\documentclass[a4paper,12pt]{article}

\usepackage{jheppub} 
   \usepackage[T1]{fontenc}                  
\usepackage{hyperref}
\hypersetup{
    bookmarks=true,         
    unicode=false,          
    pdftoolbar=true,        
    pdfmenubar=true,        
    pdffitwindow=false,     
    pdfstartview={FitH},    
    pdftitle={My title},    
    pdfauthor={Author},     
    pdfsubject={Subject},   
    pdfcreator={Creator},   
    pdfproducer={Producer}, 
    pdfkeywords={keyword1} {key2} {key3}, 
    pdfnewwindow=true,      
    colorlinks=true,       
    linkcolor=red,          
    citecolor=purple,        
    filecolor=magenta,      
    urlcolor=blue,           
    linktocpage=true
}
\usepackage{graphics}
\usepackage{rotating}
\usepackage{subfigure}
\usepackage{pstricks}
\usepackage{color}
\usepackage{amsfonts}
\usepackage{mathrsfs}
\usepackage{epsfig}
\usepackage{amsmath,amssymb,amsthm,graphicx,latexsym}
\usepackage{ulem}

\newcommand{\be}{\begin{equation}}
\newcommand{\ee}{\end{equation}}
\newcommand{\bea}{\begin{eqnarray}}
\newcommand{\eea}{\end{eqnarray}}
\newcommand{\bml}{\begin{subequations}}
\newcommand{\eml}{\end{subequations}}
\newcommand{\bfig}{\begin{figure}}
\newcommand{\efig}{\end{figure}}

\newcommand{\slashed}{\hspace{-1.1ex}/}

\begin{document}
$~~~~~~~~~~~~~~~~~~~~~~~~~~~~~~~~~~~~~~~~~~~~~~~~~~~~~~~~~~~~~~~~~~~~~~~~~~~~~~~~~~~~$\textcolor{red}{\bf TIFR/TH/15-33}
\title{\textsc{\fontsize{45}{90}\selectfont \sffamily \bfseries  COSMOS-{\it e}'-GTachyon from String Theory
		}}

\author[a]{Sayantan Choudhury
\footnote{\textcolor{purple}{\bf Presently working as a Visiting (Post-Doctoral) fellow at DTP, TIFR, Mumbai, \\$~~~~~$Alternative
 E-mail: sayanphysicsisi@gmail.com}. ${}^{}$}}
\author[b]{Sudhakar Panda}

\affiliation[a]{Department of Theoretical Physics, Tata Institute of Fundamental Research, Colaba, Mumbai - 400005, India
}
\affiliation[b]{Institute of Physics, Sachivalaya Marg, Bhubaneswar, Odisha - 751005, India
}

\emailAdd{sayantan@theory.tifr.res.in, panda@iopb.res.in  }

\abstract{In this article, our prime objective is to study the inflationary paradigm from generalized
tachyon (GTachyon) living on the world volume of a non-BPS string theory. The tachyon action is considered here 
is getting modified compared to the original action. One can
quantify the amount of the modification via a power $q$ instead of $1/2$ in the effective action.
Using this set up we study inflation from various types of tachyonic
potentials, using
which we constrain the index $q$ within, $1/2<q<2$, and a specific combination ($\propto \alpha^{'}M^4_s/g_s$) of Regge slope $\alpha^{'}$, string coupling constant $g_{s}$ and mass scale of 
tachyon $M_s$,
from the recent Planck 2015 and Planck+BICEP2/Keck
Array joint data. We explicitly study the inflationary consequences from single field, assisted
field and multi-field tachyon set up. Specifically for single field and assisted field
case we derive the results in the quasi-de-Sitter background in which we will
utilize the details of cosmological perturbations and quantum fluctuations. Also we 
derive the expressions for all inflationary observables using any arbitrary
vacuum and Bunch-Davies vacuum. For single field and assisted field
case we derive-the inflationary
flow equations, new sets of consistency
relations. Also we derive the field excursion formula for tachyon, which 
shows that assisted inflation is in more safer side compared to the single field case to validate effective field theory framework. Further we study the features of CMB Angular power
spectrum from TT, TE and EE correlations from scalar fluctuations within
the allowed range of $q$ for each potentials from single field set-up. We also put constraints
from the temperature anisotropy and polarization spectra, which
shows that our analysis is consistent with the Planck 2015 data. Finally, using
$\delta N$ formalism we derive the expressions for inflationary
observables in the context of multi-field tachyons. 
}

\maketitle
\flushbottom
\section{Introduction}
\label{aa1}
The primordial inflationary paradigm is a very sublime thought to explain various aspects of
the early universe, which creates the perturbations and the matter. For recent developments
see refs.~\cite{Martin:2010kz,Martin:2014vha,Martin:2014rqa,Martin:2014lra,Martin:2013nzq,Linde:2014nna,Lyth:1998xn,Lyth:2007qh,Baumann:2014nda,Baumann:2009ds}.
The success of primordial inflation, can be gauged by the current
observations arising from the cosmic microwave background (CMB) radiation \cite{Durrer:2001gq,Hu:2001bc,Kamionkowski:1999qc}. The
observations from Planck have put interestingly tight bounds on a number of cosmological observables related to the
perturbations, which also determines any departure from the Gaussian perturbations and the constraint on tensor-to-scalar ratio, $r_{\star}$, which can potentially unearth the scale of New Physics within any given effective
field theory set-up. But a big issue may crop up in model discrimination and also in the removal of the degeneracy of
cosmological parameters obtained from CMB observations. Non-canonical interactions in the effective
theory set-up \cite{Cheung:2007st,Weinberg:2008hq,LopezNacir:2011kk,Choudhury:2014sxa,Choudhury:2014uxa,Choudhury:2012yh,Choudhury:2015yna,Unnikrishnan:2012zu,Unnikrishnan:2013vga,Assassi:2013gxa} is one of the possibilities through which one can address this issue~\footnote{The other possibilities are- non-minimal interactions 
between the matter and gravity sector (example: SM Higgs with Einstein gravity \cite{Bezrukov:2007ep,Choudhury:2013zna}), addition of higher derivative terms in the gravity sector (example: Starobinsky model \cite{Starobinsky:1980te}),
modification of gravity sector through extra dimensions (example: 5D
membrane models \cite{Randall:1999ee,Randall:1999vf,Dvali:2000hr,Maeda:2007cb,Shtanov:2002mb,Sahni:2002dx,Sahni:2002vs,Shtanov:2002ek,Kar:2012kn,Choudhury:2013yg,Choudhury:2013dia,Choudhury:2014hna,Choudhury:2011sq,Choudhury:2011rz,Choudhury:2012ib,Choudhury:2013qza,Choudhury:2014sua,Choudhury:2015wfa,Choudhury:2013aqa,Choudhury:2013eoa}).}. The natural source for such non-canonical interactions
is string theory. The most promising example is the 
Dirac-Bonn-Infeld (DBI) effective action for tachyon field \cite{Sen:1998sm,Bergshoeff:2000dq,Sen:2003tm,Cederwall:1996uu}. In this paper,
we will only focus on tachyons in weakly coupled type IIA/ IIB string theory \cite{Maldacena:1997re}.

The phenomena of tachyon condensation was introduced in refs.~\cite{Sen:1998sm,Bergshoeff:2000dq}, where type IIA/IIB string theory
and the tachyon instability on D-branes have studied elaborately. The rolling tachyon \cite{Sen:2002nu,Leblond:2003db,Kim:2003he} in
weakly coupled type IIA/ IIB string theory may be
described in terms of an effective field theory for the
tachyon condensate 
in refs.~\cite{Sen:1999xm,Sen:2002an,Sen:2002in,Sen:2002qa,Gopakumar:2000rw,Gopakumar:2000na,Minwalla:2003hj,Basu:2010uz,David:2001vm,Aganagic:2000mh,Rastelli:2000hv,Sen:2000kd,Ohmori:2001am,Taylor:2002uv,Headrick:2004hz,Berasaluce-Gonzalez:2013sna,Das:2008af}. The cosmological implications 
of the tachyon were studied
in \cite{Sen:2003mv,Deshamukhya:2009wc,Panda:2005sg,Chingangbam:2004ng,Choudhury:2002xu,Mazumdar:2001mm,Gibbons:2003gb,Barbosa-Cendejas:2015rba,G.:2015cha,Sami:2002fs,Cremades:2005ir,Kofman:2002rh,Steer:2003yu}. For more
details see also refs.~\cite{Nozari:2013mba,Li:2013cem,Nozari:2014qba,Gibbons:2002md,Frolov:2002rr,Fairbairn:2002yp,Shiu:2002qe,Padmanabhan:2002sh,Mukohyama:2002cn}. In the context of cosmology, the detailed consequences have been studied in the following topics over many years:
\begin{enumerate}
 \item Inflationary paradigm \cite{Deshamukhya:2009wc,Panda:2005sg,Chingangbam:2004ng,Choudhury:2002xu,Mazumdar:2001mm,Gibbons:2003gb,Barbosa-Cendejas:2015rba,G.:2015cha,Sami:2002fs,Cremades:2005ir,Kofman:2002rh,Steer:2003yu,Fairbairn:2002yp},
 
 \item Primordial non-Gaussianity and CMB aspects \cite{Maldacena:2002vr,Maldacena:2011nz,Calcagni:2004bb,Enqvist:2005qu,Barnaby:2006km,Barnaby:2007yb,Dutta:2008if,Libanov:2008mk},
 
 \item Reheating scenario and particle creation \cite{Cline:2002it,Jain:2009ep},
 
 \item Late time cosmic acceleration (Dark Energy) \cite{Bagla:2002yn,Keresztes:2014jua,Copeland:2004hq,Bamba:2012cp}.
\end{enumerate}
In this work, we introduce the most generalized version of the tachyon effective action in which we are interested to study the cosmological consequences from 
the non-canonical higher dimensional 
effective field theory operators, originated from string theory. Technically we rename such a non-canonical fields 
to be generalized tachyon (GTachyon), which we will frequently use throughout the rest of the paper.  
The prime motivation of writing this paper is following~\footnote{Since Gtachyon contains non-minimal interactions, it is naturally expected 
that it will sufficiently modify the consistency relations. Also due to such non-canonical interactions it is expected that the amount of primordial 
non-Gaussianity is getting enhanced by a sufficient amount, by breaking the non-Gaussian consistency relations known in cosmological literature.
For the completeness it is important to mention here that, we have not investigated the issues related to primordial non-Gaussianity 
yet in this paper. We will report on these issues in our future work.}:
\begin{itemize}
 \item To update the present status of non-canonical interaction for stringy tachyon field appearing within the framework 
 of string theory after releasing the Planck 2015 data.
 
 \item To explicitly study the role of tachyon field and its generalized version GTachyon, to explain the observed temperature anisotropy and polarization in CMB Angular power spectrum.
 
 \item To study the specific role of the most generalized version of non-canonical higher 
 dimensional effective field theory Wilsonian operators.
 
 \item To give a broad overview of the present constraints on 
 inflationary paradigm from the most generalized version of the tachyon string theory.
 
 \item To test the explicit dynamical features of various tachyonic potentials obtained from string theory and also to know about the specific 
 structural form of effective field theory operators for a specific type of potential. 
\end{itemize}
Throughout the paper we have taken the following assumptions:
\begin{enumerate}
 \item Tachyon field $T$ is minimally coupled to the Einstein gravity sector.
 
 \item Initial condition for inflation is fixed via Bunch-Davies (BD) vacuum \cite{BD}. For the completeness we also present the results obtained from arbitrary vacuum (AV).
  For a classical initial condition the amplitude
of the primordial gravitational waves would be very tiny and practically undetectable, therefore this can
be treated as the first observable proof of quantum theory gravity, such as string theory. However, apart from the
importance and applicability of quantum version of Bunch-Davis vacuum on its theoretical and
observational ground it is still not at all well understood from the previous works in this area
whether the quantum Bunch-Davies vacuum is the only possible source of generating
large value of tensor-to-scalar ratio during inflationary epoch or not. One of the prime possibilities
coming from the deviation from quantum Bunch-Davies vacuum aka consideration
of quantum non-Bunch-Davies or arbitrary vacuum in the present non-canonical picture which may also
responsible for the generation of large tensor-to-scalar ratio and large non-Gaussianity during inflation.
Within the framework of effective field theory, such an arbitrary vacuum is commonly
identified to be the $\alpha$-vaccua \cite{Danielsson:2002mb,deBoer:2004nd,Goldstein:2003ut,Collins:2003zv,Collins:2003ze,Brunetti:2005pr,Naidu:2004ue}, which has string theoretic 
origin. 

\item We would like to point out that the tachyon mode appears
 in the quantization of the open string struck to the non-BPS brane. 
 The effective action of this tachyon field is constructed on the
 assumption that the tachyon field couples only to the graviton
 of the closed string sector with fixed vacuum expectation value
 for dilation field. Open string tachyon condensation phenomena
 fits in well with this assumption.   
 
 \item Sound speed is $c_{S}< 1$ in general for non-canonical interactions \cite{Cheung:2007st,Weinberg:2008hq,LopezNacir:2011kk,Choudhury:2014sxa,Choudhury:2014uxa,Choudhury:2012yh,Choudhury:2015yna,Unnikrishnan:2012zu,Unnikrishnan:2013vga,Assassi:2013gxa},
 which is the most promising ingredient
 for tachyonic set-up, to generate simultaneously the detectable amount of tensor-to-scalar ratio and large non-Gaussianity.
 This fact can be more clearly visualized when we go to all higher order expansion in slow-roll or more precisely by taking
 the exact solution of the mode functions for scalar and tensor fluctuations as obtained from cosmological perturbation theory
 by appropriately choosing Bunch-Davies or any arbitrary initial conditions for inflation. 
 
 \item UV cut-off of the effective field theory is fixed at $\Lambda_{UV}\sim M_p$, where $M_p=2.43\times 10^{18}~{\rm GeV}$ is the
reduced Planck mass. But in principle one can fix the scale between GUT
scale and reduced Planck scale i.e. $\Lambda_{GUT}<\Lambda_{UV}\leq M_p$. But in such a
situation $\Lambda_{UV}$ acts as a regulating parameter in the effective field theory \cite{Assassi:2013gxa,Choudhury:2014hja}.

\item Within the region of  $N_{\star}(=N_{cmb})\approx {\cal O}(50-70)$ e-foldings, we will use the following constarints 
in the background of $\Lambda$CDM model for:\\ \\
\underline{\bf \textcolor{blue}{Planck (2015)+WMAP-9+high~L(TT) data:}\cite{Ade:2015lrj}}
 \begin{eqnarray}
  \label{obscons3a}r(k_\star)&\leq&  0.11~~ ({\rm within}~ 2\sigma ~C.L.),\\
 \label{obscons3b}\ln(10^{10}P_{S})&=&3.089\pm 0.036~~ ({\rm within}~ 2\sigma ~C.L.),\\
 \label{obscons3c}n_{S}&=&0.9569 \pm 0.0077~~ ({\rm within}~ 3\sigma ~C.L.),\\
 \label{obscons3d}\alpha_{S}&=&dn_{S}/d\ln k=0.011^{+ 0.014}_{-0.013}~~({\rm within}~1.5\sigma~C.L.),\\
 \label{obscons3e}\kappa_{S}&=&d^{2}n_{S}/d\ln k^{2}=0.029^{+0.015}_{-0.016}~~({\rm within}~1.5\sigma~C.L.)\,.
 \end{eqnarray}
\underline{\bf \textcolor{blue}{Planck (2015)+BICEP2/Keck Array joint data:}\cite{Ade:2015tva}}
 \begin{eqnarray}
  \label{obscons4a}r(k_\star)&\leq&  0.12~~ ({\rm within}~ 2\sigma ~C.L.),\\
 \label{obscons4b}\ln(10^{10}P_{S})&=&3.089^{+0.024}_{-0.027}~~ ({\rm within}~ 2\sigma ~C.L.),\\
 \label{obscons4c}n_{S}&=&0.9600 \pm 0.0071~~ ({\rm within}~ 3\sigma ~C.L.),\\
 \label{obscons4d}\alpha_{S}&=&dn_{S}/d\ln k=-0.022\pm 0.010~~({\rm within}~1.5\sigma~C.L.),\\
 \label{obscons4e}\kappa_{S}&=&d^{2}n_{S}/d\ln k^{2}=0.020^{+0.016}_{-0.015}~~({\rm within}~1.5\sigma~C.L.)\,.
 \end{eqnarray}
\end{enumerate}
The plan of the paper is as follows:
\begin{itemize}
	\item  In section \ref{aa2}, we have discussed the role of tachyon in non-BPS barnes in weakly coupled
 type IIA/ IIB string theory and also introduced 
	the GTachyon in the effective action of string theory. 
	
	\item In section \ref{aa3}, we have introduced and studied the features from variants of tachyonic potentials inspired from 
	non-BPS branes in string theory.
	
	\item In section \ref{aa4}, we have studied the cosmological dynamics from GTachyons, in which we have explicitly discussed 
	the unperturbed evolution and dynamical solution in various phases of the universe.
	
	\item In section \ref{aa5}, we have explicitly studied inflationary paradigm from single, assisted and multi-field Gtachyons.
	Particularly for single field and assisted field
case in presence of GTachyon, we have derived-the inflationary
Hubble flow and potential dependent flow equations, new sets of consistency
relations, which are valid in the slow-roll regime and field excursion formula for tachyon
in terms of inflationary observables.
	
	\item In section \ref{aa6}, we have mentioned the future prospects of the present work and summarized the context of the present work.
\end{itemize}
In this paper, we have explored various cosmological consequences from GTachyonic field. We start with the basic introduction of tachyons in the context
of non-BPS string theory, where we also introduce the GTachyon
field, in presence of which the tachyon action is getting modified and one can
quantify the amount of the modification via a superscript $q$ instead of $1/2$.
This modification exactly mimics the role of effective field theory operators and
studying the various cosmological features from this theory, one of our final
objectives is to constrain the index $q$ and a specific combination ($\propto \alpha^{'}M^4_s/g_s$) of Regge slope parameter $\alpha^{'}$, string coupling constant $g_{s}$ and mass scale of 
tachyon $M_s$, from the recent Planck 2015 and Planck+BICEP2/Keck
Array joint data. To serve this purpose, we introduce various types of tachyonic
potentials-Inverse cosh, Logarithmic, Exponential and Inverse polynomial, using
which we constrain the index $q$. To explore this issue in details, we start with
the characteristic features of the each potentials.
Next we discuss the dynamics of GTachyon as well as usual tachyon for single,
assisted and multi-field scenario. Next we have explicitly studied the inflationary paradigm from single field, assisted
field and multi-field tachyon set up. Specifically for single field and assisted field
case we have derived the results in the quasi-de-Sitter background in which we have
utilized the details of- (1) cosmological perturbations and quantum fluctuations
for scalar and tensor modes, (2) Slow-roll prescription. In this context we have
derived the expressions for all inflationary observables using any arbitrary
vacuum and also for Bunch-Davies vacuum. For single field and assisted field
case in presence of GTachyon we have derived-the inflationary
Hubble flow and potential dependent flow equations, new sets of consistency
relations valid in the slow-roll regime 
and also derived the expression for the field excursion formula for tachyon
in terms of inflationary observables from both of the solutions obtained from arbitrary and Bunch-Davies initial conditions for inflation. 
Particularly the derived formula for the field excursion for GTachyon can be treated as an one of the important probes through which one can
distinguish between various tachyon models and also check the validity of effective field theory prescription
and compare the results obtained from assisted inflation as well. The results obtained in this context explicitly 
shows that assisted inflation is better compared to the single field inflation 
from the tachyon portal, provided the number of identical tachyon fields are required to be large to validate effective field theory prescription. Next using the explicit for of the
tachyonic potentials we have studied the inflationary constraints and quantify
the allowed range of the generalized index $q$ for each potentials. Hence
using the each specific form of the tachyonic potentials in the context of
single filed scenario, we have studied the features of CMB Angular power
spectrum from TT, TE and EE correlations from scalar fluctuations within
the allowed range of $q$ for each potentials. We also put the constraints
from the Planck temperature anisotropy and polarization data, which
shows that our analysis is consistent with the data. 
We have additionally studied the features of tensor contribution in the
CMB Angular power spectrum from TT, BB, TE and EE correlations, which
will give more interesting information in near future while
the signature of  primordial B-modes can be detected. Further, using
$\delta N$ formalism, we have derived the expressions for inflationary
observables in the context of multi-field tachyons and demonstrated the results for Inverse cosh potential for completeness.

\section{GTachyon in string theory}
\label{aa2}
In this section we explicitly study the world volume actions for non-BPS branes which finally govern their cosmological dynamics.
For the sake of simplicity in this discussion we neglect the contribution from the fermions and concentrate only on the massless bosonic
fields for the non-BPS branes. The world volume action for non-BPS branes is described by the sum of the Dirac-Born-Infeld (DBI)
and the Wess-Zumino (WZ) term in type IIA/IIB string theory. The effective action for DBI in a non-BPS $p$-brane is given by \cite{Sen:2004nf,Sen:2004wt}:
\bea\label{eq1}
S^{(p)}_{\rm DBI}&=&-{\cal T}_{p}\int d^{p+1}\sigma~e^{-\phi}~\sqrt{-{\rm det}\left({\cal Z}_{\mu\nu}+{\cal F}_{\mu\nu}\right)},
\eea
where the metric has signature $(-, +, +, +)$ and ${\cal Z}_{\mu\nu}$ is defined as:
\bea\label{eq1v2}
{\cal Z}_{\mu\nu}&=& G_{\mu\nu}+\alpha^{'}\partial_{\mu}T\partial_{\nu}T.
\eea
Here $T$ is the dimensionless tachyon field whose properties have been discussed later in details. Also in Eq~(\ref{eq1}) 
${\cal T}_{p}$ represents the brane tension defined as \cite{Sen:2004nf,Sen:2004wt}:
\be\begin{array}{lll}\label{gb3}
 \displaystyle {\cal T}_{p} = 
 \displaystyle  
 \sqrt{2}(2\pi)^{-p}g^{-1}_{s}
\end{array}\ee
and $\alpha^{'}$ represents the Regge slope parameter in string theory.
Here type IIA/IIB string
theory contains the non-BPS D$p$-branes \cite{Sen:1999mg}, which have precisely those dimensions which BPS D-branes do not
have explicitly. This implies that type IIA string theory has non-BPS D$p$-branes for only odd $p$ and type IIB string
theory has non-BPS D$p$-branes for only even $p$ in the present context. Additionally, it is important to note that, $g_{s}$ characterizes the string coupling constant.
Also in Eq~(\ref{eq1}), $G_{\mu\nu}$ is defined via the following transformation equation:
\be\label{eq2} G_{\mu\nu}=G_{MN}\partial_{\mu}X^{M}\partial_{\nu}X^{N}.\ee
In Eq~(\ref{eq2}) $M,N$ characterize the ten dimensional ($D=10$) indices which runs from $0,1,\cdots,9$; $\sigma^{\mu}(0\leq \mu\leq p)$ denotes
the world volume coordinates of the D$p$ brane. Also it is important to note that, in this discussion, $G_{MN}$ represents the ten-dimensional ($D=10$)
background metric for type IIA/IIB string
theory.

It is important to mention here that, in the context of non-BPS D-brane, stringy tachyon comes from only one specific sector of string theory and consequently it is
a real scalar field using which we will explain the cosmological dynamics in this article for $p=3$. Additionally it is also important to note that, Type IIA/IIB string theories contain unstable non-BPS D-branes in their spectrum.
The most easiest way to define these types of D-branes
in the context of IIA/IIB string theory is to start the computation with a coincident BPS D$p$ - $\bar{\rm D}p$-brane pair in
type IIB/IIA string theory, and then take an specific orbifold of the string theory by $(-1)^{F_L}$ , where $F_L$
signifies the specific contribution to the space-time fermion number from the left-moving sector
of the world-sheet string theory. Now in this context the Ramond Ramond (RR) fields are odd under $(-1)^{F_L}$ transformation and consequently all the Ramond Ramond (RR) fields of type
IIB/IIA string theory are projected out by the same amount via $(-1)^{F_L}$ projection. As a result the twisted sector stringy states then
give us back the Ramond Ramond (RR) fields of type IIA/IIB string theory in the present context. Most importantly, here the $(-1)^{F_L}$ projection reverses the signature of the
Ramond Ramond (RR) charge and consequently it transforms a BPS D$p$-brane to a $\bar{\rm D}p$-brane and vice versa.
This further implies that due to its operation
on the open string states on a D$p$-$\bar{\rm D}p$-brane stringy system will do the job of conjugate operation on the Chan-Paton factor
by the action of exchange operator $\sigma_{1}$ in this context. Thus technically the modding out operation on the D$p$-$\bar{\rm D}p$-brane by exactly the amount of $(-1)^{F_L}$ eliminates 
all the open string states which carry Chan-Paton factor $\sigma_{2}$ and $\sigma_{3}$, as both of them anti-commute with exchange operator $\sigma_{1}$.
Additionally it is important to note that, this operation finally keeps the open string states which are characterized via the Chan-Paton factors $I$ and $\sigma_{1}$. Finally all such operations gives us a
non-BPS D$p$-brane in the present context. Although in this discussion we are only interested in the non-BPS D-branes, the most important characteristic feature that distinguishes the physics of 
non-BPS D-branes from BPS D-branes
is that the mass spectrum of open strings on a non-BPS D-brane contains a single mode of
negative mass squared besides infinite number of other modes of positive definite mass squared. This negative mass squired mode is identified to be the tachyonic
mode which is exactly equivalent to a particular linear combination of the two tachyonic modes living on
the original brane-antibrane pair stringy system that survives the previously mentioned $(-1)^{F_L}$ projection and contains the exactly same
mass squared contribution.

In our analysis for the sake of simplicity we have neglected the contribution from the antisymmetric Kalb-Ramond $2$-form field from the effective action but
the gauge invariance of the action requires the presence of all such antisymmetric tensor contributions in the original version of the string effective action.
In the present context, it is important to note that, $G_{\mu\nu}$ can be physically interpreted as the induced metric on the membrane. Additionally it is important to note that,
the background metric $G_{MN}$ is not at all arbitrary for the present setup but the structural form of the metric is restricted in a specific sense that it has to 
satisfy the sets of background field equations in this context. Also in our discussion the transverse component of the fluctuations of the $D$-membrane is 
described by ($9-p$) number of scalar fields $X^{i}$, where the index $i$ runs from $p+1\leq i\leq 9$, and the gauge field $A_{\mu}$ describes
the fluctuations along the longitudinal direction of the membrane.

Before writing down the total effective action for non-BPS $D$-brane in string theory it is important to mention that 
the non-BPS $p$-brane has an extra tachyon field appearing in both of the Dirac-Born-Infeld (DBI) and the Wess-Zumino (WZ) stringy effective actions. The corresponding effective
actions can be written in the non-BPS string theory setup as \cite{Sen:2004nf,Sen:2004wt}:
\bea\label{eq5}
S^{(p)}_{\rm DBI}&=&-\int d^{p+1}\sigma~e^{-\phi}~\sqrt{-{\rm det}\left({\cal Z}_{\mu\nu}+{\cal F}_{\mu\nu}\right)}~\Theta(T,\partial_{\mu}T,D_{\mu}\partial_{\nu}T),\\
\label{eq6}S^{(p)}_{\rm WZ}&=& \int d^{p+1}\sigma~C\wedge dT\wedge e^{{\cal F}},
\eea
where the field $C$ contains Ramond-Ramond (RR) fields and the leading term has a ($p+1$)-differential form. 
This also mimics the role of a source term for the membrane and its presence is explicitly required for consistency of the specific version of the field theory like 
anomaly cancellation within the setup of string theory. It is important to mention here that the world volume action for a ${\rm Dp}$-brane in $(p+1)$ dimensional, where $p$ characterizes the 
spatial dimension and $1$ stands for time. The WZ term plays an important role since the BPS ${\rm Dp}$-brane is charged under a $(p+1)$ rank RR gauge field. Consequently the total action is therefore the DBI
action together with the WZ action. For the non-BPS ${\rm Dp}$-brane, the brane is charged under a rank $p$ RR gauge field. As a result the WZ action consists of wedge product of this $p$ form and additionally one form $dT$, 
where $T$ is the tachyon field. Also it is important to note that for non-BPS case ${\cal F}_{\mu\nu}$ is explicitly defined as \cite{Sen:2004nf,Sen:2004wt}:
\bea\label{eqxx3} {\cal F}_{\mu\nu}&=&B_{\mu\nu}+2\pi\alpha^{'}F_{\mu\nu}
+\partial_{\mu}Y^{I}\partial_{\nu}Y^{I}+\left(G_{IJ}+B_{IJ}\right)\partial_{\mu}Y^{I}\partial_{\nu}Y^{J}\nonumber\\&&~~~~~~~~~~~~~~~~~~~
+\left(G_{\mu I}+B_{\mu I}\right)\partial_{\nu}Y^{I}+\left(G_{I\nu}+B_{I\nu}\right)\partial_{\mu}Y^{I},\eea
and $T$ represents the dimensionless tachyon field and $\Theta$ characterizes the generalized functional in non-BPS $p$-membrane. 
In Eq~(\ref{eqxx3}), the rank-2 field strength tensor $F_{\mu\nu}$ is defined as:
\be 
F_{\mu\nu}=\partial_{[\mu}A_{\nu]}
\ee
and $B_{\mu\nu}$ represents a rank-2 Kalb-Ramond field and sometimes this can be interpreted as the pullback of $B_{MN}$ onto the D-brane world-volume.
Using this specific action mentioned in Eq~(\ref{eqxx3}), we can compute the source contributions and terms for various closed string fields produced
by the membrane. Additionally, it is important to note that, on a non-BPS D$p$-membrane world volume we have infinite
tower of massive fields, a ${\rm U}$(1) gauge field $A_{\mu} $ with the restriction on gauge indices, $0 \leq \mu \leq p$, and a set of scalar fields $Y^{I}$,
one for each direction $y^{I}$
transverse besides the tachyonic field to the D-brane. Here it is important note that, $(p + 1) \leq I \leq D$, where $D$ being $9$ for
superstring theory and $25$ for bosonic string theory.
Within the present setup the tachyon field is defined in such a way that for 
\be\label{eq6} T=0,~~F={\cal T}_{p}\ee constraint condition is explicitly satisfied. 
For non-BPS $p$ membrane the field content $C$ as appearing in the Wess-Zumino (WZ) action contains the Ramond-Ramond (RR) fields but careful observation clearly indicates that 
the leading order contribution in $C$ is characterized by $p$-form in the effective action.
Also it is important to mention here that, for constant tachyon background $T$ , the Wess-Zumino (WZ) effective action automatically 
vanishes in the present context as: \be\label{eq7} dT=0\ee and for such a specific background the generalized functional $\Theta$ can be recast in the following form as:
\bea\label{eq8}
\Theta(T,\partial_{\mu}T,D_{\mu}\partial_{\nu}T)&=& V(T),
\eea
where $V(T)$ represents the effective tachyon potential within which the contribution from the membrane tension is already taken. Consequently Eq~(\ref{eq5}) can be recast in the following simplified form as:
\bea 
S^{(p)}_{\rm DBI}&=&-\int d^{p+1}\sigma~e^{-\phi}~V(T)~\sqrt{-{\rm det}\left({\cal Z}_{\mu\nu}+{\cal F}_{\mu\nu}\right)}.
\eea
For non-BPS string theory in the constant dilaton background the purely tachyonic part of the action, after inclusion of the massless fields on the
D$p$-membrane world-volume around the tachyon vacuum is
given by:
\bea\label{eq5ccxc}
S^{(p)}_{\rm D}&=&-\int d^{p+1}\sigma~V(T)~\sqrt{-{\rm det}\left({\cal Z}_{\mu\nu}+F_{\mu\nu}+\partial_{\mu}Y^{I}\partial_{\nu}Y^{I}\right)}.
\eea
After neglecting the contribution from the massless fields for non-BPS string theory the tachyonic part of the effective action describing the 
D$p$-brane world-volume is
given by \cite{Sen:2004nf,Sen:2004wt} the following simplified form: 
\bea\label{eq5cv}
S^{(p)}_{\rm D}&=&-\int d^{p+1}\sigma~V(T)~\sqrt{-{\rm det}\left({\cal Z}_{\mu\nu}\right)}\nonumber\\
&=&-\int d^{p+1}\sigma~V(T)~\sqrt{-{\rm det}\left(G_{\mu\nu}+\alpha^{'}\partial_{\mu}T\partial_{\nu}T\right)}\nonumber\\
&=&-\int d^{p+1}\sigma~\sqrt{-g}~V(T)~\sqrt{1+\alpha^{'}g^{\mu\nu}\partial_{\mu}T\partial_{\nu}T}.
\eea
In a more generalized prescription Eq~(\ref{eq5cv}) can be modified into the following effective action as:
\bea\label{eq5cv1}
S^{(pq)}_{\rm D}&=&-\int d^{p+1}\sigma~\sqrt{-g}~V(T)~\left(1+\alpha^{'}g^{\mu\nu}\partial_{\mu}T\partial_{\nu}T\right)^{q}.
\eea
which we identified to be the most generalized Gtachyon action in string theory.
Here for $p=3$  i.e. for D$3$ brane, Eq~(\ref{eq5cv}) refers to the following crucial issues:
\begin{itemize}
 \item   Here $p=3, q=1/2$ corresponds to the exact tachyonic behavior of the effective action and it is commonly used to describe the cosmological dynamics,
 
 \item  Here $p=3, q=1$ corresponds to the single field behavior in cosmological dynamics where the kinetic term of the tachyon field is non-canonical.
        In this case the non-canonical contribution in the effective action is given by $\alpha^{'}V(T)$, where $V(T)$ is the tachyon effective potential.
        This situation is different from the usual single field models of inflation where the kinetic term is canonical within the framework of string theory.
 
 \item  For $p=3, 1/2<q<1$ or $p=3, q<1/2$ or $p=3, q>1$ contains various non-trivial features in cosmological dynamics. In this case the effective action is significantly 
 different from the usual tachyon action as appearing in the context of string theory. In this case the effective action describes a 
 huge class of effective field theory of inflationary models which can be embedded within the framework of tachyon in string theory.  In the present context,
 the generalized factor $V(T)~\left(1+\alpha^{'}g^{\mu\nu}\partial_{\mu}T\partial_{\nu}T\right)^{q}$ with exponent $q$ can be treated as Wilsonian operators as appearing in the context of 
 effective field theory. For an example let us consider a situation where we treat the Regge slope parameter $\alpha^{'}$ is small. In this case the generalized factor
 $V(T)~\left(1+\alpha^{'}g^{\mu\nu}\partial_{\mu}T\partial_{\nu}T\right)^{q}$ takes the following structure:
 \bea V(T)~\left(1+\alpha^{'}g^{\mu\nu}\partial_{\mu}T\partial_{\nu}T\right)^{q}&=& V(T)~\sum^{q}_{k=0}{}^{q}C_{k}\left(\alpha^{'}g^{\mu\nu}\partial_{\mu}T\partial_{\nu}T\right)^{k}\nonumber\\
   &=&V(T)\left[1+q\left(\alpha^{'}g^{\mu\nu}\partial_{\mu}T\partial_{\nu}T\right)\right.\nonumber\\ &&\left.+\frac{q(q-1)}{2}\left(\alpha^{'}g^{\mu\nu}\partial_{\mu}T\partial_{\nu}T\right)^2+\cdots\right],\eea
   where $\cdots$ contain higher order terms which are suppressed by the powers of Regge slope parameter $\alpha^{'}$. Here for each value of $k(=0,1,2,\cdots,q)$ the expansion 
   factor ${}^{q}C_{k}\left(\alpha^{'}\right)^{k}$ mimics the role of 
   Wilson coefficients.
\end{itemize}
Here we would like to point out that the tachyon mode appears
 in the quantization of the open string struck to the non-BPS brane. 
 The effective action of this tachyon field is constructed on the
 assumption that the tachyon field couples only to the graviton
 of the closed string sector with fixed vacuum expectation value
 for dilation field. Open string tachyon condensation phenomena
 fits in well with this assumption.  
 
For a multi-tachyonic field scenario one can generalize the tachyonic part of the non-BPS action as stated in Eq~(\ref{eq5cv}) and Eq~(\ref{eq5cv1}) as:
\bea\label{eq5cvm}
S^{(p)}_{\rm D}&=&-\sum^{N}_{i=1}\int d^{p+1}\sigma~\sqrt{-g}~V(T_i)~\sqrt{1+\alpha^{'}g^{\mu\nu}\partial_{\mu}T_i \partial_{\nu}T_i},\\
\label{eq5cv1m}
S^{(pq)}_{\rm D}&=&-\sum^{N}_{i=1}\int d^{p+1}\sigma~\sqrt{-g}~V(T_i)~\left(1+\alpha^{'}g^{\mu\nu}\partial_{\mu}T_i \partial_{\nu}T_i \right)^{q}.
\eea
In this case one can introduce a total effective potential of tachyonic fields $V_{E}(T)$, which can be expressed in terms of $N$ component tachyonic fields as:
\bea
V_{E}(T)=\sum^{N}_{i=1}V(T_i)
\eea
which is very useful to study the cosmological dynamics for $p=3$ case i.e. for D$3$ brane.

Also for assisted case one can assume all the multi-tachyonic $N$ number of fields are identical to each other i.e.
\bea T_{i}=T~~~~~\forall~~ i=1,2,3,......,N.\eea
Consequently in such a prescription Eq~(\ref{eq5cvm}) and Eq~(\ref{eq5cv1m}) can be re-written as:
\bea\label{eq5cvmas}
S^{(p)}_{\rm D}&=&-\sum^{N}_{i=1}\int d^{p+1}\sigma~\sqrt{-g}~V(T)~\sqrt{1+\alpha^{'}g^{\mu\nu}\partial_{\mu}T \partial_{\nu}T}\nonumber\\
&=&-\int d^{p+1}\sigma~\sqrt{-g}~NV(T)~\sqrt{1+\alpha^{'}g^{\mu\nu}\partial_{\mu}T \partial_{\nu}T},\\
\label{eq5cv1mas}
S^{(pq)}_{\rm D}&=&-\sum^{N}_{i=1}\int d^{p+1}\sigma~\sqrt{-g}~V(T)~\left(1+\alpha^{'}g^{\mu\nu}\partial_{\mu}T \partial_{\nu}T_i \right)^{q}\nonumber\\
&=&-\int d^{p+1}\sigma~\sqrt{-g}~NV(T)~\left(1+\alpha^{'}g^{\mu\nu}\partial_{\mu}T\partial_{\nu}T \right)^{q}.
\eea
In this case the total effective potential of tachyonic fields $V_{E}(T)$ can be recast as:
\bea
V_{E}(T)=\sum^{N}_{i=1}V(T_i)=\sum^{N}_{i=1}V(T)=NV(T).
\eea
In the next section we will discuss the various aspects of tachyonic potential $V(T)$ and also mention the various models of tachyonic potential that 
can be derived from string theory background.
\section{Variants of tachyonic models in string theory}
\label{aa3}
On general string theoretic grounds, one can argued that at the specified minimum $T_{0}$ of the effective potential $V(T)$ vanishes \cite{Sen:2004nf,Sen:2004wt} i.e.
\bea\label{eq9}
V(T_{0})=0.
\eea
Consequently the world volume action vanishes identically and in this situation the gauge field mimics the role of Lagrange multiplier field. Finally this imposes a constraint on the non-BPS setup such that the gauge current
also vanishes identically. This implies that all the states which are charged under this gauge field to disappear from the spectrum.
Also it is important to mention another important feature of the tachyonic potential in which it admits kink profile for the stringy tachyonic field. Also on an unstable non-BPS $p$-brane tachyon condensation occurs to form a kink profile 
and finally it forms a stable BPS ($p-1$) brane configuration. This kink profile for the tachyon is expected to give a $\delta$-function from $dT$- contribution and thus to reproduce the standard WZ term in the resulting $D(p-1)$-brane.
Most importantly, the kink solution effectively reduces the dimension of the world-volume by one.
Although, finding out an explicit form of the tachyonic potential is a very difficult, but string theory predicts the approximated form of the tachyonic potential.
Additionally, we also assume that the tachyonic potential $V(T)$ satisfies the following properties to describe the cosmological dynamics for non-BPS D$3$ brane setup :
\begin{enumerate}
 \item  Tachyonic potential at $T=0$ satisfies:
 \be\label{gk1} V(T=0)=\lambda=\frac{M^4_s}{(2\pi)^3 g_s},\ee
 where $g_{s}$ is the string coupling constant and $M_{s}$ signifies the mass scale of the tachyonic string theory.
 For multi tachyonic field and assisted case Eq~(\ref{gk1}) is modified as:
 \bea\label{gk2} V_{E}(T=0)&=&\sum^{N}_{i=1}V(T_{i}=0)=\sum^{N}_{i=1}\lambda_{i}=\sum^{N}_{i=1}\frac{M^4_s}{(2\pi)^3 g^{(i)}_s},\\
 \label{gk3} V_{E}(T=0)&=&\sum^{N}_{i=1}V(T_{i}=T=0)=\sum^{N}_{i=1}\lambda_{i}=N\lambda=\frac{M^4_s N}{(2\pi)^3 g_s}.
 \eea
 Here $g^{(i)}_s$ represents the string coupling constant for the ith field content, 
 which are not same for all $N$ number of tachyonic degrees of freedom. For the sake of simplicity, here we also assume that the mass scale associated 
 with $N$ number of tachyons for multi-tachyonic case and assisted case is identical for all degrees of freedom.
 On the other hand, for assisted case we assume all couplings are exactly identical and consequently we get an overall factor of $N$ multiplied with the result obtained for 
 single tachyonic field case.

 \item Inflation generally takes place at an energy scale:
 \be\label{gk4} V^{1/4}_{inf}=V^{1/4}(\tilde{T}_{0})\propto \lambda^{1/4}\ee with
 the single tachyon field fixed at $T \sim \tilde{T}_{0}$, and within the setup of string theory $\tilde{T}_{0}$ 
 is identified to be the mass scale of the tachyon by the following fashion:
 \be\label{gk5} \tilde{T}_{0} \sim M_{s}.\ee
 For multi tachyonic field case Eq~(\ref{gk4}) and Eq~(\ref{gk4}) are modified as:
 \bea\label{gk6} V^{1/4}_{inf}&=&V^{1/4}_{E}=\left\{\sum^{N}_{i=1}V(\tilde{T}_{0i})\right\}^{1/4}\propto \left\{\sum^{N}_{i=1}\lambda_{i}\right\}^{1/4}=\left\{\sum^{N}_{i=1}\frac{M^4_s}{(2\pi)^3 g^{(i)}_s}\right\}^{1/4},\\
 \label{gk5} \tilde{T}_{0} &\sim& \sum^{N}_{i=1} \tilde{T}_{0i}=M_{s}.
 \eea
 Similarly for assisted case Eq~(\ref{gk4}) and Eq~(\ref{gk4}) are modified as:
 \bea\label{gk6} V^{1/4}_{inf}&=&V^{1/4}_{E}=\left\{\sum^{N}_{i=1}V(\tilde{T}_{0i})\right\}^{1/4}\propto 
 \left\{\sum^{N}_{i=1}\lambda_{i}\right\}^{1/4}\nonumber\\&&~~~~~~~~~~~~~~~~~~~~~~~~~~~~~~~~=
 (N\lambda)^{1/4}=\left\{\frac{M^4_s N}{(2\pi)^3 g_s}\right\}^{1/4},\\
 \label{gk5} \tilde{T}_{0} &\sim& \sum^{N}_{i=1} \tilde{T}_{0i}=\sum^{N}_{i=1} \tilde{T}_{0}=N\tilde{T}_{0}=NM_{s}.
 \eea
 
 \item For $T>0$ the first derivative of the single tachyonic potential is always positive i.e. 
 \be\label{gk7} V^{'}(T>0)>0 \ee
 where $'$ represents differentiation with respect to tachyon field $T$.
 For multi tachyonic and assisted case Eq~(\ref{gk7}) can be recast as:
 \bea\label{gk8} V^{'}_{E}(T>0)&=&\sum^{N}_{i=1}\frac{dV(T_{i}>0)}{dT_i}>0,\\
 \label{gk8} V^{'}_{E}(T>0)&=&\sum^{N}_{i=1}\frac{dV(T_{i}>0)}{dT_i}=\sum^{N}_{i=1}\frac{dV(T>0)}{dT}=N V^{'}(T>0)>0.~~~~~~\eea
 
 \item At the asymptotic case of the single tachyonic field $|T|\rightarrow \infty$ the potential satisfy:
 \be\label{eq8} V(|T|\rightarrow \infty)\rightarrow 0.\ee
 For multi tachyonic and assisted case Eq~(\ref{gk8}) can be recast as:
 \bea\label{eq9} V_{E}(|T|\rightarrow \infty)&=&\sum^{N}_{i=1}V(|T_{i}|\rightarrow \infty)\rightarrow 0,\\
 \label{eq10} V_{E}(|T|\rightarrow \infty)&=&\sum^{N}_{i=1}V(|T_{i}|\rightarrow \infty)=
 NV(|T|\rightarrow \infty)\rightarrow 0.\eea
 
 \item Also one can consider that the tachyonic potential contains a global maximum at $T = 0$ i.e.
       \be\label{eq11} V^{''}(T=0)<0\ee
       for which the value of the potential is given by Eq~(\ref{gk1}). Similarly 
       for multi tachyonic and assisted case Eq~(\ref{eq11}) can be recast as:
       \bea\label{eq11} V^{''}_{E}(T=0)&=&\sum^{N}_{i=1}\frac{d^{2}V(T_{i}>0)}{dT^{2}_i}<0,\\
       \label{eq12} V^{''}_{E}(T=0)&=&\sum^{N}_{i=1}\frac{d^{2}V(T_{i}>0)}{dT^{2}_i}=\sum^{N}_{i=1}\frac{d^{2}V(T>0)}{dT^{2}}=NV^{''}(T=0)<0.~~~~~~~~\eea
\end{enumerate}
In the next subsections we mention variants of tachyonic potentials which satisfy the various above mentioned characteristics.

\subsection{Model I: Inverse cosh potential}
\label{aa3a}
For single field case the first model of tachyonic potential is given by \cite{Cremades:2005ir,Kim:2003he,Leblond:2003db,Lambert:2003zr}:
\be\label{we1}
V(T)=\frac{\lambda}{{\rm cosh}\left(\frac{T}{T_{0}}\right)},\ee
and for multi tachyonic and assisted case the total effective potential is given by:
\be\begin{array}{lll}\label{gbdf3}
 \displaystyle V_{E}(T) =\left\{\begin{array}{lll}
                    \displaystyle  
                    \sum^{N}_{i=1}\frac{\lambda_{i}}{{\rm cosh}\left(\frac{T_i}{T_{0i}}\right)} \,,~~~~~~ &
 \mbox{\small {\bf for {Multi tachyonic}}}  \\ 
 \displaystyle  
 \sum^{N}_{i=1}\frac{\lambda}{{\rm cosh}\left(\frac{T}{T_{0}}\right)}=\frac{N\lambda}{{\rm cosh}\left(\frac{T}{T_{0}}\right)} \,,~~~~~ &
 \mbox{\small {\bf for {Assisted tachyonic}}}.
          \end{array}
\right.
\end{array}\ee
Here the potential satisfies the following criteria:
\begin{itemize}
 \item  At $T=0$ for single field tachyonic potential:\be V(T=0)=\lambda \ee
 and for multi tachyonic and assisted case we have:
 \bea V_{E}(T=0)&=&\sum^{N}_{i=1}\lambda_{i},\\
 V_{E}(T=0)&=&\sum^{N}_{i=1}\lambda_{i}=N\lambda.\eea
 
 \item At $T=T_0$ for single field tachyonic potential:\be V(T=T_0)=\frac{\lambda}{{\rm cosh}(1)}\neq 0 \ee
 and for multi tachyonic and assisted case we have:
 \bea V_{E}(T=0)&=&\sum^{N}_{i=1}\frac{\lambda_{i}}{{\rm cosh}(1)}\neq 0 ,\\
 V_{E}(T=0)&=&\sum^{N}_{i=1}\frac{\lambda_{i}}{{\rm cosh}(1)}=\frac{N\lambda}{{\rm cosh}(1)}\neq 0 .\eea
 
 \item For single field tachyonic potential:\bea\label{op1} V^{'}(T)&=&-\frac{\lambda}{T_0}{\rm sech}\left(\frac{T}{T_0}\right) {\rm tanh}\left(\frac{T}{T_0}\right),\\
 \label{op2}V^{''}(T)&=&-\frac{\lambda}{T^2_0}\left[{\rm sech}^3\left(\frac{T}{T_0}\right)-{\rm sech}\left(\frac{T}{T_0}\right) {\rm tanh}^2\left(\frac{T}{T_0}\right)\right].\eea
 Now to find the extrema of the potential we substitute \be V^{'}(T)=0\ee which give rise to the follwing solutions for the the tachyonic field:
 \bea |T|=2m~ T_{0}\pi ,~~  \left(2m+1\right)~T_{0}\pi\eea
 where $m\in \mathbb{Z}$. Further substituting the solutions for tachyonic field in Eq~(\ref{op2}) we get:
 \bea\label{opp2}V^{''}(|T|=2m~ T_{0}\pi)&=&-\frac{\lambda}{T^2_0},\\
 \label{opp2}V^{''}(|T|=\left(2m+1\right)~T_{0}\pi)&=&-\frac{\lambda}{T^2_0},\eea
 and at these points the value of the potential is computed as:
 \bea\label{opp3} V(|T|=2m~ T_{0}\pi)&=&\frac{\lambda}{{\rm cosh}\left(2m~\pi\right)}=\lambda~ {\rm sech}\left(2m~\pi\right),\\
 V(|T|=\left(2m+1\right)~T_{0}\pi)&=&\frac{\lambda}{{\rm cosh}\left(\left(2m+1\right)~\pi\right)}=\lambda~ {\rm sech}\left(\left(2m+1\right)~\pi\right).~~~~~~~~\eea
 It is important to note for single tachyonic case that for $\lambda>0$, $V^{''}(|T|=2m~ T_{0}\pi ,~~  \left(2m+1\right)~T_{0}\pi)>0$ i.e. we get maxima on the potential
 and for the assisted case the results are same, provided following replacement occurs:
 \bea \lambda\rightarrow N\lambda.\eea
 and finally for multi tachyonic case we have:
\bea\label{opd1} V^{'}(T_{i})&=&-\frac{\lambda_{i}}{T_{0i}}{\rm sech}\left(\frac{T_{i}}{T_{0i}}\right) {\rm tanh}\left(\frac{T_{i}}{T_{0i}}\right),\\
 \label{opd2}V^{''}(T_{i})&=&-\frac{\lambda_{i}}{T^2_{0i}}\left[{\rm sech}^3\left(\frac{T_{i}}{T_{0i}}\right)-{\rm sech}\left(\frac{T_{i}}{T_{0i}}\right)
 {\rm tanh}^2\left(\frac{T_{i}}{T_{0i}}\right)\right].\eea
 Now to find the extrema of the potential we substitute \be V^{'}(T_{j})=0\forall j=1,2,...,N\ee which give rise to the following solutions
 for the the $j$th tachyonic field:
 \bea |T_{j}|=2m~ T_{0j}\pi ,~~  \left(2m+1\right)~T_{0j}\pi~~~ \forall j=1,2,.....,N\eea
 where $m\in \mathbb{Z}$. Further substituting the solutions for tachyonic field in Eq~(\ref{opd2}) we get:
 \bea\label{oppd2}V^{''}(|T_{j}|=2m~ T_{0j}\pi)&=&-\frac{\lambda_{j}}{T^2_{0j}},\\
 \label{oppd2}V^{''}(|T_{j}|=\left(2m+1\right)~T_{0j}\pi)&=&-\frac{\lambda_{j}}{T^2_{0j}},\eea
 and at these points the value of the total effective potential is computed as:
 \bea\label{oppd3} V^{(1)}_{E}=\sum^{N}_{j=1}V(|T_{j}|=2m~ T_{0j}\pi)&=&\sum^{N}_{j=1}\frac{\lambda_{j}}{{\rm cosh}\left(2m~\pi\right)}\nonumber\\&=&{\rm sech}\left(2m~\pi\right)\sum^{N}_{j=1}\lambda_{j},\\
V^{(2)}_{E}= \sum^{N}_{j=1}V(|T_{j}|=\left(2m+1\right)~T_{0j}\pi)&=&\sum^{N}_{j=1}\frac{\lambda_{j}}{{\rm cosh}\left(\left(2m+1\right)~\pi\right)}\nonumber\\&=&{\rm sech}\left(\left(2m+1\right)~\pi\right)\sum^{N}_{j=1}\lambda_{j}.~~~~~~~~\eea 
 It is important to note for multi tachyonic case that for $\lambda_{j}>0$, $V^{''}(|T_{j}|=2m~ T_{0j}\pi ,~~  \left(2m+1\right)~T_{0j}\pi)>0$ i.e. we get maxima on the potential $V(T_{j})$ as well as in $V_{E}(T)$.
\end{itemize}
\subsection{Model II: Logarithmic potential}
\label{aa3b}
For single field case the second model of tachyonic potential is given by \cite{Mazumdar:2001mm}:
\be\label{we2}
V(T)=\lambda\left\{ \left(\frac{T}{T_{0}}\right)^2\left[\ln\left(\frac{T}{T_{0}}\right)\right]^2+1\right\},\ee
and for multi tachyonic and assisted case the total effective potential is given by:
\be\begin{array}{lll}\label{gbdfaa3}
 \displaystyle V_{E}(T) =\left\{\begin{array}{lll}
                    \displaystyle  
                   \sum^{N}_{i=1}\lambda_{i}\left\{\left(\frac{T_i}{T_{0i}}\right)^2 \left[\ln\left(\frac{T_i}{T_{0i}}\right)\right]^2+1\right\} \,,~~~~~~ &
 \mbox{\small {\bf for {Multi tachyonic}}}  \\ 
 \displaystyle  
 \sum^{N}_{i=1}\lambda\left\{ \left(\frac{T}{T_{0}}\right)^2\left[\ln\left(\frac{T}{T_{0}}\right)\right]^2+1\right\}\\
 \displaystyle~~~~~~~~~~~~~=N\lambda\left\{ \left(\frac{T}{T_{0}}\right)^2\left[\ln\left(\frac{T}{T_{0}}\right)\right]^2+1\right\} \,,~~~~~ &
 \mbox{\small {\bf for {Assisted tachyonic}}}.
          \end{array}
\right.
\end{array}\ee
Here the potential satisfies the following criteria:
\begin{itemize}
 \item  At $T=0$ for single field tachyonic potential:\be V(T=0)=\lambda \ee
 and for multi tachyonic and assisted case we have:
 \bea V_{E}(T=0)&=&\sum^{N}_{i=1}\lambda_{i},\\
 V_{E}(T=0)&=&\sum^{N}_{i=1}\lambda_{i}=N\lambda.\eea
 
 \item At $T=T_0$ for single field tachyonic potential:\be V(T=T_0)=\lambda \ee
 and for multi tachyonic and assisted case we have:
 \bea V_{E}(T=T_0)&=&\sum^{N}_{i=1}\lambda_{i},\\
 V_{E}(T=T_0)&=&\sum^{N}_{i=1}\lambda_{i}=N\lambda.\eea
 
 \item For single field tachyonic potential:\bea\label{opvqq1} V^{'}(T)&=&\frac{2 T\lambda }{T^2_0}\ln \left(\frac{T}{T_0}\right)\left[1+\ln\left(\frac{T}{T_0}\right)\right],\\
 \label{opvqq2}V^{''}(T)&=&\frac{2\lambda}{T^2_0}\left[1+3\ln \left(\frac{T}{T_0}\right)+\ln^2 \left(\frac{T}{T_0}\right)\right].\eea
 Now to find the extrema of the potential we substitute \be V^{'}(T)=0\ee which give rise to the follwing solutions for the the tachyonic field:
 \bea T=0 ,~~T_{0},~~  \frac{T_{0}}{e}.\eea
 Further substituting the solutions for tachyonic field in Eq~(\ref{opvqq2}) we get:
 \bea\label{oppvqq2}V^{''}(T=0)&\rightarrow&\infty,\\
 \label{oppvvggqq2}V^{''}(T=T_{0})&=&\frac{2\lambda}{T^2_0},\\ 
 \label{xc}V^{''}\left(T=\frac{T_{0}}{e}\right)&=&-\frac{2\lambda}{T^2_0},\\
 \eea
 and at these points the value of the potential is computed as:
 \bea\label{oppvqq3} V(T=0)&=&\lambda,\\
 V(T=T_{0})&=&\lambda,\\
 V\left(T=\frac{T_{0}}{e}\right)&=&\lambda\left(1+\frac{1}{e^2}\right).
 \eea
 It is important to note for single tachyonic case that for $\lambda>0$, $V^{''}(T=T_{0})>0$ and $V^{''}(T=\frac{T_{0}}{e})<0$ i.e. we get both maxima and minima on the potential.
 For the assisted case the results are same, provided following replacement occurs:
 \bea \lambda\rightarrow N\lambda.\eea
 and finally for multi tachyonic case we have:
\bea\label{opvqqqx1} V^{'}(T_{i})&=&\frac{2 \lambda_{i}T_{i} }{T^2_{0i}}\ln \left(\frac{T_{i}}{T_{0i}}\right)\left[1+\ln\left(\frac{T_{i}}{T_{0i}}\right)\right],\\
 \label{opvqqqx2}V^{''}(T_{i})&=&\frac{2\lambda_{i}}{T^2_{0i}}\left[1+3\ln \left(\frac{T_{i}}{T_{0i}}\right)+\ln^2 \left(\frac{T_{i}}{T_{0i}}\right)\right].\eea
 Now to find the extrema of the potential we substitute \be V^{'}(T_{j})=0\forall j=1,2,...,N\ee which give rise to the following solutions
 for the the $j$th tachyonic field:
 \bea T_{j}=0 ,~~T_{0j},~~  \frac{T_{0j}}{e}~~~ \forall j=1,2,.....,N.\eea
  Further substituting the solutions for tachyonic field in Eq~(\ref{opvqqqx2}) we get:
 \bea\label{oppqqxxxx2}V^{''}(T_{j}=0)&\rightarrow&\infty,\\
 \label{oppqqvxx2}V^{''}(T_{j}=T_{0j})&=&\frac{2\lambda_{j}}{T^2_{0j}},\\ 
 \label{xxcqq}V^{''}\left(T_{j}=\frac{T_{0j}}{e}\right)&=&-\frac{2\lambda_{j}}{T^2_{0j}},
 \eea
 and at these points the value of the total effective potential is computed as:
 \bea\label{oppxqqq3} V^{(1)}_{E}=\sum^{N}_{j=1}V(T_{j}=0)&=&\sum^{N}_{j=1}\lambda_{j},\\
 \label{oppxvbqqq3} V^{(2)}_{E}=\sum^{N}_{j=1}V(T_{j}=T_{0j})&=&\sum^{N}_{j=1}\lambda_{j},\\
\label{sd1c} V^{(3)}_{E}= \sum^{N}_{j=1}V\left(T_{j}=\frac{T_{0j}}{e}\right)&=&\left(1+\frac{1}{e^2}\right)\sum^{N}_{j=1}\lambda_{j}.~~~~~~~~\eea 
 It is important to note for multi tachyonic case that for $\lambda_{j}>0$, $V^{''}(T=T_{0j})>0$ and $V^{''}(T=\frac{T_{0j}}{e})<0$ i.e. we get both maxima and minima on the potential $V(T_{j})$ as well as in $V_{E}(T)$.
\end{itemize}
\subsection{Model III: Exponential potential-Type I}
\label{aa3c}
For single field case the third model of tachyonic potential is given by \cite{Sami:2002fs}:
\be\label{weee1}
V(T)=\lambda\exp\left(-\frac{T}{T_{0}}\right),\ee
and for multi tachyonic and assisted case the total effective potential is given by:
\be\begin{array}{lll}\label{gbdfeee3}
 \displaystyle V_{E}(T) =\left\{\begin{array}{lll}
                    \displaystyle  
                    \sum^{N}_{i=1}\lambda_{i}\exp\left(-\frac{T_{i}}{T_{0i}}\right) \,,~~~~~~ &
 \mbox{\small {\bf for {Multi tachyonic}}}  \\ 
 \displaystyle  
 \sum^{N}_{i=1}\lambda\exp\left(-\frac{T}{T_{0}}\right)=N\lambda\exp\left(-\frac{T}{T_{0}}\right) \,,~~~~~ &
 \mbox{\small {\bf for {Assisted tachyonic}}}.
          \end{array}
\right.
\end{array}\ee
Here the potential satisfies the following criteria:
\begin{itemize}
 \item  At $T=0$ for single field tachyonic potential:\be V(T=0)=\lambda \ee
 and for multi tachyonic and assisted case we have:
 \bea V_{E}(T=0)&=&\sum^{N}_{i=1}\lambda_{i},\\
 V_{E}(T=0)&=&\sum^{N}_{i=1}\lambda_{i}=N\lambda.\eea
 
 \item At $T=T_0$ for single field tachyonic potential:\be V(T=T_0)=\frac{\lambda}{e} \ee
 and for multi tachyonic and assisted case we have:
 \bea V_{E}(T=T_0)&=&\frac{1}{e}\sum^{N}_{i=1}\lambda_{i},\\
 V_{E}(T=T_0)&=&\frac{1}{e}\sum^{N}_{i=1}\lambda_{i}=\frac{N\lambda}{e}.\eea
 
 \item For single field tachyonic potential:\bea\label{opvdf1} V^{'}(T)&=&-\frac{\lambda }{T_0}\exp\left(-\frac{T}{T_{0}}\right),\\
 \label{opvdf2}V^{''}(T)&=&\frac{\lambda }{T^2_0}\exp\left(-\frac{T}{T_{0}}\right).\eea
 Now to find the extrema of the potential we substitute \be V^{'}(T)=0\ee which give rise to the following solution for the the tachyonic field:
 \bea T\rightarrow \infty.\eea
 Further substituting the solutions for tachyonic field in Eq~(\ref{opvdf2}) we get:
 \bea\label{oppv2}V^{''}(T\rightarrow \infty)&\rightarrow& 0,
 \eea
 and also at the points $T=0,~T_{0}$ we have:
 \bea\label{oppvdf2}V^{''}(T=0)&=& \frac{\lambda}{T^2_0},\\
 \label{oppvdf2}V^{''}(T=T_{0})&=& \frac{\lambda}{eT^2_0},
 \eea
 and at these points the value of the potential is computed as:
 \bea\label{oppvdf3} V(T\rightarrow \infty)&\rightarrow& 0,\\
 \label{oppvdf3} V(T=0)&=& \lambda,\\
 \label{oppvdff3} V(T=T_{0})&=& \frac{\lambda}{e}.
 \eea
 It is important to note for single tachyonic case that for $\lambda>0$, $V^{''}(T=0,~~T_{0} )>0$ i.e. we get an asymptotic behavior of the potential. 
 For the assisted case the results are same, provided following replacement occurs:
 \bea \lambda\rightarrow N\lambda.\eea
 and finally for multi tachyonic case we have:
\bea\label{opvxwww1} V^{'}(T_{i})&=&-\frac{\lambda_{i} }{T_0}\exp\left(-\frac{T_{i}}{T_{0i}}\right),\\
 \label{opvxwww2}V^{''}(T_{i})&=&\frac{\lambda_{i} }{T^2_{0i}}\exp\left(-\frac{T_{i}}{T_{0i}}\right).\eea
 Now to find the extrema of the potential we substitute \be V^{'}(T_{j})=0\forall j=1,2,...,N\ee which give rise to the follwing solutions
 for the the $j$th tachyonic field:
 \bea T_{j}\rightarrow \infty.\eea
 Further substituting the solutions for tachyonic field in Eq~(\ref{opvxwww2}) we get:
 \bea\label{oppvwww2}V^{''}(T_{j}\rightarrow \infty)&\rightarrow& 0,
 \eea
 and also at the points $T_{j}=0,~T_{0j}$ we have:
 \bea\label{oppvwwww2}V^{''}(T_{j}=0)&=& \frac{\lambda_{j}}{T^2_{0j}},\\
 \label{oppvwwwww2}V^{''}(T_{j}=T_{0j})&=& \frac{\lambda_{j}}{eT^2_{0j}},
 \eea
 and at these points the value of the total effective potential is computed as:
 \bea\label{oppvwwwws3}  V^{(1)}_{E}=\sum^{N}_{j=1}V(T_{j}\rightarrow \infty)&\rightarrow& 0,\\
 \label{oppvwwwwst3} V^{(2)}_{E}=\sum^{N}_{j=1}V(T_{j}=0)&=& \sum^{N}_{j=1}\lambda_{j},\\
 \label{oppvwwwwstt3} V^{(3)}_{E}= \sum^{N}_{j=1}V(T_{j}=T_{0j})&=&  \frac{1}{e}\sum^{N}_{j=1}\lambda_{j}.
 \eea
 It is important to note for multi tachyonic case that for $\lambda_{j}>0$, $V^{''}(T=0,~~T_{0j} )>0$ i.e. we get an asymptotic behavior of the potential $V(T_{j})$ as well as in $V_{E}(T)$.
\end{itemize}
\subsection{Model IV: Exponential potential-Type II (Gaussian)}
\label{aa3d}
For single field case the first model of tachyonic potential is given by \cite{Raeymaekers:2004cu}:
\be\label{we1}
V(T)=\lambda\exp\left[-\left(\frac{T}{T_{0}}\right)^2\right],\ee
and for multi tachyonic and assisted case the total effective potential is given by:
\be\begin{array}{lll}\label{gbdftt3}
 \displaystyle V_{E}(T) =\left\{\begin{array}{lll}
                    \displaystyle  
                    \sum^{N}_{i=1}\lambda\exp\left[-\left(\frac{T}{T_{0}}\right)^2\right] \,,~~~~~~ &
 \mbox{\small {\bf for {Multi tachyonic}}}  \\ 
 \displaystyle  
 \sum^{N}_{i=1}\lambda\exp\left[-\left(\frac{T}{T_{0}}\right)^2\right]=N\lambda\exp\left[-\left(\frac{T}{T_{0}}\right)^2\right] \,,~~~~~ &
 \mbox{\small {\bf for {Assisted tachyonic}}}.
          \end{array}
\right.
\end{array}\ee
Here the potential satisfies the following criteria:
\begin{itemize}
 \item  At $T=0$ for single field tachyonic potential:\be V(T=0)=\lambda \ee
 and for multi tachyonic and assisted case we have:
 \bea V_{E}(T=0)&=&\sum^{N}_{i=1}\lambda_{i},\\
 V_{E}(T=0)&=&\sum^{N}_{i=1}\lambda_{i}=N\lambda.\eea
 
 \item At $T=T_0$ for single field tachyonic potential:\be V(T=T_0)=\frac{\lambda}{e} \ee
 and for multi tachyonic and assisted case we have:
 \bea V_{E}(T=T_0)&=&\frac{1}{e}\sum^{N}_{i=1}\lambda_{i},\\
 V_{E}(T=T_0)&=&\frac{1}{e}\sum^{N}_{i=1}\lambda_{i}=\frac{N\lambda}{e}.\eea
 
 \item For single field tachyonic potential:\bea\label{opvnn1} V^{'}(T)&=&-\frac{2\lambda T }{T^2_0}\exp\left[-\left(\frac{T}{T_{0}}\right)^2\right],\\
 \label{opvnn2}V^{''}(T)&=&-\frac{2\lambda  }{T^2_0}\exp\left[-\left(\frac{T}{T_{0}}\right)^2\right]\left\{1-\frac{2T^2}{T^2_0}\right\}.\eea
 Now to find the extrema of the potential we substitute \be V^{'}(T)=0\ee which give rise to the following solution for the the tachyonic field:
 \bea T=0,~~ \infty.\eea
 Further substituting the solutions for tachyonic field in Eq~(\ref{opvnn2}) we get:
 \bea\label{oppvnn2}V^{''}\left(T=0\right)&=& -\frac{2\lambda}{T^2_{0}},\\
 \label{oppvnn2}V^{''}(T\rightarrow \infty)&\rightarrow& 0
 \eea
 and also additionally for $T=T_{0}$ we have:
 \bea\label{oppvmmm2}V^{''}(T=T_0)&=& \frac{2\lambda}{eT^2_0},
 \eea
 and at these points the value of the potential is computed as:
 \bea\label{oppvmmm3} V(T\rightarrow \infty)&\rightarrow& 0,\\
 \label{oppvmmmm3} V(T=0)&=& \lambda,\\
 \label{oppvmmmmm3} V(T=T_{0})&=& \frac{\lambda}{e}.
 \eea
 It is important to note for single tachyonic case that for $\lambda>0$, $V^{''}(T=0,~~T_{0} )>0$ i.e. we get maxima on the potential. 
 For the assisted case the results are same, provided following replacement occurs:
 \bea \lambda\rightarrow N\lambda.\eea
 and finally for multi tachyonic case we have:
\bea\label{opv551} V^{'}(T_{i})&=&-\frac{2\lambda_{i} T_{i} }{T^2_{0i}}\exp\left[-\left(\frac{T_{i}}{T_{0i}}\right)^2\right],\\
 \label{opv552}V^{''}(T_{i})&=&-\frac{2\lambda_{i}  }{T^2_{0i}}\exp\left[-\left(\frac{T_{i}}{T_{0i}}\right)^2\right]\left\{1-\frac{2T^2_{i}}{T^2_{0i}}\right\}.\eea
 Now to find the extrema of the potential we substitute \be V^{'}(T_{j})=0\forall j=1,2,...,N\ee which give rise to the follwing solutions
 for the the $j$th tachyonic field:
 \bea T_{j}=0,~ \infty.\eea
 Further substituting the solutions for tachyonic field in Eq~(\ref{opv552}) we get:
 \bea\label{oppv552}V^{''}(T_{j}=0)&=& -\frac{2\lambda}{T^2_{0}},\\
 \label{oppv552}V^{''}(T_{j}\rightarrow \infty)&\rightarrow& 0,
 \eea
 Additionally for the point $T_{j}=T_{0j}$ we have:
 \bea\label{oppv452}V^{''}(T_{j}=T_{0j})&=& \frac{2\lambda_{j}}{eT^2_{0j}},
 \eea
 and at these points the value of the total effective potential is computed as:
 \bea\label{oppv453}  V^{(1)}_{E}=\sum^{N}_{j=1}V(T_{j}\rightarrow \infty)&\rightarrow& 0,\\
 \label{oppv463} V^{(2)}_{E}=\sum^{N}_{j=1}V(T_{j}=0)&=& \sum^{N}_{j=1}\lambda_{j},\\
 \label{oppv473} V^{(3)}_{E}= \sum^{N}_{j=1}V(T_{j}=T_{0j})&=&  \frac{1}{e}\sum^{N}_{j=1}\lambda_{j}.
 \eea
 It is important to note for multi tachyonic case that for $\lambda_{j}>0$, $V^{''}(T=0,~~T_{0j} )>0$ i.e. we get maxima on the potential $V(T_{j})$ as well as in $V_{E}(T)$.
\end{itemize}
\subsection{Model V: Inverse power-law potential}
\label{aa3e}
For single field case the first model of tachyonic potential is given by \cite{Steer:2003yu,Abramo:2003cp}:
\be\label{we1}
V(T)=\frac{\lambda}{\left[1+\left(\frac{T}{T_{0}}\right)^4\right]},\ee
and for multi tachyonic and assisted case the total effective potential is given by:
\be\begin{array}{lll}\label{gbdfer43}
 \displaystyle V_{E}(T) =\left\{\begin{array}{lll}
                    \displaystyle  
                    \sum^{N}_{i=1}\frac{\lambda_{i}}{\left[1+\left(\frac{T_{i}}{T_{0i}}\right)^4\right]} \,,~~~~~~ &
 \mbox{\small {\bf for {Multi tachyonic}}}  \\ 
 \displaystyle  
 \sum^{N}_{i=1}\frac{\lambda}{\left[1+\left(\frac{T}{T_{0}}\right)^4\right]}=\frac{N\lambda}{\left[1+\left(\frac{T}{T_{0}}\right)^4\right]} \,,~~~~~ &
 \mbox{\small {\bf for {Assisted tachyonic}}}.
          \end{array}
\right.
\end{array}\ee
Here the potential satisfies the following criteria:
\begin{itemize}
 \item  At $T=0$ for single field tachyonic potential:\be V(T=0)=\lambda \ee
 and for multi tachyonic and assisted case we have:
 \bea V_{E}(T=0)&=&\sum^{N}_{i=1}\lambda_{i},\\
 V_{E}(T=0)&=&\sum^{N}_{i=1}\lambda_{i}=N\lambda.\eea
 
 \item At $T=T_0$ for single field tachyonic potential:\be V(T=T_0)=\frac{\lambda}{2} \ee
 and for multi tachyonic and assisted case we have:
 \bea V_{E}(T=T_0)&=&\frac{1}{2}\sum^{N}_{i=1}\lambda_{i},\\
 V_{E}(T=T_0)&=&\frac{1}{2}\sum^{N}_{i=1}\lambda_{i}=\frac{N\lambda}{2}.\eea
 
 \item For single field tachyonic potential:\bea\label{opver1} V^{'}(T)&=&-\frac{4\lambda T^3 }{T^4_0 \left[1+\left(\frac{T}{T_{0}}\right)^4\right]^2},\\
 \label{opver2}V^{''}(T)&=&\frac{32 \lambda T^6}{T^8_0 \left[1+\left(\frac{T}{T_{0}}\right)^4\right]^3}-\frac{12 \lambda T^2}{T^4_0 \left[1+\left(\frac{T}{T_{0}}\right)^4\right]^2}.\eea
 Now to find the extrema of the potential we substitute \be V^{'}(T)=0\ee which give rise to the follwing solution for the the tachyonic field:
 \bea T=0,~~ \infty.\eea
 Further substituting the solutions for tachyonic field in Eq~(\ref{opver2}) we get:
 \bea\label{oppver2}V^{''}\left(T=0\right)&=& 0,\\
 \label{oppver2}V^{''}(T\rightarrow \infty)&\rightarrow& 0
 \eea
 and also additionally for $T=T_{0}$ we have:
 \bea\label{oppverr2}V^{''}(T=T_0)&=& \frac{\lambda}{T^2_0},
 \eea
 and at these points the value of the potential is computed as:
 \bea\label{oppverr3} V(T\rightarrow \infty)&\rightarrow& 0,\\
 \label{oppveyy3} V(T=0)&=& \lambda,\\
 \label{oppvess3} V(T=T_{0})&=& \frac{\lambda}{2}.
 \eea
 It is important to note for single tachyonic case that for $\lambda>0$, $V^{''}(T=0,~~T_{0} )>0$ i.e. we get maxima on the potential. 
 For the assisted case the results are same, provided following replacement occurs:
 \bea \lambda\rightarrow N\lambda.\eea
 and finally for multi tachyonic case we have:
\bea\label{opvqw1} V^{'}(T_{i})&=&-\frac{4\lambda_{i} T^3_{i} }{T^4_{0i} \left[1+\left(\frac{T_{i}}{T_{0i}}\right)^4\right]^2},\\
 \label{opvqw2}V^{''}(T_{i})&=&\frac{32 \lambda_{i} T^6_{i}}{T^8_{0i} \left[1+\left(\frac{T_{i}}{T_{0i}}\right)^4\right]^3}-\frac{12 \lambda_{i} T^2_{i}}{T^4_{0i} \left[1+\left(\frac{T_{i}}{T_{0i}}\right)^4\right]^2}.\eea
 Now to find the extrema of the potential we substitute: \be V^{'}(T_{j})=0\forall j=1,2,...,N\ee which give rise to the follwing solutions
 for the the $j$th tachyonic field:
 \bea T_{j}=0,~ \infty.\eea
 Further substituting the solutions for tachyonic field in Eq~(\ref{opvqw2}) we get:
 \bea\label{oppvqww2}V^{''}(T_{j}=0)&=&0,\\
 \label{oppvweee2}V^{''}(T_{j}\rightarrow \infty)&\rightarrow& 0,
 \eea
 Additionally for the point $T_{j}=T_{0j}$ we have:
 \bea\label{oppvjg2}V^{''}(T_{j}=T_{0j})&=& \frac{\lambda_{j}}{T^2_{0j}},
 \eea
 and at these points the value of the total effective potential is computed as:
 \bea\label{oppvjg3}  V^{(1)}_{E}=\sum^{N}_{j=1}V(T_{j}\rightarrow \infty)&\rightarrow& 0,\\
 \label{oppvjjk3} V^{(2)}_{E}=\sum^{N}_{j=1}V(T_{j}=0)&=& \sum^{N}_{j=1}\lambda_{j},\\
 \label{oppvkky3} V^{(3)}_{E}= \sum^{N}_{j=1}V(T_{j}=T_{0j})&=&  \frac{1}{2}\sum^{N}_{j=1}\lambda_{j}.
 \eea
 It is important to note for multi tachyonic case that for $\lambda_{j}>0$, $V^{''}(T=0,~~T_{0j} )>0$ i.e. we get maxima on the potential $V(T_{j})$ as well as in $V_{E}(T)$.
\end{itemize}
\section{Cosmological dynamics from GTachyon}
\label{aa4}

\subsection{Unperturbed evolution}
\label{aa4a}
For $p=3$ non-BPS branes the total tachyonic model action can be written as:
\be\begin{array}{lll}\label{e1}
 \displaystyle S_{T} =\left\{\begin{array}{lll}
                    \displaystyle  
                    \int d^{4}\sigma~\sqrt{-g}~\left[\frac{M^{2}_p}{2}R-~V(T)~\sqrt{1+\alpha^{'}g^{\mu\nu}\partial_{\mu}T\partial_{\nu}T}\right] \,,~~~~~~ &
 \mbox{\small {\bf for {Single}}}  \\ 
 \displaystyle  
 \int d^{4}\sigma~\sqrt{-g}~\left[\frac{M^{2}_p}{2}R-~\sum^{N}_{i=1}V(T_{i})~\sqrt{1+\alpha^{'}g^{\mu\nu}\partial_{\mu}T_{i}\partial_{\nu}T_{i}}\right] \,,~~~~~ &
 \mbox{\small {\bf for {Multi }}} \\ 
 \displaystyle  
\int d^{4}\sigma~\sqrt{-g}~\left[\frac{M^{2}_p}{2}R-~NV(T)~\sqrt{1+\alpha^{'}g^{\mu\nu}\partial_{\mu}T\partial_{\nu}T}\right] \,,~~~~~ &
 \mbox{\small {\bf for {Assisted }}}.
          \end{array}
\right.
\end{array}\ee
and in a more generalized situation Eq~(\ref{e1}) is modified as:
\be\begin{array}{lll}\label{e2}
 \displaystyle S^{(q)}_{T} =\left\{\begin{array}{lll}
                    \displaystyle  
                    \int d^{4}\sigma~\sqrt{-g}~\left[\frac{M^{2}_p}{2}R-~V(T)~\left(1+\alpha^{'}g^{\mu\nu}\partial_{\mu}T\partial_{\nu}T\right)^{q}\right] \,,~~~~~~ &
 \mbox{\small {\bf for {Single}}}  \\ 
 \displaystyle  
 \int d^{4}\sigma~\sqrt{-g}~\left[\frac{M^{2}_p}{2}R-~\sum^{N}_{i=1}V(T_{i})~\left(1+\alpha^{'}g^{\mu\nu}\partial_{\mu}T_{i}\partial_{\nu}T_{i}\right)^{q}\right] \,,~~~~~ &
 \mbox{\small {\bf for {Multi }}} \\ 
 \displaystyle  
\int d^{4}\sigma~\sqrt{-g}~\left[\frac{M^{2}_p}{2}R-~NV(T)~\left(1+\alpha^{'}g^{\mu\nu}\partial_{\mu}T\partial_{\nu}T\right)^{q}\right] \,,~~~~~ &
 \mbox{\small {\bf for {Assisted }}}
          \end{array}
\right.
\end{array}\ee
where $M_p$ is the reduced Planck mass, $M_p =2.43\times 10^{18}{\rm GeV}$.
By varying the action as stated in Eq~(\ref{e1}) and Eq~(\ref{e2}), with respect to the metric $g_{\mu\nu}$ we get the following equation of motion:
\bea G_{\mu\nu}=\frac{T_{\mu\nu}}{M^{2}_{p}} \eea
where $G_{\mu\nu}$ is defined as:
\bea G_{\nu\nu}= R_{\mu\nu}-\frac{1}{2}g_{\mu\nu}R \eea
and the energy-momentum stress tensor $T_{\mu\nu}$ is defined as:
\be T_{\mu\nu}=-\frac{2}{\sqrt{-g}}\frac{\delta(\sqrt{-g}{\cal L}_{\bf Tachyon})}{\delta g^{\mu\nu}} \ee
where ${\cal L}_{\bf Tachyon}$ be the tachyonic part of the Lagrangian for non-BPS setup. Explicitly using Eq~(\ref{e1}) the energy-momentum stress tensor can be computed as:
\be\begin{array}{lll}\label{e3}
 \displaystyle T_{\mu\nu} =\left\{\begin{array}{lll}
                    \displaystyle  
                    V(T)\left[\frac{\alpha^{'}\partial_{\mu}T\partial_{\nu}T}{\sqrt{1+\alpha^{'}g^{\mu\nu}\partial_{\mu}T\partial_{\nu}T}}-g_{\mu\nu}\sqrt{1+\alpha^{'}g^{\mu\nu}\partial_{\mu}T\partial_{\nu}T}\right] \,,~~~~~~ &
 \mbox{\small {\bf for {Single }}}  \\ 
 \displaystyle  
 \sum^{N}_{i=1}V(T_{i})\left[\frac{\alpha^{'}\partial_{\mu}T_{i}\partial_{\nu}T_{i}}{\sqrt{1+\alpha^{'}g^{\mu\nu}\partial_{\mu}T_{i}\partial_{\nu}T_{i}}}-g_{\mu\nu}\sqrt{1+\alpha^{'}g^{\mu\nu}\partial_{\mu}T_{i}\partial_{\nu}T_{i}}\right] \,,~~~~~ &
 \mbox{\small {\bf for {Multi}}} \\ 
 \displaystyle  
NV(T)\left[\frac{\alpha^{'}\partial_{\mu}T\partial_{\nu}T}{\sqrt{1+\alpha^{'}g^{\mu\nu}\partial_{\mu}T\partial_{\nu}T}}-g_{\mu\nu}\sqrt{1+\alpha^{'}g^{\mu\nu}\partial_{\mu}T\partial_{\nu}T}\right] \,,~~~~~ &
 \mbox{\small {\bf for {Assisted }}}.
          \end{array}
\right.
\end{array}\ee
and similarly in a more generalized situation using Eq~(\ref{e2}) the energy-momentum stress tensor can be computed as:
\be\begin{array}{lll}\label{e4}
 \displaystyle T^{(q)}_{\mu\nu} =\left\{\begin{array}{lll}
                    \displaystyle  
                    V(T)\left[\frac{2q\alpha^{'}\partial_{\mu}T\partial_{\nu}T}{\left(1+\alpha^{'}g^{\mu\nu}\partial_{\mu}T\partial_{\nu}T\right)^{1-q}}-g_{\mu\nu}\left(
                    1+\alpha^{'}g^{\mu\nu}\partial_{\mu}T\partial_{\nu}T\right)^{q}
                    \right] \,,~ &
 \mbox{\small {\bf for {Single }}}  \\ 
 \displaystyle  
 \sum^{N}_{i=1}V(T_{i})\left[\frac{2q\alpha^{'}\partial_{\mu}T_{i}\partial_{\nu}T_{i}}{\left(1+\alpha^{'}g^{\mu\nu}\partial_{\mu}T_{i}\partial_{\nu}T_{i}\right)^{1-q}}-g_{\mu\nu}\left(
                    1+\alpha^{'}g^{\mu\nu}\partial_{\mu}T_{i}\partial_{\nu}T_{i}\right)^{q}\right]  \,,~ &
 \mbox{\small {\bf for {Multi }}} \\ 
 \displaystyle  
NV(T)\left[\frac{2q\alpha^{'}\partial_{\mu}T\partial_{\nu}T}{\left(1+\alpha^{'}g^{\mu\nu}\partial_{\mu}T\partial_{\nu}T\right)^{1-q}}-g_{\mu\nu}\left(
                    1+\alpha^{'}g^{\mu\nu}\partial_{\mu}T\partial_{\nu}T\right)^{q}\right]  \,,~ &
 \mbox{\small {\bf for {Assisted }}}.
          \end{array}
\right.
\end{array}\ee
It clearly appears that, for $q=1/2$, the result is perfectly consistent with the Eq~(\ref{e3}).
Further using the perfect fluid assumption the energy-momentum stress tensor can be written as:
\be\begin{array}{lll}\label{e5}
 \displaystyle T_{\mu\nu} =\left\{\begin{array}{lll}
                    \displaystyle  
                    (\rho +p)u_{\mu}u_{\nu}+ g_{\mu\nu} p \,,~~~~ &
 \mbox{\small {\bf for {Single }}}  \\ 
 \displaystyle  
 \sum^{N}_{i=1}(\rho_{i} +p_{i})u^{(i)}_{\mu}u^{(i)}_{\nu}+ g_{\mu\nu} p_{i}  \,,~~~ &
 \mbox{\small {\bf for {Multi }}} \\
 \displaystyle  
N\left[(\rho +p)u_{\mu}u_{\nu}+ g_{\mu\nu} p\right]  \,,~~~ &
 \mbox{\small {\bf for {Assisted }}}.
          \end{array}
\right.
\end{array}\ee
where for the assisted case we assume that the density and pressure of all identical tachyonic modes are same.
Here $u_{\mu}$ signifies the four velocity of the fluid, which is defined as:
\be\begin{array}{lll}\label{e6}
 \displaystyle u_{\mu} =\left\{\begin{array}{lll}
                    \displaystyle  
                    \frac{\partial_{\mu}T}{\sqrt{-g^{\alpha\beta}\partial_{\alpha}T\partial_{\beta}T}} \,,~~~~ &
 \mbox{\small {\bf for {Single }}}  \\ 
 \displaystyle  
 \sum^{N}_{i=1} \frac{\partial_{\mu}T_{i}}{\sqrt{-g^{\alpha\beta}\partial_{\alpha}T_{i}\partial_{\beta}T_{i}}}  \,,~~~ &
 \mbox{\small {\bf for {Multi }}} \\ 
 \displaystyle  
\frac{N\partial_{\mu}T}{\sqrt{-g^{\alpha\beta}\partial_{\alpha}T\partial_{\beta}T}} \,,~~~ &
 \mbox{\small {\bf for {Assisted }}}.
          \end{array}
\right.
\end{array}\ee
Further comparing Eq~(\ref{e3}) and Eq~(\ref{e5}) the density $\rho$ and pressure $p$ for tachyonic field can be computed as:
\be\begin{array}{lll}\label{e7}
 \displaystyle \rho =\left\{\begin{array}{lll}
                    \displaystyle  
                    \frac{V(T)}{\sqrt{1-\alpha^{'}\dot{T}^2}} \,,~~~~ &
 \mbox{\small {\bf for {Single }}}  \\ 
 \displaystyle  
 \sum^{N}_{i=1} \frac{V(T)}{\sqrt{1-\alpha^{'}\dot{T}^2}}  \,,~~~ &
 \mbox{\small {\bf for {Multi }}} \\ 
 \displaystyle  
\frac{NV(T)}{\sqrt{1-\alpha^{'}\dot{T}^2}} \,,~~~ &
 \mbox{\small {\bf for {Assisted }}}.
          \end{array}
\right.
\end{array}\ee
and
\be\begin{array}{lll}\label{e8}
 \displaystyle p =\left\{\begin{array}{lll}
                    \displaystyle  
                    -V(T)\sqrt{1-\alpha^{'}\dot{T}^2} \,,~~~~ &
 \mbox{\small {\bf for {Single }}}  \\ 
 \displaystyle  
 -\sum^{N}_{i=1} V(T)\sqrt{1-\alpha^{'}\dot{T}^2} \,,~~~ &
 \mbox{\small {\bf for {Multi }}} \\ 
 \displaystyle  
-NV(T)\sqrt{1-\alpha^{'}\dot{T}^2} \,,~~~ &
 \mbox{\small {\bf for {Assisted }}}.
          \end{array}
\right.
\end{array}\ee
Similarly for the generalized situation comparing Eq~(\ref{e4}) and Eq~(\ref{e5}) the density $\rho$ and pressure $p$ for tachyonic field can be computed as:
\be\begin{array}{lll}\label{e9}
 \displaystyle \rho =\left\{\begin{array}{lll}
                    \displaystyle  
                    \frac{V(T)}{\left(1-\alpha^{'}\dot{T}^2\right)^{1-q}}\left[1-\alpha^{'}(1-2q)\dot{T}^2\right] \,,~~~~ &
 \mbox{\small {\bf for {Single }}}  \\ 
 \displaystyle  
 \sum^{N}_{i=1}  \frac{V(T_{i})}{\left(1-\alpha^{'}\dot{T}^2_{i}\right)^{1-q}}\left[1-\alpha^{'}(1-2q)\dot{T}^2_{i}\right]  \,,~~~ &
 \mbox{\small {\bf for {Multi }}} \\ 
 \displaystyle  
 \frac{NV(T)}{\left(1-\alpha^{'}\dot{T}^2\right)^{1-q}}\left[1-\alpha^{'}(1-2q)\dot{T}^2\right] \,,~~~ &
 \mbox{\small {\bf for {Assisted }}}.
          \end{array}
\right.
\end{array}\ee
and
\be\begin{array}{lll}\label{e10}
 \displaystyle p =\left\{\begin{array}{lll}
                    \displaystyle  
                    -V(T)\left(1-\alpha^{'}\dot{T}^2\right)^{q} \,,~~~~ &
 \mbox{\small {\bf for {Single }}}  \\ 
 \displaystyle  
 -\sum^{N}_{i=1} V(T_{i})\left(1-\alpha^{'}\dot{T}^2_{i}\right)^{q} \,,~~~ &
 \mbox{\small {\bf for {Multi }}} \\ 
 \displaystyle  
-NV(T)\left(1-\alpha^{'}\dot{T}^2\right)^{q} \,,~~~ &
 \mbox{\small {\bf for {Assisted }}}.
          \end{array}
\right.
\end{array}\ee
Next using Eq~(\ref{e7}) and Eq~(\ref{e8}) one can write down the expression for the equation of state parameter:
\be\begin{array}{lll}\label{e11}
 \displaystyle w =\left\{\begin{array}{lll}
                    \displaystyle  
                    \frac{p}{\rho}=\left(\alpha^{'}\dot{T}^2-1\right) \,,~~~~ &
 \mbox{\small {\bf for {Single }}}  \\ 
 \displaystyle  
 \sum^{N}_{i=1} \frac{p_{i}}{\rho_{i}}=\sum^{N}_{i=1}\left(\alpha^{'}\dot{T}^2_{i}-1\right) \,,~~~ &
 \mbox{\small {\bf for {Multi }}} \\ 
 \displaystyle \frac{Np}{\rho}= 
N\left(\alpha^{'}\dot{T}^2-1\right) \,,~~~ &
 \mbox{\small {\bf for {Assisted }}}.
          \end{array}
\right.
\end{array}\ee
and for the generalized case using Eq~(\ref{e9}) and Eq~(\ref{e10}) one can write down the expression for the equation of state parameter:
\be\begin{array}{lll}\label{e12}
 \displaystyle w =\left\{\begin{array}{lll}
                    \displaystyle  
                    \frac{p}{\rho}=\frac{\left(\alpha^{'}\dot{T}^2-1\right)}{\left[1-\alpha^{'}(1-2q)\dot{T}^2\right] } \,,~~~~ &
 \mbox{\small {\bf for {Single }}}  \\ 
 \displaystyle  
 \sum^{N}_{i=1} \frac{p_{i}}{\rho_{i}}=\sum^{N}_{i=1}\frac{\left(\alpha^{'}\dot{T}^2_{i}-1\right)}{\left[1-\alpha^{'}(1-2q)\dot{T}^2_{i}\right] } \,,~~~ &
 \mbox{\small {\bf for {Multi }}} \\ 
 \displaystyle \frac{Np}{\rho}= 
\frac{N\left(\alpha^{'}\dot{T}^2-1\right)}{\left[1-\alpha^{'}(1-2q)\dot{T}^2\right] } \,,~~~ &
 \mbox{\small {\bf for {Assisted }}}.
          \end{array}
\right.
\end{array}\ee
Further using the spatially flat $k=0$ FLRW metric defined through the following line element:
\be ds^2=-dt^2+a^{2}(t)d{\bf x}^2\ee
the Friedmann equations can be written as:
\bea\label{ku1} H^{2}&=&\left(\frac{\dot{a}}{a}\right)^2=\frac{\rho}{3M^{2}_{p}},\\
\label{ku2}\dot{H}+ H^{2} &=& \frac{\ddot{a}}{a}=-\frac{(\rho+3p)}{6M^{2}_{p}}.\eea
where the density $\rho$ and pressure $p$ is computed in Eq~(\ref{e7})-Eq~(\ref{eq10}). Also $H$ is the Hubble parameter, defined as:
\be H=\frac{1}{a(t)}\frac{da(t)}{dt}=\frac{\dot{a}}{a}. \ee
On the other hand, varying the action as stated in Eq~(\ref{e1}) with respect to the tachyon field the equation of motion can be written as:
\be\begin{array}{lll}\label{e13}
 \displaystyle 0 =\left\{\begin{array}{lll}
                    \displaystyle  
                    \frac{1}{\sqrt{-g}}\partial_{\mu}\left(\sqrt{-g}V(T)\sqrt{1+g^{\alpha\beta}\partial_{\alpha}T\partial_{\beta}T}\right)
                    =\frac{\alpha^{'}\ddot{T}}{\left(1-\alpha^{'}\dot{T}^2\right)}
+3H\alpha^{'}\dot{T}+\frac{dV(T)}{V(T)dT} \,, &
 \mbox{\small {\bf for {Single }}}  \\ 
 \displaystyle  
 \frac{1}{\sqrt{-g}}\partial_{\mu}\left(\sqrt{-g}V(T_{i})\sqrt{1+g^{\alpha\beta}\partial_{\alpha}T_{i}\partial_{\beta}T_{i}}\right)
                    =\frac{\alpha^{'}\ddot{T}_{i}}{\left(1-\alpha^{'}\dot{T}^2_{i}\right)}
+3H\alpha^{'}\dot{T}_{i}+\frac{dV(T_{i})}{V(T_{i})dT_{i}}  \,, &
 \mbox{\small {\bf for {Multi }}} \\ 
 \displaystyle \frac{1}{\sqrt{-g}}\partial_{\mu}\left(\sqrt{-g}V(T)\sqrt{1+g^{\alpha\beta}\partial_{\alpha}T\partial_{\beta}T}\right)
                    =\frac{\alpha^{'}\ddot{T}}{\left(1-\alpha^{'}\dot{T}^2\right)}
+3H\alpha^{'}\dot{T}+\frac{dV(T)}{V(T)dT}  \,, &
 \mbox{\small {\bf for {Assisted }}}.
          \end{array}
\right.
\end{array}\ee
Similarly, in the most generalized case, varying the action as stated in Eq~(\ref{e2}) with respect to the tachyon field the equation of motion can be written as:
\be\begin{array}{lll}\label{e14}
 \displaystyle 0 =\left\{\begin{array}{lll}
                    \displaystyle  
                    \frac{1}{\sqrt{-g}}\partial_{\mu}\left(\sqrt{-g}V(T)\left(1+g^{\alpha\beta}\partial_{\alpha}T\partial_{\beta}T\right)^q\right)\\
                    \displaystyle~~~~~~~=\frac{2q\alpha^{'}\ddot{T}}{\left(1-\alpha^{'}\dot{T}^2\right)}
+6q\alpha^{'}H\frac{\dot{T}}{1-\alpha^{'}(1-2q)\dot{T}^2}+\frac{dV(T)}{V(T)dT} \,,~~~~ &
 \mbox{\small {\bf for {Single }}}  \\ 
 \displaystyle  
 \frac{1}{\sqrt{-g}}\partial_{\mu}\left(\sqrt{-g}V(T_{i})\left(1+g^{\alpha\beta}\partial_{\alpha}T_{i}\partial_{\beta}T_{i}\right)^q\right)\\
                    \displaystyle~~~~~~~=\frac{2q\alpha^{'}\ddot{T}_{i}}{\left(1-\alpha^{'}\dot{T}^2_{i}\right)}
+6q\alpha^{'}H\frac{\dot{T}_{i}}{1-\alpha^{'}(1-2q)\dot{T}^2_{i}}+\frac{dV(T_{i})}{V(T_{i})dT_{i}}  \,,~~~ &
 \mbox{\small {\bf for {Multi }}} \\ 
 \displaystyle \frac{1}{\sqrt{-g}}\partial_{\mu}\left(\sqrt{-g}V(T)\left(1+g^{\alpha\beta}\partial_{\alpha}T\partial_{\beta}T\right)^q\right)\\
                    \displaystyle~~~~~~~=\frac{2q\alpha^{'}\ddot{T}}{\left(1-\alpha^{'}\dot{T}^2\right)}
+6q\alpha^{'}H\frac{\dot{T}}{1-\alpha^{'}(1-2q)\dot{T}^2}+\frac{dV(T)}{V(T)dT}  \,,~~~ &
 \mbox{\small {\bf for {Assisted }}}.
          \end{array}
\right.
\end{array}\ee
Further using Eq~(\ref{e7}), Eq~(\ref{e8}) and Eq~(\ref{e13}) the expression for the adiabatic sound speed $c_{A}$ turns out to be:
\be\begin{array}{lll}\label{e15}
 \displaystyle c_{A} =\left\{\begin{array}{lll}
                    \displaystyle  
                    \sqrt{\frac{\dot{p}}{\dot{\rho}}}=\sqrt{-w\left[1+\frac{2}{3H\alpha^{'}\dot{T}}\frac{dV(T)}{V(T)dT}\right]} \,,~~~~ &
 \mbox{\small {\bf for {Single }}}  \\ 
 \displaystyle  
 \sqrt{\sum^{N}_{i=1} \frac{\dot{p}_{i}}{\dot{\rho}_{i}}}=\sqrt{\sum^{N}_{i=1}\left\{-w_{i}\left[1+\frac{2}{3H\alpha^{'}\dot{T}_{i}}\frac{dV(T_{i})}{V(T_{i})dT_{i}}\right]\right\}} \,,~~~ &
 \mbox{\small {\bf for {Multi }}} \\ 
 \displaystyle    \sqrt{\frac{N\dot{p}}{\dot{\rho}}}=\sqrt{-wN\left[1+\frac{2}{3H\alpha^{'}\dot{T}}\frac{dV(T)}{V(T)dT}\right]} \,,~~~ &
 \mbox{\small {\bf for {Assisted }}}.
          \end{array}
\right.
\end{array}\ee
and for the generalized case using Eq~(\ref{e7}), Eq~(\ref{e8}) and Eq~(\ref{e14}) the expression for the adiabatic sound speed turns out to be:
\be\begin{array}{lll}\label{e16}
 \displaystyle \tiny\tiny c_{A} =\left\{\begin{array}{lll}\tiny
                    \displaystyle  
                    \sqrt{\frac{-w\left[1+\frac{\left(1-\alpha^{'}(1-2q)\dot{T}^2\right)}{3qH\alpha^{'}\dot{T}}
                    \frac{dV(T)}{V(T)dT}\right]}{\left[\left(\frac{1}{q}-1\right)\left\{1-\frac{(1-2q)}{(1-q)}w\right\}
                    \left\{1+\frac{\left(1-\alpha^{'}(1-2q)\dot{T}^2\right)}{6qH\alpha^{'}\dot{T}}\frac{dV(T)}{V(T)dT}\right\}
                    -\frac{\left(1-\alpha^{'}(1-2q)\dot{T}^2\right)}{6qH\alpha^{'}\dot{T}}\frac{dV(T)}{V(T)dT}\right]}} \,,~~~~ &
 \mbox{\small {\bf for {Single }}}  \\ 
 \displaystyle  
 \sqrt{\sum^{N}_{i=1}\left\{\frac{-w_{i}\left[1+\frac{\left(1-\alpha^{'}(1-2q)\dot{T}^2_{i}\right)}{3qH\alpha^{'}\dot{T}_{i}}
                    \frac{dV(T_{i})}{V(T_{i})dT_{i}}\right]}{\left[\left(\frac{1}{q}-1\right)\left\{1-\frac{(1-2q)}{(1-q)}w_{i}\right\}
                    \left\{1+\frac{\left(1-\alpha^{'}(1-2q)\dot{T}^2_{i}\right)}{6qH\alpha^{'}\dot{T}_{i}}\frac{dV(T_{i})}{V(T_{i})dT_{i}}\right\}
                    -\frac{\left(1-\alpha^{'}(1-2q)\dot{T}^2_{i}\right)}{6qH\alpha^{'}\dot{T}_{i}}\frac{dV(T_{i})}{V(T_{i})dT_{i}}\right]}\right\}} \,,~~~ &
 \mbox{\small {\bf for {Multi }}} \\ 
 \displaystyle   \sqrt{\frac{-wN\left[1+\frac{\left(1-\alpha^{'}(1-2q)\dot{T}^2\right)}{3qH\alpha^{'}\dot{T}}
                    \frac{dV(T)}{V(T)dT}\right]}{\left[\left(\frac{1}{q}-1\right)\left\{1-\frac{(1-2q)}{(1-q)}w\right\}
                    \left\{1+\frac{\left(1-\alpha^{'}(1-2q)\dot{T}^2\right)}{6qH\alpha^{'}\dot{T}}\frac{dV(T)}{V(T)dT}\right\}
                    -\frac{\left(1-\alpha^{'}(1-2q)\dot{T}^2\right)}{6qH\alpha^{'}\dot{T}}\frac{dV(T)}{V(T)dT}\right]}} \,,~~~ &
 \mbox{\small {\bf for {Assisted }}}.
          \end{array}
\right.
\end{array}\ee
It is important to mention here that, substituting $q=1/2$ in Eq~(\ref{e16}) one can get back the result obtained in Eq~(\ref{e15}).
Similarly in the present context the effective sound speed $c_{S}$ is defined as:
\be\begin{array}{lll}\label{e17}
 \displaystyle c_{S} =\left\{\begin{array}{lll}\tiny
                    \displaystyle  \tiny
                    \sqrt{\frac{\frac{\partial p}{\partial \dot{T}^2}}{\frac{\partial \rho}{\partial \dot{T}^2}}}=\sqrt{-w} \,,~~~~ &
 \mbox{\small {\bf for {Single }}}  \\ 
 \displaystyle  
 \sqrt{\sum^{N}_{i=1} \frac{\frac{\partial p_{i}}{\partial \dot{T}^2_{i}}}{\frac{\partial \rho_{i}}{\partial \dot{T}^2_{i}}}}
 =\sqrt{-\sum^{N}_{i=1}w_{i}} \,,~~~ &
 \mbox{\small {\bf for {Multi }}} \\ 
 \displaystyle    \sqrt{N\frac{\frac{\partial p}{\partial \dot{T}^2}}{\frac{\partial \rho}{\partial \dot{T}^2}}}=\sqrt{-wN} \,,~~~ &
 \mbox{\small {\bf for {Assisted }}}.
          \end{array}
\right.
\end{array}\ee
and for the generalized case the effective sound speed $c_{S}$ is defined as:
\be\begin{array}{lll}\label{e18}
 \displaystyle c_{S} =\left\{\begin{array}{lll}\footnotesize
                    \displaystyle  
                    \sqrt{\frac{\frac{\partial p}{\partial \dot{T}^2}}{\frac{\partial \rho}{\partial \dot{T}^2}}}=\sqrt{\left[1+\frac{2(q-1)(1+w)}{1+(1-2q)(2w+1)}\right]} \,,~~~~ &
 \mbox{\small {\bf for {Single }}}  \\ 
 \displaystyle  
 \sqrt{\sum^{N}_{i=1} \frac{\frac{\partial p_{i}}{\partial \dot{T}^2_{i}}}{\frac{\partial \rho_{i}}{\partial \dot{T}^2_{i}}}}
 =\sqrt{\sum^{N}_{i=1}\left[1+\frac{2(q-1)(1+w_{i})}{1+(1-2q)(2w_{i}+1)}\right]} \,,~~~ &
 \mbox{\small {\bf for {Multi }}} \\ 
 \displaystyle    \sqrt{N\frac{\frac{\partial p}{\partial \dot{T}^2}}{\frac{\partial \rho}{\partial \dot{T}^2}}}=\sqrt{N\left[1+\frac{2(q-1)(1+w)}{1+(1-2q)(2w+1)}\right]} \,,~~~ &
 \mbox{\small {\bf for {Assisted }}}.
          \end{array}
\right.
\end{array}\ee
Finally comparing Eq~(\ref{e15}), Eq~(\ref{e16}), Eq~(\ref{e17}) and Eq~(\ref{e18}) we get the following relationship between adiabatic and effective sound speed in tachyonic
field theory:
\be\begin{array}{lll}\label{e19}
 \displaystyle c_{A} =\left\{\begin{array}{lll}\footnotesize
                    \displaystyle  
                    \sqrt{\frac{\dot{p}}{\dot{\rho}}}=c_{S}\sqrt{\left[1+\frac{2}{3H\alpha^{'}\dot{T}}\frac{dV(T)}{V(T)dT}\right]} \,,~~~~ &
 \mbox{\small {\bf for {Single }}}  \\ 
 \displaystyle  
 \sqrt{\sum^{N}_{i=1} \frac{\dot{p}_{i}}{\dot{\rho}_{i}}}=\sqrt{\sum^{N}_{i=1}\left\{c^{2}_{S,i}\left[1+\frac{2}{3H\alpha^{'}\dot{T}_{i}}\frac{dV(T_{i})}{V(T_{i})dT_{i}}\right]\right\}} \,,~~~ &
 \mbox{\small {\bf for {Multi }}} \\ 
 \displaystyle    \sqrt{\frac{N\dot{p}}{\dot{\rho}}}=c_{S}\sqrt{\left[1+\frac{2}{3H\alpha^{'}\dot{T}}\frac{dV(T)}{V(T)dT}\right]} \,,~~~ &
 \mbox{\small {\bf for {Assisted }}}.
          \end{array}
\right.
\end{array}\ee
and for the generalized case we get:
\be\begin{array}{lll}\label{e20}
 \displaystyle \tiny\tiny c_{A} =\left\{\begin{array}{lll}\tiny
                    \displaystyle  
                    \sqrt{\frac{\frac{c^{2}_{S}}{\left[1+\frac{(1-2q)}{(1-q)}\left(c^{2}_{S}-1\right)\right]}
                    \left[1+\frac{\left(1-\alpha^{'}(1-2q)\dot{T}^2\right)}{3qH\alpha^{'}\dot{T}}
                    \frac{dV(T)}{V(T)dT}\right]}{\left[\left(\frac{1}{q}-1\right)\left\{1+\frac{\frac{(1-2q)}{(1-q)}c^{2}_{S}}{\left[1+\frac{(1-2q)}{(1-q)}\left(c^{2}_{S}-1\right)\right]}\right\}
                    \left\{1+\frac{\left(1-\alpha^{'}(1-2q)\dot{T}^2\right)}{6qH\alpha^{'}\dot{T}}\frac{dV(T)}{V(T)dT}\right\}
                    -\frac{\left(1-\alpha^{'}(1-2q)\dot{T}^2\right)}{6qH\alpha^{'}\dot{T}}\frac{dV(T)}{V(T)dT}\right]}} \,,~~~~ &
 \mbox{\small {\bf for {Single }}}  \\ 
 \displaystyle  
 \sqrt{\sum^{N}_{i=1}\left\{\frac{\frac{c^{2}_{S,i}}{\left[1+\frac{(1-2q)}{(1-q)}\left(c^{2}_{S,i}-1\right)\right]}
 \left[1+\frac{\left(1-\alpha^{'}(1-2q)\dot{T}^2_{i}\right)}{3qH\alpha^{'}\dot{T}_{i}}
                    \frac{dV(T_{i})}{V(T_{i})dT_{i}}\right]}{\left[\left(\frac{1}{q}-1\right)\left\{1+\frac{\frac{(1-2q)}{(1-q)}c^{2}_{S,i}}{\left[1+\frac{(1-2q)}{(1-q)}\left(c^{2}_{S,i}-1\right)\right]}\right\}
                    \left\{1+\frac{\left(1-\alpha^{'}(1-2q)\dot{T}^2_{i}\right)}{6qH\alpha^{'}\dot{T}_{i}}\frac{dV(T_{i})}{V(T_{i})dT_{i}}\right\}
                    -\frac{\left(1-\alpha^{'}(1-2q)\dot{T}^2_{i}\right)}{6qH\alpha^{'}\dot{T}_{i}}\frac{dV(T_{i})}{V(T_{i})dT_{i}}\right]}\right\}} \,,~~~ &
 \mbox{\small {\bf for {Multi }}} \\ 
 \displaystyle   \sqrt{\frac{\frac{c^{2}_{S}}{\left[1+\frac{(1-2q)}{(1-q)}\left(\frac{c^{2}_{S}}{N}-1\right)\right]}
 \left[1+\frac{\left(1-\alpha^{'}(1-2q)\dot{T}^2\right)}{3qH\alpha^{'}\dot{T}}
                    \frac{dV(T)}{V(T)dT}\right]}{\left[\left(\frac{1}{q}-1\right)\left\{1+\frac{\frac{(1-2q)}{(1-q)}c^{2}_{S}}{N\left[1+\frac{(1-2q)}{(1-q)}\left(\frac{c^{2}_{S}}{N}-1\right)\right]}\right\}
                    \left\{1+\frac{\left(1-\alpha^{'}(1-2q)\dot{T}^2\right)}{6qH\alpha^{'}\dot{T}}\frac{dV(T)}{V(T)dT}\right\}
                    -\frac{\left(1-\alpha^{'}(1-2q)\dot{T}^2\right)}{6qH\alpha^{'}\dot{T}}\frac{dV(T)}{V(T)dT}\right]}} \,,~~~ &
 \mbox{\small {\bf for {Assisted }}}.
          \end{array}
\right.
\end{array}\ee
Let us mention other crucial issues which we use throughout the analysis performed in this paper:
\begin{enumerate}
 \item {\bf At early times} the tachyonic field satisfy the following small field criteria to validate Effective Field Theory prescription within the framework of tachyonic string theory:
 \bea \label{g1}\frac{|T|}{M_{p}}<<1,\\
 \label{g2}|\dot{T}|<<\frac{1}{\sqrt{\alpha^{'}}}\eea
 and for the generalized $q\neq 1/2$ case additionally we have to satisfy another constraint:
 \bea \label{g3}\sqrt{(1-2q)}|\dot{T}|<<\frac{1}{\sqrt{\alpha^{'}}}.\eea
 
 \item {\bf At early times} using Eq~(\ref{g1}), Eq~(\ref{g2}) and Eq~(\ref{g3}) in Eq~(\ref{e11}) and Eq~(\ref{e12}), the equation of state parameter $w$ can be approximated as:
     \bea w\approx -1. \eea
 \item {\bf At late times} the tachyonic field satisfy the following small field criteria within the framework of tachyonic string theory:
 \bea \label{g4}\frac{|T|}{M_{p}}\sim 1,\\
 \label{g5}\frac{|\dot{T}|}{M^{2}_{p}}\sim \frac{1}{\sqrt{\alpha^{'}}}\eea
 and for the generalized $q\neq 1/2$ case additionally we have to satisfy another constraint:
 \bea \label{g6}\sqrt{(1-2q)}|\dot{T}|\sim\frac{1}{\sqrt{\alpha^{'}}}.\eea
 \item {\bf At late times} using Eq~(\ref{g4}), Eq~(\ref{g5}) and Eq~(\ref{g6}) in Eq~(\ref{e11}) and Eq~(\ref{e12}), the equation of state parameter $w$ can be approximated as:
 \bea w\approx 0.\eea
 \item There might be another interesting possibility appear in the present context, where the tachyonic modes satisfy the large field criteria, represented by the following constraint:
 \bea \label{g7}\frac{|T|}{M_{p}}>>1,\\
 \label{g8}|\dot{T}|>>\frac{1}{\sqrt{\alpha^{'}}}\eea
 and for the generalized $q\neq 1/2$ case additionally we have to satisfy another constraint:
 \bea \label{g9}\sqrt{(1-2q)}|\dot{T}|>>\frac{1}{\sqrt{\alpha^{'}}}.\eea
 \item Further using Eq~(\ref{g7}), Eq~(\ref{g8}) and Eq~(\ref{g9}) in Eq~(\ref{e11}) and Eq~(\ref{e12}), the equation of state parameter $w$ can be approximated as:
     \be\begin{array}{lll}\label{g10}
 \displaystyle w =\left\{\begin{array}{lll}
                    \displaystyle  
                    \frac{p}{\rho}=\alpha^{'}\dot{T}^2 \,,~~~~ &
 \mbox{\small {\bf for {Single }}}  \\ 
 \displaystyle  
 \sum^{N}_{i=1} \frac{p_{i}}{\rho_{i}}=\sum^{N}_{i=1}\alpha^{'}\dot{T}^2_{i} \,,~~~ &
 \mbox{\small {\bf for {Multi }}} \\ 
 \displaystyle \frac{Np}{\rho}= 
N\alpha^{'}\dot{T}^2 \,,~~~ &
 \mbox{\small {\bf for {Assisted }}}.
          \end{array}
\right.
\end{array}\ee
and for the generalized $q\neq 1/2$ case we have:
 \be\begin{array}{lll}\label{g11}
 \displaystyle w =\left\{\begin{array}{lll}
                    \displaystyle  
                    \frac{p}{\rho}=\frac{1}{(2q-1) } \,,~~~~ &
 \mbox{\small {\bf for {Single }}}  \\ 
 \displaystyle  
 \sum^{N}_{i=1} \frac{p_{i}}{\rho_{i}}=\sum^{N}_{i=1}\frac{1}{(2q-1) } \,,~~~ &
 \mbox{\small {\bf for {Multi }}} \\ 
 \displaystyle \frac{Np}{\rho}= 
\frac{N}{(2q-1) } \,,~~~ &
 \mbox{\small {\bf for {Assisted }}}.
          \end{array}
\right.
\end{array}\ee
Here one can control the parameter $q$ and $N$ to get the desired value of equation of state parameter $w$, which is necessarily required to explain the cosmological dynamics.
 \item
Additionally, it is important to note that, within the setup of string theory the tachyonic modes can form cluster on small cosmological scales. Consequently tachyonic string theory
can be treated as a unified prescription to explain the inflationary paradigm and dark matter.
\item Reheating and creation of matter particles in a class of specific models
where the minimum of the tachyon potential
$V(T)$ is at $T\rightarrow \infty$, which is a pathological issue in the present context
because the
tachyon field in such type of string theories does not participate in oscillations~\footnote{Here it is important to note that the 
oscillations are necessarily required to explain reheating phenomena.}. To solve this crucial pathological problem in the present context, one can think about a particular physical situation 
where the universe is initially dominated by a inflationary epoch and this can be explained via
the energy density of the tachyon condensate as mentioned in the introduction of this article and according to this proposal the setup will always remain dominated
by the tachyons. To resolve this pathological issue it may happen that the tachyon condensation phenomena is potentially responsible for
a short period of inflationary epoch prior, which occurs at 
a Planckian mass scale, $M_p\sim 2.43\times 10^{18}{\rm GeV}$ and also one need to require
a second stage of inflationary epoch just followed by the previous one which occurs at the vicinity of the GUT scale ($10^{16}{\rm GeV}$). This directly implies
that the tachyon serves no crucial purpose in the post inflationary epoch
till at the very later stages of its cosmological evolution on time scales.
All these crucial pathological problems do not
appear in the context of well known hybrid inflationary setup where the complex tachyon field
has a specific minimum value at the sub-Planckian ($<M_{p}$) regime given by the list of constraint equations as stated in Eq~(\ref{g1}), Eq~(\ref{g2}) and in Eq~(\ref{g3}).
In this paper, we have studied the cosmological consequences from different class of tachyonic potentials appearing in the context of string theory which have no connection to the hybrid inflationary model, but to explain
CMB constraints tachyon condensation phenomena plays important role. Instead of studying the tachyon condensation phenomena, in this paper we have studied the cosmological perturbation theory and its physical 
consequences in detail in later sections.

\item The energy density of tachyons after inflation should
be fine tuned to be sub-dominant until the
very later stages of the cosmological evolution of the universe.

\end{enumerate}

\subsection{Dynamical solution for various phases}
\label{aa4b}
In this section our prime objective is to study the dynamical behavior of the tachyonic field in the background of spatially flat FLRW metric and in presence of Einstein Hilbert term in the gravity sector.
Below we explicitly show that the solution for the tachyonic field can explain various phases of the universe starting from inflation to the dust formation. To study the cosmological dynamics from 
the tachyonic string field theoretic setup let us start with the following solution ansatz of tachyon field:
\be\begin{array}{lll}\label{g12}
 \displaystyle \dot{T} =\left\{\begin{array}{lll}
                    \displaystyle  
                    \frac{1}{\sqrt{\alpha^{'}}}\left(\frac{\exp\left(\frac{2t}{bT_0}\right)-1}{\exp\left(\frac{2t}{bT_0}\right)+1}\right)=\frac{1}{\sqrt{\alpha^{'}}}{\rm tanh}\left(\frac{t}{b T_0}\right) \,,~~~~ &
 \mbox{\small {\bf for {Single }}}  \\ 
 \displaystyle  
 \sum^{N}_{i=1} \dot{T}_{i}=\sum^{N}_{i=1}\frac{1}{\sqrt{\alpha^{'}}}\left(\frac{\exp\left(\frac{2t}{bT_{0i}}\right)-1}{\exp\left(\frac{2t}{bT_{0i}}\right)+1}\right)=\sum^{N}_{i=1}\frac{1}{\sqrt{\alpha^{'}}}{\rm tanh}\left(\frac{t}{b T_{0i}}\right)\,,~~~ &
 \mbox{\small {\bf for {Multi }}} \\ 
 \displaystyle \frac{N}{\sqrt{\alpha^{'}}}\left(\frac{\exp\left(\frac{2t}{bT_0}\right)-1}{\exp\left(\frac{2t}{bT_0}\right)+1}\right)=\frac{N}{\sqrt{\alpha^{'}}}{\rm tanh}\left(\frac{t}{b T_0}\right) \,,~~~ &
 \mbox{\small {\bf for {Assisted }}}.
          \end{array}
\right.
\end{array}\ee
which will satisfy the equation of motion of tachyon field as stated in Eq~(\ref{e13}) and Eq~(\ref{e14}) respectively. 
Here $b$ is a new parameter of the theory which has inverse square mass dimension i.e. $[M]^{-1}$. Consequently the argument of the hyperbolic functions i.e. $(\frac{t}{b T_0})$ is dimensionless.
From various cosmological observations it is possible to put stringent constraint on the newly introduced parameter $b$.
Further integrating Eq~(\ref{g12}) we get the following solutions for the tachyonic field:
\be\begin{array}{lll}\label{g14}
 \displaystyle T(t) =\left\{\begin{array}{lll}
                    \displaystyle  
                    \frac{b T_{0}}{\sqrt{\alpha^{'}}}\ln\left[{\rm cosh}\left(\frac{t}{b T_0}\right)\right] \,,~~~~ &
 \mbox{\small {\bf for {Single }}}  \\ 
 \displaystyle  
 \sum^{N}_{i=1}T_{i}(t)=\sum^{N}_{i=1}\frac{b T_{0i}}{\sqrt{\alpha^{'}}}\ln\left[{\rm cosh}\left(\frac{t}{b T_{0i}}\right)\right]\,,~~~ &
 \mbox{\small {\bf for {Multi }}} \\ 
 \displaystyle \frac{Nb T_{0}}{\sqrt{\alpha^{'}}}\ln\left[{\rm cosh}\left(\frac{t}{b T_0}\right)\right] \,,~~~ &
 \mbox{\small {\bf for {Assisted }}}.
          \end{array}
\right.
\end{array}\ee
For both of the cases we fix the boundary condition in such a way that the tachyonic field $T$ satisfy the constraint:
$T(t=0)=T(0)=0.$
Further using Eq~(\ref{ku1}) and Eq~(\ref{e13}) we get the following constraint condition for the cosmological time dependent potential $V(t)$:
\be\begin{array}{lll}\label{g16}
 \displaystyle 0 =\left\{\begin{array}{lll}
                    \displaystyle  
                    \left[\frac{1}{b T_{0}}+\frac{\dot{V}}{V}{\rm coth}\left(\frac{t}{b T_0}\right)+\sqrt{\frac{3V}{{\rm cosh}
                    \left(\frac{t}{b T_0}\right)}}\frac{{\rm sinh}\left(\frac{t}{b T_0}\right)}{M_p }\right] \,,~~~~ &
 \mbox{\small {\bf for {Single }}}  \\ 
 \displaystyle  
 \sum^{N}_{i=1} \left[\frac{1}{b T_{0i}}+\frac{\dot{V}}{V}{\rm coth}\left(\frac{t}{b T_{0i}}\right)+\sqrt{\frac{3V}{
 {\rm cosh}\left(\frac{t}{b T_{0i}}\right)}}\frac{{\rm sinh}\left(\frac{t}{b T_{0i}}\right)}{M_p}\right] \,,~~~ &
 \mbox{\small {\bf for {Multi }}} \\ 
 \displaystyle N\left[\frac{1}{b T_{0}}+\frac{\dot{V}}{V}{\rm coth}\left(\frac{t}{b T_0}\right)+\sqrt{\frac{3V}
 {{\rm cosh}\left(\frac{t}{b T_0}\right)}}\frac{{\rm sinh}\left(\frac{t}{b T_0}\right)}{ M_p}\right]  \,,~~~ &
 \mbox{\small {\bf for {Assisted }}}
          \end{array}
\right.
\end{array}\ee
and for the generalized case we get:
\be\begin{array}{lll}\label{g17}
 \displaystyle 0 =\left\{\begin{array}{lll}
                    \displaystyle  
                    \left[\frac{1}{b T_{0}}+\frac{1}{2q}\frac{\dot{V}}{V}{\rm coth}\left(\frac{t}{b T_0}\right)+\frac{\sqrt{\frac{3V}{{\rm cosh}\left(\frac{t}{b T_0}\right)}}{\rm sinh}
                    \left(\frac{t}{b T_0}\right)}{ M_p \left[1-(1-2q){\rm tanh}^2\left(\frac{t}{b T_0}\right)\right]}\right] \,,~ &
 \mbox{\small {\bf for {Single }}}  \\ 
 \displaystyle  
 \sum^{N}_{i=1} \left[\frac{1}{b T_{0i}}+\frac{1}{2q}\frac{\dot{V}}{V}{\rm coth}\left(\frac{t}{b T_{0i}}\right)+\frac{\sqrt{\frac{3V}{{\rm cosh}
 \left(\frac{t}{b T_{0i}}\right)}}{\rm sinh}\left(\frac{t}{b T_{0i}}\right)}{ M_p \left[1-(1-2q){\rm tanh}^2\left(\frac{t}{b T_{0i}}\right)\right]}\right] \,,~ &
 \mbox{\small {\bf for {Multi }}} \\ 
 \displaystyle N\left[\frac{1}{b T_{0}}+\frac{1}{2q}\frac{\dot{V}}{V}{\rm coth}\left(\frac{t}{b T_0}\right)+\frac{\sqrt{\frac{3V}{{\rm cosh}\left(\frac{t}{b T_0}\right)}}{\rm sinh}
 \left(\frac{t}{b T_0}\right)}{ M_p \left[1-(1-2q){\rm tanh}^2\left(\frac{t}{b T_0}\right)\right]}\right]  \,,~ &
 \mbox{\small {\bf for {Assisted }}}
          \end{array}
\right.
\end{array}\ee
Solutions of Eq~(\ref{g16}) and Eq~(\ref{g17}) is given by:
\be\begin{array}{lll}\label{g18}
 \displaystyle V(t) =\left\{\begin{array}{lll}
                    \displaystyle  
                    \frac{\lambda}{{\rm cosh}\left(\frac{t}{b T_0}\right)}\left[\frac{1}{1+\frac{\sqrt{3\lambda}bT_{0}}{2M_{p}}\left\{\frac{t}{b T_{0}}-{\rm tanh}\left(\frac{t}{b T_0}\right)\right\}}\right]^2 \,, &
 \mbox{\small {\bf for {Single }}}  \\ 
 \displaystyle  
 \sum^{N}_{i=1}V_{i}(t)=\sum^{N}_{i=1}\frac{\lambda_{i}}{{\rm cosh}\left(\frac{t}{b T_{0i}}\right)}\left[\frac{1}{1+\frac{\sqrt{3\lambda}bT_{0i}}{2M_{p}}\left\{\frac{t}{b T_{0i}}-{\rm tanh}\left(\frac{t}
 {b T_{0i}}\right)\right\}}\right]^2 \,,~ &
 \mbox{\small {\bf for {Multi }}} \\ 
 \displaystyle \frac{N\lambda}{{\rm cosh}\left(\frac{t}{b T_0}\right)}\left[\frac{1}{1+\frac{\sqrt{3\lambda}bT_{0}}{2M_{p}}\left\{\frac{t}{b T_{0}}-{\rm tanh}\left(\frac{t}{b T_0}\right)\right\}}\right]^2  \,,~ &
 \mbox{\small {\bf for {Assisted }}}
          \end{array}
\right.
\end{array}\ee
and for the generalized case we get:
\be\begin{array}{lll}\label{g19}
 \displaystyle V(t) =\left\{\begin{array}{lll}
                    \displaystyle  
                    \frac{\lambda}{\left[{\rm cosh}\left(\frac{t}{b T_0}\right)\right]^{2q}}\left[\frac{1}{1+\frac{\sqrt{3\lambda}bT_{0}}{2M_{p}}\left\{\frac{t}{b T_{0}}-{\rm tanh}\left(\frac{t}{b T_0}\right)\right\}}\right]^2 \,, &
 \mbox{\small {\bf for {Single }}}  \\ 
 \displaystyle  
 \sum^{N}_{i=1}V_{i}(t)=\sum^{N}_{i=1}\frac{\lambda_{i}}{\left[{\rm cosh}\left(\frac{t}{b T_{0i}}\right)\right]^{2q}}\left[\frac{1}{1+\frac{\sqrt{3\lambda}bT_{0i}}{2M_{p}}\left\{\frac{t}{b T_{0i}}-{\rm tanh}\left(\frac{t}
 {b T_{0i}}\right)\right\}}\right]^2 \,,~ &
 \mbox{\small {\bf for {Multi }}} \\ 
 \displaystyle \frac{N\lambda}{\left[{\rm cosh}\left(\frac{t}{b T_0}\right)\right]^{2q}}\left[\frac{1}{1+\frac{\sqrt{3\lambda}bT_{0}}{2M_{p}}\left\{\frac{t}{b T_{0}}-{\rm tanh}\left(\frac{t}{b T_0}\right)\right\}}\right]^2  \,,~ &
 \mbox{\small {\bf for {Assisted }}}
          \end{array}
\right.
\end{array}\ee
where to get the analytical solution from the generalized case we assume that time scale of consideration satisfy the following constraint:
\bea\label{opiu} t<<b T_{0}~{\rm tanh}^{-1}\left[\sqrt{\frac{1}{2q-1}}\right] \eea
which is valid for all values of $q$ except $q\neq 1/2$. For usual tachyonic case and for the generalized situation we use the following normalization condition:
\bea \label{iopq}
V(t=0)=V(0)=\lambda
\eea
to fix the value of arbitrary integration constant. Further using the explicit solution for the tachyonic field as appearing in Eq~(\ref{g14}), we can write the time as a function of tachyonic field as:
\be\begin{array}{lll}\label{g14xx}
 \displaystyle t =\left\{\begin{array}{lll}
                    \displaystyle  
                    b T_0 \left[{\rm cosh}^{-1}\left\{\exp\left(\frac{\sqrt{\alpha^{'}}T}{b T_{0}}\right)\right\}\right] \,,~~~~ &
 \mbox{\small {\bf for {Single }}}  \\ 
 \displaystyle  
 \sum^{N}_{i=1}b T_{0i} \left[{\rm cosh}^{-1}\left\{\exp\left(\frac{\sqrt{\alpha^{'}}T}{b T_{0i}}\right)\right\}\right]\,,~~~ &
 \mbox{\small {\bf for {Multi }}} \\ 
 \displaystyle Nb T_0 \left[{\rm cosh}^{-1}\left\{\exp\left(\frac{\sqrt{\alpha^{'}}T}{b T_{0}}\right)\right\}\right] \,,~~~ &
 \mbox{\small {\bf for {Assisted }}}.
          \end{array}
\right.
\end{array}\ee
and further substituting Eq~(\ref{g14xx}) in Eq~(\ref{g18}) and Eq~(\ref{g19})
we get the following expression for the potential as a function of tachyonic field:
\be\begin{array}{lll}\label{g20}\tiny
 \displaystyle V(T) =\left\{\begin{array}{lll}
                    \displaystyle  
                    \lambda\exp\left(-\frac{\sqrt{\alpha^{'}}T}{b T_{0}}\right)\left[\frac{1}{1+\frac{\sqrt{3\lambda}bT_{0}}{2M_{p}}\left\{{\rm cosh}^{-1}\left\{\exp\left(\frac{\sqrt{\alpha^{'}}T}{b T_{0}}\right)\right\}
                    -{\rm tanh}\left({\rm cosh}^{-1}\left\{\exp\left(\frac{\sqrt{\alpha^{'}}T}{b T_{0}}\right)\right\}\right)\right\}}\right]^2 \,, &
 \mbox{\small {\bf for {Single }}}  \\ 
 \displaystyle  
 \sum^{N}_{i=1}\lambda_{i}\exp\left(-\frac{\sqrt{\alpha^{'}}T_{i}}{b T_{0i}}\right)\left[\frac{1}{1+\frac{\sqrt{3\lambda}bT_{0i}}{2M_{p}}\left\{{\rm cosh}^{-1}\left\{\exp\left(\frac{\sqrt{\alpha^{'}}T_{i}}{b T_{0i}}\right)\right\}
                    -{\rm tanh}\left({\rm cosh}^{-1}\left\{\exp\left(\frac{\sqrt{\alpha^{'}}T_{i}}{b T_{0i}}\right)\right\}\right)\right\}}\right]^2 \,,~ &
 \mbox{\small {\bf for {Multi }}} \\ 
 \displaystyle N\lambda\exp\left(-\frac{\sqrt{\alpha^{'}}T}{b T_{0}}\right)\left[\frac{1}{1+\frac{\sqrt{3\lambda}bT_{0}}{2M_{p}}\left\{{\rm cosh}^{-1}\left\{\exp\left(\frac{\sqrt{\alpha^{'}}T}{b T_{0}}\right)\right\}
                    -{\rm tanh}\left({\rm cosh}^{-1}\left\{\exp\left(\frac{\sqrt{\alpha^{'}}T}{b T_{0}}\right)\right\}\right)\right\}}\right]^2  \,,~ &
 \mbox{\small {\bf for {Assisted }}}
          \end{array}
\right.
\end{array}\ee
and for the generalized case we get:
\be\begin{array}{lll}\label{g21}\tiny
 \displaystyle V(T) =\left\{\begin{array}{lll}
                    \displaystyle  
                    \lambda\exp\left(-\frac{2q\sqrt{\alpha^{'}}T}{b T_{0}}\right)\left[\frac{1}{1+\frac{\sqrt{3\lambda}bT_{0}}{2M_{p}}\left\{{\rm cosh}^{-1}\left\{\exp\left(\frac{2q\sqrt{\alpha^{'}}T}{b T_{0}}\right)\right\}
                    -{\rm tanh}\left({\rm cosh}^{-1}\left\{\exp\left(\frac{2q\sqrt{\alpha^{'}}T}{b T_{0}}\right)\right\}\right)\right\}}\right]^2 \,, &
 \mbox{\small {\bf for {Single }}}  \\ 
 \displaystyle  
 \sum^{N}_{i=1}\lambda_{i}\exp\left(-\frac{2q\sqrt{\alpha^{'}}T_{i}}{b T_{0i}}\right)\left[\frac{1}{1+\frac{\sqrt{3\lambda}bT_{0i}}{2M_{p}}\left\{{\rm cosh}^{-1}\left\{\exp\left(\frac{2q\sqrt{\alpha^{'}}T_{i}}{b T_{0i}}\right)\right\}
                    -{\rm tanh}\left({\rm cosh}^{-1}\left\{\exp\left(\frac{2q\sqrt{\alpha^{'}}T_{i}}{b T_{0i}}\right)\right\}\right)\right\}}\right]^2 \,,~ &
 \mbox{\small {\bf for {Multi }}} \\ 
 \displaystyle N\lambda\exp\left(-\frac{2q\sqrt{\alpha^{'}}T}{b T_{0}}\right)\left[\frac{1}{1+\frac{\sqrt{3\lambda}bT_{0}}{2M_{p}}\left\{{\rm cosh}^{-1}\left\{\exp\left(\frac{2q\sqrt{\alpha^{'}}T}{b T_{0}}\right)\right\}
                    -{\rm tanh}\left({\rm cosh}^{-1}\left\{\exp\left(\frac{2q\sqrt{\alpha^{'}}T}{b T_{0}}\right)\right\}\right)\right\}}\right]^2  \,,~ &
 \mbox{\small {\bf for {Assisted }}}.
          \end{array}
\right.
\end{array}\ee
Next we use the following redefinition in the tachyonic field :
\bea
\frac{\sqrt{\alpha^{'}}T}{b T_{0}}&\rightarrow& \frac{T}{T_{0}},\\
\frac{\sqrt{\alpha^{'}}T}{b T_{0i}}&\rightarrow& \frac{T}{T_{0i}}.
\eea
Hence using the redefinition, the potential as stated in Eq~(\ref{g20}) and Eq~(\ref{g21}) can be re-expressed as:
\be\begin{array}{lll}\label{g22}\tiny
 \displaystyle V(T) =\left\{\begin{array}{lll}
                    \displaystyle  
                    \lambda\exp\left(-\frac{T}{T_{0}}\right)\left[\frac{1}{1+\frac{\sqrt{3\lambda}bT_{0}}{2M_{p}}\left\{{\rm cosh}^{-1}\left\{\exp\left(\frac{\sqrt{\alpha^{'}}T}{b T_{0}}\right)\right\}
                    -{\rm tanh}\left({\rm cosh}^{-1}\left\{\exp\left(\frac{\sqrt{\alpha^{'}}T}{b T_{0}}\right)\right\}\right)\right\}}\right]^2 \,, &
 \mbox{\small {\bf for {Single }}}  \\ 
 \displaystyle  
 \sum^{N}_{i=1}\lambda_{i}\exp\left(-\frac{T_{i}}{T_{0i}}\right)\left[\frac{1}{1+\frac{\sqrt{3\lambda}bT_{0i}}{2M_{p}}\left\{{\rm cosh}^{-1}\left\{\exp\left(\frac{\sqrt{\alpha^{'}}T_{i}}{b T_{0i}}\right)\right\}
                    -{\rm tanh}\left({\rm cosh}^{-1}\left\{\exp\left(\frac{\sqrt{\alpha^{'}}T_{i}}{b T_{0i}}\right)\right\}\right)\right\}}\right]^2 \,,~ &
 \mbox{\small {\bf for {Multi }}} \\  
 \displaystyle N\lambda\exp\left(-\frac{T}{T_{0}}\right)\left[\frac{1}{1+\frac{\sqrt{3\lambda}bT_{0}}{2M_{p}}\left\{{\rm cosh}^{-1}\left\{\exp\left(\frac{\sqrt{\alpha^{'}}T}{b T_{0}}\right)\right\}
                    -{\rm tanh}\left({\rm cosh}^{-1}\left\{\exp\left(\frac{\sqrt{\alpha^{'}}T}{b T_{0}}\right)\right\}\right)\right\}}\right]^2  \,,~ &
 \mbox{\small {\bf for {Assisted }}}
          \end{array}
\right.
\end{array}\ee
and for the generalized case we get:
\be\begin{array}{lll}\label{g23}\tiny
 \displaystyle V(T) =\left\{\begin{array}{lll}
                    \displaystyle  
                    \lambda\exp\left(-\frac{2qT}{T_{0}}\right)\left[\frac{1}{1+\frac{\sqrt{3\lambda}bT_{0}}{2M_{p}}\left\{{\rm cosh}^{-1}\left\{\exp\left(\frac{2q\sqrt{\alpha^{'}}T}{b T_{0}}\right)\right\}
                    -{\rm tanh}\left({\rm cosh}^{-1}\left\{\exp\left(\frac{2q\sqrt{\alpha^{'}}T}{b T_{0}}\right)\right\}\right)\right\}}\right]^2 \,, &
 \mbox{\small {\bf for {Single }}}  \\
 \displaystyle  
 \sum^{N}_{i=1}\lambda_{i}\exp\left(-\frac{2qT_{i}}{ T_{0i}}\right)\left[\frac{1}{1+\frac{\sqrt{3\lambda}bT_{0i}}{2M_{p}}\left\{{\rm cosh}^{-1}\left\{\exp\left(\frac{2q\sqrt{\alpha^{'}}T_{i}}{b T_{0i}}\right)\right\}
                    -{\rm tanh}\left({\rm cosh}^{-1}\left\{\exp\left(\frac{2q\sqrt{\alpha^{'}}T_{i}}{b T_{0i}}\right)\right\}\right)\right\}}\right]^2 \,,~ &
 \mbox{\small {\bf for {Multi }}} \\ 
 \displaystyle N\lambda\exp\left(-\frac{2qT}{ T_{0}}\right)\left[\frac{1}{1+\frac{\sqrt{3\lambda}bT_{0}}{2M_{p}}\left\{{\rm cosh}^{-1}\left\{\exp\left(\frac{2q\sqrt{\alpha^{'}}T}{b T_{0}}\right)\right\}
                    -{\rm tanh}\left({\rm cosh}^{-1}\left\{\exp\left(\frac{2q\sqrt{\alpha^{'}}T}{b T_{0}}\right)\right\}\right)\right\}}\right]^2  \,,~ &
 \mbox{\small {\bf for {Assisted }}}.
          \end{array}
\right.
\end{array}\ee
Now let us explicitly study the limiting behavior of the potential $V(T)$ in detail, which is appended below:
\begin{itemize}
 \item  At $T<<T_{0}$ and $T_{i}<<T_{0i}$ limiting case one can use the following approximation for the usual tachyonic case:
 \bea \ln\left[{\rm cosh}\left(\frac{t}{b T_0}\right)\right]&<<&1,\\
  \ln\left[{\rm cosh}\left(\frac{t}{b T_{0i}}\right)\right]&<<&1.\eea
 Using this approximation one can use the following expansion:
 \bea 
 \ln\left[{\rm cosh}\left(\frac{t}{b T_0}\right)\right]&\approx& \frac{1}{2}\left(\frac{t}{b T_{0}}\right)^2,\\
 \ln\left[{\rm cosh}\left(\frac{t}{b T_{0i}}\right)\right]&\approx& \frac{1}{2}\left(\frac{t}{b T_{0i}}\right)^2.
 \eea
Hence using the solution obtained for tachyonic field as stated in Eq~(\ref{g14}) we get:
\be\begin{array}{lll}\label{g24}
 \displaystyle T(t) =\left\{\begin{array}{lll}
                    \displaystyle  
                    \frac{T_{0}}{2}\left(\frac{t}{b T_{0}}\right)^2 \,,~~~~ &
 \mbox{\small {\bf for {Single }}}  \\ 
 \displaystyle  
 \sum^{N}_{i=1}T_{i}(t)=\sum^{N}_{i=1}\frac{T_{0i}}{2}\left(\frac{t}{b T_{0i}}\right)^2 \,,~~~ &
 \mbox{\small {\bf for {Multi }}} \\ 
 \displaystyle \frac{NT_{0}}{2}\left(\frac{t}{b T_{0}}\right)^2  \,,~~~ &
 \mbox{\small {\bf for {Assisted }}}.
          \end{array}
\right.
\end{array}\ee
and by inverting Eq~(\ref{g24}) the associated time scale can be computed as:
\be\begin{array}{lll}\label{g25}
 \displaystyle t =\left\{\begin{array}{lll}
                    \displaystyle  
                    b T_{0}\sqrt{\frac{2T}{ T_{0}}} \,,~~~~ &
 \mbox{\small {\bf for {Single }}}  \\ 
 \displaystyle  
 \sum^{N}_{i=1}b T_{0i}\sqrt{\frac{2T}{ T_{0i}}} \,,~~~ &
 \mbox{\small {\bf for {Multi }}} \\ 
 \displaystyle Nb T_{0}\sqrt{\frac{2T}{ T_{0}}}  \,,~~~ &
 \mbox{\small {\bf for {Assisted }}}.
          \end{array}
\right.
\end{array}\ee
Consequently we have:
\bea {\bf Single: } ~~~~~~~~{\rm tanh}\left(\frac{t}{b T_{0}}\right)\approx{\rm tanh}\left(\sqrt{\frac{2T}{ T_{0}}}\right)\approx \sqrt{\frac{2T}{ T_{0}}},~~~~~~\\ 
{\bf Multi: } ~~~~~~~~{\rm tanh}\left(\frac{t}{b T_{0i}}\right)\approx{\rm tanh}\left(\sqrt{\frac{2T}{ T_{0i}}}\right)\approx \sqrt{\frac{2T}{ T_{0i}}},~~~~~~\\
{\bf Assisted: } ~~~~~~~~{\rm tanh}\left(\frac{t}{b T_{0}}\right)\approx{\rm tanh}\left(\sqrt{\frac{2T}{ T_{0}}}\right)\approx \sqrt{\frac{2T}{ T_{0}}}.~~~~~~
\eea
and finally for the usual tachyonic case the potential $V(T)$ can be approximated as:
\be\begin{array}{lll}\label{g26}
 \displaystyle V(T) =\left\{\begin{array}{lll}
                    \displaystyle  
                    \lambda\exp\left(-\frac{T}{T_{0}}\right) \,, &
 \mbox{\small {\bf for {Single }}}  \\ 
 \displaystyle  
 \sum^{N}_{i=1}V(T_{i})=\sum^{N}_{i=1}\lambda_{i}\exp\left(-\frac{T_{i}}{T_{0i}}\right) \,,~ &
 \mbox{\small {\bf for {Multi }}} \\  
 \displaystyle N\lambda\exp\left(-\frac{T}{T_{0}}\right)  \,,~ &
 \mbox{\small {\bf for {Assisted }}}
          \end{array}
\right.
\end{array}\ee
and for the generalized case we have:
\be\begin{array}{lll}\label{g27}
 \displaystyle V(T) =\left\{\begin{array}{lll}
                    \displaystyle  
                    \lambda\exp\left(-\frac{2qT}{T_{0}}\right) \,, &
 \mbox{\small {\bf for {Single }}}  \\ 
 \displaystyle  
 \sum^{N}_{i=1}V(T_{i})=\sum^{N}_{i=1}\lambda_{i}\exp\left(-\frac{2qT_{i}}{T_{0i}}\right) \,,~ &
 \mbox{\small {\bf for {Multi }}} \\ 
 \displaystyle N\lambda\exp\left(-\frac{2qT}{T_{0}}\right)  \,,~ &
 \mbox{\small {\bf for {Assisted }}}.
          \end{array}
\right.
\end{array}\ee
where the behavior of these types of potentials has been elaborately mentioned in the earlier section.

 \item
 At $T>>T_{0}$ and $T_{i}>>T_{0i}$ limiting case one can use the following approximation for the usual tachyonic case:
 \bea \ln\left[{\rm cosh}\left(\frac{t}{b T_0}\right)\right]&>>&1,\\
  \ln\left[{\rm cosh}\left(\frac{t}{b T_{0i}}\right)\right]&>>&1.\eea
 Using this approximation one can use the following expansion:
 \bea 
 \ln\left[{\rm cosh}\left(\frac{t}{b T_0}\right)\right]&\approx& \ln\left[\frac{1}{2}{\exp}\left(\frac{t}{b T_0}\right)\right]\approx\frac{t}{b T_0},\\
 \ln\left[{\rm cosh}\left(\frac{t}{b T_{0i}}\right)\right]&\approx& \ln\left[\frac{1}{2}{\exp}\left(\frac{t}{b T_{0i}}\right)\right]\approx\frac{t}{b T_{0i}}.
 \eea
Hence using the solution obtained for tachyonic field as stated in Eq~(\ref{g14}) we get:
\be\begin{array}{lll}\label{g28}
 \displaystyle T(t) =\left\{\begin{array}{lll}
                    \displaystyle  
                    \frac{t}{b} \,,~~~~ &
 \mbox{\small {\bf for {Single }}}  \\ 
 \displaystyle  
 \sum^{N}_{i=1}T_{i}(t)=\sum^{N}_{i=1} \frac{t}{2b}=\frac{Nt}{b}  \,,~~~ &
 \mbox{\small {\bf for {Multi }}} \\ 
 \displaystyle  \frac{Nt}{b}   \,,~~~ &
 \mbox{\small {\bf for {Assisted }}}.
          \end{array}
\right.
\end{array}\ee
and by inverting Eq~(\ref{g28}) the associated time scale can be computed as:
\be\begin{array}{lll}\label{g29}
 \displaystyle t =\left\{\begin{array}{lll}
                    \displaystyle  
                    b T \,,~~~~ &
 \mbox{\small {\bf for {Single }}}  \\ 
 \displaystyle  
 \sum^{N}_{i=1}b T_{i} \,,~~~ &
 \mbox{\small {\bf for {Multi }}} \\ 
 \displaystyle Nb T  \,,~~~ &
 \mbox{\small {\bf for {Assisted }}}.
          \end{array}
\right.
\end{array}\ee
Consequently we have:
\bea {\bf Single: } ~~~~~~~~{\rm tanh}\left(\frac{t}{b T_{0}}\right)\approx 1,~~~~~~\\ 
{\bf Multi: } ~~~~~~~~{\rm tanh}\left(\frac{t}{b T_{0i}}\right)\approx 1,~~~~~~\\
{\bf Assisted: } ~~~~~~~~{\rm tanh}\left(\frac{t}{b T_{0}}\right)\approx 1.~~~~~~
\eea
and finally for the usual tachyonic case the potential $V(T)$ can be approximated as:
\be\begin{array}{lll}\label{g26}
 \displaystyle V(T) =\left\{\begin{array}{lll}
                    \displaystyle  
                    \frac{4M^{2}_{p}}{3 b^2 T^{2}_{0}}\left(\frac{T_{0}}{T}\right)^2 \exp\left(-\frac{T}{T_{0}}\right) \,, &
 \mbox{\small {\bf for {Single }}}  \\ 
 \displaystyle  
 \sum^{N}_{i=1}V(T_{i})=\sum^{N}_{i=1}\frac{4M^{2}_{p}}{3 b^2 T^{2}_{0i}}\left(\frac{T_{0i}}{T_{i}}\right)^2\exp\left(-\frac{T_{i}}{T_{0i}}\right) \,,~ &
 \mbox{\small {\bf for {Multi }}} \\ 
 \displaystyle \frac{4M^{2}_{p}N}{3 b^2 T^{2}_{0}}\left(\frac{T_{0}}{T}\right)^2\exp\left(-\frac{T}{T_{0}}\right)  \,,~ &
 \mbox{\small {\bf for {Assisted }}}
          \end{array}
\right.
\end{array}\ee
and for the generalized case we have:
\be\begin{array}{lll}\label{g27}
 \displaystyle V(T) =\left\{\begin{array}{lll}
                    \displaystyle  
                    \frac{4M^{2}_{p}}{3 b^2 T^{2}_{0}}\left(\frac{T_{0}}{T}\right)^2\exp\left(-\frac{2qT}{T_{0}}\right) \,, &
 \mbox{\small {\bf for {Single }}}  \\ 
 \displaystyle  
 \sum^{N}_{i=1}V(T_{i})=\sum^{N}_{i=1}\frac{4M^{2}_{p}}{3 b^2 T^{2}_{0i}}\left(\frac{T_{0i}}{T_{i}}\right)^2\exp\left(-\frac{2qT_{i}}{T_{0i}}\right) \,,~ &
 \mbox{\small {\bf for {Multi }}} \\ 
 \displaystyle \frac{4M^{2}_{p}N}{3 b^2 T^{2}_{0}}\left(\frac{T_{0}}{T}\right)^2\exp\left(-\frac{2qT}{T_{0}}\right)  \,,~ &
 \mbox{\small {\bf for {Assisted }}}.
          \end{array}
\right.
\end{array}\ee
where the scale of inflation for the usual tachyonic case is fixed by:
\be\begin{array}{lll}\label{g28}
 \displaystyle  V^{1/4}_{inf}\propto \left\{\begin{array}{lll}
                    \displaystyle  
                    \lambda^{1/4} \,, &
 \mbox{\small {\bf for {Single }}}  \\ 
 \displaystyle  
 \left\{\sum^{N}_{i=1}\lambda_{i}\right\}^{1/4} \,,~ &
 \mbox{\small {\bf for {Multi }}} \\ 
 \displaystyle (N\lambda)^{1/4}  \,,~ &
 \mbox{\small {\bf for {Assisted }}}.
          \end{array}
\right.
\end{array}\ee
and for the generalized case it is fixed by:
\be\begin{array}{lll}\label{g29}
 \displaystyle  V^{1/4}_{inf}\propto \left\{\begin{array}{lll}
                    \displaystyle  
                    \left(\frac{4M^{2}_{p}}{3 b^2 T^{2}_{0}}\right)^{1/4} \,, &
 \mbox{\small {\bf for {Single }}}  \\ 
 \displaystyle  
 \left\{\sum^{N}_{i=1}\frac{4M^{2}_{p}}{3 b^2 T^{2}_{0i}}\right\}^{1/4} \,,~ &
 \mbox{\small {\bf for {Multi }}} \\ 
 \displaystyle \left(\frac{4M^{2}_{p}N}{3 b^2 T^{2}_{0}}\right)^{1/4}  \,,~ &
 \mbox{\small {\bf for {Assisted }}}.
          \end{array}
\right.
\end{array}\ee
Here the potential satisfy the following criteria:
\begin{itemize}
 \item  At $T=0$ for single field tachyonic potential:\be V(T=0)\rightarrow \infty \ee
 and for multi tachyonic and assisted case we have:
 \bea V_{E}(T=0)&\rightarrow& \infty.\eea
 
 \item At $T=T_0$ for single field tachyonic potential:\be V(T=T_0)=\frac{4M^{2}_{p}}{3e b^2 T^{2}_{0}} \ee
 and also for multi tachyonic and assisted case we have:
 \bea V_{E}(T=T_0)&=&\sum^{N}_{i=1}\frac{4M^{2}_{p}}{3e b^2 T^{2}_{0i}},\\
 V_{E}(T=T_0)&=&\frac{4M^{2}_{p}N}{3e b^2 T^{2}_{0}}.\eea
 For the generalized case one can repeat the same computation with the following redefinition of the 
 $b$ parameter of tachyonic field theory:
 \bea b^{-2}\exp(-2q)\rightarrow b^{-2}.\eea
 
 \item For single field tachyonic potential:\bea\label{r1} V^{'}(T)&=&-\frac{4 M^{2}_{p}}{3 b^2  T^3}\exp\left(-\frac{T}{T_{0}}\right)
 \left(2
 +\frac{T}{T_{0}}\right),\\
 \label{r2}V^{''}(T)&=&\frac{4 M^{2}_{p}}{3 b^2 T^4}\exp\left(-\frac{T}{T_{0}}\right)
 \left\{6+4\left(\frac{T}{T_{0}}\right)
 +\left(\frac{T}{T_{0}}\right)^2\right\}.\eea
 Now to find the extrema of the potential we substitute \be V^{'}(T)=0\ee which give rise to the following solution for the the tachyonic field:
 \bea T=-2T_{0},~~ \infty.\eea
 Further substituting the solutions for tachyonic field in Eq~(\ref{r2}) we get:
 \bea\label{r4}V^{''}\left(T=-2T_{0}\right)&=& \frac{e^2 M^{2}_{p}}{6 b^2 T^4_{0}},\\
 \label{r6}V^{''}(T\rightarrow \infty)&\rightarrow& 0
 \eea
 and also additionally for $T=T_{0}$ we have:
 \bea\label{r7}V^{''}(T=T_0)&=& \frac{44 M^{2}_{p}}{3e  b^2 T^4_{0}},
 \eea
 and at these points the value of the potential is computed as:
 \bea\label{r8} V(T\rightarrow \infty)&\rightarrow& 0,\\
 \label{r9} V(T=-2T_{0})&=& \frac{e^2 M^{2}_{p}}{3 b^2 T^2_{0}},\\
 \label{r10} V(T=T_{0})&=& \frac{4 M^{2}_{p}}{3 b^2 T^2_{0}}.
 \eea
 It is important to note for single tachyonic case that:
 \begin{itemize}
 \item for $b^2>0$, $V^{''}(T=-2T_{0},~~T_{0} )>0$ i.e. we get maxima on the potential.
 
 \item for $b^2<0$, $V^{''}(T=-2T_{0},~~T_{0} )<0$ i.e. we get minima on the potential.
 \end{itemize}
 and for the assisted case the results are same, provided following replacement occurs:
 \bea \lambda\rightarrow N\lambda.\eea
 and finally for multi tachyonic case we have:
\bea\label{ur1} V^{'}(T_{i})&=&-\frac{4 M^{2}_{p}}{3 b^2  T^3_{i}}\exp\left(-\frac{T_{i}}{T_{0i}}\right)
 \left(2
 +\frac{T_{i}}{T_{0i}}\right),\\
 \label{ur2}V^{''}(T_{i})&=&\frac{4 M^{2}_{p}}{3 b^2 T^4_{i}}\exp\left(-\frac{T_{i}}{T_{0i}}\right)
 \left\{6+4\left(\frac{T_{i}}{T_{0i}}\right)
 +\left(\frac{T_{i}}{T_{0i}}\right)^2\right\}.\eea
 Now to find the extrema of the potential we substitute: \be V^{'}(T_{j})=0\forall j=1,2,...,N\ee which give rise to the follwing solutions
 for the the $j$th tachyonic field:
 \bea T_{j}=-2T_{0j},~ \infty.\eea
 Further substituting the solutions for tachyonic field in Eq~(\ref{ur2}) we get:
 \bea\label{ur3}V^{''}\left(T_{j}=-2T_{0j}\right)&=& \frac{e^2 M^{2}_{p}}{6 b^2 T^4_{0j}},\\
 \label{ur4}V^{''}(T_{j}\rightarrow \infty)&\rightarrow& 0
 \eea
 Additionally for the point $T_{j}=T_{0j}$ we have:
 \bea\label{ur5}V^{''}(T_{j}=T_{0j})&=& \frac{44 M^{2}_{p}}{3e  b^2 T^4_{0j}},
 \eea
 and at these points the value of the total effective potential is computed as:
 \bea\label{ur6}  V^{(1)}_{E}=\sum^{N}_{j=1}V(T_{j}\rightarrow \infty)&\rightarrow& 0,\\
 \label{ur7} V^{(2)}_{E}=\sum^{N}_{j=1}V(T_{j}=-2T_{0j})&=& \sum^{N}_{j=1}\frac{e^2 M^{2}_{p}}{3 b^2 T^2_{0j}},\\
 \label{ur8} V^{(3)}_{E}= \sum^{N}_{j=1}V(T_{j}=T_{0j})&=& \frac{4N M^{2}_{p}}{3 b^2 T^2_{0}}.
 \eea
 It is important to note for multi tachyonic case that:
 \begin{itemize}
 \item for $b^2>0$, $V^{''}(T=-2T_{0j},~~T_{0j} )>0$ i.e. we get maxima on the potential $V_{j}(T)$ as well in $V_{E}(T)$.
 
 \item for $b^2<0$, $V^{''}(T=-2T_{0j},~~T_{0j} )<0$ i.e. we get minima on the potential $V_{j}(T)$ as well in $V_{E}(T)$.
 
 \end{itemize}
\end{itemize}
\end{itemize}

Next using Eq~(\ref{ku1}) Hubble parameter can be expressed in terms of the usual tachyonic potential $V(T)$ as:
\be\begin{array}{lll}\label{g30}
 \displaystyle  H^{2}(t)= \left\{\begin{array}{lll}
                    \displaystyle  
                    \frac{\lambda}{3M^{2}_{p}}\left[\frac{1}{1+\frac{\sqrt{3\lambda}bT_{0}}{2M_{p}}\left\{\frac{t}{b T_{0}}-{\rm tanh}\left(\frac{t}{b T_0}\right)\right\}}\right]^2 \,, &
 \mbox{\small {\bf for {Single }}}  \\ 
 \displaystyle  
 \sum^{N}_{i=1}H_{i}(t)=\sum^{N}_{i=1}\frac{\lambda_{i}}{3M^{2}_{p}}\left[\frac{1}{1+\frac{\sqrt{3\lambda}bT_{0i}}{2M_{p}}\left\{\frac{t}{b T_{0i}}-{\rm tanh}\left(\frac{t}
 {b T_{0i}}\right)\right\}}\right]^2 \,,~ &
 \mbox{\small {\bf for {Multi }}} \\  
 \displaystyle \frac{N\lambda}{3M^{2}_{p}}\left[\frac{1}{1+\frac{\sqrt{3\lambda}bT_{0}}{2M_{p}}\left\{\frac{t}{b T_{0}}-{\rm tanh}\left(\frac{t}{b T_0}\right)\right\}}\right]^2  \,,~ &
 \mbox{\small {\bf for {Assisted }}}
          \end{array}
\right.
\end{array}\ee
and for the generalized case we have:
\be\begin{array}{lll}\label{g31}
 \displaystyle  H^{2}(t)= \left\{\begin{array}{lll}
                    \displaystyle  
                    \frac{\lambda}{3M^{2}_{p}\left[{\rm cosh}\left(\frac{t}{b T_0}\right)\right]^{2(2q-1)}}\left[\frac{1}{1+\frac{\sqrt{3\lambda}bT_{0}}{2M_{p}}\left\{\frac{t}{b T_{0}}-{\rm tanh}\left(\frac{t}{b T_0}\right)\right\}}\right]^2 \,, &
 \mbox{\small {\bf for {Single }}}  \\ 
 \displaystyle  
 \sum^{N}_{i=1}\frac{\lambda_{i}}{3M^{2}_{p}\left[{\rm cosh}\left(\frac{t}{b T_{0i}}\right)\right]^{2(2q-1)}}\left[\frac{1}{1+\frac{\sqrt{3\lambda}bT_{0i}}{2M_{p}}\left\{\frac{t}{b T_{0i}}-{\rm tanh}\left(\frac{t}
 {b T_{0i}}\right)\right\}}\right]^2 \,,~ &
 \mbox{\small {\bf for {Multi }}} \\ 
 \displaystyle \frac{N\lambda}{3M^{2}_{p}\left[{\rm cosh}\left(\frac{t}{b T_0}\right)\right]^{2(2q-1)}}\left[\frac{1}{1+\frac{\sqrt{3\lambda}bT_{0}}{2M_{p}}\left\{\frac{t}{b T_{0}}-{\rm tanh}\left(\frac{t}{b T_0}\right)\right\}}\right]^2  \,,~ &
 \mbox{\small {\bf for {Assisted }}}
          \end{array}
\right.
\end{array}\ee
Let us study the limiting situation from the expression obtained for Hubble parameter explicitly:
\begin{itemize}
 \item In the $T<<T_{0}$ and $T_{j}<<T_{0j}$ limiting situation the Hubble parameter can be approximated for the usual tachyonic case as:
 \be\begin{array}{lll}\label{g32}
 \displaystyle  H^{2}(t)= \left\{\begin{array}{lll}
                    \displaystyle  
                    \frac{\lambda}{3M^{2}_{p}}\,, &
 \mbox{\small {\bf for {Single }}}  \\ 
 \displaystyle  
 \sum^{N}_{i=1}\frac{\lambda_{i}}{3M^{2}_{p}} \,,~ &
 \mbox{\small {\bf for {Multi }}} \\ 
 \displaystyle \frac{N\lambda}{3M^{2}_{p}}  \,,~ &
 \mbox{\small {\bf for {Assisted }}}
          \end{array}
\right.
\end{array}\ee
and for the generalized case we have:
\be\begin{array}{lll}\label{g33}
 \displaystyle  H^{2}(t)= \left\{\begin{array}{lll}
                    \displaystyle  
                    \frac{\lambda}{3M^{2}_{p}\left[{\rm cosh}\left(\frac{t}{b T_0}\right)\right]^{2(2q-1)}}\,, &
 \mbox{\small {\bf for {Single }}}  \\ 
 \displaystyle  
 \sum^{N}_{i=1}\frac{\lambda_{i}}{3M^{2}_{p}\left[{\rm cosh}\left(\frac{t}{b T_{0i}}\right)\right]^{2(2q-1)}} \,,~ &
 \mbox{\small {\bf for {Multi }}} \\  
 \displaystyle \frac{N\lambda}{3M^{2}_{p}\left[{\rm cosh}\left(\frac{t}{b T_0}\right)\right]^{2(2q-1)}}  \,,~ &
 \mbox{\small {\bf for {Assisted }}}.
          \end{array}
\right.
\end{array}\ee
 In this limiting situation the solution for the scale factor $a(t)$ from Eq~(\ref{g32}) can be expressed as:
 \be\begin{array}{lll}\label{g33}
 \displaystyle  a(t)= \left\{\begin{array}{lll}
                    \displaystyle  
                    a_{inf}\exp\left[\frac{\sqrt{\lambda}}{\sqrt{3}M_{p}}\left(t-t_{inf}\right)\right]\,, &
 \mbox{\small {\bf for {Single }}}  \\ 
 \displaystyle  
 \sum^{N}_{i=1}a_{inf}\exp\left[\sqrt{\frac{\lambda_{i}}{3M^{2}_{p}}}\left(t-t_{inf}\right)\right] \,,~ &
 \mbox{\small {\bf for {Multi }}} \\  
 \displaystyle a_{inf}\exp\left[\frac{\sqrt{N\lambda}}{\sqrt{3}M_{p}}\left(t-t_{inf}\right)\right] \,,~ &
 \mbox{\small {\bf for {Assisted }}}
          \end{array}
\right.
\end{array}\ee
which is exactly the de-Sitter solution required for inflation. Here at the inflationary time scale $t=t_{inf}$ the scale factor is given by:
\be a_{inf}=a(t=t_{inf}). \ee
On the other hand for the generalized case we have the following solution for the scale factor $a(t)$:
\be\begin{array}{lll}\label{g34}\small
 \displaystyle  a(t)= \left\{\begin{array}{lll}
                    \displaystyle  
                    a_{inf}\exp\left[\frac{\sqrt{\lambda}b T_{0}}{\sqrt{3}M_{p}}\left\{{\rm sinh}\left(\frac{t}{b T_0}\right) 
                     \, _2F_1\left[\frac{1}{2},q;\frac{3}{2};-\sinh ^2\left(\frac{t}{b T_0}\right)\right]\right\}^{t}_{t_{inf}}\right]\,, &
 \mbox{\small {\bf for {Single }}}  \\ 
 \displaystyle  
\sum^{N}_{i=1} a_{inf}\exp\left[\frac{\sqrt{\lambda_{i}}b T_{0i}}{\sqrt{3}M_{p}}\left\{{\rm sinh}\left(\frac{t}{b T_{0i}}\right) 
                     \, _2F_1\left[\frac{1}{2},q;\frac{3}{2};-\sinh ^2\left(\frac{t}{b T_{0i}}\right)\right]\right\}^{t}_{t_{inf}}\right] \,,~ &
 \mbox{\small {\bf for {Multi }}} \\  
 \displaystyle a_{inf}\exp\left[\frac{\sqrt{N\lambda}b T_{0}}{\sqrt{3}M_{p}}\left\{{\rm sinh}\left(\frac{t}{b T_0}\right) 
                     \, _2F_1\left[\frac{1}{2},q;\frac{3}{2};-\sinh ^2\left(\frac{t}{b T_0}\right)\right]\right\}^{t}_{t_{inf}}\right] \,,~ &
 \mbox{\small {\bf for {Assisted }}}
          \end{array}
\right.
\end{array}\ee
which replicates the behaviour of quasi de-Sitter solution during inflation. For $q=1/2$ case it exactly follows the de-Sitter behavior.
 \item In the $T>>T_{0}$ and $T_{j}>>T_{0j}$ limiting situation the Hubble parameter can be approximated for the usual tachyonic case as:
 \be\begin{array}{lll}\label{g35}
 \displaystyle  H^{2}(t)= \left\{\begin{array}{lll}
                    \displaystyle  
                    \frac{4}{9t^2}\,, &
 \mbox{\small {\bf for {Single }}}  \\  
 \displaystyle  
 \sum^{N}_{i=1}\frac{4}{9t^2} \,,~ &
 \mbox{\small {\bf for {Multi }}} \\  
 \displaystyle  \frac{4N}{9t^2}  \,,~ &
 \mbox{\small {\bf for {Assisted }}}
          \end{array}
\right.
\end{array}\ee
and for the generalized case we have:
\be\begin{array}{lll}\label{g36}
 \displaystyle  H^{2}(t)= \left\{\begin{array}{lll}
                    \displaystyle  
                    \frac{4}{9t^2\left[\frac{1}{2}\exp\left(\frac{t}{b T_0}\right)\right]^{2(2q-1)}}\,, &
 \mbox{\small {\bf for {Single }}}  \\  
 \displaystyle  
 \sum^{N}_{i=1}\frac{4}{9t^2\left[\frac{1}{2}\exp\left(\frac{t}{b T_{0i}}\right)\right]^{2(2q-1)}} \,,~ &
 \mbox{\small {\bf for {Multi }}} \\ 
 \displaystyle  \frac{4N}{9t^2\left[\frac{1}{2}\exp\left(\frac{t}{b T_0}\right)\right]^{2(2q-1)}}  \,,~ &
 \mbox{\small {\bf for {Assisted }}}
          \end{array}
\right.
\end{array}\ee
 In this limiting situation the solution for the scale factor $a(t)$ from Eq~(\ref{g35}) can be expressed as:
 \be\begin{array}{lll}\label{g37}
 \displaystyle  a(t)= \left\{\begin{array}{lll}
                    \displaystyle  
                    a_{dust}\left(\frac{t}{t_{dust}}\right)^{2/3}\,, &
 \mbox{\small {\bf for {Single }}}  \\  
 \displaystyle  
 \sum^{N}_{i=1}a_{dust}\left(\frac{t}{t_{dust}}\right)^{2/3} \,,~ &
 \mbox{\small {\bf for {Multi }}} \\  
 \displaystyle a_{dust}\left(\frac{t}{t_{dust}}\right)^{2/3} \,,~ &
 \mbox{\small {\bf for {Assisted }}}
          \end{array}
\right.
\end{array}\ee
which is exactly the dust like solution required for the formation of dark matter. Here at the inflationary time scale $t=t_{dust}$ the scale factor is given by:
\be a_{dust}=a(t=t_{dust}). \ee
On the other hand for the generalized case we have the following solution for the scale factor $a(t)$:
\be\begin{array}{lll}\label{g38}
 \displaystyle  a(t)= \left\{\begin{array}{lll}
                    \displaystyle  
                    a_{dust}\exp\left[\frac{4^q}{3}   \left\{\text{Ei}\left[\left(\frac{t}{b T_{0}}\right)-2q\right]-\text{Ei}\left[\left(\frac{t_{dust}}{b T_{0}}\right)-2q\right]\right\}\right]\,, &
 \mbox{\small {\bf for {Single }}}  \\  
 \displaystyle  
\sum^{N}_{i=1} a_{dust}\exp\left[\frac{4^q}{3}   \left\{\text{Ei}\left[\left(\frac{t}{b T_{0i}}\right)-2q\right]-\text{Ei}\left[\left(\frac{t_{dust}}{b T_{0i}}\right)-2q\right]\right\}\right] \,,~ &
 \mbox{\small {\bf for {Multi }}} \\ 
 \displaystyle a_{dust}\exp\left[\frac{N4^q}{3}   \left\{\text{Ei}\left[\left(\frac{t}{b T_{0}}\right)-2q\right]-\text{Ei}\left[\left(\frac{t_{dust}}{b T_{0}}\right)-2q\right]\right\}\right] \,,~ &
 \mbox{\small {\bf for {Assisted }}}
          \end{array}
\right.
\end{array}\ee
which replicates the behaviour of quasi dust like solution. For $q=1/2$ case it exactly follows the dust like behavior.
 
\end{itemize}

\section{Inflationary paradigm from GTachyon}
\label{aa5}
It is a very well known fact that during cosmological inflation, quantum fluctuations are
stretched on the scales larger than the size of the horizon. Consequently, they are frozen until they re-enter
the horizon at the end of inflationary phase. In the present context a single tachyonic field drives inflationary paradigm, which finally giving rise to 
the large scale
perturbations with a quasi scale invariant primordial power spectrum corresponding to the scalar and tensor modes.
Deviations from the scale invariance in the primordial power spectrum can be measured in terms of the slow-roll parameters
which we will explicitly discuss in the next subsections. By detailed computation we explicitly
show that at lowest order of the primordial spectrum the scalar perturbations is exactly
same as that obtained for usual single field inflationary setup. For completeness the next to leading order corrections to the cosmological perturbations 
are also derived, which finally giving rise to the sufficient
change in the cosmological consistency relations with respect to the results obtained for the usual single field inflationary setup. Hence we 
apply all the derived results for the tachyonic inflationary models as explicitly mentioned in the previous section and study the CMB constraints 
by applying the recent Planck 2015 data.

\subsection{Computation for Single field inflation}
\label{aa5a}
\subsubsection{Condition for inflation}

For single field tachyonic inflation, the prime condition for inflation is given by:
\bea
\dot{H}+H^{2}&=&\left(\frac{\ddot{a}}{{a}}\right)=-\frac{(\rho+3p)}{6 M^2_p}>0
\eea
which can be re-expressed in terms of the following constraint condition in the context of single field tachyonic inflation:
\be\begin{array}{lll}\label{g39}
 \displaystyle \frac{V(T)}{3M^{2}_{p}\sqrt{1-\alpha^{'}\dot{T}^2}}\left(1-\frac{3}{2}\alpha^{'}\dot{T}^2\right)>0.
\end{array}\ee
Here Eq~(\ref{g39}) implies that to satisfy inflationary constraints in the slow-roll regime the following constraint always holds good:
\bea \label{wa1}\dot{T}&<& \sqrt{\frac{2}{3\alpha^{'}}},\\ 
\label{wa2}\ddot{T}&<&  3H \dot{T}<\sqrt{\frac{6}{\alpha^{'}}}H.\eea
Consequently the field equations are approximated as`:
\be\begin{array}{lll}\label{e13x}
 \displaystyle 
3H\alpha^{'}\dot{T}+\frac{dV(T)}{V(T)dT}\approx 0 \,,
\end{array}\ee
Similarly, in the most generalized case, 
\be\begin{array}{lll}\label{g39}
 \displaystyle \frac{V(T)}{3M^{2}_{p}\left(1-\alpha^{'}\dot{T}^2\right)^{1-q}}\left(1-(1+q)\alpha^{'}\dot{T}^2\right)>0.
\end{array}\ee
Here Eq~(\ref{g39}) implies that to satisfy inflationary constraints in the slow-roll regime the following constraint always holds good:
\bea \dot{T}&<& \sqrt{\frac{1}{\alpha^{'}(1+q)}},\\ 
\ddot{T}&<&  3H \dot{T}<\sqrt{\frac{9}{\alpha^{'}(1+q)}}H.\eea
Consequently the field equations are approximated as`:
\be\begin{array}{lll}\label{e14xx}
 \displaystyle 
6q\alpha^{'}H\dot{T}+\frac{dV(T)}{V(T)dT}\approx 0 \,
\end{array}\ee
Also for both the cases in the slow-roll regime the Friedmann equation is modified as:
\bea\label{qxcc1} H^{2}&\approx& \frac{V(T)}{3M^{2}_{p}}.\eea
Further substituting Eq~(\ref{qxcc1}) in Eq~(\ref{e13x}) and Eq~(\ref{e14xx}) we get:
\bea\label{e13gvx}
 \displaystyle 
\frac{\sqrt{3V(T)}}{M_{p}}\alpha^{'}\dot{T}+\frac{dV(T)}{V(T)dT}\approx 0 \,,\\
\label{e13gvvx} 6q\frac{\sqrt{V(T)}}{\sqrt{3}M_{p}}\alpha^{'}\dot{T}+\frac{dV(T)}{V(T)dT}\approx 0 \,.
\eea
Finally the general solution for both the cases can be expressed in terms of the single field tachyonic potential $V(T)$ as:
\bea\label{e13gvxc1}
 \displaystyle 
\label{bw1}t-t_{i}\approx -\frac{\sqrt{3}\alpha^{'}}{M_{p}}\int^{T}_{T_{i}}dT~\frac{V^{3/2}(T)}{V^{'}(T)} \,,\\
\label{bw2} t-t_{i}\approx -\frac{6q\alpha^{'}}{\sqrt{3}M_{p}}\int^{T}_{T_{i}}dT~\frac{V^{3/2}(T)}{V^{'}(T)}.
\eea
Let us now re-write Eq~(\ref{bw1}) and Eq~(\ref{bw2}), in terms of the string theoretic tachyonic potentials as already mentioned in the last section.
For $q=1/2$ situation we get:
\be\begin{array}{lll}\label{bw3}\footnotesize
 \displaystyle  t-t_{i}\approx \left\{\begin{array}{lll}\tiny
                    \displaystyle  
                    \frac{\alpha^{'}T^{2}_{0}}{2M_{p}}\sqrt{3\lambda} \left\{\ln \left[\frac{\left(\sqrt{\text{sech}\left(\frac{T_{i}}{T_{0}}\right)}+1\right) \left(1-\sqrt{\text{sech}\left(\frac{T}{T_{0}}\right)}\right)}
                    {\left(\sqrt{\text{sech}\left(\frac{T}{T_{0}}\right)}+1\right)\left(1-\sqrt{\text{sech}\left(\frac{T_{i}}{T_{0}}\right)}\right)}\right]\right.\\ \left.
                    \displaystyle +2 \tan ^{-1}\left[\frac{\sqrt{\text{sech}\left(\frac{T_{i}}{T_{0}}\right)}-\sqrt{\text{sech}\left(\frac{T}{T_{0}}\right)}}{1+\sqrt{\text{sech}\left(\frac{T_{i}}{T_{0}}\right)\text{sech}\left(\frac{T}{T_{0}}\right)}}
                    \right]\right\}\,, &
 \mbox{\small {\bf for {Model~1 }}}  \\ 
 \displaystyle  
                    -\frac{\alpha^{'}T^{2}_{0}}{2M_{p}}\sqrt{3\lambda}~ \ln \left(\frac{\ln \left(\frac{T}{T_{0}}\right)\left(\ln \left(\frac{T_{i}}{T_{0}}\right)+1\right)}{\ln \left(\frac{T_{i}}{T_{0}}\right)
                    \left(\ln \left(\frac{T}{T_{0}}\right)+1\right)}\right)
                   \,, &
 \mbox{\small {\bf for {Model~2 }}}  \\  
 \displaystyle  
                    -\frac{2\alpha^{'} T^2_{0}}{M_{p}}\sqrt{3\lambda} \left\{\exp\left[-\frac{T}{2T_{0}}\right]-\exp\left[-\frac{T_{i}}{2T_{0}}\right]\right\}\,, &
 \mbox{\small {\bf for {Model~3 }}}  \\ 
 \displaystyle  
                    -\frac{\alpha^{'}T^{2}_{0}}{4M_{p}}\sqrt{3\lambda}\left\{\text{Ei}\left(-\frac{T^2_{i}}{2 T^2_{0}}\right) 
                    -\text{Ei}\left(-\frac{T^2}{2 T^2_{0}}\right)\right\}\,, &
 \mbox{\small {\bf for {Model~4 }}}  \\  
 \displaystyle  
                    \frac{\alpha^{'}T^{2}_{0}}{8 M_{p}} \sqrt{3\lambda}  \left\{\ln\left(\frac{\sqrt{T^4_{0}+T^4}+T^2}{\sqrt{T^4_{0}+T^4_{i}}+T^2_{i}}\right)
                    +\sqrt{1+\left(\frac{T_{0}}{T_{i}}\right)^4}-\sqrt{1+\left(\frac{T_{0}}{T}\right)^4}\right\}\,, &
 \mbox{\small {\bf for {Model~5 }}}
          \end{array}
\right.
\end{array}\ee
and for any arbitrary $q$ we get the following generalized result:
\be\begin{array}{lll}\label{bw4}\footnotesize
 \displaystyle  t-t_{i}\approx \left\{\begin{array}{lll}\tiny
                    \displaystyle  
                    \frac{3q\alpha^{'}T^{2}_{0}}{M_{p}}\sqrt{\frac{\lambda}{3}} \left\{\ln \left[\frac{\left(\sqrt{\text{sech}\left(\frac{T_{i}}{T_{0}}\right)}+1\right) \left(1-\sqrt{\text{sech}\left(\frac{T}{T_{0}}\right)}\right)}
                    {\left(\sqrt{\text{sech}\left(\frac{T}{T_{0}}\right)}+1\right)\left(1-\sqrt{\text{sech}\left(\frac{T_{i}}{T_{0}}\right)}\right)}\right]\right.\\ \left.
                    \displaystyle ~~~~~~~~~~~~~~~~~~+2 \tan ^{-1}\left[\frac{\sqrt{\text{sech}\left(\frac{T_{i}}{T_{0}}\right)}-\sqrt{\text{sech}\left(\frac{T}{T_{0}}\right)}}{1+\sqrt{\text{sech}\left(\frac{T_{i}}{T_{0}}\right)\text{sech}\left(\frac{T}{T_{0}}\right)}}
                    \right]\right\}\,, &
 \mbox{\small {\bf for {Model~1 }}}  \\ 
 \displaystyle  
                    -\frac{3q\alpha^{'}T^{2}_{0}}{M_{p}}\sqrt{\frac{\lambda}{3}}~ \ln \left(\frac{\ln \left(\frac{T}{T_{0}}\right)\left(\ln \left(\frac{T_{i}}{T_{0}}\right)+1\right)}{\ln \left(\frac{T_{i}}{T_{0}}\right)
                    \left(\ln \left(\frac{T}{T_{0}}\right)+1\right)}\right)
                   \,, &
 \mbox{\small {\bf for {Model~2 }}}  \\  
 \displaystyle  
                    -\frac{12q\alpha^{'} T^2_{0}}{M_{p}}\sqrt{\frac{\lambda}{3}} \left\{\exp\left[-\frac{T}{2T_{0}}\right]-\exp\left[-\frac{T_{i}}{2T_{0}}\right]\right\}\,, &
 \mbox{\small {\bf for {Model~3 }}}  \\  
 \displaystyle  
                    -\frac{3q\alpha^{'}T^{2}_{0}}{2M_{p}}\sqrt{\frac{\lambda}{3}}\left\{\text{Ei}\left(-\frac{T^2_{i}}{2 T^2_{0}}\right) 
                    -\text{Ei}\left(-\frac{T^2}{2 T^2_{0}}\right)\right\}\,, &
 \mbox{\small {\bf for {Model~4 }}}  \\  
 \displaystyle  
                    \frac{3q\alpha^{'}T^{2}_{0}}{4 M_{p}} \sqrt{\frac{\lambda}{3}}  \left\{\ln\left(\frac{\sqrt{T^4_{0}+T^4}+T^2}{\sqrt{T^4_{0}+T^4_{i}}+T^2_{i}}\right)
                   +\sqrt{1+\left(\frac{T_{0}}{T_{i}}\right)^4}-\sqrt{1+\left(\frac{T_{0}}{T}\right)^4}\right\}\,. &
 \mbox{\small {\bf for {Model~5 }}}
          \end{array}
\right.
\end{array}\ee
Further using Eq~(\ref{bw3}), Eq~(\ref{bw4}) and Eq~(\ref{qxcc1}) we get the following solution for the scale factor 
in terms of the tachyonic field for usual $q=1/2$ and for generalized value of $q$ as:
\be\begin{array}{lll}\label{scss1}
 \displaystyle  a= a_{i}\times \left\{\begin{array}{lll}
                    \displaystyle  
                    \exp\left[-\frac{\alpha^{'}}{M^2_p}\int^{T}_{T_{i}}dT~\frac{V^{2}(T)}{V^{'}(T)}\right]\,,~~~~~~ &
 \mbox{\small {\bf for {$q=1/2$ }}}  \\  
 \displaystyle   
                    \exp\left[-\frac{2q\alpha^{'}}{M^2_p}\int^{T}_{T_{i}}dT~\frac{V^{2}(T)}{V^{'}(T)}\right]\,.~~~~~~ &
 \mbox{\small {\bf for {~any~arbitrary~ $q$ }}} 
          \end{array}
\right.
\end{array}\ee
Finally re-writing Eq~(\ref{scss1}), in terms of the string theoretic tachyonic potentials as already mentioned in the last section
for $q=1/2$ we get: 
\be\begin{array}{lll}\label{bwst3}\tiny
 \displaystyle  a\approx a_{i}\times\left\{\begin{array}{lll}
                    \displaystyle  
                    \exp\left[\frac{\alpha^{'}T^{2}_{0}\lambda}{M^{2}_{p}}\ln \left(\frac{\tanh \left(\frac{T}{2 T_{0}}\right)}{\tanh \left(\frac{T_{i}}{2 T_{0}}\right)}\right)\right]\,, &
 \mbox{\small {\bf for {Model~1 }}}  \\ 
 \displaystyle  
                     \exp\left[-\frac{\alpha^{'}T^{2}_{0}\lambda}{2M^{2}_{p}}~ \ln \left(\frac{\ln \left(\frac{T}{T_{0}}\right)\left(\ln \left(\frac{T_{i}}{T_{0}}\right)+1\right)}{\ln \left(\frac{T_{i}}{T_{0}}\right)
                    \left(\ln \left(\frac{T}{T_{0}}\right)+1\right)}\right)\right]
                   \,, &
 \mbox{\small {\bf for {Model~2 }}}  \\ 
 \displaystyle  
                   \exp\left[\frac{\alpha^{'}T^2_{0}\lambda}{M^2_{p}} \left(\exp\left[-\frac{T_{i}}{T_{0}}\right]-\exp\left[-\frac{T}{T_{0}}\right]\right)\right]\,, &
 \mbox{\small {\bf for {Model~3 }}}  \\ 
 \displaystyle  
                    \exp\left[-\frac{\alpha^{'}T^{2}_{0}\lambda}{4M^2_{p}}\left\{\text{Ei}\left(-\frac{T^2_{i}}{T^2_{0}}\right)-\text{Ei}\left(-\frac{T^2}{T^2_{0}}\right)\right\}\right]\,, &
 \mbox{\small {\bf for {Model~4 }}}  \\ 
 \displaystyle  
                    \exp\left[-\frac{\alpha^{'}T^4_{0}\lambda}{8M^2_{p}} \left(\frac{1}{T^2}-\frac{1}{T^2_{i}}\right)\right]\,, &
 \mbox{\small {\bf for {Model~5 }}}
          \end{array}
\right.
\end{array}\ee
 and for any arbitrary $q$ we get:
\be\begin{array}{lll}\label{bwst3x}\tiny
 \displaystyle  a\approx a_{i}\times\left\{\begin{array}{lll}
                    \displaystyle  
                    \exp\left[\frac{2q\alpha^{'}T^{2}_{0}\lambda}{3M^{2}_{p}}\ln \left(\frac{\tanh \left(\frac{T}{2 T_{0}}\right)}{\tanh \left(\frac{T_{i}}{2 T_{0}}\right)}\right)\right]\,, &
 \mbox{\small {\bf for {Model~1 }}}  \\ 
 \displaystyle  
                     \exp\left[-\frac{q\alpha^{'}T^{2}_{0}\lambda}{M^{2}_{p}}~ \ln \left(\frac{\ln \left(\frac{T}{T_{0}}\right)\left(\ln \left(\frac{T_{i}}{T_{0}}\right)+1\right)}{\ln \left(\frac{T_{i}}{T_{0}}\right)
                    \left(\ln \left(\frac{T}{T_{0}}\right)+1\right)}\right)\right]
                   \,, &
 \mbox{\small {\bf for {Model~2 }}}  \\  
 \displaystyle  
                   \exp\left[\frac{2q\alpha^{'}T^2_{0}\lambda}{M^2_{p}} \left(\exp\left[-\frac{T_{i}}{T_{0}}\right]-\exp\left[-\frac{T}{T_{0}}\right]\right)\right]\,, &
 \mbox{\small {\bf for {Model~3 }}}  \\ 
 \displaystyle  
                    \exp\left[-\frac{q\alpha^{'}T^{2}_{0}\lambda}{2M^2_{p}}\left\{\text{Ei}\left(-\frac{T^2_{i}}{T^2_{0}}\right)-\text{Ei}\left(-\frac{T^2}{T^2_{0}}\right)\right\}\right]\,, &
 \mbox{\small {\bf for {Model~4 }}}  \\ 
 \displaystyle  
                    \exp\left[-\frac{q\alpha^{'}T^4_{0}\lambda}{4M^2_{p}} \left(\frac{1}{T^2}-\frac{1}{T^2_{i}}\right)\right]\,. &
 \mbox{\small {\bf for {Model~5 }}}
          \end{array}
\right.
\end{array}\ee
Hence using Eq~(\ref{bw3}), Eq~(\ref{bw4}), Eq~(\ref{bwst3}) and Eq~(\ref{bwst3x}) one can study the parametric behaviour of the scale factor $a(t)$ with respect to time $t$ and expected to be as like exact de-Sitter
or quasi de-Sitter solution during inflationary slow-roll phase, as explicitly derived in the previous section.

\subsubsection{Analysis using Slow-roll formalism}
Here our prime objective is to define slow-roll parameters for tachyon inflation in terms of the Hubble parameter and the single field tachyonic inflationary potential. Using the slow-roll 
approximation one can expand various cosmological observables in terms of small dynamical quantities derived from the appropriate derivatives of the Hubble parameter and of the inflationary potential. 
To start with here we use the horizon-flow parameters based on derivatives of
Hubble parameter with respect to the number of e-foldings $N$, defined as:
\bea\label{e-fold}
N(t)&=&\int^{t_{end}}_{t}H(t)~dt,\eea
where $t_{end}$ signifies the end of inflation. Further using Eq~(\ref{e13x}), Eq~(\ref{e14xx}), Eq~(\ref{qxcc1}) and eq~(\ref{e-fold}) we get:
\be\begin{array}{lll}\label{vcv}
 \displaystyle  \frac{dT}{dN}=\frac{\dot{T}}{H}=\left\{\begin{array}{lll}
                    \displaystyle  
                    -\frac{2H^{'}}{3\alpha^{'}H^{3}}\,,~~~~~~ &
 \mbox{\small {\bf for {$q=1/2$ }}}  \\ 
 \displaystyle   
                    -\frac{2H^{'}}{3\alpha^{'}H^{3}}\left(\frac{1-\alpha^{'}(1-2q)\dot{T}^2}{2q}\right)\,.~~~~~~ &
 \mbox{\small {\bf for {~any~arbitrary~ $q$ }}} 
          \end{array}
\right.
\end{array}\ee
where $H^{'}>0$ which makes always $\dot{T}>0$ during inflationary phase.
Further using Eq~(\ref{vcv}) we get the following differential operator identity for tachyonic inflation:
\be\begin{array}{lll}\label{vcv2}
 \displaystyle  \frac{1}{H}\frac{d}{dt}=\frac{d}{dN}=\frac{d}{d\ln k}=\left\{\begin{array}{lll}
                    \displaystyle  
                   -\frac{2H^{'}}{3\alpha^{'}H^{3}}\frac{d}{dT}\,,~~~~~~ &
 \mbox{\small {\bf for {$q=1/2$ }}}  \\  
 \displaystyle   
                -\frac{2H^{'}}{3\alpha^{'}H^{3}}\left(\frac{1-\alpha^{'}(1-2q)\dot{T}^2}{2q}\right)\frac{d}{dT}\,.~~~~~~ &
 \mbox{\small {\bf for {~any~arbitrary~ $q$ }}} 
          \end{array}
\right.
\end{array}\ee
Next we define the following Hubble slow roll parameters:
\bea \label{h1} \epsilon_{0}&=&\frac{H_{\star}}{H},\\
                \epsilon_{i+1}&=&\frac{d\ln|\epsilon_{i}|}{dN},~~~~~~i\geq 1
                \eea
                where $H_{\star}$ be the Hubble parameter at the pivot scale. Further using the differential operator identity as mentioned in Eq~(\ref{vcv2}) we get the following Hubble flow equation for tachyonic inflation for $i\geq 0$ :
   \be\begin{array}{lll}\label{vcv4}
 \displaystyle  \frac{1}{H}\frac{d\epsilon_{i}}{dt}=\frac{d\epsilon_{i}}{dN}=\epsilon_{i+1}\epsilon_{i}=\left\{\begin{array}{lll}
                    \displaystyle  
                   -\frac{2H^{'}}{3\alpha^{'}H^{3}}\frac{d\epsilon_{i}}{dT}\,,~~~~~~ &
 \mbox{\small {\bf for {$q=1/2$ }}}  \\ 
 \displaystyle   
                -\frac{2H^{'}}{3\alpha^{'}H^{3}}\left(\frac{1-\alpha^{'}(1-2q)\dot{T}^2}{2q}\right)\frac{d\epsilon_{i}}{dT}\,.~~~~~~ &
 \mbox{\small {\bf for {~any~arbitrary~ $q$ }}} 
          \end{array}
\right.
\end{array}\ee             
For realistic estimate from the single field tachyonic inflationary model substituting the free index $i$ to $i=0,1,2$ in Eq~(\ref{h1}) and Eq~(\ref{vcv4}) we get the contributions from the first three Hubble slow-roll parameter,
which can be depicted as:
\bea\label{bp1}
 \displaystyle  \epsilon_{1}=\frac{d\ln|\epsilon_{0}|}{dN}&=&-\frac{\dot{H}}{H^2}=\left\{\begin{array}{lll}
                    \displaystyle  
                   \frac{2}{3\alpha^{'}}\left(\frac{H^{'}}{H^2}\right)^{2}=\frac{3}{2}\alpha^{'}\dot{T}^{2}\,,~~~~ &
 \mbox{\small {\bf for {$q=1/2$ }}}  \\ 
 \displaystyle   
                \frac{2}{3\alpha^{'}}\left(\frac{H^{'}}{H^2}\right)^{2}\left(\frac{1-\alpha^{'}(1-2q)\dot{T}^2}{2q}\right)\\
                \displaystyle~~~~~~~~~~~~=\frac{3}{2}\alpha^{'}\dot{T}^{2}\left(\frac{2q}{1-\alpha^{'}(1-2q)\dot{T}^2}\right)\,.~~~~ &
 \mbox{\small {\bf for {~any~arbitrary~ $q$ }}} 
          \end{array}
\right.\\ \nonumber\\  \label{bp2}
 \displaystyle  \epsilon_{2}=\frac{d\ln|\epsilon_{1}|}{dN}&=&\frac{\ddot{H}}{H\dot{H}}+2\epsilon_{1}=\left\{\begin{array}{lll}
                    \displaystyle  
                   \sqrt{\frac{2}{3\alpha^{'}\epsilon_{1}}}\frac{\epsilon^{'}_{1}}{H}=2\frac{\ddot{T}}{H\dot{T}}\,,~~~~ &
 \mbox{\small {\bf for {$q=1/2$ }}}  \\ 
 \displaystyle   
                \sqrt{\frac{2}{3\alpha^{'}\epsilon_{1}}\left(\frac{1-\alpha^{'}(1-2q)\dot{T}^2}{2q}\right)}\frac{\epsilon^{'}_{1}}{H}\\
                \displaystyle~~~~~~~=\frac{2\ddot{T}}{H\dot{T}\left(1-\alpha^{'}(1-2q)\dot{T}^2\right)}\,.~~~~ &
 \mbox{\small {\bf for {~any~arbitrary~ $q$ }}} 
          \end{array}
\right.\\ \nonumber \\ 
\label{bp3}
 \displaystyle  \epsilon_{3}=\frac{d\ln|\epsilon_{2}|}{dN}&=&\frac{1}{\epsilon_{2}}\left[\frac{\dddot{H}}{H^2 \dot{H}}-3\frac{\ddot{H}}{H^3}-\frac{\ddot{H}^2}{H^2 \dot{H}^2}+4\frac{\dot{H}^2}{H^4}\right]\nonumber 
 \\\nonumber \\&=&\left\{\begin{array}{lll}
                    \displaystyle  
                   \sqrt{\frac{2\epsilon_{1}}{3\alpha^{'}}}\frac{\epsilon^{'}_{2}}{H}=\left[\frac{2\dddot{T}}{H^2\dot{T}\epsilon_{2}}+\epsilon_{1}-\frac{\epsilon_{2}}{2}\right]\,,~~~~ &
 \mbox{\small {\bf for {$q=1/2$ }}}  \\
 \displaystyle   
               \sqrt{\frac{2\epsilon_{1}}{3\alpha^{'}}\left(\frac{1-\alpha^{'}(1-2q)\dot{T}^2}{2q}\right)}\frac{\epsilon^{'}_{2}}{H}\\
                \displaystyle~~~~=\frac{\left[\frac{2\dddot{T}}{H^2\dot{T}\epsilon_{2}}+\epsilon_{1}-\frac{\epsilon_{2}}{2}\right]}{\left(1-\alpha^{'}(1-2q)\dot{T}^2\right)}
                +\frac{\frac{4\alpha^{'}(1-2q)\ddot{T}^2}{H} }{\left(1-\alpha^{'}(1-2q)\dot{T}^2\right)^2}\,.~~~~ &
 \mbox{\small {\bf for {~any~arbitrary~ $q$ }}} 
          \end{array}
\right.
\eea
It is important to note that:
\begin{itemize}
 \item In the present context $\epsilon_{1}$ is characterized by the part of the total tachyonic energy density $\dot{T}^2$. Inflation occurs when $\epsilon_{1}<1$ and ends when $\epsilon_{1}=1$ which is 
 exactly same like other single field slow-roll inflationary paradigm.
 \item The slow-roll parameter $\epsilon_{2}$ characterizes the ratio
of the field acceleration relative to the frictional contribution acting on it due to the expansion.
\item The third slow-roll parameter $\epsilon_{3}$ is made up of both $\epsilon_{1}$ and $\epsilon_{2}$. More precisely, $\epsilon_{3}$ is made up of $\dot{T}$, $\ddot{T}$ and $\dddot{T}$. This clearly implies that 
the third slow-roll parameter $\epsilon_{3}$ carries the contribution from the part of total tachyonic energy density, field acceleration relative to the frictional contribution and rate of change of field acceleration.
\item The slow-roll conditions stated in Eq~(\ref{wa1}) and Eq~(\ref{wa2}) are satisfied when the slow-roll parameters satisfy $\epsilon_{1}<<1$, $\epsilon_{2}<<1$ and $\epsilon_{3}<<1$. This also implies that in the 
slow-roll regime of the tachyonic inflation product of the two slow-roll parameters are also less than unity. For an example from Eq~(\ref{bp3}) it is clearly observed that to satisfy the slow-roll condition we need to have additionally
$\epsilon_{2}\epsilon_{3}<<1$.
\end{itemize}
Now for the sake of clarity, using Hamilton-Jacobi formalism, the Friedman equations and conservation equation can be re-written as:
\be\begin{array}{lll}\label{vcbb2}
 \displaystyle  0\approx\left\{\begin{array}{lll}
                    \displaystyle  
                   \left[H^{'}(T)\right]^2 -\frac{9\alpha^{'}}{4}H^{4}(T)+\frac{\alpha^{'}}{4M^{4}_{p}}V^{2}(T)\,,~~~~~~ &
 \mbox{\small {\bf for {$q=1/2$ }}}  \\  
 \displaystyle   
                H^{2}(T)\left[1-\frac{4\left[H^{'}(T)\right]^2}{9\alpha^{'} H^{4}(T)}\left(\frac{1-\alpha^{'}(1-2q)\dot{T}^2}{2q}\right)^2\right]^{1-q}\\
                \displaystyle ~~~~~~~-\frac{V(T)}{3M^{2}_{p}}\left[1-\frac{4(1-2q)\left[H^{'}(T)\right]^2}{9\alpha^{'} H^{4}(T)}\left(\frac{1-\alpha^{'}(1-2q)\dot{T}^2}{2q}\right)^2\right]\,.~~~~~~ &
 \mbox{\small {\bf for {~any~arbitrary~ $q$ }}} 
          \end{array}
\right.
\end{array}\ee
and 
\be\begin{array}{lll}\label{vcbb3}
 \displaystyle  H^{'}(T)\approx\left\{\begin{array}{lll}
                    \displaystyle  
                   -\frac{3\alpha^{'}}{2}H^{2}(T)\dot{T}\,,~~~~~~ &
 \mbox{\small {\bf for {$q=1/2$ }}}  \\  
 \displaystyle   
                -\frac{3\alpha^{'}}{2}H^{2}(T)\dot{T}\left(\frac{2q}{1-\alpha^{'}(1-2q)\dot{T}^2}\right)\,.~~~~~~ &
 \mbox{\small {\bf for {~any~arbitrary~ $q$ }}} 
          \end{array}
\right.
\end{array}\ee
Further using the definition of first Hubble slow-roll parameter $\epsilon_{1}$ in Eq~(\ref{vcbb2}) we get:
\be\begin{array}{lll}\label{vcbb4}
 \displaystyle  \left\{\begin{array}{lll}
                    \displaystyle  
                   H^{2}(T)\left[1-\frac{2}{3}\epsilon_{1}(T)\right]^{1/2}=\frac{V(T)}{3M^{2}_{p}}\,,~~~~~~ &
 \mbox{\small {\bf for {$q=1/2$ }}}  \\  
 \displaystyle   
                H^{2}(T)\left[1-\frac{1}{3q}\epsilon_{1}(T)\right]^{1-q}\approx\frac{V(T)}{3M^{2}_{p}}\left[1-\frac{(1-2q)}{3q}\epsilon_{1}(T)\right]\,.~~~~~~ &
 \mbox{\small {\bf for {~any~arbitrary~ $q$ }}} 
          \end{array}
\right.
\end{array}\ee
where for the arbitrary $q$ we have used the following constraint condition:
\bea 1-\underbrace{\alpha^{'}(1-2q)\dot{T}^{2}}_{<<1}\approx 1 .\eea
Now as in the slow-roll regime of tachyonic inflation $\epsilon_{1}(T)<<1$, consequently one can expand the exponents appearing in the left hand side of Eq~(\ref{vcbb4}), which leads to the following simplified expression:
\be\begin{array}{lll}\label{vcbb5}
 \displaystyle  \left\{\begin{array}{lll}
                    \displaystyle  
                   H^{2}(T)\left[1-\frac{1}{3}\epsilon_{1}(T)\right]+{\cal O}(\epsilon^2_{1} (T))=\frac{V(T)}{3M^{2}_{p}}\,,~~~~~~ &
 \mbox{\small {\bf for {$q=1/2$ }}}  \\  
 \displaystyle   
                H^{2}(T)\left[1-\frac{1-q}{3q}\epsilon_{1}(T)\right]+{\cal O}(\epsilon^2_{1} (T))\approx\frac{V(T)}{3M^{2}_{p}}\left[1-\frac{(1-2q)}{3q}\epsilon_{1}(T)\right]\,.~~~~~~ &
 \mbox{\small {\bf for {~any~arbitrary~ $q$ }}} 
          \end{array}
\right.
\end{array}\ee
It is important to mention here that:
\begin{itemize}
 \item The result for $q=1/2$ implies that except for the second order correction term in slow-roll i.e. ${\cal O}(\epsilon^2_{1} (T))$ the rest of the contribution exactly matches with the known result for 
 the single field slow-roll inflationary models. But in the non slow-roll limiting situation truncating at second order in slow-roll is not allowed and in that case for correct computation 
 one need to consider the full binomial series expansion.
 
 \item For $q\neq \frac{1}{2}$ i.e. for any other arbitrary $q$ in the slow-roll regime of tachyonic inflation we have allowed the second order correction term in slow-roll i.e. ${\cal O}(\epsilon^2_{1} (T))$ as appearing for $q\neq \frac{1}{2}$.
 But the final result implies significant deviation from the result that is well known for single field slow-roll inflationary models in the slow-roll regime. As mentioned earlier in the non slow-roll limiting
 situation truncating at a certain order in slow-roll is not allowed and in that case for correct computation 
 one need to consider the full binomial series expansion. Also for $q\neq \frac{1}{2}$ case, the right hand side of Eq~(\ref{vcbb5}) gets modified in presence of slow-roll parameter $\epsilon_{1}$.
 
\end{itemize}
Our next objective is to express the Hubble slow-roll parameters in therms of the tachyon potential dependent slow-roll parameters. To serve this purpose let us start with writing the expression for the derivatives of the 
potential in terms of the Hubble slow-roll parameters. Allowing upto the second order contribution in Hubble slow-roll parameters we get \cite{Steer:2003yu}:
\bea 
\label{tep1} M^{2}_{p}\frac{V^{'}(T)}{V(T)H(T)}&=&-\sqrt{6\epsilon_{1}}\frac{\left(1-\frac{2}{3}\epsilon_{1}+\frac{\epsilon_{2}}{6}\right)}{\left(1-\frac{2}{3}\epsilon_{1}\right)},\\
\label{tep2} M^{4}_{p}\frac{V^{''}(T)}{V(T)H^2(T)}&=&6\epsilon_{1}\frac{\left(1-\frac{2}{3}\epsilon_{1}+\frac{\epsilon_{2}}{6}\right)\left(1-\frac{2}{3}\epsilon_{1}-\frac{\epsilon_{2}}{3}\right)}
{\left(1-\frac{2}{3}\epsilon_{1}\right)^2}\nonumber \\&&~~~~~~~~~~~~~~~+\frac{\epsilon_{2}}{2}\frac{\left(5\epsilon_{1}-\frac{\epsilon_{2}}{3}-\epsilon_{3}\right)}{\left(1-\frac{2}{3}\epsilon_{1}\right)}
+3\left(\epsilon_{1}-\frac{\epsilon_{2}}{2}\right),
\eea
Further using Eq~(\ref{bp1}), Eq~(\ref{bp2}), Eq~(\ref{bp3}), Eq~(\ref{tep1}) and Eq~(\ref{tep2}) one can re-express the Hubble slow-roll parameters in terms of the potential dependent slow-roll parameter as:
\bea\label{bp1}\footnotesize
 \displaystyle  \epsilon_{1}&\approx&\left\{\begin{array}{lll}
                    \displaystyle  
                   \frac{M^{2}_{p}}{2\alpha^{'}}\frac{V^{'2}(T)}{V^{3}(T)}=\frac{\epsilon_{V}}{V(T)\alpha^{'}}\equiv\bar{\epsilon}_{V}\,,~~~~~~~~~~~~~~~~~~~~~~~~~~~~~~~~~~~~ &
 \mbox{\small {\bf for {$q=1/2$ }}}  \\ 
 \displaystyle   
                 \frac{M^{2}_{p}}{4q\alpha^{'}}\frac{V^{'2}(T)}{V^{3}(T)}=\frac{\epsilon_{V}}{2qV(T)\alpha^{'}}\equiv\frac{\bar{\epsilon}_{V}}{2q}\,.~~~~ &
 \mbox{\small {\bf for {~any~ $q$ }}} 
          \end{array}
\right.\\  \label{bp2}
 \displaystyle  \epsilon_{2}&\approx&\left\{\begin{array}{lll}
                    \displaystyle  
                   \frac{M^2_{p}}{\alpha^{'}}\left(3\frac{V^{'2}(T)}{V^{3}(T)}-2\frac{V^{''}(T)}{V^{2}(T)}\right)
                   =\frac{2\left(3\epsilon_{V}-\eta_{V}\right)}{V(T)\alpha^{'}}=2\left(3\bar{\epsilon}_{V}-\bar{\eta}_{V}\right)\,,~ &
 \mbox{\small {\bf for {$q=1/2$ }}}  \\ 
 \displaystyle   
                 \frac{M^2_{p}}{\sqrt{2q}\alpha^{'}}\left(3\frac{V^{'2}(T)}{V^{3}(T)}-2\frac{V^{''}(T)}{V^{2}(T)}\right)
                   =\frac{\sqrt{\frac{2}{q}}\left(3\epsilon_{V}-\eta_{V}\right)}{V(T)\alpha^{'}}=\sqrt{\frac{2}{q}}\left(3\bar{\epsilon}_{V}-\bar{\eta}_{V}\right)\,. &
 \mbox{\small {\bf for {~any~ $q$ }}} 
          \end{array}
\right.\\ 
\label{bp3}
 \displaystyle  \epsilon_{3}\epsilon_{2}&\approx&\left\{\begin{array}{lll}
                    \displaystyle  
                   \frac{M^{4}_{p}}{V^{2}(T)\alpha^{'2}}\left(2\frac{V^{'''}(T)V^{'}(T)}{V^{2}(T)}-10\frac{V^{''}(T)V^{'2}(T)}{V^{3}(T)}+9\frac{V^{'4}(T)}{V^{4}(T)}\right)\\
                   \displaystyle~~~~~ =\frac{\left(2\xi^2_{V}-5\eta_{V}\epsilon_{V}+36\epsilon^2_{V}\right)}{V^{2}(T)\alpha^{'2}}=\left(2\bar{\xi}^2_{V}-5\bar{\eta}_{V}\bar{\epsilon}_{V}+36\bar{\epsilon}^2_{V}\right)\,,~~~~ &
 \mbox{\small {\bf for {$q=1/2$ }}}  \\ 
 \displaystyle   
              \frac{M^{4}_{p}}{\sqrt{2q}V^{2}(T)\alpha^{'2}}\left(2\frac{V^{'''}(T)V^{'}(T)}{V^{2}(T)}-10\frac{V^{''}(T)V^{'2}(T)}{V^{3}(T)}+9\frac{V^{'4}(T)}{V^{4}(T)}\right)\\
                   \displaystyle ~~~~~=\frac{\left(2\xi^2_{V}-5\eta_{V}\epsilon_{V}+36\epsilon^2_{V}\right)}{\sqrt{2q} V^{2}(T)\alpha^{'2}}=\frac{\left(2\bar{\xi}^2_{V}-5\bar{\eta}_{V}\bar{\epsilon}_{V}+36\bar{\epsilon}^2_{V}\right)}{\sqrt{2q}}\,.~~~~ &
 \mbox{\small {\bf for {~any~ $q$ }}} 
          \end{array}
\right.
\eea
where the potential dependent slow-roll parameters $\epsilon_{V}, \eta_{V}, \xi^{2}_{V}, \sigma^{3}_{V}$ are defined as:
\bea \label{wzq1} \epsilon_{V}&=&\frac{M^{2}_{p}}{2}\left(\frac{V^{'}(T)}{V(T)}\right)^2,\\
\label{wzq2} \eta_{V}&=& M^{2}_{p}\left(\frac{V^{''}(T)}{V(T)}\right),\\
\label{wzq3} \xi^{2}_{V}&=& M^{4}_{p}\left(\frac{V^{'}(T)V^{'''}(T)}{V^2(T)}\right),\\
\label{wzq4} \sigma^{3}_{V}&=& M^{6}_{p}\left(\frac{V^{'2}(T)V^{''''}(T)}{V^3(T)}\right),
\eea
which is exactly similar to the expression for the slow-roll parameter as appearing in the context of single field slow-roll inflationary models.
However, for the sake of clarity here we introduce new sets of potential dependent slow-roll parameters for tachyonic inflation by rescaling with the appropriate powers of $\alpha^{'}V(T)$:
\bea \label{wzq11} \bar{\epsilon}_{V}&=&\frac{\epsilon_{V}}{\alpha^{'}V(T)}=\frac{M^{2}_{p}}{2\alpha^{'}V(T)}\left(\frac{V^{'}(T)}{V(T)}\right)^2,\\
\label{wzq22} \bar{\eta}_{V}&=& \frac{{\eta}_{V}}{\alpha^{'}V(T)}=\frac{M^{2}_{p}}{\alpha^{'}V(T)}\left(\frac{V^{''}(T)}{V(T)}\right),\\
\label{wzq33} \bar{\xi}^{2}_{V}&=& \frac{\xi^{2}_{V}}{\alpha^{'2}V^2(T)}=\frac{M^{4}_{p}}{\alpha^{'2}V^2(T)}\left(\frac{V^{'}(T)V^{'''}(T)}{V^2(T)}\right),\\
\label{wzq44} \bar{\sigma}^{3}_{V}&=& \frac{\sigma^{3}_{V}}{\alpha^{'3}V^3(T)}=\frac{M^{6}_{p}}{\alpha^{'3}V^3(T)}\left(\frac{V^{'2}(T)V^{''''}(T)}{V^3(T)}\right).
\eea
Further using Eq~(\ref{wzq1})-Eq~(\ref{wzq44}) we get the following operator identity for tachyonic inflation:
\be\begin{array}{lll}\label{vchj2}
 \displaystyle  \frac{1}{H}\frac{d}{dt}=\frac{d}{dN}=\frac{d}{d\ln k}\approx\left\{\begin{array}{lll}
                    \displaystyle  
                   \sqrt{\frac{2\bar{\epsilon}_{V}}{V(T)\alpha^{'}}}M_{p}\left(1-\frac{2}{3}\bar{\epsilon}_{V}\right)^{1/4}\frac{d}{dT}\,,~~~~~~ &
 \mbox{\small {\bf for {$q=1/2$ }}}  \\ 
 \displaystyle   
                \sqrt{\frac{2\bar{\epsilon}_{V}}{V(T)\alpha^{'}}}\frac{M_{p}}{2q}\left(1-\frac{1}{3q}\bar{\epsilon}_{V}\right)^{1/4}\frac{d}{dT}\,.~~~~~~ &
 \mbox{\small {\bf for {~any~arbitrary~ $q$ }}} 
          \end{array}
\right.
\end{array}\ee
Finally using Eq~(\ref{vchj2}) we get the following sets of flow equations in the context of tachyoinc inflation:
\bea\label{bp11}
 \displaystyle  \frac{d\epsilon_{1}}{dN}&=&\left\{\begin{array}{lll}
                    \displaystyle  
                   \frac{d\bar{\epsilon}_{V}}{dN}=2\bar{\epsilon}_{V}\left(\bar{\eta}_{V}-3\bar{\epsilon}_{V}\right)\left(1-\frac{2}{3}\bar{\epsilon}_{V}\right)^{1/4}\,,~~~~~~~~~~~~~~~~~~~~~~~~~~~~~~~ &
 \mbox{\small {\bf for {$q=1/2$ }}}  \\
 \displaystyle   
                \frac{1}{2q}\frac{d\bar{\epsilon}_{V}}{dN}=\frac{\bar{\epsilon}_{V}}{q}\left(\bar{\eta}_{V}-3\bar{\epsilon}_{V}\right)\left(1-\frac{1}{3q}\bar{\epsilon}_{V}\right)^{1/4}\,.~~~~ &
 \mbox{\small {\bf for {~any~ $q$ }}} 
          \end{array}
\right.\\   \label{bp22}
 \displaystyle   \frac{d\epsilon_{2}}{dN}&=&\left\{\begin{array}{lll}
                    \displaystyle  
                   2\left(10\bar{\epsilon}_{V}\bar{\eta}_{V}-18\bar{\epsilon}^2_{V}-\bar{\xi}^2_{V}\right)\left(1-\frac{2}{3}\bar{\epsilon}_{V}\right)^{1/4}\,,~~~~~~~~~~~~~~~~~~~~~~~~~~~~~~~~~~ &
 \mbox{\small {\bf for {$q=1/2$ }}}  \\ 
 \displaystyle   
                 \sqrt{\frac{2}{q}}\left(10\bar{\epsilon}_{V}\bar{\eta}_{V}-18\bar{\epsilon}^2_{V}-\bar{\xi}^2_{V}\right)\left(1-\frac{1}{3q}\bar{\epsilon}_{V}\right)^{1/4}\,.~~~~ &
 \mbox{\small {\bf for {~any~ $q$ }}} 
          \end{array}
\right.\\ 
\label{bp33}
 \displaystyle  \frac{d(\epsilon_{2}\epsilon_{3})}{dN}&=&\left\{\begin{array}{lll}
                    \displaystyle  
                   \left(2\bar{\sigma}^{3}_{V}-216\bar{\epsilon}^{3}_{V}+2\bar{\xi}^{2}_{V}\bar{\eta}_{V}-7\bar{\xi}^{2}_{V}\bar{\epsilon}_{V}+194\bar{\epsilon}^2_{V}\bar{\eta}_{V}-10\bar{\eta}^2_{V}\bar{\epsilon}_{V}
                   \right)\left(1-\frac{2}{3}\bar{\epsilon}_{V}\right)^{1/4}\,,~~~~ &
 \mbox{\small {\bf for {$q=1/2$ }}}  \\ 
 \displaystyle   
               \frac{\left(2\bar{\sigma}^{3}_{V}-216\bar{\epsilon}^{3}_{V}+2\bar{\xi}^{2}_{V}\bar{\eta}_{V}-7\bar{\xi}^{2}_{V}\bar{\epsilon}_{V}+194\bar{\epsilon}^2_{V}\bar{\eta}_{V}-10\bar{\eta}^2_{V}\bar{\epsilon}_{V}
                   \right)}{\sqrt{2q}}\left(1-\frac{1}{3q}\bar{\epsilon}_{V}\right)^{1/4}\,.~~~~ &
 \mbox{\small {\bf for {~any~ $q$ }}} 
          \end{array}
\right.
\eea
where we use the following consistency conditions for re-scaled potential dependent slow-roll parameters:
\bea\label{bp1aas}
 \displaystyle  \frac{d\bar{\epsilon}_{V}}{dN}&=&\left\{\begin{array}{lll}
                    \displaystyle  
                   2\bar{\epsilon}_{V}\left(\bar{\eta}_{V}-3\bar{\epsilon}_{V}\right)\left(1-\frac{2}{3}\bar{\epsilon}_{V}\right)^{1/4}\,,~~~~~~~~~~~~~~~~~~~~~~~~~~~~~~~ &
 \mbox{\small {\bf for {$q=1/2$ }}}  \\ 
 \displaystyle   
                2\bar{\epsilon}_{V}\left(\bar{\eta}_{V}-3\bar{\epsilon}_{V}\right)\left(1-\frac{1}{3q}\bar{\epsilon}_{V}\right)^{1/4}\,.~~~~ &
 \mbox{\small {\bf for {~any~ $q$ }}} 
          \end{array}
\right.
\eea 
\bea\label{bp2aas}
 \displaystyle  \frac{d\bar{\eta}_{V}}{dN}&=&\left\{\begin{array}{lll}
                    \displaystyle  
                   \left(\bar{\xi}^2_{V}-4\bar{\epsilon}_{V}\bar{\eta}_{V}\right)\left(1-\frac{2}{3}\bar{\epsilon}_{V}\right)^{1/4}\,,~~~~~~~~~~~~~~~~~~~~~~~~~~~~~~~ &
 \mbox{\small {\bf for {$q=1/2$ }}}  \\ 
 \displaystyle   
                \left(\bar{\xi}^2_{V}-4\bar{\epsilon}_{V}\bar{\eta}_{V}\right)\left(1-\frac{1}{3q}\bar{\epsilon}_{V}\right)^{1/4}\,.~~~~ &
 \mbox{\small {\bf for {~any~ $q$ }}} 
          \end{array}
\right.
\\ \label{bp3aas}
 \displaystyle  \frac{d\bar{\xi}^2_{V}}{dN}&=&\left\{\begin{array}{lll}
                    \displaystyle  
                   \left(\bar{\sigma}^3_{V}+\bar{\xi}^2_{V}\bar{\eta}_{V}-\bar{\xi}^2_{V}\bar{\epsilon}_{V}\right)\left(1-\frac{2}{3}\bar{\epsilon}_{V}\right)^{1/4}\,,~~~~~~~~~~~~~~~~~~~~~~~~~~~~~~~ &
 \mbox{\small {\bf for {$q=1/2$ }}}  \\ 
 \displaystyle   
                \left(\bar{\sigma}^3_{V}+\bar{\xi}^2_{V}\bar{\eta}_{V}-\bar{\xi}^2_{V}\bar{\epsilon}_{V}\right)\left(1-\frac{1}{3q}\bar{\epsilon}_{V}\right)^{1/4}\,.~~~~ &
 \mbox{\small {\bf for {~any~ $q$ }}} 
          \end{array}
\right.
\\ \label{bp4aas}
 \displaystyle  \frac{d\bar{\sigma}^3_{V}}{dN}&=&\left\{\begin{array}{lll}
                    \displaystyle  
                   \bar{\sigma}^3_{V}\left(\bar{\eta}_{V}-12\bar{\epsilon}_{V}\right)\left(1-\frac{2}{3}\bar{\epsilon}_{V}\right)^{1/4}\,,~~~~~~~~~~~~~~~~~~~~~~~~~~~~~~~ &
 \mbox{\small {\bf for {$q=1/2$ }}}  \\ 
 \displaystyle   
                \bar{\sigma}^3_{V}\left(\bar{\eta}_{V}-12\bar{\epsilon}_{V}\right)\left(1-\frac{1}{3q}\bar{\epsilon}_{V}\right)^{1/4}\,.~~~~ &
 \mbox{\small {\bf for {~any~ $q$ }}} 
          \end{array}
\right.
\eea 
In terms of the slow-roll parameters, the number of e-foldings can be re-expressed as:
\be\begin{array}{lll}\label{e-fold2}\footnotesize
 \displaystyle  N(T)=\left\{\begin{array}{lll}\footnotesize
                    \displaystyle  
                   \sqrt{\frac{3\alpha^{'}}{2}}\int^{T_{end}}_{T}\frac{H(T)}{\sqrt{\epsilon_{1}}}~dT=\sqrt{\frac{3\alpha^{'}}{2}}\int^{T_{end}}_{T}\frac{H(T)}{\sqrt{\bar{\epsilon}_{V}}}~dT
\approx\frac{\alpha^{'}}{M^{2}_{p}}\int^{T}_{T_{end}}
\frac{V^{2}(T)}{V^{'}(T)}~dT\,, &
 \mbox{\small {\bf for {$q=1/2$ }}}  \\ 
 \displaystyle   
               2q\sqrt{\frac{3\alpha^{'}}{2}}\int^{T_{end}}_{T}\frac{H(T)}{\sqrt{\epsilon_{1}}}~dT= \sqrt{3\alpha^{'}q}\int^{T_{end}}_{T}\frac{H(T)}{\sqrt{\bar{\epsilon}_{V}}}~dT
 \approx\frac{ \sqrt{2q}\alpha^{'}}{M^{2}_{p}}\int^{T}_{T_{end}}
\frac{V^{2}(T)}{V^{'}(T)}~dT\,. &
 \mbox{\small {\bf for {~any~ $q$ }}} 
          \end{array}
\right.
\end{array}\ee
where $T_{end}$ characterizes the tachyonic field value at the end of inflation $t=t_{end}$. It is important to mention here that in the single field tachyoinc inflationary paradigm the field value of the 
tachyon at the end of inflation is computed from the following condition:
\be {\rm max}_{\phi=\phi_{e}}\left[\epsilon_{V},|\eta_{V}|,|\xi^{2}_{V}|,|\sigma^{3}_{V}|\right]\equiv 1.\ee
Let $T_{\star}$ denote the value of tachyonic field $T$ at which a length scale or more precisely the modes crosses the Hubble radius
during inflation, which is given by the momentum at pivot scale $k_{\star} = a_{\star}H_{\star}$. Here $a_{\star}$ and $H_{\star}$ signify the 
scale factor and Hubble parameter at the horizon crossing scale or at the pivot scale. Then the definition of the number of e-foldings as stated in Eq~(\ref{e-fold}) and Eq~(\ref{e-fold2}) gives:
\bea\label{ui1} N_{\star}=N(T_{\star})=\ln\left(\frac{a_{end}}{a_{\star}}\right),\eea
where $a_{end}$ be the scale factor at the end of inflation. Then using Eq~(\ref{ui1}) the corresponding horizon crossing momentum scale or the the pivot scale can be computed as:
\bea c_{S}k_{\star}=a_{\star}H_{\star}=a_{end}H_{\star}\exp\left(-N_{\star}\right).\eea
Now at any arbitrary momentum scale the number of e-foldings, $N(k)$, between the Hubble
exit of the relevant modes and the end of inflation can be expressed as:
\bea \label{nk} N(k)\approx 71.21 -\ln\left(\frac{k}{k_{0}}\right)+\frac{1}{4}\ln\left(\frac{V(T_{\star})}{M^{4}_{p}}\right)+\frac{1}{4}\ln\left(\frac{V(T_{\star})}{\rho_{end}}\right)+\frac{1-3w_{int}}{12(1+w_{int})}
\ln\left(\frac{\rho_{reh}}{\rho_{end}}\right),~~~~~~~
\eea
where $\rho_{end}$ is the energy density at the end of inflation, $\rho_{reh}$ is an energy scale during
reheating, $c_{S}k_{0}= a_{0}H_{0}$ is the present Hubble scale, $V(T_{\star})$ corresponds to the potential
energy when the relevant modes left the Hubble patch during inflation corresponding
to the momentum scale $c_{S}k_{\star} = a_{\star}H_{\star}=c_{S}k_{cmb}$, and $w_{int}$ characterizes the effective equation of state
parameter between the end of inflation and the energy scale during reheating. 
Further using Eq~(\ref{nk}) in Eq~(\ref{ui1}) we get the following expression:
\bea \label{nk1} N_{\star}\approx 71.21 -\ln\left(\frac{k_{\star}}{k_{0}}\right)+\frac{1}{4}\ln\left(\frac{V(T_{\star})}{M^{4}_{p}}\right)+\frac{1}{4}\ln\left(\frac{V(T_{\star})}{\rho_{end}}\right)+\frac{1-3w_{int}}{12(1+w_{int})}
\ln\left(\frac{\rho_{reh}}{\rho_{end}}\right),~~~~~~~
\eea
 which is very useful to fix the number of e-foldings within $50<N_{\star}<70$ for tachyonic inflation.
\subsubsection{Basics of tachyonic perturbations}
In this subsection we explicitly discuss about the cosmological linear perturbation theory within the framework of tachyonic inflation.
Let us clearly mention that here we have various ways of characterizing cosmological perturbations in the context of inflation,
which finally depend on the choice of gauge. Let us do the computation in the longitudinal guage, where the scalar metric perturbations of the 
FLRW background are given by the following infinitesimal line element:
\bea \label{line}
ds^2 &=& -\left(1+2{\bf \Phi}(t,{\bf x})\right)dt^2+a^{2}(t)\left(1-2{\bf \Psi}(t,{\bf x})\right)\delta_{ij}dx^{i}dx^{j},
\eea
where $a(t)$ is the scale factor, ${\bf \Phi}(t,{\bf x})$ and ${\bf \Psi}(t,{\bf x})$ characterizes the gauge invariant metric perturbations.
Specifically, the perturbation of the FLRW metric leads to the perturbation in the energy-momentum stress tensor via the Einstein field equation or equivalently through the Friedmann equations.
For the perturbed metric as mentioned in Eq~(\ref{line}), the perturbed Einstein field equations can be expressed for $q=1/2$ case of tachyonic inflationary setup as:
\bea \label{l1} 3H\left(H{\bf \Phi}(t,{\bf k})+\dot{\bf \Psi}(t,{\bf k})\right)+\frac{k^{2}}{a^{2}(t)}&=&-\frac{1}{2M^{2}_{p}}\delta \rho,\\
\label{l2} \ddot{\bf \Psi}(t,{\bf k})+3H\left(H{\bf \Phi}(t,{\bf k})+\dot{\bf \Psi}(t,{\bf k})\right)+H\dot{\bf \Phi}(t,{\bf k})\nonumber~~~~\\~~~~~~~~~~+2\dot{H}{\bf \Phi}(t,{\bf k})+\frac{k^{2}}{3a^{2}(t)}
\left({\bf \Phi}(t,{\bf k})-{\bf \Psi}(t,{\bf k})\right)&=&\frac{1}{2M^{2}_{p}}\delta p,\\
\label{l3} \dot{\bf \Psi}(t,{\bf k})+H{\bf \Phi}(t,{\bf k})&=&-\frac{\alpha^{'}V(T)}{\sqrt{1-\alpha^{'}\dot{T}^{2}}}\frac{\dot{T}}{M^{2}_{p}}\delta T,~~~~\\
\label{l4}{\bf \Psi}(t,{\bf k})-{\bf \Phi}(t,{\bf k})&=&0.
\eea
Similarly, for any arbitrary $q$ the perturbed Einstein field equations can be expressed as:
\bea \label{l11} 3H\left(H{\bf \Phi}(t,{\bf k})+\dot{\bf \Psi}(t,{\bf k})\right)+\frac{k^{2}}{a^{2}(t)}&=&-\frac{1}{2M^{2}_{p}}\delta \rho,\\
\label{l22} \ddot{\bf \Psi}(t,{\bf k})+3H\left(H{\bf \Phi}(t,{\bf k})+\dot{\bf \Psi}(t,{\bf k})\right)+H\dot{\bf \Phi}(t,{\bf k})\nonumber~~~~\\~~~~~~~~~~+2\dot{H}{\bf \Phi}(t,{\bf k})+\frac{k^{2}}{3a^{2}(t)}
\left({\bf \Phi}(t,{\bf k})-{\bf \Psi}(t,{\bf k})\right)&=&\frac{1}{2M^{2}_{p}}\delta p,\\
\label{l33} \dot{\bf \Psi}(t,{\bf k})+H{\bf \Phi}(t,{\bf k})&=&-\frac{\alpha^{'}V(T)\left[1-\alpha^{'}(1-2q)\dot{T}^2\right]}{\left(1-\alpha^{'}\dot{T}^2\right)^{1-q}} \frac{\dot{T}}{M^{2}_{p}}\delta T,~~~~~~~~~~\\
\label{l44}{\bf \Psi}(t,{\bf k})-{\bf \Phi}(t,{\bf k})&=&0.
\eea
Here ${\bf \Phi}(t,{\bf k})$ and ${\bf \Psi}(t,{\bf k})$ are the two gauge invariant metric perturbations in the Fourier space are defined via the following transformation:
\bea \label{f1} {\bf \Phi}(t,{\bf x})=\int d^{3}k~ {\bf \Phi}(t,{\bf k})~\exp(i {\bf k} . {\bf x}),\\
\label{f2} {\bf \Psi}(t,{\bf x})=\int d^{3}k~ {\bf \Psi}(t,{\bf k})~\exp(i {\bf k} . {\bf x}),
\eea
Additionally, it is important to note that in Eq~(\ref{l4}), the two gauge invariant metric perturbations ${\bf \Phi}(t,{\bf k})$ and ${\bf \Psi}(t,{\bf k})$ are equal in the context of minimally coupled tachyonic string field theoretic model 
with Einstein gravity sector. In Eq~(\ref{l1}) and Eq~(\ref{l2}) the perturbed energy density $\delta\rho$ and pressure $\delta p$ are given by:
\be\begin{array}{lll}\label{e7aa}\footnotesize
 \displaystyle \delta\rho =\left\{\begin{array}{lll}\footnotesize
                    \displaystyle  
                    \frac{V^{'}(T)\delta T}{\sqrt{1-\alpha^{'}\dot{T}^2}}
                    +\frac{\alpha^{'}V(T)\left(\dot{T}\delta\dot{T}+\dot{T}^2{\bf \Phi}(t,{\bf k})\right)}{\left(1-\alpha^{'}\dot{T}^2\right)^{3/2}} \,,~~~~ &
 \mbox{\small {\bf for $q=1/2$}}  \\
 \displaystyle  
  \frac{\left\{V^{'}(T)\left[1-\alpha^{'}(1-2q)\dot{T}^2\right]\delta T-4\alpha^{'}(1-2q)V(T)\dot{T}\delta \dot{T}\right\}}
  {\left(1-\alpha^{'}\dot{T}^2\right)^{1-q}}\\ \displaystyle ~~~~~
  +\frac{2\alpha^{'}(1-q)V(T)\left[1-\alpha^{'}(1-2q)\dot{T}^2\right]\left(\dot{T}\delta\dot{T}+\dot{T}^2{\bf \Phi}(t,{\bf k})\right)}
  {\left(1-\alpha^{'}\dot{T}^2\right)^{2-q}} \,,~~~ &
 \mbox{\small {\bf for~any~arbitrary~$q$}}.
          \end{array}
\right.
\end{array}\ee
and
\be\begin{array}{lll}\label{e8aa}\footnotesize
 \displaystyle \delta p =\left\{\begin{array}{lll}\footnotesize
                    \displaystyle  
                    -V^{'}(T)\sqrt{1-\alpha^{'}\dot{T}^2}\delta T +\frac{\alpha^{'}V(T)\left(\dot{T}\delta\dot{T}+\dot{T}^2{\bf \Phi}(t,{\bf k})\right)}{\sqrt{1-\alpha^{'}\dot{T}^2}} \,,~~~~ &
 \mbox{\small {\bf for $q=1/2$}}  \\ 
 \displaystyle  
 -V^{'}(T)\left(1-\alpha^{'}\dot{T}^2\right)^{q}\delta T+\frac{2q\alpha^{'}V(T)\left(\dot{T}\delta\dot{T}+\dot{T}^2{\bf \Phi}(t,{\bf k})\right)}{\left(1-\alpha^{'}\dot{T}^2\right)^{1-q}} \,,~~~ &
 \mbox{\small {\bf for ~any~arbitrary~$q$}}.
          \end{array}
\right.
\end{array}\ee
Similarly after the variation of the tachyoinic field equation motion
we get the following expressions for the perturbed equation of motion: 
\be\begin{array}{lll}\label{e8aa}\footnotesize
 \displaystyle 0 \approx\left\{\begin{array}{lll}\footnotesize
                    \displaystyle  
                    \delta \ddot{T}+3H\delta\dot{T}+\frac{2\alpha^{'}\ddot{T}\left(\dot{T}\delta\dot{T}+\dot{T}^2 {\bf \Phi}(t,{\bf k})\right)}{\left(1-\alpha^{'}\dot{T}^2\right)}
                    \\ \displaystyle ~~~~~~~+\frac{M_{p}\sqrt{1-\alpha^{'}\dot{T}^2}}{\alpha^{'}V(T)}\left[\left(\frac{k^2}{a^2} - 3\dot{H}\right){\bf \Phi}(t,{\bf k})-\frac{2k^2}{a^2}{\bf \Psi}(t,{\bf k})
                    \right.\\ \left. \displaystyle~~~~-3\left(\ddot{\bf \Psi}(t,{\bf k})+4H\dot{\bf \Psi}(t,{\bf k})+H\dot{\bf \Phi}(t,{\bf k})+\dot{H}{\bf \Phi}(t,{\bf k})+4H^2 {\bf \Phi}(t,{\bf k})\right)\right]\\
                    \displaystyle -\left\{6H\alpha^{'}\dot{T}^3 - \frac{2V^{'}(T)}{\alpha^{'}V(T)}\left(1-\alpha^{'}\dot{T}^2\right)\right\}{\bf \Phi}(t,{\bf k})-\left(\dot{\bf \Psi}(t,{\bf k})+3\dot{\bf \Psi}(t,{\bf k})\right)\dot{T}\\
                    \displaystyle ~~~~~~-\frac{M_{p}\left(1-\alpha^{'}\dot{T}^2\right)}{\alpha^{'}}\left(\frac{V^{''}(T)}{V(T)}-\frac{V^{'2}(T)}{V^{2}(T)}\right)\,,~~~~ &
 \mbox{\small {\bf for $q=1/2$}}  \\ 
 \displaystyle  
                    \delta \ddot{T}+3H\delta\dot{T}+\frac{2\alpha^{'}\ddot{T}\left(\dot{T}\delta\dot{T}+\dot{T}^2 {\bf \Phi}(t,{\bf k})\right)}{\left(1-\alpha^{'}\dot{T}^2\right)^{2(1-q)}}
                    \\ \displaystyle ~~~~~~~+\frac{M_{p}\left(1-\alpha^{'}\dot{T}^2\right)^{1-q}}{\alpha^{'}V(T)\left[1-\alpha^{'}(1-2q)\dot{T}^2\right]}\left[\left(\frac{k^2}{a^2} - 3\dot{H}\right){\bf \Phi}(t,{\bf k})-\frac{2k^2}{a^2}{\bf \Psi}(t,{\bf k})
                    \right.\\ \left. \displaystyle~~~~-3\left(\ddot{\bf \Psi}(t,{\bf k})+4H\dot{\bf \Psi}(t,{\bf k})+H\dot{\bf \Phi}(t,{\bf k})+\dot{H}{\bf \Phi}(t,{\bf k})+4H^2 {\bf \Phi}(t,{\bf k})\right)\right]\\
                    \displaystyle -\left\{6H\alpha^{'}\dot{T}^3 - \sqrt{\frac{2}{q}}\frac{V^{'}(T)}{\alpha^{'}V(T)}\left(1-\alpha^{'}\dot{T}^2\right)^{2(1-q)}\right\}{\bf \Phi}(t,{\bf k})-\left(\dot{\bf \Psi}(t,{\bf k})+3\dot{\bf \Psi}(t,{\bf k})\right)\dot{T}\\
                    \displaystyle ~~~~~~-\frac{M_{p}\left(1-\alpha^{'}\dot{T}^2\right)^{2(1-q)}}{\sqrt{2q}\alpha^{'}}\left(\frac{V^{''}(T)}{V(T)}-\frac{V^{'2}(T)}{V^{2}(T)}\right)\,,~~~ &
 \mbox{\small {\bf for ~any~$q$}}.
          \end{array}
\right.
\end{array}\ee 
Further we will perform the following steps throughout the next part of the computation:
\begin{itemize}
 \item First of all we decompose the scalar perturbations into two components-(1) {\bf entropic or isocurvature perturbations} which can be usually treated as the orthogonal projective part to the trajectory and (2) 
 {\bf adiabatic or curvature perturbations} which can be usually treated as the parallel projective part to the trajectory.
 \item If inflation is governed by a single scalar field then we deal with {\bf adiabatic or curvature perturbations}. On the other hand for multiple scalar fields we deal with {\bf entropic or isocurvature perturbations}.
 \item In the present context the inflationary dynamics is governed by a single tachyonic field, which implies the surviving part of the cosmological perturbations are governed by the {\bf adiabatic} contribution.
 \item Within the framework of first order cosmological perturbation theory we define a gauge invariant primordial curvature perturbation on the scales outside the horizon:
       \bea \zeta= {\bf \Psi}-\frac{H}{\dot{\rho}}\delta\rho. \eea
 \item Next we consider the uniform density hypersurface in which 
 \bea \delta \rho=0.\eea
 Consequently the curvature perturbation is governed by:
  \bea \zeta={\bf \Psi}.\eea
  \item Further, the time evolution of the curvature perturbation can be expressed as:
  \bea\label{pqw1} \dot{\zeta}=H\left(\frac{\delta \bar{p}}{\rho+p}\right),\eea
  where $\delta \bar{p}$ characterizes the non-adiabatic or entropic contribution in the first order linearized cosmological perturbation. In the present context $\delta \bar{p}$ can be expressed as:
  \bea \delta \bar{p}= \Gamma \dot{p},\eea
  where $\Gamma$ characterizes the relative displacement between hypersurfaces of uniform pressure and density. Additionally, it is important to note that Eq~(\ref{pqw1}) signifies the change in the {\bf curvature perturbation}
  on the uniform density hypersurfaces on the large scales. Also from Eq~(\ref{pqw1}) it is clearly observed that the contribution from the time evolution of the {\bf adiabatic or curvature perturbations} are directly proportional to 
  the non-adiabatic contribution which comes from significantly from the isocurvature part of pressure perturbation $\delta p$ and are completely independent of the specific mathematical structure of the gravitational field equations
  in the context of Einstein gravity framework. In generalized prescription
  the pressure perturbation in arbitrary gauge can be decomposed into the following two contributions:
  \bea \delta p = c^{2}_{S}\delta \rho + \delta \bar{p},\eea
  where $c^{2}_{S}$ is the effective sound speed, which is mentioned in the earlier section of the paper. 
  \item Now let us consider a situation where the pressure perturbation is completely made up of adiabatic contribution from the cosmological
  perturbation on large cosmological scales. Consequently we get:
  \bea \delta \bar{p}= \Gamma \dot{p}=0 ~~~~\Rightarrow~~~~\zeta={\bf Constant},\eea
  which is consistent with the single field slow-roll conditions in the context of tachyonic inflationary setup. Finally, in the uniform density hypersurfaces, the curvature perturbation can be written in terms of the tachyonic field 
  fluctuations on spatially flat hypersurfaces as: 
  \bea \zeta=-H\left(\frac{\delta T}{\dot{T}}\right).\eea
   \end{itemize}
\subsubsection{Computation of scalar power spectrum}
In this subsection our prime objective is to compute the primordial power spectra of scalar quantum fluctuations from tachyonic inflation and study the cosmological consequences from previously mentioned string theory originated 
tachyonic potentials in the light of Planck 2015 data. To serve this purpose let us start with the following canonical variable $v_{\bf k}$, which can be quantized with the standard techniques:
\bea v_{\bf k}\equiv z~M_{p}~\zeta_{\bf k}, \eea
where $\zeta_{k}$ is the curvature perturbation in the momentum space, which can be expressed in terms of the curvature perturbation in position space through the following Fourier transformation:
\bea \zeta(t,{\bf x})=\int d^{3}k~\zeta_{\bf k}(t)~\exp(i{\bf k}.{\bf x}). \eea
Also $z$ is defined as:
\be\begin{array}{lll}\label{e8aat}
 \displaystyle z =\frac{a(t)}{c_{S} H M_{p}}\sqrt{\rho +p}=\left\{\begin{array}{lll}
                    \displaystyle  
                    \frac{\sqrt{3\alpha^{'}}a(t)\dot{T}}{\sqrt{1-\alpha^{'}\dot{T}^2}}=\frac{a(t)}{c_{S}}\sqrt{2\epsilon_{1}}= \frac{a(t)}{c_{S}}\sqrt{2\bar{\epsilon}_{V}}\,,~~~~ &
 \mbox{\small {\bf for $q=1/2$}}  \\ 
 \displaystyle  
 \frac{\sqrt{6q\alpha^{'}}a(t)\dot{T}}{\sqrt{1-\alpha^{'}\dot{T}^2}}\sqrt{\frac{1+(1-2q)\alpha^{'}\dot{T}^2}{1-(1-2q)\alpha^{'}\dot{T}^2}}
=\frac{a(t)\sqrt{2\epsilon_{1}}}{c_{S}}
 =\frac{a(t)\sqrt{2\bar{\epsilon}_{V}}}{\sqrt{2q}c_{S}} \,,~~~ &
 \mbox{\small {\bf for ~any~$q$}}.
          \end{array}
\right.
\end{array}\ee
Next we use conformal time $\eta$ instead of using the time $t$, which is defined via the following infinitesimal transformation:
\bea 
dt= a~d\eta
\eea
using which one can redefine the Hubble parameter in conformal coordinate system as:
\bea {\cal H}(\eta)=\frac{1}{a(\eta)}\frac{da(\eta)}{d\eta}=a~H(t)
\eea
Further we derive the equation of motion of the scalar fluctuation by extremizing the tachyonic model action as:
\bea \label{diff}
\left[\frac{d^{2}}{d\eta^2}+\left(c^{2}_{S}k^{2}-\frac{1}{z}\frac{d^{2}z}{d\eta^2}\right)\right]v_{\bf k}=0
\eea
where 
\be\begin{array}{lll}\label{rte}\footnotesize
 \displaystyle \frac{1}{z}\frac{d^{2}z}{d\eta^2} = 2a^2H^2+\overbrace{\frac{2}{3}\frac{\left(\frac{1}{z}\frac{dz}{d\eta}\frac{d\epsilon_{1}}{d\eta}+\frac{1}{2}\frac{d^2\epsilon_{1}}{d\eta^2}\right)}
 {\left(1-\frac{2}{3}\epsilon_{1}\right)}
                    +\frac{1}{9}\frac{\left(\frac{d\epsilon_{1}}{d\eta}\right)^2}{\left(1-\frac{2}{3}\epsilon_{1}\right)^2}}^{Higher~order~slow-roll~correction}\\
                    \displaystyle~~~~~~~=\left\{\begin{array}{lll}\footnotesize
                    \displaystyle  
                    2a^2H^2+\frac{2}{3}\frac{\left(\frac{1}{z}\frac{dz}{d\eta}\frac{d\bar{\epsilon}_{V}}{d\eta}+\frac{1}{2}\frac{d^2\bar{\epsilon}_{V}}{d\eta^2}\right)}{\left(1-\frac{2}{3}\bar{\epsilon}_{V}\right)}
                    +\frac{1}{9}\frac{\left(\frac{d\bar{\epsilon}_{V}}{d\eta}\right)^2}{\left(1-\frac{2}{3}\bar{\epsilon}_{V}\right)^2}\\
                    \displaystyle~~~~~=a^2H^2\left[2+8\bar{\epsilon}_{V}-3\bar{\eta}_{V}+\left(3\bar{\epsilon}_{V}-\bar{\eta}_{V}\right)^2+\bar{\epsilon}_{V}\left(3\bar{\epsilon}_{V}-\bar{\eta}_{V}\right)
                    \right.\\ \left.~~~~~~~~~~~~~~\displaystyle +
                    \frac{1}{2}\left(2\bar{\xi}^2_{V}-5\bar{\eta}_{V}\bar{\epsilon}_{V}+36\bar{\epsilon}^2_{V}\right)\right]+\cdots \,,~~~~ &
 \mbox{\small {\bf for $q=1/2$}}  \\ 
 \displaystyle  
  2a^2H^2+\frac{1}{3q}\frac{\left(\frac{1}{z}\frac{dz}{d\eta}\frac{d\bar{\epsilon}_{V}}{d\eta}+\frac{1}{2}\frac{d^2\bar{\epsilon}_{V}}{d\eta^2}\right)}{\left(1-\frac{1}{3q}\bar{\epsilon}_{V}\right)}
                    +\frac{1}{36q^2}\frac{\left(\frac{d\bar{\epsilon}_{V}}{d\eta}\right)^2}{\left(1-\frac{1}{3q}\bar{\epsilon}_{V}\right)^2}\\
                    \displaystyle~~~~~=a^2H^2\left[2+\left(\frac{9}{\sqrt{2q}}-\frac{1}{2q}\right)\bar{\epsilon}_{V}-\frac{3}{\sqrt{2q}}\bar{\eta}_{V}+\frac{1}{2q}\left(3\bar{\epsilon}_{V}-\bar{\eta}_{V}\right)^2
                    
                    \right.\\ \left.~~~~~~~~~~~~~~\displaystyle +\frac{1}{(2q)^{3/2}}\bar{\epsilon}_{V}\left(3\bar{\epsilon}_{V}-\bar{\eta}_{V}\right)+
                    \frac{1}{2\sqrt{2q}}\left(2\bar{\xi}^2_{V}-5\bar{\eta}_{V}\bar{\epsilon}_{V}+36\bar{\epsilon}^2_{V}\right)\right]+\cdots \,,~~~ &
 \mbox{\small {\bf for ~any~$q$}}.
          \end{array}
\right.
\end{array}\ee
and the factor $aH$ can be expressed in terms of conformal time $\eta$ as:
\be\begin{array}{lll}\label{e8aate}\footnotesize
 \displaystyle \eta \approx\left\{\begin{array}{lll}\footnotesize
                    \displaystyle  
                    -\frac{1}{aH}\left(1+\bar{\epsilon}_{V}\right)+\cdots \,,~~~~ &
 \mbox{\small {\bf for $q=1/2$}}  \\ 
 \displaystyle  
 -\frac{1}{aH}\left(1+\frac{\bar{\epsilon}_{V}}{2q}\right)+\cdots \,,~~~ &
 \mbox{\small {\bf for ~any~arbitrary~$q$}}.
          \end{array}
\right.
\end{array}\ee
Further replacing the factor $aH$ in Eq~(\ref{rte}), we finally get the following simplified expression:
\be\begin{array}{lll}\label{rteq}\footnotesize
 \displaystyle \frac{1}{z}\frac{d^{2}z}{d\eta^2} 
                  \approx\left\{\begin{array}{lll}\footnotesize
                    \displaystyle  
                   \frac{1}{\eta^2}\left[\left(\frac{3}{2}+4\bar{\epsilon}_{V}-\bar{\eta}_{V}\right)^2-\frac{1}{4}\right]+\cdots \,,~~~~ &
 \mbox{\small {\bf for $q=1/2$}}  \\ 
 \displaystyle  
   \frac{1}{\eta^2}\left[\left\{\frac{3}{2}+\left(\frac{1}{2q}+\frac{3}{\sqrt{2q}}\right)\bar{\epsilon}_{V}-\frac{1}{\sqrt{2q}}\bar{\eta}_{V}\right\}^2-\frac{1}{4}\right]+\cdots  \,,~~~ &
 \mbox{\small {\bf for ~any~$q$}}.
          \end{array}
\right.
\end{array}\ee
 Now for further simplification in the computation of scalar power spectrum we introduce a new factor $\nu$ which is defined as:
 \be\begin{array}{lll}\label{rtex}\footnotesize
 \displaystyle \nu
                  \approx\left\{\begin{array}{lll}\footnotesize
                    \displaystyle  
                   \left(\frac{3}{2}+4\bar{\epsilon}_{V}-\bar{\eta}_{V}\right)+\cdots\,,~~~~ &
 \mbox{\small {\bf for $q=1/2$}}  \\ 
 \displaystyle  
  \left\{\frac{3}{2}+\left(\frac{1}{2q}+\frac{3}{\sqrt{2q}}\right)\bar{\epsilon}_{V}-\frac{1}{\sqrt{2q}}\bar{\eta}_{V}\right\}+\cdots  \,,~~~ &
 \mbox{\small {\bf for ~any~$q$}}.
          \end{array}
\right.
\end{array}\ee
Hence using Eq~(\ref{rtex}) in Eq~(\ref{diff}), we get the following simplified form of the Equation of motion:
\bea \label{diff2}
\left[\frac{d^{2}}{d\eta^2}+\left(c^{2}_{S}k^{2}-\frac{\left(\nu^2-\frac{1}{4}\right)}{\eta^2}\right)\right]v_{\bf k}(\eta)=0
\eea
and the most general solution of Eq~(\ref{diff2}) is given by:
\bea \label{solve} 
v_{\bf k}(\eta)=\sqrt{-\eta}\left[C_{1}H^{(1)}_{\nu}\left(-kc_{S}\eta\right)+C_{2}H^{(2)}_{\nu}\left(-kc_{S}\eta\right)\right].
\eea
where $C_{1}$ and $C_{2}$ are two arbitrary integration constants, which can be fixed from the appropriate choice of the boundary conditions.
Additionally $H^{(1)}_{\nu}$ and $H^{(2)}_{\nu}$ represent the Hankel function of the first and second kind with rank $\nu$. Now to impose the well known 
Bunch-Davies boundary condition at early times we have used:
\bea \label{lima}
\lim_{kc_{S}\eta\rightarrow-\infty} H^{(1)}_{\nu}\left(-kc_{S}\eta\right)&=&\sqrt{\frac{2}{\pi}}\frac{1}{\sqrt{-\eta}}\exp\left(ikc_{S}\eta\right)\exp\left(i\frac{\pi}{2}\left(\nu+\frac{1}{2}\right)\right),\\
\lim_{kc_{S}\eta\rightarrow-\infty} H^{(2)}_{\nu}\left(-kc_{S}\eta\right)&=&\sqrt{\frac{2}{\pi}}\frac{1}{\sqrt{-\eta}}\exp\left(-ikc_{S}\eta\right)\exp\left(-i\frac{\pi}{2}\left(\nu+\frac{1}{2}\right)\right).
\eea
As a result the previously mentioned integration constants are fixed at the  values:$ C_{1}=\sqrt{\frac{\pi}{2}},~~~~  C_{2}= 0.$
Consequently the solution of the mode function for scalar fluctuations takes the following form:
\bea \label{solve2} 
v_{\bf k}(\eta)=\sqrt{-\frac{\eta\pi}{2}}H^{(1)}_{\nu}\left(-kc_{S}\eta\right).
\eea
On the other hand, the solution stated in Eq~(\ref{solve}) determines the future evolution of the mode including its super-horizon dynamics at
$c_{S}k<<aH$ or $|k c_{S}\eta|<<1$  or $k c_{S}\eta\rightarrow 0$ and this is due to:
\bea \label{lima2}
\lim_{kc_{S}\eta\rightarrow 0} H^{(1)}_{\nu}\left(-kc_{S}\eta\right)&=&\frac{i}{\pi}\Gamma(\nu)\left(\frac{-kc_{S}\eta}{2}\right)^{-\nu}.
\eea
Consequently the solution of the mode function for scalar fluctuations takes the following form:
\bea \label{solve3} 
v_{\bf k}(\eta)=\sqrt{-\frac{\eta\pi}{2}}\frac{i}{\pi}\Gamma(\nu)\left(\frac{-kc_{S}\eta}{2}\right)^{-\nu}.
\eea
Finally combining the results obtained in Eq~(\ref{solve}), Eq~(\ref{solve2}) and Eq~(\ref{solve3}) we get:
\be\begin{array}{lll}\label{rtex}\footnotesize
 \displaystyle v_{\bf k}(\eta)
                  =\left\{\begin{array}{lll}\footnotesize
                    \displaystyle  
                  \sqrt{-\eta}\left[C_{1}H^{(1)}_{\nu}\left(-kc_{S}\eta\right)+C_{2}H^{(2)}_{\nu}\left(-kc_{S}\eta\right)\right],~~~~ &
 \mbox{\small {\bf for AV}}  \\ 
 \displaystyle  
  \sqrt{-\frac{\eta\pi}{2}}H^{(1)}_{\nu}\left(-kc_{S}\eta\right)  \,,~~~ &
 \mbox{\small {\bf for BD + $|k c_{S}\eta|>>1$}} \\ 
 \displaystyle  
  \sqrt{-\frac{\eta\pi}{2}}\frac{i}{\pi}\Gamma(\nu)\left(\frac{-kc_{S}\eta}{2}\right)^{-\nu} \,,~~~ &
 \mbox{\small {\bf for BD + $|k c_{S}\eta|<<1$}}.
          \end{array}
\right.
\end{array}\ee
where ${\bf AV}$ and ${\bf BD}$ signify arbitrary vacuum and Bunch-Davies vacuum respectively.
Finally the two point function from scalar fluctuation for both ${\bf AV}$ and ${\bf BD}$ can be expressed as:
\begin{eqnarray}\footnotesize
 \label{sc2}\langle \zeta_{\bf k}\zeta_{\bf k^{'}}\rangle =\left(\frac{H}{\dot{T}}\right)^{2}\langle \delta T_{\bf k}
 \delta T_{\bf k^{'}}\rangle =\frac{1}{z^{2}M^{2}_{p}}\langle v_{\bf k}
 v_{\bf k^{'}}\rangle =(2\pi)^{3}\delta^{3}({\bf k}+{\bf k^{'}})\frac{2\pi^{2}}{k^{3}}\Delta_{\zeta}(k),~~~~~~
\end{eqnarray}
where the primordial power spectrum for scalar modes at any
arbitrary momentum scale $k$ can be written for both ${\bf AV}$ and ${\bf BD}$ with $q=1/2$ as:
\begin{eqnarray}
\label{ps2} 
\label{ps1} \footnotesize\Delta_{\zeta}(k)&\equiv& \frac{k^{3}P_{\zeta}(k)}{2\pi^2}=\frac{k^{3}|v_{k}|^{2}}{2\pi^{2}z^{2}M^{2}_{p}}\nonumber\\ 
&=&\left\{\begin{array}{lll}\footnotesize
                    \displaystyle  \left\{\begin{array}{lll}
                    \displaystyle  
                 \frac{2^{2\nu-3}\left(-k\eta c_{S}\right)^{3-2\nu} H^2}{8c_{S}\bar{\epsilon}_{V}(1+\bar{\epsilon}_{V})^2 \pi^{2}M^{2}_{p}}\left|\frac{\Gamma(\nu)}{\Gamma\left(\frac{3}{2}\right)}\right|^{2}\,,~~~~ &
 \mbox{\small {\bf for $|k c_{S}\eta|<<1$}}  \\ 
 \displaystyle  
                \frac{2^{2\nu-3}c^{2-2\nu}_{S}\left(1+\bar{\epsilon}_{V}\right)^{1-2\nu} H^2}{8\bar{\epsilon}_{V} \pi^{2}M^{2}_{p}}\left|\frac{\Gamma(\nu)}{\Gamma\left(\frac{3}{2}\right)}\right|^{2}\,,~~~~ &
 \mbox{\small {\bf for $|k c_{S}\eta|=1$}}  \\ 
 \displaystyle  
  \frac{\left(-k\eta c_{S}\right)^3 H^2|H^{(1)}_{\nu}\left(-kc_{S}\eta\right)|^{2}}{8c_{S}\bar{\epsilon}_{V}(1+\bar{\epsilon}_{V})^2 \pi M^{2}_{p}}\,,~~~ &
 \mbox{\small {\bf for $|k c_{S}\eta|>>1$}}.
          \end{array}
\right.~~~~~~~~~ &
 \mbox{\small \underline{\bf for BD}}  \\ 
 \displaystyle  
 \frac{\left(-k\eta c_{S}\right)^3 H^2\left|C_{1}H^{(1)}_{\nu}\left(-kc_{S}\eta\right)+C_{2}H^{(2)}_{\nu}
 \left(-kc_{S}\eta\right)\right|^2}{4c_{S}\bar{\epsilon}_{V}(1+\bar{\epsilon}_{V})^2 \pi^{2}M^{2}_{p}}\,,~~~~~~~~~ &
 \mbox{\small \underline{\bf for AV}}.
 \end{array}
\right.\end{eqnarray}
and similarly the primordial power spectrum for scalar modes at any
arbitrary momentum scale $k$ can be written for both ${\bf AV}$ and ${\bf BD}$ with any arbitrary $q$ as:
\begin{eqnarray}
\label{ps1} \footnotesize\Delta_{\zeta}(k)&\equiv& \frac{k^{3}P_{\zeta}(k)}{2\pi^2}=\frac{k^{3}|v_{k}|^{2}}{2\pi^{2}z^{2}M^{2}_{p}}\nonumber\\ 
&=&\left\{\begin{array}{lll}\footnotesize
                    \displaystyle  \left\{\begin{array}{lll}
                    \displaystyle  
                 \frac{2^{2\nu-3}q\left(-k\eta c_{S}\right)^{3-2\nu} H^2}{4c_{S}\bar{\epsilon}_{V}\left(1+\frac{1}{2q}\bar{\epsilon}_{V}\right)^2  \pi^{2}M^{2}_{p}}\left|\frac{\Gamma(\nu)}{\Gamma\left(\frac{3}{2}\right)}\right|^{2}\,,~~~~ &
 \mbox{\small {\bf for $|k c_{S}\eta|<<1$}}  \\ 
 \displaystyle  
                \frac{2^{2\nu-3}q ~c^{2-2\nu}_{S}\left(1+\frac{1}{2q}\bar{\epsilon}_{V}\right)^{1-2\nu} H^2}{4\bar{\epsilon}_{V} \pi^{2}M^{2}_{p}}\left|\frac{\Gamma(\nu)}{\Gamma\left(\frac{3}{2}\right)}\right|^{2}\,,~~~~ &
 \mbox{\small {\bf for $|k c_{S}\eta|=1$}}  \\ 
 \displaystyle  
  \frac{q\left(-k\eta c_{S}\right)^3 H^2|H^{(1)}_{\nu}\left(-kc_{S}\eta\right)|^{2}}{4c_{S}\bar{\epsilon}_{V}\left(1+\frac{1}{2q}\bar{\epsilon}_{V}\right)^2 \pi M^{2}_{p}}\,,~~~ &
 \mbox{\small {\bf for $|k c_{S}\eta|>>1$}}.
          \end{array}
\right.~~~~~~~~~ &
 \mbox{\small \underline{\bf for BD}}  \\ 
 \displaystyle  
 \frac{q\left(-k\eta c_{S}\right)^3 H^2\left|C_{1}H^{(1)}_{\nu}\left(-kc_{S}\eta\right)+C_{2}H^{(2)}_{\nu}
 \left(-kc_{S}\eta\right)\right|^2}{2c_{S}\bar{\epsilon}_{V}\left(1+\frac{1}{2q}\bar{\epsilon}_{V}\right)^2 \pi^{2}M^{2}_{p}}\,,~~~~~~~~~ &
 \mbox{\small \underline{\bf for AV}}.
 \end{array}
\right.\end{eqnarray}
where the effective sound speed $c_{S}$ is given by:
\be\begin{array}{lll}\label{rteqxc}\footnotesize
 \displaystyle c_{S} 
                  =\left\{\begin{array}{lll}\footnotesize
                    \displaystyle  
                   \sqrt{1-\frac{2}{3} {\epsilon}_{1}}=\sqrt{1-\frac{2}{3}\bar{\epsilon}_{V}}\,,~~~~ &
 \mbox{\small {\bf for $q=1/2$}}  \\
 \displaystyle  
  \sqrt{\frac{1-\frac{2}{3} {\epsilon}_{1}}{1+\frac{2(1-2q)}{3q} {\epsilon}_{1}}}=\sqrt{\frac{1-\frac{1}{3q} \bar{\epsilon}_{V}}{1+\frac{(1-2q)}{3q^2}  \bar{\epsilon}_{V}}} \,,~~~ &
 \mbox{\small {\bf for ~any~$q$}}.
          \end{array}
\right.
\end{array}\ee
Now starting from the expression for primordial power spectrum for scalar modes one can compute the spectral tilt at any
 arbitrary momentum scale $k$ for both ${\bf AV}$ and ${\bf BD}$ with $q=1/2$ as:
\begin{eqnarray}\footnotesize
\label{ps11} n_{\zeta}(k)-1 &\equiv& \frac{d\ln \Delta_{\zeta}(k)}{d\ln k} = \frac{d\ln \Delta_{\zeta}(k)}{d N}\nonumber\\ 
&\approx&\left\{\begin{array}{lll}\footnotesize
                    \displaystyle \footnotesize \left\{\begin{array}{lll}
                    \displaystyle \footnotesize
                 (3-2\nu)\left[1-\frac{2}{3}\bar{\epsilon}_{V}(\bar{\eta}_{V}-3\bar{\epsilon}_{V})\left(1-\frac{2}{3}\bar{\epsilon}_{V}\right)^{1/4}\right]+\cdots, &
 \mbox{\small {\bf for $|k c_{S}\eta|<<1$}}  \\ 
 \displaystyle \small 
                (3-2\nu)\left[1-\frac{2}{3}\bar{\epsilon}_{V}(\bar{\eta}_{V}-3\bar{\epsilon}_{V})\left(1-\frac{2}{3}\bar{\epsilon}_{V}\right)^{1/4}\right]+\cdots\,,&
 \mbox{\small {\bf for $|k c_{S}\eta|=1$}}  \\ 
 \displaystyle  \small
  \bar{\epsilon}_{V}-\left[\frac{2}{3}\bar{\epsilon}_{V}(\bar{\eta}_{V}-3\bar{\epsilon}_{V}) +2(\bar{\eta}_{V}-3\bar{\epsilon}_{V})\right]\left(1-\frac{2}{3}\bar{\epsilon}_{V}\right)^{1/4}\\
  \displaystyle +\frac{\left(-c_{S}\eta\right)\left[H^{(1)}_{\nu-1}\left(-kc_{S}\eta\right)-H^{(1)}_{\nu+1}
 \left(-kc_{S}\eta\right)\right]}{H^{(1)}_{\nu}\left(-kc_{S}\eta\right)}+\cdots\,, &
 \mbox{\small {\bf for $|k c_{S}\eta|>>1$}}.
          \end{array}
\right. &
 \mbox{\small \underline{\bf for BD}}  \\ 
 \displaystyle  
 \bar{\epsilon}_{V}-\left[\frac{2}{3}\bar{\epsilon}_{V}(\bar{\eta}_{V}-3\bar{\epsilon}_{V}) +2(\bar{\eta}_{V}-3\bar{\epsilon}_{V})\right]\left(1-\frac{2}{3}\bar{\epsilon}_{V}\right)^{1/4}\\
  \displaystyle +\frac{\left(-c_{S}\eta\right)C_{1}\left[H^{(1)}_{\nu-1}\left(-kc_{S}\eta\right)-H^{(1)}_{\nu+1}
 \left(-kc_{S}\eta\right)\right]}{\left[C_{1}H^{(1)}_{\nu}\left(-kc_{S}\eta\right)+C_{2}H^{(2)}_{\nu}
 \left(-kc_{S}\eta\right)\right]}\\
 \displaystyle +\frac{\left(-c_{S}\eta\right)C_{2}\left[H^{(2)}_{\nu-1}\left(-kc_{S}\eta\right)-H^{(2)}_{\nu+1}
 \left(-kc_{S}\eta\right)\right]}{\left[C_{1}H^{(1)}_{\nu}\left(-kc_{S}\eta\right)+C_{2}H^{(2)}_{\nu}
 \left(-kc_{S}\eta\right)\right]}+\cdots
 \,, &
 \mbox{\small \underline{\bf for AV}}.
 \end{array}
\right.\end{eqnarray}
and for both ${\bf AV}$ and ${\bf BD}$ with any arbitrary $q$ as:
\begin{eqnarray}
\label{ps111} n_{\zeta}(k)-1 &\equiv& \frac{d\ln \Delta_{\zeta}(k)}{d\ln k} = \frac{d\ln \Delta_{\zeta}(k)}{d N}\nonumber\\ 
&=&\left\{\begin{array}{lll}
                    \displaystyle \footnotesize \left\{\begin{array}{lll}\footnotesize
                    \displaystyle  
                  (3-2\nu)\left[1-\frac{1}{3q}\bar{\epsilon}_{V}(\bar{\eta}_{V}-3\bar{\epsilon}_{V})\left(1-\frac{1}{3q}\bar{\epsilon}_{V}\right)^{1/4}\right]+\cdots\,,&
 \mbox{\small {\bf for $|k c_{S}\eta|<<1$}}  \\ 
 \displaystyle  
                 (3-2\nu)\left[1-\frac{1}{3q}\bar{\epsilon}_{V}(\bar{\eta}_{V}-3\bar{\epsilon}_{V})\left(1-\frac{1}{3q}\bar{\epsilon}_{V}\right)^{1/4}\right]+\cdots\,, &
 \mbox{\small {\bf for $|k c_{S}\eta|=1$}}  \\ 
 \displaystyle  
  \small
  \frac{\bar{\epsilon}_{V}}{2q}-\left[\frac{1}{3q}\bar{\epsilon}_{V}(\bar{\eta}_{V}-3\bar{\epsilon}_{V}) +\frac{1}{q}(\bar{\eta}_{V}-3\bar{\epsilon}_{V})\right]\left(1-\frac{2}{3}\bar{\epsilon}_{V}\right)^{1/4}\\
  \displaystyle +\frac{\left(-c_{S}\eta\right)\left[H^{(1)}_{\nu-1}\left(-kc_{S}\eta\right)-H^{(1)}_{\nu+1}
 \left(-kc_{S}\eta\right)\right]}{H^{(1)}_{\nu}\left(-kc_{S}\eta\right)}+\cdots\,, &
 \mbox{\small {\bf for $|k c_{S}\eta|>>1$}}.
          \end{array}
\right. &
 \mbox{\small \underline{\bf for BD}}  \\ 
 \displaystyle  
 \frac{\bar{\epsilon}_{V}}{2q}-\left[\frac{1}{3q}\bar{\epsilon}_{V}(\bar{\eta}_{V}-3\bar{\epsilon}_{V})
  \displaystyle +\frac{1}{q}(\bar{\eta}_{V}-3\bar{\epsilon}_{V})\right]\left(1-\frac{2}{3}\bar{\epsilon}_{V}\right)^{1/4}\\
  \displaystyle +\frac{\left(-c_{S}\eta\right)C_{1}\left[H^{(1)}_{\nu-1}\left(-kc_{S}\eta\right)-H^{(1)}_{\nu+1}
 \left(-kc_{S}\eta\right)\right]}{\left[C_{1}H^{(1)}_{\nu}\left(-kc_{S}\eta\right)+C_{2}H^{(2)}_{\nu}
 \left(-kc_{S}\eta\right)\right]}\\
 \displaystyle +\frac{\left(-c_{S}\eta\right)C_{1}\left[H^{(2)}_{\nu-1}\left(-kc_{S}\eta\right)-H^{(2)}_{\nu+1}
 \left(-kc_{S}\eta\right)\right]}{\left[C_{1}H^{(1)}_{\nu}\left(-kc_{S}\eta\right)+C_{2}H^{(2)}_{\nu}
 \left(-kc_{S}\eta\right)\right]}+\cdots\,, &
 \mbox{\small \underline{\bf for AV}}.
 \end{array}
\right.\end{eqnarray}
One can also consider the following approximations to simplify the final derived form of the primordial scalar power spectrum for {\bf BD} vacuum with $|k c_{S}\eta|=1$ case:
\begin{enumerate}
 \item We start with the {\it Laurent expansion} of the Gamma function:
       \bea \label{r1zax} \Gamma(\nu) &=& \frac{1}{\nu}-\gamma+\frac{1}{2}\left(\gamma^2+\frac{\pi^2}{6}\right)\nu-\frac{1}{6}\left(\gamma^3+\frac{\gamma \pi^2}{2}+2\zeta(3)\right)\nu^2 +{\cal O}(\nu^3).~~~~~~~~~~~~
       \eea
       where $\gamma$ being the Euler Mascheroni constant and $\zeta(3)$ characterizing the Reimann zeta function of 
       order $3$ originating in the expansion of the gamma function. 
 \item  Hence using the result of Eq~(\ref{r1zax}) for $q=1/2$ and for arbitrary $q$ we can write:
 \be\begin{array}{lll}\label{r2x}
 \displaystyle \Gamma(\nu) 
                  =\left\{\begin{array}{lll}
                    \displaystyle  
                   \frac{1}{\left(\frac{3}{2}+4\bar{\epsilon}_{V}-\bar{\eta}_{V}\right)}-\gamma+\frac{1}{2}
                   \left(\gamma^2+\frac{\pi^2}{6}\right)\left(\frac{3}{2}+4\bar{\epsilon}_{V}-\bar{\eta}_{V}\right)\\
                   \displaystyle~~~~~~~~-\frac{1}{6}\left(\gamma^3+\frac{\gamma \pi^2}{2}+2\zeta(3)\right)
                   \left(\frac{3}{2}+4\bar{\epsilon}_{V}-\bar{\eta}_{V}\right)^2 +\cdots\,,~~~~ &
 \mbox{\small {\bf for $q=1/2$}}  \\
 \displaystyle  
  \frac{1}{ \left\{\frac{3}{2}+\left(\frac{1}{2q}+\frac{3}{\sqrt{2q}}\right)\bar{\epsilon}_{V}-\frac{1}{\sqrt{2q}}\bar{\eta}_{V}\right\}}
  -\gamma \\ \displaystyle ~~~+\frac{1}{2}\left(\gamma^2+\frac{\pi^2}{6}\right) \left\{\frac{3}{2}+\left(\frac{1}{2q}+\frac{3}{\sqrt{2q}}\right)\bar{\epsilon}_{V}-\frac{1}{\sqrt{2q}}\bar{\eta}_{V}\right\}
  \\ \displaystyle~-\frac{1}{6}\left(\gamma^3+\frac{\gamma \pi^2}{2}+2\zeta(3)\right) \left\{\frac{3}{2}+\left(\frac{1}{2q}+\frac{3}{\sqrt{2q}}\right)\bar{\epsilon}_{V}-\frac{1}{\sqrt{2q}}\bar{\eta}_{V}\right\}^2
  +\cdots \,,~~~ &
 \mbox{\small {\bf for ~any~$q$}}.
          \end{array}
\right.
\end{array}\ee
\item In the slow-roll regime of inflation all the slow-roll parameters satisfy the following constraint:
          \bea \bar{\epsilon}_{V}&<<& 1,\\
               |\bar{\eta}_{V}|&<<& 1,\\
               |\bar{\xi}^{2}_{V}|&<<& 1,\\ 
               |\bar{\sigma}^{3}_{V}|&<<& 1.\eea
               Using these approximations the primordial scalar power spectrum can be expressed as:
               \begin{eqnarray} 
\label{psxcc1} \Delta_{\zeta, \star}
&\approx&\left\{\left[1-({\cal C}_{E}+1){\epsilon}_{1}-\frac{{\cal C}_{E}}{2}\epsilon_{2}\right]^2 \frac{H^{2}}{8\pi^2M^{2}_{p}c_{S}{\epsilon}_{1}}\right\}_{k_{\star}=a_{\star}H_{\star}}\nonumber\\
&=&\left\{\begin{array}{lll}
                    \displaystyle 
 \displaystyle  \left\{\left[1-({\cal C}_{E}+1)\bar{\epsilon}_{V}-{\cal C}_{E}\left(3\bar{\epsilon}_{V}-\bar{\eta}_{V}\right)\right]^2 \frac{H^{2}}{8\pi^2M^{2}_{p}c_{S}\bar{\epsilon}_{V}}\right\}_{k_{\star}=a_{\star}H_{\star}}\,,~~~~~~~~~ &
 \mbox{\small {\bf for $q=1/2$}}  \\ 
 \displaystyle \left\{\left[1-({\cal C}_{E}+1)\frac{\bar{\epsilon}_{V}}{2q}-\frac{{\cal C}_{E}}{\sqrt{2q}}\left(3\bar{\epsilon}_{V}-\bar{\eta}_{V}\right)\right]^2 \frac{q H^{2}}{4\pi^2M^{2}_{p}c_{S}\bar{\epsilon}_{V}}\right\}_{k_{\star}=a_{\star}H_{\star}} \,,~~~~~~~~~ &
 \mbox{\small {\bf for any $q$}}.
 \end{array}
\right.\end{eqnarray}
where ${\cal C}_{E}$ is given by:
\bea  
{\cal C}_{E}= -2 + \ln 2 + \gamma \approx -0.72. 
\eea
\item Using the slow-roll approximations one can further approximate the expression for sound speed as:
\be\begin{array}{lll}\label{rteqxccc}
 \displaystyle c^2_{S} 
                  =\left\{\begin{array}{lll}
                    \displaystyle  
                  1-\frac{2}{3}\bar{\epsilon}_{V}+{\cal O}(\bar{\epsilon}^2_{V})+\cdots \,,~~~~ &
 \mbox{\small {\bf for $q=1/2$}}  \\ 
 \displaystyle  
 1-\frac{(1-q)}{3q^2}\bar{\epsilon}_{V}+{\cal O}(\bar{\epsilon}^2_{V})+\cdots \,,~~~ &
 \mbox{\small {\bf for ~any~$q$}}.
          \end{array}
\right.
\end{array}\ee
\item Hence using the result in Eq~(\ref{psxcc1}) we get the following simplified expression for the primordial scalar power spectrum:
\begin{eqnarray} 
\label{psxcc1v0} \Delta_{\zeta, \star}
&\approx&\left\{\begin{array}{lll}
                    \displaystyle 
 \displaystyle  
                \left\{\left[1-\left({\cal C}_{E}+\frac{5}{6}\right)\bar{\epsilon}_{V}-{\cal C}_{E}
                \left(3\bar{\epsilon}_{V}-\bar{\eta}_{V}\right)\right]^2 \frac{H^{2}}{8\pi^2M^{2}_{p}\bar{\epsilon}_{V}}\right\}_{k_{\star}=a_{\star}H_{\star}} \,,~~~~ &
 \mbox{\small {\bf for $q=1/2$}}  \\ 
 \displaystyle  
                \left\{\left[1-({\cal C}_{E}+1-\Sigma)\frac{\bar{\epsilon}_{V}}{2q}-\frac{{\cal C}_{E}}{\sqrt{2q}}\left(3\bar{\epsilon}_{V}-\bar{\eta}_{V}\right)\right]^2 \frac{q H^{2}}{4\pi^2M^{2}_{p}
                \bar{\epsilon}_{V}}\right\}_{k_{\star}=a_{\star}H_{\star}} \,,~~~~ &
 \mbox{\small {\bf for any $q$}}.
 \end{array}
\right.\end{eqnarray}
where the factor $\Sigma$ is defined as:
\begin{eqnarray} 
\label{psxcc1v1} \Sigma
&=&\left\{\begin{array}{lll}
                    \displaystyle 
 \displaystyle  
               \frac{1}{6} \,,~~~~ &
 \mbox{\small {\bf for $q=1/2$}}  \\ 
 \displaystyle  
               \frac{1-q}{6q} \,,~~~~ &
 \mbox{\small {\bf for any $q$}}.
 \end{array}
\right.\end{eqnarray}
\item Next one can compute the scalar spectral tilt ($n_{S}$) of the primordial scalar power spectrum as:
\begin{eqnarray} 
\label{psxcc1v2} n_{\zeta, \star}-1 
&\approx&\left\{\begin{array}{lll}
                    \displaystyle 
 \displaystyle  -2{\epsilon}_{1}-{\epsilon}_{2}-2{\epsilon}^{2}_{1}-\left(2{\cal C}_{E}+\frac{8}{3}\right){\epsilon}_{1}{\epsilon}_{2}-{\cal C}_{E}{\epsilon}_{2}{\epsilon}_{3}+\cdots\\
               \displaystyle
                =2\bar{\eta}_{V}-8\bar{\epsilon}_{V}-2\bar{\epsilon}^{2}_{V}-2\left(2{\cal C}_{E}+\frac{8}{3}\right)\bar{\epsilon}_{V}\left(3\bar{\epsilon}_{V}-\bar{\eta}_{V}\right)\\
               \displaystyle~~~~~~~-{\cal C}_{E}\left(2\bar{\xi}^{2}_{V}-5\bar{\eta}_{V}\bar{\epsilon}_{V}+36\bar{\epsilon}^{2}_{V}\right)+\cdots,~~ &
 \mbox{\small {\bf for $q=1/2$}}  \\ 
 \displaystyle  
               \displaystyle  -2{\epsilon}_{1}-{\epsilon}_{2}-2{\epsilon}^{2}_{1}-\left(2{\cal C}_{E}+3-2\Sigma\right){\epsilon}_{1}{\epsilon}_{2}-{\cal C}_{E}{\epsilon}_{2}{\epsilon}_{3}+\cdots\\
               \displaystyle
                =\sqrt{\frac{2}{q}}\bar{\eta}_{V}-\left(\frac{1}{q}+3\sqrt{\frac{2}{q}}\right)\bar{\epsilon}_{V}-\frac{\bar{\epsilon}^{2}_{V}}{2q^2}\\
                \displaystyle -\frac{2}{(2q)^{3/2}}\left(2{\cal C}_{E}+3-2\Sigma\right)\bar{\epsilon}_{V}\left(3\bar{\epsilon}_{V}-\bar{\eta}_{V}\right)\\
               \displaystyle~~~~~~~-\frac{{\cal C}_{E}}{\sqrt{2q}}\left(2\bar{\xi}^{2}_{V}-5\bar{\eta}_{V}\bar{\epsilon}_{V}+36\bar{\epsilon}^{2}_{V}\right)+\cdots\,,~~ &
 \mbox{\small {\bf for any $q$}}.
 \end{array}
\right.\end{eqnarray}
\item Next one can compute the running of the scalar spectral tilt ($\alpha_{S}$) of the primordial scalar power spectrum as:
\begin{eqnarray}
\label{psxcc1v3} \alpha_{\zeta, \star}&=&\left(\frac{dn_{\zeta}(k)}{d\ln k}\right)_{k_{\star}=a_{\star}H_{\star}}=\left(\frac{dn_{\zeta}(k)}{dN}\right)_{k_{\star}=a_{\star}H_{\star}}\\
&\approx&\small\left\{ \small\begin{array}{lll}
                    \displaystyle 
\displaystyle -\left\{\left[4\bar{\epsilon}_{V}(1+\bar{\epsilon}_{V})(\bar{\eta}_{V}-3\bar{\epsilon}_{V})+2\left(10\bar{\epsilon}_{V}\bar{\eta}_{V}
                -18\bar{\epsilon}^2_{V}-\bar{\xi}^2_{V}\right)\right]\right.\nonumber\\ \left.
                \displaystyle -{\cal C}_{E}\left(2\bar{\sigma}^{3}_{V}-216\bar{\epsilon}^{3}_{V}+2\bar{\xi}^{2}_{V}\bar{\eta}_{V}-7\bar{\xi}^{2}_{V}\bar{\epsilon}_{V}
                +194\bar{\epsilon}^2_{V}\bar{\eta}_{V}-10\bar{\eta}_{V}\bar{\epsilon}_{V}
                   \right)\displaystyle\right.\\ \left.
               \displaystyle-\left(2{\cal C}_{E}+\frac{8}{3}\right)\left[2\bar{\epsilon}_{V}\left(10\bar{\epsilon}_{V}\bar{\eta}_{V}-18\bar{\epsilon}^2_{V}
               -\bar{\xi}^2_{V}\right)\right.\right.\\ \left.\left.
               \displaystyle -4\bar{\epsilon}_{V}\left(3\bar{\epsilon}_{V}-\bar{\eta}_{V}\right)^2\right]\right\}\displaystyle\left(1-\frac{2}{3}\bar{\epsilon}_{V}\right)^{1/4}+\cdots,~~ &
 \mbox{\small {\bf for $q=1/2$}}  \\ 
 \displaystyle  
               -\left\{\left[\sqrt{\frac{2}{q}}\frac{\bar{\epsilon}_{V}}{q}\left(1+\frac{\bar{\epsilon}_{V}}{2q}\right)(\bar{\eta}_{V}-3\bar{\epsilon}_{V})+\sqrt{\frac{2}{q}}\left(10\bar{\epsilon}_{V}\bar{\eta}_{V}
                -18\bar{\epsilon}^2_{V}-\bar{\xi}^2_{V}\right)\right]\right.\nonumber\\ \left.
                \displaystyle -\frac{{\cal C}_{E}}{\sqrt{2q}}\left(2\bar{\sigma}^{3}_{V}-216\bar{\epsilon}^{3}_{V}+2\bar{\xi}^{2}_{V}\bar{\eta}_{V}-7\bar{\xi}^{2}_{V}\bar{\epsilon}_{V}
                +194\bar{\epsilon}^2_{V}\bar{\eta}_{V}-10\bar{\eta}_{V}\bar{\epsilon}_{V}
                   \right)\right.\nonumber\\ \left.
               \displaystyle-\left(2{\cal C}_{E}+\frac{8}{3}\right)\left[\sqrt{\frac{2}{q}}\bar{\epsilon}_{V}\left(10\bar{\epsilon}_{V}\bar{\eta}_{V}-18\bar{\epsilon}^2_{V}
               -\bar{\xi}^2_{V}\right)\right.\right.\\ \left.\left.\displaystyle
               -\frac{4}{(2q)^{3/2}}\bar{\epsilon}_{V}\left(3\bar{\epsilon}_{V}-\bar{\eta}_{V}\right)^2\right]\right\}\displaystyle\left(1-\frac{1}{3q}\bar{\epsilon}_{V}\right)^{1/4}+\cdots\,,~~ &
 \mbox{\small {\bf for any $q$}}.
 \end{array}
\right.\end{eqnarray}
\item Finally, one can also compute the running of the running of scalar spectral tilt ($\kappa_{S}$) of the primordial scalar power spectrum as:
\begin{eqnarray}
\label{psxcc1v4} \kappa_{\zeta, \star}&=&\left(\frac{d^2n_{\zeta}(k)}{d\ln k^2}\right)_{k_{\star}=a_{\star}H_{\star}}
=\left(\frac{d^2n_{\zeta}(k)}{dN^2}\right)_{k_{\star}=a_{\star}H_{\star}}\nonumber\\
&\approx&\left\{ \small\begin{array}{lll}
                    \displaystyle 
\displaystyle -\left[8\bar{\epsilon}_{V}(1+\bar{\epsilon}_{V})(\bar{\eta}_{V}-3\bar{\epsilon}_{V})^2+8\bar{\epsilon}^2_{V}(\bar{\eta}_{V}-3\bar{\epsilon}_{V})^2\right.\\ \left. \displaystyle 
+8\bar{\epsilon}_{V}(1+\bar{\epsilon}_{V})\left(\bar{\xi}^2_{V}-10\bar{\epsilon}_{V}\bar{\eta}_{V}
+18\bar{\epsilon}^{2}_{V}\right)\right.\\ \left. \displaystyle  +2\left(20\bar{\epsilon}_{V}\bar{\eta}_{V}
\left(\bar{\eta}_{V}-3\bar{\epsilon}_{V}\right)+10\bar{\epsilon}_{V}\left(\bar{\xi}^2_{V}-4\bar{\epsilon}_{V}\bar{\eta}_{V}\right)
                \right.\right.\\ \left.\left.\displaystyle -72\bar{\epsilon}^2_{V}\left(\bar{\eta}_{V}-3\bar{\epsilon}_{V}\right)-
                \left(\bar{\sigma}^3_{V}+\bar{\xi}^2_{V}\bar{\eta}_{V}-\bar{\xi}^2_{V}\bar{\epsilon}_{V}\right)\right)\right]\displaystyle
                \left(1-\frac{2}{3}\bar{\epsilon}_{V}\right)^{1/4}+\cdots,~~ &
 \mbox{\small {\bf for $q=1/2$}}  \\ 
 \displaystyle  
              
                \displaystyle 
\displaystyle -\left[\left(\frac{2}{q}\right)^{3/2}\bar{\epsilon}_{V}\left(1+\frac{\bar{\epsilon}_{V}}{2q}\right)(\bar{\eta}_{V}
-3\bar{\epsilon}_{V})^2+\sqrt{\frac{2}{q}}\frac{2}{q^2}\bar{\epsilon}^2_{V}(\bar{\eta}_{V}-3\bar{\epsilon}_{V})^2\right.\\ \left. \displaystyle 
+\sqrt{\frac{2}{q}}\frac{2}{q}\bar{\epsilon}_{V}\left(1+\frac{\bar{\epsilon}_{V}}{2q}\right)\left(\bar{\xi}^2_{V}-10\bar{\epsilon}_{V}\bar{\eta}_{V}
+18\bar{\epsilon}^{2}_{V}\right)\right.\\ \left. \displaystyle  +\sqrt{\frac{2}{q}}\left(\frac{10}{q}\bar{\epsilon}_{V}\bar{\eta}_{V}
\left(\bar{\eta}_{V}-3\bar{\epsilon}_{V}\right)+10\bar{\epsilon}_{V}\left(\bar{\xi}^2_{V}-4\bar{\epsilon}_{V}\bar{\eta}_{V}\right)
                \right.\right.\\ \left.\left.\displaystyle -\frac{36}{q}\bar{\epsilon}^2_{V}\left(\bar{\eta}_{V}-3\bar{\epsilon}_{V}\right)-
                \left(\bar{\sigma}^3_{V}+\bar{\xi}^2_{V}\bar{\eta}_{V}-\bar{\xi}^2_{V}\bar{\epsilon}_{V}\right)\right)\right]\displaystyle
                \left(1-\frac{1}{3q}\bar{\epsilon}_{V}\right)^{1/4}+\cdots,~~ &
 \mbox{\small {\bf for any $q$}}.
 \end{array}
\right.\end{eqnarray}
\end{enumerate}

\subsubsection{Computation of tensor power spectrum}

In this subsection our prime objective is to compute the primordial power spectra of tensor quantum fluctuations from tachyonic inflation and study the cosmological consequences from previously mentioned string theory originated 
tachyonic potentials in the light of Planck 2015 data. To serve this purpose let us start with the following canonical variable $u_{\bf k}$, which can be quantized with the standard techniques:
\bea u^{\gamma}_{\bf k}\equiv \frac{a}{\sqrt{2}}~M_{p}~h^{\gamma}_{\bf k}, \eea
where $h^{\gamma}_{\bf k}$ is the curvature perturbation in the momentum space, which can be expressed in terms of the curvature perturbation in position space through the following Fourier transformation:
\bea h^{\gamma}(t,{\bf x})=\int d^{3}k~h^{\gamma}_{\bf k}(t)~\exp(i{\bf k}.{\bf x}). \eea
Here the superscript $\gamma$ stands for the helicity index for the transverse and traceless spin-2 graviton degrees of freedom. In general
tensor modes can be written in terms of the two orthogonal polarization basis vectors.

Further we derive the equation of motion of the tensor fluctuation by extremizing the tachyonic model action as:
\bea \label{diff2}
\left[\frac{d^{2}}{d\eta^2}+\left(k^{2}-\frac{1}{a}\frac{d^{2}a}{d\eta^2}\right)\right]u_{\bf k}=0
\eea
where the helicity index $\gamma$ is summed over in the Fourier modes for the tensor contribution $u_{\bf k}$ and 
\be\begin{array}{lll}\label{rte2}
 \displaystyle \frac{1}{a}\frac{d^{2}a}{d\eta^2} =\left\{\begin{array}{lll}
                    \displaystyle  
                    a^2H^2\left[2-\bar{\epsilon}_{V}\right]+\cdots \,,~~~~ &
 \mbox{\small {\bf for $q=1/2$}}  \\ 
 \displaystyle  
   a^2H^2\left[2-\frac{1}{2q}\bar{\epsilon}_{V}\right]+\cdots\,,~~~ &
 \mbox{\small {\bf for ~any~$q$}}.
          \end{array}
\right.
\end{array}\ee
Further replacing the factor $aH$ in Eq~(\ref{rte}), we finally get the following simplified expression:
\be\begin{array}{lll}\label{rteq2}
 \displaystyle \frac{1}{a}\frac{d^{2}a}{d\eta^2} 
                  \approx\left\{\begin{array}{lll}
                    \displaystyle  
                   \frac{1}{\eta^2}\left[\left(\frac{3}{2}+\bar{\epsilon}_{V}\right)^2-\frac{1}{4}\right]+\cdots \,,~~~~ &
 \mbox{\small {\bf for $q=1/2$}}  \\ 
 \displaystyle  
   \frac{1}{\eta^2}\left[\left\{\frac{3}{2}+\frac{1}{2q}\bar{\epsilon}_{V}\right\}^2-\frac{1}{4}\right]+\cdots  \,,~~~ &
 \mbox{\small {\bf for ~any~$q$}}.
          \end{array}
\right.
\end{array}\ee
 Now for further simplification in the computation of tensor power spectrum we introduce a new factor $\mu$ which is defined as:
 \be\begin{array}{lll}\label{rtex2}
 \displaystyle \mu
                  \approx\left\{\begin{array}{lll}
                    \displaystyle  
                   \left(\frac{3}{2}+\bar{\epsilon}_{V}\right)+\cdots\,,~~~~ &
 \mbox{\small {\bf for $q=1/2$}}  \\ 
 \displaystyle  
  \left\{\frac{3}{2}+\frac{1}{2q}\bar{\epsilon}_{V}\right\}+\cdots  \,,~~~ &
 \mbox{\small {\bf for ~any~$q$}}.
          \end{array}
\right.
\end{array}\ee
Hence using Eq~(\ref{rtex2}) in Eq~(\ref{diff2}), we get the following simplified form of the Equation of motion:
\bea \label{diff3}
\left[\frac{d^{2}}{d\eta^2}+\left(k^{2}-\frac{\left(\mu^2-\frac{1}{4}\right)}{\eta^2}\right)\right]u_{\bf k}(\eta)=0
\eea
and the most general solution of Eq~(\ref{diff3}) is given by:
\bea \label{solve3} 
u_{\bf k}(\eta)=\sqrt{-\eta}\left[D_{1}H^{(1)}_{\mu}\left(-k\eta\right)+D_{2}H^{(2)}_{\mu}\left(-k\eta\right)\right].
\eea
where $D_{1}$ and $D_{2}$ are two arbitrary integration constants, which can be fixed from the appropriate choice of the boundary conditions.
Additionally $H^{(1)}_{\mu}$ and $H^{(2)}_{\mu}$ represent the Hankel function of the first and second kind with rank $\mu$. Now to impose the well known 
Bunch-Davies boundary condition at early times we have used:
\bea \label{lima3}
\lim_{k\eta\rightarrow-\infty} H^{(1)}_{\mu}\left(-k\eta\right)&=&\sqrt{\frac{2}{\pi}}\frac{1}{\sqrt{-\eta}}\exp\left(ik\eta\right)\exp\left(i\frac{\pi}{2}\left(\mu+\frac{1}{2}\right)\right),\\
\lim_{k\eta\rightarrow-\infty} H^{(2)}_{\mu}\left(-k\eta\right)&=&\sqrt{\frac{2}{\pi}}\frac{1}{\sqrt{-\eta}}\exp\left(-ik\eta\right)\exp\left(-i\frac{\pi}{2}\left(\mu+\frac{1}{2}\right)\right).
\eea
As a result the previously mentioned integration constants are fixed at the following values:
\bea \label{nu13} D_{1}&=& \sqrt{\frac{\pi}{2}},\\
\label{nu23}  D_{2}&=& 0.
\eea
Consequently the solution of the mode function for tensor fluctuations takes the following form:
\bea \label{solve4} 
u_{\bf k}(\eta)=\sqrt{-\frac{\eta\pi}{2}}H^{(1)}_{\mu}\left(-k\eta\right).
\eea
On the other hand, the solution stated in Eq~(\ref{solve3}) determines the future evolution of the mode including its super-horizon dynamics at
$k<<aH$ or $|k \eta|<<1$  or $k \eta\rightarrow 0$ and this is due to:
\bea \label{lima4}
\lim_{k\eta\rightarrow 0} H^{(1)}_{\mu}\left(-k\eta\right)&=&\frac{i}{\pi}\Gamma(\mu)\left(\frac{-k\eta}{2}\right)^{-\mu}.
\eea
Consequently the solution of the mode function for tensor fluctuations takes the following form:
\bea \label{solve5} 
u_{\bf k}(\eta)=\sqrt{-\frac{\eta\pi}{2}}\frac{i}{\pi}\Gamma(\mu)\left(\frac{-k\eta}{2}\right)^{-\mu}.
\eea
Finally combining the results obtained in Eq~(\ref{solve}), Eq~(\ref{solve2}) and Eq~(\ref{solve3}) we get:
\be\begin{array}{lll}\label{rtex5}
 \displaystyle u_{\bf k}(\eta)
                  =\left\{\begin{array}{lll}
                    \displaystyle  
                  \sqrt{-\eta}\left[D_{1}H^{(1)}_{\mu}\left(-k\eta\right)+D_{2}H^{(2)}_{\mu}\left(-k\eta\right)\right],~~~~ &
 \mbox{\small {\bf for AV}}  \\ 
 \displaystyle  
  \sqrt{-\frac{\eta\pi}{2}}H^{(1)}_{\mu}\left(-k\eta\right)  \,,~~~ &
 \mbox{\small {\bf for BD + $|k \eta|>>1$}} \\ 
 \displaystyle  
  \sqrt{-\frac{\eta\pi}{2}}\frac{i}{\pi}\Gamma(\mu)\left(\frac{-k\eta}{2}\right)^{-\mu} \,,~~~ &
 \mbox{\small {\bf for BD + $|k \eta|<<1$}}.
          \end{array}
\right.
\end{array}\ee
where ${\bf AV}$ and ${\bf BD}$ signify arbitrary vacuum and Bunch-Davies vacuum respectively.
Finally the two point function from tensor fluctuation for both ${\bf AV}$ and ${\bf BD}$ can be expressed as:
\begin{eqnarray}
 \label{sc6}\langle h_{\bf k}h_{\bf k^{'}}\rangle =\left(\frac{4}{M^2_{p}}\right)\langle \delta \psi_{\bf k}
 \delta \psi_{\bf k^{'}}\rangle =\frac{2}{a^{2}M^{2}_{p}}\langle u_{\bf k}
 u_{\bf k^{'}}\rangle =(2\pi)^{3}\delta^{3}({\bf k}+{\bf k^{'}})\frac{2\pi^{2}}{k^{3}}\Delta_{h}(k),~~~~~~
\end{eqnarray}
where the primordial power spectrum for tensor modes at any
arbitrary momentum scale $k$ can be written for both ${\bf AV}$ and ${\bf BD}$ with $q=1/2$ as:
\begin{eqnarray}
\label{ps7} \Delta_{h}(k)&\equiv& \overbrace{2}^{Due~to~graviton~helicity}\times\frac{k^{3}P_{h}(k)}{2\pi^2}=\frac{k^{3}|u_{k}|^{2}}{\pi^{2}a^{2}M^{2}_{p}}\nonumber\\ 
&=& \small\left\{\begin{array}{lll}
                    \displaystyle  \small \left\{\begin{array}{lll}
                    \displaystyle   \small
                 \frac{2^{2\mu-2}\left(-k\eta\right)^{3-2\mu} H^2}{(1+\bar{\epsilon}_{V})^2 \pi^{2}M^{2}_{p}}\left|\frac{\Gamma(\mu)}{\Gamma\left(\frac{3}{2}\right)}\right|^{2}\,,~~~~ &
 \mbox{\small {\bf for $|k \eta|<<1$}}  \\ 
 \displaystyle  
                \frac{2^{2\mu-2}\left(1+\bar{\epsilon}_{V}\right)^{1-2\mu} H^2}{\pi^{2}M^{2}_{p}}\left|\frac{\Gamma(\mu)}{\Gamma\left(\frac{3}{2}\right)}\right|^{2}\,,~~~~ &
 \mbox{\small {\bf for $|k \eta|=1$}}  \\ 
 \displaystyle  
  \frac{\left(-k\eta \right)^3 H^2|H^{(1)}_{\mu}\left(-k\eta\right)|^{2}}{(1+\bar{\epsilon}_{V})^2 \pi M^{2}_{p}}\,,~~~ &
 \mbox{\small {\bf for $|k \eta|>>1$}}.
          \end{array}
\right.~~~~~~~~~ &
 \mbox{\small \underline{\bf for BD}}  \\ 
 \displaystyle  
 \frac{2\left(-k\eta \right)^3 H^2\left|D_{1}H^{(1)}_{\mu}\left(-k\eta\right)+D_{2}H^{(2)}_{\mu}
 \left(-k\eta\right)\right|^2}{(1+\bar{\epsilon}_{V})^2 \pi^{2}M^{2}_{p}}\,,~~~~~~~~~ &
 \mbox{\small \underline{\bf for AV}}.
 \end{array}
\right.\end{eqnarray}
and similarly the primordial power spectrum for tensor modes at any
arbitrary momentum scale $k$ can be written for both ${\bf AV}$ and ${\bf BD}$ with any arbitrary $q$ as:
\begin{eqnarray}
\label{ps8} 
  \Delta_{h}(k)&\equiv& \overbrace{2}^{Due~to~graviton~helicity}\times\frac{k^{3}P_{h}(k)}{2\pi^2}=\frac{k^{3}|u_{k}|^{2}}{\pi^{2}a^{2}M^{2}_{p}}\nonumber\\ 
&=& \small\left\{\begin{array}{lll}
                    \displaystyle  \small \left\{\begin{array}{lll}
                    \displaystyle  
                 \frac{2^{2\mu-1}q\left(-k\eta \right)^{3-2\mu} H^2}{\left(1+\frac{1}{2q}\bar{\epsilon}_{V}\right)^2  \pi^{2}M^{2}_{p}}\left|\frac{\Gamma(\mu)}{\Gamma\left(\frac{3}{2}\right)}\right|^{2}\,,~~~~ &
 \mbox{\small {\bf for $|k \eta|<<1$}}  \\ 
 \displaystyle  
                \frac{2^{2\mu-1}q \left(1+\frac{1}{2q}\bar{\epsilon}_{V}\right)^{1-2\mu} H^2}{\pi^{2}M^{2}_{p}}\left|\frac{\Gamma(\mu)}{\Gamma\left(\frac{3}{2}\right)}\right|^{2}\,,~~~~ &
 \mbox{\small {\bf for $|k \eta|=1$}}  \\ 
 \displaystyle  
  \frac{2q\left(-k\eta \right)^3 H^2|H^{(1)}_{\mu}\left(-k\eta\right)|^{2}}{\bar{\epsilon}_{V}\left(1+\frac{1}{2q}\bar{\epsilon}_{V}\right)^2 \pi M^{2}_{p}}\,,~~~ &
 \mbox{\small {\bf for $|k \eta|>>1$}}.
          \end{array}
\right.~~~~~~~~~ &
 \mbox{\small \underline{\bf for BD}}  \\ 
 \displaystyle  
 \frac{4q\left(-k\eta \right)^3 H^2\left|C_{1}H^{(1)}_{\mu}\left(-k\eta\right)+C_{2}H^{(2)}_{\mu}
 \left(-k\eta\right)\right|^2}{\left(1+\frac{1}{2q}\bar{\epsilon}_{V}\right)^2\pi^{2}M^{2}_{p}}\,,~~~~~~~~~ &
 \mbox{\small \underline{\bf for AV}}.
 \end{array}
\right.\end{eqnarray}
Now starting from the expression for primordial power spectrum for tensor modes one can compute the spectral tilt at any
 arbitrary momentum scale $k$ for both ${\bf AV}$ and ${\bf BD}$ with $q=1/2$ as:
\begin{eqnarray}
\label{ps9} n_{h}(k) &\equiv& \frac{d\ln \Delta_{h}(k)}{d\ln k} = \frac{d\ln \Delta_{h}(k)}{d N}\nonumber\\ 
&\approx&\left\{\begin{array}{lll}
                    \displaystyle \small \left\{\begin{array}{lll}
                    \displaystyle  \small
                 (3-2\mu)\left[1-\frac{2}{3}\bar{\epsilon}_{V}(\bar{\eta}_{V}-3\bar{\epsilon}_{V})\left(1-\frac{2}{3}\bar{\epsilon}_{V}\right)^{1/4}\right]+\cdots, &
 \mbox{\small {\bf for $|k \eta|<<1$}}  \\ 
 \displaystyle \small 
                (3-2\mu)\left[1-\frac{2}{3}\bar{\epsilon}_{V}(\bar{\eta}_{V}-3\bar{\epsilon}_{V})\left(1-\frac{2}{3}\bar{\epsilon}_{V}\right)^{1/4}\right]+\cdots\,,&
 \mbox{\small {\bf for $|k \eta|=1$}}  \\ 
 \displaystyle  \small
  \bar{\epsilon}_{V}-\frac{2}{3}\bar{\epsilon}_{V}(\bar{\eta}_{V}-3\bar{\epsilon}_{V})\left(1-\frac{2}{3}\bar{\epsilon}_{V}\right)^{1/4}\\
  \displaystyle -2(\bar{\eta}_{V}-3\bar{\epsilon}_{V})\left(1-\frac{2}{3}\bar{\epsilon}_{V}\right)^{1/4}\\
  \displaystyle \frac{\left(-\eta\right)\left[H^{(1)}_{\mu-1}\left(-k\eta\right)-H^{(1)}_{\mu+1}
 \left(-k\eta\right)\right]}{H^{(1)}_{\mu}\left(-k\eta\right)}+\cdots\,, &
 \mbox{\small {\bf for $|k \eta|>>1$}}.
          \end{array}
\right. &
 \mbox{\small \underline{\bf for BD}}  \\ 
 \displaystyle  
 \bar{\epsilon}_{V}-\frac{2}{3}\bar{\epsilon}_{V}(\bar{\eta}_{V}-3\bar{\epsilon}_{V})\left(1-\frac{2}{3}\bar{\epsilon}_{V}\right)^{1/4}\\
  \displaystyle -2(\bar{\eta}_{V}-3\bar{\epsilon}_{V})\left(1-\frac{2}{3}\bar{\epsilon}_{V}\right)^{1/4}\\
  \displaystyle \frac{\left(-\eta\right)C_{1}\left[H^{(1)}_{\mu-1}\left(-k\eta\right)-H^{(1)}_{\mu+1}
 \left(-k\eta\right)\right]}{\left[C_{1}H^{(1)}_{\mu}\left(-k\eta\right)+C_{2}H^{(2)}_{\mu}
 \left(-k\eta\right)\right]}\\
 \displaystyle \frac{\left(-\eta\right)C_{1}\left[H^{(2)}_{\mu-1}\left(-k\eta\right)-H^{(2)}_{\mu+1}
 \left(-k\eta\right)\right]}{\left[C_{1}H^{(1)}_{\mu}\left(-k\eta\right)+C_{2}H^{(2)}_{\mu}
 \left(-k\eta\right)\right]}+\cdots
 \,, &
 \mbox{\small \underline{\bf for AV}}.
 \end{array}
\right.\end{eqnarray}
and for both ${\bf AV}$ and ${\bf BD}$ with any arbitrary $q$ as:
\begin{eqnarray}
\label{ps10} n_{h}(k) &\equiv& \frac{d\ln \Delta_{h}(k)}{d\ln k} = \frac{d\ln \Delta_{h}(k)}{d N}\nonumber\\ 
&=&\left\{\begin{array}{lll}
                    \displaystyle \small \left\{\begin{array}{lll}
                    \displaystyle  
                  (3-2\mu)\left[1-\frac{1}{3q}\bar{\epsilon}_{V}(\bar{\eta}_{V}-3\bar{\epsilon}_{V})\left(1-\frac{1}{3q}\bar{\epsilon}_{V}\right)^{1/4}\right]+\cdots\,,&
 \mbox{\small {\bf for $|k \eta|<<1$}}  \\ 
 \displaystyle  
                 (3-2\mu)\left[1-\frac{1}{3q}\bar{\epsilon}_{V}(\bar{\eta}_{V}-3\bar{\epsilon}_{V})\left(1-\frac{1}{3q}\bar{\epsilon}_{V}\right)^{1/4}\right]+\cdots\,, &
 \mbox{\small {\bf for $|k \eta|=1$}}  \\ 
 \displaystyle  
  \small
  \frac{\bar{\epsilon}_{V}}{2q}-\frac{1}{3q}\bar{\epsilon}_{V}(\bar{\eta}_{V}-3\bar{\epsilon}_{V})\left(1-\frac{1}{3q}\bar{\epsilon}_{V}\right)^{1/4}\\
  \displaystyle -\frac{1}{q}(\bar{\eta}_{V}-3\bar{\epsilon}_{V})\left(1-\frac{2}{3}\bar{\epsilon}_{V}\right)^{1/4}\\
  \displaystyle \frac{\left(-\eta\right)\left[H^{(1)}_{\mu-1}\left(-k\eta\right)-H^{(1)}_{\mu+1}
 \left(-k\eta\right)\right]}{H^{(1)}_{\mu}\left(-k\eta\right)}+\cdots\,, &
 \mbox{\small {\bf for $|k \eta|>>1$}}.
          \end{array}
\right. &
 \mbox{\small \underline{\bf for BD}}  \\ 
 \displaystyle  
 \frac{\bar{\epsilon}_{V}}{2q}-\frac{1}{3q}\bar{\epsilon}_{V}(\bar{\eta}_{V}-3\bar{\epsilon}_{V})\left(1-\frac{1}{3q}\bar{\epsilon}_{V}\right)^{1/4}\\
  \displaystyle -\frac{1}{q}(\bar{\eta}_{V}-3\bar{\epsilon}_{V})\left(1-\frac{2}{3}\bar{\epsilon}_{V}\right)^{1/4}\\
  \displaystyle \frac{\left(-\eta\right)C_{1}\left[H^{(1)}_{\mu-1}\left(-k\eta\right)-H^{(1)}_{\mu+1}
 \left(-k\eta\right)\right]}{\left[C_{1}H^{(1)}_{\mu}\left(-k\eta\right)+C_{2}H^{(2)}_{\mu}
 \left(-k\eta\right)\right]}\\
 \displaystyle \frac{\left(-\eta\right)C_{1}\left[H^{(2)}_{\mu-1}\left(-k\eta\right)-H^{(2)}_{\mu+1}
 \left(-k\eta\right)\right]}{\left[C_{1}H^{(1)}_{\mu}\left(-k\eta\right)+C_{2}H^{(2)}_{\mu}
 \left(-k\eta\right)\right]}+\cdots\,, &
 \mbox{\small \underline{\bf for AV}}.
 \end{array}
\right.\end{eqnarray}
One can also consider the following approximations to simplify the final derived form of the primordial scalar power spectrum for {\bf BD} vacuum with $|k c_{S}\eta|=1$ case:
\begin{enumerate}
 \item We start with the {\it Laurent expansion} of the Gamma function:
       \bea \label{r1x11} \Gamma(\mu) &=& \frac{1}{\mu}-\gamma+\frac{1}{2}\left(\gamma^2+\frac{\pi^2}{6}\right)\mu-\frac{1}{6}\left(\gamma^3+\frac{\gamma \pi^2}{2}+2\zeta(3)\right)\mu^2 +{\cal O}(\mu^3).~~~~~~~~~~~~
       \eea
       where $\gamma$ being the Euler Mascheroni constant and $\zeta(3)$ characterizing the Reimann zeta function of 
       order $3$ originating in the expansion of the gamma function. 
 \item  Hence using the result of Eq~(\ref{r1x11}) for $q=1/2$ and for arbitrary $q$ we can write:
 \be\begin{array}{lll}\label{r2x11}
 \displaystyle \Gamma(\mu) 
                  =\left\{\begin{array}{lll}
                    \displaystyle  
                   \frac{1}{\left(\frac{3}{2}+\bar{\epsilon}_{V}\right)}-\gamma+\frac{1}{2}
                   \left(\gamma^2+\frac{\pi^2}{6}\right)\left(\frac{3}{2}+\bar{\epsilon}_{V}\right)\\
                   \displaystyle~~~~~~~~-\frac{1}{6}\left(\gamma^3+\frac{\gamma \pi^2}{2}+2\zeta(3)\right)
                   \left(\frac{3}{2}+\bar{\epsilon}_{V}\right)^2 +\cdots\,,~~~~ &
 \mbox{\small {\bf for $q=1/2$}}  \\ 
 \displaystyle  
  \frac{1}{ \left\{\frac{3}{2}+\frac{1}{2q}\bar{\epsilon}_{V}\right\}}
  -\gamma 
  +\frac{1}{2}\left(\gamma^2+\frac{\pi^2}{6}\right) \left\{\frac{3}{2}+\frac{1}{2q}\bar{\epsilon}_{V}\right\}
  \\ \displaystyle~-\frac{1}{6}\left(\gamma^3+\frac{\gamma \pi^2}{2}+2\zeta(3)\right) \left\{\frac{3}{2}+\frac{1}{2q}\bar{\epsilon}_{V}\right\}^2
  +\cdots \,,~~~ &
 \mbox{\small {\bf for ~any~$q$}}.
          \end{array}
\right.
\end{array}\ee
\item In the slow-roll regime of inflation all the slow-roll parameters satisfy the following constraint:
          \bea \bar{\epsilon}_{V}&<<& 1,\\
               |\bar{\eta}_{V}|&<<& 1,\\
               |\bar{\xi}^{2}_{V}|&<<& 1,\\ 
               |\bar{\sigma}^{3}_{V}|&<<& 1.\eea
               Using these approximations the primordial scalar power spectrum can be expressed as:
               \begin{eqnarray} 
\label{psxcc122} \Delta_{h, \star}
&\approx&\left\{\left[1-({\cal C}_{E}+1){\epsilon}_{1}\right]^2 \frac{2H^{2}}{\pi^2M^{2}_{p}}\right\}_{k_{\star}=a_{\star}H_{\star}}\nonumber\\
&=&\left\{\begin{array}{lll}
                    \displaystyle 
 \displaystyle  \left\{\left[1-({\cal C}_{E}+1)\bar{\epsilon}_{V}\right]^2 \frac{2H^{2}}{\pi^2M^{2}_{p}}\right\}_{k_{\star}=a_{\star}H_{\star}}\,,~~~~~~~~~ &
 \mbox{\small {\bf for $q=1/2$}}  \\ \\
 \displaystyle \left\{\left[1-({\cal C}_{E}+1)\frac{\bar{\epsilon}_{V}}{2q}\right]^2 \frac{2H^{2}}{\pi^2M^{2}_{p}}\right\}_{k_{\star}=a_{\star}H_{\star}} \,,~~~~~~~~~ &
 \mbox{\small {\bf for any $q$}}.
 \end{array}
\right.\end{eqnarray}
where ${\cal C}_{E}$ is given by:
\bea  
{\cal C}_{E}= -2 + \ln 2 + \gamma \approx -0.72. 
\eea

\item Next one can compute the scalar spectral tilt ($n_{S}$) of the primordial scalar power spectrum as:
\begin{eqnarray} 
\label{psxcc133} n_{h, \star} 
&\approx&\left\{\begin{array}{lll}
                    \displaystyle 
 \displaystyle  -2{\epsilon}_{1}\left[1+{\epsilon}_{1}+\left({\cal C}_{E}+1\right){\epsilon}_{2}\right]+\cdots\\
               \displaystyle
                =-2\bar{\epsilon}_{V}\left[1+\bar{\epsilon}_{V}+2\left({\cal C}_{E}+1\right)\left(3\bar{\epsilon}_{V}-\bar{\eta}_{V}\right)\right]+\cdots,~~ &
 \mbox{\small {\bf for $q=1/2$}}  \\ 
 \displaystyle  
               \displaystyle  -2{\epsilon}_{1}\left[1+{\epsilon}_{1}+\left({\cal C}_{E}+1\right){\epsilon}_{2}\right]+\cdots\\
               \displaystyle
                =-\frac{\bar{\epsilon}_{V}}{q}\left[1+\frac{\bar{\epsilon}_{V}}{2q}+\sqrt{\frac{2}{q}}\left({\cal C}_{E}+1\right)\left(3\bar{\epsilon}_{V}-\bar{\eta}_{V}\right)\right]+\cdots\,,~~ &
 \mbox{\small {\bf for any $q$}}.
 \end{array}
\right.\end{eqnarray}
\item Next one can compute the running of the tensor spectral tilt ($\alpha_{h}$) of the primordial scalar power spectrum as:
\begin{eqnarray}
\label{psxcc144} \alpha_{h, \star}&=&\left(\frac{dn_{h}(k)}{d\ln k}\right)_{k_{\star}=a_{\star}H_{\star}}=\left(\frac{dn_{h}(k)}{dN}\right)_{k_{\star}=a_{\star}H_{\star}}\\
&\approx&\left\{ \small\begin{array}{lll}
                    \displaystyle 
\displaystyle -\left[4\bar{\epsilon}_{V}(1+\bar{\epsilon}_{V})(\bar{\eta}_{V}-3\bar{\epsilon}_{V})+4\bar{\epsilon}^2_{V}(\bar{\eta}_{V}-3\bar{\epsilon}_{V})
\right.\\ \left. \displaystyle ~~~~~~~~~~~-8\left({\cal C}_{E}+1\right)\bar{\epsilon}_{V}(\bar{\eta}_{V}-3\bar{\epsilon}_{V})^2 \right.\\ \left. \displaystyle~~~~~~~~~~~~-2\bar{\epsilon}_{V}\left(10\bar{\epsilon}_{V}\bar{\eta}_{V}
                -18\bar{\epsilon}^2_{V}-\bar{\xi}^2_{V}\right)\right]_{\star}\displaystyle\left(1-\frac{2}{3}\bar{\epsilon}_{V}\right)^{1/4}_{\star}+\cdots,~~ &
 \mbox{\small {\bf for $q=1/2$}}  \\ 
 \displaystyle  
               -\left[2\frac{\bar{\epsilon}_{V}}{q}\left(1+\frac{\bar{\epsilon}_{V}}{2q}\right)(\bar{\eta}_{V}-3\bar{\epsilon}_{V})+\frac{1}{q^2}\bar{\epsilon}^2_{V}(\bar{\eta}_{V}-3\bar{\epsilon}_{V})
\right.\\ \left. \displaystyle ~~~~~~~~~~~-\frac{8}{(2q)^{5/2}}\left({\cal C}_{E}+1\right)\bar{\epsilon}_{V}(\bar{\eta}_{V}-3\bar{\epsilon}_{V})^2 \right.\\ \left. \displaystyle~~~~~~~~~~~~-
\sqrt{\frac{2}{q}}\bar{\epsilon}_{V}\left(10\bar{\epsilon}_{V}\bar{\eta}_{V}
                -18\bar{\epsilon}^2_{V}-\bar{\xi}^2_{V}\right)\right]_{\star}\displaystyle\left(1-\frac{1}{3q}\bar{\epsilon}_{V}\right)^{1/4}_{\star}+\cdots\,,~~ &
 \mbox{\small {\bf for any $q$}}.
 \end{array}
\right.\end{eqnarray}
\item Finally, one can also compute the running of the running of scalar spectral tilt ($\kappa_{S}$) of the primordial scalar power spectrum as:
\begin{eqnarray}
\label{psxcc155} \kappa_{h, \star}&=&\left(\frac{d^2n_{h}(k)}{d\ln k^2}\right)_{k_{\star}=a_{\star}H_{\star}}
=\left(\frac{d^2n_{h}(k)}{dN^2}\right)_{k_{\star}=a_{\star}H_{\star}}\nonumber\\
&\approx&\left\{ \small\begin{array}{lll}
                    \displaystyle 
\displaystyle \displaystyle -\left[8\bar{\epsilon}_{V}(1+\bar{\epsilon}_{V})(\bar{\eta}_{V}-3\bar{\epsilon}_{V})^2+8\bar{\epsilon}^2_{V}(\bar{\eta}_{V}-3\bar{\epsilon}_{V})^2
\right.\\ \left. \displaystyle-16\left({\cal C}_{E}+1\right)\bar{\epsilon}_{V}\left\{
(\bar{\eta}_{V}-3\bar{\epsilon}_{V})^3 -(\bar{\eta}_{V}-3\bar{\epsilon}_{V})\left(10\bar{\epsilon}_{V}\bar{\eta}_{V}
                -18\bar{\epsilon}^2_{V}-\bar{\xi}^2_{V}\right)\right\}\right.\\ 
\left. \displaystyle~~~~~~~~~~~~-4\bar{\epsilon}_{V}(1+\bar{\epsilon}_{V})\left(10\bar{\epsilon}_{V}\bar{\eta}_{V}
                -18\bar{\epsilon}^2_{V}-\bar{\xi}^2_{V}\right)\right]_{\star}\displaystyle\left(1-\frac{2}{3}\bar{\epsilon}_{V}\right)^{1/4}_{\star}+\cdots,~~ &
 \mbox{\small {\bf for $q=1/2$}}  \\ 
 \displaystyle  
              
                \displaystyle 
\displaystyle -\left[\frac{4}{q}\bar{\epsilon}_{V}\left(1+\frac{\bar{\epsilon}_{V}}{2q}\right)(\bar{\eta}_{V}
-3\bar{\epsilon}_{V})^2+\frac{2}{q^2}\bar{\epsilon}^2_{V}(\bar{\eta}_{V}-3\bar{\epsilon}_{V})^2\right.\\ \left. \displaystyle 
+\frac{2}{q}\bar{\epsilon}_{V}\left(1+\frac{\bar{\epsilon}_{V}}{2q}\right)\left(\bar{\xi}^2_{V}-10\bar{\epsilon}_{V}\bar{\eta}_{V}
+18\bar{\epsilon}^{2}_{V}\right)\right.\\ \left. \displaystyle -\frac{8}{q}\left({\cal C}_{E}+1\right)\bar{\epsilon}_{V}\left\{
(\bar{\eta}_{V}-3\bar{\epsilon}_{V})^3 -(\bar{\eta}_{V}-3\bar{\epsilon}_{V})\left(10\bar{\epsilon}_{V}\bar{\eta}_{V}
                -18\bar{\epsilon}^2_{V}-\bar{\xi}^2_{V}\right)\right\}\right]_{\star}\\
                \displaystyle~~~~~~~~~~~~~~~~~~~~~~~~~~~~\times\left(1-\frac{1}{3q}\bar{\epsilon}_{V}\right)^{1/4}_{\star}+\cdots,~~ &
 \mbox{\small {\bf for any $q$}}.
 \end{array}
\right.\end{eqnarray}
\end{enumerate}

\subsubsection{Modified consistency relations}
In this subsection we derive the new (modified) consistency relations for single tachyonic field inflation:
\begin{enumerate}
 \item Let us first start with tenor-to-scalar ratio $r$, which can be defined at any arbitrary momentum scale $k$ for $q=1/2$ case as:
 \begin{eqnarray}
\label{ps166} r(k)&\equiv& \frac{\Delta_{h}(k)}{\Delta_{\zeta}(k)}=2\frac{P_{h}(k)}{P_{\zeta}(k)}=2\frac{|u_{k}|^{2}}{|v_{k}|^{2}}\left(\frac{z}{a}\right)^2\nonumber\\ 
&=& \small\left\{\begin{array}{lll}
                    \displaystyle  \small \left\{\begin{array}{lll}\small
                    \displaystyle   \small
                 16\times 2^{2(\mu-\nu)} \bar{\epsilon}_{V}\left(-k\eta c_{S}\right)^{2(\nu-\mu)}c^{2\mu-2}_{S}\left|\frac{\Gamma(\mu)}{\Gamma\left(\nu\right)}\right|^{2}\,,~~~~ &
 \mbox{\small {\bf for $|k c_{S}\eta|<<1$}}  \\ 
 \displaystyle  
               16\times 2^{2(\mu-\nu)} \bar{\epsilon}_{V}\left(1+\bar{\epsilon}_{V}\right)^{2(\nu-\mu)}c^{2\nu-2}_{S}\left|\frac{\Gamma(\mu)}{\Gamma\left(\nu\right)}\right|^{2}\,,~~~~ &
 \mbox{\small {\bf for $|k c_{S}\eta|=1$}}  \\ 
 \displaystyle  
  \frac{8\bar{\epsilon}_{V}}{c^{2}_{S}}\times\frac{|H^{(1)}_{\mu}\left(-k\eta\right)|^{2}}{|H^{(1)}_{\nu}\left(-kc_{S}\eta\right)|^{2}}\,,~~~ &
 \mbox{\small {\bf for $|k c_{S}\eta|>>1$}}.
          \end{array}
\right.~~~~~ &
 \mbox{\small \underline{\bf for BD}}  \\ 
 \displaystyle  
  \frac{8\bar{\epsilon}_{V}}{c^{2}_{S}}\times\frac{\left|D_{1}H^{(1)}_{\mu}\left(-k\eta\right)+D_{2}H^{(2)}_{\mu}
 \left(-k\eta\right)\right|^2}{\left|C_{1}H^{(1)}_{\nu}\left(-kc_{S}\eta\right)+C_{2}H^{(2)}_{\nu}
 \left(-kc_{S}\eta\right)\right|^2}\,,~~~~ &
 \mbox{\small \underline{\bf for AV}}.
 \end{array}
\right.\end{eqnarray}
Similarly for arbitrary $q$ one can write the following expression for tenor-to-scalar ratio $r$ at any arbitrary momentum scale as:
\begin{eqnarray}
\label{ps177} r(k)&\equiv& \frac{\Delta_{h}(k)}{\Delta_{\zeta}(k)}=2\frac{P_{h}(k)}{P_{\zeta}(k)}=2\frac{|u_{k}|^{2}}{|v_{k}|^{2}}\left(\frac{z}{a}\right)^2\nonumber\\ 
&=& \small\left\{\begin{array}{lll}
                    \displaystyle  \small \left\{\begin{array}{lll}\small
                    \displaystyle   \small
                 16\times 2^{2(\mu-\nu)} \bar{\epsilon}_{V}\left(-k\eta c_{S}\right)^{2(\nu-\mu)}c^{2\mu-2}_{S}\left|\frac{\Gamma(\mu)}{\Gamma\left(\nu\right)}\right|^{2}\,,~~~~ &
 \mbox{\small {\bf for $|k c_{S}\eta|<<1$}}  \\ 
 \displaystyle  
               16\times 2^{2(\mu-\nu)} \bar{\epsilon}_{V}\left(1+\frac{\bar{\epsilon}_{V}}{2q}\right)^{2(\nu-\mu)}c^{2\nu-2}_{S}\left|\frac{\Gamma(\mu)}{\Gamma\left(\nu\right)}\right|^{2}\,,~~~~ &
 \mbox{\small {\bf for $|k c_{S}\eta|=1$}}  \\ 
 \displaystyle  
  \frac{8\bar{\epsilon}_{V}}{c^{2}_{S}}\times\frac{|H^{(1)}_{\mu}\left(-kc_{S}\eta\right)|^{2}}{|H^{(1)}_{\nu}\left(-kc_{S}\eta\right)|^{2}}\,,~~~ &
 \mbox{\small {\bf for $|k c_{S}\eta|>>1$}}.
          \end{array}
\right.~~~~~ &
 \mbox{\small \underline{\bf for BD}}  \\ 
 \displaystyle  
  \frac{8\bar{\epsilon}_{V}}{c^{2}_{S}}\times\frac{\left|D_{1}H^{(1)}_{\mu}\left(-kc_{S}\eta\right)+D_{2}H^{(2)}_{\mu}
 \left(-kc_{S}\eta\right)\right|^2}{\left|C_{1}H^{(1)}_{\nu}\left(-kc_{S}\eta\right)+C_{2}H^{(2)}_{\nu}
 \left(-kc_{S}\eta\right)\right|^2}\,,~~~~ &
 \mbox{\small \underline{\bf for AV}}.
 \end{array}
\right.\end{eqnarray}
\item Next for {\bf BD} vacuum with $|k c_{S}\eta|=1$ case within slow-roll regime we can approximately write the following expression for tensor-to-scalar ratio:
\begin{eqnarray}
\label{psxcc188} r_{\star}&=&\frac{\Delta_{h}(k_{\star})}{\Delta_{\zeta}(k_{\star})}=2\frac{P_{h}(k_{\star})}{P_{\zeta}(k_{\star})}=2\frac{|u_{k_{\star}}|^{2}}{|v_{k_{\star}}|^{2}}\left(\frac{z}{a}\right)^2_{\star}\\ 
&=& \left[16{\epsilon}_{1}c_{S}\frac{\left[1-({\cal C}_{E}+1){\epsilon}_{1}\right]^2}{\left[1-({\cal C}_{E}+1){\epsilon}_{1}-\frac{{\cal C}_{E}}{2}{\epsilon}_{2}\right]^2}\right]_{k_{\star}=a_{\star}H_{\star}}\nonumber\\
&\approx&\left\{ \small\begin{array}{lll}
                    \displaystyle 
\displaystyle \left[16\bar{\epsilon}_{V}c_{S}\frac{\left[1-({\cal C}_{E}+1)\bar{\epsilon}_{V}\right]^2}{\left[1-({\cal C}_{E}+1)\bar{\epsilon}_{V}-{\cal C}_{E}\left(3\bar{\epsilon}_{V}-\bar{\eta}_{V}\right)\right]^2}\right]_{k_{\star}=a_{\star}H_{\star}}\\
\displaystyle =\left[16\bar{\epsilon}_{V}\frac{\left[1-({\cal C}_{E}+1)\bar{\epsilon}_{V}\right]^2}{\left[1-\left({\cal C}_{E}+\frac{5}{6}\right)\bar{\epsilon}_{V}-{\cal C}_{E}
                \left(3\bar{\epsilon}_{V}-\bar{\eta}_{V}\right)\right]^2}\right]_{k_{\star}=a_{\star}H_{\star}},~~ &
 \mbox{\small {\bf for $q=1/2$}}  \\ 
 \displaystyle  
                \left[\frac{8}{q}\bar{\epsilon}_{V}c_{S}\frac{\left[1-({\cal C}_{E}+1)\frac{\bar{\epsilon}_{V}}{2q}\right]^2}{\left[1-({\cal C}_{E}+1)\frac{\bar{\epsilon}_{V}}{2q}-\frac{{\cal C}_{E}}{\sqrt{2q}}\left(3\bar{\epsilon}_{V}-\bar{\eta}_{V}\right)\right]^2}\right]_{k_{\star}=a_{\star}H_{\star}}\\
\displaystyle =\left[\frac{8}{q}\bar{\epsilon}_{V}\frac{\left[1-({\cal C}_{E}+1)\frac{\bar{\epsilon}_{V}}{2q}\right]^2}{\left[1-({\cal C}_{E}+1-\Sigma)\frac{\bar{\epsilon}_{V}}{2q}-\frac{{\cal C}_{E}}{\sqrt{2q}}\left(3\bar{\epsilon}_{V}-\bar{\eta}_{V}\right)\right]^2}\right]_{k_{\star}=a_{\star}H_{\star}}\,,~~ &
 \mbox{\small {\bf for any $q$}}.
 \end{array}
\right.\end{eqnarray}
\item Hence the consistency relation between the tensor-to-scalar ratio $r$ and spectral tilt $n_{h}$ for tensor modes for {\bf BD} vacuum with $|k c_{S}\eta|=1$ case can be written as: 
\begin{eqnarray}
\label{psxcc199} r_{\star}
&=& -8n_{h,\star}\times\left[c_{S}\frac{\left[1-({\cal C}_{E}+1){\epsilon}_{1}\right]^2}{\left[1-({\cal C}_{E}+1)
{\epsilon}_{1}-\frac{{\cal C}_{E}}{2}{\epsilon}_{2}\right]^2\left[1+{\epsilon}_{1}+\left({\cal C}_{E}
+1\right){\epsilon}_{2}\right]}\right]_{k_{\star}=a_{\star}H_{\star}}\nonumber\\ \nonumber \\
&\approx& -8n_{h,\star}\times\underbrace{\left\{ \tiny\begin{array}{lll}
                    \displaystyle 
\displaystyle\left[c_{S}\frac{\left[1-({\cal C}_{E}+1)\bar{\epsilon}_{V}\right]^2}{\left[1-({\cal C}_{E}+1)\bar{\epsilon}_{V}-{\cal C}_{E}\left(3\bar{\epsilon}_{V}-\bar{\eta}_{V}\right)\right]^2\left[1+\bar{\epsilon}_{V}+2\left({\cal C}_{E}+1\right)\left(3\bar{\epsilon}_{V}-\bar{\eta}_{V}\right)\right]}\right]_{k_{\star}=a_{\star}H_{\star}}\\
\displaystyle =\left[\frac{\left[1-({\cal C}_{E}+1)\bar{\epsilon}_{V}\right]^2}{\left[1-\left({\cal C}_{E}+\frac{5}{6}\right)\bar{\epsilon}_{V}-{\cal C}_{E}
                \left(3\bar{\epsilon}_{V}-\bar{\eta}_{V}\right)\right]^2\left[1+\bar{\epsilon}_{V}+2\left({\cal C}_{E}+1\right)\left(3\bar{\epsilon}_{V}-\bar{\eta}_{V}\right)\right]}\right]_{k_{\star}=a_{\star}H_{\star}},~~ &
 \mbox{\small {\bf for $q=1/2$}}  \\ 
 \displaystyle  
                \left[c_{S}\frac{\left[1-({\cal C}_{E}+1)\frac{\bar{\epsilon}_{V}}{2q}\right]^2}{\left[1-({\cal C}_{E}+1)\frac{\bar{\epsilon}_{V}}{2q}-\frac{{\cal C}_{E}}
                {\sqrt{2q}}\left(3\bar{\epsilon}_{V}-\bar{\eta}_{V}\right)\right]^2\left[1+\frac{\bar{\epsilon}_{V}}{2q}+\sqrt{\frac{2}{q}}\left({\cal C}_{E}+1\right)\left(3\bar{\epsilon}_{V}-\bar{\eta}_{V}\right)\right]}\right]_{k_{\star}=a_{\star}H_{\star}}\\
\displaystyle =\left[\frac{\left[1-({\cal C}_{E}+1)\frac{\bar{\epsilon}_{V}}{2q}\right]^2}{\left[1-({\cal C}_{E}+1-\Sigma)\frac{\bar{\epsilon}_{V}}{2q}-\frac{{\cal C}_{E}}{\sqrt{2q}}
\left(3\bar{\epsilon}_{V}-\bar{\eta}_{V}\right)\right]^2\left[1+\frac{\bar{\epsilon}_{V}}{2q}+\sqrt{\frac{2}{q}}\left({\cal C}_{E}+1\right)\left(3\bar{\epsilon}_{V}-\bar{\eta}_{V}\right)\right]}\right]_{k_{\star}=a_{\star}H_{\star}}\,,~~ &
 \mbox{\small {\bf for any $q$}}.
 \end{array}
\right.}_{Correction~factor}\end{eqnarray}
\item Next one can express the first two slow-roll parameters $\bar{\epsilon}_{V}$ and $\bar{\eta}_{V}$ in terms of the inflationary observables as:
\begin{eqnarray}
\label{psxcc1101} \bar{\epsilon}_{V}
&\approx&\left\{ \begin{array}{lll}
                    \displaystyle 
\displaystyle\epsilon_{1}\approx -\frac{n_{h,\star}}{2}+\cdots\approx \frac{r_{\star}}{16}+\cdots,~~~~~~~~~~~~~~~~~~~~~~~~~~~~~~~~~~~~~~~~~~ &
 \mbox{\small {\bf for $q=1/2$}}  \\ 
 \displaystyle  
                2q\epsilon_{1}\approx -qn_{h,\star}+\cdots\approx \frac{qr_{\star}}{8}+\cdots\,,~~ &
 \mbox{\small {\bf for any $q$}}.
 \end{array}
\right. \nonumber\\ 
\label{psxcc1} \bar{\eta}_{V}
&\approx&\left\{ \begin{array}{lll}
                    \displaystyle 
\displaystyle 3\epsilon_{1}-\frac{\epsilon_{2}}{2}\approx \frac{1}{2}\left(n_{\zeta,\star}-1+\frac{r_{\star}}{2}\right)+\cdots\approx \frac{1}{2}\left(n_{\zeta,\star}-1-4n_{h,\star}\right)+\cdots,~~ &
 \mbox{\small {\bf for $q=1/2$}}  \\ 
 \displaystyle  
                6q\epsilon_{1}-\sqrt{\frac{q}{2}}\epsilon_{2}\approx \sqrt{\frac{q}{2}}\left(n_{\zeta,\star}-1+\left(\frac{1}{q}+3\sqrt{\frac{2}{q}}\right)\frac{qr_{\star}}{8}\right)+\cdots\\
                \displaystyle~~~~~~~~~~~~~~~~~\approx \sqrt{\frac{q}{2}}\left(n_{\zeta,\star}-1-q\left(\frac{1}{q}+3\sqrt{\frac{2}{q}}\right)n_{h,\star}\right)+\cdots\,,~~ &
 \mbox{\small {\bf for any $q$}}.
 \end{array}
\right.\end{eqnarray}
\item Then the connecting consistency relation between tensor and scalar spectral tilt and tensor-to-scalar ratio can be expressed as:
\begin{eqnarray} 
\label{psxcc112} n_{h, \star} 
&\approx&\left\{\begin{array}{lll}
                    \displaystyle 
               \displaystyle
                -\frac{r_{\star}}{8c_{S}}\left[1-\frac{r_{\star}}{16}+\left(1-n_{\zeta,\star}\right)-{\cal C}_{E}\left\{\frac{r_{\star}}{8}
                +\left(n_{\zeta,\star}-1\right)\right\}\right]+\cdots,~~ &
 \mbox{\small {\bf for $q=1/2$}}  \\ 
 \displaystyle  
               \displaystyle
                -\frac{r_{\star}}{8c_{S}}\left[1+\left\{\left(\frac{3q}{8}\sqrt{\frac{2}{q}}
                -\left(\sqrt{\frac{2}{q}}+5\right)\frac{1}{16}\right)r_{\star}
                +\frac{\left(1-n_{\zeta,\star}\right)}{\sqrt{2q}}\right\}
                \right.\\ \left.\displaystyle+\sqrt{\frac{2}{q}}{\cal C}_{E}\left(\frac{3qr_{\star}}{8}-\frac{1}{2}\left\{n_{\zeta,\star}-1+\left(\frac{1}{q}+3\sqrt{\frac{2}{q}}\right)\frac{qr_{\star}}{8}\right\}\right)\right]+\cdots\,,~~ &
 \mbox{\small {\bf for any $q$}}.
 \end{array}
\right. \end{eqnarray}
Finally using the approximated version of the expression for $c_{S}$ in terms of slow-roll parameters one can recast this consistency condition as: 
\begin{eqnarray} 
\label{psxcc113} n_{h, \star} 
&\approx&\left\{\begin{array}{lll}
                    \displaystyle 
               \displaystyle
                -\frac{r_{\star}}{8}\left[1-\frac{r_{\star}}{24}+\left(1-n_{\zeta,\star}\right)-{\cal C}_{E}\left\{\frac{r_{\star}}{8}
                +\left(n_{\zeta,\star}-1\right)\right\}\right]+\cdots,~~ &
 \mbox{\small {\bf for $q=1/2$}}  \\ 
 \displaystyle  
               \displaystyle
                -\frac{r_{\star}}{8}\left[1+\left\{\left(\frac{3q}{8}\sqrt{\frac{2}{q}}
                -\left(\sqrt{\frac{2}{q}}+5\right)\frac{1}{16}+\frac{\Sigma}{8}\right)r_{\star}
                +\frac{\left(1-n_{\zeta,\star}\right)}{\sqrt{2q}}\right\}
                \right.\\ \left.\displaystyle+\sqrt{\frac{2}{q}}{\cal C}_{E}\left(\frac{3qr_{\star}}{8}-\frac{1}{2}\left\{n_{\zeta,\star}-1+\left(\frac{1}{q}+3\sqrt{\frac{2}{q}}\right)\frac{qr_{\star}}{8}\right\}\right)\right]+\cdots\,,~~ &
 \mbox{\small {\bf for any $q$}}.
 \end{array}
\right. \end{eqnarray}
\item Next the running of the sound speed $c_{S}$ can be written in terms of slow-roll parameters as:
\be\begin{array}{lll}\label{rteqxc14}
 \displaystyle S=\frac{\dot{c}_{S}}{Hc_{S}}=\frac{d\ln c_{S}}{dN}= \frac{d\ln c_{S}}{d\ln k}\\ 
                 ~~=\left\{\begin{array}{lll}
                    \displaystyle  
                   -\frac{2}{3}\bar{\epsilon}_{V}\left(\bar{\eta}_{V}-3\bar{\epsilon}_{V}\right)\left(1-\frac{2}{3}\bar{\epsilon}_{V}\right)^{1/4}+\cdots \,,~~~~ &
 \mbox{\small {\bf for $q=1/2$}}  \\ 
 \displaystyle  
 -\frac{(1-q)}{3q^2}\bar{\epsilon}_{V}\left(\bar{\eta}_{V}-3\bar{\epsilon}_{V}\right)\left(1-\frac{1}{3q}\bar{\epsilon}_{V}\right)^{1/4}+\cdots \,,~~~ &
 \mbox{\small {\bf for ~any~$q$}}.
          \end{array}
\right.
\end{array}\ee
which can be treated as another slow-roll parameter in the present context. One can also recast the slow-roll parameter $S$ in terms of the inflationary observables as:
\be\begin{array}{lll}\label{rteqxc15}
 \displaystyle S=\frac{\dot{c}_{S}}{Hc_{S}}=\frac{d\ln c_{S}}{dN}= \frac{d\ln c_{S}}{d\ln k}\\ 
                 ~~=\left\{\begin{array}{lll}
                    \displaystyle  
                   -\frac{r_{\star}}{48}\left(n_{\zeta,\star}-1+\frac{r_{\star}}{8}\right)\left(1-\frac{r_{\star}}{24}\right)^{1/4}+\cdots \,,~~~~ &
 \mbox{\small {\bf for $q=1/2$}}  \\ 
 \displaystyle  
 -\frac{(1-q)}{24q^2}\sqrt{\frac{q}{2}}r_{\star}\left(n_{\zeta,\star}-1+\frac{r_{\star}}{8}\right)\left(1-\frac{r_{\star}}{24}\right)^{1/4}+\cdots \,,~~~ &
 \mbox{\small {\bf for ~any~$q$}}.
          \end{array}
\right.
\end{array}\ee
\item Further the running of tensor spectral tilt can be written in terms of the inflationary observables as:
\begin{eqnarray}
\label{psxcc116} \alpha_{h, \star}&=&\left(\frac{dn_{h}(k)}{d\ln k}\right)_{k_{\star}=a_{\star}H_{\star}}=\left(\frac{dn_{h}(k)}{dN}\right)_{k_{\star}=a_{\star}H_{\star}}\\
&\approx&\left\{ \small\begin{array}{lll}
                    \displaystyle 
\displaystyle -\left[\frac{r_{\star}}{8}\left(1+\frac{r_{\star}}{8}\right)\left(n_{\zeta,\star}-1+\frac{r_{\star}}{8}\right)-\frac{r_{\star}}{8}\left(n_{\zeta,\star}-1+\frac{r_{\star}}{8}\right)^2
\right.\\ \left. \displaystyle ~~~~~~~~~~~-{\cal C}_{E}\frac{r_{\star}}{8}\left(n_{\zeta,\star}-1+\frac{r_{\star}}{8}\right)^2 \right]\displaystyle\left(1-\frac{r_{\star}}{24}\right)^{1/4}+\cdots, &
 \mbox{\small {\bf for $q=1/2$}}  \\ 
 \displaystyle  
               --\left[\frac{r_{\star}}{4}\sqrt{\frac{q}{2}}\left(1+\frac{r_{\star}}{8}\right)\left(n_{\zeta,\star}-1+\frac{r_{\star}}{8}\right)
\right.\\ \left. \displaystyle -\frac{1}{(2q)^{1/2}}\left({\cal C}_{E}+1\right)\frac{r_{\star}}{8}\left(n_{\zeta,\star}-1+\frac{r_{\star}}{8}\right)^2 \right]\displaystyle
\left(1-\frac{r_{\star}}{24}\right)^{1/4}+\cdots\,, &
 \mbox{\small {\bf for any $q$}}.
 \end{array}
\right.\end{eqnarray}
\item Next the scalar power spectrum can be expressed in terms of the other inflationary observables as:
\begin{eqnarray} 
\label{psxcc117} \Delta_{\zeta, \star}
&\approx&\left\{\begin{array}{lll}
                    \displaystyle 
 \displaystyle  
                \left[1-\left({\cal C}_{E}+\frac{5}{6}\right)\frac{r_{\star}}{16}+\frac{{\cal C}_{E}}{2}
                \left(n_{\zeta,\star}-1+\frac{r_{\star}}{8}\right)\right]^2 \frac{2H^{2}_{\star}}{\pi^2M^{2}_{p}r_{\star}}+\cdots \,,~~~~ &
 \mbox{\small {\bf for $q=1/2$}}  \\ 
 \displaystyle  
                \left[1-({\cal C}_{E}+1-\Sigma)\frac{r_{\star}}{16}+\frac{{\cal C}_{E}}{2}\left(n_{\zeta,\star}-1
                +\frac{r_{\star}}{8}\right)\right]^2 \displaystyle\frac{2 H^{2}_{\star}}{\pi^2M^{2}_{p}
                r_{\star}}+\cdots \,,~~~~ &
 \mbox{\small {\bf for any $q$}}.
 \end{array}
\right.\end{eqnarray}
\item Further the tensor power spectrum can be expressed in terms of the other inflationary observables as:
\begin{eqnarray} 
\label{psxcc118} \Delta_{h, \star}
&=&\left\{\begin{array}{lll}
                    \displaystyle 
 \displaystyle  \left[1-({\cal C}_{E}+1)\frac{r_{\star}}{16}\right]^2 \frac{2H^{2}_{\star}}{\pi^2M^{2}_{p}}+\cdots\,,~~~~~~~~~ &
 \mbox{\small {\bf for $q=1/2$}}  \\ 
 \displaystyle \left[1-({\cal C}_{E}+1)\frac{r_{\star}}{16}\right]^2 \frac{2H^{2}_{\star}}{\pi^2M^{2}_{p}}+\cdots \,,~~~~~~~~~ &
 \mbox{\small {\bf for any $q$}}.
 \end{array}
\right.\end{eqnarray}
\item Next the running of the tensor to scalar ratio can be expressed in terms of inflationary observables as:
\begin{eqnarray}
\label{psxcc119} \alpha_{r,\star}&=&\left(\frac{d r}{d\ln k}\right)_{\star}=\left(\frac{d r}{d N}\right)_{\star}=-8\alpha_{h, \star}+\cdots\nonumber\\
&\approx&\left\{ \small\begin{array}{lll}
                    \displaystyle 
\displaystyle \left[r_{\star}\left(1+\frac{r_{\star}}{8}\right)\left(n_{\zeta,\star}-1+\frac{r_{\star}}{8}\right)
-r_{\star}\left(n_{\zeta,\star}-1+\frac{r_{\star}}{8}\right)^2
\right.\\ \left. \displaystyle ~~~~~~~~~~~-{\cal C}_{E}r_{\star}\left(n_{\zeta,\star}-1+\frac{r_{\star}}{8}\right)^2 \right]\displaystyle\left(1-\frac{r_{\star}}{24}\right)^{1/4}+\cdots, &
 \mbox{\small {\bf for $q=1/2$}}  \\ 
 \displaystyle  
               \left[2r_{\star}\sqrt{\frac{q}{2}}\left(1+\frac{r_{\star}}{8}\right)\left(n_{\zeta,\star}-1+\frac{r_{\star}}{8}\right)
\right.\\ \left. \displaystyle -\frac{1}{(2q)^{1/2}}\left({\cal C}_{E}+1\right)r_{\star}\left(n_{\zeta,\star}-1
+\frac{r_{\star}}{8}\right)^2 \right]\displaystyle
\left(1-\frac{r_{\star}}{24}\right)^{1/4}+\cdots\,, &
 \mbox{\small {\bf for any $q$}}.
 \end{array}
\right.\end{eqnarray}
\item Finally the scale of single field tachyonic inflation can be expressed in terms of the Hubble parameter and the other inflationary observables as: 
\begin{eqnarray} \label{p20}
H_{inf}&=&\label{psxcc1} H_{\star}\nonumber\\
&\approx&\left\{\begin{array}{lll}
                    \displaystyle 
 \displaystyle  
               \frac{\sqrt{\frac{\Delta_{\zeta, \star}r_{\star}}{2}}\pi M_{p}}{ \left[1-\left({\cal C}_{E}+\frac{5}{6}\right)\frac{r_{\star}}{16}+\frac{{\cal C}_{E}}{2}
                \left(n_{\zeta,\star}-1+\frac{r_{\star}}{8}\right)\right]} +\cdots \,,~~~~ &
 \mbox{\small {\bf for $q=1/2$}}  \\ 
 \displaystyle \frac{\sqrt{\frac{\Delta_{\zeta, \star}r_{\star}}{2}}\pi M_{p}}{ \left[1-({\cal C}_{E}+1-\Sigma)
 \frac{r_{\star}}{16}+\frac{{\cal C}_{E}}{2}\left(n_{\zeta,\star}-1
                +\frac{r_{\star}}{8}\right)\right]} 
                +\cdots \,,~~~~ &
 \mbox{\small {\bf for any $q$}}.
 \end{array}
\right.\end{eqnarray}
One can recast this statement in terms of inflationary potential as:
\begin{eqnarray} 
\label{psxcc12200} \sqrt[4]{V_{inf}}&=&\sqrt[4]{V_{\star}}\nonumber\\
&\approx&\left\{\begin{array}{lll}
                    \displaystyle 
 \displaystyle  
               \frac{\sqrt[4]{\frac{3\Delta_{\zeta, \star}r_{\star}}{2}}\sqrt{\pi} M_{p}}{ \sqrt{\left[1-\left({\cal C}_{E}+
               \frac{5}{6}\right)\frac{r_{\star}}{16}+\frac{{\cal C}_{E}}{2}
                \left(n_{\zeta,\star}-1+\frac{r_{\star}}{8}\right)\right]}} +\cdots \,, &
 \mbox{\small {\bf for $q=1/2$}}  \\ 
 \displaystyle \frac{\sqrt[4]{\frac{3\Delta_{\zeta, \star}r_{\star}}{2}}\sqrt{\pi} M_{p}}{ \sqrt{\left[1-({\cal C}_{E}+1-\Sigma)
 \frac{r_{\star}}{16}+\frac{{\cal C}_{E}}{2}\left(n_{\zeta,\star}-1
                +\frac{r_{\star}}{8}\right)\right]}} 
                +\cdots \,, &
 \mbox{\small {\bf for any $q$}}.
 \end{array}
\right.\end{eqnarray}
\end{enumerate}

\subsubsection{Field excursion for tachyon}
In this subsection we explicitly derive the expression for the field excursion~\footnote{In the context of effective field theory,
with minimally coupled scalar field with Einstein gravity we compute the field excursion formula in refs.~\cite{Choudhury:2014sua,Choudhury:2015pqa,Choudhury:2014wsa,Choudhury:2014kma,Choudhury:2013iaa}. } for tachyonic inflation defined as:
\bea |\Delta T| &=& |T_{cmb}-T_{end}|=|T_{\star}-T_{end}| \eea
where $T_{cmb}$, $T_{end}$ and $T_{\star}$ signify the tachyon field value at the time of horizon exit, at end of inflation and at pivot scale respectively.
Here we perform the computation for both {\bf AV} and {\bf BD} vacuum.
For for the sake of simplicity the pivot scale is fixed at the horizon exit scale. To compute the expression for the field excursion we perform the following steps:
\begin{enumerate}
 \item We start with the operator identity for single field tachyon using which one can write expression for the tachyon field variation with respect to the momentum scale ($k$) or number of e-foldings ($N$) 
 in terms of the inflationary observables as:
 \be\begin{array}{lll}\label{vchj2xc}
 \displaystyle  \frac{1}{H}\frac{dT}{dt}=\frac{dT}{dN}=\frac{dT}{d\ln k}\approx\left\{\begin{array}{lll}
                    \displaystyle  
                   \sqrt{\frac{r}{8V(T)\alpha^{'}}}M_{p}\left(1-\frac{r}{24}\right)^{1/4}+\cdots\,,~~~~~~ &
 \mbox{\small {\bf for {$q=1/2$ }}}  \\ 
 \displaystyle   
                \sqrt{\frac{qr}{4V(T)\alpha^{'}}}\frac{M_{p}}{2q}\left(1-\frac{r}{24}\right)^{1/4}+\cdots\,.~~~~~~ &
 \mbox{\small {\bf for {~any~arbitrary~ $q$ }}} 
          \end{array}
\right.
\end{array}\ee
where the tensor-to-scalar ratio $r$ is function of $k$ or $N$.
 
 \item Next using Eq~(\ref{vchj2xc}) we can write the following integral equation:
 \be\begin{array}{lll}\label{vchj2xc22zx}
 \displaystyle  \int^{T_{\star}}_{T_{end}} dT~\sqrt{V(T)} \approx\left\{\begin{array}{lll}
                    \displaystyle  
                   \int^{k_{\star}}_{k_{end}} d\ln k~\sqrt{\frac{r}{8\alpha^{'}}}M_{p}\left(1-\frac{r}{24}\right)^{1/4}+\cdots\,\\ \displaystyle  
                  \displaystyle =\int^{N_{\star}}_{N_{end}} dN~\sqrt{\frac{r}{8\alpha^{'}}}M_{p}\left(1-\frac{r}{24}\right)^{1/4}+\cdots\,,~~~~~~ &
 \mbox{\small {\bf for {$q=1/2$ }}}  \\ 
 \displaystyle   
                \int^{k_{\star}}_{k_{end}} d\ln k~\sqrt{\frac{qr}{4\alpha^{'}}}\frac{M_{p}}{2q}\left(1-\frac{r}{24}\right)^{1/4}+\cdots\,\\
                \displaystyle   
                =\int^{N_{\star}}_{N_{end}} dN~\sqrt{\frac{qr}{4\alpha^{'}}}\frac{M_{p}}{2q}\left(1-\frac{r}{24}\right)^{1/4}+\cdots\,.~~~~~~ &
 \mbox{\small {\bf for {~any~arbitrary~ $q$ }}} 
          \end{array}
\right.
\end{array}\ee
 \item Next we parametrize the form of tensor-to-scalar ratio for $q=1/2$ and for any arbitrary $q$ at any arbitrary scale as:
 \be\begin{array}{lll}\label{rk1az}
  \displaystyle r(k)\displaystyle =\left\{\begin{array}{ll}\left\{\begin{array}{ll}
                    \displaystyle  r_{\star} &
 \mbox{ {\bf for \underline{Case I}}}  \\ 
         \displaystyle  r_{\star}\left(\frac{k}{k_{\star}}\right)^{n_{h,\star}-n_{\zeta,\star}+1} & \mbox{ {\bf for \underline{Case II}}}\\ 
\displaystyle  r_{\star}\left(\frac{k}{k_{*}}\right)^{n_{h,\star}-n_{\zeta,\star}+1+\frac{\alpha_{h,\star}-\alpha_{\zeta,\star}}{2!}\ln\left(\frac{k}{k_{\star}}\right)+\cdots} & \mbox{ {\bf for \underline{Case III}}}.
          \end{array}
\right.& \mbox{ {\bf for {BD}}} \\ 
\displaystyle  
  \frac{qr_{\star}}{2c^{2}_{S}}\times\frac{\left|D_{1}H^{(1)}_{\mu}\left(-kc_{S}\eta\right)+D_{2}H^{(2)}_{\mu}
 \left(-kc_{S}\eta\right)\right|^2}{\left|C_{1}H^{(1)}_{\nu}\left(-kc_{S}\eta\right)+C_{2}H^{(2)}_{\nu}
 \left(-kc_{S}\eta\right)\right|^2}\, & \mbox{ {\bf for {AV}}} 
\end{array}
\right.
\end{array}\ee
where $k_{*}$ be the pivot scale of momentum. One can also express Eq~(\ref{rk1az}) in terms of number of e-foldings ($N$) as:
\be\begin{array}{lll}\label{rk1az3}
  \displaystyle r(N)\displaystyle =\left\{\begin{array}{ll}\left\{\begin{array}{ll}
                    \displaystyle  r_{\star} &
 \mbox{ {\bf for \underline{Case I}}}  \\ 
         \displaystyle  r_{\star}\exp\left[(N-N_{\star})(n_{h,\star}-n_{\zeta,\star}+1)\right] & \mbox{ {\bf for \underline{Case II}}}\\ 
\displaystyle  r_{\star}\exp\left[(N-N_{\star})\left\{(n_{h,\star}-n_{\zeta,\star}+1)\right.\right.\\ \left. \left. \displaystyle\displaystyle~~~~~~~~~~+\frac{\alpha_{h,\star}-\alpha_{\zeta,\star}}{2!}(N-N_{\star})+\cdots\right\}\right] & \mbox{ {\bf for \underline{Case III}}}.
          \end{array}
\right.& \mbox{ {\bf for {BD}}} \\ 
\displaystyle  
  \frac{qr_{\star}}{2c^{2}_{S}}\times\frac{\left|D_{1}H^{(1)}_{\mu}\left(-k_{\star}c_{S}\eta\exp\left[N-N_{\star}\right]\right)+D_{2}H^{(2)}_{\mu}
 \left(-k_{\star}c_{S}\eta\exp\left[N-N_{\star}\right]\right)\right|^2}{\left|C_{1}H^{(1)}_{\nu}\left(-k_{\star}c_{S}\eta\exp\left[N-N_{\star}\right]\right)+C_{2}H^{(2)}_{\nu}
 \left(-k_{\star}c_{S}\eta\exp\left[N-N_{\star}\right]\right)\right|^2}\, & \mbox{ {\bf for {AV}}} 
\end{array}
\right.
\end{array}\ee
where $k$ and $N$ is connected through the following expression:
$\frac{k}{k_{\star}}= \exp\left[(N-N_{\star})\right].$
Here the three possibilities for {\bf BD} vacuum are:-
\begin{itemize}
                        \item \underline{\bf Case I} stands for a situation where the spectrum is scale invariant. This is the similar situation 
                                            as considered in case of Lyth bound. This possibility also surmounts to the Harrison \& Zeldovich
spectrum, which is completely ruled out by Planck 2015+WMAP9 data within 5$\sigma$ C.L.

\item  \underline{\bf Case II} stands for a situation where spectrum follows power law feature 
through the spectral tilt $(n_{\zeta},n_{h})$. This possibility is also tightly constrained by
the WMAP9 and Planck 2015+WMAP9 data within 2$\sigma$ C.L.,
\item \underline{\bf Case III} signifies a situation where the spectrum shows deviation from power low in presence of running of the
 spectral tilt $(\alpha_{\zeta},\alpha_{h})$ along with logarithmic correction in the momentum scale as appearing in the exponent.
 This possibility is favoured by WMAP9 data
and tightly constrained within 2$\sigma$ window by Planck+WMAP9 data.
\end{itemize}
\item For any value of $q$ including $q=1/2$ we need to compute the following integral:
\be\begin{array}{lll}\label{raqq1}
  \displaystyle \int^{k_{\star}}_{k_{end}} d\ln k~\sqrt{\frac{qr}{4\alpha^{'}}}\frac{M_{p}}{2q}\left(1-\frac{r}{24}\right)^{1/4}\,\\  \approx\small\left\{\begin{array}{ll}\tiny
                    \displaystyle  \sqrt{\frac{qr_{\star}}{4\alpha^{'}}}\frac{M_{p}}{2q}\left(1-\frac{r_{\star}}{24}\right)^{1/4}\ln\left(\frac{k_{\star}}{k_{end}}\right) &
 \mbox{ {\bf for \underline{Case I}}}  \\ \\
         \displaystyle \frac{\sqrt{\frac{qr_{\star}}{4\alpha^{'}}}\frac{M_{p}}{q}}{3(n_{h,\star}-n_{\zeta,\star}+1)}\left[  \left\{\, _2F_1\left[\frac{1}{2},\frac{3}{4};\frac{3}{2};\frac{r_{\star}}{24} \right]+2 \left(1-\frac{r_{\star}}{24}
         \right)^{1/4}\right\}\right.\\ \left.
         \displaystyle -\left(\frac{k_{end}}{k_{\star}}\right)^{\frac{n_{h,\star}-n_{\zeta,\star}+1}{2}} \left\{\, _2F_1\left[\frac{1}{2},\frac{3}{4};\frac{3}{2};\frac{r_{\star}}{24}
         \left(\frac{k_{end}}{k_{\star}}\right)^{n_{h,\star}-n_{\zeta,\star}+1}\right]\right.\right.\\ \left.\left.\displaystyle~~~~~~~~+2 \left(1-\frac{r_{\star}}{24}\left(\frac{k_{end}}{k_{\star}}\right)^{n_{h,\star}
         -n_{\zeta,\star}+1}
         \right)^{1/4}\right\}\right] & \mbox{ {\bf for \underline{Case II}}}\\ \\
\displaystyle  \sqrt{\frac{\pi q r_{\star}}{\alpha^{'}(\alpha_{h,\star}-\alpha_{\zeta,\star})}}\frac{M_{p}}{48q}
e^{-\frac{3 (n_{h,\star}-n_{\zeta,\star}+1)^2}{4 (\alpha_{h,\star}-\alpha_{\zeta,\star})}}\\
\displaystyle ~~\left[12 e^{\frac{(n_{h,\star}-n_{\zeta,\star}+1)^2}{ 2(\alpha_{h,\star}-\alpha_{\zeta,\star})}}\left\{ \text{erfi}\left(\frac{n_{h,\star}-n_{\zeta,\star}+1}{2 \sqrt{\alpha_{h,\star}-\alpha_{\zeta,\star}}}
\right)\right.\right.\\ \left.\left. 
\displaystyle~~~~~-
\text{erfi}\left(\frac{n_{h,\star}-n_{\zeta,\star}+1}{2 \sqrt{\alpha_{h,\star}-\alpha_{\zeta,\star}}}+\frac{\sqrt{\alpha_{h,\star}-\alpha_{\zeta,\star}}}{2}\ln\left(\frac{k_{end}}{k_{\star}}\right)
\right)\right\}\right.\\ \left. \displaystyle~~~~~~-\frac{\sqrt{3}r_{\star}}{24} \left\{\text{erfi}
\left(\frac{\sqrt{3} (n_{h,\star}-n_{\zeta,\star}+1)}{2 \sqrt{\alpha_{h,\star}-\alpha_{\zeta,\star}}}\right)\right.\right.\\ \left.\left. 
\displaystyle-\text{erfi}
\left(\frac{\sqrt{3} (n_{h,\star}-n_{\zeta,\star}+1)}{2 \sqrt{\alpha_{h,\star}-\alpha_{\zeta,\star}}}+\frac{\sqrt{3(\alpha_{h,\star}-\alpha_{\zeta,\star})}}{2}
\ln\left(\frac{k_{end}}{k_{\star}}\right)\right)\right\}\right] & \mbox{ {\bf for \underline{Case III}}}.
          \end{array}
\right.& \mbox{ \underline{\bf for {BD}}} 
\end{array}\ee
Similarly for {\bf AV} we get the following result:
\be\begin{array}{lll}\label{raqq2}
  \displaystyle \int^{k_{\star}}_{k_{end}} d\ln k~\sqrt{\frac{qr}{4\alpha^{'}}}\frac{M_{p}}{2q}\left(1-\frac{r}{24}\right)^{1/4}\,\\  \approx\small\left\{\begin{array}{ll}
                    \displaystyle  \displaystyle  
  \sqrt{\frac{r_{\star}}{8\alpha^{'}}}\frac{M_{p}}{2c_{S}}\left(1-\frac{qr_{\star}}{48c^{2}_{S}}\right)^{1/4}\ln\left(\frac{k_{\star}}{k_{end}}\right) &
 \mbox{ {\bf for \underline{Case I}}}  \\ \\
         \displaystyle \displaystyle  
  \sqrt{\frac{r_{\star}}{8\alpha^{'}}}\frac{M_{p}\left|D\right|}{2c_{S}\left|C\right|}\left(1-\frac{qr_{\star}}{48c^{2}_{S}}\frac{\left|D\right|^2}
  {\left|C\right|^2}\right)^{1/4}\ln\left(\frac{k_{\star}}{k_{end}}\right) & \mbox{ {\bf for \underline{Case II}}}.
          \end{array}
\right.& \mbox{ \underline{\bf for {AV}}} 
\end{array}\ee
In terms number of e-foldings $N$ one can re-express Eq~(\ref{raqq1}) and Eq~(\ref{raqq2}) as:
\be\begin{array}{lll}\label{raqg1}
  \displaystyle \int^{k_{\star}}_{k_{end}} d\ln k~\sqrt{\frac{qr}{4\alpha^{'}}}\frac{M_{p}}{2q}\left(1-\frac{r}{24}\right)^{1/4}\,\\  \approx\small\left\{\begin{array}{ll}
                    \displaystyle  \sqrt{\frac{qr_{\star}}{4\alpha^{'}}}\frac{M_{p}}{2q}\left(1-\frac{r_{\star}}{24}\right)^{1/4}\left(N_{\star}-N_{end}\right) &
 \mbox{ {\bf for \underline{Case I}}}  \\ \\
         \displaystyle \frac{\sqrt{\frac{qr_{\star}}{4\alpha^{'}}}\frac{M_{p}}{q}}{3(n_{h,\star}-n_{\zeta,\star}+1)}\left[  
         \left\{\, _2F_1\left[\frac{1}{2},\frac{3}{4};\frac{3}{2};\frac{r_{\star}}{24} \right]+2 \left(1-\frac{r_{\star}}{24}
         \right)^{1/4}\right\}\right.\\ \left.
         \displaystyle -e^{\frac{n_{h,\star}-n_{\zeta,\star}+1}{2}(N_{end}-N_{\star})} \left\{\, _2F_1\left[\frac{1}{2},\frac{3}{4};\frac{3}{2};\frac{r_{\star}}{24}
         e^{(n_{h,\star}-n_{\zeta,\star}+1)(N_{end}-N_{\star})}\right]\right.\right.\\ \left.\left.\displaystyle~~~~~~~~+2 \left(1-\frac{r_{\star}}{24}e^{(n_{h,\star}
         -n_{\zeta,\star}+1)(N_{end}-N_{\star})}
         \right)^{1/4}\right\}\right] & \mbox{ {\bf for \underline{Case II}}}\\ \\
\displaystyle  \sqrt{\frac{\pi q r_{\star}}{\alpha^{'}(\alpha_{h,\star}-\alpha_{\zeta,\star})}}\frac{M_{p}}{48q}
e^{-\frac{3 (n_{h,\star}-n_{\zeta,\star}+1)^2}{4 (\alpha_{h,\star}-\alpha_{\zeta,\star})}}\\
\displaystyle ~~\left[12 e^{\frac{(n_{h,\star}-n_{\zeta,\star}+1)^2}{ 2(\alpha_{h,\star}-\alpha_{\zeta,\star})}}\left\{ \text{erfi}\left(\frac{n_{h,\star}-n_{\zeta,\star}+1}{2 \sqrt{\alpha_{h,\star}-\alpha_{\zeta,\star}}}
\right)\right.\right.\\ \left.\left. 
\displaystyle~~~~~-
\text{erfi}\left(\frac{n_{h,\star}-n_{\zeta,\star}+1}{2 \sqrt{\alpha_{h,\star}-\alpha_{\zeta,\star}}}+\frac{\sqrt{\alpha_{h,\star}-\alpha_{\zeta,\star}}}{2}(N_{end}-N_{\star})
\right)\right\}\right.\\ \left. \displaystyle~~~~~~-\frac{\sqrt{3}r_{\star}}{24} \left\{\text{erfi}
\left(\frac{\sqrt{3} (n_{h,\star}-n_{\zeta,\star}+1)}{2 \sqrt{\alpha_{h,\star}-\alpha_{\zeta,\star}}}\right)\right.\right.\\ \left.\left. 
\displaystyle-\text{erfi}
\left(\frac{\sqrt{3} (n_{h,\star}-n_{\zeta,\star}+1)}{2 \sqrt{\alpha_{h,\star}-\alpha_{\zeta,\star}}}+\frac{\sqrt{3(\alpha_{h,\star}-\alpha_{\zeta,\star})}}{2}
(N_{end}-N_{\star})\right)\right\}\right] & \mbox{ {\bf for \underline{Case III}}}.
          \end{array}
\right.& \mbox{ \underline{\bf for {BD}}} 
\end{array}\ee
\be\begin{array}{lll}\label{raqg2}
  \displaystyle \int^{k_{\star}}_{k_{end}} d\ln k~\sqrt{\frac{qr}{4\alpha^{'}}}\frac{M_{p}}{2q}\left(1-\frac{r}{24}\right)^{1/4}\,\\ \\ \approx\small\left\{\begin{array}{ll}
                    \displaystyle  \displaystyle  
  \sqrt{\frac{r_{\star}}{8\alpha^{'}}}\frac{M_{p}}{2c_{S}}\left(1-\frac{qr_{\star}}{48c^{2}_{S}}\right)^{1/4}(N_{\star}-N_{end}) &
 \mbox{ {\bf for \underline{Case I}}}  \\ 
         \displaystyle \displaystyle  
  \sqrt{\frac{r_{\star}}{8\alpha^{'}}}\frac{M_{p}\left|D\right|}{2c_{S}\left|C\right|}\left(1-\frac{qr_{\star}}{48c^{2}_{S}}\frac{\left|D\right|^2}
  {\left|C\right|^2}\right)^{1/4}(N_{\star}-N_{end}) & \mbox{ {\bf for \underline{Case II}}}.
          \end{array}
\right.& \mbox{ \underline{\bf for {AV}}} 
\end{array}\ee

Here the two possibilities for {\bf AV} vacuum are:-
\begin{itemize}
                        \item \underline{\bf Case I} stands for a situation where the spectrum is characterized by the constraint i) $D_{1}=D_{2}=C_{1}=C_{2}\neq 0$, ii) $D_{1}=D_{2}$, $C_{1}=C_{2}=0$, iii) $D_{1}=D_{2}=0$,
                        $C_{1}=C_{2}$.

\item  \underline{\bf Case II} stands for a situation where the spectrum is characterized by the constraint i) $\mu\approx \nu$, $D_{1}=D_{2}=D\neq 0$ and $C_{1}=C_{2}=C\neq 0$, ii) $\mu\approx \nu$, $D_{1}=D\neq 0$, $D_{2}=0$ and $C_{1}=C\neq 0$, $C_{2}=0
$, iii) $\mu\approx \nu$, $D_{2}=D\neq 0$, $D_{1}=0$ and $C_{2}=C\neq 0$, $C_{1}=0$.
\end{itemize} 
\item Next we assume that the generic tachyonic potential $V(T)$ can be expressed as:
\bea V(T)&=& V_{0}+\sum^{\infty}_{n=1}\frac{1}{n!}\left(\frac{d^{n}V(T)}{dT^{n}}\right)_{T=T_{0}}\left(T-T_{0}\right)^{n}.\eea
where the contribution from $V_{0}$ fix the scale of potential and the higher order Taylor expansion co-efficients characterize the shape of potential.
\item Further we need to compute the following integral:
\be\begin{array}{lll}\label{vde1}
 \displaystyle  \int^{T_{\star}}_{T_{end}} dT~\sqrt{V(T)}\approx \int^{T_{\star}}_{T_{end}} dT~\sqrt{V_{0}+\sum^{\infty}_{n=1}\frac{1}{n!}\left(\frac{d^{n}V(T)}{dT^{n}}\right)_{T=T_{0}}\left(T-T_{0}\right)^{n}}\\
 \displaystyle~~~~~~~~~~~~~~~~~~~~~~\approx \int^{T_{\star}}_{T_{end}} dT~\sqrt{V_{0}}\left\{1+\sum^{\infty}_{n=1}\frac{1}{2 n!V_{0}}\left(\frac{d^{n}V(T)}{dT^{n}}\right)_{T=T_{0}}\left(T-T_{0}\right)^{n}+\cdots\right\}\\
 \displaystyle~~~~~~~~~~~~~~~~~~~~~~\approx \sqrt{V_{0}}\Delta T\left\{1+\sum^{\infty}_{n=1}\frac{\left(\frac{d^{n}V(T)}{dT^{n}}\right)_{T=T_{0}}}{2(n+1) n!V_{0}}\frac{\left[\left(T_{\star}-T_{0}\right)^{n+1}-
 \left(T_{end}-T_{0}\right)^{n+1}\right]}{\Delta T}+\cdots\right\}
\end{array}\ee
\item Next we assume that:
 \be \frac{1}{V_{0}}\left(\frac{d^{n}V(T)}{dT^{n}}\right)_{T=T_{0}}\frac{\left[\left(T_{\star}-T_{0}\right)^{n+1}-
 \left(T_{end}-T_{0}\right)^{n+1}\right]}{\Delta T}<<1.\ee
 Consequently from Eq~(\ref{vde1}) we get:
\be\label{vde2}
  \int^{T_{\star}}_{T_{end}} dT~\sqrt{V(T)}\approx\sqrt{V_{0}}\Delta T
\ee
\item Finally using Eq~(\ref{raqg1}), Eq~(\ref{raqg2}), Eq~(\ref{vde2}) and Eq~(\ref{vchj2xc22zx}) we get:
\be\begin{array}{lll}\label{raqggg1}
  \displaystyle \frac{\Delta T}{M_{p}}\,  \approx\small\left\{\begin{array}{ll}
                    \displaystyle  \sqrt{\frac{qr_{\star}}{4\alpha^{'}V_{0}}}\frac{1}{2q}\left(1-\frac{r_{\star}}{24}\right)^{1/4}\left(N_{\star}-N_{end}\right) &
 \mbox{ {\bf for \underline{Case I}}}  \\ \\
         \displaystyle \frac{\sqrt{\frac{qr_{\star}}{4\alpha^{'}V_{0}}}\frac{1}{q}}{3(n_{h,\star}-n_{\zeta,\star}+1)}\left[  
         \left\{\, _2F_1\left[\frac{1}{2},\frac{3}{4};\frac{3}{2};\frac{r_{\star}}{24} \right]+2 \left(1-\frac{r_{\star}}{24}
         \right)^{1/4}\right\}\right.\\ \left.
         \displaystyle -e^{\frac{n_{h,\star}-n_{\zeta,\star}+1}{2}(N_{end}-N_{\star})} \left\{\, _2F_1\left[\frac{1}{2},\frac{3}{4};\frac{3}{2};\frac{r_{\star}}{24}
         e^{(n_{h,\star}-n_{\zeta,\star}+1)(N_{end}-N_{\star})}\right]\right.\right.\\ \left.\left.\displaystyle~~~~~~~~+2 \left(1-\frac{r_{\star}}{24}e^{(n_{h,\star}
         -n_{\zeta,\star}+1)(N_{end}-N_{\star})}
         \right)^{1/4}\right\}\right] & \mbox{ {\bf for \underline{Case II}}}\\ \\
\displaystyle  \sqrt{\frac{\pi q r_{\star}}{\alpha^{'}(\alpha_{h,\star}-\alpha_{\zeta,\star})V_{0}}}\frac{1}{48q}
e^{-\frac{3 (n_{h,\star}-n_{\zeta,\star}+1)^2}{4 (\alpha_{h,\star}-\alpha_{\zeta,\star})}}\\
\displaystyle ~~\left[12 e^{\frac{(n_{h,\star}-n_{\zeta,\star}+1)^2}{ 2(\alpha_{h,\star}-\alpha_{\zeta,\star})}}\left\{ \text{erfi}\left(\frac{n_{h,\star}-n_{\zeta,\star}+1}{2 \sqrt{\alpha_{h,\star}-\alpha_{\zeta,\star}}}
\right)\right.\right.\\ \left.\left. 
\displaystyle~~~~~-
\text{erfi}\left(\frac{n_{h,\star}-n_{\zeta,\star}+1}{2 \sqrt{\alpha_{h,\star}-\alpha_{\zeta,\star}}}+\frac{\sqrt{\alpha_{h,\star}-\alpha_{\zeta,\star}}}{2}(N_{end}-N_{\star})
\right)\right\}\right.\\ \left. \displaystyle~~~~~~-\frac{\sqrt{3}r_{\star}}{24} \left\{\text{erfi}
\left(\frac{\sqrt{3} (n_{h,\star}-n_{\zeta,\star}+1)}{2 \sqrt{\alpha_{h,\star}-\alpha_{\zeta,\star}}}\right)\right.\right.\\ \left.\left. 
\displaystyle-\text{erfi}
\left(\frac{\sqrt{3} (n_{h,\star}-n_{\zeta,\star}+1)}{2 \sqrt{\alpha_{h,\star}-\alpha_{\zeta,\star}}}+\frac{\sqrt{3(\alpha_{h,\star}-\alpha_{\zeta,\star})}}{2}
(N_{end}-N_{\star})\right)\right\}\right] & \mbox{ {\bf for \underline{Case III}}}.
          \end{array}
\right.& \mbox{ \small\underline{\bf for {BD}}} 
\\\end{array}\ee
\be\begin{array}{lll}\label{raqggg2}
  \displaystyle \frac{\Delta T}{M_{p}} \approx\small\left\{\begin{array}{ll}
                    \displaystyle  \displaystyle  
  \sqrt{\frac{r_{\star}}{8\alpha^{'}V_{0}}}\frac{1}{2c_{S}}\left(1-\frac{qr_{\star}}{48c^{2}_{S}}\right)^{1/4}(N_{\star}-N_{end}) &
 \mbox{ {\bf for \underline{Case I}}}  \\ \\
         \displaystyle \displaystyle  
  \sqrt{\frac{r_{\star}}{8\alpha^{'}V_{0}}}\frac{\left|D\right|}{2c_{S}\left|C\right|}\left(1-\frac{qr_{\star}}{48c^{2}_{S}}\frac{\left|D\right|^2}
  {\left|C\right|^2}\right)^{1/4}(N_{\star}-N_{end}) & \mbox{ {\bf for \underline{Case II}}}.
          \end{array}
\right.& \mbox{ \underline{\bf for {AV}}} 
\end{array}\ee
\item Next using Eq~(\ref{psxcc12200}) in eq~(\ref{raqggg1}) and Eq~(\ref{raqggg2}) we get:
\be\begin{array}{lll}\label{raqgggg1}
  \displaystyle \frac{\Delta T}{M_{p}}\,  \approx\small\left\{\begin{array}{ll}
                    \displaystyle  \sqrt{\frac{qr_{\star}}{4\alpha^{'}}}\frac{1}{2q\sqrt{\frac{3\Delta_{\zeta, \star}r_{\star}}{2}}\pi M^2_{p}}\left(1-\frac{r_{\star}}{24}\right)^{1/4}\left(N_{\star}-N_{end}\right)\\
                    \displaystyle   \left[1-({\cal C}_{E}+1-\Sigma)
 \frac{r_{\star}}{16}+\frac{{\cal C}_{E}}{2\sqrt{2q}}\left(n_{\zeta,\star}-1
                \right.\right.\\ \left.\left.~~~~~~~~~~~~~\displaystyle+\left(1+3\sqrt{2q}-6q\right)\frac{r_{\star}}{8}\right)\right]&
 \mbox{ {\bf for \underline{Case I}}}  \\ \\
         \displaystyle \frac{\sqrt{\frac{qr_{\star}}{4\alpha^{'}}}\frac{1}{q\sqrt{\frac{3\Delta_{\zeta, \star}r_{\star}}{2}}\pi M^2_{p}}}{3(n_{h,\star}-n_{\zeta,\star}+1)}\left[  
         \left\{\, _2F_1\left[\frac{1}{2},\frac{3}{4};\frac{3}{2};\frac{r_{\star}}{24} \right]+2 \left(1-\frac{r_{\star}}{24}
         \right)^{1/4}\right\}\right.\\ \left.
         \displaystyle -e^{\frac{n_{h,\star}-n_{\zeta,\star}+1}{2}(N_{end}-N_{\star})} \left\{\, _2F_1\left[\frac{1}{2},\frac{3}{4};\frac{3}{2};\frac{r_{\star}}{24}
         e^{(n_{h,\star}-n_{\zeta,\star}+1)(N_{end}-N_{\star})}\right]\right.\right.\\ \left.\left.\displaystyle~~~~~~~~+2 \left(1-\frac{r_{\star}}{24}e^{(n_{h,\star}
         -n_{\zeta,\star}+1)(N_{end}-N_{\star})}
         \right)^{1/4}\right\}\right]\\
         \displaystyle   \left[1-({\cal C}_{E}+1-\Sigma)
 \frac{r_{\star}}{16}+\frac{{\cal C}_{E}}{2\sqrt{2q}}\left(n_{\zeta,\star}-1
                \right.\right.\\ \left.\left.~~~~~~~~~~~~~\displaystyle+\left(1+3\sqrt{2q}-6q\right)\frac{r_{\star}}{8}\right)\right]& \mbox{ {\bf for \underline{Case II}}}\\ \\
\displaystyle \frac{ \sqrt{\frac{\pi q r_{\star}}{\alpha^{'}(\alpha_{h,\star}-\alpha_{\zeta,\star})}}}{48q\sqrt{\frac{3\Delta_{\zeta, \star}r_{\star}}{2}}\pi M^2_{p}}
e^{-\frac{3 (n_{h,\star}-n_{\zeta,\star}+1)^2}{4 (\alpha_{h,\star}-\alpha_{\zeta,\star})}}\\
\displaystyle ~~\left[12 e^{\frac{(n_{h,\star}-n_{\zeta,\star}+1)^2}{ 2(\alpha_{h,\star}-\alpha_{\zeta,\star})}}\left\{ \text{erfi}\left(\frac{n_{h,\star}-n_{\zeta,\star}+1}{2 \sqrt{\alpha_{h,\star}-\alpha_{\zeta,\star}}}
\right)\right.\right.\\ \left.\left. 
\displaystyle~~~~~-
\text{erfi}\left(\frac{n_{h,\star}-n_{\zeta,\star}+1}{2 \sqrt{\alpha_{h,\star}-\alpha_{\zeta,\star}}}+\frac{\sqrt{\alpha_{h,\star}-\alpha_{\zeta,\star}}}{2}(N_{end}-N_{\star})
\right)\right\}\right.\\ \left. \displaystyle~~~~~~-\frac{\sqrt{3}r_{\star}}{24} \left\{\text{erfi}
\left(\frac{\sqrt{3} (n_{h,\star}-n_{\zeta,\star}+1)}{2 \sqrt{\alpha_{h,\star}-\alpha_{\zeta,\star}}}\right)\right.\right.\\ \left.\left. 
\displaystyle-\text{erfi}
\left(\frac{\sqrt{3} (n_{h,\star}-n_{\zeta,\star}+1)}{2 \sqrt{\alpha_{h,\star}-\alpha_{\zeta,\star}}}+\frac{\sqrt{3(\alpha_{h,\star}-\alpha_{\zeta,\star})}}{2}
(N_{end}-N_{\star})\right)\right\}\right]\\
\displaystyle   \left[1-({\cal C}_{E}+1-\Sigma)
 \frac{r_{\star}}{16}+\frac{{\cal C}_{E}}{2\sqrt{2q}}\left(n_{\zeta,\star}-1
                \right.\right.\\ \left.\left.~~~~~~~~~~~~~\displaystyle+\left(1+3\sqrt{2q}-6q\right)\frac{r_{\star}}{8}\right)\right]& \mbox{ {\bf for \underline{Case III}}}.
          \end{array}
\right.& \mbox{ \small\underline{\bf for {BD}}} 
\end{array}\ee
\be\begin{array}{lll}\label{raqgggg2}
  \displaystyle \frac{\Delta T}{M_{p}} \approx\small\left\{\begin{array}{ll}
                    \displaystyle  \displaystyle  
  \sqrt{\frac{r_{\star}}{8\alpha^{'}}}\frac{\left(1-\frac{qr_{\star}}{48c^{2}_{S}}\right)^{1/4}(N_{\star}-N_{end})}{2c_{S}\sqrt{\frac{3\Delta_{\zeta, \star}r_{\star}}{2}}\pi M^2_{p}}\\ \displaystyle
                 \left[1-({\cal C}_{E}+1-\Sigma)
 \frac{r_{\star}}{16}+\frac{{\cal C}_{E}}{2\sqrt{2q}}\left(n_{\zeta,\star}-1
                +\left(1+3\sqrt{2q}-6q\right)\frac{r_{\star}}{8}\right)\right]&
 \mbox{ {\bf for \underline{Case I}}}  \\ 
         \displaystyle \displaystyle  
  \sqrt{\frac{r_{\star}}{8\alpha^{'}}}\frac{\left|D\right|\left(1-\frac{qr_{\star}}{48c^{2}_{S}}\frac{\left|D\right|^2}
  {\left|C\right|^2}\right)^{1/4}(N_{\star}-N_{end})}{2c_{S}\left|C\right|\sqrt{\frac{3\Delta_{\zeta, \star}r_{\star}}{2}}\pi M^2_{p}}\\ \displaystyle
                \left[1-({\cal C}_{E}+1-\Sigma)
 \frac{r_{\star}}{16}+\frac{{\cal C}_{E}}{2\sqrt{2q}}\left(n_{\zeta,\star}-1
                +\left(1+3\sqrt{2q}-6q\right)\frac{r_{\star}}{8}\right)\right] & \mbox{ {\bf for \underline{Case II}}}.
          \end{array}
\right.& \mbox{ \underline{\bf \small for {AV}}} 
\end{array}\ee
Further using the approximated form of the sound speed $c_{S}$ the expression for the field excursion for {\bf AV} can be re-written as:
\be\begin{array}{lll}\label{raqggggg2}
  \displaystyle \frac{\Delta T}{M_{p}} \approx\small\left\{\begin{array}{ll}
                    \displaystyle  \displaystyle  
  \sqrt{\frac{r_{\star}}{8\alpha^{'}}}\frac{\left(1-\frac{qr_{\star}}{48\left[1-\frac{(1-q)}{3q}\frac{r_{\star}}{8}\right]}\right)^{1/4}(N_{\star}-N_{end})}{2\sqrt{1-\frac{(1-q)}{3q}\frac{r_{\star}}{8}}\sqrt{\frac{3\Delta_{\zeta, \star}r_{\star}}{2}}\pi M^2_{p}}\\ \displaystyle
                 \left[1-({\cal C}_{E}+1-\Sigma)
 \frac{r_{\star}}{16}+\frac{{\cal C}_{E}}{2\sqrt{2q}}\left(n_{\zeta,\star}-1
                +\left(1+3\sqrt{2q}-6q\right)\frac{r_{\star}}{8}\right)\right]&
 \mbox{ {\bf for \underline{Case I}}}  \\ 
         \displaystyle \displaystyle  
  \sqrt{\frac{r_{\star}}{8\alpha^{'}}}\frac{\left|D\right|\left(1-\frac{qr_{\star}}{48\left[1-\frac{(1-q)}{3q}\frac{r_{\star}}{8}\right]}\frac{\left|D\right|^2}
  {\left|C\right|^2}\right)^{1/4}(N_{\star}-N_{end})}{2\sqrt{1-\frac{(1-q)}{3q}\frac{r_{\star}}{8}}\left|C\right|\sqrt{\frac{3\Delta_{\zeta, \star}r_{\star}}{2}}\pi M^2_{p}}\\ \displaystyle
                \left[1-({\cal C}_{E}+1-\Sigma)
 \frac{r_{\star}}{16}+\frac{{\cal C}_{E}}{2\sqrt{2q}}\left(n_{\zeta,\star}-1
                +\left(1+3\sqrt{2q}-6q\right)\frac{r_{\star}}{8}\right)\right] & \mbox{ {\bf for \underline{Case II}}}.
          \end{array}
\right.& \mbox{ \underline{\bf \small for {AV}}} 
\end{array}\ee
\end{enumerate}

\subsubsection{Semi analytical study and Cosmological parameter estimation}
\begin{figure}[htb]
	\centering
	\includegraphics[width=12cm,height=7cm]{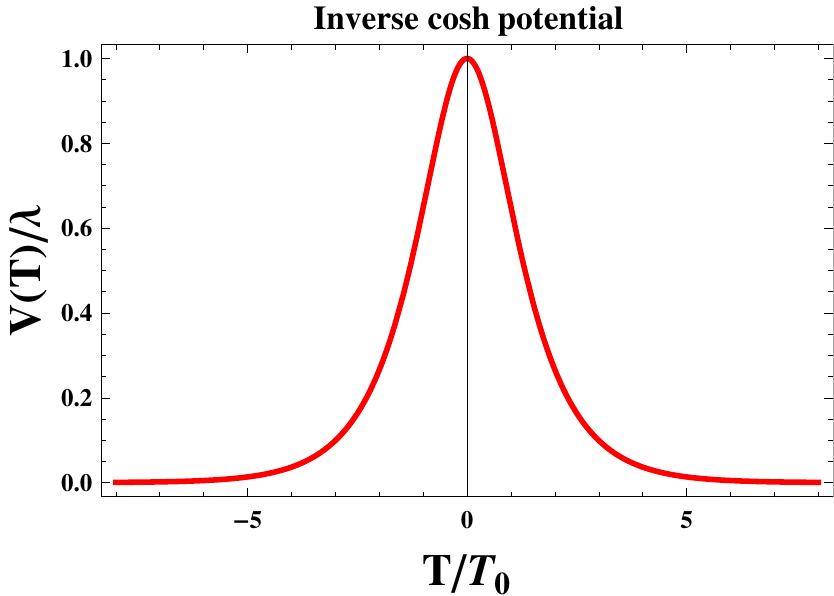}
	\caption{Variation of the Inverse cosh potential $V(T)/\lambda$ with field $T/T_{0}$ in dimensionless units. 
	}
	\label{fig1}
\end{figure}

In this subsection our prime objective are:
\begin{itemize}
 \item  To compute various inflationary observables from variants of tachyonic single field potentials as mentioned earlier in this paper,
 
 \item  Estimate the relevant cosmological parameters from the proposed models,
 
 \item Next
to compare the effectiveness of all of these models in the light of recent Planck 2015 data alongwith other combined constraints.

\item Finally we will check the compatibility of all of these models with the CMB TT, TE and EE angular power spectra as observed by Planck 2015.
\end{itemize}
\underline{\bf Model I: Inverse cosh potential}\\
For single field case the first model of tachyonic potential is given by:
\be\label{webbbb1}
V(T)=\frac{\lambda}{{\rm cosh}\left(\frac{T}{T_{0}}\right)},\ee
where $\lambda$ characterize the scale of inflation and $T_{0}$ is the parameter of the model.
In Fig.~(\ref{fig1}) we have depicted the symmetric behavior of the Inverse cosh potential with respect to scaled field coordinate $T/T_{0}$ in dimensionless units around the origin fixed at $T/T_{0}=0$.
In this case the tachyon field started rolling down from the top hight of the potential and take part in inflationary dynamics.  


Next using specified form of the potential the potential dependent slow-roll parameters are computed as:
\bea \bar{\epsilon}_{V}&=&\frac{1}{2g}\frac{\sinh^2\left(\frac{T}{T_{0}}\right) }{\cosh\left(\frac{T}{T_{0}}\right) },\\
\bar{\eta}_{V}&=&\frac{1}{g}\left[\frac{\sinh^2\left(\frac{T}{T_{0}}\right) }{\cosh\left(\frac{T}{T_{0}}\right) }-{\rm sech}\left(\frac{T}{T_{0}}\right)\right],\\
\bar{\xi}^2_{V}&=&\frac{1}{g^2}\frac{\sinh^2\left(\frac{T}{T_{0}}\right) }{\cosh\left(\frac{T}{T_{0}}\right) }\left[\frac{\sinh^2\left(\frac{T}{T_{0}}\right) }{\cosh\left(\frac{T}{T_{0}}\right) }-5~{\rm sech}\left(\frac{T}{T_{0}}\right)\right],~~~~~\\
\bar{\sigma}^3_{V}&=&\frac{1}{g^3}\frac{\sinh^2\left(\frac{T}{T_{0}}\right) }{\cosh\left(\frac{T}{T_{0}}\right) }\left[\frac{\sinh^2\left(\frac{T}{T_{0}}\right) }{\cosh\left(\frac{T}{T_{0}}\right) }\left\{\frac{\sinh^2\left(\frac{T}{T_{0}}\right) }{\cosh\left(\frac{T}{T_{0}}\right) }-18~{\rm sech}\left(\frac{T}{T_{0}}\right)\right\}
+5~{\rm sech}^2\left(\frac{T}{T_{0}}\right)\right],~~~~~~~~~~~~~~
\eea
where the factor $g$ is defined as:
\be g= \frac{\alpha^{'}\lambda T^2_0}{M^2_p}=\frac{M^4_s}{(2\pi)^3 g_s}\frac{\alpha^{'} T^2_0}{M^2_p}.\ee
Next we compute the number of e-foldings from this model:
\be\begin{array}{lll}\label{e-fold5}
 \displaystyle  N(T)=\left\{\begin{array}{lll}
                    \displaystyle  
                   g~\ln\left[\frac{\tanh\left(\frac{T_{end}}{2T_{0}}\right)}{\tanh\left(\frac{T}{2T_{0}}\right)}\right]\,,~~~~~~ &
 \mbox{\small {\bf for {$q=1/2$ }}}  \\ 
 \displaystyle   
             \sqrt{2q}~g~\ln\left[\frac{\tanh\left(\frac{T_{end}}{2T_{0}}\right)}{\tanh\left(\frac{T}{2T_{0}}\right)}\right]\,.~~~~~~ &
 \mbox{\small {\bf for {~any~arbitrary~ $q$ }}} 
          \end{array}
\right.
\end{array}\ee
Further using the condition to end inflation:
\bea \bar{\epsilon}_{V}(T_{end})=1,\\
      |\bar{\eta}_{V}(T_{end})|=1,\eea
      we get the following field value at the end of inflation:
      \be T_{end}=T_{0}~{\rm sech}^{-1}(g).\ee
Next using $N=N_{cmb}=N_{\star}$ and $T=T_{cmb}=T_{\star}$ at the horizon crossing we get,
\be\begin{array}{lll}\label{e-fold6}
 \displaystyle  T_{\star}=2T_{0}\times\left\{\begin{array}{lll}
                    \displaystyle  
                   \tanh^{-1}\left[\exp\left(-\frac{N_{\star}}{g}\right)\tanh\left(\frac{1}{2}{\rm sech}^{-1}(g)\right)\right]\,,~~~~~~ &
 \mbox{\small {\bf for {$q=1/2$ }}}  \\ 
 \displaystyle   
              \tanh^{-1}\left[\exp\left(-\frac{N_{\star}}{\sqrt{2q}g}\right)\tanh\left(\frac{1}{2}{\rm sech}^{-1}(g)\right)\right]\,.~~~~~~ &
 \mbox{\small {\bf for {~any~arbitrary~ $q$ }}} 
          \end{array}
\right.
\end{array}\ee
Consequently the field excursion can be computed as:
\be\begin{array}{lll}\label{e-fold6ex}
 \displaystyle  |\Delta T|=T_{0}\times\left\{\begin{array}{lll}
                    \displaystyle  
                   \left|2\tanh^{-1}\left[\exp\left(-\frac{N_{\star}}{g}\right)\tanh\left(\frac{1}{2}{\rm sech}^{-1}(g)\right)\right]-{\rm sech}^{-1}(g)\right|\,,~~~~~~ &
 \mbox{\small {\bf for {$q=1/2$ }}}  \\ 
 \displaystyle   
              \left|2\tanh^{-1}\left[\exp\left(-\frac{N_{\star}}{\sqrt{2q}g}\right)\tanh\left(\frac{1}{2}{\rm sech}^{-1}(g)\right)\right]-{\rm sech}^{-1}(g)\right|\,.~~~~~~ &
 \mbox{\small {\bf for {~any~ $q$ }}} 
          \end{array}
\right.
\end{array}\ee
In the slow-roll regime of inflation the following approximations holds good:
\be\begin{array}{lll}\label{e-fold7}
 \displaystyle   \cosh\left(\frac{T_{\star}}{T_{0}}\right)\approx\left\{\begin{array}{lll}
                    \displaystyle  
                   \frac{1}{\tanh\left(\frac{N_{\star}}{g}\right)}\,,~~~~~~ &
 \mbox{\small {\bf for {$q=1/2$ }}}  \\ 
 \displaystyle   
             \frac{1}{\tanh\left(\frac{N_{\star}}{\sqrt{2q}g}\right)}\,.~~~~~~ &
 \mbox{\small {\bf for {~any~arbitrary~ $q$ }}} 
          \end{array}
\right.
\end{array}\ee
\be\begin{array}{lll}\label{e-fold8}
 \displaystyle   \sinh\left(\frac{T_{\star}}{T_{0}}\right)\approx\left\{\begin{array}{lll}
                    \displaystyle  
                   \frac{1}{\sinh\left(\frac{N_{\star}}{g}\right)}\,,~~~~~~ &
 \mbox{\small {\bf for {$q=1/2$ }}}  \\ 
 \displaystyle   
             \frac{1}{\sinh\left(\frac{N_{\star}}{\sqrt{2q}g}\right)}\,.~~~~~~ &
 \mbox{\small {\bf for {~any~arbitrary~ $q$ }}} 
          \end{array}
\right.
\end{array}\ee
Using Eq~(\ref{e-fold7}) and Eq~(\ref{e-fold8}) in the definition of potential dependent slow-roll parameter 
finally we compute the  following inflationary observables:
\be\begin{array}{lll}\label{e-fold9}
 \displaystyle   \Delta_{\zeta,\star}\approx\frac{g\lambda}{12\pi^2M^{4}_{p}}\times\left\{\begin{array}{lll}
                    \displaystyle  
                   \sinh^2\left(\frac{N_{\star}}{g}\right)\,,~~~~~~ &
 \mbox{\small {\bf for {$q=1/2$ }}}  \\ 
 \displaystyle   
             2q~\sinh^2\left(\frac{N_{\star}}{\sqrt{2q}g}\right)\,.~~~~~~ &
 \mbox{\small {\bf for {~any~arbitrary~ $q$ }}} 
          \end{array}
\right.
\end{array}\ee
\be\begin{array}{lll}\label{e-fold10}
 \displaystyle   n_{\zeta,\star}-1\approx-\frac{2}{g}\times\left\{\begin{array}{lll}
                    \displaystyle  
                   \frac{1}{\tanh\left(\frac{N_{\star}}{g}\right)}\,,~~~~~~ &
 \mbox{\small {\bf for {$q=1/2$ }}}  \\ 
 \displaystyle   
             \frac{1}{\sqrt{2q}\tanh\left(\frac{N_{\star}}{\sqrt{2q}g}\right)}\,.~~~~~~ &
 \mbox{\small {\bf for {~any~arbitrary~ $q$ }}} 
          \end{array}
\right.
\end{array}\ee
\be\begin{array}{lll}\label{e-fold11}
 \displaystyle   \alpha_{\zeta,\star}\approx-\frac{2}{g^2}\times\left\{\begin{array}{lll}
                    \displaystyle  
                   \frac{1}{\sinh^2\left(\frac{N_{\star}}{g}\right)}\,,~~~~~~ &
 \mbox{\small {\bf for {$q=1/2$ }}}  \\ 
 \displaystyle   
             \frac{1}{2q\sinh^2\left(\frac{N_{\star}}{\sqrt{2q}g}\right)}\,.~~~~~~ &
 \mbox{\small {\bf for {~any~arbitrary~ $q$ }}} 
          \end{array}
\right.
\end{array}\ee
\be\begin{array}{lll}\label{e-fold12}
 \displaystyle   \kappa_{\zeta,\star}\approx-\frac{4}{g^3}\times\left\{\begin{array}{lll}
                    \displaystyle  
                   \frac{\cosh\left(\frac{N_{\star}}{g}\right)}{\sinh^3\left(\frac{N_{\star}}{g}\right)}\,,~~~~~~ &
 \mbox{\small {\bf for {$q=1/2$ }}}  \\ 
 \displaystyle   
             \frac{\cosh\left(\frac{N_{\star}}{2qg}\right)}{(2q)^{3/2}\sinh^3\left(\frac{N_{\star}}{\sqrt{2q}g}\right)}\,.~~~~~~ &
 \mbox{\small {\bf for {~any~arbitrary~ $q$ }}} 
          \end{array}
\right.
\end{array}\ee
\be\begin{array}{lll}\label{e-fold13}
 \displaystyle   r_{\star}\approx\frac{16}{g}\times\left\{\begin{array}{lll}
                    \displaystyle  
                   \frac{1}{\sinh\left(\frac{2N_{\star}}{g}\right)}\,,~~~~~~ &
 \mbox{\small {\bf for {$q=1/2$ }}}  \\ 
 \displaystyle   
             \frac{1}{2q\sinh\left(\frac{2N_{\star}}{\sqrt{2q}g}\right)}\,.~~~~~~ &
 \mbox{\small {\bf for {~any~arbitrary~ $q$ }}} 
          \end{array}
\right.
\end{array}\ee
\begin{figure*}[htb]
\centering
\subfigure[$r$ vs $n_{\zeta}$.]{
    \includegraphics[width=12.2cm,height=6.5cm] {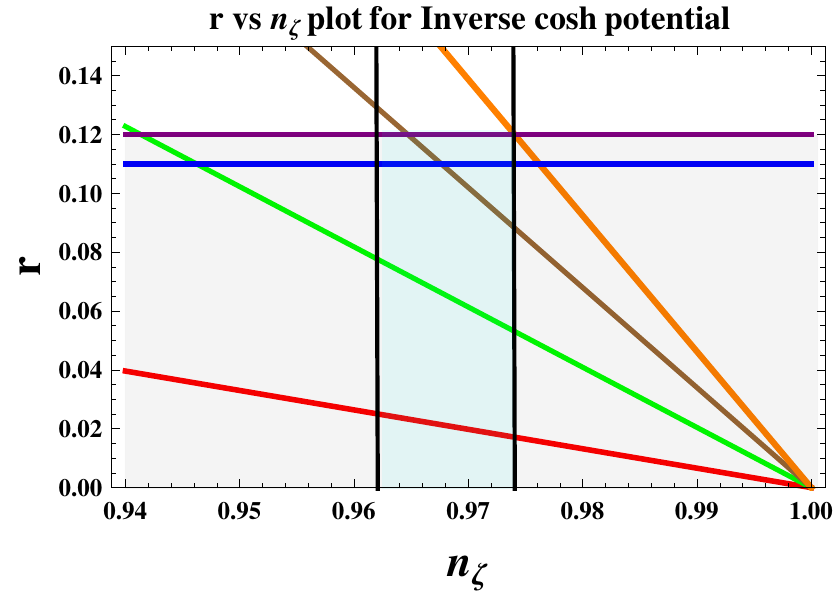}
    \label{fig2a}
}
\subfigure[$r$ vs $g$.]{
    \includegraphics[width=12.2cm,height=6.5cm] {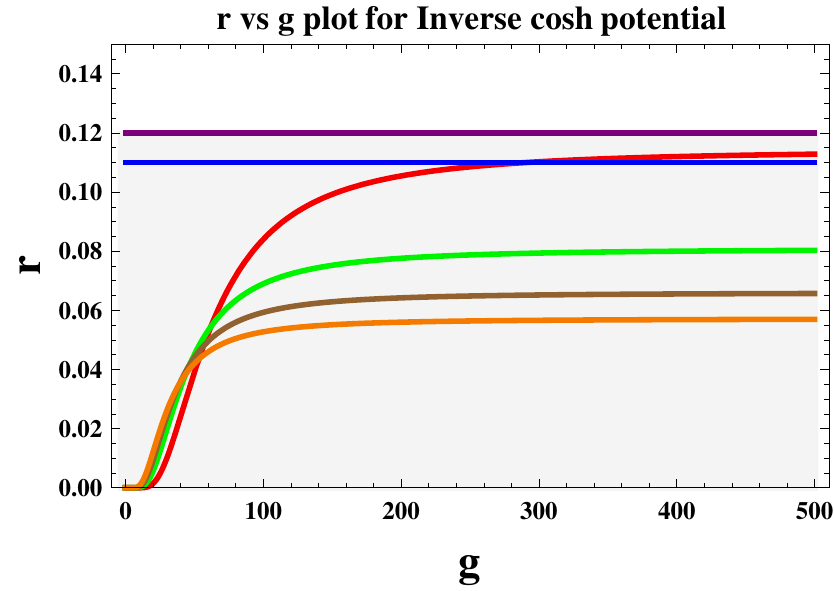}
    \label{fig2b}
}
\caption[Optional caption for list of figures]{Behaviour of the tensor-to-scalar ratio $r$ with respect to \ref{fig2a} the scalar spectral index $n_{\zeta}$
and \ref{fig2b} the parameter $g$ for Inverse cosh potential.
The purple and blue coloured line represent the upper bound of tensor-to-scalar ratio allowed by Planck+ BICEP2+Keck Array joint constraint and only Planck 2015 data respectively. For both the figures 
 \textcolor{red}{red},~\textcolor{green}{green},
~\textcolor{brown}{brown},~\textcolor{orange}{orange} colored curve represent $q=1/2$, $q=1$, $q=3/2$ and $q=2$ respectively.
The \textcolor{cyan}{cyan} color shaded region bounded by two vertical black coloured lines in \ref{fig2a} represent the Planck $2\sigma$ allowed region and the rest of the light gray shaded region is
showing the $1\sigma$ allowed range, which is at present 
disfavoured by the Planck data and Planck+ BICEP2+Keck Array joint constraint. From \ref{fig2a} and \ref{fig2b}, it 
is also observed that, within $50<N_{\star}<70$ the Inverse cosh potential is favoured only for the characteristic index $1/2<q<2$, by Planck 2015 data and Planck+ BICEP2+Keck Array joint analysis.
In \ref{fig2b}, we have explicitly shown that in $r-g$ plane the observationally favoured lower bound for the characteristic index is $q\geq 1/2$. } 
\label{fig2}
\end{figure*}
\begin{figure*}[htb]
\centering
\subfigure[$\Delta_{\zeta}$ vs $g$.]{
    \includegraphics[width=7.2cm,height=8cm] {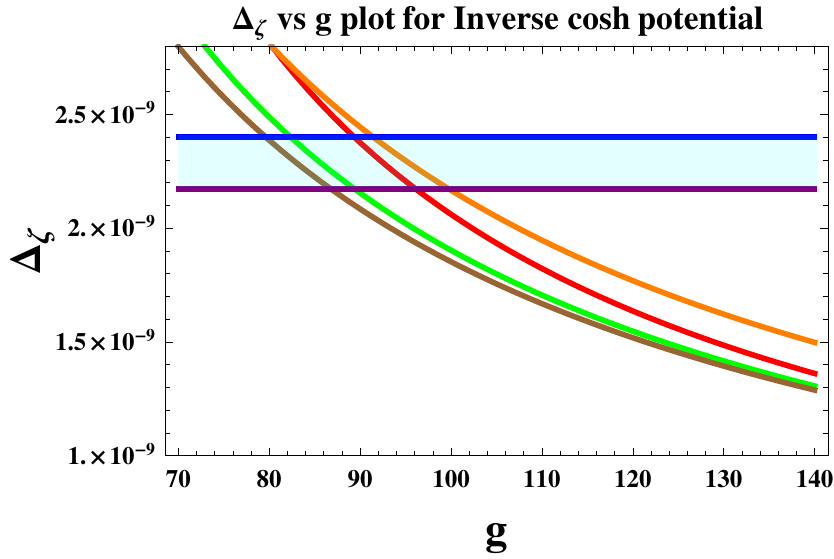}
    \label{fig3a}
}
\subfigure[$n_{\zeta}$ vs $g$.]{
    \includegraphics[width=7.2cm,height=8cm] {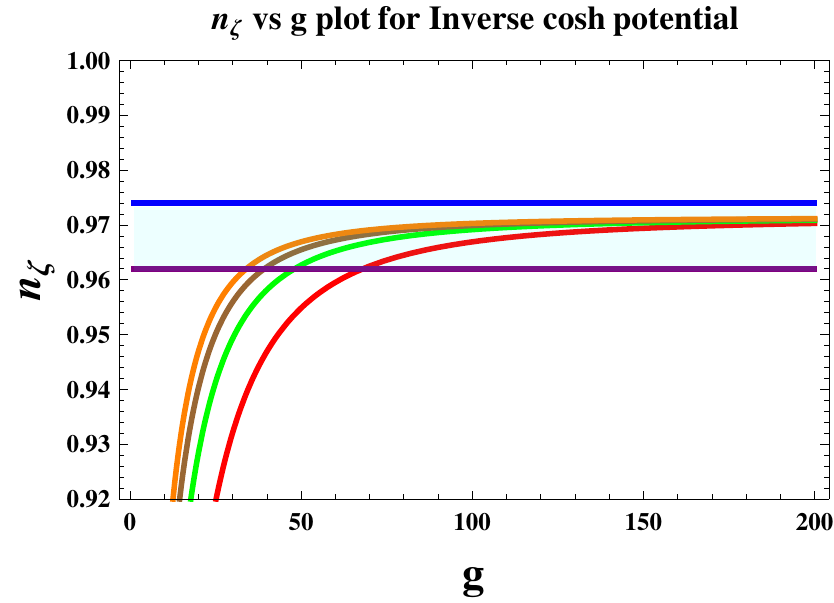}
    \label{fig3b}
}
\subfigure[$\Delta_{\zeta}$ vs $n_{\zeta}$.]{
    \includegraphics[width=11.2cm,height=8cm] {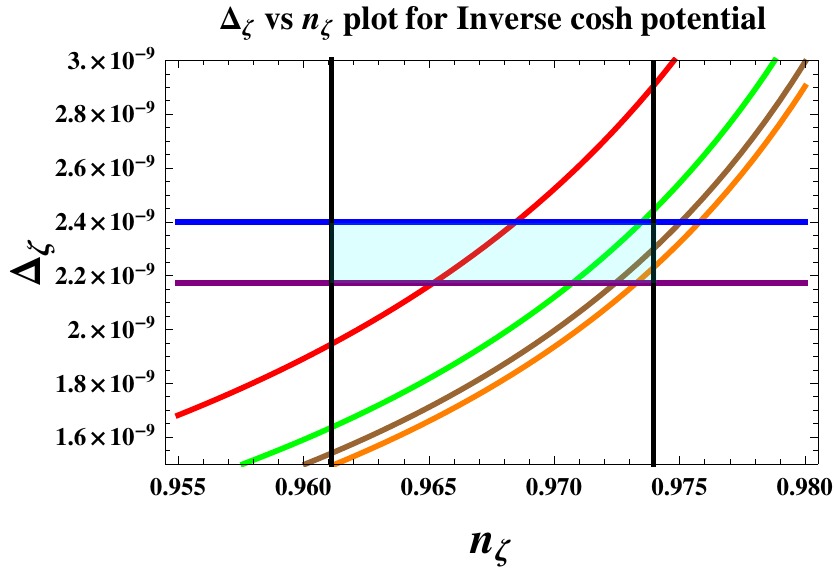}
    \label{fig3c}
}
\caption[Optional caption for list of figures]{Variation of the \ref{fig3a} scalar power spectrum $\Delta_{\zeta}$
vs scalar spectral index $n_{\zeta}$, \ref{fig3b} scalar power spectrum $\Delta_{\zeta}$ vs the stringy parameter $g$
and \ref{fig3c} scalar spectral tilt $n_{\zeta}$ vs the stringy parameter $g$. The purple and blue coloured line represent the upper and lower bound allowed by WMAP+Planck 2015 data respectively. 
The green dotted region bounded by two vertical black coloured lines represent the Planck $2\sigma$ allowed region and the rest of the light gray shaded region is disfavoured by the Planck+WMAP constraint.} 
\label{fig3}
\end{figure*}
\begin{figure*}[htb]
\centering
\subfigure[$\alpha_{\zeta}$ vs $n_{\zeta}$.]{
    \includegraphics[width=12.2cm,height=7.3cm] {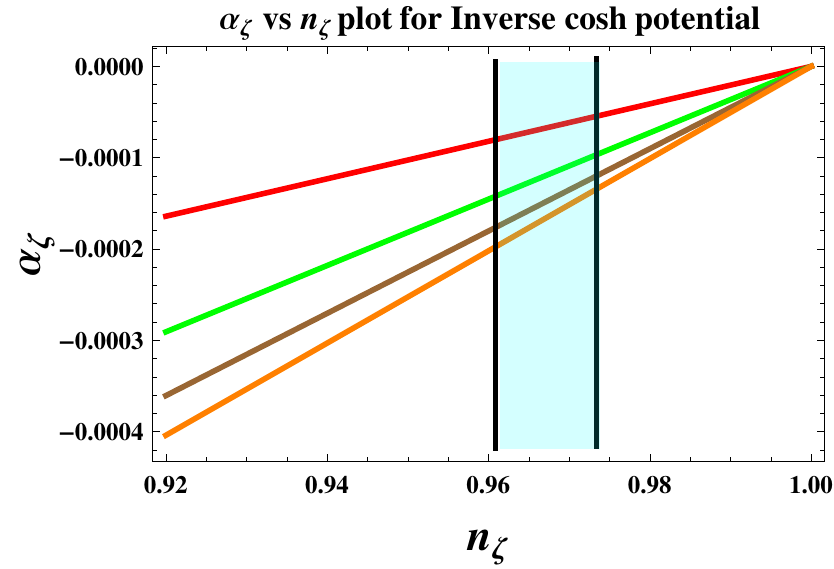}
    \label{fig4a}
}
\subfigure[$\kappa_{\zeta}$ vs $n_{\zeta}$.]{
    \includegraphics[width=12.2cm,height=7.3cm] {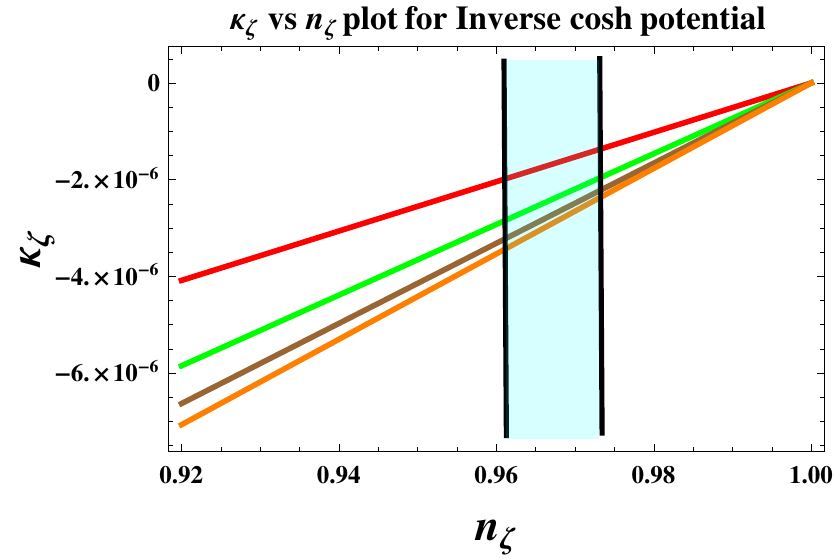}
    \label{fig4b}
}
\caption[Optional caption for list of figures]{Behaviour of the \ref{fig4a} running of the scalar spectral tilt $\alpha_{\zeta}$ and \ref{fig4b} running of the 
running of the scalar spectral tilt $\kappa_{\zeta}$ with respect to the scalar spectral index $n_{\zeta}$ for Inverse cosh potential with $g=88$.
For both the figures 
 \textcolor{red}{red},~\textcolor{green}{green},
~\textcolor{brown}{brown},~\textcolor{orange}{orange} colored curve represent $q=1/2$, $q=1$, $q=3/2$ and $q=2$ respectively.
The \textcolor{cyan}{cyan} color shaded region bounded by two vertical black coloured lines in \ref{fig4a} and \ref{fig4a} represent the Planck $2\sigma$ allowed region and the rest of the light gray shaded region is
showing the $1\sigma$ allowed range, which is at present 
disfavoured by the Planck data and Planck+ BICEP2+Keck Array joint constraint. From \ref{fig4a} and \ref{fig4b}, it 
is also observed that, within $50<N_{\star}<70$ the Inverse cosh potential is favoured for the characteristic index $1/2<q<2$, by Planck 2015 data and Planck+ BICEP2+Keck Array joint analysis.} 
\label{fig4}
\end{figure*}
For inverse cosh potential we get the following consistency relations:
\be\begin{array}{lll}\label{e-fold14}
 \displaystyle   r_{\star}\approx 4(1-n_{\zeta,\star})\times\left\{\begin{array}{lll}
                    \displaystyle  
                   \frac{1}{\cosh^2\left(\frac{N_{\star}}{g}\right)}\,,~~~~~~ &
 \mbox{\small {\bf for {$q=1/2$ }}}  \\ 
 \displaystyle   
             \frac{\sqrt{2q}}{\cosh^2\left(\frac{N_{\star}}{2qg}\right)}\,.~~~~~~ &
 \mbox{\small {\bf for {~any~arbitrary~ $q$ }}} 
          \end{array}
\right.
\end{array}\ee
\be\begin{array}{lll}\label{e-fold15}
 \displaystyle   \Delta_{\zeta,\star}\approx\frac{\lambda}{12\pi^2M^{4}_{p}(1-n_{\zeta,\star})}\times\left\{\begin{array}{lll}
                    \displaystyle  
                   \sinh\left(\frac{2N_{\star}}{g}\right)\,,~~~~~~ &
 \mbox{\small {\bf for {$q=1/2$ }}}  \\ 
 \displaystyle   
             \sqrt{2q}\sinh\left(\frac{2N_{\star}}{\sqrt{2q}g}\right)\,.~~~~~~ &
 \mbox{\small {\bf for {~any~arbitrary~ $q$ }}} 
          \end{array}
\right.
\end{array}\ee
\be\begin{array}{lll}\label{e-fold16}
 \displaystyle   \Delta_{\zeta,\star}\approx\frac{2\lambda}{3\pi^2M^{4}_{p}r_{\star}}\times\left\{\begin{array}{lll}
                    \displaystyle  
                   \tanh\left(\frac{N_{\star}}{g}\right)\,,~~~~~~ &
 \mbox{\small {\bf for {$q=1/2$ }}}  \\ 
 \displaystyle   
             \tanh\left(\frac{N_{\star}}{\sqrt{2q}g}\right)\,.~~~~~~ &
 \mbox{\small {\bf for {~any~arbitrary~ $q$ }}} 
          \end{array}
\right.
\end{array}\ee
\be\begin{array}{lll}\label{e-fold17}
 \displaystyle   \alpha_{\zeta,\star}\approx\frac{2}{g}\times\left(n_{\zeta,\star}-1\right)\times\left\{\begin{array}{lll}
                    \displaystyle  
                  \frac{1}{ \sinh\left(\frac{2N_{\star}}{g}\right)}\,,~~~~~~ &
 \mbox{\small {\bf for {$q=1/2$ }}}  \\ 
 \displaystyle   
            \frac{1}{\sqrt{2q}\sinh\left(\frac{2N_{\star}}{\sqrt{2q}g}\right)}\,.~~~~~~ &
 \mbox{\small {\bf for {~any~arbitrary~ $q$ }}} 
          \end{array}
\right.
\end{array}\ee
\be\begin{array}{lll}\label{e-fold18}
 \displaystyle   \kappa_{\zeta,\star}\approx\frac{2}{g^2}\times\left(n_{\zeta,\star}-1\right)\times\left\{\begin{array}{lll}
                    \displaystyle  
                  \frac{1}{ \sinh^2\left(\frac{N_{\star}}{g}\right)}\,,~~~~~~ &
 \mbox{\small {\bf for {$q=1/2$ }}}  \\ 
 \displaystyle   
            \frac{1}{2q\sinh^2\left(\frac{N_{\star}}{\sqrt{2q}g}\right)}\,.~~~~~~ &
 \mbox{\small {\bf for {~any~arbitrary~ $q$ }}} 
          \end{array}
\right.
\end{array}\ee
Let us now discuss the general constraints on the parameters of tachyonic string theory including the factor $q$ 
and on the parameters appearing in the 
expression for Inverse cosh potential. In Fig.~(\ref{fig2a}) and Fig.~(\ref{fig2b}), we have shown the behavior of the tensor-to-scalar ratio $r$ with respect to the scalar spectral index $n_{\zeta}$
and the model parameter $g$ for Inverse cosh potential respectively. In both the figures the \textcolor{purple}{purple} and \textcolor{blue}{blue}
coloured line represent the upper bound of tensor-to-scalar ratio allowed by Planck+ BICEP2+Keck Array
joint constraint and only Planck 2015 data respectively. For both the figures 
 \textcolor{red}{red},~\textcolor{green}{green},
~\textcolor{brown}{brown},~\textcolor{orange}{orange} colored curve represent $q=1/2$, $q=1$, $q=3/2$ and $q=2$ respectively.
The \textcolor{cyan}{cyan} color shaded region bounded by two vertical black coloured lines in Fig.~(\ref{fig2a}) represent the
Planck $2\sigma$ allowed region and the rest of the light gray shaded region
is showing the $1\sigma$ allowed range, which is at present 
disfavoured by the Planck 2015 data and Planck+ BICEP2+Keck Array joint constraint. The rest of the region 
is completely ruled out by the present observational constraints. From Fig.~(\ref{fig2a}) and Fig.~(\ref{fig2b}), it 
is also observed that, within $50<N_{\star}<70$, the Inverse cosh potential is favoured only for the characteristic
index $1/2<q<2$, by Planck 2015 data and Planck+ BICEP2+Keck Array joint analysis. Also in Fig.~(\ref{fig2a}) for $q=1/2$, $q=1$, $q=3/2$ and $q=2$ we fix $N_{\star}/g\sim 0.8$. This implies that for $50<N_{\star}<70$,
the prescribed window for $g$ from $r-n_{\zeta}$ plot is given by, $63<g<88$.
In Fig.~(\ref{fig2b}), we have explicitly shown that the in $r-g$ plane the observationally
favoured lower bound for the characteristic index is $q\geq 1/2$. It is additionally important to note that, for $q>>2$, the 
tensor-to-scalar ratio computed from the model is negligibly small for Inverse cosh potential.
This implies that if the inflationary tensor mode is detected near to its present upper bound on tensor-to-scalar ratio then all $q>>2$ 
possibilities for tachyonic inflation can be discarded for Inverse cosh potential. On the contrary,
if inflationary tensor modes are never detected by any of the future observational probes then $q>>2$
possibilities for tachyonic inflation in case of Inverse cosh potential is highly 
prominent. Also it is important to mention that, in Fig.~(\ref{fig2b}) within the window $0<g<100$, 
if we smaller the value of $g$, then the inflationary tensor-to-scalar ratio also gradually decreases.
To analyze the results more clearly let us describe the cosmological features from Fig.~(\ref{fig2a}) in detail.
Let us first start with the $q=1/2$ situation, in which the $2\sigma$ constraint on the scalar spectral tilt is satisfied
within the window of tensor-to-scalar ratio, $0.015<r<0.025$ for $50<N_{\star}<70$. Next for $q=1$ case, the same constraint 
is satisfied within the window of tensor-to-scalar ratio, $ 0.050<r<0.075$ for $50<N_{\star}<70$. Further for $q=3/2$ case, the same constraint 
on scalar spectral tilt is satisfied within the window of tensor-to-scalar ratio, $ 0.090<r<0.12$ for $50<N_{\star}<70$.
Finally, for $q=2$ situation, the value for the tensor-to-scalar ratio is $r<0.12$, which is tightly constrained from the upper bound of 
spectral tilt from Planck 2015 observational data. 

In Fig.~(\ref{fig3a}), Fig.~(\ref{fig3b}) and Fig.~(\ref{fig3c}), we have depicted the behavior of the 
scalar power spectrum $\Delta_{\zeta}$ vs the stringy parameter $g$, 
scalar spectral tilt $n_{\zeta}$ vs the stringy parameter $g$ and scalar power spectrum $\Delta_{\zeta}$
vs scalar spectral index $n_{\zeta}$ for Inverse cosh potential respectively. It is important to note that,
for all of the figures 
 \textcolor{red}{red},~\textcolor{green}{green},
~\textcolor{brown}{brown},~\textcolor{orange}{orange} colored curve represent $q=1/2$, $q=1$, $q=3/2$ and $q=2$ respectively.
The \textcolor{purple}{purple} and \textcolor{blue}{blue} coloured line represent the upper and lower bound allowed by WMAP+Planck 2015 data respectively. 
The \textcolor{cyan}{cyan} color shaded region bounded by two vertical black coloured lines represent the Planck
$2\sigma$ allowed region and the rest of the light gray shaded region is the $1\sigma$ region, which is presently 
disfavoured by the joint Planck+WMAP constraints. The rest of the region 
is completely ruled out by the present observational constraints. From Fig.~(\ref{fig3a}) it is clearly 
observed that the observational constraints on the amplitude of the scalar mode fluctuations satisfy within the window $80<g<100$. For $g>100$ the 
corresponding amplitude falls down in a non-trivial fashion by following the exact functional form as stated in Eq~(\ref{e-fold9}). Next using the behavior as
shown in Fig.~(\ref{fig3b}), the lower bound on the stringy parameter $g$ is constrained as, $g>50$ by using the non-trivial
relationship as stated in Eq~(\ref{e-fold10}). But at this lower bound of the parameter $g$ 
the amplitude of the scalar power spectrum is larger compared to the present observational constraints. This implies that, to satisfy both of the constraint from 
the amplitude of the scalar power spectrum and its spectral tilt within $2\sigma$ CL the constrained numerical value of the stringy parameter is
lying within the window $80<g<100$. Also in Fig.~(\ref{fig3c}) for $q=1/2$, $q=1$, $q=3/2$ and $q=2$ we fix $N_{\star}/g\sim 0.8$, which further implies that for $50<N_{\star}<70$,
the prescribed window for $g$ from $\Delta_{\zeta}-n_{\zeta}$ plot is given by, $63<g<88$. If we additionally impose the constraint from the upper bound on tensor-to-scalar ratio then also the allowed parameter 
range is lying within the almost similar window i.e. $88<g<100$.

\begin{figure}[htb]
	\centering
	\includegraphics[width=12cm,height=8cm]{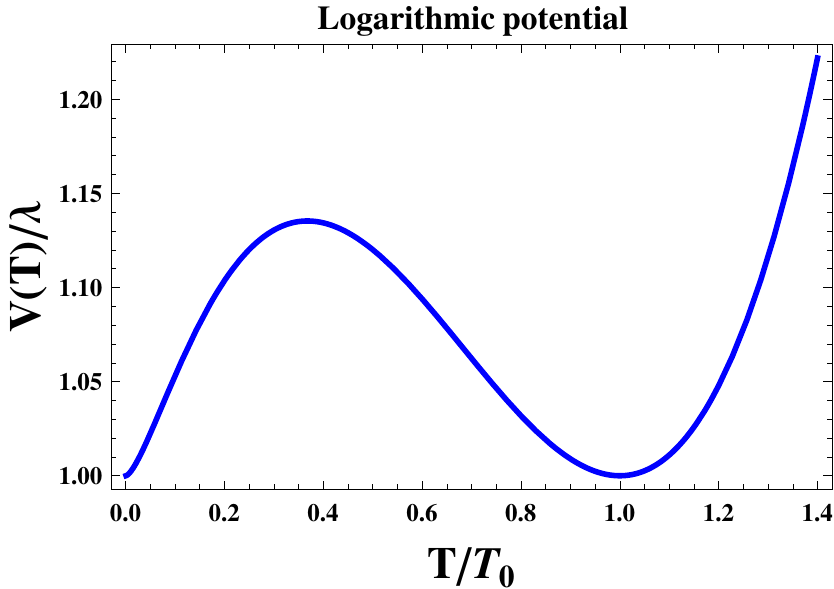}
	\caption{Variation of the Logarithmic potential $V(T)/\lambda$ with field $T/T_{0}$ in dimensionless units.
	}
	\label{fig5}
\end{figure}

In Fig.~(\ref{fig4a}) and Fig.~(\ref{fig4b}), we have shown the behaviour
of the running of the scalar spectral tilt $\alpha_{\zeta}$ and running of the 
running of the scalar spectral tilt $\kappa_{\zeta}$ with respect to the scalar spectral index $n_{\zeta}$ for Inverse cosh potential with $g=88$ respectively.
For both the figures 
 \textcolor{red}{red},~\textcolor{green}{green},
~\textcolor{brown}{brown},~\textcolor{orange}{orange} colored curve represent $q=1/2$, $q=1$, $q=3/2$ and $q=2$ respectively.
The \textcolor{cyan}{cyan} color shaded region bounded by two vertical black coloured lines in both the plots represent the Planck $2\sigma$ allowed region and the rest of the light gray shaded region is
showing the $1\sigma$ allowed range, which is at present 
disfavoured by the Planck data and Planck+ BICEP2+Keck Array joint constraint. From both of these figures, it 
is also observed that, within $50<N_{\star}<70$ the Inverse cosh potential is favoured for the characteristic index $1/2<q<2$, by Planck
2015 data and Planck+ BICEP2+Keck Array joint analysis. From Fig.~(\ref{fig4a}) and Fig.~(\ref{fig4b}), it is observed that within the $2\sigma$ 
observed range of the scalar spectral tilt $n_{\zeta}$, as the value of the characteristic parameter $q$ increases,
the value of the running $\alpha_{\zeta}$
and running of the running $\kappa_{\zeta}$ decreases for Inverse cosh potential. It is also important to note that for $1/2<q<2$, the numerical value 
of the running $\alpha_{\zeta}\sim {\cal O}(-10^{-4})$
and running of the running $\kappa_{\zeta}\sim {\cal O}(-10^{-6})$, which are perfectly 
consistent with the $1.5\sigma$ constraints on running and running of the running as obtained from Planck 2015 data.\\ \\
\underline{\bf Model II: Logarithmic potential}\\
For single field case the second model of tachyonic potential is given by:
\be\label{webbbb2b}
V(T)=\lambda\left\{ \left(\frac{T}{T_{0}}\right)^2\left[\ln\left(\frac{T}{T_{0}}\right)\right]^2+1\right\},\ee
where $\lambda$ characterize the scale of inflation and $T_{0}$ is the parameter of the model.
In Fig.~(\ref{fig5}) we have depicted the behavior of the Logarithmic potential with respect to scaled field coordinate $T/T_{0}$ in dimensionless units.
In this case the tachyon field started rolling down from the top hight of the potential either from the left or right hand side and take part in inflationary dynamics.  


Next using specified form of the potential the potential dependent slow-roll parameters are computed as:
\bea \bar{\epsilon}_{V}&=&\frac{1}{2g}\frac{4\left(\frac{T}{T_{0}}\right)^2\ln^2\left(\frac{T}{T_{0}}\right)\left[1+\ln\left(\frac{T}{T_{0}}\right)\right]^2}{\left[1
+\left(\frac{T}{T_{0}}\right)^2\ln^2\left(\frac{T}{T_{0}}\right)\right]^3},\\
\bar{\eta}_{V}&=&\frac{1}{g}\frac{2\left\{\ln\left(\frac{T}{T_{0}}\right)+\left[1+\ln\left(\frac{T}{T_{0}}\right)\right]^2\right\}}{\left[1
+\left(\frac{T}{T_{0}}\right)^2\ln^2\left(\frac{T}{T_{0}}\right)\right]^2},\\
\bar{\xi}^2_{V}&=&\frac{1}{g^2}\frac{4\ln\left(\frac{T}{T_{0}}\right)\left[1+\ln\left(\frac{T}{T_{0}}\right)\right]\left[3+2\ln\left(\frac{T}{T_{0}}\right)\right]}{\left[1
+\left(\frac{T}{T_{0}}\right)^2\ln^2\left(\frac{T}{T_{0}}\right)\right]^4},~~~~~\\
\bar{\sigma}^3_{V}&=&-\frac{1}{g^3}\frac{4\ln^2\left(\frac{T}{T_{0}}\right)\left[1+\ln\left(\frac{T}{T_{0}}\right)\right]^2\left[1+2\ln\left(\frac{T}{T_{0}}\right)\right]}{\left[1
+\left(\frac{T}{T_{0}}\right)^2\ln^2\left(\frac{T}{T_{0}}\right)\right]^6},~~~~~~~~~~~~~~
\eea
where the factor $g$ is defined as:
\be g= \frac{\alpha^{'}\lambda T^2_0}{M^2_p}=\frac{M^4_s}{(2\pi)^3 g_s}\frac{\alpha^{'} T^2_0}{M^2_p}.\ee
Next we compute the number of e-foldings from this model:
\be\begin{array}{lll}\label{e-fold5b}
 \displaystyle  N(T)=\left\{\begin{array}{lll}
                    \displaystyle  
                   \frac{g}{2}~\left\{-\frac{2 \text{Ei}\left[2 \left(\ln \left(\frac{T}{T_{0}}\right)+1\right)\right]}{e^2}-\frac{\text{Ei}\left[4 \left(\ln \left(\frac{T}{T_{0}}\right)+1\right)\right]}{e^4}\right.\\ \left.
                   \displaystyle +\left(\frac{T}{T_{0}}\right)^4\left[\frac{11 }{32}-\frac{3}{8}  \ln \left(\frac{T}{T_{0}}\right)+\frac{1}{4}  \ln ^2\left(\frac{T}{T_{0}}\right)
                   \right]\right.\\ \left.
                   \displaystyle +\left(\frac{T}{T_{0}}\right)^2+\ln \left(\frac{\ln \left(\frac{T}{T_{0}}\right)}{1+\ln \left(\frac{T}{T_{0}}\right)}\right)\right\}^{T}_{T_{end}}\,,~~~~~~ &
 \mbox{\small {\bf for {$q=1/2$ }}}  \\ 
 \displaystyle   
             \sqrt{2q}\frac{g}{2}~\left\{-\frac{2 \text{Ei}\left[2 \left(\ln \left(\frac{T}{T_{0}}\right)+1\right)\right]}{e^2}-\frac{\text{Ei}\left[4 \left(\ln \left(\frac{T}{T_{0}}\right)+1\right)\right]}{e^4}\right.\\ \left.
                   \displaystyle +\left(\frac{T}{T_{0}}\right)^4\left[\frac{11 }{32}-\frac{3}{8}  \ln \left(\frac{T}{T_{0}}\right)+\frac{1}{4}  \ln ^2\left(\frac{T}{T_{0}}\right)
                   \right]\right.\\ \left.
                   \displaystyle +\left(\frac{T}{T_{0}}\right)^2+\ln \left(\frac{\ln \left(\frac{T}{T_{0}}\right)}{1+\ln \left(\frac{T}{T_{0}}\right)}\right)\right\}^{T}_{T_{end}}\,.~~~~~~ &
 \mbox{\small {\bf for {~any~arbitrary~ $q$ }}} 
          \end{array}
\right.
\end{array}\ee
Further using the condition to end inflation:
\bea \bar{\epsilon}_{V}(T_{end})=1,\\
      |\bar{\eta}_{V}(T_{end})|=1,\eea
      we get the following transcendental equation:
      \be \left[1+\ln \left(\frac{T_{end}}{T_{0}}\right)\right]^2\left\{\left(\frac{T_{end}}{T_{0}}\right)^2\ln^2 \left(\frac{T_{end}}{T_{0}}\right)-1\right\}=\ln \left(\frac{T_{end}}{T_{0}}\right),\ee
      from which we get following sets of real solutions for the field value:
      \bea T_{end}&=& \left(0.07~T_{0},~ 0.69~T_{0},~1.83~T_{0}\right).\eea
      Then using this result we need to numerically solve the transcendental equation of $T_{\star}$ which involves $N_{\star}$ explicitly. However, in the slow-roll regime of inflation 
we get the following simplified expression for the field value $T_{\star}$ in terms of $N_{\star}$, $T_{end}$ and $T_{0}$ as:
\be\begin{array}{lll}\label{e-fold6b}
 \displaystyle  T_{\star}\approx T_{0}\times\left\{\begin{array}{lll}
                    \displaystyle  
                   \sqrt{\frac{2N_{\star}}{g}+\left(\frac{T_{end}}{T_{0}}\right)^2}\,,~~~~~~ &
 \mbox{\small {\bf for {$q=1/2$ }}}  \\ 
 \displaystyle   
              \sqrt{\frac{2N_{\star}}{\sqrt{2q}g}+\left(\frac{T_{end}}{T_{0}}\right)^2}\,.~~~~~~ &
 \mbox{\small {\bf for {~any~arbitrary~ $q$ }}} 
          \end{array}
\right.
\end{array}\ee
where we have explicitly used that fact that in the slow-roll regime the quadratic term gives the dominant contribution in $N_{\star}$.

Also the field excursion can be computed as:
\be\begin{array}{lll}\label{e-fold6bex}
 \displaystyle  |\Delta T|=T_{0}\times\left\{\begin{array}{lll}
                    \displaystyle  
                   \left|\sqrt{\frac{2N_{\star}}{g}+c^2}-c\right|\,,~~~~~~ &
 \mbox{\small {\bf for {$q=1/2$ }}}  \\ 
 \displaystyle   
              \left|\sqrt{\frac{2N_{\star}}{\sqrt{2q}g}+c^2}-c\right|\,.~~~~~~ &
 \mbox{\small {\bf for {~any~arbitrary~ $q$ }}} 
          \end{array}
\right.
\end{array}\ee
where $c=0.07,0.69,1.83$. During numerical estimation we have taken $c=0.07$ as it is compatible with the observational constraints from Planck 2015 data.

Finally using the previously mentioned definition of potential dependent slow-roll parameter 
 we compute the  following inflationary observables:
\be\begin{array}{lll}\label{e-fold9b}\tiny
 \displaystyle   \Delta_{\zeta,\star}\approx\frac{g\lambda}{48\pi^2 M^{4}_{p}}\times\left\{\begin{array}{lll}
                    \displaystyle  
                  \frac{\left[1
+\frac{1}{4}\left(\frac{2N_{\star}}{g}+\left(\frac{T_{end}}{T_{0}}\right)^2\right)\ln^2\left(\frac{2N_{\star}}{g}+\left(\frac{T_{end}}{T_{0}}\right)^2\right)\right]^4}{\left(\frac{2N_{\star}}{g}
+\left(\frac{T_{end}}{T_{0}}\right)^2\right)\ln^2\left(\frac{2N_{\star}}{g}+\left(\frac{T_{end}}{T_{0}}\right)^2\right)\left[1+\frac{1}{2}\ln\left(\frac{2N_{\star}}{g}+\left(\frac{T_{end}}{T_{0}}\right)^2\right)\right]^2}\,, &
 \mbox{\small {\bf for {$q=1/2$ }}}  \\ 
 \displaystyle   
             \frac{2q\left[1
+\frac{1}{4}\left(\frac{2N_{\star}}{\sqrt{2q}g}+\left(\frac{T_{end}}{T_{0}}\right)^2\right)\ln^2\left(\frac{2N_{\star}}{\sqrt{2q}g}+\left(\frac{T_{end}}{T_{0}}\right)^2\right)\right]^4}{\left(\frac{2N_{\star}}{\sqrt{2q}g}
+\left(\frac{T_{end}}{T_{0}}\right)^2\right)\ln^2\left(\frac{2N_{\star}}{\sqrt{2q}g}+\left(\frac{T_{end}}{T_{0}}\right)^2\right)\left[1+\frac{1}{2}\ln\left(\frac{2N_{\star}}{\sqrt{2q}g}+\left(\frac{T_{end}}{T_{0}}\right)^2\right)\right]^2}\,\,. &
 \mbox{\small {\bf for {~any~ $q$ }}} 
          \end{array}
\right.
\end{array}\ee
\be\begin{array}{lll}\label{e-fold10b}\tiny
 \displaystyle   n_{\zeta,\star}-1\approx\frac{4}{g}\times\left\{\begin{array}{lll}
                    \displaystyle  
                  \frac{ \vartheta_{\star}}{\left[1
+\frac{1}{4}\left(\frac{2N_{\star}}{g}+\left(\frac{T_{end}}{T_{0}}\right)^2\right)\ln^2\left(\frac{2N_{\star}}{g}+\left(\frac{T_{end}}{T_{0}}\right)^2\right)\right]^3}\,, &
 \mbox{\small {\bf for {$q=1/2$ }}}  \\ 
 \displaystyle   
            \displaystyle  
                  \frac{ \vartheta_{\star}}{\left[1
+\frac{1}{4}\left(\frac{2N_{\star}}{\sqrt{2q}g}+\left(\frac{T_{end}}{T_{0}}\right)^2\right)\ln^2\left(\frac{2N_{\star}}{\sqrt{2q}g}+\left(\frac{T_{end}}{T_{0}}\right)^2\right)\right]^3}\,. &
 \mbox{\small {\bf for {~any~ $q$ }}} 
          \end{array}
\right.
\end{array}\ee
\begin{figure*}[htb]
\centering
\subfigure[$r$ vs $n_{\zeta}$.]{
    \includegraphics[width=10.2cm,height=6.5cm] {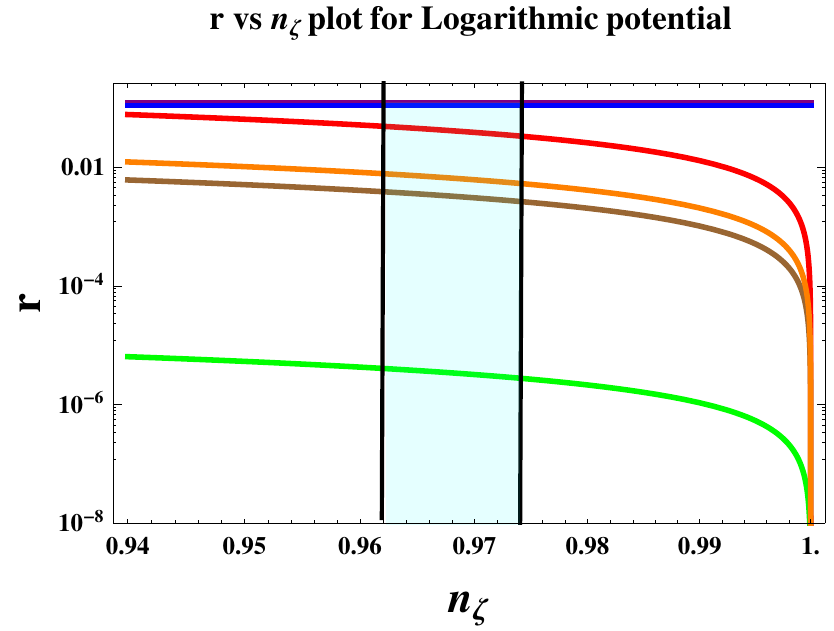}
    \label{fig6a}
}
\subfigure[$r$ vs $g$.]{
    \includegraphics[width=10.2cm,height=6.5cm] {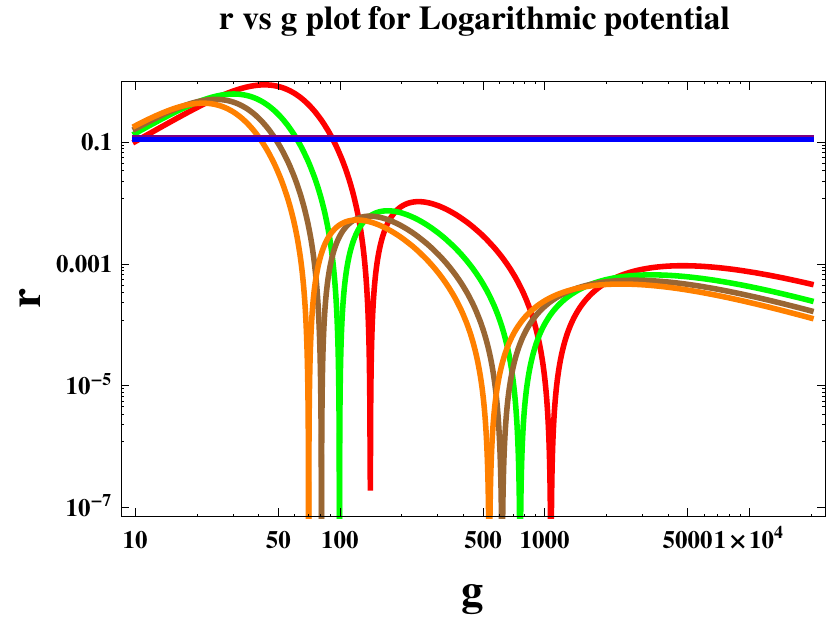}
    \label{fig6b}
}
\caption[Optional caption for list of figures]{Behaviour of the tensor-to-scalar ratio $r$ with respect to \ref{fig6a} the scalar spectral index $n_{\zeta}$
and \ref{fig6b} the parameter $g$ for Logarithmic potential.
The purple and blue coloured line represent the upper bound of tensor-to-scalar ratio allowed by Planck+ BICEP2+Keck Array joint constraint and only Planck 2015 data respectively. For both the figures 
 \textcolor{red}{red},~\textcolor{green}{green},
~\textcolor{brown}{brown},~\textcolor{orange}{orange} colored curve represent $q=1/2$, $q=1$, $q=3/2$ and $q=2$ respectively.
The \textcolor{cyan}{cyan} color shaded region bounded by two vertical black coloured lines in \ref{fig6a} represent the Planck $2\sigma$ allowed region and the rest of the light gray shaded region is
showing the $1\sigma$ allowed range, which is at present 
disfavoured by the Planck data and Planck+ BICEP2+Keck Array joint constraint. From \ref{fig6a} and \ref{fig6b}, it 
is also observed that, within $50<N_{\star}<70$ the Logarithmic potential is favoured for the characteristic index $1/2<q<2$, by Planck 2015 data and Planck+ BICEP2+Keck Array joint analysis.
In \ref{fig6a}, we have explicitly shown that in $r-n_{\zeta}$ plane the observationally favoured lower bound for the characteristic index is $q\geq 1/2$. } 
\label{fig6}
\end{figure*}
\begin{figure*}[htb]
\centering
\subfigure[$\Delta_{\zeta}$ vs $g$.]{
    \includegraphics[width=7.2cm,height=8cm] {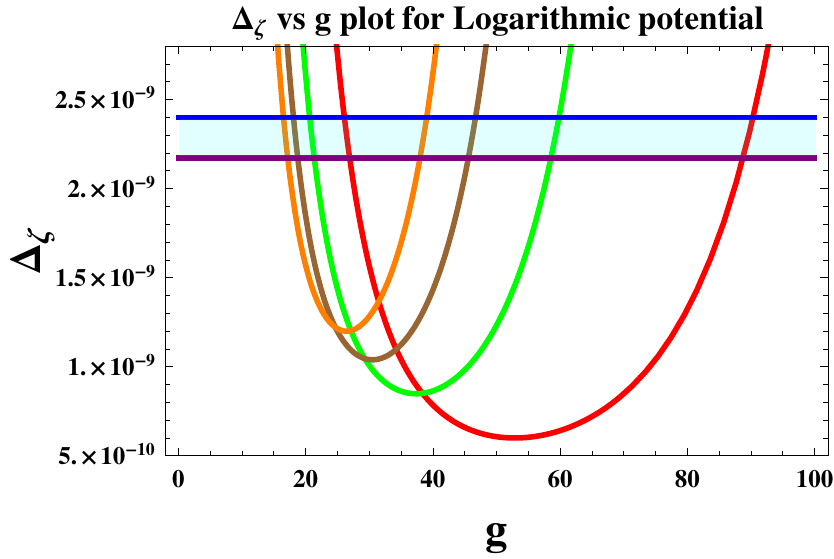}
    \label{fig7a}
}
\subfigure[$n_{\zeta}$ vs $g$.]{
    \includegraphics[width=7.2cm,height=8cm] {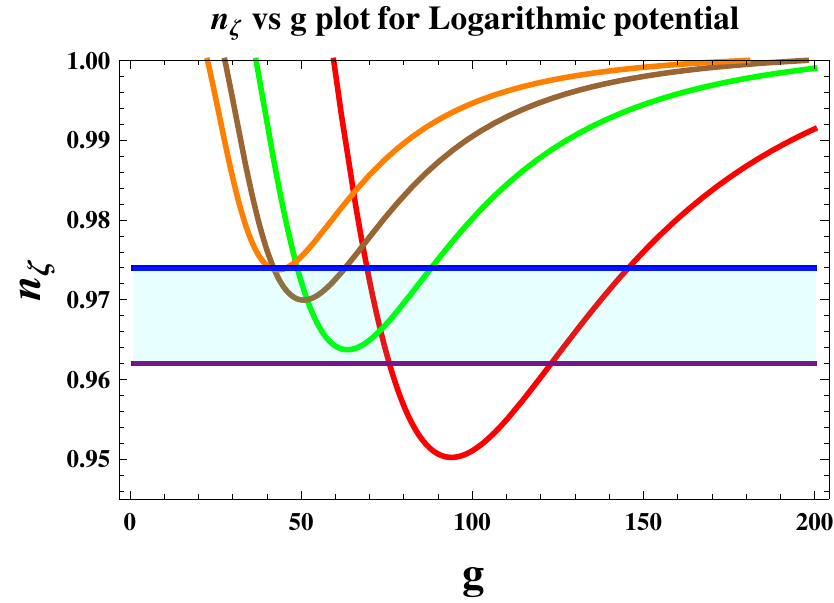}
    \label{fig7b}
}
\subfigure[$\Delta_{\zeta}$ vs $n_{\zeta}$.]{
    \includegraphics[width=11.2cm,height=8cm] {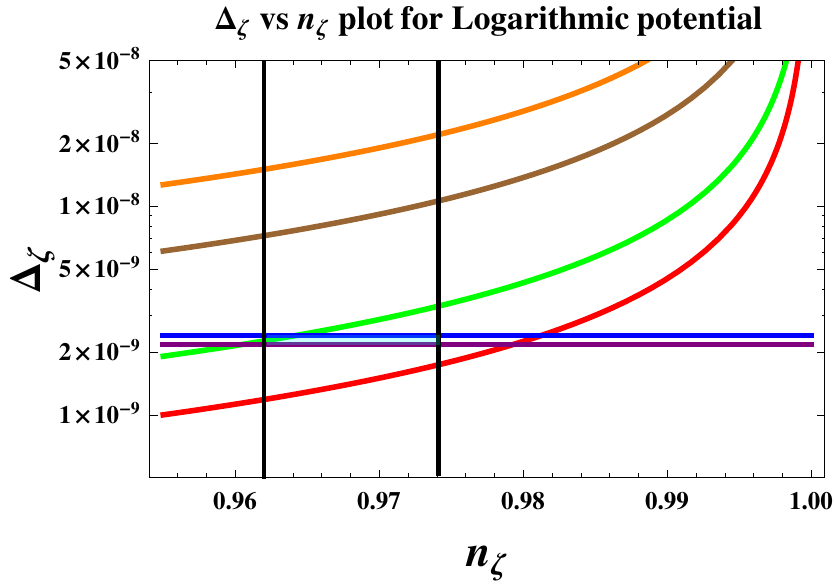}
    \label{fig7c}
}
\caption[Optional caption for list of figures]{Variation of the \ref{fig7a} scalar power spectrum $\Delta_{\zeta}$
vs scalar spectral index $n_{\zeta}$, \ref{fig7b} scalar power spectrum $\Delta_{\zeta}$ vs the stringy parameter $g$
and \ref{fig7c} scalar spectral tilt $n_{\zeta}$ vs the stringy parameter $g$. The purple and blue coloured line represent the upper and lower bound allowed by WMAP+Planck 2015 data respectively. 
The green dotted region bounded by two vertical black coloured lines represent the Planck $2\sigma$ allowed region and the rest of the light gray shaded region is disfavoured by the Planck+WMAP constraint.} 
\label{fig7}
\end{figure*}
\be\begin{array}{lll}\label{e-fold13b}\tiny
 \displaystyle   r_{\star}\approx\frac{32}{g}\times\left\{\begin{array}{lll}
                    \displaystyle  
                  \frac{\left(\frac{2N_{\star}}{g}+\left(\frac{T_{end}}{T_{0}}\right)^2\right)\ln^2\left(\frac{2N_{\star}}{g}+\left(\frac{T_{end}}{T_{0}}\right)^2\right)
                  \left[1+\frac{1}{2}\ln\left(\frac{2N_{\star}}{g}+\left(\frac{T_{end}}{T_{0}}\right)^2\right)\right]^2}{\left[1
+\frac{1}{4}\left(\frac{2N_{\star}}{g}+\left(\frac{T_{end}}{T_{0}}\right)^2\right)\ln^2\left(\frac{2N_{\star}}{g}+\left(\frac{T_{end}}{T_{0}}\right)^2\right)\right]^3}\,,~~~~~~ &
 \mbox{\small {\bf for {$q=1/2$ }}}  \\ 
 \displaystyle   
            \frac{1}{2q}\frac{\left(\frac{2N_{\star}}{\sqrt{2q}g}+\left(\frac{T_{end}}{T_{0}}\right)^2\right)\ln^2\left(\frac{2N_{\star}}{\sqrt{2q}g}+\left(\frac{T_{end}}{T_{0}}\right)^2\right)\left[1+\frac{1}{2}
            \ln\left(\frac{2N_{\star}}{\sqrt{2q}g}+\left(\frac{T_{end}}{T_{0}}\right)^2\right)\right]^2}{\left[1
+\frac{1}{4}\left(\frac{2N_{\star}}{\sqrt{2q}g}+\left(\frac{T_{end}}{T_{0}}\right)^2\right)\ln^2\left(\frac{2N_{\star}}{\sqrt{2q}g}+\left(\frac{T_{end}}{T_{0}}\right)^2\right)\right]^3}\,.~~~~~~ &
 \mbox{\small {\bf for {~any~arbitrary~ $q$ }}} 
          \end{array}
\right.
\end{array}\ee
where 
\be\begin{array}{lll}\label{e-fold13bss}\tiny
 \displaystyle   \vartheta_{\star}\approx\left\{\begin{array}{lll}
                    \displaystyle  
                  \left[1-\frac{3}{4}\left(\frac{2N_{\star}}{g}+\left(\frac{T_{end}}{T_{0}}\right)^2
\right)\ln^2\left(\frac{2N_{\star}}{g}+\left(\frac{T_{end}}{T_{0}}\right)^2\right)\right]\left[1+\frac{1}{2}\ln\left(\frac{2N_{\star}}{g}
+\left(\frac{T_{end}}{T_{0}}\right)^2\right)\right]^2\\  \displaystyle
+\frac{1}{2}\ln\left(\frac{2N_{\star}}{g}+\left(\frac{T_{end}}{T_{0}}\right)^2\right)\left[1
+\frac{1}{4}\left(\frac{2N_{\star}}{g}+\left(\frac{T_{end}}{T_{0}}\right)^2\right)\ln^2\left(\frac{2N_{\star}}{g}+\left(\frac{T_{end}}{T_{0}}\right)^2\right)\right]\,,~~~~~~ &
 \mbox{\small {\bf for {$q=1/2$ }}}  \\ 
 \displaystyle   
            \left[1-\frac{G}{4}\left(\frac{N_{\star}}{qg}
\right)\ln^2\left(\frac{2N_{\star}}{\sqrt{2q}g}+\left(\frac{T_{end}}{T_{0}}\right)^2\right)\right]\left[1+\frac{1}{2}\ln\left(\frac{2N_{\star}}{\sqrt{2q}g}+\left(\frac{T_{end}}{T_{0}}\right)^2\right)\right]^2\\  \displaystyle
+\frac{P}{2}\ln\left(\frac{2N_{\star}}{\sqrt{2q}g}+\left(\frac{T_{end}}{T_{0}}\right)^2\right)\left[1
+\frac{1}{4}\left(\frac{2N_{\star}}{\sqrt{2q}g}+\left(\frac{T_{end}}{T_{0}}\right)^2\right)\ln^2\left(\frac{2N_{\star}}{\sqrt{2q}g}+\left(\frac{T_{end}}{T_{0}}\right)^2\right)\right]\,.~~~~~~ &
 \mbox{\small {\bf for {~any~arbitrary~ $q$ }}} 
          \end{array}
\right.
\end{array}\ee
Here the constants $G$ and $P$ are defined as:
\bea G&=&\frac{1}{2q}\left(1+2\sqrt{2q}\right),\\
P&=&\frac{1}{\sqrt{2q}}.\eea
For inverse cosh potential we get the following consistency relations:
\be\begin{array}{lll}\label{e-fold14b}\tiny
 \displaystyle   r_{\star}\approx 8(1-n_{\zeta,\star})\times\left\{\begin{array}{lll}
                    \displaystyle  
                  \frac{\left(\frac{2N_{\star}}{g}+\left(\frac{T_{end}}{T_{0}}\right)^2\right)\ln^2\left(\frac{2N_{\star}}{g}+\left(\frac{T_{end}}{T_{0}}\right)^2\right)\left[1+\frac{1}{2}\ln\left(\frac{2N_{\star}}{g}
                  +\left(\frac{T_{end}}{T_{0}}\right)^2\right)\right]^2}{ |\vartheta_{\star}|}\,,~~~~~~ &
 \mbox{\small {\bf for {$q=1/2$ }}}  \\ 
 \displaystyle   
            \frac{\left(\frac{2N_{\star}}{\sqrt{2q}g}+\left(\frac{T_{end}}{T_{0}}\right)^2\right)\ln^2\left(\frac{2N_{\star}}{\sqrt{2q}g}+\left(\frac{T_{end}}{T_{0}}\right)^2\right)
            \left[1+\frac{1}{2}\ln\left(\frac{2N_{\star}}{\sqrt{2q}g}+\left(\frac{T_{end}}{T_{0}}\right)^2\right)\right]^2}{2q |\vartheta_{\star}|}\,.~~~~~~ &
 \mbox{\small {\bf for {~any~ $q$ }}} 
          \end{array}
\right.
\end{array}\ee
\be\begin{array}{lll}\label{e-fold15b}\tiny
\displaystyle   \Delta_{\zeta,\star}\approx\frac{\lambda}{12\pi^2 M^{4}_{p}}\times\left\{\begin{array}{lll}
                    \displaystyle  
                  \frac{ |\vartheta_{\star}|}{(n_{\zeta,\star}-1)\left(\frac{2N_{\star}}{g}+\left(\frac{T_{end}}{T_{0}}\right)^2
\right)\ln^2\left(\frac{2N_{\star}}{g}+\left(\frac{T_{end}}{T_{0}}\right)^2\right)\left[1+\frac{1}{2}\ln\left(\frac{2N_{\star}}{g}+\left(\frac{T_{end}}{T_{0}}\right)^2\right)\right]^2}\,, &
 \mbox{\small {\bf for {$q=1/2$ }}}  \\ 
 \displaystyle   
             \frac{2q |\vartheta_{\star}|}{(n_{\zeta,\star}-1)\left(\frac{2N_{\star}}{\sqrt{2q}g}+\left(\frac{T_{end}}{T_{0}}\right)^2
\right)\ln^2\left(\frac{2N_{\star}}{\sqrt{2q}g}+\left(\frac{T_{end}}{T_{0}}\right)^2\right)\left[1+\frac{1}{2}\ln\left(\frac{2N_{\star}}{\sqrt{2q}g}+\left(\frac{T_{end}}{T_{0}}\right)^2\right)\right]^2}\,. &
 \mbox{\small {\bf for {~any~ $q$ }}} 
          \end{array}
\right.
\end{array}\ee
\be\begin{array}{lll}\label{e-fold16b}\tiny
 \displaystyle   \Delta_{\zeta,\star}\approx\frac{2\lambda}{3\pi^2M^{4}_{p}r_{\star}}\times\left\{\begin{array}{lll}
                    \displaystyle  
                   \left[1
+\frac{1}{4}\left(\frac{2N_{\star}}{g}+\left(\frac{T_{end}}{T_{0}}\right)^2\right)\ln^2\left(\frac{2N_{\star}}{g}+\left(\frac{T_{end}}{T_{0}}\right)^2\right)\right]\,,~~~~~~ &
 \mbox{\small {\bf for {$q=1/2$ }}}  \\ 
 \displaystyle   
             \left[1
+\frac{1}{4}\left(\frac{2N_{\star}}{\sqrt{2q}g}+\left(\frac{T_{end}}{T_{0}}\right)^2\right)\ln^2\left(\frac{2N_{\star}}{\sqrt{2q}g}+\left(\frac{T_{end}}{T_{0}}\right)^2\right)\right]\,.~~~~~~ &
 \mbox{\small {\bf for {~any~arbitrary~ $q$ }}} 
          \end{array}
\right.
\end{array}\ee
Let us now discuss the general constraints on the parameters of tachyonic string theory including the factor $q$ 
and on the paramters appearing in the 
expression for Logarithmic potential. In Fig.~(\ref{fig6a}) and Fig.~(\ref{fig6b}), we have shown the behavior of the tensor-to-scalar ratio $r$ with respect to the scalar spectral index $n_{\zeta}$
and the model parameter $g$ for Logarithmic potential respectively. In both the figures the \textcolor{purple}{purple} and \textcolor{blue}{blue}
coloured line represent the upper bound of tensor-to-scalar ratio allowed by Planck+ BICEP2+Keck Array
joint constraint and only Planck 2015 data respectively. For both the figures 
 \textcolor{red}{red},~\textcolor{green}{green},
~\textcolor{brown}{brown},~\textcolor{orange}{orange} colored curve represent $q=1/2$, $q=1$, $q=3/2$ and $q=2$ respectively.
The \textcolor{cyan}{cyan} color shaded region bounded by two vertical black coloured lines in Fig.~(\ref{fig6a}) represent the
Planck $2\sigma$ allowed region and the rest of the light gray shaded region
is showing the $1\sigma$ allowed range, which is at present 
disfavoured by the Planck 2015 data and Planck+ BICEP2+Keck Array joint constraint. The rest of the region 
is completely ruled out by the present observational constraints. From Fig.~(\ref{fig6a}) and Fig.~(\ref{fig6b}), it 
is also observed that, within $50<N_{\star}<70$, the Logarithmic potential is favoured for the characteristic
index $1/2<q<2$, by Planck 2015 data and Planck+ BICEP2+Keck Array joint analysis. Also in Fig.~(\ref{fig6a}) for $q=1/2$, $q=1$, $q=3/2$ and $q=2$ we fix $N_{\star}/g\sim 0.7$. This implies that for $50<N_{\star}<70$,
the prescribed window for $g$ from $r-n_{\zeta}$ plot is given by, $71.4<g<100$.
In Fig.~(\ref{fig6b}), we have explicitly shown that the in $r-g$ plane the observationally
favoured lower bound for the characteristic index is $q\geq 1/2$. It is additionally important to note that, for $q>>2$, the 
tensor-to-scalar ratio computed from the model is negligibly small for Logarithmic potential.
This implies that if the inflationary tensor mode is detected near to its present upper bound on tensor-to-scalar ratio then all $q>>2$ 
possibilities for tachyonic inflation can be discarded for Logarithmic potential. On the contrary,
if inflationary tensor modes are never detected by any of the future observational probes then $q>>2$
possibilities for tachyonic inflation in case of Logarithmic potential is highly 
prominent. Also it is important to mention that, in Fig.~(\ref{fig6b}) within the window $40<g<150$, 
if we increase the value of $g$, then the inflationary tensor-to-scalar ratio also gradually decreases. After that within $71<g<300$ the value of tensor-to-scalar ratio slightly increases and again it falls down to lower value within the 
the interval $120<g<1000$. 

\begin{figure}[htb]
	\centering
	\includegraphics[width=12cm,height=8cm]{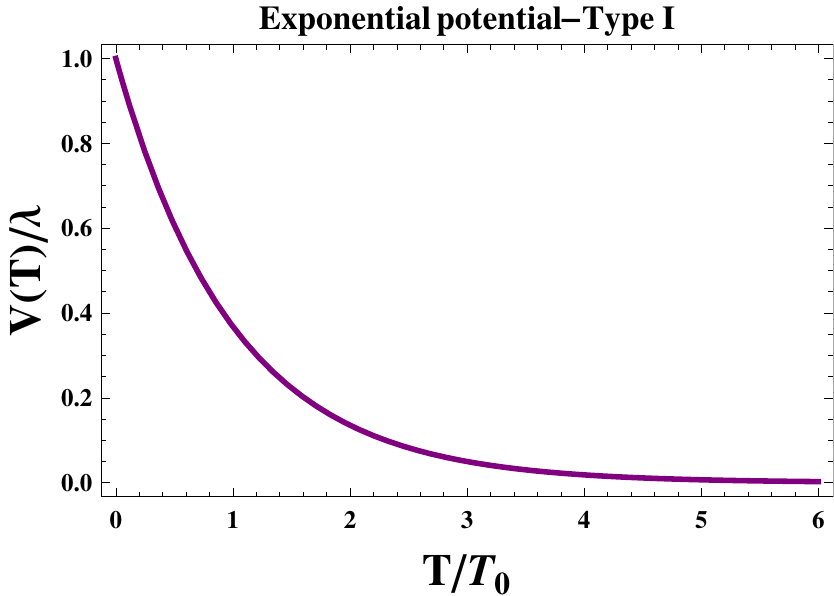}
	\caption{Variation of the Exponential potential-Type I $V(T)/\lambda$ with field $T/T_{0}$ in dimensionless units.
	}
	\label{fig8}
\end{figure}

In Fig.~(\ref{fig7a}), Fig.~(\ref{fig7b}) and Fig.~(\ref{fig7c}), we have depicted the behavior of the 
scalar power spectrum $\Delta_{\zeta}$ vs the stringy parameter $g$, 
scalar spectral tilt $n_{\zeta}$ vs the stringy parameter $g$ and scalar power spectrum $\Delta_{\zeta}$
vs scalar spectral index $n_{\zeta}$ for Logarithmic potential respectively. It is important to note that,
for all of the figures 
 \textcolor{red}{red},~\textcolor{green}{green},
~\textcolor{brown}{brown},~\textcolor{orange}{orange} colored curve represent $q=1/2$, $q=1$, $q=3/2$ and $q=2$ respectively.
The \textcolor{purple}{purple} and \textcolor{blue}{blue} coloured line represent the upper and lower bound allowed by WMAP+Planck 2015 data respectively. 
The \textcolor{cyan}{cyan} color shaded region bounded by two vertical black coloured lines represent the Planck
$2\sigma$ allowed region and the rest of the light gray shaded region is the $1\sigma$ region, which is presently 
disfavoured by the joint Planck+WMAP constraints. The rest of the region 
is completely ruled out by the present observational constraints. From Fig.~(\ref{fig7a}) it is clearly 
observed that the observational constraints on the amplitude of the scalar mode fluctuations satisfy within the window
$g=(26,90)$ for $q=1/2$, $g=(20,60)$ for $q=1$, $g=(18,48)$ for $q=3/2$ and $g=(16,38)$ for $q=2$. For all the cases outside the 
mentioned window for the parameter $g$ the 
corresponding amplitude gradually increases in a non-trivial fashion by following the exact functional form as stated in Eq~(\ref{e-fold9b}). Next using the behavior as
shown in Fig.~(\ref{fig7b}), the lower bound on the stringy parameter $g$ is constrained as, $g>45$ by using the non-trivial
relationship as stated in Eq~(\ref{e-fold10b}) for $q=2$. Similarly by observing the Fig.~(\ref{fig7b}) one can find out the other lower bounds on $g$ from different value of $q$. But at this lower bound of the parameter $g$ 
the amplitude of the scalar power spectrum is very larger compared to the present observational constraints. This implies that, to satisfy both of the constraint from 
the amplitude of the scalar power spectrum and its spectral tilt within $2\sigma$ CL the constrained numerical value of the stringy parameter is
lying within the window $48<g<90$ in which $q=2$ case is slightly disfavoured compared to the other lower values of $q$ studied in the present context. Also in Fig.~(\ref{fig7c}) for $q=1/2$, and $q=1$
we fix $N_{\star}/g\sim 0.7$, which further implies that for $50<N_{\star}<70$,
the prescribed window for $g$ from $\Delta_{\zeta}-n_{\zeta}$ plot is given by, $71.4<g<100$. If we additionally impose the constraint from the upper bound on tensor-to-scalar ratio then also the allowed parameter 
range is lying within the almost similar window i.e. $71.4<g<90$ and combining all the constraints the allowed value of the characteristic index is lying within $1/2<q<1$. However, $q=1/2$ is tightly constrained 
as depicted in  Fig.~(\ref{fig7c}).
\\ \\
\underline{\bf Model III: Exponential potential-Type I}\\
For single field case the third model of tachyonic potential is given by:
\be\label{weeeee1c}
V(T)=\lambda\exp\left(-\frac{T}{T_{0}}\right),\ee
where $\lambda$ characterize the scale of inflation and $T_{0}$ is the parameter of the model.
In Fig.~(\ref{fig8}) we have depicted the behavior of the Exponential potential-Type I with respect to scaled field coordinate $T/T_{0}$ in dimensionless units.
In this case the tachyon field started rolling down from the top hight of the potential from the left hand side and take part in inflationary dynamics.  


Next using specified form of the potential the potential dependent slow-roll parameters are computed as:
\bea \bar{\epsilon}_{V}&=&\frac{1}{2g}\exp\left(\frac{T}{T_{0}}\right),\\
\bar{\eta}_{V}&=&\frac{1}{g}\exp\left(\frac{T}{T_{0}}\right),\\
\bar{\xi}^2_{V}&=&\frac{1}{g^2}\exp\left(\frac{2T}{T_{0}}\right),~~~~~\\
\bar{\sigma}^3_{V}&=&\frac{1}{g^3}\exp\left(\frac{3T}{T_{0}}\right),~~~~~~~~~~~~~~
\eea
where the factor $g$ is defined as:
\be g= \frac{\alpha^{'}\lambda T^2_0}{M^2_p}=\frac{M^4_s}{(2\pi)^3 g_s}\frac{\alpha^{'} T^2_0}{M^2_p}.\ee
Next we compute the number of e-foldings from this model:
\be\begin{array}{lll}\label{e-fold5c}
 \displaystyle  N(T)=\left\{\begin{array}{lll}
                    \displaystyle  
                   g~\left[\exp\left(-\frac{T}{T_{0}}\right)-\exp\left(-\frac{T_{end}}{T_{0}}\right)\right]\,,~~~~~~ &
 \mbox{\small {\bf for {$q=1/2$ }}}  \\ 
 \displaystyle   
             \sqrt{2q}g~\left[\exp\left(-\frac{T}{T_{0}}\right)-\exp\left(-\frac{T_{end}}{T_{0}}\right)\right]\,.~~~~~~ &
 \mbox{\small {\bf for {~any~arbitrary~ $q$ }}} 
          \end{array}
\right.
\end{array}\ee
Further using the condition to end inflation:
\bea \bar{\epsilon}_{V}(T_{end})=1.\eea
      we get the following field value at the end of inflation:
      \be T_{end}=T_{0}~\ln(2g).\ee
Next using $N=N_{cmb}=N_{\star}$ and $T=T_{cmb}=T_{\star}$ at the horizon crossing we get,
\be\begin{array}{lll}\label{e-fold6c}
 \displaystyle  T_{\star}=T_{0}\times\left\{\begin{array}{lll}
                    \displaystyle  
                    \ln\left[\frac{g}{\frac{1}{2}+N_{\star}}\right]\,,~~~~~~ &
 \mbox{\small {\bf for {$q=1/2$ }}}  \\ 
 \displaystyle   
              \ln\left[\frac{g}{\frac{1}{2}+\frac{N_{\star}}{\sqrt{2q}}}\right]\,.~~~~~~ &
 \mbox{\small {\bf for {~any~arbitrary~ $q$ }}} 
          \end{array}
\right.
\end{array}\ee
Also the field excursion can be computed as:
\be\begin{array}{lll}\label{e-fold6cex}
 \displaystyle  |\Delta T|=T_{0}\times\left\{\begin{array}{lll}
                    \displaystyle  
                    \left|\ln\left[\frac{1}{1+2N_{\star}}\right]\right|\,,~~~~~~ &
 \mbox{\small {\bf for {$q=1/2$ }}}  \\ 
 \displaystyle   
              \left|\ln\left[\frac{1}{1+\frac{2N_{\star}}{\sqrt{2q}}}\right]\right|\,.~~~~~~ &
 \mbox{\small {\bf for {~any~arbitrary~ $q$ }}} 
          \end{array}
\right.
\end{array}\ee
\begin{figure*}[htb]
\centering
\subfigure[$r$ vs $n_{\zeta}$.]{
    \includegraphics[width=7.2cm,height=9.2cm] {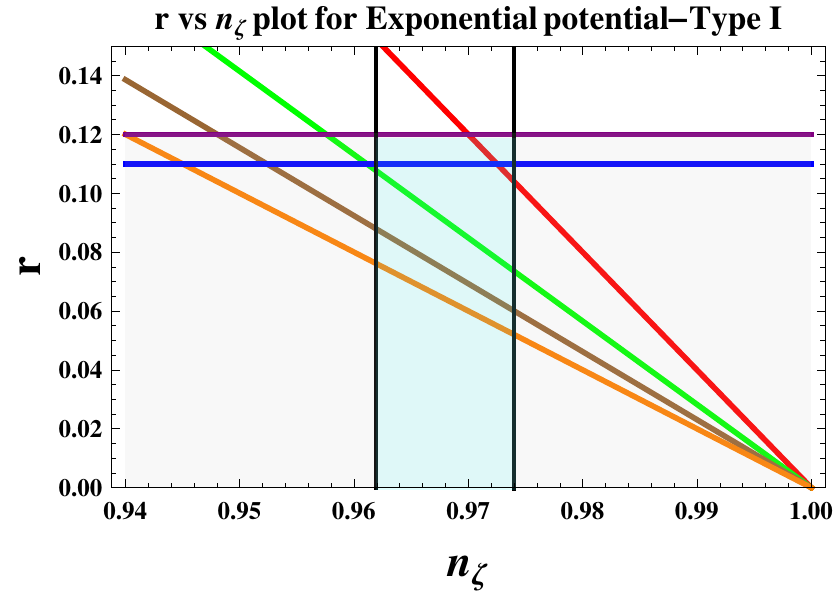}
    \label{fig9a}
}
\subfigure[$\Delta_{\zeta}$ vs $n_{\zeta}$.]{
    \includegraphics[width=7.2cm,height=9.2cm] {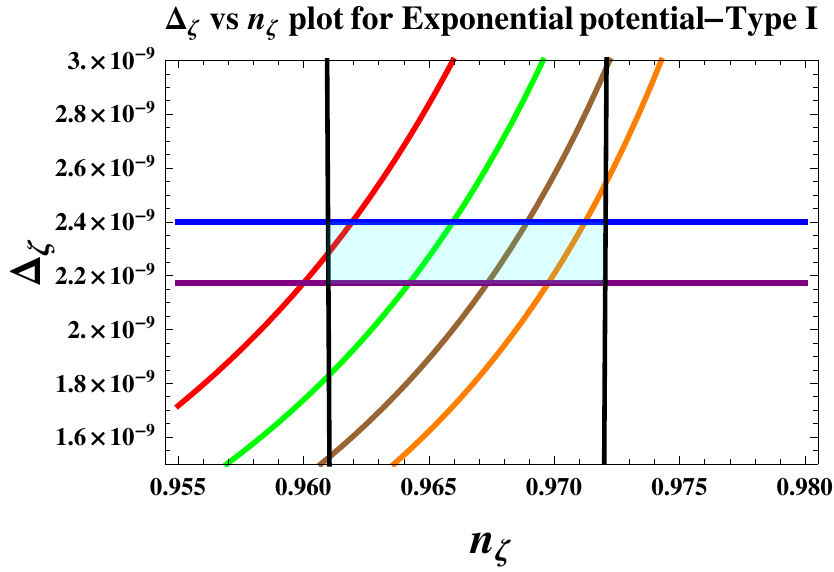}
    \label{fig9b}
}
\caption[Optional caption for list of figures]{In \ref{fig9a} the behaviour of the tensor-to-scalar ratio $r$ with respect to the scalar spectral index $n_{\zeta}$
and \ref{fig9b} the parameter $g$ for Exponential potential-Type I.
The purple and blue coloured line represent the upper bound of tensor-to-scalar ratio allowed by Planck+ BICEP2+Keck Array joint constraint and only Planck 2015 data respectively. For both the figures 
 \textcolor{red}{red},~\textcolor{green}{green},
~\textcolor{brown}{brown},~\textcolor{orange}{orange} colored curve represent $q=1/2$, $q=1$, $q=3/2$ and $q=2$ respectively.
The \textcolor{cyan}{cyan} color shaded region bounded by two vertical black coloured lines in \ref{fig9a} represent the Planck $2\sigma$ allowed region and the rest of the light gray shaded region is
showing the $1\sigma$ allowed range, which is at present 
disfavoured by the Planck data and Planck+ BICEP2+Keck Array joint constraint. 
In \ref{fig9a}, we have explicitly shown that in $r-n_{\zeta}$ plane the observationally favoured lower bound for the characteristic index is $q\geq 1/2$.  Variation of the scalar power spectrum $\Delta_{\zeta}$
vs scalar spectral index $n_{\zeta}$ is shown in \ref{fig9b}. The purple and blue coloured line represent the upper and lower bound allowed by WMAP+Planck 2015 data respectively. 
The green dotted region bounded by two vertical black coloured lines represent the Planck $2\sigma$ allowed region and the rest of the light gray shaded region is disfavoured by the Planck+WMAP constraint.
From \ref{fig9a} and \ref{fig9b}, it 
is also observed that, within $50<N_{\star}<70$ the Exponential potential-Type I is favoured for the characteristic index $1<q<2$, by Planck 2015 data and Planck+ BICEP2+Keck Array joint analysis.} 
\label{fig9}
\end{figure*}
\begin{figure*}[htb]
\centering
\subfigure[$\alpha_{\zeta}$ vs $n_{\zeta}$.]{
    \includegraphics[width=12.2cm,height=7.3cm] {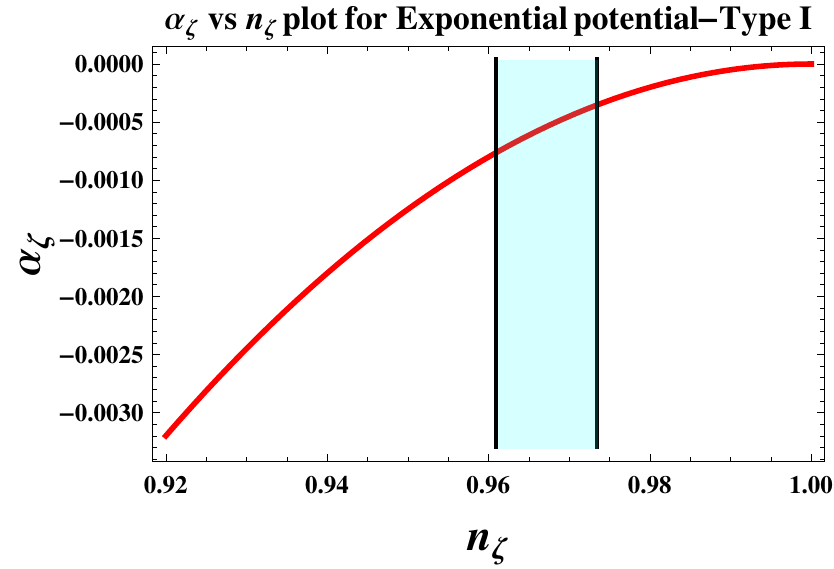}
    \label{fig10a}
}
\subfigure[$\kappa_{\zeta}$ vs $n_{\zeta}$.]{
    \includegraphics[width=12.2cm,height=7.3cm] {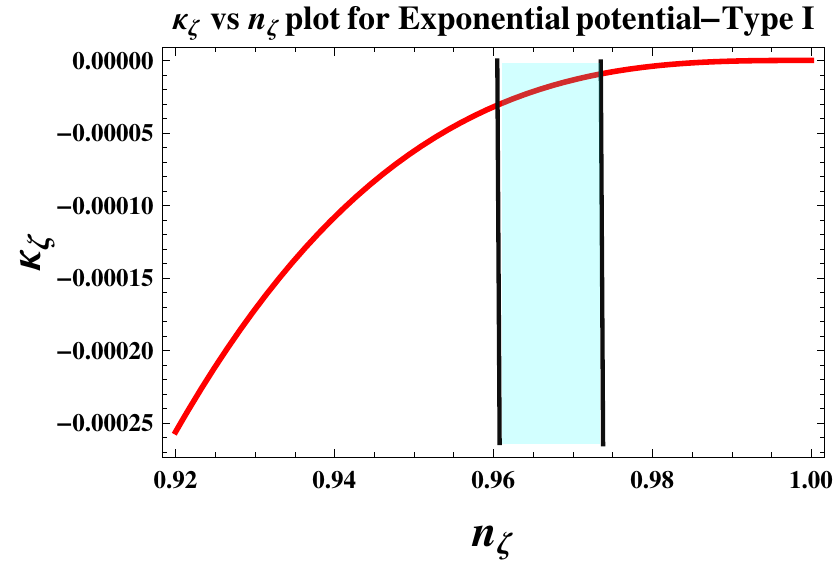}
    \label{fig10b}
}
\caption[Optional caption for list of figures]{Behaviour of the \ref{fig10a} running of the scalar spectral tilt $\alpha_{\zeta}$ and \ref{fig10b} running of the 
running of the scalar spectral tilt $\kappa_{\zeta}$ with respect to the scalar spectral index $n_{\zeta}$ for Exponential potential-Type I.
For both the figures 
 \textcolor{red}{red} colored curve represent for any value of $q$ respectively.
The \textcolor{cyan}{cyan} color shaded region bounded by two vertical black coloured lines in \ref{fig10a} and \ref{fig10a} represent the Planck $2\sigma$ allowed region and the rest of the light gray shaded region is
showing the $1\sigma$ allowed range, which is at present 
disfavoured by the Planck data and Planck+ BICEP2+Keck Array joint constraint. From \ref{fig10a} and \ref{fig10b}, it 
is also observed that, within $50<N_{\star}<70$ the Exponential potential-Type I is favoured for the characteristic index $1<q<2$, by Planck 2015 data and Planck+ BICEP2+Keck Array joint analysis.} 
\label{fig10}
\end{figure*}
Finally we compute the following inflationary observables:
\be\begin{array}{lll}\label{e-fold9c}
 \displaystyle   \Delta_{\zeta,\star}\approx\frac{g\lambda}{12\pi^2M^{4}_{p}}\times\left\{\begin{array}{lll}
                    \displaystyle  
                   \left(\frac{N_{\star}+\frac{1}{2}}{g}\right)^2\,,~~~~~~ &
 \mbox{\small {\bf for {$q=1/2$ }}}  \\ 
 \displaystyle   
            2q~ \left(\frac{\frac{N_{\star}}{\sqrt{2q}}+\frac{1}{2}}{g}\right)^2\,.~~~~~~ &
 \mbox{\small {\bf for {~any~arbitrary~ $q$ }}} 
          \end{array}
\right.
\end{array}\ee
\be\begin{array}{lll}\label{e-fold10c}
 \displaystyle   n_{\zeta,\star}-1\approx \left\{\begin{array}{lll}
                    \displaystyle  
                  -\frac{2}{N_{\star}+\frac{1}{2}}\,,~~~~~~ &
 \mbox{\small {\bf for {$q=1/2$ }}}  \\ 
 \displaystyle   
             -\frac{2}{\sqrt{2q}\left(\frac{N_{\star}}{\sqrt{2q}}+\frac{1}{2}\right)}\,.~~~~~~ &
 \mbox{\small {\bf for {~any~arbitrary~ $q$ }}} 
          \end{array}
\right.
\end{array}\ee
\be\begin{array}{lll}\label{e-fold11c}
 \displaystyle   \alpha_{\zeta,\star}\approx\left\{\begin{array}{lll}
                    \displaystyle  
                   -\frac{2}{\left(N_{\star}+\frac{1}{2}\right)^2}\,,~~~~~~ &
 \mbox{\small {\bf for {$q=1/2$ }}}  \\ 
 \displaystyle   
             -\frac{2}{2q\left(\frac{N_{\star}}{\sqrt{2q}}+\frac{1}{2}\right)^2}\,.~~~~~~ &
 \mbox{\small {\bf for {~any~arbitrary~ $q$ }}} 
          \end{array}
\right.
\end{array}\ee
\be\begin{array}{lll}\label{e-fold12c}
 \displaystyle   \kappa_{\zeta,\star}\approx\left\{\begin{array}{lll}
                    \displaystyle  
                  -\frac{4}{\left(N_{\star}+\frac{1}{2}\right)^3}\,,~~~~~~ &
 \mbox{\small {\bf for {$q=1/2$ }}}  \\ 
 \displaystyle   
             -\frac{4}{(2q)^{3/2}\left(\frac{N_{\star}}{\sqrt{2q}}+\frac{1}{2}\right)^3}\,.~~~~~~ &
 \mbox{\small {\bf for {~any~arbitrary~ $q$ }}} 
          \end{array}
\right.
\end{array}\ee
\be\begin{array}{lll}\label{e-fold13c}
 \displaystyle   r_{\star}\approx\left\{\begin{array}{lll}
                    \displaystyle  
                   \frac{8}{\left(N_{\star}+\frac{1}{2}\right)}\,,~~~~~~ &
 \mbox{\small {\bf for {$q=1/2$ }}}  \\ 
 \displaystyle   
             \frac{8}{2q\left(\frac{N_{\star}}{\sqrt{2q}}+\frac{1}{2}\right)}\,.~~~~~~ &
 \mbox{\small {\bf for {~any~arbitrary~ $q$ }}} 
          \end{array}
\right.
\end{array}\ee
For inverse cosh potential we get the following consistency relations:
\be\begin{array}{lll}\label{e-fold14c}
 \displaystyle   r_{\star}\approx 4(1-n_{\zeta,\star})\times\left\{\begin{array}{lll}
                    \displaystyle  
                   1\,,~~~~~~ &
 \mbox{\small {\bf for {$q=1/2$ }}}  \\ 
 \displaystyle   
             \frac{1}{\sqrt{2q}}\,.~~~~~~ &
 \mbox{\small {\bf for {~any~arbitrary~ $q$ }}} 
          \end{array}
\right.
\end{array}\ee
\be\begin{array}{lll}\label{e-fold15c}
 \displaystyle   \Delta_{\zeta,\star}\approx\frac{\lambda}{3g\pi^2M^{4}_{p}(1-n_{\zeta,\star})^2}\times\left\{\begin{array}{lll}
                    \displaystyle  
                   1\,,~~~~~~ &
 \mbox{\small {\bf for {$q=1/2$ }}}  \\ 
 \displaystyle   
             1\,.~~~~~~ &
 \mbox{\small {\bf for {~any~arbitrary~ $q$ }}} 
          \end{array}
\right.
\end{array}\ee
\be\begin{array}{lll}\label{e-fold16c}
 \displaystyle   \Delta_{\zeta,\star}\approx\frac{2\lambda}{3g\pi^2M^{4}_{p}r_{\star}}\times\left\{\begin{array}{lll}
                    \displaystyle  
                   \left(N_{\star}+\frac{1}{2}\right)\,,~~~~~~ &
 \mbox{\small {\bf for {$q=1/2$ }}}  \\ 
 \displaystyle   
             \left(\frac{N_{\star}}{\sqrt{2q}}+\frac{1}{2}\right)\,.~~~~~~ &
 \mbox{\small {\bf for {~any~arbitrary~ $q$ }}} 
          \end{array}
\right.
\end{array}\ee
\be\begin{array}{lll}\label{e-fold17c}
 \displaystyle   \alpha_{\zeta,\star}\approx-\frac{1}{2}\left(n_{\zeta,\star}-1\right)^2\times\left\{\begin{array}{lll}
                    \displaystyle  
                   1\,,~~~~~~ &
 \mbox{\small {\bf for {$q=1/2$ }}}  \\ 
 \displaystyle   
             1\,.~~~~~~ &
 \mbox{\small {\bf for {~any~arbitrary~ $q$ }}} 
          \end{array}
\right.
\end{array}\ee
\be\begin{array}{lll}\label{e-fold18c}
 \displaystyle   \kappa_{\zeta,\star}\approx\frac{1}{2}\left(n_{\zeta,\star}-1\right)^3\times\left\{\begin{array}{lll}
                    \displaystyle  
                   1\,,~~~~~~ &
 \mbox{\small {\bf for {$q=1/2$ }}}  \\ 
 \displaystyle   
             1\,.~~~~~~ &
 \mbox{\small {\bf for {~any~arbitrary~ $q$ }}} 
          \end{array}
\right.
\end{array}\ee
Let us now discuss the general constraints on the parameters of tachyonic string theory including the factor $q$ 
and on the parameters appearing in the 
expression for Exponential potential-Type I. In Fig.~(\ref{fig9a}), we have shown the behavior of the tensor-to-scalar ratio $r$ with respect to the scalar spectral index $n_{\zeta}$
 for Exponential potential-Type I respectively. In both the figures the \textcolor{purple}{purple} and \textcolor{blue}{blue}
coloured line represent the upper bound of tensor-to-scalar ratio allowed by Planck+ BICEP2+Keck Array
joint constraint and only Planck 2015 data respectively. For both the figures 
 \textcolor{red}{red},~\textcolor{green}{green},
~\textcolor{brown}{brown},~\textcolor{orange}{orange} colored curve represent $q=1/2$, $q=1$, $q=3/2$ and $q=2$ respectively.
The \textcolor{cyan}{cyan} color shaded region bounded by two vertical black coloured lines in Fig.~(\ref{fig9a}) represent the
Planck $2\sigma$ allowed region and the rest of the light gray shaded region
is showing the $1\sigma$ allowed range, which is at present 
disfavoured by the Planck 2015 data and Planck+ BICEP2+Keck Array joint constraint. The rest of the region 
is completely ruled out by the present observational constraints. From Fig.~(\ref{fig9a}), it 
is also observed that, within $50<N_{\star}<70$, the Exponential potential-Type I is favoured for the characteristic
index $1/2<q<2$, by Planck 2015 data and Planck+ BICEP2+Keck Array joint analysis. Also in Fig.~(\ref{fig9a}) for $q=1/2$, $q=1$, $q=3/2$ and $q=2$ we fix $N_{\star}/g\sim 0.8$. This implies that for $50<N_{\star}<70$,
the prescribed window for $g$ from $r-n_{\zeta}$ plot is given by, $63<g<88$.
To analyze the results more clearly let us describe the cosmological features from Fig.~(\ref{fig9a}) in detail.
Let us first start with the $q=1/2$ situation, in which the $2\sigma$ constraint on the scalar spectral tilt is satisfied
within the window of tensor-to-scalar ratio, $0.10<r<0.12$ for $50<N_{\star}<70$. Next for $q=1$ case, the same constraint 
is satisfied within the window of tensor-to-scalar ratio, $ 0.07<r<0.11$ for $50<N_{\star}<70$. Further for $q=3/2$ case, the same constraint 
on scalar spectral tilt is satisfied within the window of tensor-to-scalar ratio, $ 0.06<r<0.085$ for $50<N_{\star}<70$.
Finally, for $q=2$ situation, the value for the tensor-to-scalar ratio is $0.05<r<0.075$.

In Fig.~(\ref{fig9b}), we have depicted the behavior of the 
scalar power spectrum $\Delta_{\zeta}$ vs 
scalar spectral tilt $n_{\zeta}$ for Exponential potential-Type I respectively. It is important to note that,
for all of the figures 
 \textcolor{red}{red},~\textcolor{green}{green},
~\textcolor{brown}{brown},~\textcolor{orange}{orange} colored curve represent $q=1/2$, $q=1$, $q=3/2$ and $q=2$ respectively.
The \textcolor{purple}{purple} and \textcolor{blue}{blue} coloured line represent the upper and lower bound allowed by WMAP+Planck 2015 data respectively. 
The \textcolor{cyan}{cyan} color shaded region bounded by two vertical black coloured lines represent the Planck
$2\sigma$ allowed region and the rest of the light gray shaded region is the $1\sigma$ region, which is presently 
disfavoured by the joint Planck+WMAP constraints. The rest of the region 
is completely ruled out by the present observational constraints. Also in Fig.~(\ref{fig9b}) for $q=1/2$, $q=1$, $q=3/2$ and $q=2$
we fix $N_{\star}\sim 70$. But plots can be reproduced for $50<N_{\star} <70$ also considering the present observational constraints from Planck 2015 data.
It is important to mention here that if we combine the constraints obtained from Fig.~(\ref{fig9a}) and Fig.~(\ref{fig9b}), then by comparing the behaviour of the GTachyon in 
$r-n_{\zeta}$ and $\Delta_{\zeta}-n_{\zeta}$ we clearly observe that the $q=1/2$ case is almost discarded as it is not consistent with both of the $2\sigma$ constraints simultaneously.

\begin{figure}[htb]
	\centering
	\includegraphics[width=12cm,height=10cm]{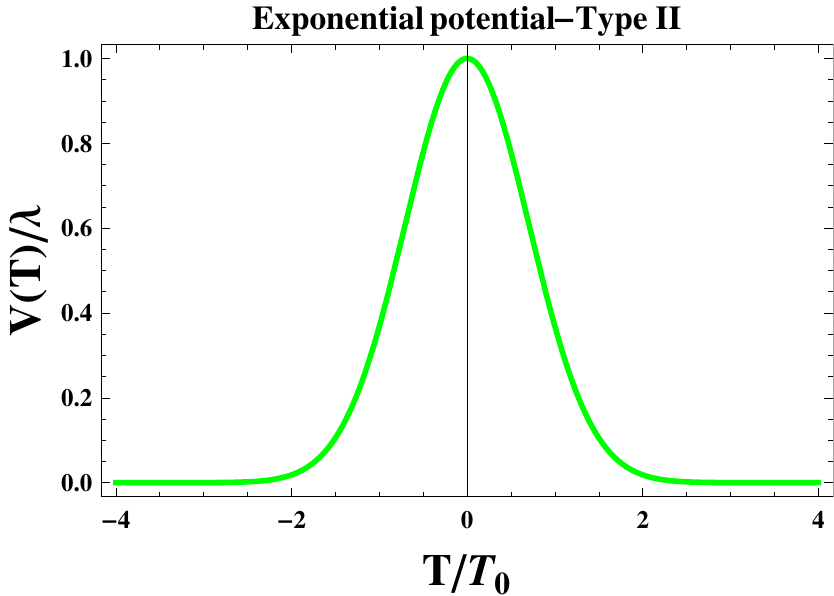}
	\caption{Variation of the Exponential potential-Type II $V(T)/\lambda$ with field $T/T_{0}$ in dimensionless units.
	}
	\label{fig11}
\end{figure}

In Fig.~(\ref{fig10a}) and Fig.~(\ref{fig10b}), we have shown the behaviour
of the running of the scalar spectral tilt $\alpha_{\zeta}$ and running of the 
running of the scalar spectral tilt $\kappa_{\zeta}$ with respect to the scalar spectral index $n_{\zeta}$ for Exponential potential-Type I respectively.
For both the figures 
 \textcolor{red}{red} colored curve represent the behavior for any arbitrary values of $q$.
The \textcolor{cyan}{cyan} color shaded region bounded by two vertical black coloured lines in both the plots represent the Planck $2\sigma$ allowed region and the rest of the light gray shaded region is
showing the $1\sigma$ allowed range, which is at present 
disfavoured by the Planck data and Planck+ BICEP2+Keck Array joint constraint. From both of these figures, it 
is also observed that, within $50<N_{\star}<70$ the Exponential potential-Type I is favoured for the characteristic index $1<q<2$, by Planck
2015 data and Planck+ BICEP2+Keck Array joint analysis. It is also important to note that for any values of $q$, the numerical value 
of the running $\alpha_{\zeta}\sim {\cal O}(-10^{-4})$
and running of the running $\kappa_{\zeta}\sim {\cal O}(-10^{-5})$, which are perfectly 
consistent with the $1.5\sigma$ constraints on running and running of the running as obtained from Planck 2015 data. For $q=1/2$, $q=1$, $q=3/2$ and $q=2$ $g$ is not explicitly appearing in the various inflationary observables except the amplitude of scalar power spectrum in 
this case. To produce the correct value of the amplitude of the scalar power spectra we fix the parameter $360<g<400$.\\
\underline{\bf Model IV: Exponential potential-Type II}\\
For single field case the first model of tachyonic potential is given by:
\be\label{weeeee1d}
V(T)=\lambda\exp\left[-\left(\frac{T}{T_{0}}\right)^2\right],\ee
where $\lambda$ characterize the scale of inflation and $T_{0}$ is the parameter of the model.
In Fig.~(\ref{fig11}) we have depicted the symmetric behavior of the Exponential potential-Type I with respect to scaled field coordinate $T/T_{0}$ in dimensionless units.
In this case the tachyon field started rolling down from the top hight of the potential situated at the origin and take part in inflationary dynamics.  


Next using specified form of the potential the potential dependent slow-roll parameters are computed as:
\bea \bar{\epsilon}_{V}&=&\frac{1}{2g}4\left(\frac{T}{T_{0}}\right)^2\exp\left[\left(\frac{T}{T_{0}}\right)^2\right],\\
\bar{\eta}_{V}&=&\frac{1}{g}\left[2\left(\frac{T}{T_{0}}\right)^2-1\right]\exp\left[\left(\frac{T}{T_{0}}\right)^2\right],\\
\bar{\xi}^2_{V}&=&\frac{1}{g^2}8\left(\frac{T}{T_{0}}\right)^2\left[2\left(\frac{T}{T_{0}}\right)^2-3\right]\exp\left[2\left(\frac{T}{T_{0}}\right)^2\right],~~~~~\\
\bar{\sigma}^3_{V}&=&\frac{1}{g^3}16\left(\frac{T}{T_{0}}\right)^2\left[4\left(\frac{T}{T_{0}}\right)^4-12\left(\frac{T}{T_{0}}\right)^2+3\right]\exp\left[3\left(\frac{T}{T_{0}}\right)^2\right],~~~~~~~~~~~~~~
\eea
where the factor $g$ is defined as:
\be g= \frac{\alpha^{'}\lambda T^2_0}{M^2_p}=\frac{M^4_s}{(2\pi)^3 g_s}\frac{\alpha^{'} T^2_0}{M^2_p}.\ee
Next we compute the number of e-foldings from this model~\footnote{It is important to note that here we have used the following approximation:
\be {\rm Ei}\left[-\left(\frac{T}{T_{0}}\right)^2\right]\approx \gamma +2\ln\left(\frac{T}{T_{0}}\right)+\cdots \ee
where all the terms represented via $\cdots$ are negligibly small in the series expansion. Here $\gamma$ represents the \text{is Euler{'}s constant}, with numerical value $\gamma\simeq 0.577216$.}:
\be\begin{array}{lll}\label{e-fold5d}
 \displaystyle  N(T)\approx\left\{\begin{array}{lll}
                    \displaystyle  
                   g~\left[\ln\left(\frac{T}{T_{0}}\right)-\ln\left(\frac{T_{end}}{T_{0}}\right)\right]\,,~~~~~~ &
 \mbox{\small {\bf for {$q=1/2$ }}}  \\ 
 \displaystyle   
             \sqrt{2q}~g~\left[\ln\left(\frac{T}{T_{0}}\right)-\ln\left(\frac{T_{end}}{T_{0}}\right)\right]\,.~~~~~~ &
 \mbox{\small {\bf for {~any~arbitrary~ $q$ }}} 
          \end{array}
\right.
\end{array}\ee
Further using the condition to end inflation:
\bea \bar{\epsilon}_{V}(T_{end})=1.\eea
      we get the following transcendental equation to determine the field value at the end of inflation:
      \be \left(\frac{T_{end}}{T_{0}}\right)^2\exp\left[\left(\frac{T_{end}}{T_{0}}\right)^2\right]=\frac{g}{2}\ee
      which we need to solve numerically for a given value of $g$. For the sake of simplicity in the further computation one can consider the following possibilities:
      \be\begin{array}{lll}\label{e-fold6d}
 \displaystyle  T_{end}\approx\left\{\begin{array}{lll}
                    \displaystyle  
                   T_{0}\,,~~~~~~ &
 \mbox{\small {\bf for {$\frac{g}{2}\sim e$ }}}  \\ 
 \displaystyle   
             \sqrt{\frac{g}{2}}~T_{0}\,.~~~~~~ &
 \mbox{\small {\bf for {$\frac{g}{2}<<1$ }}} 
          \end{array}
\right.
\end{array}\ee
      Next using $N=N_{cmb}=N_{\star}$ and $T=T_{cmb}=T_{\star}$ at the horizon crossing we get,
\be\begin{array}{lll}\label{e-fold7d}
 \displaystyle  T_{\star}=T_{0}\times\left\{\begin{array}{lll}
                    \displaystyle  
                    \exp\left[\frac{N_{\star}}{g}+\ln\left(\frac{T_{end}}{T_{0}}\right)\right]\,,~~~~~~ &
 \mbox{\small {\bf for {$q=1/2$ }}}  \\ 
 \displaystyle   
              \exp\left[\frac{N_{\star}}{\sqrt{2q}g}+\ln\left(\frac{T_{end}}{T_{0}}\right)\right]\,.~~~~~~ &
 \mbox{\small {\bf for {~any~arbitrary~ $q$ }}} 
          \end{array}
\right.
\end{array}\ee
Also the field excursion can be computed as:
\be\begin{array}{lll}\label{e-fold7dex}
 \displaystyle  |\Delta T|=T_{0}\times\left\{\begin{array}{lll}
                    \displaystyle  
                    \left|\exp\left[\frac{N_{\star}}{g}+\ln p\right]-p\right|\,,~~~~~~ &
 \mbox{\small {\bf for {$q=1/2$ }}}  \\ 
 \displaystyle   
              \left|\exp\left[\frac{N_{\star}}{\sqrt{2q}g}+\ln p\right]-p\right|\,.~~~~~~ &
 \mbox{\small {\bf for {~any~arbitrary~ $q$ }}} 
          \end{array}
\right.
\end{array}\ee
where $p=1, \sqrt{\frac{g}{2}}$. But during numerical estimation we only take $p= \sqrt{\frac{g}{2}}$ because $p=1$ is disfavoured by the Planck 2015 data.

Finally we compute the following inflationary observables:
\be\begin{array}{lll}\label{e-fold9d}
 \displaystyle   \Delta_{\zeta,\star}\approx\frac{g\lambda}{48\pi^2M^{4}_{p}}\times\left\{\begin{array}{lll}
                    \displaystyle  
                  \frac{ \exp\left(-2\exp\left[\frac{2N_{\star}}{g}+2\ln\left(\frac{T_{end}}{T_{0}}\right)\right]\right)}{ \exp\left[\frac{2N_{\star}}{g}+2\ln\left(\frac{T_{end}}{T_{0}}\right)\right]}\,,~~~~~~ &
 \mbox{\small {\bf for {$q=1/2$ }}}  \\ 
 \displaystyle   
            2q~ \frac{ \exp\left(-2\exp\left[\frac{2N_{\star}}{\sqrt{2q}g}+2\ln\left(\frac{T_{end}}{T_{0}}\right)\right]\right)}{ \exp\left[\frac{2N_{\star}}{\sqrt{2q}g}+2\ln\left(\frac{T_{end}}{T_{0}}\right)\right]}\,.~~~~~~ &
 \mbox{\small {\bf for {~any~arbitrary~ $q$ }}} 
          \end{array}
\right.
\end{array}\ee
\be\begin{array}{lll}\label{e-fold10d}
 \displaystyle   n_{\zeta,\star}-1\approx \left\{\begin{array}{lll}
                    \displaystyle  
                 -\frac{2}{g}\left\{1+2\left(\frac{T_{end}}{T_{0}}\right)^2~ \exp\left[\frac{2N_{\star}}{g}\right]\right\}\,,~~~~~~ &
 \mbox{\small {\bf for {$q=1/2$ }}}  \\ 
 \displaystyle   
            -\frac{2}{\sqrt{2q}~g}\left\{1+2\left(\frac{T_{end}}{T_{0}}\right)^2~ \exp\left[\frac{2N_{\star}}{\sqrt{2q}g}\right]\right\}\,.~~~~~~ &
 \mbox{\small {\bf for {~any~arbitrary~ $q$ }}} 
          \end{array}
\right.
\end{array}\ee
\begin{figure*}[htb]
\centering
\subfigure[$r$ vs $n_{\zeta}$.]{
    \includegraphics[width=10.2cm,height=7cm] {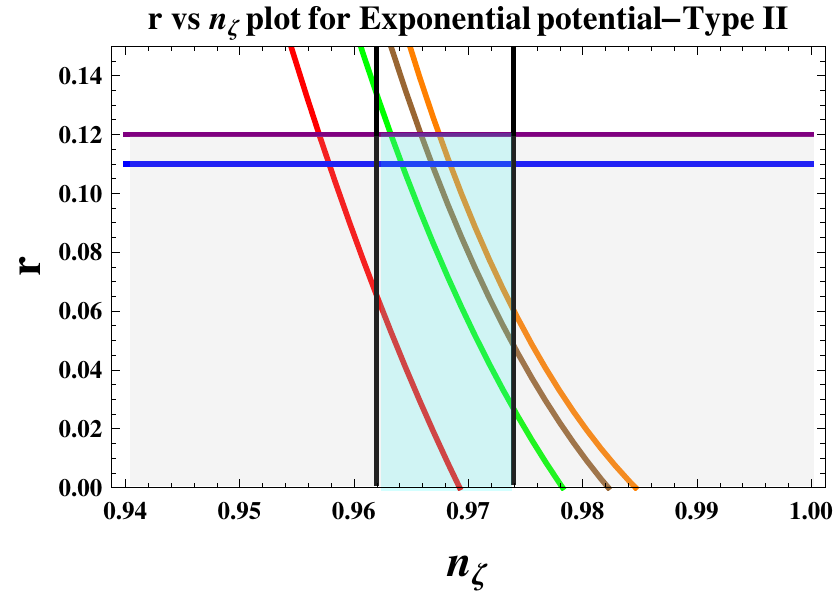}
    \label{fig12a}
}
\subfigure[$r$ vs $g$.]{
    \includegraphics[width=10.2cm,height=7cm] {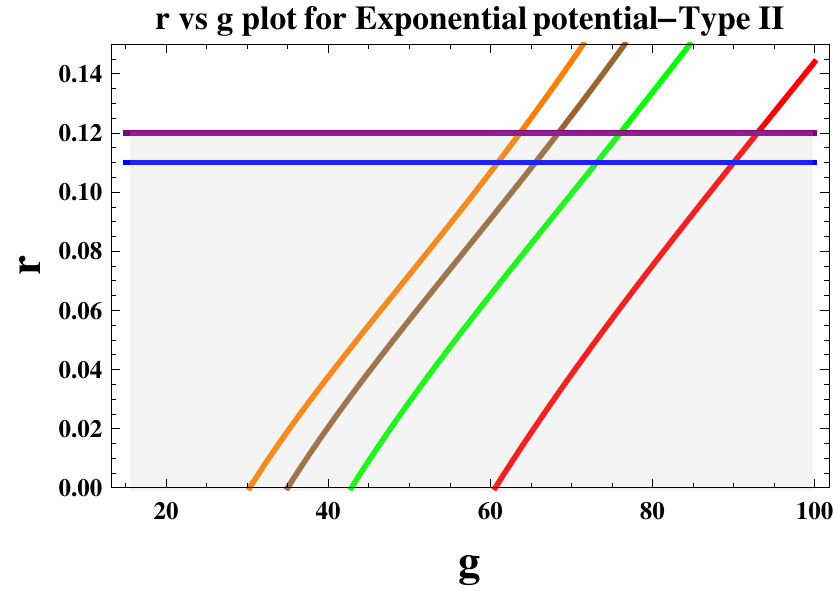}
    \label{fig12b}
}
\caption[Optional caption for list of figures]{Behaviour of the tensor-to-scalar ratio $r$ with respect to \ref{fig12a} the scalar spectral index $n_{\zeta}$
and \ref{fig12b} the parameter $g$ for Exponential potential-Type II.
The purple and blue coloured line represent the upper bound of tensor-to-scalar ratio allowed by Planck+ BICEP2+Keck Array joint constraint and only Planck 2015 data respectively. For both the figures 
 \textcolor{red}{red},~\textcolor{green}{green},
~\textcolor{brown}{brown},~\textcolor{orange}{orange} colored curve represent $q=1/2$, $q=1$, $q=3/2$ and $q=2$ respectively.
The \textcolor{cyan}{cyan} color shaded region bounded by two vertical black coloured lines in \ref{fig12a} represent the Planck $2\sigma$ allowed region and the rest of the light gray shaded region is
showing the $1\sigma$ allowed range, which is at present 
disfavoured by the Planck data and Planck+ BICEP2+Keck Array joint constraint. From \ref{fig12a} and \ref{fig12b}, it 
is also observed that, within $50<N_{\star}<70$ the Logarithmic potential is favoured
for the characteristic index $1/2<q<2$, by Planck 2015 data and Planck+ BICEP2+Keck Array joint analysis.} 
\label{fig12}
\end{figure*}
\begin{figure*}[htb]
\centering
\subfigure[$\Delta_{\zeta}$ vs $g$.]{
    \includegraphics[width=7.2cm,height=8cm] {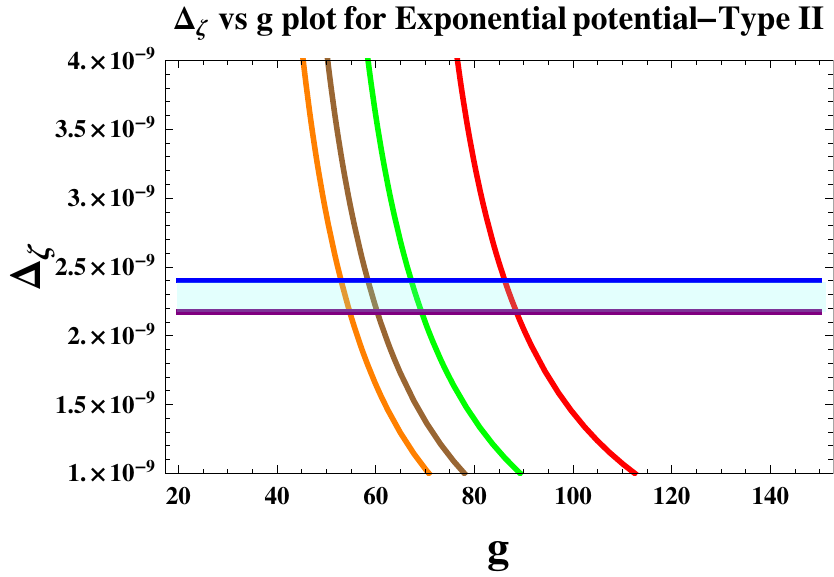}
    \label{fig13a}
}
\subfigure[$n_{\zeta}$ vs $g$.]{
    \includegraphics[width=7.2cm,height=8cm] {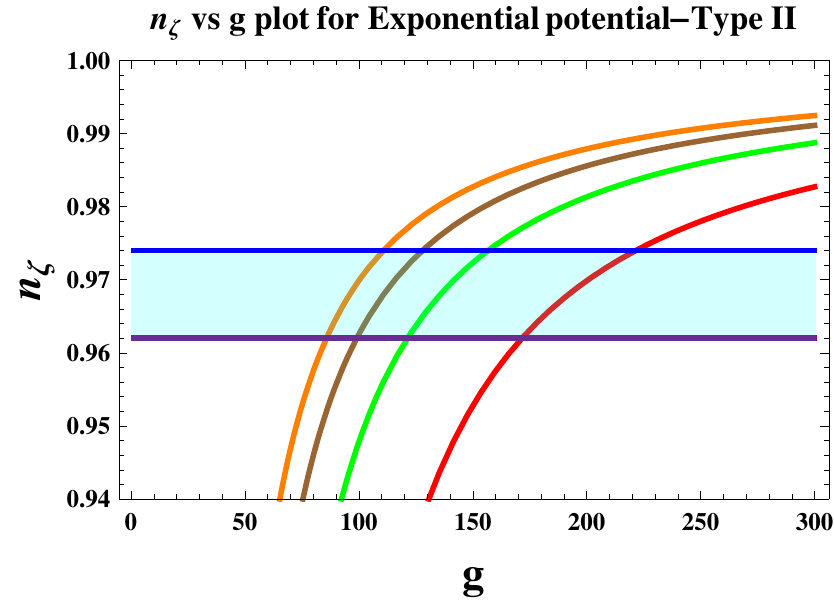}
    \label{fig13b}
}
\subfigure[$\Delta_{\zeta}$ vs $n_{\zeta}$.]{
    \includegraphics[width=11.2cm,height=8cm] {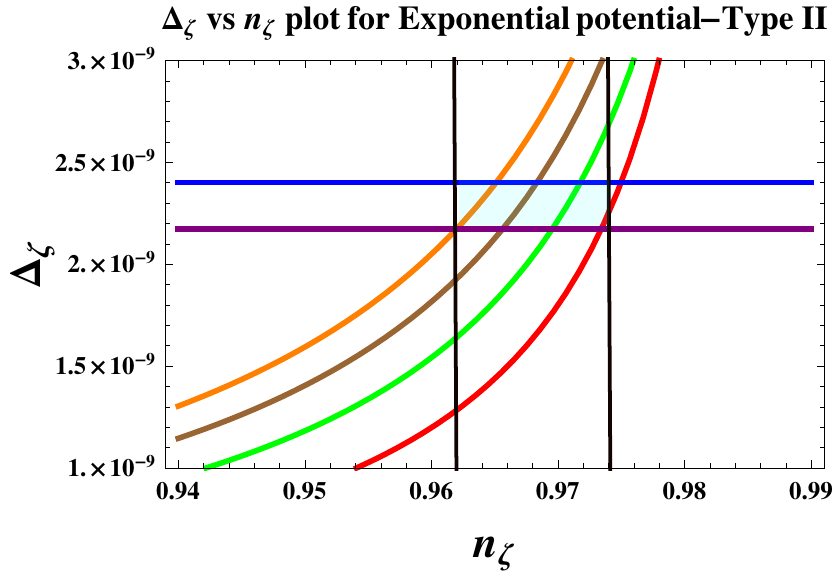}
    \label{fig13c}
}
\caption[Optional caption for list of figures]{Variation of the \ref{fig13a} scalar power spectrum $\Delta_{\zeta}$
vs scalar spectral index $n_{\zeta}$, \ref{fig13b} scalar power spectrum $\Delta_{\zeta}$ vs the stringy parameter $g$
and \ref{fig13c} scalar spectral tilt $n_{\zeta}$ vs the stringy parameter $g$. The purple and blue coloured line represent the upper and lower bound allowed by WMAP+Planck 2015 data respectively. 
The green dotted region bounded by two vertical black coloured lines represent the Planck $2\sigma$ allowed region and the rest of the light gray shaded region is disfavoured by the Planck+WMAP constraint.} 
\label{fig13}
\end{figure*}
\begin{figure*}[htb]
\centering
\subfigure[$\alpha_{\zeta}$ vs $n_{\zeta}$.]{
    \includegraphics[width=12.2cm,height=7.3cm] {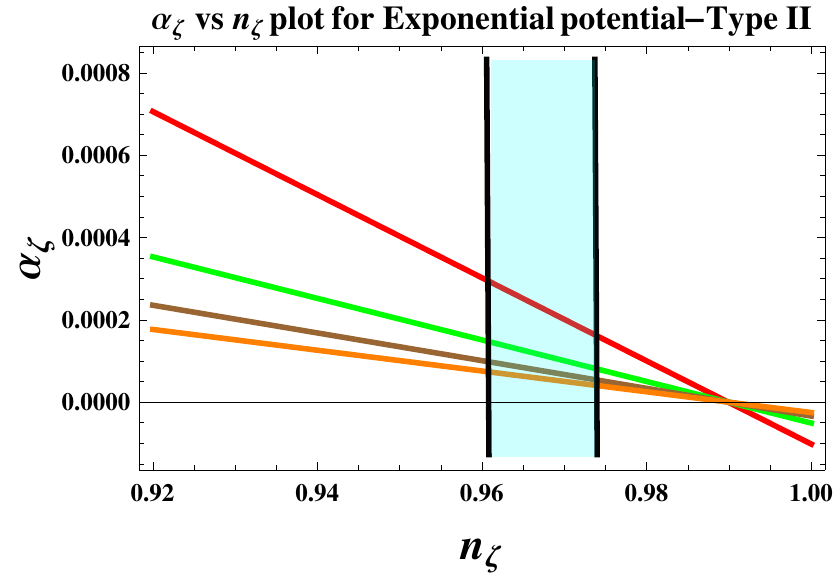}
    \label{fig14a}
}
\subfigure[$\kappa_{\zeta}$ vs $n_{\zeta}$.]{
    \includegraphics[width=12.2cm,height=7.3cm] {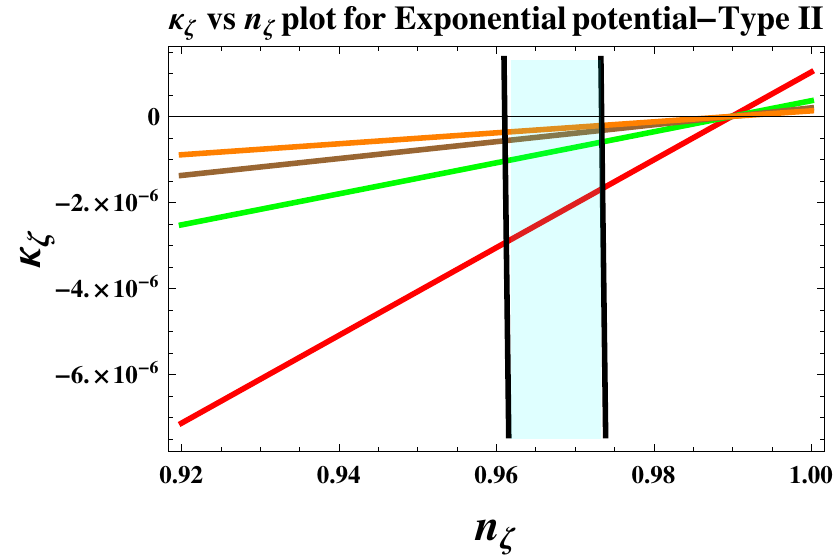}
    \label{fig14b}
}
\caption[Optional caption for list of figures]{Behaviour of the \ref{fig14a} running of the scalar spectral tilt $\alpha_{\zeta}$ and \ref{fig14b} running of the 
running of the scalar spectral tilt $\kappa_{\zeta}$ with respect to the scalar spectral index $n_{\zeta}$ for Inverse cosh potential with $g=88$.
For both the figures 
 \textcolor{red}{red},~\textcolor{green}{green},
~\textcolor{brown}{brown},~\textcolor{orange}{orange} colored curve represent $q=1/2$, $q=1$, $q=3/2$ and $q=2$ respectively.
The \textcolor{cyan}{cyan} color shaded region bounded by two vertical black coloured lines in \ref{fig14a} and \ref{fig14a} represent the Planck $2\sigma$ allowed region and the rest of the light gray shaded region is
showing the $1\sigma$ allowed range, which is at present 
disfavoured by the Planck data and Planck+ BICEP2+Keck Array joint constraint. From \ref{fig14a} and \ref{fig14b}, it 
is also observed that, within $50<N_{\star}<70$ the Inverse cosh potential is favoured for the characteristic index $1/2<q<2$, by Planck 2015 data and Planck+ BICEP2+Keck Array joint analysis.} 
\label{fig14}
\end{figure*}
\be\begin{array}{lll}\label{e-fold11d}
 \displaystyle   \alpha_{\zeta,\star}\approx \frac{8}{g^2}\times\left\{\begin{array}{lll}
                    \displaystyle  
                  \left(\frac{T_{end}}{T_{0}}\right)^2~ \exp\left[\frac{2N_{\star}}{g}\right]\,,~~~~~~ &
 \mbox{\small {\bf for {$q=1/2$ }}}  \\ 
 \displaystyle   
             \frac{1}{2q}\left(\frac{T_{end}}{T_{0}}\right)^2~ \exp\left[\frac{2N_{\star}}{\sqrt{2q}g}\right]\,.~~~~~~ &
 \mbox{\small {\bf for {~any~arbitrary~ $q$ }}} 
          \end{array}
\right.
\end{array}\ee
\be\begin{array}{lll}\label{e-fold12d}
 \displaystyle   \kappa_{\zeta,\star}\approx -\frac{16}{g^3}\times\left\{\begin{array}{lll}
                    \displaystyle  
                  \left(\frac{T_{end}}{T_{0}}\right)^2~ \exp\left[\frac{2N_{\star}}{g}\right]\,,~~~~~~ &
 \mbox{\small {\bf for {$q=1/2$ }}}  \\ 
 \displaystyle   
             \frac{1}{(2q)^{3/2}}\left(\frac{T_{end}}{T_{0}}\right)^2~ \exp\left[\frac{2N_{\star}}{\sqrt{2q}g}\right]\,.~~~~~~ &
 \mbox{\small {\bf for {~any~arbitrary~ $q$ }}} 
          \end{array}
\right.
\end{array}\ee
\be\begin{array}{lll}\label{e-fold13d}
 \displaystyle   r_{\star}\approx\frac{32}{g}\times\left\{\begin{array}{lll}
                    \displaystyle  
                   \left(\frac{T_{end}}{T_{0}}\right)^2 \exp\left[\frac{2N_{\star}}{g}+\left(\frac{T_{end}}{T_{0}}\right)^2\exp\left[\frac{2N_{\star}}{g}\right]\right]\,,~~~~~~ &
 \mbox{\small {\bf for {$q=1/2$ }}}  \\ 
 \displaystyle   
             \frac{1}{2q}\left(\frac{T_{end}}{T_{0}}\right)^2 \exp\left[\frac{2N_{\star}}{\sqrt{2q}g}+\left(\frac{T_{end}}{T_{0}}\right)^2\exp\left[\frac{2N_{\star}}{\sqrt{2q}g}\right]\right]\,.~~~~~~ &
 \mbox{\small {\bf for {~any~arbitrary~ $q$ }}} 
          \end{array}
\right.
\end{array}\ee
For inverse cosh potential we get the following consistency relations:
\be\begin{array}{lll}\label{e-fold14d}
  \displaystyle   r_{\star}\approx\frac{16}{g}\times\left\{\begin{array}{lll}
                    \displaystyle  
                   \left[\frac{g}{2}\left(1-n_{\zeta,\star}\right)-1\right] \exp\left[\frac{1}{2}\left(\frac{g}{2}\left(1-n_{\zeta,\star}
                   \right)-1\right)\right]\,,~~~~~~ &
 \mbox{\small {\bf for {$q=1/2$ }}}  \\ 
 \displaystyle   
             \frac{1}{2q}\left[\frac{\sqrt{2q}g}{2}\left(1-n_{\zeta,\star}
             \right)-1\right] \exp\left[\frac{1}{2}\left(\frac{\sqrt{2q}g}{2}\left(1-n_{\zeta,\star}\right)-1\right)\right]\,.~~~~~~ &
 \mbox{\small {\bf for {~any~arbitrary~ $q$ }}} 
          \end{array}
\right.
\end{array}\ee
\be\begin{array}{lll}\label{e-fold15d}
 \displaystyle   \Delta_{\zeta,\star}\approx\frac{g\lambda}{24\pi^2M^{4}_{p}}\times\left\{\begin{array}{lll}
                    \displaystyle  
                  \frac{ \exp\left(-\left[\frac{g}{2}\left(1-n_{\zeta,\star}
                  \right)-1\right]\right)}{\left[\frac{g}{2}\left(1-n_{\zeta,\star}\right)-1\right]}\,,~~~~~~ &
 \mbox{\small {\bf for {$q=1/2$ }}}  \\ 
 \displaystyle   
            2q~ \frac{ \exp\left(-\left[\frac{\sqrt{2q}g}{2}\left(1-n_{\zeta,\star}
            \right)-1\right]\right)}{\left[\frac{\sqrt{2q}g}{2}\left(1-n_{\zeta,\star}\right)-1\right]}\,.~~~~~~ &
 \mbox{\small {\bf for {~any~arbitrary~ $q$ }}} 
          \end{array}
\right.
\end{array}\ee
\be\begin{array}{lll}\label{e-fold16d}
 \displaystyle   \Delta_{\zeta,\star}\approx\frac{2\lambda}{3\pi^2M^{4}_{p}r_{\star}}\times\left\{\begin{array}{lll}
                    \displaystyle  
                   \exp\left[-\left(\frac{T_{end}}{T_{0}}\right)^2~ \exp\left[\frac{2N_{\star}}{g}\right]\right]\,,~~~~~~ &
 \mbox{\small {\bf for {$q=1/2$ }}}  \\ 
 \displaystyle   
            \exp\left[-\left(\frac{T_{end}}{T_{0}}\right)^2~ \exp\left[\frac{2N_{\star}}{\sqrt{2q}g}\right]\right]\,.~~~~~~ &
 \mbox{\small {\bf for {~any~arbitrary~ $q$ }}} 
          \end{array}
\right.
\end{array}\ee
\be\begin{array}{lll}\label{e-fold17d}
 \displaystyle   \alpha_{\zeta,\star}\approx \frac{4}{g^2}\times\left\{\begin{array}{lll}
                    \displaystyle  
                  \left[\frac{g}{2}\left(1-n_{\zeta,\star}
                  \right)-1\right]\,,~~~~~~ &
 \mbox{\small {\bf for {$q=1/2$ }}}  \\ 
 \displaystyle   
             \frac{1}{2q}\left[\frac{\sqrt{2q}g}{2}\left(1-n_{\zeta,\star}
            \right)-1\right]\,.~~~~~~ &
 \mbox{\small {\bf for {~any~arbitrary~ $q$ }}} 
          \end{array}
\right.
\end{array}\ee
\be\begin{array}{lll}\label{e-fold18d}
 \displaystyle   \kappa_{\zeta,\star}\approx -\frac{8}{g^3}\times\left\{\begin{array}{lll}
                    \displaystyle  
                  \left[\frac{g}{2}\left(1-n_{\zeta,\star}
                  \right)-1\right]\,,~~~~~~ &
 \mbox{\small {\bf for {$q=1/2$ }}}  \\ 
 \displaystyle   
             \frac{1}{(2q)^{3/2}}\left[\frac{\sqrt{2q}g}{2}\left(1-n_{\zeta,\star}
            \right)-1\right]\,.~~~~~~ &
 \mbox{\small {\bf for {~any~arbitrary~ $q$ }}} 
          \end{array}
\right.
\end{array}\ee
Let us now discuss the general constraints on the parameters of tachyonic string theory including the factor $q$ 
and on the parameters appearing in the 
expression for Exponential potential-Type II. In Fig.~(\ref{fig12a}) and Fig.~(\ref{fig12b}), we have shown the behavior of the tensor-to-scalar ratio $r$ with respect to the scalar spectral index $n_{\zeta}$
and the model parameter $g$ for Exponential potential-Type II respectively. In both the figures the \textcolor{purple}{purple} and \textcolor{blue}{blue}
coloured line represent the upper bound of tensor-to-scalar ratio allowed by Planck+ BICEP2+Keck Array
joint constraint and only Planck 2015 data respectively. For both the figures 
 \textcolor{red}{red},~\textcolor{green}{green},
~\textcolor{brown}{brown},~\textcolor{orange}{orange} colored curve represent $q=1/2$, $q=1$, $q=3/2$ and $q=2$ respectively.
The \textcolor{cyan}{cyan} color shaded region bounded by two vertical black coloured lines in Fig.~(\ref{fig12a}) represent the
Planck $2\sigma$ allowed region and the rest of the light gray shaded region
is showing the $1\sigma$ allowed range, which is at present 
disfavoured by the Planck 2015 data and Planck+ BICEP2+Keck Array joint constraint. The rest of the region 
is completely ruled out by the present observational constraints. From Fig.~(\ref{fig12a}) and Fig.~(\ref{fig12b}), it 
is also observed that, within $50<N_{\star}<70$, the Exponential potential-Type II is favoured only for the characteristic
index $1/2<q<2$, by Planck 2015 data and Planck+ BICEP2+Keck Array joint analysis.
Also in Fig.~(\ref{fig12a}) for $q=1/2$, $q=1$, $q=3/2$ and $q=2$ we fix $N_{\star}/g\sim 0.85$. This implies that for $50<N_{\star}<70$,
the prescribed window for $g$ from $r-n_{\zeta}$ plot is given by, $59<g<82.3$ considering $1/2<q<2$.
It is additionally important to note that, for $q<<1/2$, the 
tensor-to-scalar ratio computed from the model is negligibly small for Exponential potential-Type II.
This implies that if the inflationary tensor mode is detected near to its present upper bound on tensor-to-scalar ratio then all $q<<1/2$ 
possibilities for tachyonic inflation can be discarded for Exponential potential-Type II. On the contrary,
if inflationary tensor modes are never detected by any of the future observational probes then $q<<1/2$
possibilities for tachyonic inflation in case of Exponential potential-Type II is highly 
prominent. Also it is important to mention that, in Fig.~(\ref{fig12b}) within the window $30<g<93$, 
if we smaller the value of $g$, then the inflationary tensor-to-scalar ratio also gradually decreases.
To analyze the results more clearly let us describe the cosmological features from Fig.~(\ref{fig12a}) in detail.
Let us first start with the $q=1/2$ situation, in which the $2\sigma$ constraint on the scalar spectral tilt is satisfied
within the window of tensor-to-scalar ratio, $0<r<0.065$ for $50<N_{\star}<70$. Next for $q=1$ case, the same constraint 
is satisfied within the window of tensor-to-scalar ratio, $0.030<r<0.12$ for $50<N_{\star}<70$. Further for $q=3/2$ case, the same constraint 
on scalar spectral tilt is satisfied within the window of tensor-to-scalar ratio, $0.045<r<0.12$ for $50<N_{\star}<70$.
Finally, for $q=2$ situation, the value for the tensor-to-scalar ratio is $0.060<r<0.12$, which is tightly constrained from the upper bound of 
spectral tilt from Planck 2015 observational data. 

In Fig.~(\ref{fig13a}), Fig.~(\ref{fig13b}) and Fig.~(\ref{fig13c}), we have depicted the behavior of the 
scalar power spectrum $\Delta_{\zeta}$ vs the stringy parameter $g$, 
scalar spectral tilt $n_{\zeta}$ vs the stringy parameter $g$ and scalar power spectrum $\Delta_{\zeta}$
vs scalar spectral index $n_{\zeta}$ for Exponential potential-Type-II respectively. It is important to note that,
for all of the figures 
 \textcolor{red}{red},~\textcolor{green}{green},
~\textcolor{brown}{brown},~\textcolor{orange}{orange} colored curve represent $q=1/2$, $q=1$, $q=3/2$ and $q=2$ respectively.
The \textcolor{purple}{purple} and \textcolor{blue}{blue} coloured line represent the upper and lower bound allowed by WMAP+Planck 2015 data respectively. 
The \textcolor{cyan}{cyan} color shaded region bounded by two vertical black coloured lines represent the Planck
$2\sigma$ allowed region and the rest of the light gray shaded region is the $1\sigma$ region, which is presently 
disfavoured by the joint Planck+WMAP constraints. The rest of the region 
is completely ruled out by the present observational constraints. From Fig.~(\ref{fig13a}) it is clearly 
observed that the observational constraints on the amplitude of the scalar mode
fluctuations satisfy within the window $55<g<93$ considering $1/2<q<2$.
For $g>93$ the 
corresponding amplitude falls down in a non-trivial fashion by following the exact functional form as stated in Eq~(\ref{e-fold9d}). Next using
the behavior as
shown in Fig.~(\ref{fig13b}), the lower bound on the stringy parameter $g$ is constrained as, $g>73$ by using the non-trivial
relationship as stated in Eq~(\ref{e-fold10d}). But at this lower bound of the parameter $g$ 
the amplitude of the scalar power spectrum is slightly outside to the present observational constraints. This implies that, to satisfy
both of the constraint from 
the amplitude of the scalar power spectrum and its spectral tilt within $2\sigma$ CL the
constrained numerical value of the stringy parameter is
lying within the window $73<g<93$. Also in Fig.~(\ref{fig13c}) for $q=1/2$, $q=1$, $q=3/2$ and $q=2$ we
fix $N_{\star}/g\sim 0.85$, which further implies that for $50<N_{\star}<70$,
the prescribed window for $g$ from $\Delta_{\zeta}-n_{\zeta}$ plot is given by, $59<g<82.3$. If we additionally impose the constraint from the upper bound on tensor-to-scalar ratio then also the allowed parameter 
range is lying within the window i.e. $73<g<82.3$.

\begin{figure}[htb]
	\centering
	\includegraphics[width=12cm,height=8cm]{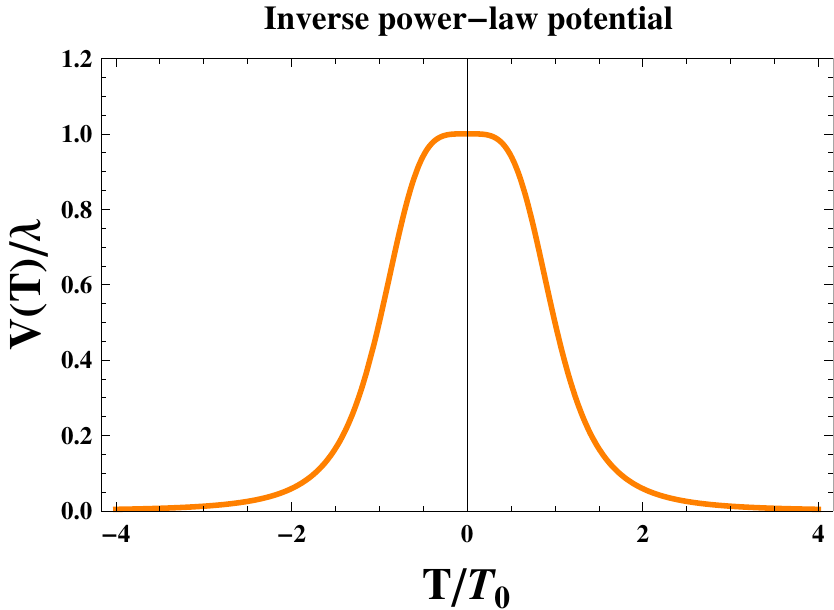}
	\caption{Variation of the Inverse power-law potential $V(T)/\lambda$ with field $T/T_{0}$ in dimensionless units.
	}
	\label{fig15}
\end{figure}

In Fig.~(\ref{fig14a}) and Fig.~(\ref{fig14b}), we have shown the behaviour
of the running of the scalar spectral tilt $\alpha_{\zeta}$ and running of the 
running of the scalar spectral tilt $\kappa_{\zeta}$ with respect to the scalar spectral index $n_{\zeta}$ for Exponential potential-Type II with $g=76$ respectively.
For both the figures 
 \textcolor{red}{red},~\textcolor{green}{green},
~\textcolor{brown}{brown},~\textcolor{orange}{orange} colored curve represent $q=1/2$, $q=1$, $q=3/2$ and $q=2$ respectively.
The \textcolor{cyan}{cyan} color shaded region bounded by two vertical black coloured lines in both the plots represent the Planck $2\sigma$ allowed region and the rest of the light gray shaded region is
showing the $1\sigma$ allowed range, which is at present 
disfavoured by the Planck data and Planck+ BICEP2+Keck Array joint constraint. From both of these figures, it 
is also observed that, within $50<N_{\star}<70$ the Exponential potential-Type II is favoured for the characteristic index $1/2<q<2$, by Planck
2015 data and Planck+ BICEP2+Keck Array joint analysis. From Fig.~(\ref{fig14a}) and Fig.~(\ref{fig14b}), it is observed that within the $2\sigma$ 
observed range of the scalar spectral tilt $n_{\zeta}$, as the value of the characteristic parameter $q$ increases,
the value of the running $\alpha_{\zeta}$ decreases
and running of the running $\kappa_{\zeta}$ increases for Exponential potential-Type II. It is also important to note that for $1/2<q<2$, the numerical value 
of the running $\alpha_{\zeta}\sim {\cal O}(10^{-4})$
and running of the running $\kappa_{\zeta}\sim {\cal O}(-10^{-6})$, which are perfectly 
consistent with the $1.5\sigma$ constraints on running and running of the running as obtained from Planck 2015 data.\\ \\ 
\underline{\bf Model V: Inverse power-law potential}\\
For single field case the first model of tachyonic potential is given by:
\be\label{weeeeeee1}
V(T)=\frac{\lambda}{\left[1+\left(\frac{T}{T_{0}}\right)^4\right]},\ee
where $\lambda$ characterize the scale of inflation and $T_{0}$ is the parameter of the model.
In Fig.~(\ref{fig15}) we have depicted the symmetric behavior of the Inverse power-law potential with respect to scaled field coordinate $T/T_{0}$ in dimensionless units.
In this case the tachyon field started rolling down from the top hight of the potential situated at the origin and take part in inflationary dynamics.

Next using specified form of the potential the potential dependent slow-roll parameters are computed as:
\bea \bar{\epsilon}_{V}&=&\frac{1}{2g}\frac{16\left(\frac{T}{T_{0}}\right)^6}{\left[1+\left(\frac{T}{T_{0}}\right)^4\right]},~~~~~~~~~~~~~~~~~~~~~~~~~~~~~~~~~~~~~~~~~~~~~~~~~~~~~~\\
 \bar{\eta}_{V}&=&\frac{1}{g}\frac{4\left[5\left(\frac{T}{T_{0}}\right)^4-3\right]}{\left[1+\left(\frac{T}{T_{0}}\right)^4\right]},~~~~~~~~~~~~~~~~~~~~~~\eea
\bea \bar{\xi}^2_{V}&=&\frac{1}{g^2}\frac{96\left(\frac{T}{T_{0}}\right)^2\left[5\left(\frac{T}{T_{0}}\right)^8-10\left(\frac{T}{T_{0}}\right)^4 +1\right]}{\left[1+\left(\frac{T}{T_{0}}\right)^4\right]^2},\\
 \bar{\sigma}^3_{V}&=&\frac{1}{g^3}\frac{384\left(\frac{T}{T_{0}}\right)^6\left[5\left(\frac{T}{T_{0}}\right)^4\left(7\left(\frac{T}{T_{0}}\right)^8-31\left(\frac{T}{T_{0}}\right)^4 +13\right)-1\right]}
{\left[1+\left(\frac{T}{T_{0}}\right)^4\right]^3},~~~~~~~~~~~~~~
\eea
where the factor $g$ is defined as:
\be g= \frac{\alpha^{'}\lambda T^2_0}{M^2_p}=\frac{M^4_s}{(2\pi)^3 g_s}\frac{\alpha^{'} T^2_0}{M^2_p}.\ee
Next we compute the number of e-foldings from this model:
\be\begin{array}{lll}\label{e-fold5e}
 \displaystyle  N(T)=\left\{\begin{array}{lll}
                    \displaystyle  
                   \frac{g}{8}~\left[\frac{1}{\left(\frac{T}{T_{0}}\right)^2}-\frac{1}{\left(\frac{T_{end}}{T_{0}}\right)^2}\right]\,,~~~~~~ &
 \mbox{\small {\bf for {$q=1/2$ }}}  \\ 
 \displaystyle   
             \sqrt{2q}~\frac{g}{8}~\left[\frac{1}{\left(\frac{T}{T_{0}}\right)^2}-\frac{1}{\left(\frac{T_{end}}{T_{0}}\right)^2}\right]\,.~~~~~~ &
 \mbox{\small {\bf for {~any~arbitrary~ $q$ }}} 
          \end{array}
\right.
\end{array}\ee
Further using the condition to end inflation:
\bea \bar{\epsilon}_{V}(T_{end})=1,\eea
      we get the following field values at the end of inflation:
      \be T_{end}\sim \sqrt{\frac{g}{8}}~T_{0}.\ee
Next using $N=N_{cmb}=N_{\star}$ and $T=T_{cmb}=T_{\star}$ at the horizon crossing we get,
\be\begin{array}{lll}\label{e-fold6e}
 \displaystyle  T_{\star}=T_{0}\times\left\{\begin{array}{lll}
                    \displaystyle  
                    \frac{1}{\sqrt{\frac{8}{g}\left(N_{\star}+1\right)}}\,,~~~~~~ &
 \mbox{\small {\bf for {$q=1/2$ }}}  \\ 
 \displaystyle   
             \frac{1}{\sqrt{\frac{8}{g}\left(\frac{N_{\star}}{\sqrt{2q}}+1\right)}}\,.~~~~~~ &
 \mbox{\small {\bf for {~any~arbitrary~ $q$ }}} 
          \end{array}
\right.
\end{array}\ee
Also the field excursion can be computed as:
\be\begin{array}{lll}\label{e-fold6eex}
 \displaystyle  |\Delta T|=T_{0}\times\left\{\begin{array}{lll}
                    \displaystyle  
                    \left|\frac{1}{\sqrt{\frac{8}{g}\left(N_{\star}+1\right)}}-\sqrt{\frac{g}{8}}\right|\,,~~~~~~ &
 \mbox{\small {\bf for {$q=1/2$ }}}  \\ 
 \displaystyle   
             \left|\frac{1}{\sqrt{\frac{8}{g}\left(\frac{N_{\star}}{\sqrt{2q}}+1\right)}}-\sqrt{\frac{g}{8}}\right|\,.~~~~~~ &
 \mbox{\small {\bf for {~any~arbitrary~ $q$ }}} 
          \end{array}
\right.
\end{array}\ee

Finally we compute the following inflationary observables:
\be\begin{array}{lll}\label{e-fold9e}
 \displaystyle   \Delta_{\zeta,\star}\approx\frac{g\lambda}{12\pi^2M^{4}_{p}}\times\left\{\begin{array}{lll}
                    \displaystyle  
                   \frac{1}{16\left(\frac{8}{g}\left(N_{\star}+1\right)\right)^3}\,,~~~~~~ &
 \mbox{\small {\bf for {$q=1/2$ }}}  \\ 
 \displaystyle   
             \frac{2q}{16\left(\frac{8}{g}\left(\frac{N_{\star}}{\sqrt{2q}}+1\right)\right)^3}\,.~~~~~~ &
 \mbox{\small {\bf for {~any~arbitrary~ $q$ }}} 
          \end{array}
\right.
\end{array}\ee
\be\begin{array}{lll}\label{e-fold10e}
 \displaystyle   n_{\zeta,\star}-1\approx\left\{\begin{array}{lll}
                    \displaystyle  
                   -\frac{3}{\left(N_{\star}+1\right)}\,,~~~~~~ &
 \mbox{\small {\bf for {$q=1/2$ }}}  \\ 
 \displaystyle   
            -\frac{3}{\left(\frac{N_{\star}}{\sqrt{2q}}+1\right)}\,.~~~~~~ &
 \mbox{\small {\bf for {~any~arbitrary~ $q$ }}} 
          \end{array}
\right.
\end{array}\ee
\begin{figure*}[htb]
\centering
\subfigure[$r$ vs $n_{\zeta}$.]{
    \includegraphics[width=7.2cm,height=7.7cm] {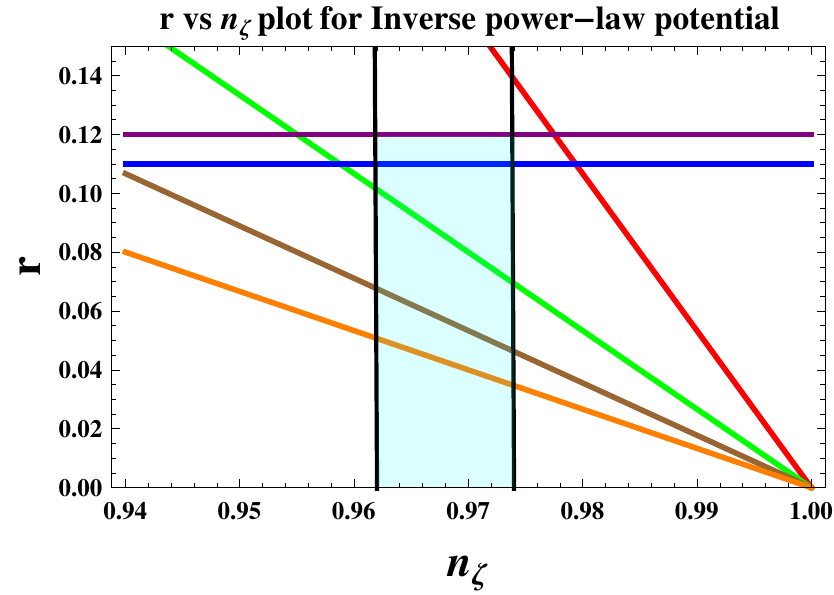}
    \label{fig16a}
}
\subfigure[$\Delta_{\zeta}$ vs $n_{\zeta}$.]{
    \includegraphics[width=7.2cm,height=7.7cm] {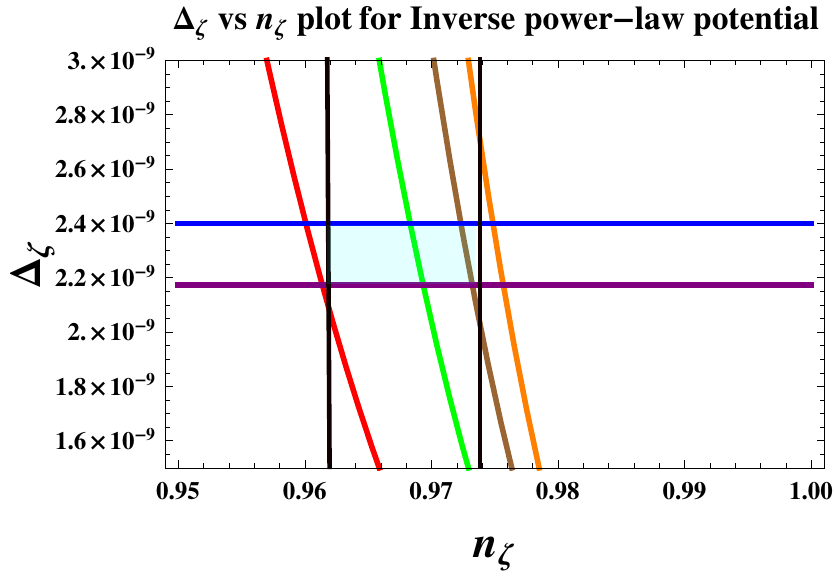}
    \label{fig16b}
}
\caption[Optional caption for list of figures]{In \ref{fig16a} the behaviour of the tensor-to-scalar ratio $r$ with respect to the scalar spectral index $n_{\zeta}$
	and \ref{fig16b} the parameter $g$ for Inverse power-law potential.
	The purple and blue coloured line represent the upper bound of tensor-to-scalar ratio allowed by Planck+ BICEP2+Keck Array joint constraint and only Planck 2015 data respectively. For both the figures 
	\textcolor{red}{red},~\textcolor{green}{green},
	~\textcolor{brown}{brown},~\textcolor{orange}{orange} colored curve represent $q=1/2$, $q=1$, $q=3/2$ and $q=2$ respectively.
	The \textcolor{cyan}{cyan} color shaded region bounded by two vertical black coloured lines in \ref{fig16a} represent the Planck $2\sigma$ allowed region and the rest of the light gray shaded region is
	showing the $1\sigma$ allowed range, which is at present 
	disfavoured by the Planck data and Planck+ BICEP2+Keck Array joint constraint. 
	In \ref{fig16a}, we have explicitly shown that in $r-n_{\zeta}$ plane the observationally disfavoured value of the characteristic index is $q\geq 1/2$.  Variation of the scalar power spectrum $\Delta_{\zeta}$
	vs scalar spectral index $n_{\zeta}$ is shown in \ref{fig16b}. The purple and blue coloured line represent the upper and lower bound allowed by WMAP+Planck 2015 data respectively. 
	The green dotted region bounded by two vertical black coloured lines represent the Planck $2\sigma$ allowed region and the rest of the light gray shaded region is disfavoured by the Planck+WMAP constraint.
	In \ref{fig16b}, we have explicitly shown that in $\Delta_{\zeta}-n_{\zeta}$ plane the observationally favoured value of the characteristic index is $q=1$ and $q=3/2$.
	From \ref{fig16a} and \ref{fig16b}, it 
	is also observed that, within $50<N_{\star}<70$ the Inverse power-law potential is favoured for the characteristic index $3/2<q<1$, by Planck 2015 data and Planck+ BICEP2+Keck Array joint analysis.} 
\label{fig16}
\end{figure*}
\begin{figure*}[htb]
\centering
\subfigure[$\alpha_{\zeta}$ vs $n_{\zeta}$.]{
    \includegraphics[width=12.2cm,height=7.3cm] {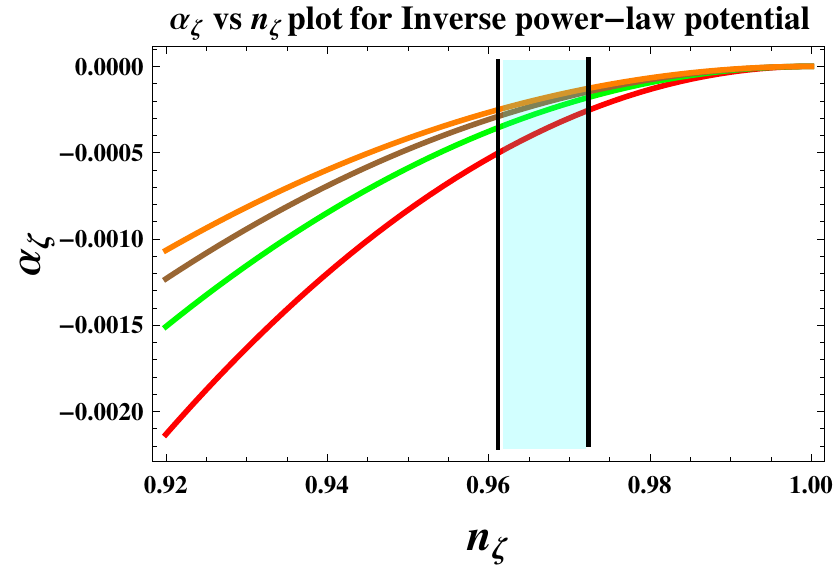}
    \label{fig17a}
}
\subfigure[$\kappa_{\zeta}$ vs $n_{\zeta}$.]{
    \includegraphics[width=12.2cm,height=7.3cm] {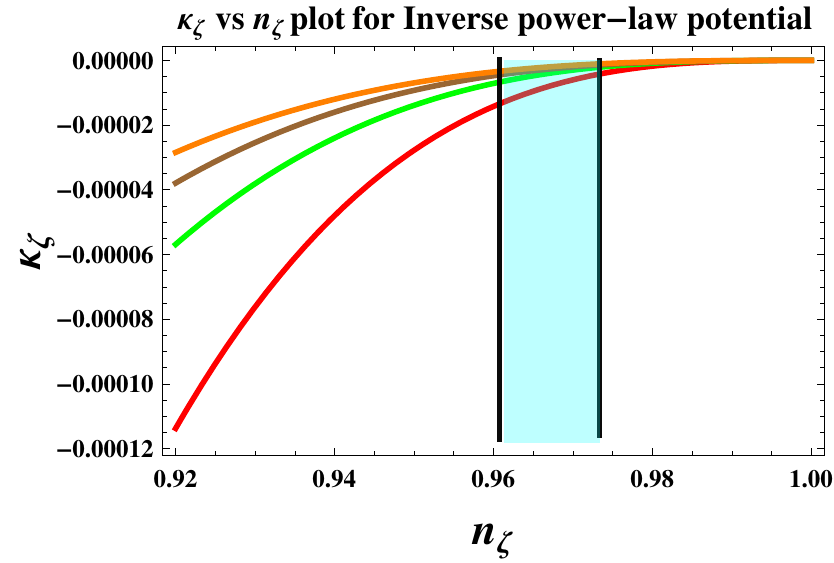}
    \label{fig17b}
}
\caption[Optional caption for list of figures]{Behaviour of the \ref{fig17a} running of the scalar spectral tilt $\alpha_{\zeta}$ and \ref{fig17b} running of the 
running of the scalar spectral tilt $\kappa_{\zeta}$ with respect to the scalar spectral index $n_{\zeta}$ for Inverse cosh potential with $g=88$.
For both the figures 
 \textcolor{red}{red},~\textcolor{green}{green},
~\textcolor{brown}{brown},~\textcolor{orange}{orange} colored curve represent $q=1/2$, $q=1$, $q=3/2$ and $q=2$ respectively.
The \textcolor{cyan}{cyan} color shaded region bounded by two vertical black coloured lines in \ref{fig17a} and \ref{fig17a} represent the Planck $2\sigma$ allowed region and the rest of the light gray shaded region is
showing the $1\sigma$ allowed range, which is at present 
disfavoured by the Planck data and Planck+ BICEP2+Keck Array joint constraint. From \ref{fig17a} and \ref{fig17b}, it 
is also observed that, within $50<N_{\star}<70$ the Inverse cosh potential is favoured for the characteristic index $1/2<q<2$, by Planck 2015 data and Planck+ BICEP2+Keck Array joint analysis.} 
\label{fig17}
\end{figure*}
\be\begin{array}{lll}\label{e-fold11e}
 \displaystyle   \alpha_{\zeta,\star}\approx\left\{\begin{array}{lll}
                    \displaystyle  
                   -\frac{3}{\left(N_{\star}+1\right)^2}\,,~~~~~~ &
 \mbox{\small {\bf for {$q=1/2$ }}}  \\ 
 \displaystyle   
            -\frac{3}{\sqrt{2q}\left(\frac{N_{\star}}{\sqrt{2q}}+1\right)^2}\,.~~~~~~ &
 \mbox{\small {\bf for {~any~arbitrary~ $q$ }}} 
          \end{array}
\right.
\end{array}\ee
\be\begin{array}{lll}\label{e-fold12e}
 \displaystyle   \kappa_{\zeta,\star}\approx\left\{\begin{array}{lll}
                    \displaystyle  
                   -\frac{6}{\left(N_{\star}+1\right)^3}\,,~~~~~~ &
 \mbox{\small {\bf for {$q=1/2$ }}}  \\ 
 \displaystyle   
              -\frac{6}{2q\left(\frac{N_{\star}}{\sqrt{2q}}+1\right)^3}\,.~~~~~~ &
 \mbox{\small {\bf for {~any~arbitrary~ $q$ }}} 
          \end{array}
\right.
\end{array}\ee
\be\begin{array}{lll}\label{e-fold13e}
 \displaystyle   r_{\star}\approx\left\{\begin{array}{lll}
                    \displaystyle  
                   \frac{16}{\left(N_{\star}+1\right)}\,,~~~~~~ &
 \mbox{\small {\bf for {$q=1/2$ }}}  \\ 
 \displaystyle   
             \frac{16}{2q\left(\frac{N_{\star}}{\sqrt{2q}}+1\right)}\,.~~~~~~ &
 \mbox{\small {\bf for {~any~arbitrary~ $q$ }}} 
          \end{array}
\right.
\end{array}\ee
For inverse cosh potential we get the following consistency relations:
\be\begin{array}{lll}\label{e-fold14e}
 \displaystyle   r_{\star}\approx \frac{16}{3}(1-n_{\zeta,\star})\times\left\{\begin{array}{lll}
                    \displaystyle  
                   1\,,~~~~~~ &
 \mbox{\small {\bf for {$q=1/2$ }}}  \\ 
 \displaystyle   
             \frac{1}{2q}\,.~~~~~~ &
 \mbox{\small {\bf for {~any~arbitrary~ $q$ }}} 
          \end{array}
\right.
\end{array}\ee
\be\begin{array}{lll}\label{e-fold15e}
 \displaystyle   \Delta_{\zeta,\star}\approx\frac{g^4 \lambda(1-n_{\zeta,\star})^3}{12\pi^2M^{4}_{p}}\times\left\{\begin{array}{lll}
                    \displaystyle  
                   \frac{1}{221184}\,,~~~~~~ &
 \mbox{\small {\bf for {$q=1/2$ }}}  \\ 
 \displaystyle   
             \frac{2q}{221184}\,.~~~~~~ &
 \mbox{\small {\bf for {~any~arbitrary~ $q$ }}} 
          \end{array}
\right.
\end{array}\ee
\be\begin{array}{lll}\label{e-fold16e}
 \displaystyle   \Delta_{\zeta,\star}\approx\frac{g^4 \lambda r^3_{\star}}{12\pi^2M^{4}_{p}}\times\left\{\begin{array}{lll}
                    \displaystyle  
                  \frac{1}{33554432}\,,~~~~~~ &
 \mbox{\small {\bf for {$q=1/2$ }}}  \\ 
 \displaystyle   
              \frac{(2q)^3}{33554432}\,.~~~~~~ &
 \mbox{\small {\bf for {~any~arbitrary~ $q$ }}} 
          \end{array}
\right.
\end{array}\ee
\be\begin{array}{lll}\label{e-fold17e}
 \displaystyle   \alpha_{\zeta,\star}\approx-\frac{1}{3}\left(n_{\zeta,\star}-1\right)^2 \times\left\{\begin{array}{lll}
                    \displaystyle  
                   1\,,~~~~~~ &
 \mbox{\small {\bf for {$q=1/2$ }}}  \\ 
 \displaystyle   
            \frac{1}{\sqrt{2q}}\,.~~~~~~ &
 \mbox{\small {\bf for {~any~arbitrary~ $q$ }}} 
          \end{array}
\right.
\end{array}\ee
\be\begin{array}{lll}\label{e-fold18e}
 \displaystyle   \kappa_{\zeta,\star}\approx\frac{2}{9}\left(n_{\zeta,\star}-1\right)^3 \times\left\{\begin{array}{lll}
                    \displaystyle  
                   1\,,~~~~~~ &
 \mbox{\small {\bf for {$q=1/2$ }}}  \\ 
 \displaystyle   
              \frac{1}{2q}\,.~~~~~~ &
 \mbox{\small {\bf for {~any~arbitrary~ $q$ }}} 
          \end{array}
\right.
\end{array}\ee
Let us now discuss the general constraints on the parameters of tachyonic string theory including the factor $q$ 
and on the parameters appearing in the 
expression for Inverse power-law potential. In Fig.~(\ref{fig16a}), we have shown the behavior of the tensor-to-scalar ratio $r$ with respect to the scalar spectral index $n_{\zeta}$
 for Inverse power-law potential respectively. In both the figures the \textcolor{purple}{purple} and \textcolor{blue}{blue}
coloured line represent the upper bound of tensor-to-scalar ratio allowed by Planck+ BICEP2+Keck Array
joint constraint and only Planck 2015 data respectively. For both the figures 
 \textcolor{red}{red},~\textcolor{green}{green},
~\textcolor{brown}{brown},~\textcolor{orange}{orange} colored curve represent $q=1/2$, $q=1$, $q=3/2$ and $q=2$ respectively.
The \textcolor{cyan}{cyan} color shaded region bounded by two vertical black coloured lines in Fig.~(\ref{fig16a}) represent the
Planck $2\sigma$ allowed region and the rest of the light gray shaded region
is showing the $1\sigma$ allowed range, which is at present 
disfavoured by the Planck 2015 data and Planck+ BICEP2+Keck Array joint constraint. The rest of the region 
is completely ruled out by the present observational constraints. From Fig.~(\ref{fig16a}), it 
is also observed that, within $50<N_{\star}<70$, the Inverse power-law potential is favoured for the characteristic
index $1<q<2$, by Planck 2015 data and Planck+ BICEP2+Keck Array joint analysis. Also in Fig.~(\ref{fig16a}) for $q=1/2$, $q=1$, $q=3/2$ and $q=2$ we fix $N_{\star}\sim 70$. One can draw the similar characteristic curves for 
any value of the e-foldings lying within the window $50<N_{\star}<70$ also. 
To analyze the results more clearly let us describe the cosmological features from Fig.~(\ref{fig16a}) in detail.
Let us first start with the $q=1/2$ situation, in which the $2\sigma$ constraint on the scalar spectral tilt is disfavoured
 for $50<N_{\star}<70$. Next for $q=1$ case, the same constraint 
is satisfied within the window of tensor-to-scalar ratio, $ 0.07<r<0.10$ for $50<N_{\star}<70$. Further for $q=3/2$ case, the same constraint 
on scalar spectral tilt is satisfied within the window of tensor-to-scalar ratio, $ 0.05<r<0.07$ for $50<N_{\star}<70$.
Finally, for $q=2$ situation, the value for the tensor-to-scalar ratio is $0.035<r<0.05$.

In Fig.~(\ref{fig16b}), we have depicted the behavior of the 
scalar power spectrum $\Delta_{\zeta}$ vs 
scalar spectral tilt $n_{\zeta}$ for Inverse power-law potential respectively. It is important to note that,
for all of the figures 
 \textcolor{red}{red},~\textcolor{green}{green},
~\textcolor{brown}{brown},~\textcolor{orange}{orange} colored curve represent $q=1/2$, $q=1$, $q=3/2$ and $q=2$ respectively.
The \textcolor{purple}{purple} and \textcolor{blue}{blue} coloured line represent the upper and lower bound allowed by WMAP+Planck 2015 data respectively. 
The \textcolor{cyan}{cyan} color shaded region bounded by two vertical black coloured lines represent the Planck
$2\sigma$ allowed region and the rest of the light gray shaded region is the $1\sigma$ region, which is presently 
disfavoured by the joint Planck+WMAP constraints. The rest of the region 
is completely ruled out by the present observational constraints. Also in Fig.~(\ref{fig16b}) for $q=1/2$, $q=1$, $q=3/2$ and $q=2$
we fix $N_{\star}\sim 70$. But plots can be reproduced for $50<N_{\star} <70$ also considering the present observational constraints from Planck 2015 data.
It is clearly observed from Fig.~(\ref{fig16b}) that the Planck 2015 constraint on amplitude of the scalar power spectrum $\Delta_{\zeta}$ and the scalar spectral tilt $n_{\zeta}$ 
disfavour $q=1/2$ and $q=2$ values for Inverse power-law potential.

In Fig.~(\ref{fig17a}) and Fig.~(\ref{fig17b}), we have shown the behaviour
of the running of the scalar spectral tilt $\alpha_{\zeta}$ and running of the 
running of the scalar spectral tilt $\kappa_{\zeta}$ with respect to the scalar spectral index $n_{\zeta}$ for Inverse power-law potential respectively.
For both the figures 
 \textcolor{red}{red},~\textcolor{green}{green},
~\textcolor{brown}{brown},~\textcolor{orange}{orange} colored curve represent $q=1/2$, $q=1$, $q=3/2$ and $q=2$ respectively.
The \textcolor{cyan}{cyan} color shaded region bounded by two vertical black coloured lines in both the plots represent the Planck $2\sigma$ allowed region and the rest of the light gray shaded region is
showing the $1\sigma$ allowed range, which is at present 
disfavoured by the Planck data and Planck+ BICEP2+Keck Array joint constraint. From both of these figures, it 
is also observed that, within $50<N_{\star}<70$ the Inverse power-law potential is favoured for the characteristic index $1/2<q<2$, by Planck
2015 data and Planck+ BICEP2+Keck Array joint analysis. If we additionally impose the constraints from $r-n_{\zeta}$ and $\Delta_{\zeta}-n_{\zeta}$ plane then the stringent window on the characteristic index 
of GTachyon is lying within $1<q<3/2$. It is also important to note that for any values of $q$, the numerical value 
of the running $\alpha_{\zeta}\sim {\cal O}(-10^{-4})$
and running of the running $\kappa_{\zeta}\sim {\cal O}(-10^{-5})$, which are perfectly 
consistent with the $1.5\sigma$ constraints on running and running of the running as obtained from Planck 2015 data. For $q=1/2$, $q=1$, $q=3/2$ and $q=2$ $g$ is not explicitly appearing in the various inflationary observables except the amplitude of scalar power spectrum in 
this case. To produce the correct value of the amplitude of the scalar power spectra we fix the parameter $600<g<700$.

\subsubsection{Analyzing CMB power spectrum}

In this section we explicitly study the cosmological consequences from the CMB TT, TE, EE, BB-angular power spectrum computed from all the proposed models of inflation
~\footnote{In this work we have not consider possibility of other cross correlators i.e. TB, EB as 
there is no observational evidence of such contributions in the CMB map. Also till date there is no observational evidence for 
inflationary origin of BB angular power spectrum except from CMB lensing \cite{Hanson:2013hsb}. However, for the sake of completeness 
in this paper we show the theoretical BB angular power 
spectra from the various models of tachyonic potential. }. 
The angular power spectra or equivalently the two point correlator of $X$ and $Y$ fields are defined as \cite{Baumann:2009ds}:
\be
C_\ell^{XY} \equiv \frac{1}{2\ell + 1} \sum^{l}_{m=-l} \langle a_{X,\ell m}^* a_{Y,\ell m} \rangle \, , \qquad X, Y = T, E, B\, .
\ee
Here $l$ characterizes the CMB multipole and $m$ signifies the magnetic quantum number which runs from $m=-l,\cdots, +l$. In general,
the field $X(\hat{n})$ and $Y(\hat{n})$ can be expressed in terms of the following harmonic expansion:
\be\begin{array}{lll}\label{e-fold15}
 \displaystyle   X(\hat{n}), Y(\hat{n})=\left\{\begin{array}{lll}
                    \displaystyle  
                   \sum^{\infty}_{l=0}\sum^{+l}_{m=-l}a_{T,lm}V_{lm}(\hat{n})\,,~~~~~~ &
 \mbox{\small {\bf for {$X=T$ }}}  \\ 
 \displaystyle   
             \sum^{\infty}_{l=0}\sum^{+l}_{m=-l}a_{E,lm}V_{lm}(\hat{n})\,,~~~~~~ &
 \mbox{\small {\bf for {$X=E$ }}} 
 \\ 
 \displaystyle   
             \sum^{\infty}_{l=0}\sum^{+l}_{m=-l}a_{B,lm}V_{lm}(\hat{n})\,,~~~~~~ &
 \mbox{\small {\bf for {$X=B$ }}} 
          \end{array}
\right.
\end{array}\ee
where $\hat{n}$ is the arbitrary directional unit vector in CMB map. 
Here $V_{lm}(\hat{n})$ are the spherical harmonics which are chosen to be the basis of the harmonic expansion of the temperature anisotropy and $E$ and $B$ polarization.
Further using the inflationary power spectra at any momentum scale $k$: \be \Delta(k) \equiv \{ \Delta_\zeta (k), \Delta_h (k) \}\ee  the angular power spectra for CMB temperature fluctuations and polarization can 
be expressed as \cite{Baumann:2009ds}:
\be
\label{equ:CXY}
\displaystyle
C_\ell^{XY} = \frac{2}{\pi} \int k^2 dk \underbrace{\Delta(k)}_{\rm Inflation}\,  \underbrace{\Theta_{X \ell}(k) \Theta_{Y \ell}(k)}_{\rm Anisotropies} \, ,
\ee
where the anisotropic integral kernel can be written as \cite{Baumann:2009ds}:
\be
\label{equ:source1}
\Theta_{X \ell}(k) = \int_0^{\eta_0} d\eta\, \underbrace{S_X(k, \eta)}_{\rm Sources}\, \underbrace{P_{X\ell}(k[\eta_0-\eta])}_{\rm Projection}\, .
\ee

\begin{figure}[t]
\centering
\subfigure[$l(l+1)C^{TT}_{l}/2\pi$~vs~$l$~(scalar)]{
    \includegraphics[width=7.2cm, height=5.7cm] {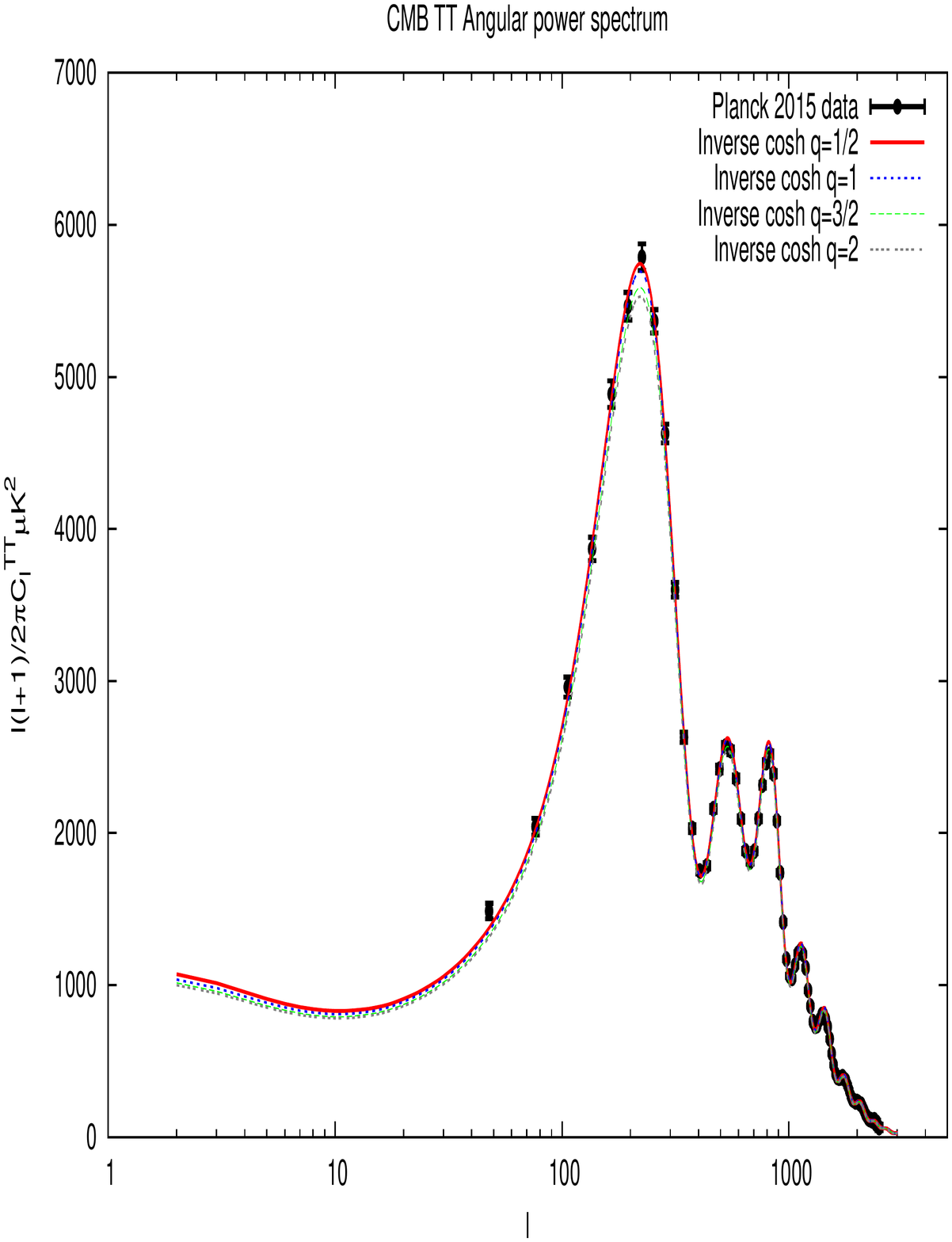}
    \label{TTm1}
}
\subfigure[$l(l+1)C^{TT}_{l}/2\pi$~vs~$l$~(scalar)]{
    \includegraphics[width=7.2cm, height=5.7cm] {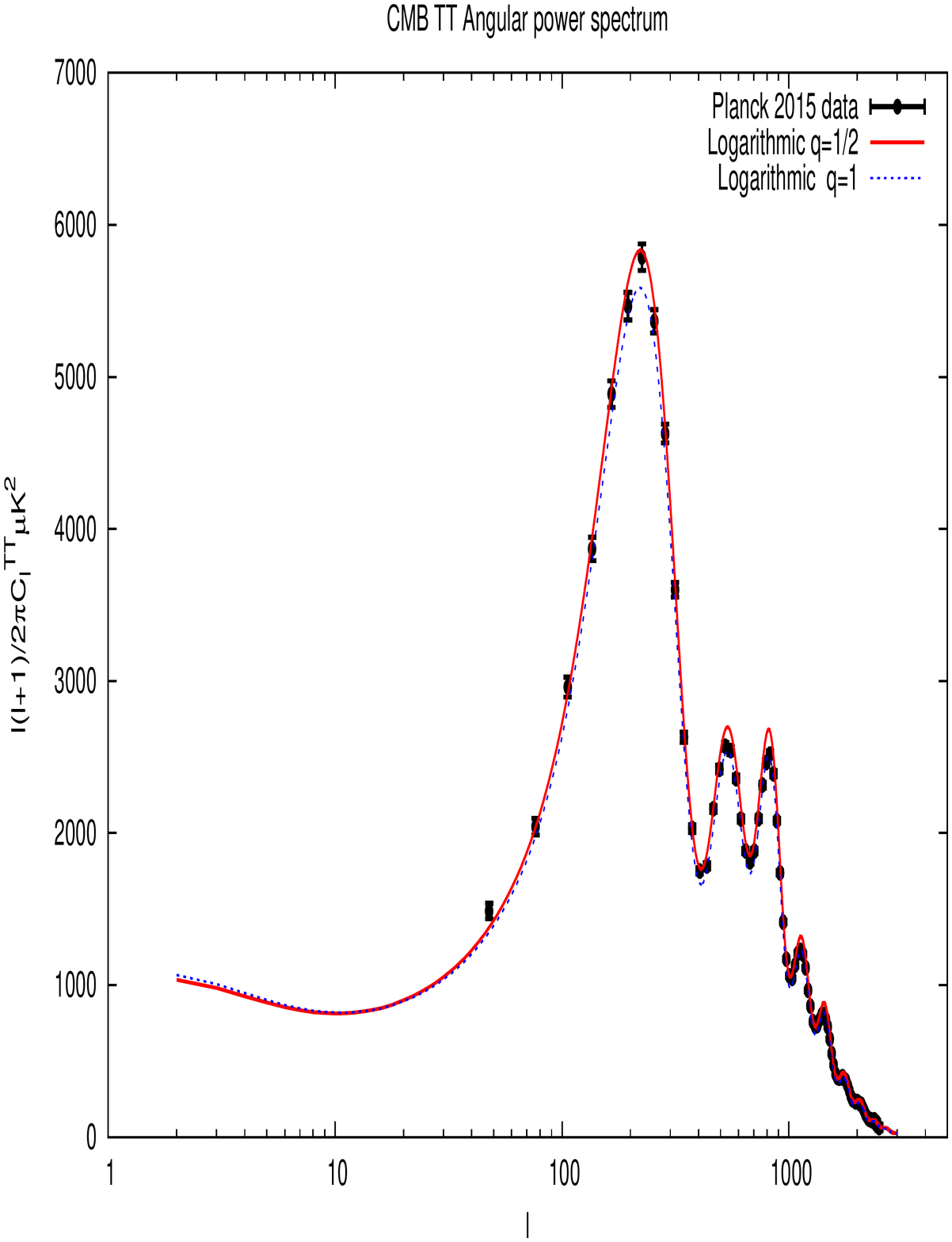}
    \label{TTm2}
}
\subfigure[$l(l+1)C^{TT}_{l}/2\pi$~vs~$l$~(scalar)]{
    \includegraphics[width=7.2cm, height=5.7cm] {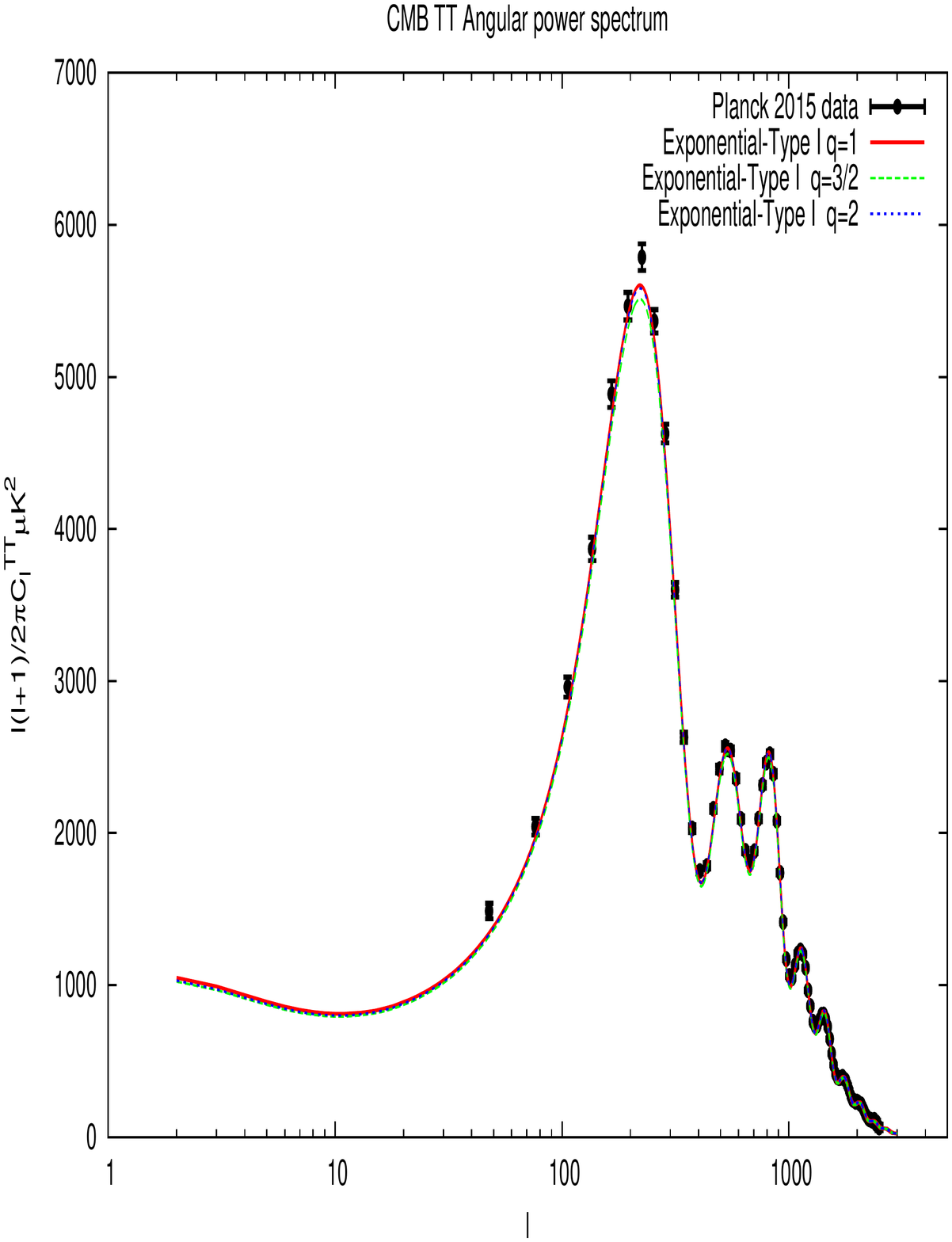}
    \label{TTm3}
}
\subfigure[$l(l+1)C^{TT}_{l}/2\pi$~vs~$l$~(scalar)]{
    \includegraphics[width=7.2cm, height=5.7cm] {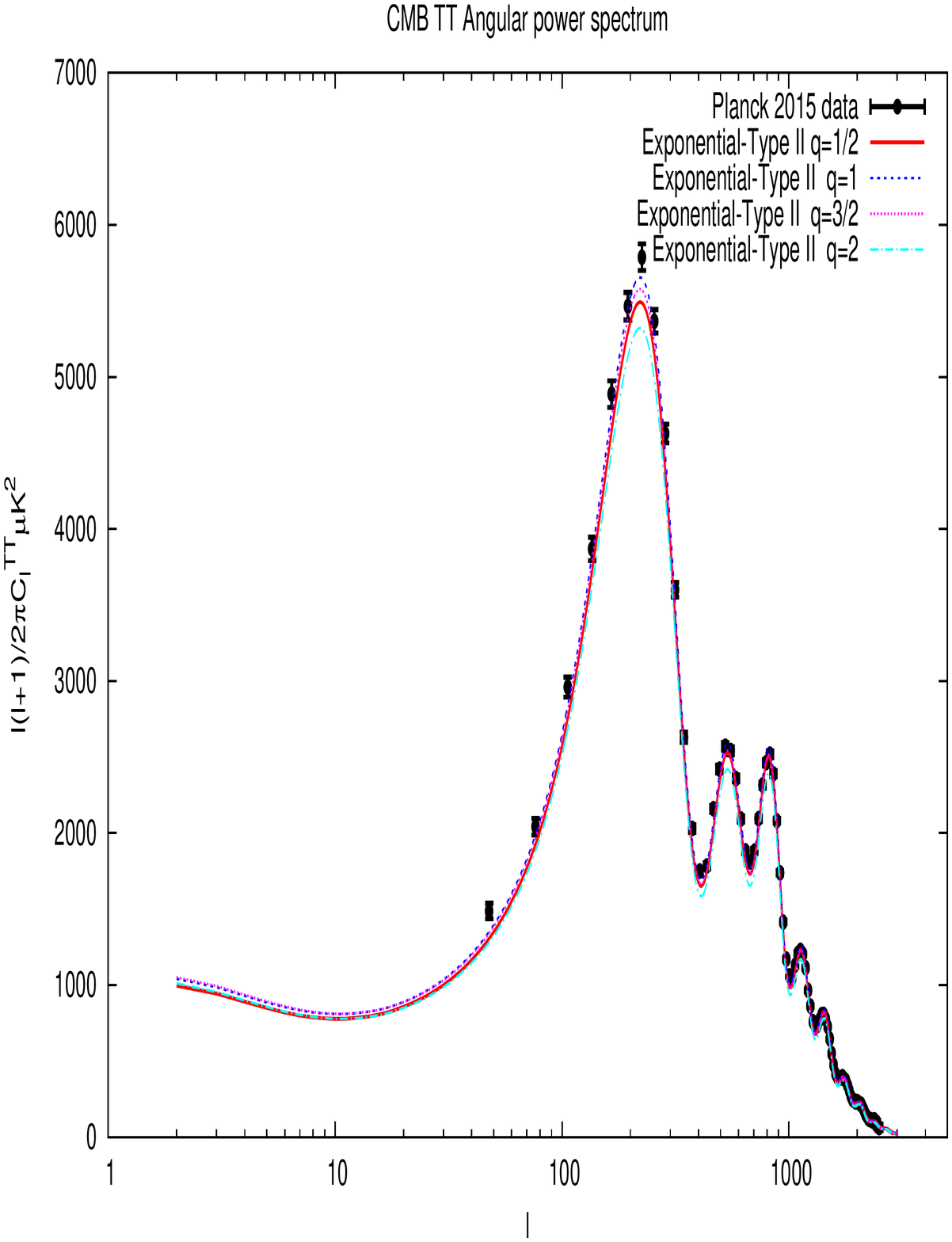}
    \label{TTm4}
}
\subfigure[$l(l+1)C^{TT}_{l}/2\pi$~vs~$l$~(scalar)]{
    \includegraphics[width=7.2cm, height=5.7cm] {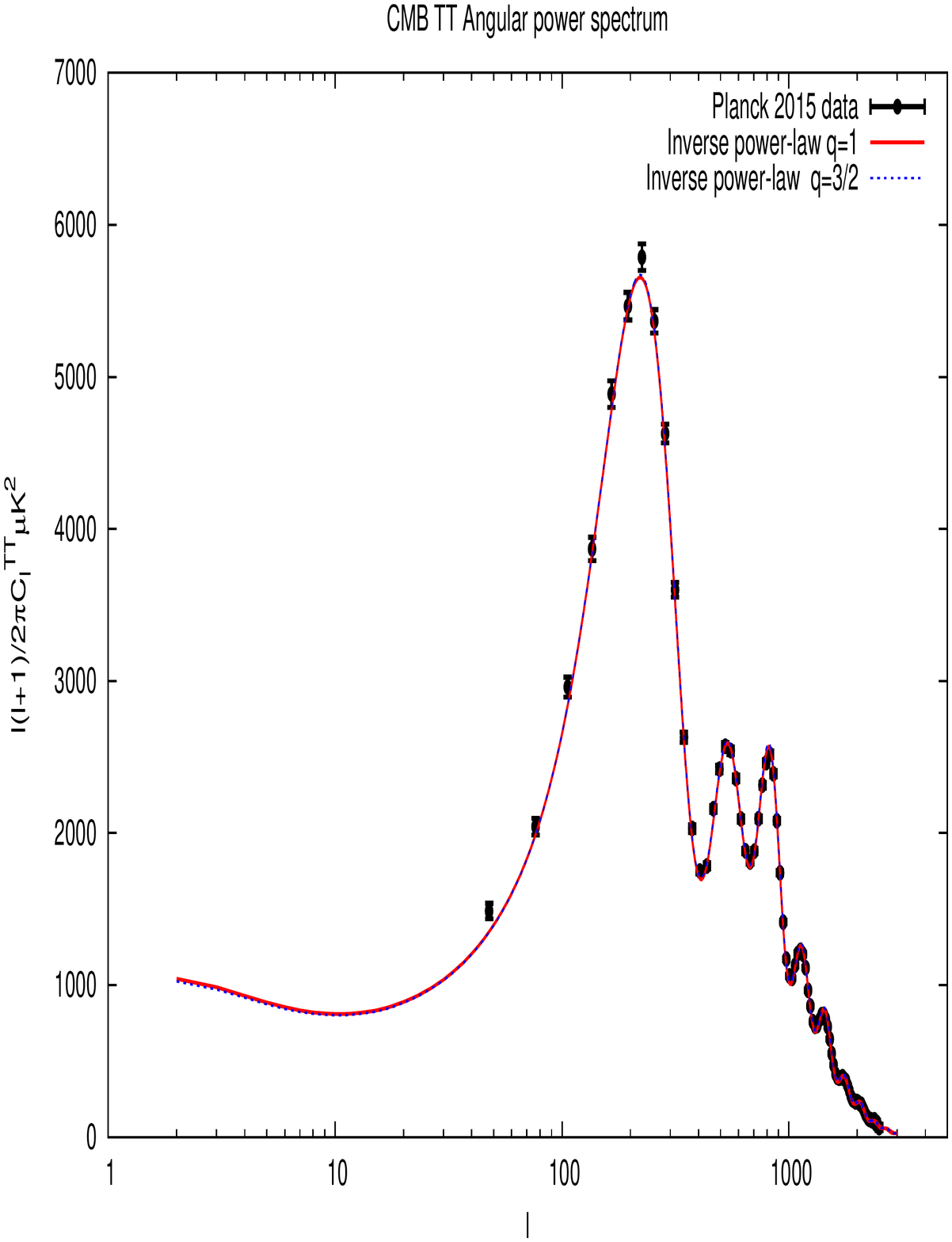}
    \label{TTm5}
}
\caption[Optional caption for list of figures]{We show the variation of CMB TT Angular power spectrum with respect to the multipole, $l$ for scalar modes for all five tachyonic models.
}
\label{fig18}
\end{figure}

\begin{figure}[t]
\centering
\subfigure[$l(l+1)C^{TE}_{l}/2\pi$~vs~$l$~(scalar)]{
    \includegraphics[width=7.2cm, height=5.7cm] {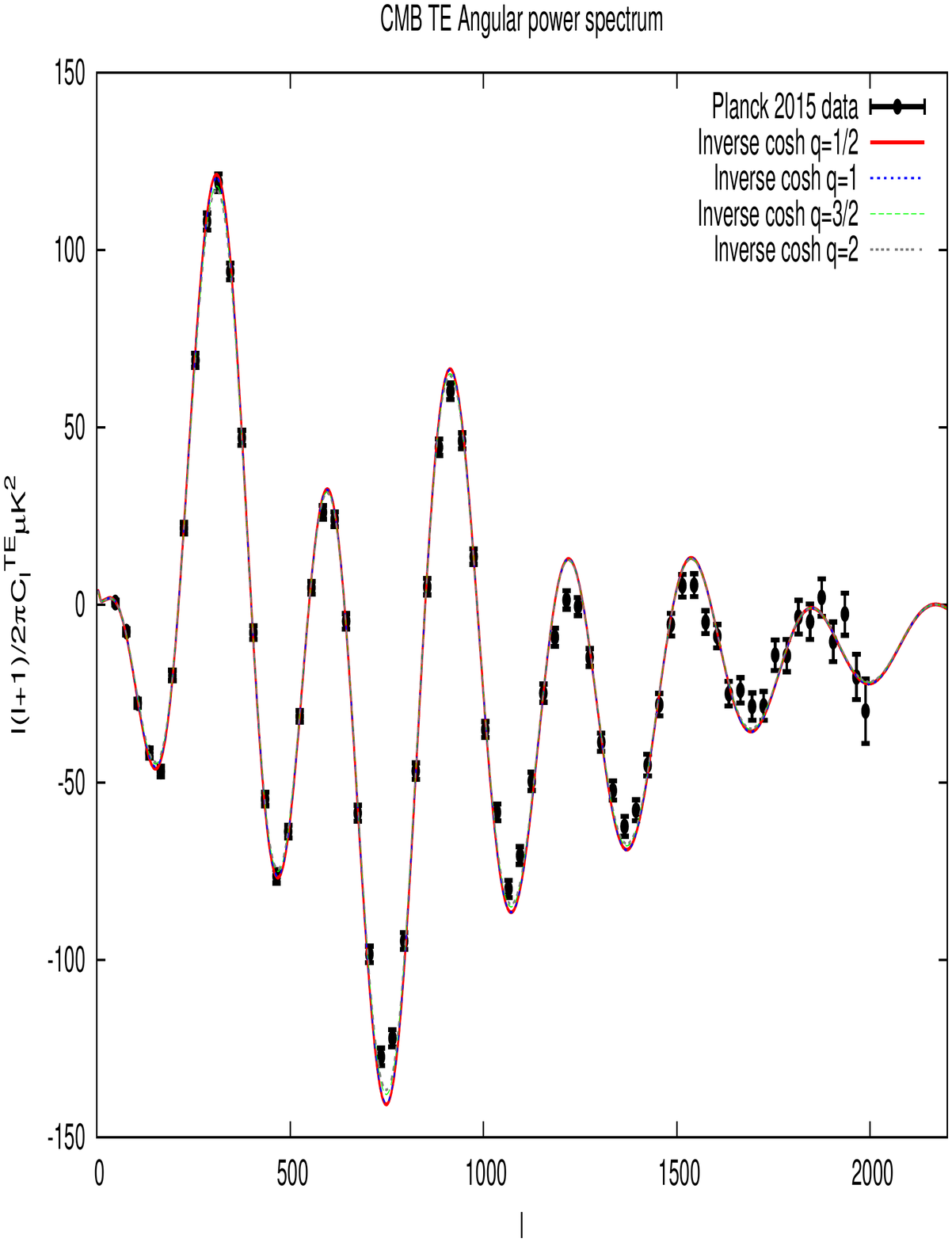}
    \label{TEm1}
}
\subfigure[$l(l+1)C^{TE}_{l}/2\pi$~vs~$l$~(scalar)]{
    \includegraphics[width=7.2cm, height=5.7cm] {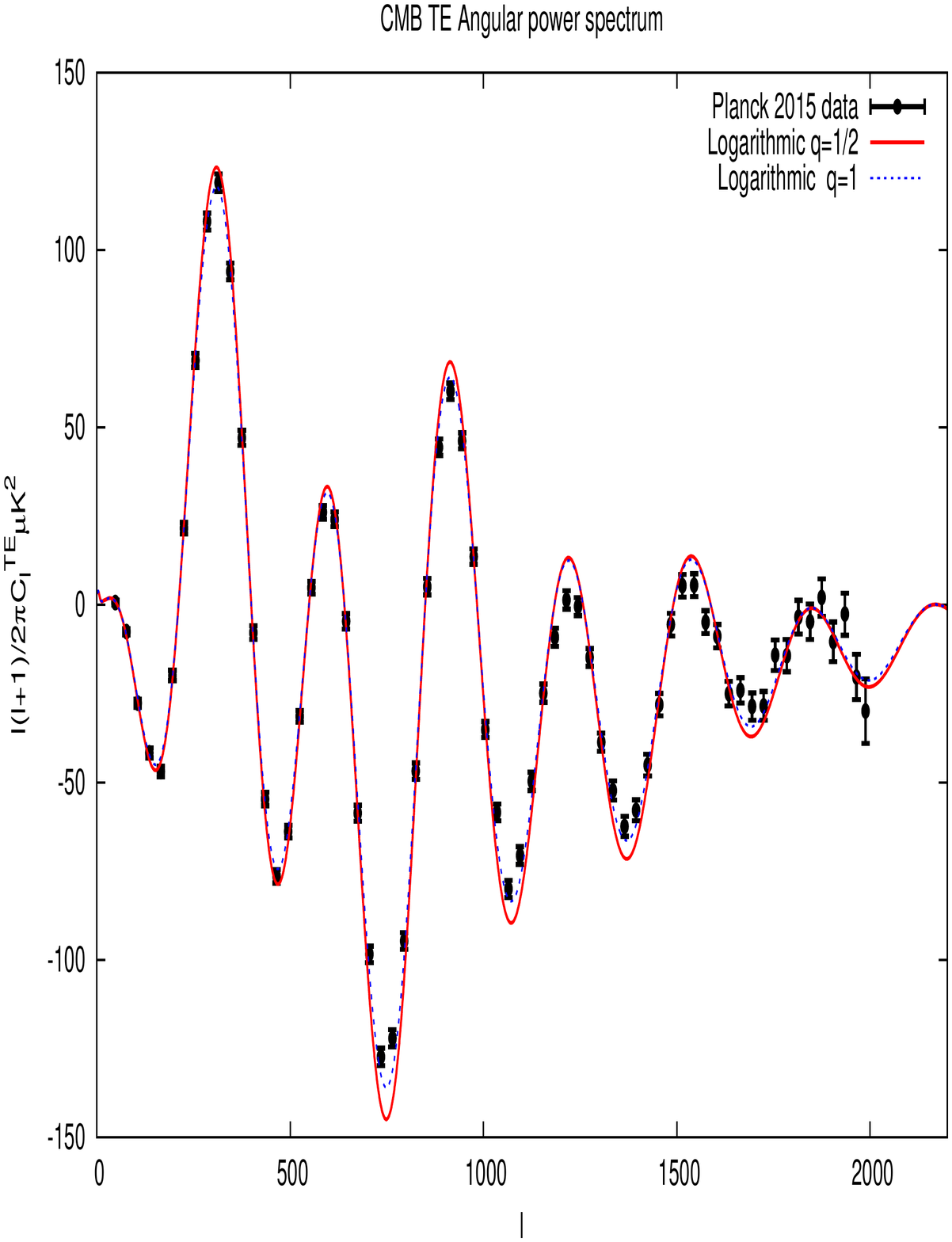}
    \label{TEm2}
}
\subfigure[$l(l+1)C^{TE}_{l}/2\pi$~vs~$l$~(scalar)]{
    \includegraphics[width=7.2cm, height=5.7cm] {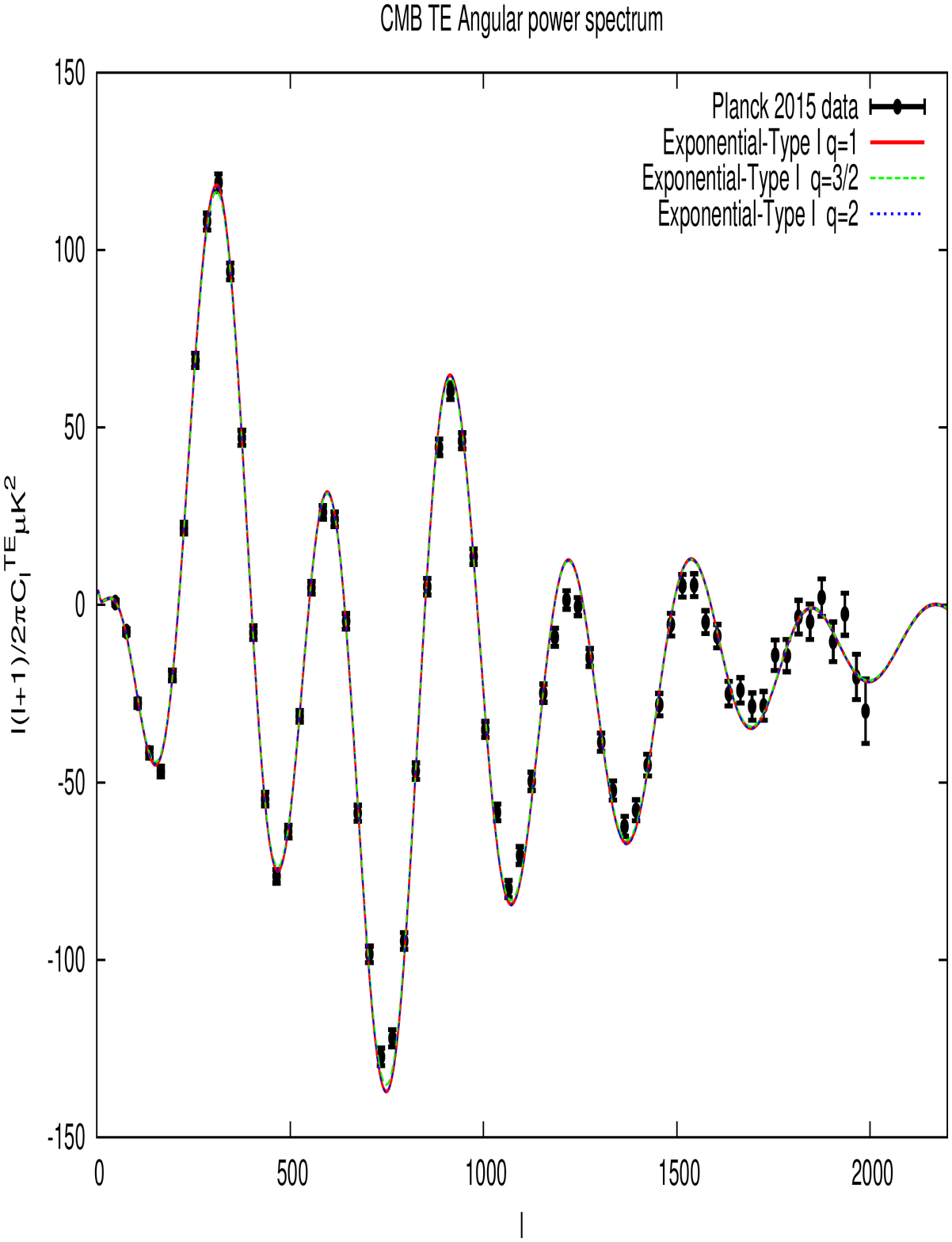}
    \label{TEm3}
}
\subfigure[$l(l+1)C^{TE}_{l}/2\pi$~vs~$l$~(scalar)]{
    \includegraphics[width=7.2cm, height=5.7cm] {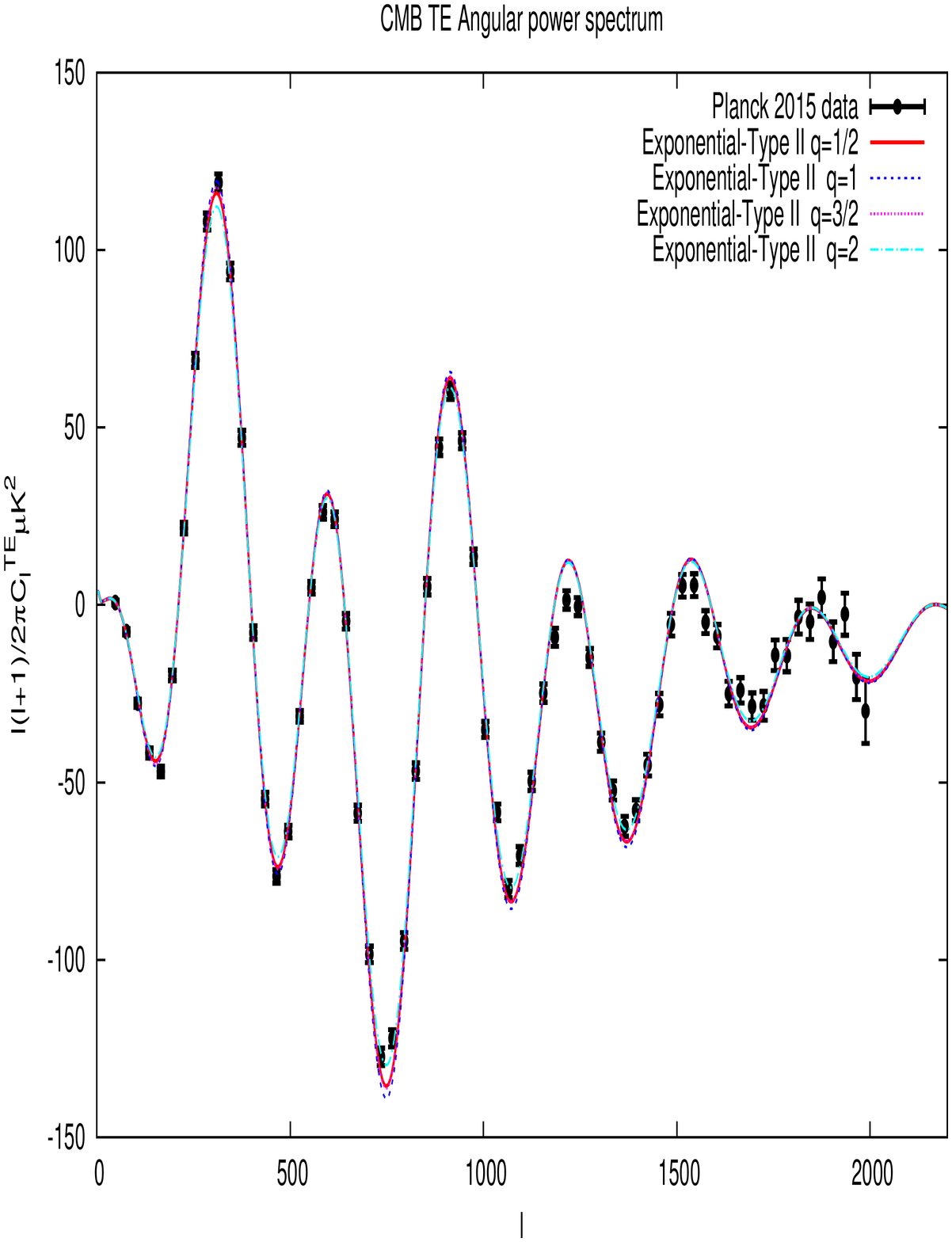}
    \label{TEm4}
}
\subfigure[$l(l+1)C^{TE}_{l}/2\pi$~vs~$l$~(scalar)]{
    \includegraphics[width=7.2cm, height=5.7cm] {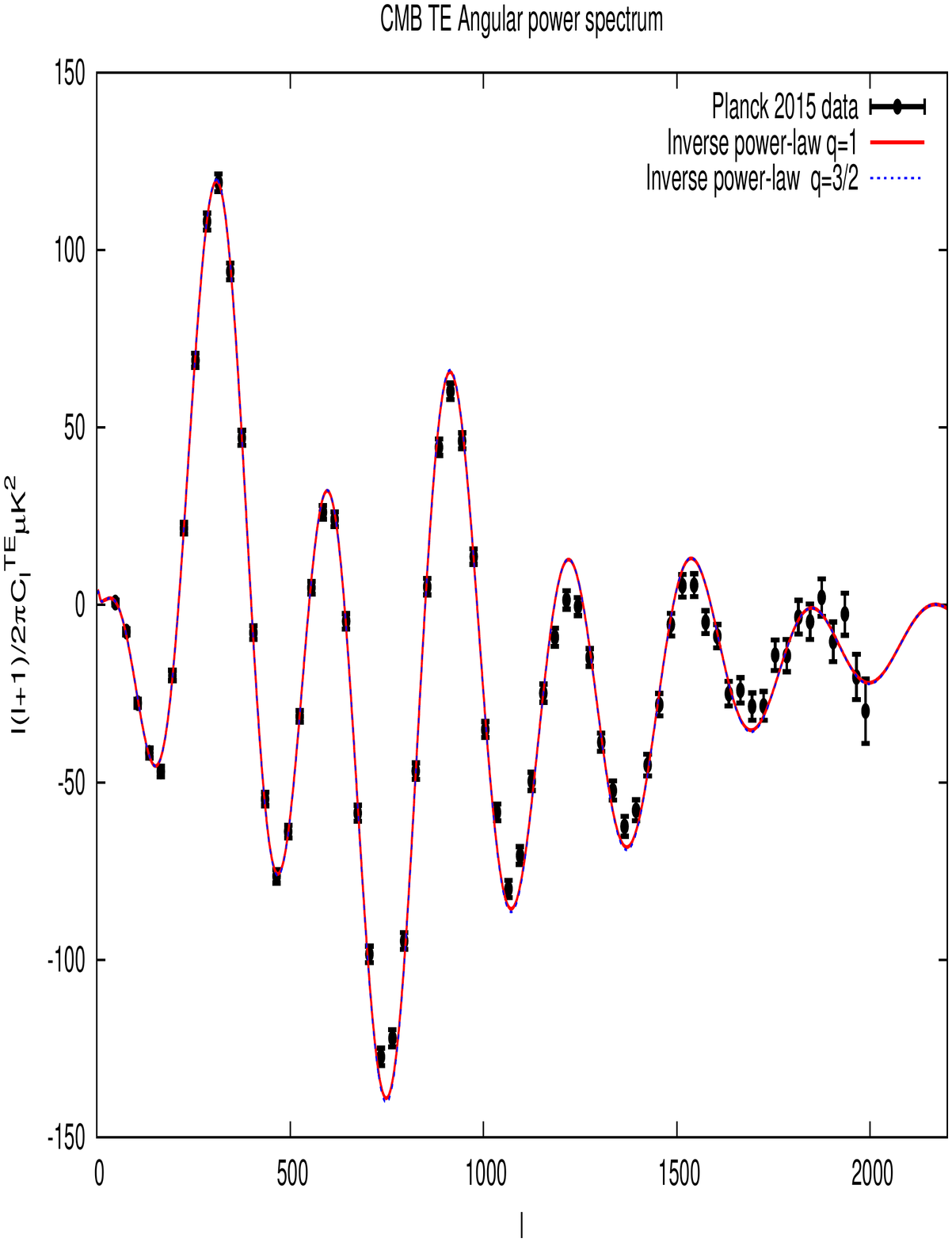}
    \label{TEm5}
}
\caption[Optional caption for list of figures]{We show the variation of CMB TE Angular power spectrum with respect to the multipole, $l$ for scalar modes for all five tachyonic models.
}
\label{fig19}
\end{figure}

\begin{figure}[t]
\centering
\subfigure[$l(l+1)C^{EE}_{l}/2\pi$~vs~$l$~(scalar)]{
    \includegraphics[width=7.2cm, height=5.7cm] {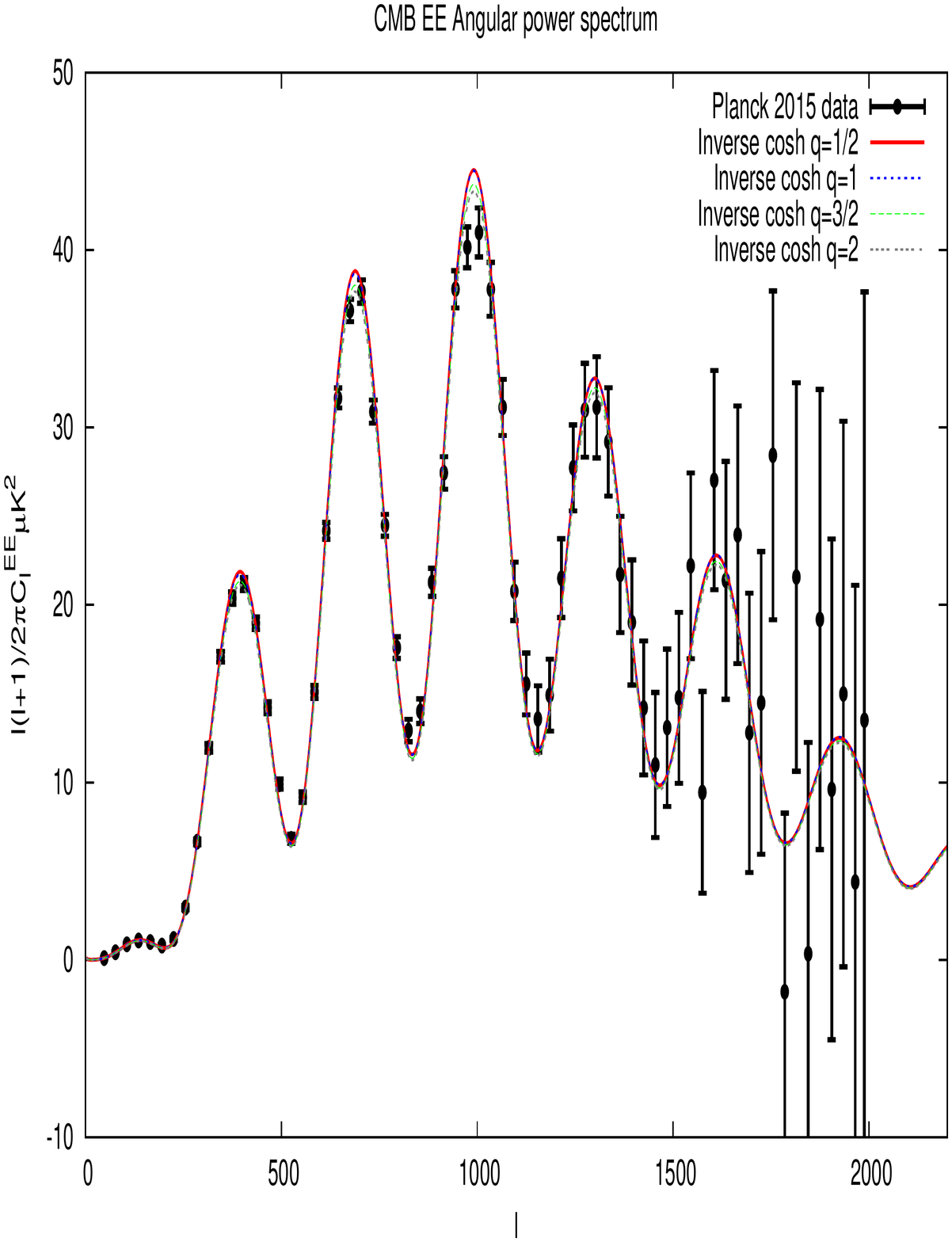}
    \label{EEm1}
}
\subfigure[$l(l+1)C^{EE}_{l}/2\pi$~vs~$l$~(scalar)]{
    \includegraphics[width=7.2cm, height=5.7cm] {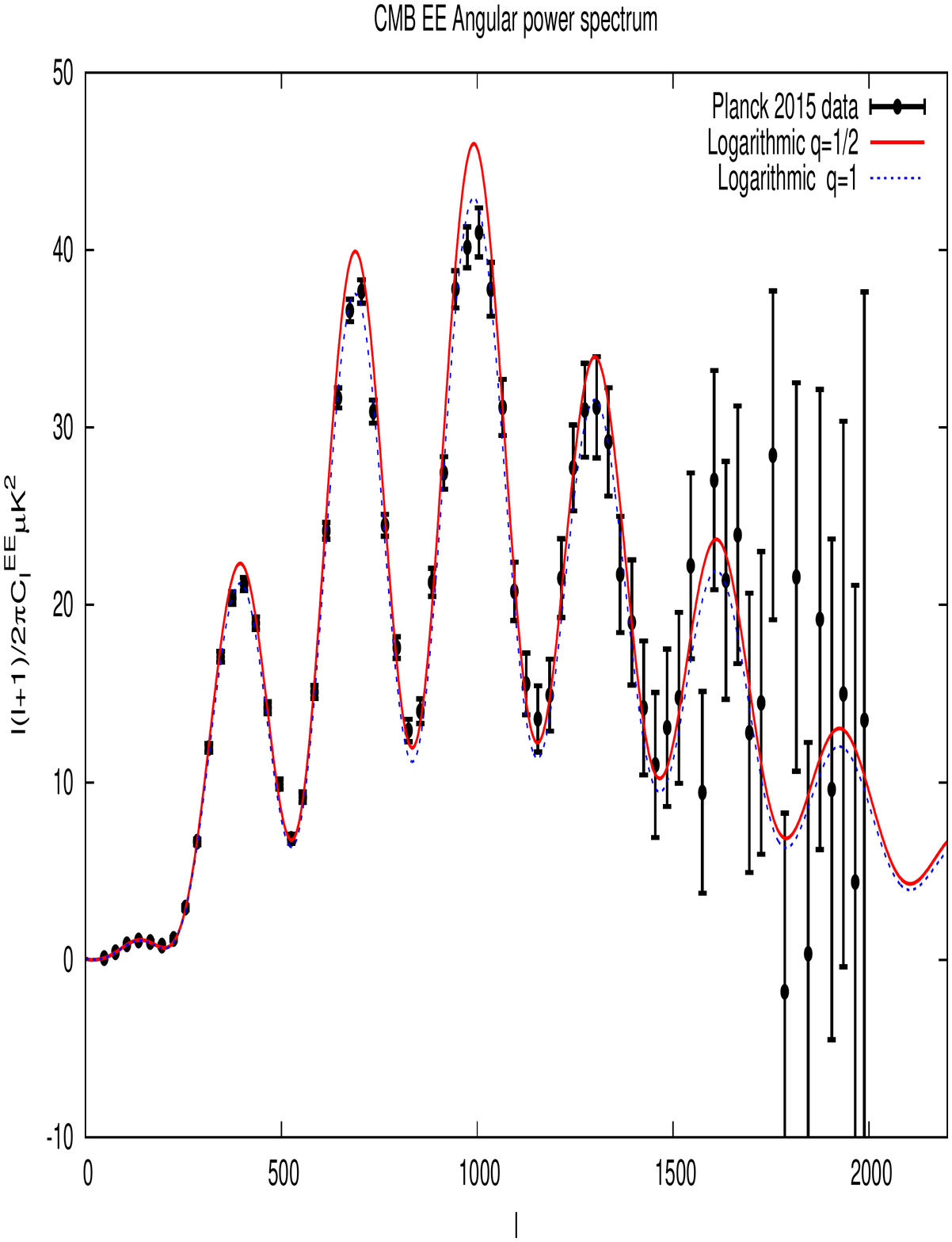}
    \label{EEm2}
}
\subfigure[$l(l+1)C^{EE}_{l}/2\pi$~vs~$l$~(scalar)]{
    \includegraphics[width=7.2cm, height=5.7cm] {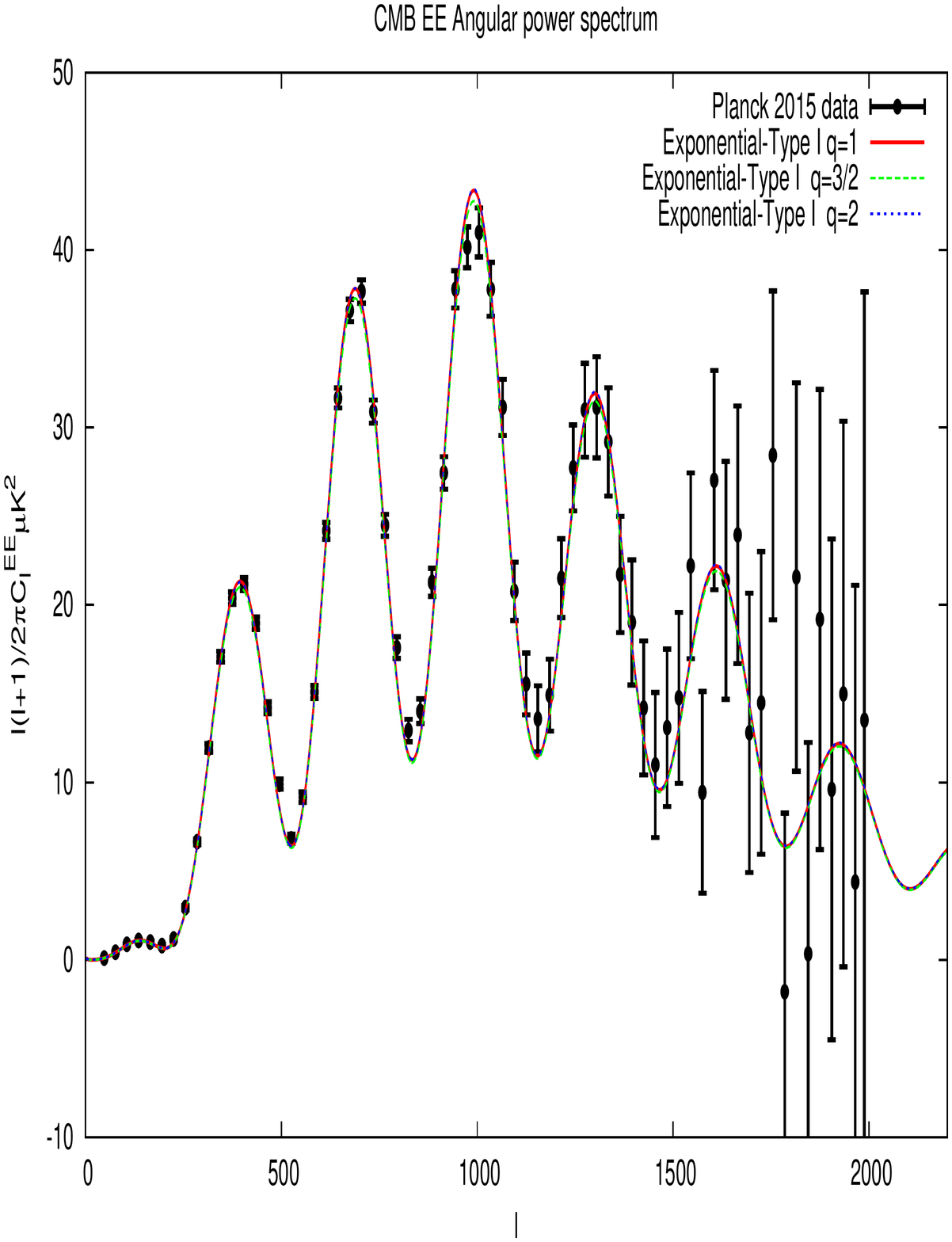}
    \label{EEm3}
}
\subfigure[$l(l+1)C^{EE}_{l}/2\pi$~vs~$l$~(scalar)]{
    \includegraphics[width=7.2cm, height=5.7cm] {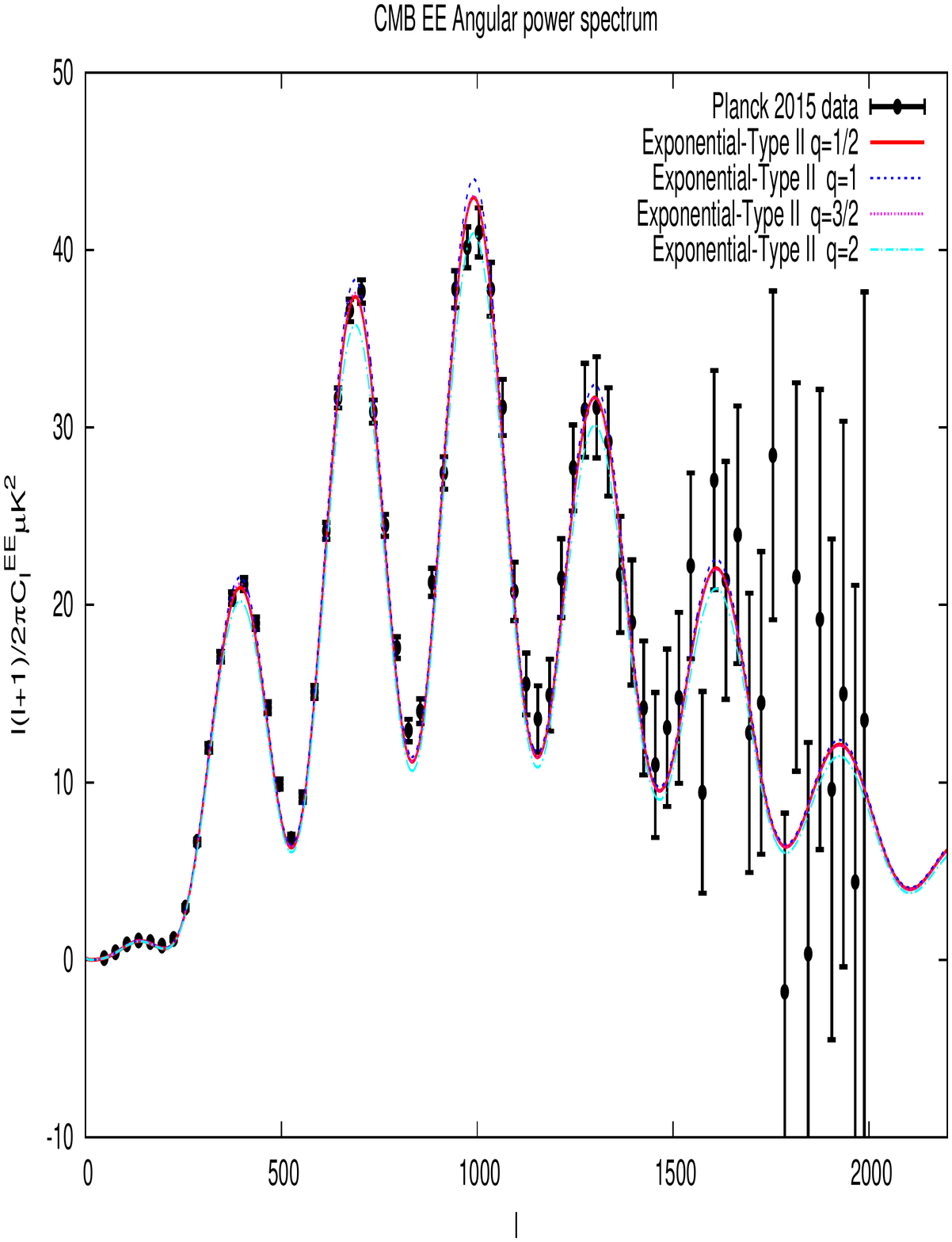}
    \label{EEm4}
}
\subfigure[$l(l+1)C^{EE}_{l}/2\pi$~vs~$l$~(scalar)]{
    \includegraphics[width=7.2cm, height=5.7cm] {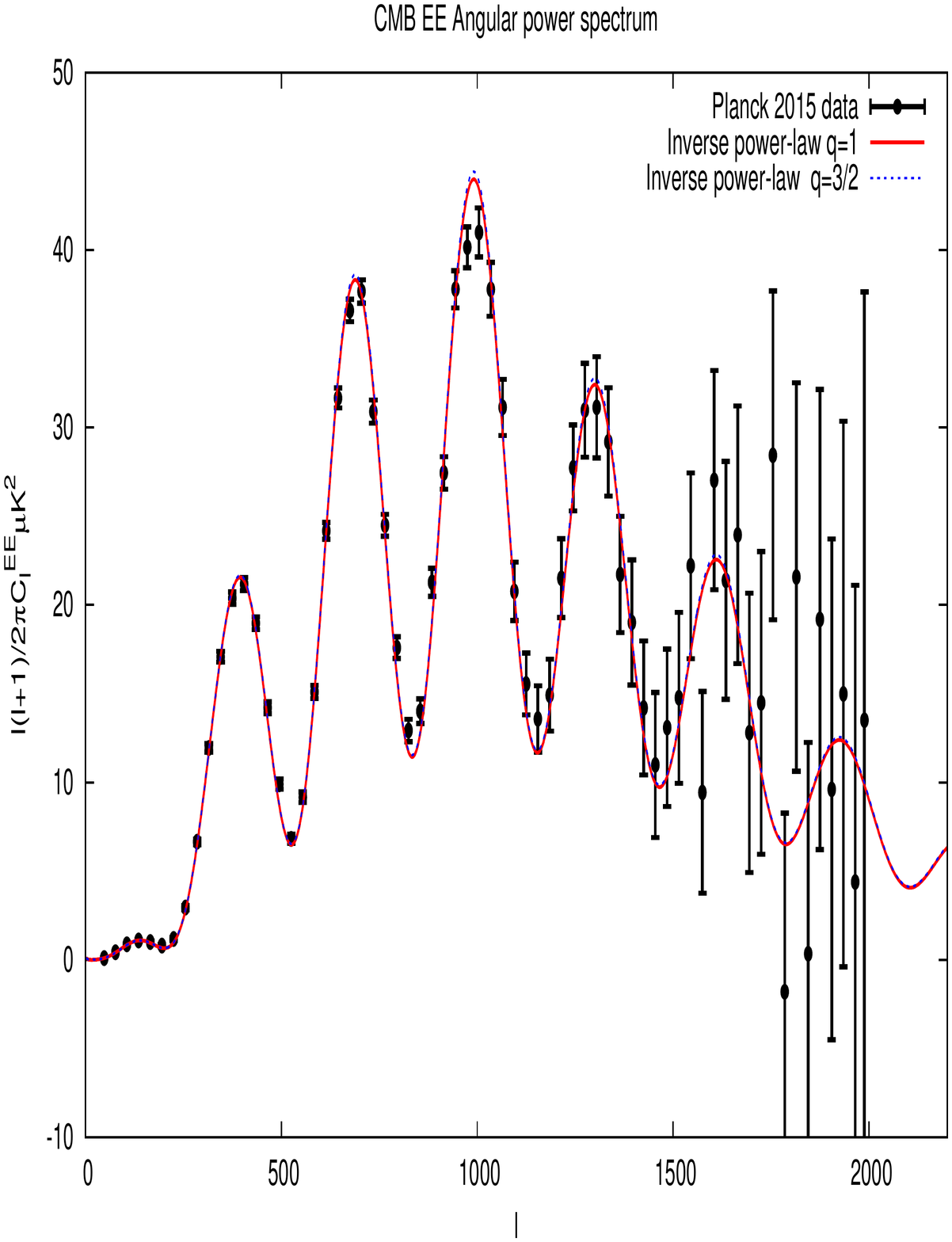}
    \label{EEm5}
}
\caption[Optional caption for list of figures]{We show the variation of CMB EE Angular power spectrum with respect to the multipole, $l$ for scalar modes for all five tachyonic models.
}
\label{fig20}
\end{figure}

\begin{figure}[t]
\centering
\subfigure[$l(l+1)C^{BB}_{l}/2\pi$~vs~$l$~(tensor)]{
    \includegraphics[width=7.2cm, height=5.7cm] {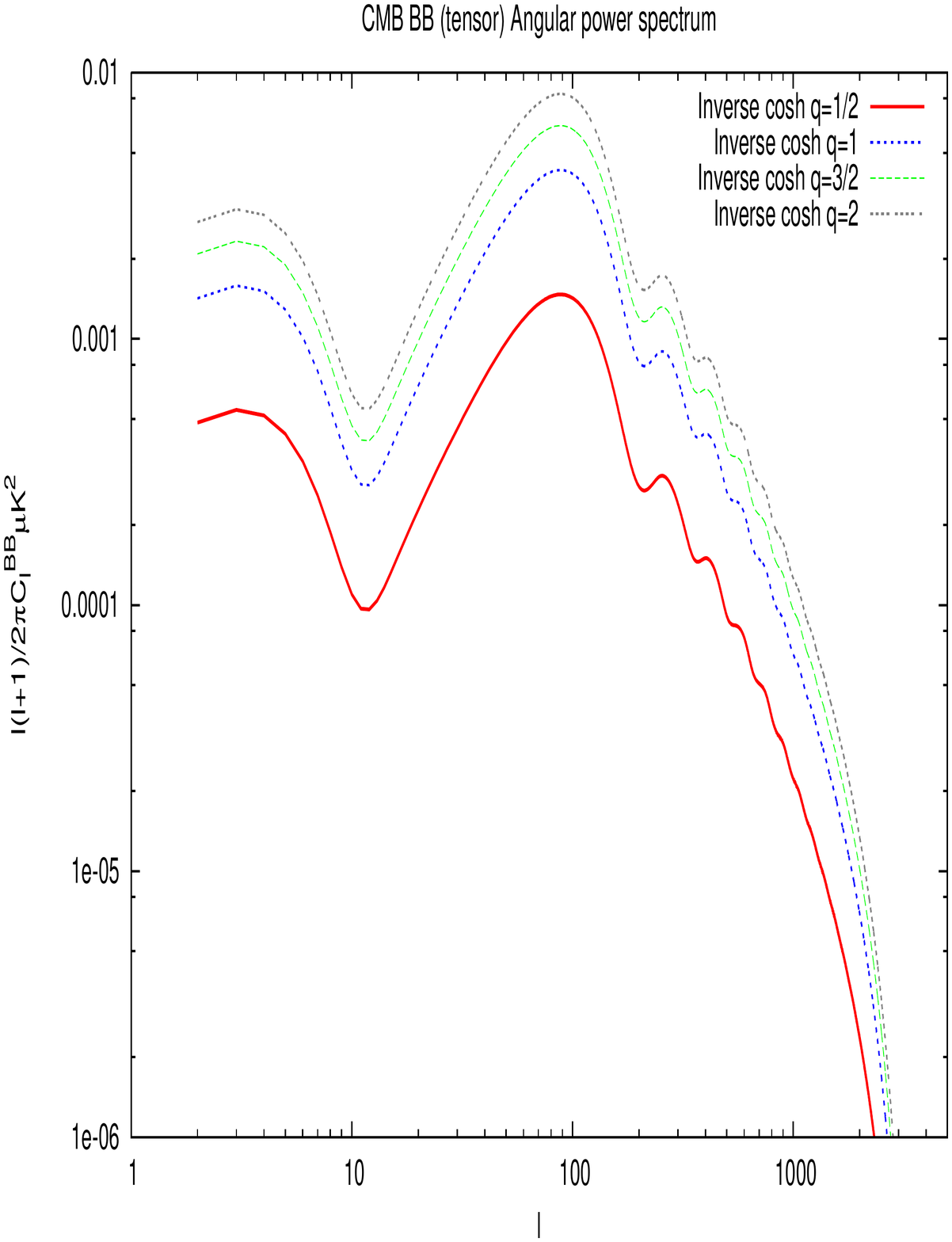}
    \label{BBtm1}
}
\subfigure[$l(l+1)C^{BB}_{l}/2\pi$~vs~$l$~(tensor)]{
    \includegraphics[width=7.2cm, height=5.7cm] {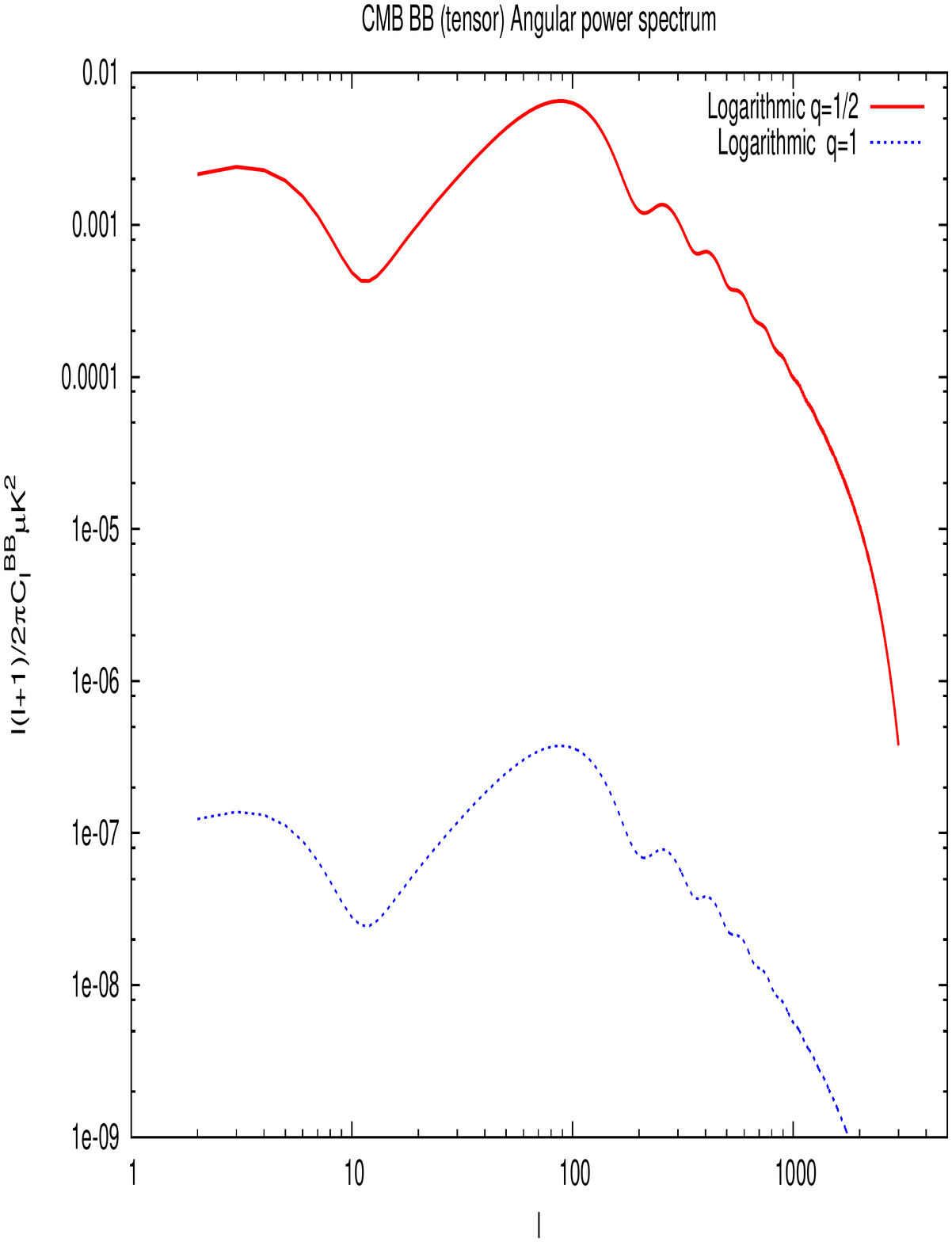}
    \label{BBtm2}
}
\subfigure[$l(l+1)C^{BB}_{l}/2\pi$~vs~$l$~(tensor)]{
    \includegraphics[width=7.2cm, height=5.7cm] {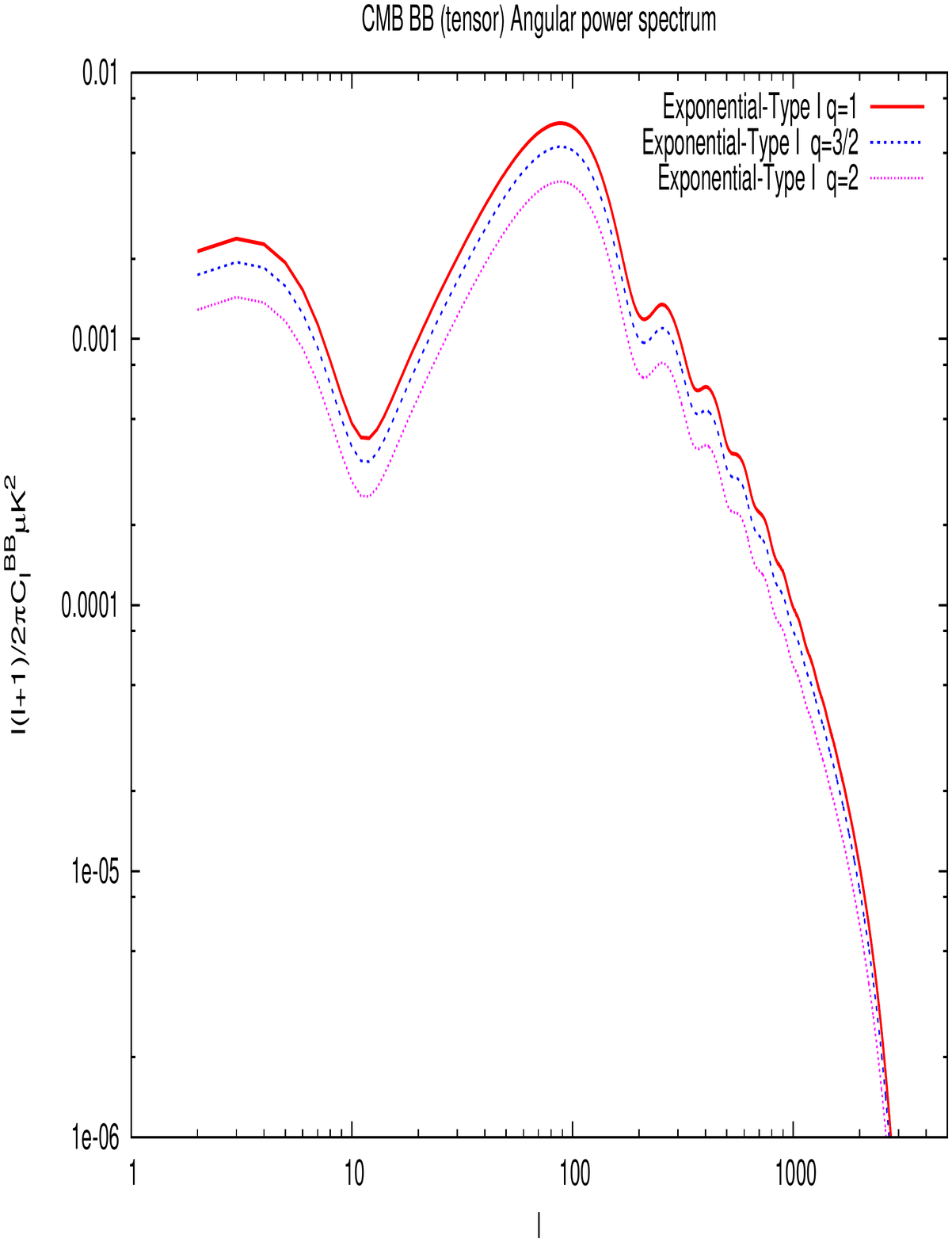}
    \label{BBtm3}
}
\subfigure[$l(l+1)C^{BB}_{l}/2\pi$~vs~$l$~(tensor)]{
    \includegraphics[width=7.2cm, height=5.7cm] {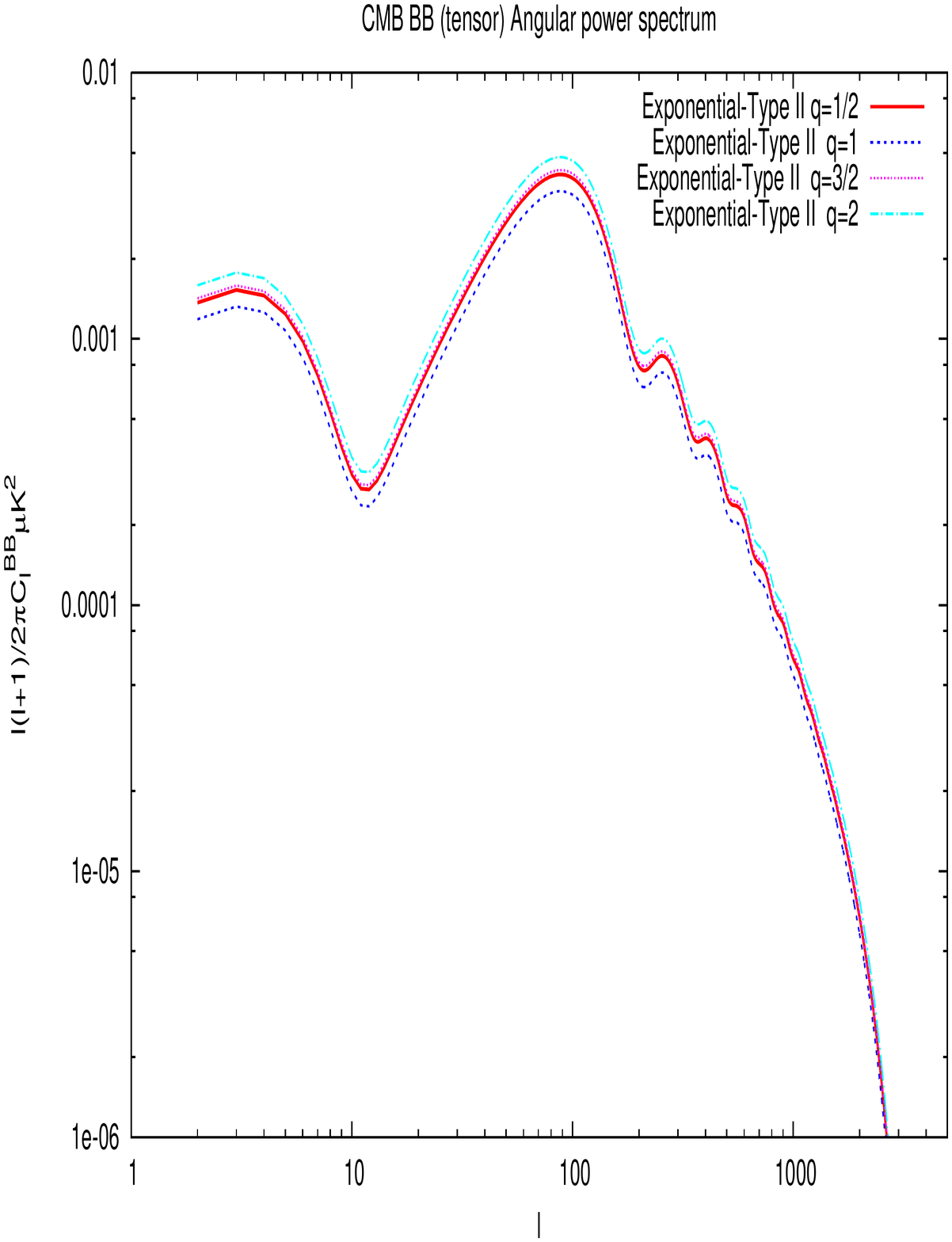}
    \label{BBtm4}
}
\subfigure[$l(l+1)C^{BB}_{l}/2\pi$~vs~$l$~(tensor)]{
    \includegraphics[width=7.2cm, height=5.7cm] {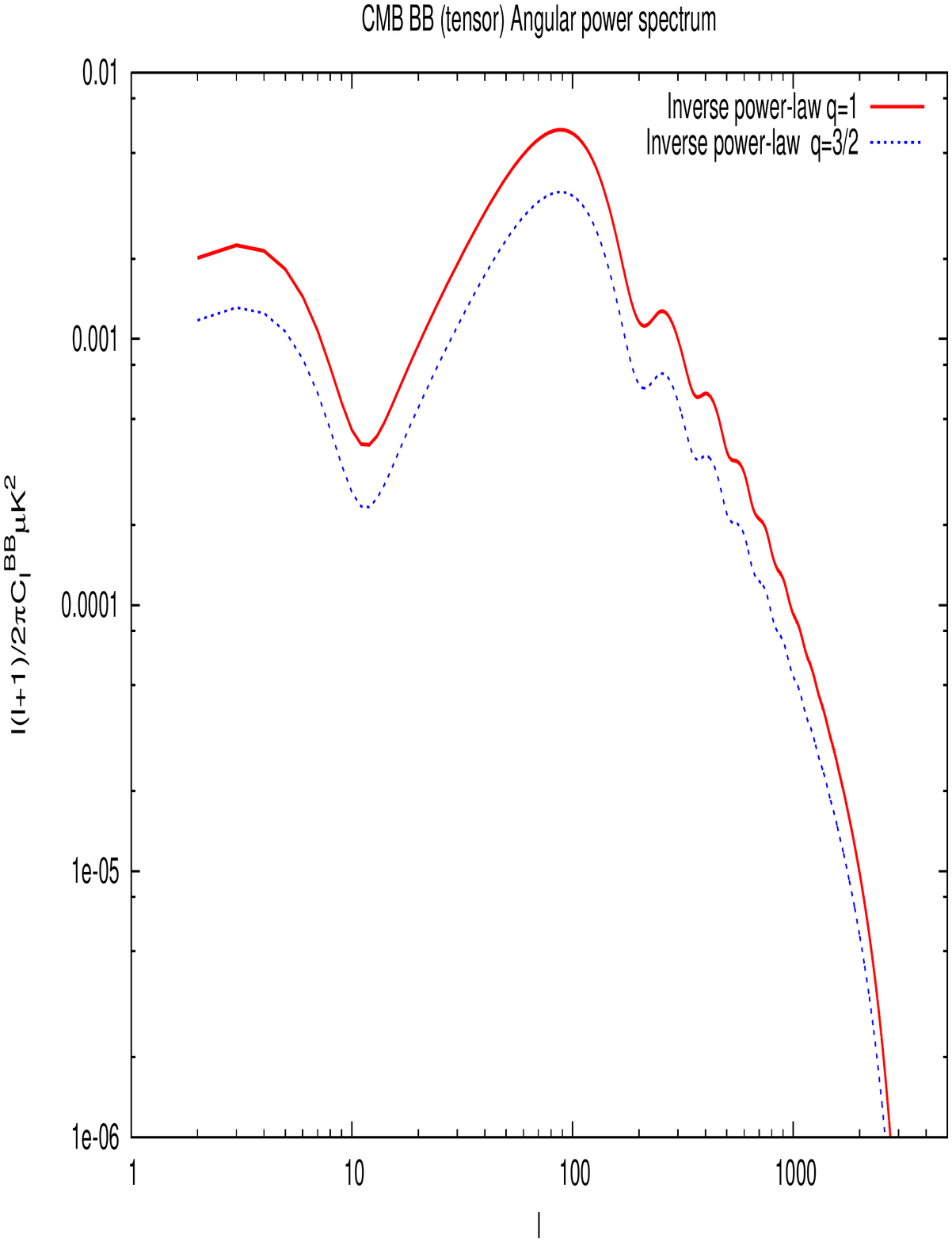}
    \label{BBtm5}
}
\caption[Optional caption for list of figures]{We show the variation of CMB BB Angular power spectrum with respect to the multipole, $l$ for tensor modes for all five tachyonic models.
}
\label{fig21}
\end{figure}

\begin{figure}[t]
\centering
\subfigure[$l(l+1)C^{TT}_{l}/2\pi$~vs~$l$~(tensor)]{
    \includegraphics[width=7.2cm, height=5.7cm] {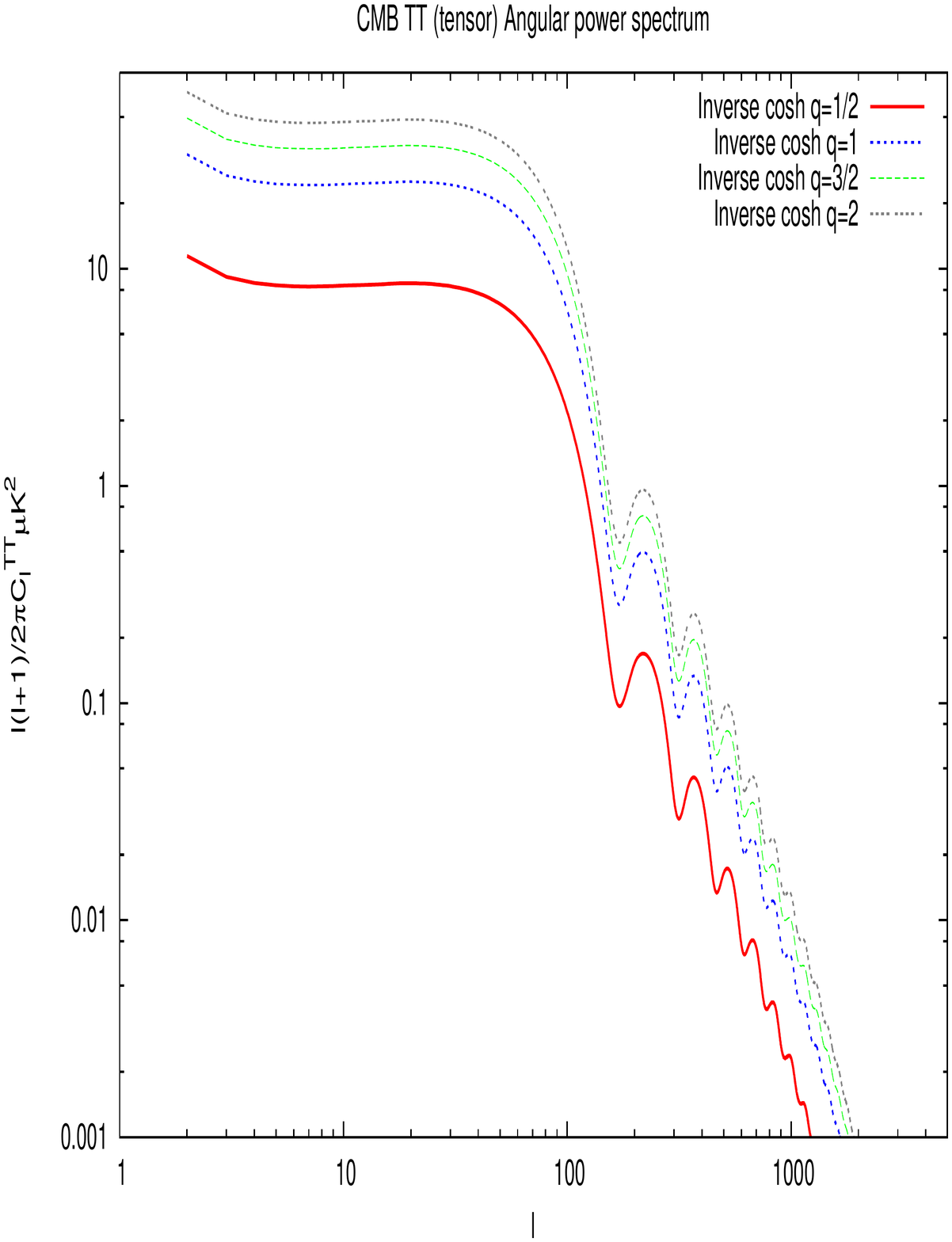}
    \label{TTtm1}
}
\subfigure[$l(l+1)C^{TT}_{l}/2\pi$~vs~$l$~(tensor)]{
    \includegraphics[width=7.2cm, height=5.7cm] {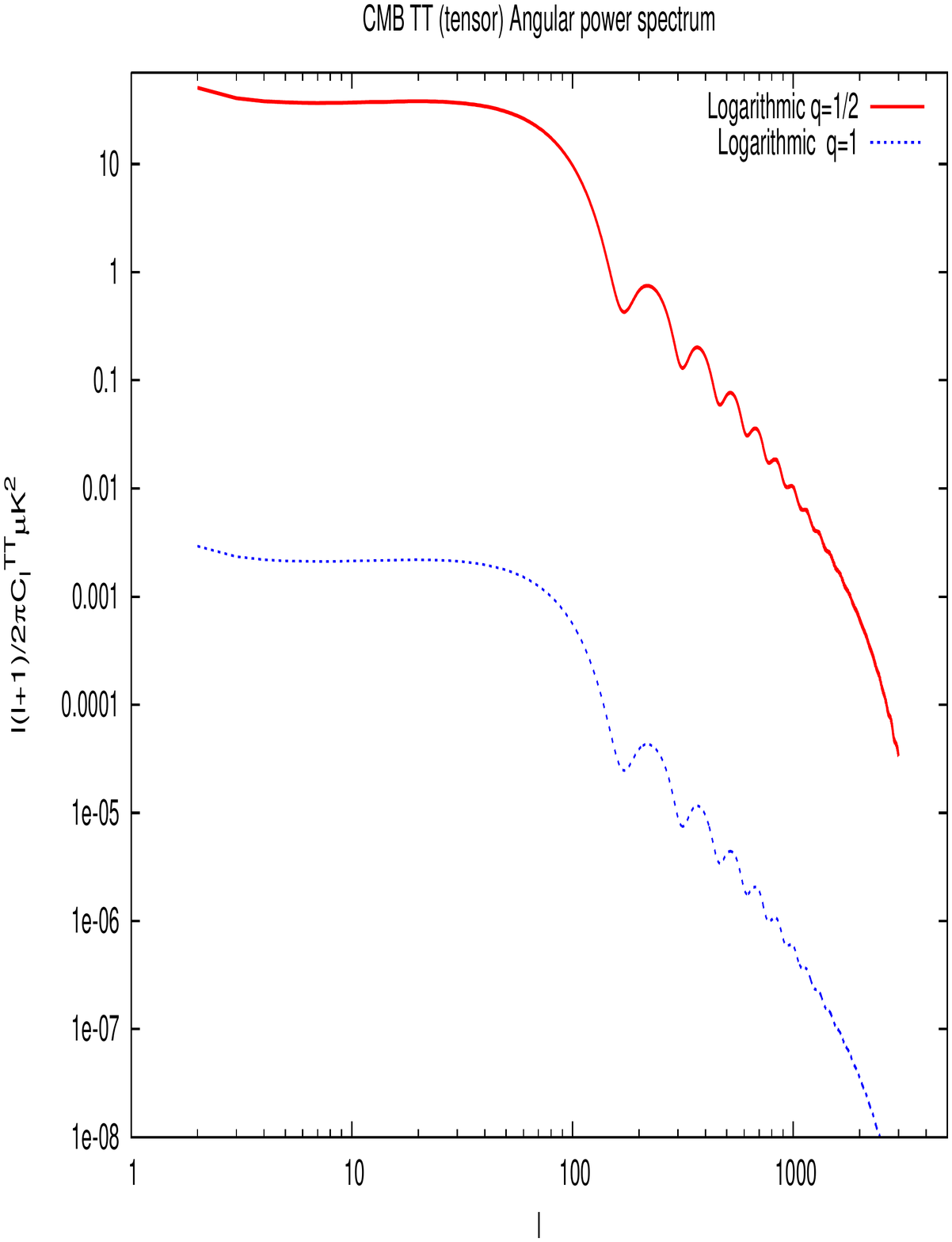}
    \label{TTtm2}
}
\subfigure[$l(l+1)C^{TT}_{l}/2\pi$~vs~$l$~(tensor)]{
    \includegraphics[width=7.2cm, height=5.7cm] {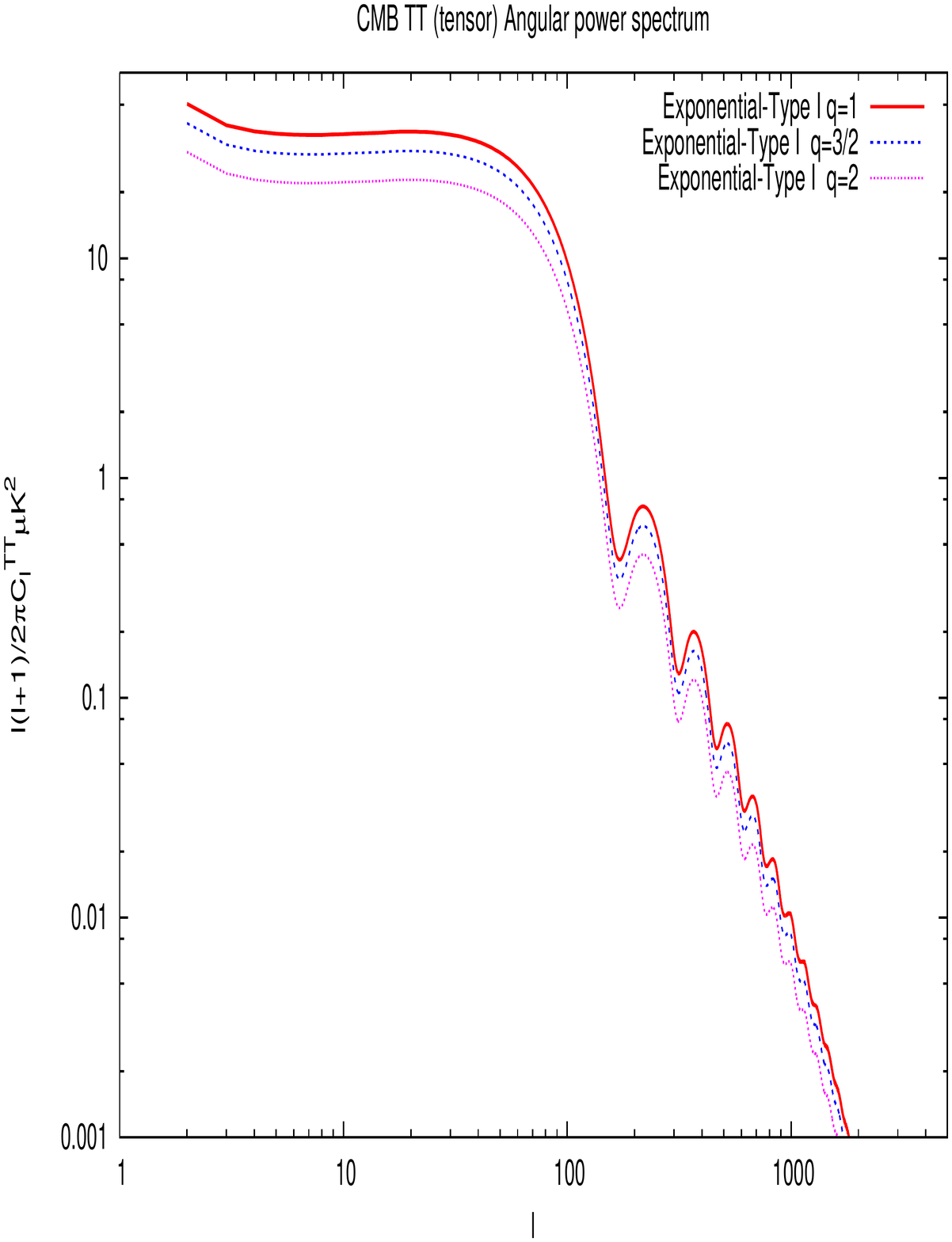}
    \label{TTtm3}
}
\subfigure[$l(l+1)C^{TT}_{l}/2\pi$~vs~$l$~(tensor)]{
    \includegraphics[width=7.2cm, height=5.7cm] {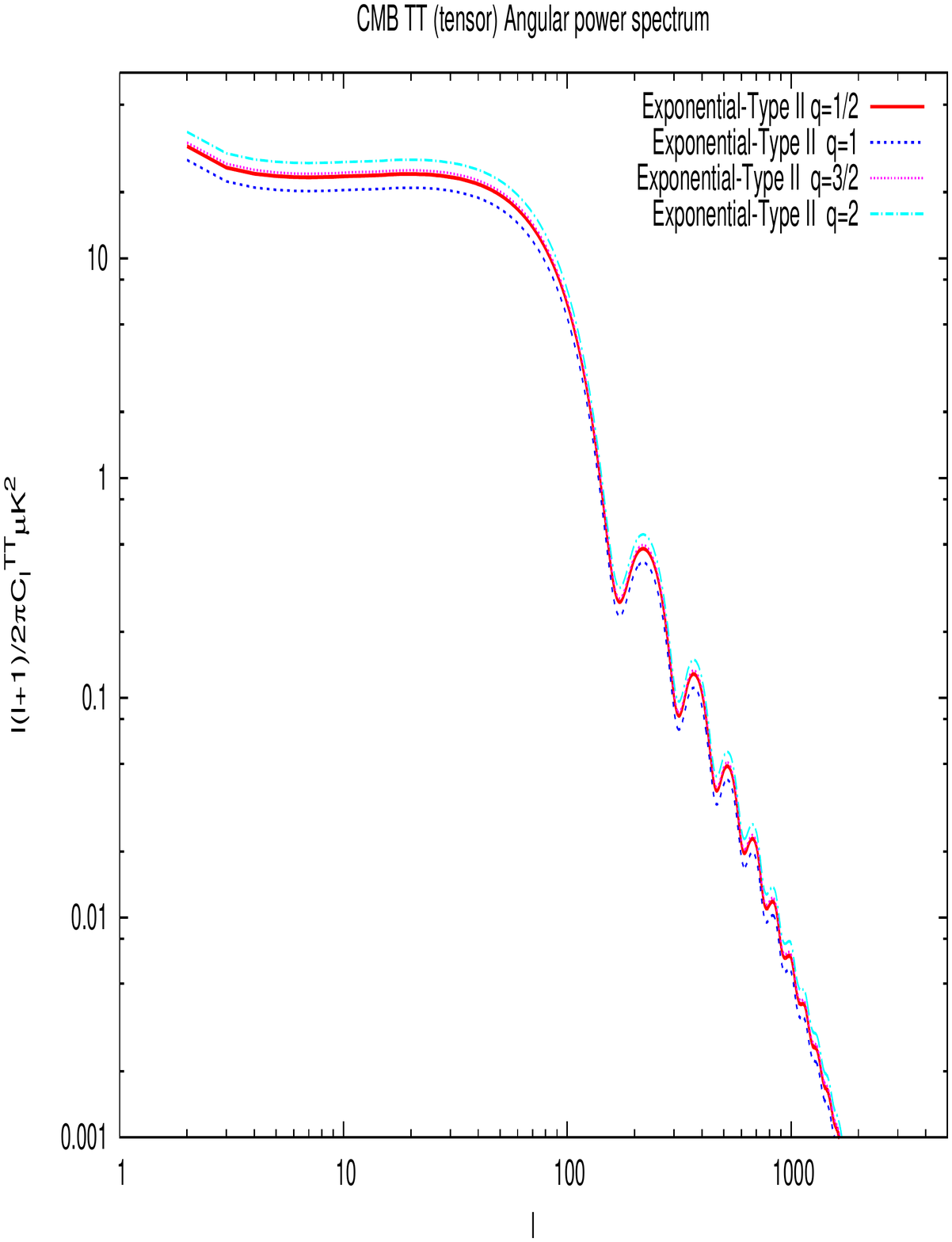}
    \label{TTtm4}
}
\subfigure[$l(l+1)C^{TT}_{l}/2\pi$~vs~$l$~(tensor)]{
    \includegraphics[width=7.2cm, height=5.7cm] {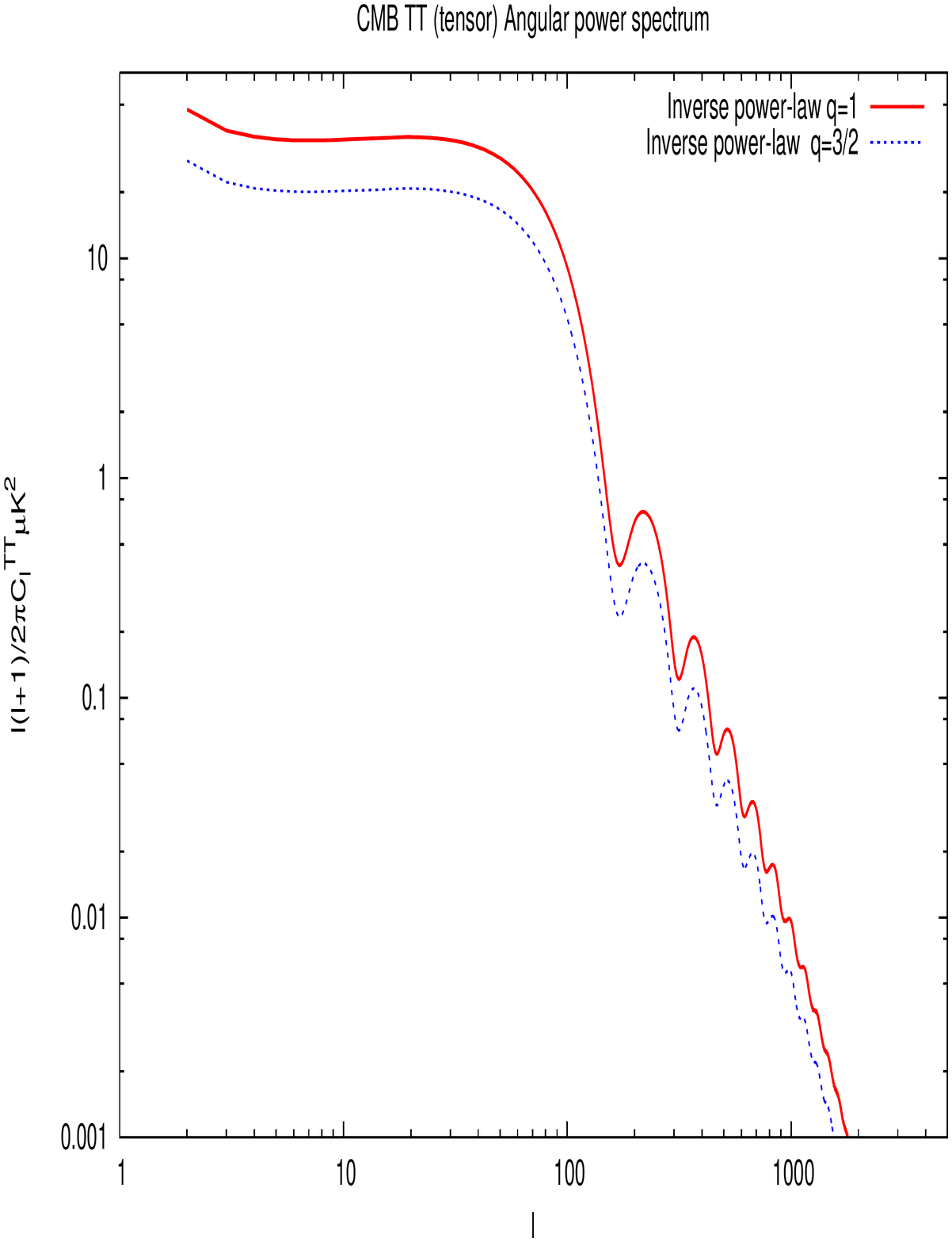}
    \label{TTtm5}
}
\caption[Optional caption for list of figures]{We show the variation of CMB TT Angular power spectrum with respect to the multipole, $l$ for tensor modes for all five tachyonic models.
}
\label{fig22}
\end{figure}

\begin{figure}[t]
\centering
\subfigure[$l(l+1)C^{TE}_{l}/2\pi$~vs~$l$~(tensor)]{
    \includegraphics[width=7.2cm, height=5.7cm] {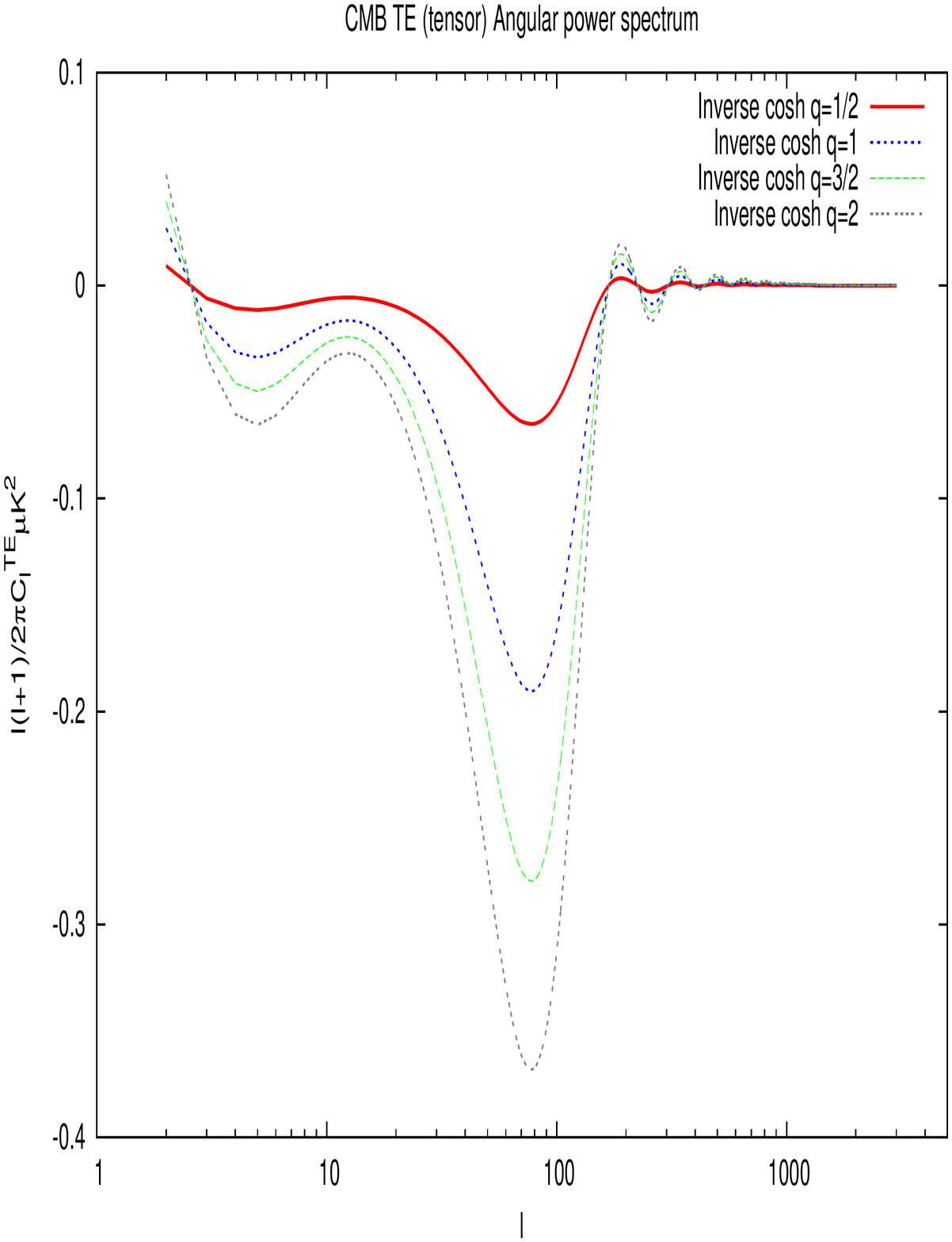}
    \label{TEtm1}
}
\subfigure[$l(l+1)C^{TE}_{l}/2\pi$~vs~$l$~(tensor)]{
    \includegraphics[width=7.2cm, height=5.7cm] {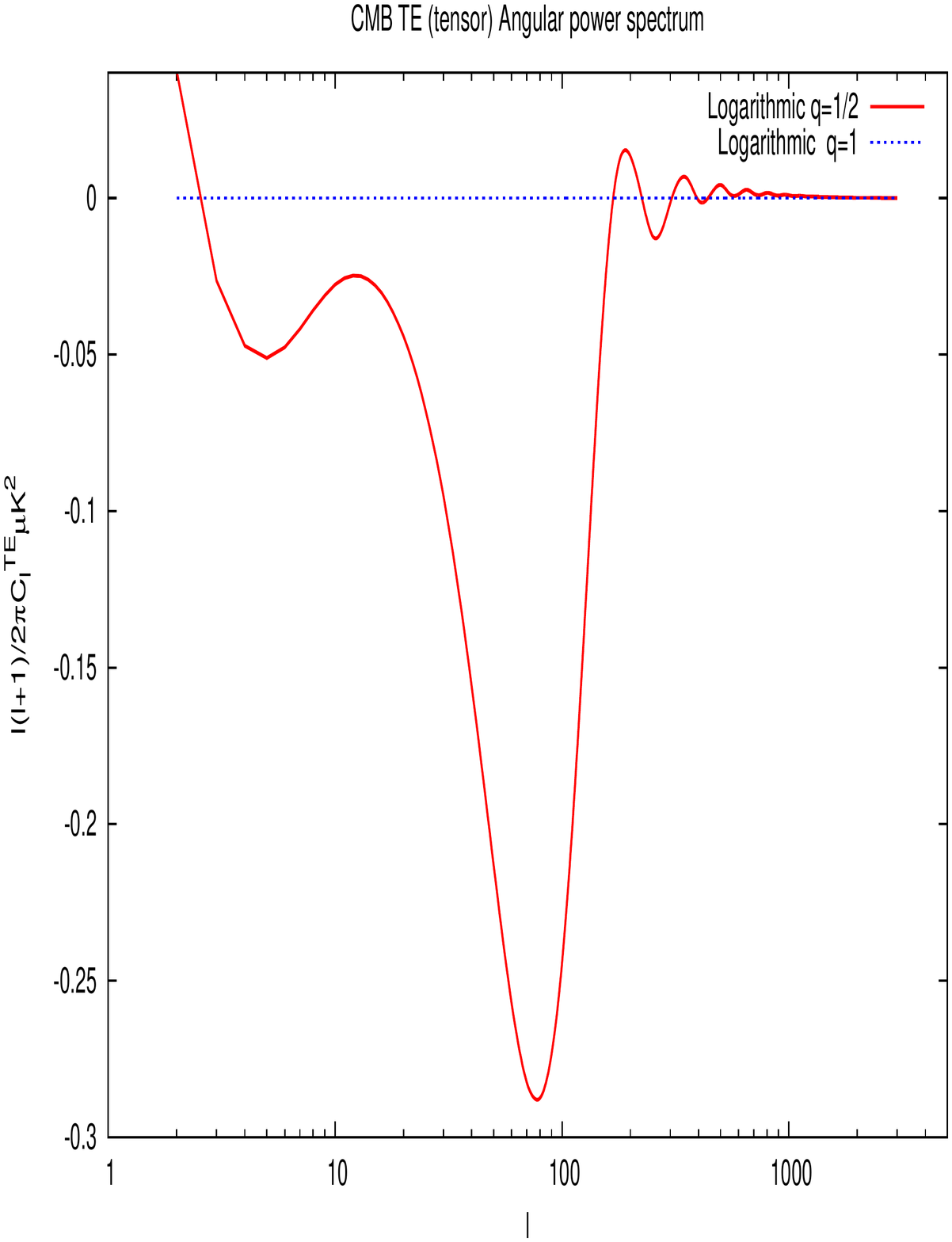}
    \label{TEtm2}
}
\subfigure[$l(l+1)C^{TE}_{l}/2\pi$~vs~$l$~(tensor)]{
    \includegraphics[width=7.2cm, height=5.7cm] {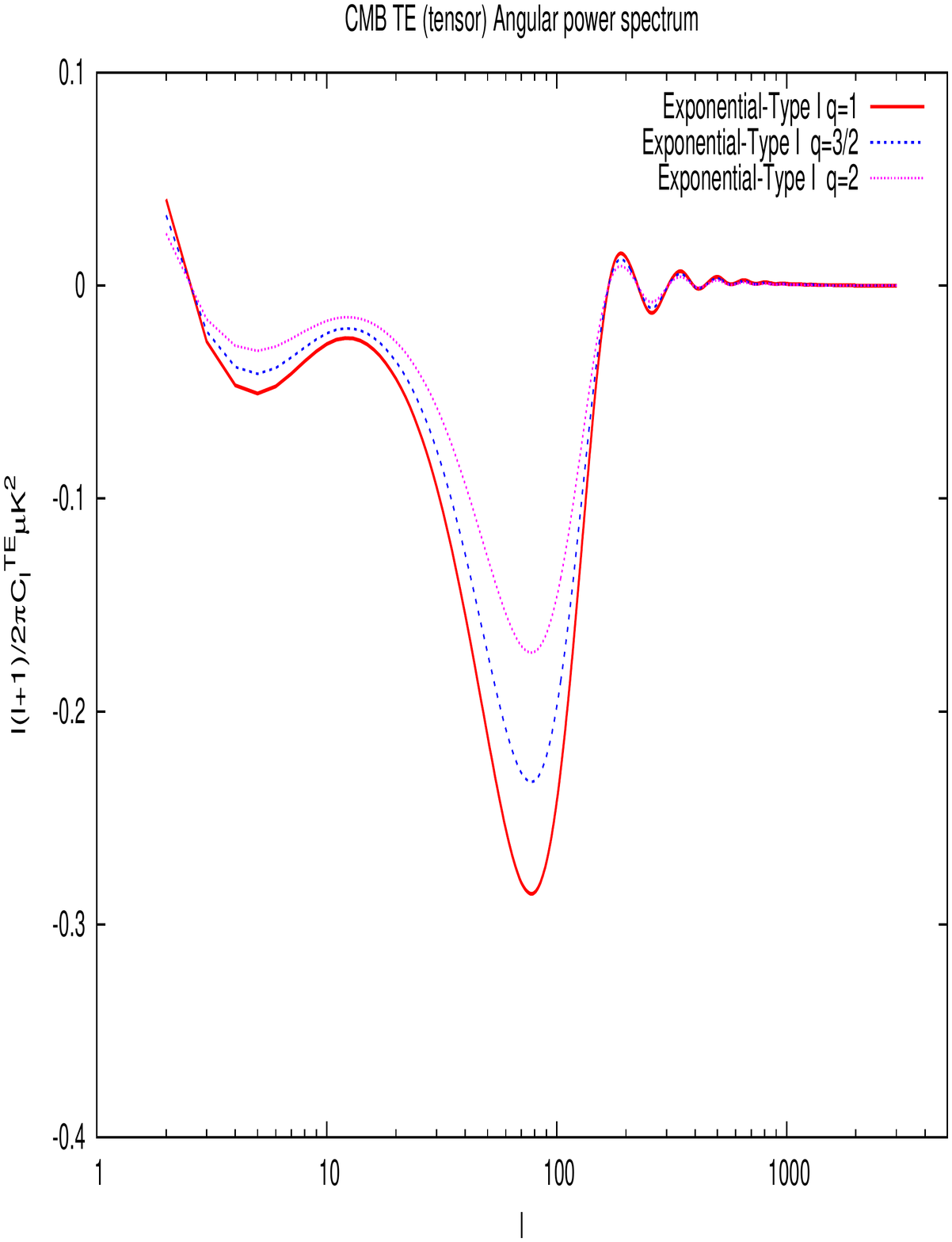}
    \label{TEtm3}
}
\subfigure[$l(l+1)C^{TE}_{l}/2\pi$~vs~$l$~(tensor)]{
    \includegraphics[width=7.2cm, height=5.7cm] {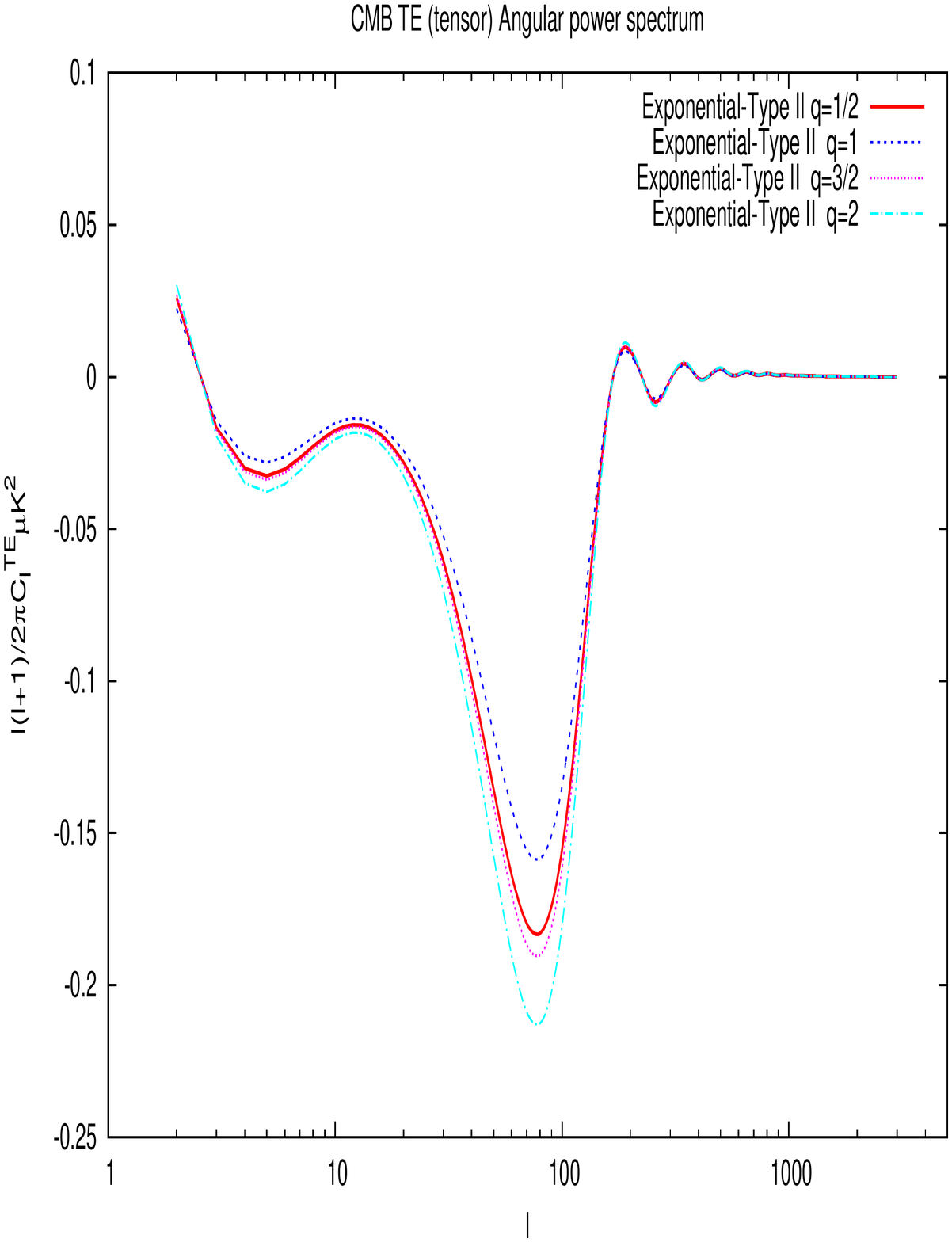}
    \label{TEtm4}
}
\subfigure[$l(l+1)C^{TE}_{l}/2\pi$~vs~$l$~(tensor)]{
    \includegraphics[width=7.2cm, height=5.7cm] {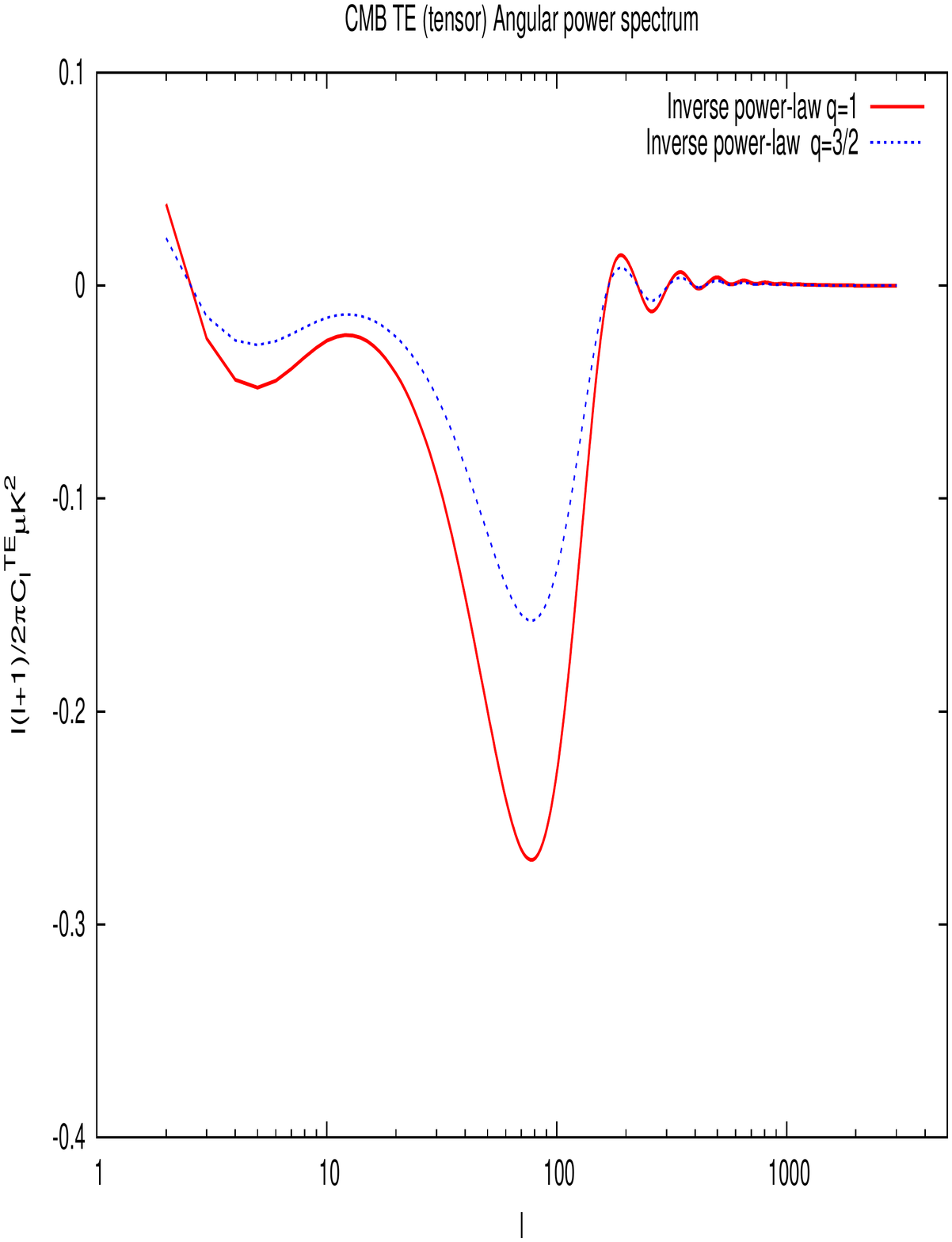}
    \label{TEtm5}
}
\caption[Optional caption for list of figures]{We show the variation of CMB TE Angular power spectrum with respect to the multipole, $l$ for tensor modes for all five tachyonic models.
}
\label{fig23}
\end{figure}

\begin{figure}[t]
\centering
\subfigure[$l(l+1)C^{EE}_{l}/2\pi$~vs~$l$~(tensor)]{
    \includegraphics[width=7.2cm, height=5.7cm] {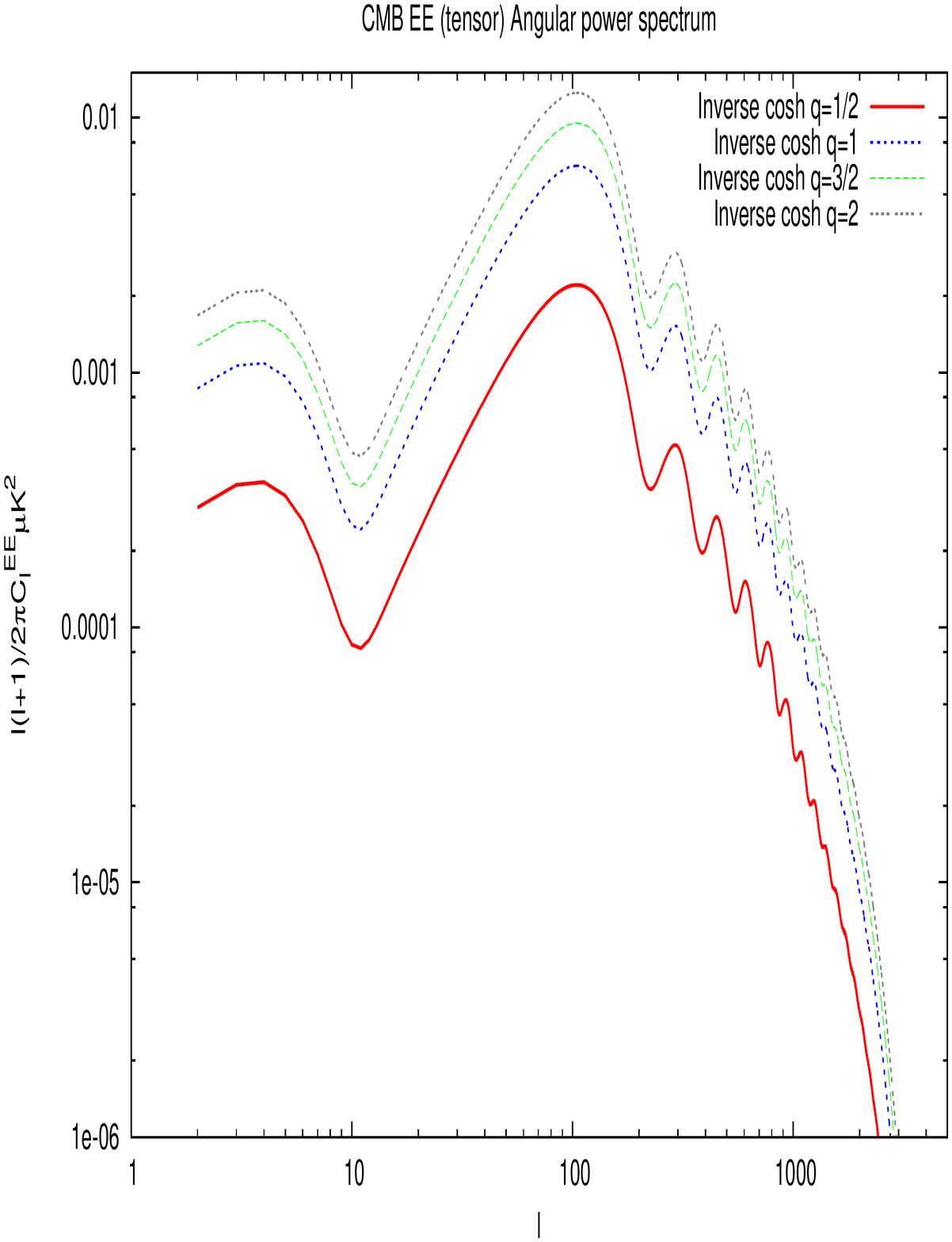}
    \label{EEtm1}
}
\subfigure[$l(l+1)C^{EE}_{l}/2\pi$~vs~$l$~(tensor)]{
    \includegraphics[width=7.2cm, height=5.7cm] {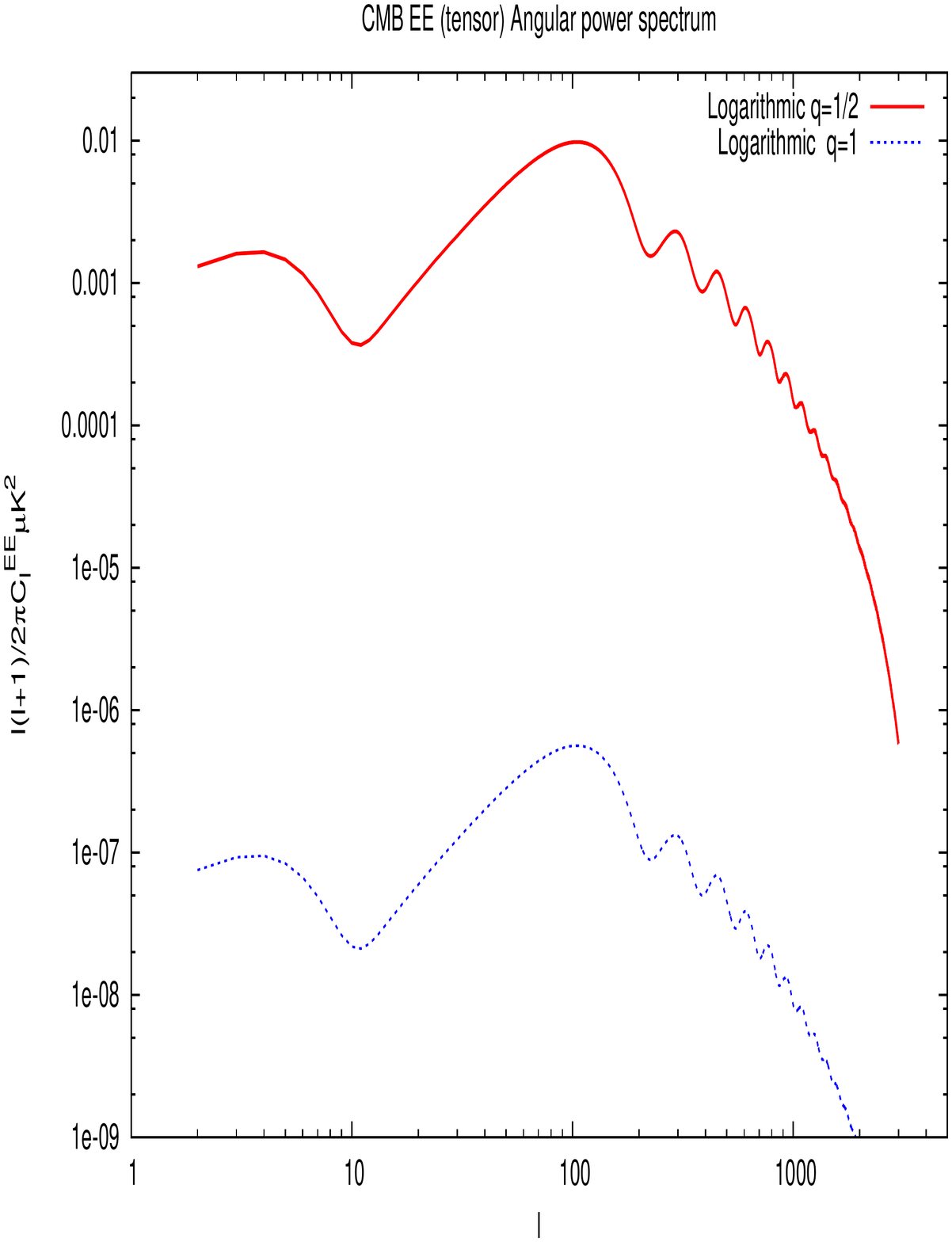}
    \label{EEtm2}
}
\subfigure[$l(l+1)C^{EE}_{l}/2\pi$~vs~$l$~(tensor)]{
    \includegraphics[width=7.2cm, height=5.7cm] {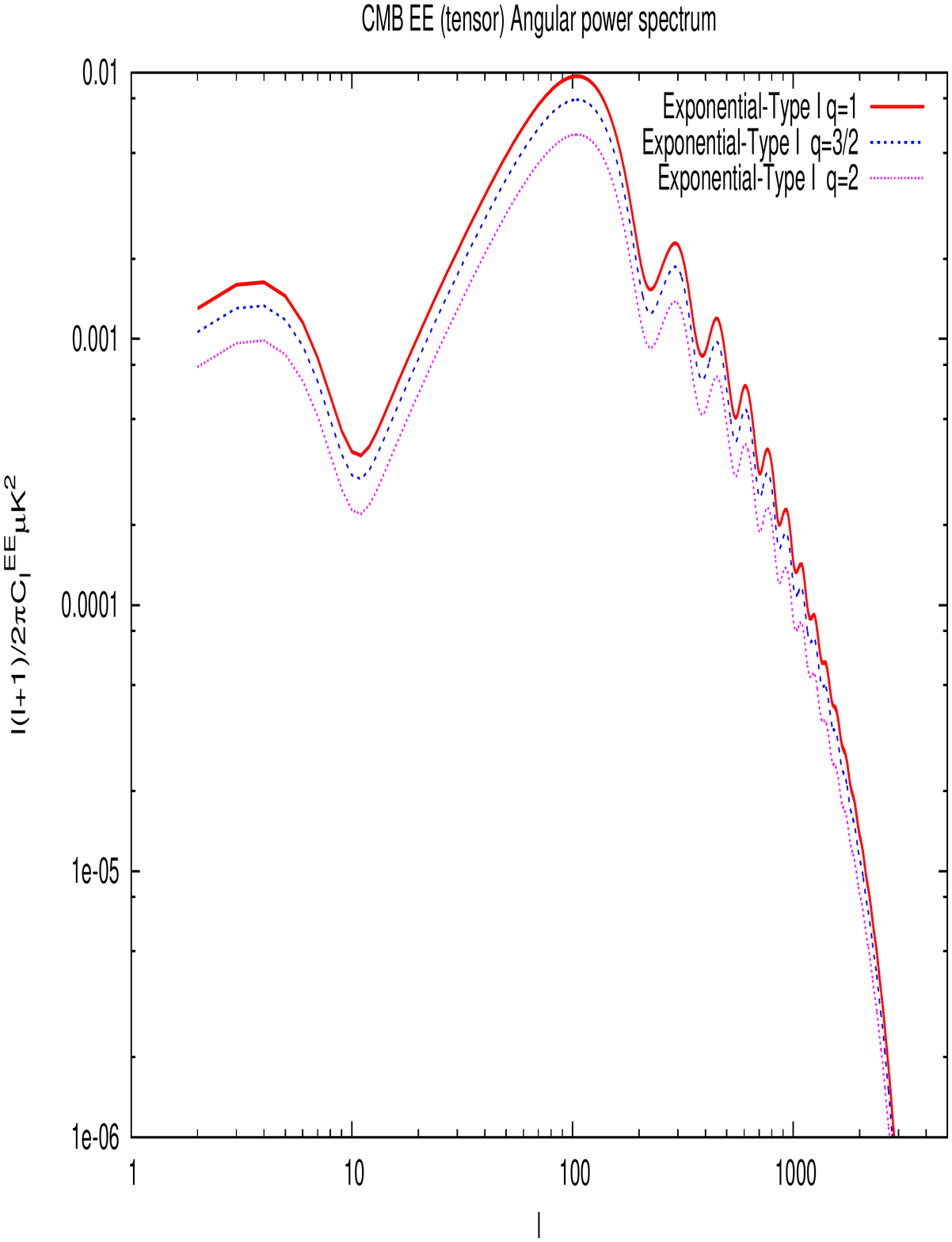}
    \label{EEtm3}
}
\subfigure[$l(l+1)C^{EE}_{l}/2\pi$~vs~$l$~(tensor)]{
    \includegraphics[width=7.2cm, height=5.7cm] {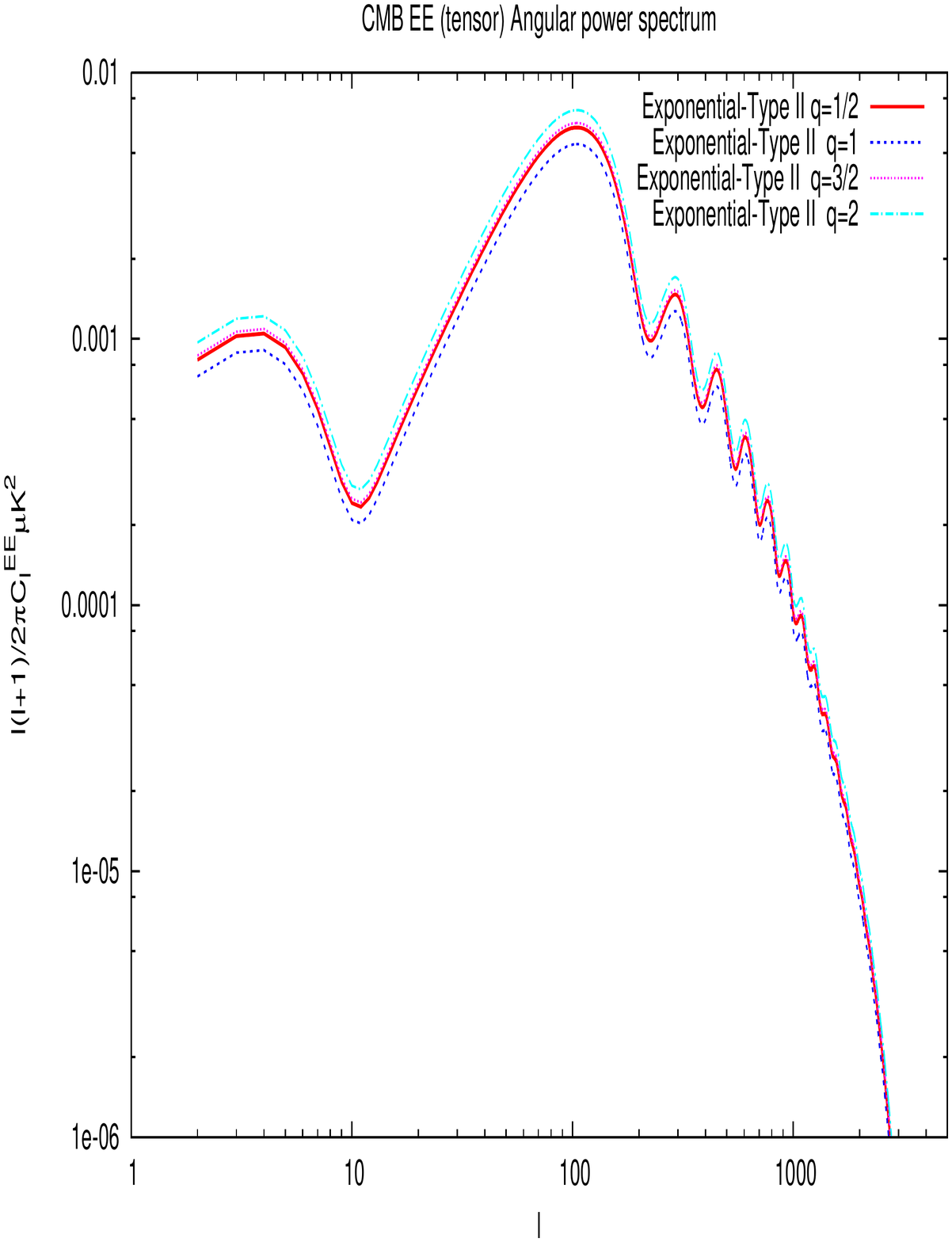}
    \label{EEtm4}
}
\subfigure[$l(l+1)C^{EE}_{l}/2\pi$~vs~$l$~(tensor)]{
    \includegraphics[width=7.2cm, height=5.7cm] {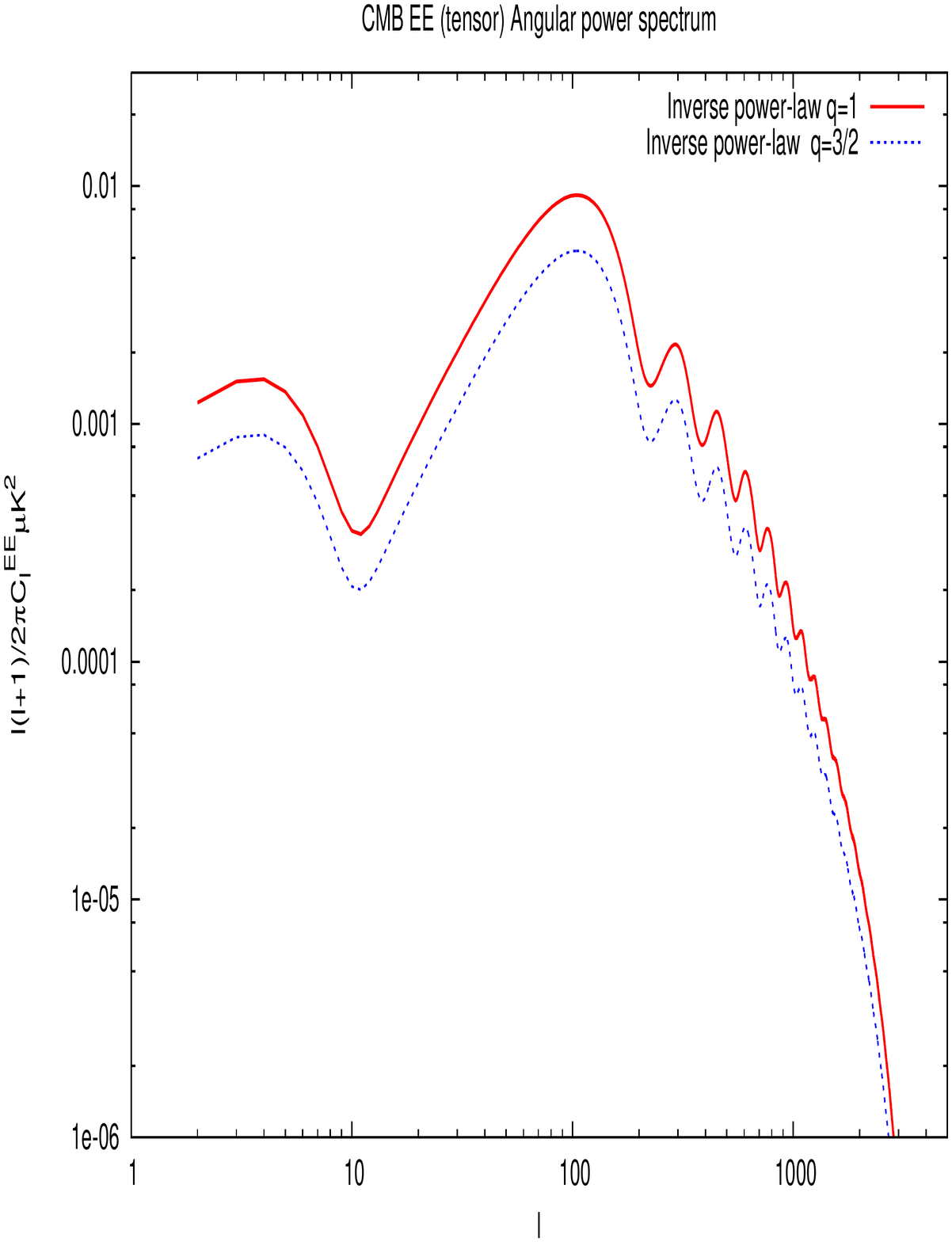}
    \label{EEtm5}
}
\caption[Optional caption for list of figures]{We show the variation of CMB EE Angular power spectrum with respect to the multipole, $l$ for tensor modes for all five tachyonic models.
}
\label{fig24}
\end{figure}

The integral (\ref{equ:CXY}) relates the inhomogeneities predicted by inflation, $\Delta(k)$, to the anisotropies observed in the CMB, $C_\ell^{XY}$. The correlations between the different $X$ and $Y$ modes
are related by the transfer functions $\Theta_{X\ell}(k)$ and $\Theta_{Y\ell}(k)$.
The transfer functions may be written as the line-of-sight integral Eq~(\ref{equ:source1}) which factorizes into physical source terms $S_X(k, \eta)$ and
 geometric projection factors $P_{X\ell}(k[\eta_0-\eta])$ through combinations of Bessel functions.

Further expressing Eq~(\ref{equ:CXY}) in terms of TT, TE, EE, BB correlation I get:
\begin{eqnarray}
 \underline{\bf For~ curvature~perturbation}:\nonumber\\
 C_\ell^{TT} &=& \frac{2}{\pi} \int k^2 dk ~{\Delta_{S}(k)}\,  {\Theta_{T \ell}(k) \Theta_{T \ell}(k)} \, ,~~~~~~~~~~~~\\
C_\ell^{TE} &=& (4\pi)^2 \int k^2 dk ~{\Delta_{S}(k)}\,  {\Theta_{T \ell}(k) \Theta_{E \ell}(k)} \, ,\\
C_\ell^{EE} &=& (4\pi)^2 \int k^2 dk ~{\Delta_{S}(k)}\,  {\Theta_{E \ell}(k) \Theta_{E \ell}(k)} \, ,\\
\underline{\bf For~ tensor~perturbation}:\nonumber\\
C_\ell^{BB} &=& (4\pi)^2 \int k^2 dk ~{\Delta_{h}(k)}\,  {\Theta_{B \ell}(k) \Theta_{B \ell}(k)} \,~~~~~~~~~~~~\\
C_\ell^{TT} &=& \frac{2}{\pi} \int k^2 dk ~{\Delta_{h}(k)}\,  {\Theta_{T \ell}(k) \Theta_{T \ell}(k)} \, ,\\
C_\ell^{TE} &=& (4\pi)^2 \int k^2 dk ~{\Delta_{h}(k)}\,  {\Theta_{T \ell}(k) \Theta_{E \ell}(k)} \, ,\\
C_\ell^{EE} &=& (4\pi)^2 \int k^2 dk ~{\Delta_{h}(k)}\,  {\Theta_{E \ell}(k) \Theta_{E \ell}(k)} \,.
\end{eqnarray}

where the inflationary power spectra $\{ \Delta_\zeta (k), \Delta_h (k) \}$ are parametrized at any arbitrary momentum scale $k$ as~\footnote{However, it is important to mention here that 
if we know the mode scalar and tensor mode functions, which are obtained from the the exact solution of Mukhanov-Sasaki equation for exactly Bunch-Davies vacuum or for any arbitrary vacuum then one can implement the exact structural form of the 
primordial power spectra. But within the slow-roll regime of inflation the presented version of the parametrization of the power spectra is exactly compatible with the exact power 
spectrum from the solution of the Mukhanov-Sasaki equation.}:
\begin{eqnarray}
 \Delta_{\zeta}(k)&=& \Delta_{\zeta,\star}\left(\frac{k}{c_{S}k_{\star}}\right)^{n_{\zeta,\star}-1+\frac{\alpha_{\zeta,\star}}{2}\ln\left(\frac{k}{c_{S}k_{\star}}\right)
+\frac{\kappa_{\zeta,\star}}{6}\ln^{2}\left(\frac{k}{c_{S}k_{\star}}\right)+....}\,,\\
 \Delta_{h}(k)&=& r(k)\Delta_{\zeta}(k)=\Delta_{h,\star}\left(\frac{k}{k_{\star}}\right)^{n_{h,\star}+\frac{\alpha_{h,\star}}{2}\ln\left(\frac{k}{k_{\star}}\right)
+\frac{\kappa_{h,\star}}{6}\ln^{2}\left(\frac{k}{k_{\star}}\right)+....}\,.
\end{eqnarray}
It is important to note that the cosmological significance of the $E$ and $B$ decomposition of CMB
polarization carries the following significant features:
\begin{itemize}
\item Curvature (density) perturbations create only polarizing $E$-modes.
\item Vector (vorticity) perturbations create mainly $B$-modes. However, it is important note that here that the contributions of vector modes decay
 with the expansion of the universe and are therefore sub-dominant at the epoch of recombination. For this reason we have neglected such sub-dominant 
effects from our rest of the analysis.
\item Tensor (gravitational wave) perturbations create both $E$-modes and $B$-modes. But to quantify the exact contribution of the primordial gravitational waves, 
specifically in the inflationary $B$ modes, one need to separate all the other significant contributions in the $B$ modes i.e. primordial magnetic field, gravitational lensing, non-Gaussianity etc.
But at present in cosmological literature no such sophisticated techniques or algorithms are available using which one can separate all of these significant contributions completely. 
\end{itemize}

To compute the momentum integrals numerically and to analyze the various features of CMB angular power spectra from all of the inflationary models mentioned in the last section, here we use a semi-analytical code ``CAMB''. 
For the numerical analysis we use here the best fit model parameters corresponding to all of the five tachyonic models, which are compatible with Planck 2015 data. 
Additionally we take $\Lambda$CDM background. Further we put all these inputs to ``CAMB'' and modify the
 inbuilt parameterization of power spectrum for scalar and tensor 
modes accordingly. After performing all the numerical computations via ``CAMB'' finally from our analysis we have
generated all the theoretical CMB angular power spectra from which we observed the following significant features:-
\begin{itemize}
\item In Fig.~(\ref{fig18}),
at low $\ell$ region $(2<l<49)$ the contributions from the running ($\alpha_{\zeta},\alpha_{h}$), and running of running ($\kappa_{\zeta},\kappa_{h}$) are very small.
Their additional contribution to the CMB power spectrum for scalar and tensor modes becomes unity $({\cal O}(1))$ within low-$l$ region
 and the original
 power spectrum becomes unchanged. As a result the tachyonic models will be well fitted with the CMB TT spectrum at low-$l$ region within 
 high cosmic variance as observed by Planck except for a few outliers according to the Planck 2013 data release~\footnote{From Fig.~(\ref{fig18}) it is observed that for Inverse cosh and Logarithmic model $q=1/2$,  Exponential
 Type-I and Type II 
 model $q=1$, Inverse power-law model $q=1$ and $q=3/2$ generalized characteristic indices are well fitted with Planck 2015 data.}. 
 But to update our analysis with the latest Planck 2015 data set we have not shown such high cosmic variance explicitly. In case of WMAP9 data
 for CMB TT spectrum the cosmic variance in the low-$l$ region is not very large compared to the 
Planck low-$l$ data. But our tachyonic models is also well fitted with WMAP9 low-$l$ data also which we have not shown explicitly in the plot to update the analysis using Planck 2015 data.
It is also important to mention that if we incorporate the uncertainties 
in the scanning multipole $l$ for the measurement of CMB TT spectrum, then also our prescribed analysis is pretty consistent with the 
Planck 2015 data. Further if we move
 towards high $\ell$ regime ($47<l<2500$) the contribution of running and running of running become stronger and
 this will enhance the power spectrum to a permissible value such that it will accurately fit high-$l$ 
data within very small cosmic variance as observed by Planck. In this way one can easily scan over all the multipoles starting from low-$l$ to high-$l$ using the
same momentum dependent parameterization of tensor-to-scalar ratio.

\item From Fig.~(\ref{fig18}), we see that the Sachs-Wolfe plateau obtained
from our proposed tachyonic models is non flat, confirming the appearance of running, and running of the running in the spectrum 
observed for low $l$ region ($l<47$).
 For larger value of the multipole ($47<l<2500$),
CMB anisotropy spectrum is dominated by the Baryon
Acoustic Oscillations (BAO), giving rise to several ups and
downs in the CMB TT spectrum. In the low $l$ region due to the presence of very 
large cosmic variance there may be other pre-inflationary scenarios which might be able to describe the TT-power spectrum better. 
In my study we have considered only the possibility for which the behaviour of tachyonic models is analyzed for both 
low and high $l$ regions. 

\item From Fig.~(\ref{fig18}), Fig.~(\ref{fig19}), Fig.~(\ref{fig20}), we observe that if we include the uncertainties in multipole $l$ as well in the observed CMB angular power spectra 
then the proposed tachyonic models is pretty consistent with the CMB TT, TE, EE for scalar mode
from Planck 2015 data. 

\item In Fig.~(\ref{fig21}), Fig.~(\ref{fig22}), Fig.~(\ref{fig23}) and Fig.~(\ref{fig24}) we have explicitly shown the 
theoretical CMB BB, TT, TE and EE angular power spectrum from tensor mode. Most importantly, if inflationary paradigm is responsible
for the nearly de-Sitter expansion of the early universe then the CMB BB spectra for tensor modes is one of the prime components trough which one can detect the 
contribution for primordial gravitational waves via tensor-to-scalar ratio. But till date only the contribution from the lensing B-modes are detected via South Pole Telescope and Planck 2015
+BICEP2/Keck Array joint mission.
But confirming the sole inflationary origin from the detection of the de-lensed version of the signal is not sufficient enough to draw any final conclusion~\footnote{In this respect one may consider the alternative frameworks of 
inflationary paradigm as well.}. There are other possibilities as
well through which it is possible to generate 
CMB B-modes, those components are, Primordial Magnetic Field, Gravitational Lensing, CP asymmetry in the lepton sector 
of particle physics etc. 
  \end{itemize}
\subsection{Computation for Assisted inflation}
\label{aa5b}
In case of assisted inflation \cite{Liddle:1998jc,Copeland:1999cs,Malik:1998gy,Piao:2002vf} all the tachyons would follow a similar 
trajectory with a unique late time attractor~\footnote{We also suggest the readers to read ref.~\cite{Dimopoulos:2005ac} for completeness.}. In more technical language one can 
state that:
\bea T_{1}\sim T_{2}\sim\cdots T_{M}\equiv T_{i}=T ~~\forall i=1,2,\cdots,M.\eea
Most importantly the detailed computation of the density and tensor perturbation
responsible for the anisotropy of CMB depends on this late time attractor behavior of the fields.
\subsubsection{Condition for inflation}
For assisted tachyonic inflation, the prime condition for inflation is given by:
\bea
\dot{H}+H^{2}&=&\left(\frac{\ddot{a}}{{a}}\right)=-\sum^{M}_{i=1}\frac{(\rho_{i}+3p_{i})}{6 M^2_p}=\frac{(\rho+3p)}{6 M^2_p}>0
\eea
which can be re-expressed in terms of the following constraint condition in the context of assisted tachyonic inflation:
\be\begin{array}{lll}\label{g39v2}
	\displaystyle \sum^{M}_{i=1}\frac{V(T_i)}{3M^{2}_{p}\sqrt{1-\alpha^{'}\dot{T}^2_i}}\left(1-\frac{3}{2}\alpha^{'}\dot{T}^2_i\right)=\frac{MV(T)}{3M^{2}_{p}\sqrt{1-\alpha^{'}\dot{T}^2}}\left(1-\frac{3}{2}\alpha^{'}\dot{T}^2\right)
	>0.
\end{array}\ee
Here Eq~(\ref{g39v2}) implies that to satisfy inflationary constraints in the slow-roll regime the following constraint always holds good:
\bea \label{wa1v2}\dot{T}&<& \sqrt{\frac{2}{3\alpha^{'}}}~~\forall i=1,2,\cdots,M,\\ 
\label{wa2v2}\ddot{T}&<&  3H \dot{T}<\sqrt{\frac{6}{\alpha^{'}}}H~~\forall i=1,2,\cdots,M.\eea
Consequently the field equations are approximated as:
\be\begin{array}{lll}\label{e13xv2}
	\displaystyle 
	3H\alpha^{'}\dot{T}_{i}+\frac{dV(T_{i})}{V(T_{i})dT_{i}}\approx 0~~\forall i=1,2,\cdots,M \,,
\end{array}\ee
Similarly, in the most generalized case, 
\be\begin{array}{lll}\label{g39v3}
	\displaystyle \sum^{M}_{i=1}\frac{V(T_{i})}{3M^{2}_{p}\left(1-\alpha^{'}\dot{T}^2_{i}
		\right)^{1-q}}\left(1-(1+q)\alpha^{'}\dot{T}^2_{i}\right)=\frac{MV(T)}{3M^{2}_{p}\left(1-\alpha^{'}\dot{T}^2
		\right)^{1-q}}\left(1-(1+q)\alpha^{'}\dot{T}^2\right)>0.
\end{array}\ee
Here Eq~(\ref{g39v3}) implies that to satisfy inflationary constraints in the slow-roll regime the following constraint always holds good:
\bea \dot{T}&<& \sqrt{\frac{1}{\alpha^{'}(1+q)}}~~\forall i=1,2,\cdots,M,\\ 
\ddot{T}&<&  3H \dot{T}<\sqrt{\frac{9}{\alpha^{'}(1+q)}}H~~\forall i=1,2,\cdots,M.\eea
Consequently the field equations are approximated as`:
\be\begin{array}{lll}\label{e14xxvv3}
	\displaystyle 
	6q\alpha^{'}H\dot{T}_{i}+\frac{dV(T_{i})}{V(T_{i})dT_{i}}\approx 0~~\forall i=1,2,\cdots,M \,
\end{array}\ee
Also for both the cases in the slow-roll regime the Friedmann equation is modified as:
\bea\label{qxcc1v3} H^{2}&\approx& \sum^{M}_{i=1}\frac{V(T_{i})}{3M^{2}_{p}}=\frac{MV(T)}{3M^2_p}.\eea
The equation of motions can be mapped to the equations of a model with single tachyonic field using the following redefinitions:
\bea \tilde{T}&=&\sqrt{M} T,\\
\tilde{V}&=&V(\tilde{T})=M V(T)=\sum^{M}_{i=1}V(T_{i}).\eea
Now to show the late time attractor behaviour of the solution of scalar fields we keep $T_{1}$, but replace the rest of the part with the redefined fields:
\bea\label{gtk1} \chi_{i}&=& T_{i}-T_{1}~~~~~\forall i=2,\cdots,M.\eea
Using Eq~(\ref{gtk1}), at late times the equation of motion can be recast as:
\be\begin{array}{lll}\label{e13}
 \displaystyle 0 =\left\{\begin{array}{lll}
                    \displaystyle  
                    \frac{1}{\sqrt{-g}}\partial_{\mu}\left(\sqrt{-g}V_{eff}(T_{1},\chi_{i})\sqrt{1+g^{\alpha\beta}\partial_{\alpha}\chi_{i}\partial_{\beta}\chi_{i}}\right)\\
                    \displaystyle~~~~~~~=\frac{\alpha^{'}\ddot{\chi}_{i}}{\left(1-\alpha^{'}\dot{\chi}^2_{i}\right)}
+3H\alpha^{'}\dot{\chi}_{i}+\frac{dV_{eff}(T_{1},\chi_{i})}{V_{eff}(T_{1},\chi_{i})d\chi_{i}} \,,~~~~ &
 \mbox{\small {\bf for {$q=1/2$ }}}  \\ 
 \displaystyle  
 \frac{1}{\sqrt{-g}}\partial_{\mu}\left(\sqrt{-g}V_{eff}(T_{1},\chi_{i})\left(1+g^{\alpha\beta}\partial_{\alpha}\chi_{i}\partial_{\beta}\chi_{i}\right)^q\right)\\
                    \displaystyle~~~~~~~=\frac{2q\alpha^{'}\ddot{\chi}_{i}}{\left(1-\alpha^{'}\dot{\chi}^2_{i}\right)}
+6q\alpha^{'}H\frac{\dot{\chi}_{i}}{1-\alpha^{'}(1-2q)\dot{\chi}^2_{i}}+\frac{dV_{eff}(T_{1},\chi_{i})}{V_{eff}(T_{1},\chi_{i})d\chi_{i}} \,,~~~ &
 \mbox{\small {\bf for {any $q$ }}}
          \end{array}
\right.
\end{array}\ee
where $V_{eff}(T_{1},\chi_{i})$ represents the effective potential. The minimum in the $\chi_{i}$ direction is always appearing at $\chi_{i}=0$, regardless of the explicit behaviour of the 
tachyon field $T_{1}$. Consequently the late time solutions has all the $T_{i}$ equal. From this analysis it is also observed that the length of the time interval to reach this attractor behaviour will
depend on the initial separation i.e. the tachyon field value $\chi_{i}$ and the extent of frictional contribution coming from the expansion rate $H$.

It is important to mention that, for assisted inflation the equation of motion for each tachyon fields follows the following simple relationship:
\bea \frac{d\ln T_{i}}{dt}\simeq \frac{d\ln T}{dt}~~\forall i=1,2,\cdots,M.\eea
Further substituting Eq~(\ref{qxcc1v3}) in Eq~(\ref{e13xv2}) and Eq~(\ref{e14xxvv3}) we get:
\bea\label{e13gvxv4}
\displaystyle 
\frac{\sqrt{3MV(T)}}{M_{p}}\alpha^{'}\dot{T}_{i}+\frac{dV(T_{i})}{V(T_{i})dT_{i}}\approx 0 \,,\\
\label{e13gvvx} 6q\frac{\sqrt{MV(T)}}{\sqrt{3}M_{p}}\alpha^{'}\dot{T}_{i}+\frac{dV(T_{i})}{V(T_{i})dT_{i}}\approx 0 \,.
\eea
Finally the general solution for both the cases can be expressed in terms of the single field tachyonic potential $V(T)$ as:
\bea\label{e13gvxc1v5}
\displaystyle 
\label{bw1}t-t_{in}\approx -\frac{\sqrt{3M}\alpha^{'}}{M_{p}}\int^{T_{i}}_{T_{in}}dT_{i}~\frac{\sqrt{V(T)}V(T_{i})}{V^{'}(T_{i})} \,,\\
\label{bw2} t-t_{in}\approx -\frac{6q\sqrt{M}\alpha^{'}}{\sqrt{3}M_{p}}\int^{T_{i}}_{T_{in}}dT_{i}~\frac{\sqrt{V(T)}V(T_{i})}{V^{'}(T_{i})}.
\eea
Further using Eq~(\ref{bw3}), Eq~(\ref{bw4}) and Eq~(\ref{qxcc1v3}) we get the following solution for the scale factor 
in terms of the tachyonic field for usual $q=1/2$ and for generalized value of $q$ as:
\be\begin{array}{lll}\label{scss1}
	\displaystyle  a= a_{in}\times \left\{\begin{array}{lll}
		\displaystyle  
		\exp\left[-\frac{\alpha^{'}M}{M^2_p}\int^{T_{i}}_{T_{in}}dT_{i}~\frac{V(T)V(T_{i})}{V^{'}(T_{i})}\right]\,,~~~~~~ &
		\mbox{\small {\bf for {$q=1/2$ }}}  \\ 
		\displaystyle   
		\exp\left[-\frac{2q\alpha^{'}M}{M^2_p}\int^{T_{i}}_{T_{in}}dT_{i}~\frac{V(T)V(T_i)}{V^{'}(T_{i})}\right]\,.~~~~~~ &
		\mbox{\small {\bf for {~any~arbitrary~ $q$ }}} 
	\end{array}
	\right.
\end{array}\ee
\subsubsection{Analysis using Slow-roll formalism}
Here our prime objective is to define slow-roll parameters for tachyon inflation in terms of the Hubble parameter and the assisted tachyonic inflationary potential. Using the slow-roll 
approximation one can expand various cosmological observables in terms of small dynamical quantities derived from the appropriate derivatives of the Hubble parameter and of the inflationary potential. 
To start with here we use the horizon-flow parameters based on derivatives of
Hubble parameter with respect to the number of e-foldings $N$, defined as:
\bea\label{e-fold}
N(t)&=&\int^{t_{end}}_{t}H(t)~dt,\eea
where $t_{end}$ signifies the end of inflation. Further using Eq~(\ref{e13x}), Eq~(\ref{e14xx}), Eq~(\ref{qxcc1}) and eq~(\ref{e-fold}) we get:
\be\begin{array}{lll}\label{vcvv23}
	\displaystyle  \frac{dT_i}{dN}=\frac{\dot{T}_i}{H}=\left\{\begin{array}{lll}
		\displaystyle  
		-\frac{2H^{'}}{3\alpha^{'}H^{3}}\,,~~~~~~ &
		\mbox{\small {\bf for {$q=1/2$ }}}  \\ 
		\displaystyle   
		-\frac{2H^{'}}{3\alpha^{'}H^{3}}\left(\frac{1-\alpha^{'}(1-2q)\dot{T}^2_i}{2q}\right)\,.~~~~~~ &
		\mbox{\small {\bf for {~any~arbitrary~ $q$ }}} 
	\end{array}
	\right.
\end{array}\ee
where $H^{'}>0$ which makes always $\dot{T}_i>0$ during inflationary phase.
Further using Eq~(\ref{vcvv23}) we get the following differential operator identity for tachyonic inflation:
\be\begin{array}{lll}\label{vcv2v2}
	\displaystyle  \frac{1}{H}\frac{d}{dt}=\frac{d}{dN}=\frac{d}{d\ln k}=\left\{\begin{array}{lll}
		\displaystyle  
		-\frac{2H^{'}}{3\alpha^{'}H^{3}}\frac{d}{dT_i}\,,~~~~~~ &
		\mbox{\small {\bf for {$q=1/2$ }}}  \\ 
		\displaystyle   
		-\frac{2H^{'}}{3\alpha^{'}H^{3}}\left(\frac{1-\alpha^{'}(1-2q)\dot{T}^2_i}{2q}\right)\frac{d}{dT_i}\,.~~~~~~ &
		\mbox{\small {\bf for {~any~arbitrary~ $q$ }}} 
	\end{array}
	\right.
\end{array}\ee
Further using the differential operator identity as mentioned in Eq~(\ref{vcv2v2}) we get the following Hubble flow equation for tachyonic inflation for $j\geq 0$ :
\be\begin{array}{lll}\label{vcv4v2}
	\displaystyle  \frac{1}{H}\frac{d\epsilon_{i}}{dt}=\frac{d\epsilon_{j}}{dN}=\epsilon_{j+1}\epsilon_{j}=\left\{\begin{array}{lll}
		\displaystyle  
		-\frac{2H^{'}}{3\alpha^{'}H^{3}}\frac{d\epsilon_{j}}{dT_i}\,,~~~~~~ &
		\mbox{\small {\bf for {$q=1/2$ }}}  \\ 
		\displaystyle   
		-\frac{2H^{'}}{3\alpha^{'}H^{3}}\left(\frac{1-\alpha^{'}(1-2q)\dot{T}^2_i}{2q}\right)\frac{d\epsilon_{j}}{dT_i}\,.~~~~~~ &
		\mbox{\small {\bf for {~any~arbitrary~ $q$ }}} 
	\end{array}
	\right.
\end{array}\ee             
For realistic estimate from the single field tachyonic inflationary model substituting the free index $j$ to $j=0,1,2$ in Eq~(\ref{h1}) and Eq~(\ref{vcv4v2}) we get the contributions from the first three Hubble slow-roll parameter,
which can be depicted as:
\bea\label{bp1v2}
\displaystyle  \epsilon_{1}=\frac{d\ln|\epsilon_{0}|}{dN}&=&-\frac{\dot{H}}{H^2}=\left\{\begin{array}{lll}
	\displaystyle  
	\frac{2}{3\alpha^{'}}\left(\frac{H^{'}}{H^2}\right)^{2}=\frac{3}{2}\alpha^{'}\dot{T}^{2}_i\,,~~~~ &
	\mbox{\small {\bf for {$q=1/2$ }}}  \\ 
	\displaystyle   
	\frac{2}{3\alpha^{'}}\left(\frac{H^{'}}{H^2}\right)^{2}\left(\frac{1-\alpha^{'}(1-2q)\dot{T}^2_i}{2q}\right)\\
	\displaystyle~~~~~~~~~~~~=\frac{3}{2}\alpha^{'}\dot{T}^{2}_i \left(\frac{2q}{1-\alpha^{'}(1-2q)\dot{T}^2_i}\right)\,.~~~~ &
	\mbox{\small {\bf for {~any~arbitrary~ $q$ }}} 
\end{array}
\right. \eea \bea \label{bp2v2}
\displaystyle  \epsilon_{2}=\frac{d\ln|\epsilon_{1}|}{dN}&=&\frac{\ddot{H}}{H\dot{H}}+2\epsilon_{1}=\left\{\begin{array}{lll}
	\displaystyle  
	\sqrt{\frac{2}{3\alpha^{'}\epsilon_{1}}}\frac{\epsilon^{'}_{1}}{H}=2\frac{\ddot{T}_i}{H\dot{T}_i}\,,~~~~ &
	\mbox{\small {\bf for {$q=1/2$ }}}  \\ 
	\displaystyle   
	\sqrt{\frac{2}{3\alpha^{'}\epsilon_{1}}\left(\frac{1-\alpha^{'}(1-2q)\dot{T}^2_i}{2q}\right)}\frac{\epsilon^{'}_{1}}{H}\\
	\displaystyle~~~~~~~=\frac{2\ddot{T}_i}{H\dot{T}\left(1-\alpha^{'}(1-2q)\dot{T}^2_i\right)}\,.~~~~ &
	\mbox{\small {\bf for {~any~arbitrary~ $q$ }}} 
\end{array}
\right. \eea \bea
\label{bp3v2}
\displaystyle  \epsilon_{3}=\frac{d\ln|\epsilon_{2}|}{dN}&=&\frac{1}{\epsilon_{2}}\left[\frac{\dddot{H}}{H^2 \dot{H}}-3\frac{\ddot{H}}{H^3}-\frac{\ddot{H}^2}{H^2 \dot{H}^2}+4\frac{\dot{H}^2}{H^4}\right]\nonumber 
\\\nonumber \\&=&\left\{\begin{array}{lll}
	\displaystyle  
	\sqrt{\frac{2\epsilon_{1}}{3\alpha^{'}}}\frac{\epsilon^{'}_{2}}{H}=\left[\frac{2\dddot{T}_i}{H^2\dot{T}_i\epsilon_{2}}+\epsilon_{1}-\frac{\epsilon_{2}}{2}\right]\,,~~~~ &
	\mbox{\small {\bf for {$q=1/2$ }}}  \\  
	\displaystyle   
	\sqrt{\frac{2\epsilon_{1}}{3\alpha^{'}}\left(\frac{1-\alpha^{'}(1-2q)\dot{T}^2_i}{2q}\right)}\frac{\epsilon^{'}_{2}}{H}\\
	\displaystyle~~~~=\frac{\left[\frac{2\dddot{T}_i}{H^2\dot{T}_i\epsilon_{2}}+\epsilon_{1}-\frac{\epsilon_{2}}{2}\right]}{\left(1-\alpha^{'}(1-2q)\dot{T}^2_i\right)}
	+\frac{\frac{4\alpha^{'}(1-2q)\ddot{T}^2_i}{H} }{\left(1-\alpha^{'}(1-2q)\dot{T}^2_i\right)^2}\,.~~~~ &
	\mbox{\small {\bf for {~any~arbitrary~ $q$ }}} 
\end{array}
\right.
\eea
Further using Eq~(\ref{bp1v2}), Eq~(\ref{bp2v2}), Eq~(\ref{bp3v3}), Eq~(\ref{tep1}) and Eq~(\ref{tep2}) one can re-express the Hubble slow-roll parameters in terms of the potential dependent slow-roll parameter as:
\bea\label{bp1v3}\small
\displaystyle  \epsilon_{1}&\approx&\left\{\begin{array}{lll}
	\displaystyle  
	\frac{M^{2}_{p}}{2\alpha^{'}}\frac{V^{'2}(T_i)}{MV(T)V^{2}(T_i)}=\frac{\epsilon_{V}(T_i)}{MV(T)\alpha^{'}}\equiv\bar{\epsilon}_{V}(T_i)\,,~~~~~~~~~~~~~~~~~~~~~~~~~~~~~~~~~~~~ &
	\mbox{\small {\bf for {$q=1/2$ }}}  \\ 
	\displaystyle   
	\frac{M^{2}_{p}}{4q\alpha^{'}}\frac{V^{'2}(T_i)}{MV(T)V^{2}(T_i)}=\frac{\epsilon_{V}(T_i)}{2qMV(T)\alpha^{'}}\equiv\frac{\bar{\epsilon}_{V}(T_i)}{2q}\,.~~~~ &
	\mbox{\small {\bf for {~any~ $q$ }}} 
\end{array}
\right.\\ \nonumber \label{bp2v3}
\displaystyle  \epsilon_{2}&\approx&\left\{\begin{array}{lll}
	\displaystyle  
	\frac{M^2_{p}}{\alpha^{'}M}\left(3\frac{V^{'2}(T_i)}{V(T)V^{2}(T_i)}-2\frac{V^{''}(T_i)}{V(T)V(T_i)}\right)\\
	\displaystyle~~~~~~~~~
	=\frac{2\left(3\epsilon_{V}(T_i)-\eta_{V}(T_i)\right)}{MV(T)\alpha^{'}}=2\left(3\bar{\epsilon}_{V}(T_i)-\bar{\eta}_{V}(T_i)\right)\,,~~~~~~~~~~~~~~~~~~ &
	\mbox{\small {\bf for {$q=1/2$ }}}  \\ 
	\displaystyle   
	\frac{M^2_{p}}{\sqrt{2q}\alpha^{'}M}\left(3\frac{V^{'2}(T_i)}{V(T)V^{2}(T_i)}-2\frac{V^{''}(T_i)}{V(T)V(T_i)}\right)\\
	\displaystyle~~~~~~~~~
	=\frac{\sqrt{\frac{2}{q}}\left(3\epsilon_{V}(T_i)-\eta_{V}(T_i)\right)}{MV(T)\alpha^{'}}=\sqrt{\frac{2}{q}}\left(3\bar{\epsilon}_{V}(T_i)-\bar{\eta}_{V}(T_i)\right)\,.~~~~~~~~~~~~~~~~~~ &
	\mbox{\small {\bf for {~any~ $q$ }}} 
\end{array}
\right.\eea\bea
\label{bp3v3}
\displaystyle  \epsilon_{3}\epsilon_{2}&\approx&\left\{\begin{array}{lll}
	\displaystyle  
	\frac{M^{4}_{p}}{M^2V^{2}(T)\alpha^{'2}}\left(2\frac{V^{'''}(T_i)V^{'}(T_i)}{V^{2}(T_i)}-10\frac{V^{''}(T_i)V^{'2}(T_i)}{V^{3}(T_i)}+9\frac{V^{'4}(T_i)}{V^{4}(T_i)}\right)\\
	\displaystyle~~~~~ =\frac{\left(2\xi^2_{V}(T_i)-5\eta_{V}(T_i)\epsilon_{V}(T_i)+36\epsilon^2_{V}(T_i)\right)}{M^2V^{2}(T)\alpha^{'2}}\\ \displaystyle~~~~~=\left(2\bar{\xi}^2_{V}(T_i)-5\bar{\eta}_{V}(T_i)\bar{\epsilon}_{V}(T_i)+36\bar{\epsilon}^2_{V}(T_i)\right)\,,~~~~ &
	\mbox{\small {\bf for {$q=1/2$ }}}  \\ 
	\displaystyle   
	\frac{M^{4}_{p}}{M^2\sqrt{2q}V^{2}(T)\alpha^{'2}}\left(2\frac{V^{'''}(T_i)V^{'}(T_i)}{V^{2}(T_i)}-10\frac{V^{''}(T_i)V^{'2}(T_i)}{V^{3}(T_i)}+9\frac{V^{'4}(T_i)}{V^{4}(T_i)}\right)\\
	\displaystyle ~~~~~=\frac{\left(2\xi^2_{V}(T_i)-5\eta_{V}(T_i)\epsilon_{V}(T_i)+36\epsilon^2_{V}(T_i)\right)}{M^2\sqrt{2q} V^{2}(T)\alpha^{'2}}\\ \displaystyle~~~~~=\frac{\left(2\bar{\xi}^2_{V}(T_i)-5\bar{\eta}_{V}(T_i)\bar{\epsilon}_{V}(T_i)+36\bar{\epsilon}^2_{V}(T_i)\right)}{\sqrt{2q}}\,.~~~~ &
	\mbox{\small {\bf for {~any~ $q$ }}} 
\end{array}
\right.
\eea
where $'=d/dT_{i}$ and the potential dependent slow-roll parameters $\epsilon_{V}, \eta_{V}, \xi^{2}_{V}, \sigma^{3}_{V}$ are defined as:
\bea \label{wzq1v2} \epsilon_{V}(T_{i})&=&\frac{M^{2}_{p}}{2}\left(\frac{V^{'}(T_{i})}{V(T_{i})}\right)^2,\\
\label{wzq2v2} \eta_{V}(T_{i})&=& M^{2}_{p}\left(\frac{V^{''}(T_{i})}{V(T_{i})}\right),\\
\label{wzq3v2} \xi^{2}_{V}(T_{i})&=& M^{4}_{p}\left(\frac{V^{'}(T_{i})V^{'''}(T_i)}{V^2(T_{i})}\right),\\
\label{wzq4v2} \sigma^{3}_{V}(T_{i})&=& M^{6}_{p}\left(\frac{V^{'2}(T_{i})V^{''''}(T_{i})}{V^3(T_{i})}\right),
\eea
which is exactly similar to the expression for the slow-roll parameter as appearing in the context of single field slow-roll inflationary models.
However, for the sake of clarity here we introduce new sets of potential dependent slow-roll parameters for tachyonic inflation by rescaling with the appropriate powers of $\alpha^{'}V(T)$:
\bea \label{wzq11v2} \bar{\epsilon}_{V}(T_{i})&=&\frac{\epsilon_{V}(T_{i})}{M\alpha^{'}V(T)}=\frac{M^{2}_{p}}{2M\alpha^{'}V(T)}\left(\frac{V^{'}(T_{i})}{V(T_{i})}\right)^2=\frac{\bar{\epsilon}_{V}}{M},\\
\label{wzq22v2} \bar{\eta}_{V}(T_{i})&=& \frac{{\eta}_{V}(T_{i})}{M\alpha^{'}V(T)}=\frac{M^{2}_{p}}{M\alpha^{'}V(T)}\left(\frac{V^{''}(T_{i})}{V(T_{i})}\right)=\frac{\bar{\eta}_{V}}{M},\\
\label{wzq33v2} \bar{\xi}^{2}_{V}(T_{i})&=& \frac{\xi^{2}_{V}(T_{i})}{M^2\alpha^{'2}V^2(T)}=\frac{M^{4}_{p}}{M^2\alpha^{'2}V^2(T)}\left(\frac{V^{'}(T_{i})V^{'''}(T_{i})}{V^2(T_{i})}\right)=\frac{\bar{\xi}^{2}_{V}}{M^2},\\
\label{wzq44v2} \bar{\sigma}^{3}_{V}(T_{i})&=& \frac{\sigma^{3}_{V}(T_{i})}{M^3\alpha^{'3}V^3(T)}=\frac{M^{6}_{p}}{M^2\alpha^{'3}V^3(T)}\left(\frac{V^{'2}(T_{i})V^{''''}(T_{i})}{V^3(T_{i})}\right)=\frac{\bar{\sigma}^{3}_{V}}{M^3}.
\eea
where $\bar{\epsilon}_{V}$, $\bar{\eta}_{V}$, $\bar{\xi}^{2}_{V}$ and $\bar{\sigma}^{3}_{V}$ are the single field tachyonic slow-roll parameters.
Further using Eq~(\ref{wzq1v2})-Eq~(\ref{wzq44v2}) we get the following operator identity for tachyonic inflation:
\be\begin{array}{lll}\label{vchj2v2}
	\displaystyle  \frac{1}{H}\frac{d}{dt}=\frac{d}{dN}=\frac{d}{d\ln k}\approx\left\{\begin{array}{lll}
		\displaystyle  
		\sqrt{\frac{2\bar{\epsilon}_{V}(T_i)}{MV(T)\alpha^{'}}}M_{p}\left(1-\frac{2}{3}\bar{\epsilon}_{V}(T_i)\right)^{1/4}\frac{d}{dT_i}\,,~~~~~~ &
		\mbox{\small {\bf for {$q=1/2$ }}}  \\ 
		\displaystyle   
		\sqrt{\frac{2\bar{\epsilon}_{V}(T_i)}{MV(T)\alpha^{'}}}\frac{M_{p}}{2q}\left(1-\frac{1}{3q}\bar{\epsilon}_{V}(T_i)\right)^{1/4}\frac{d}{dT_i}\,.~~~~~~ &
		\mbox{\small {\bf for {~any~arbitrary~ $q$ }}} 
	\end{array}
	\right.
\end{array}\ee
Finally using Eq~(\ref{vchj2v2}) we get the following sets of flow equations in the context of tachyoinc inflation:
\bea\label{bp11v2}
\displaystyle  \frac{d\epsilon_{1}}{dN}&=&\left\{\begin{array}{lll}
	\displaystyle  
	\frac{d\bar{\epsilon}_{V}(T_i)}{dN}=\frac{2}{M^2}\bar{\epsilon}_{V}\left(\bar{\eta}_{V}-3\bar{\epsilon}_{V}\right)\left(1-\frac{2}{3}\frac{\bar{\epsilon}_{V}}{M}\right)^{1/4}\,,~~~~~~~~~~~~~~~~~~~~ &
	\mbox{\small {\bf for {$q=1/2$ }}}  \\ 
	\displaystyle   
	\frac{1}{2q}\frac{d\bar{\epsilon}_{V}(T_i)}{dN}=\frac{\bar{\epsilon}_{V}}{qM^2}\left(\bar{\eta}_{V}-3\bar{\epsilon}_{V}\right)\left(1-\frac{1}{3q}\frac{\bar{\epsilon}_{V}}{M}\right)^{1/4}\,.~~~~ &
	\mbox{\small {\bf for {~any~ $q$ }}} 
\end{array}
\right.\eea \bea \label{bp22v2}
\displaystyle   \frac{d\epsilon_{2}}{dN}&=&\left\{\begin{array}{lll}
	\displaystyle  
	\frac{2}{M^2}\left(10\bar{\epsilon}_{V}\bar{\eta}_{V}-18\bar{\epsilon}^2_{V}-\bar{\xi}^2_{V}\right)\left(1-\frac{2}{3}\frac{\bar{\epsilon}_{V}}{M}\right)^{1/4}\,,~~~~~~~~~~~~~~~~~~~ &
	\mbox{\small {\bf for {$q=1/2$ }}}  \\ 
	\displaystyle   
	\frac{\sqrt{\frac{2}{q}}}{M^2}\left(10\bar{\epsilon}_{V}\bar{\eta}_{V}-18\bar{\epsilon}^2_{V}-\bar{\xi}^2_{V}\right)\left(1-\frac{1}{3q}\frac{\bar{\epsilon}_{V}}{M}\right)^{1/4}\,.~~~~ &
	\mbox{\small {\bf for {~any~ $q$ }}} 
\end{array}
\right.\eea \bea
\label{bp33v2}
\displaystyle  \frac{d(\epsilon_{2}\epsilon_{3})}{dN}&=&\left\{\begin{array}{lll}
	\displaystyle  
	\frac{1}{M^3}\left(2\bar{\sigma}^{3}_{V}-216\bar{\epsilon}^{3}_{V}+2\bar{\xi}^{2}_{V}\bar{\eta}_{V}-7\bar{\xi}^{2}_{V}\bar{\epsilon}_{V}\right.\\ \left.~~\displaystyle +194\bar{\epsilon}^2_{V}
	\bar{\eta}_{V}-10\bar{\eta}_{V}\bar{\epsilon}_{V}
	\right)\displaystyle \left(1-\frac{2}{3}\bar{\epsilon}_{V}\frac{\bar{\epsilon}_{V}}{M}\right)^{1/4}\,,~~~~ &
	\mbox{\small {\bf for {$q=1/2$ }}}  \\ 
	\displaystyle   
	\frac{1}{M^3\sqrt{2q}}\left(2\bar{\sigma}^{3}_{V}-216\bar{\epsilon}^{3}_{V}+2\bar{\xi}^{2}_{V}\bar{\eta}_{V}-7\bar{\xi}^{2}_{V}\bar{\epsilon}_{V}\right.\\ \left.~~\displaystyle +194\bar{\epsilon}^2_{V}
	\bar{\eta}_{V}-10\bar{\eta}_{V}\bar{\epsilon}_{V}
	\right)\displaystyle \left(1-\frac{1}{3q}\frac{\bar{\epsilon}_{V}}{M}\right)^{1/4}\,.~~~~ &
	\mbox{\small {\bf for {~any~ $q$ }}} 
\end{array}
\right.
\eea
where we use the following consistency conditions for re-scaled potential dependent slow-roll parameters:
\bea\label{bp1aasv2}
\displaystyle  \frac{d\bar{\epsilon}_{V}}{dN}&=&\left\{\begin{array}{lll}
	\displaystyle  
	\frac{2}{M^2}\bar{\epsilon}_{V}\left(\bar{\eta}_{V}-3\bar{\epsilon}_{V}\right)\left(1-\frac{2}{3}\frac{\bar{\epsilon}_{V}}{M}\right)^{1/4}\,,~~~~~~~~~~~~~~~~~~~~~~~~~~~~~~~ &
	\mbox{\small {\bf for {$q=1/2$ }}}  \\ 
	\displaystyle   
	\frac{2}{M^2}\bar{\epsilon}_{V}\left(\bar{\eta}_{V}-3\bar{\epsilon}_{V}\right)\left(1-\frac{1}{3q}\frac{\bar{\epsilon}_{V}}{M}\right)^{1/4}\,.~~~~ &
	\mbox{\small {\bf for {~any~ $q$ }}} 
\end{array}
\right.
\eea 
\bea\label{bp2aasv2}
\displaystyle  \frac{d\bar{\eta}_{V}}{dN}&=&\left\{\begin{array}{lll}
	\displaystyle  
	\frac{1}{M^2}\left(\bar{\xi}^2_{V}-4\bar{\epsilon}_{V}\bar{\eta}_{V}\right)\left(1-\frac{2}{3}\frac{\bar{\epsilon}_{V}}{M}\right)^{1/4}\,,~~~~~~~~~~~~~~~~~~~~~~~~~~~~~~~ &
	\mbox{\small {\bf for {$q=1/2$ }}}  \\ 
	\displaystyle   
	\frac{1}{M^2}\left(\bar{\xi}^2_{V}-4\bar{\epsilon}_{V}\bar{\eta}_{V}\right)\left(1-\frac{1}{3q}\frac{\bar{\epsilon}_{V}}{M}\right)^{1/4}\,.~~~~ &
	\mbox{\small {\bf for {~any~ $q$ }}} 
\end{array}
\right.
\eea 
\bea\label{bp3aasv2}
\displaystyle  \frac{d\bar{\xi}^2_{V}}{dN}&=&\left\{\begin{array}{lll}
	\displaystyle  
	\frac{1}{M^3}\left(\bar{\sigma}^3_{V}+\bar{\xi}^2_{V}\bar{\eta}_{V}-\bar{\xi}^2_{V}\bar{\epsilon}_{V}\right)\left(1-\frac{2}{3}\frac{\bar{\epsilon}_{V}}{M}\right)^{1/4}\,,~~~~~~~~~~~~~~~~~~~ &
	\mbox{\small {\bf for {$q=1/2$ }}}  \\ 
	\displaystyle   
	\frac{1}{M^3}\left(\bar{\sigma}^3_{V}+\bar{\xi}^2_{V}\bar{\eta}_{V}-\bar{\xi}^2_{V}\bar{\epsilon}_{V}\right)\left(1-\frac{1}{3q}\frac{\bar{\epsilon}_{V}}{M}\right)^{1/4}\,.~~~~ &
	\mbox{\small {\bf for {~any~ $q$ }}} 
\end{array}
\right.
\eea 
\bea\label{bp4aasv2}
\displaystyle  \frac{d\bar{\sigma}^3_{V}}{dN}&=&\left\{\begin{array}{lll}
	\displaystyle  
	\frac{1}{M^4}\bar{\sigma}^3_{V}\left(\bar{\eta}_{V}-12\bar{\epsilon}_{V}\right)\left(1-\frac{2}{3}\frac{\bar{\epsilon}_{V}}{M}\right)^{1/4}\,,~~~~~~~~~~~~~~~~~~~~~~~~~~~~~~~ &
	\mbox{\small {\bf for {$q=1/2$ }}}  \\ 
	\displaystyle   
	\frac{1}{M^4}\bar{\sigma}^3_{V}\left(\bar{\eta}_{V}-12\bar{\epsilon}_{V}\right)\left(1-\frac{1}{3q}\frac{\bar{\epsilon}_{V}}{M}\right)^{1/4}\,.~~~~ &
	\mbox{\small {\bf for {~any~ $q$ }}} 
\end{array}
\right.
\eea 
In terms of the slow-roll parameters, the number of e-foldings can be re-expressed as:
\be\begin{array}{lll}\label{e-fold2v7v2}
	\displaystyle  N=\left\{\begin{array}{lll}
		\displaystyle  
		\sqrt{\frac{3\alpha^{'}}{2}}\int^{T_{i,end}}_{T_i}\frac{H(T)}{\sqrt{\epsilon_{1}}}~dT_i=\sqrt{\frac{3\alpha^{'}}{2}}\int^{T_{end}}_{T_i}\frac{H(T)}{\sqrt{\bar{\epsilon}_{V}(T_i)}}~dT
		\nonumber\\ ~~~~~~~~~~~~~~~~~~~~~~~~~~~~~\displaystyle \approx\frac{\alpha^{'}M}{M^{2}_{p}}\int^{T_i}_{T_{i,end}}
		\frac{V(T)V(T_i)}{V^{'}(T_i)}~dT_i\,,~~~~~~ &
		\mbox{\small {\bf for {$q=1/2$ }}}  \\ 
		\displaystyle   
		2q\sqrt{\frac{3\alpha^{'}}{2}}\int^{T_{i,end}}_{T_i}\frac{H(T)}{\sqrt{\epsilon_{1}}}~dT_i= \sqrt{3\alpha^{'}q}\int^{T_{i,end}}_{T_i}\frac{H(T)}{\sqrt{\bar{\epsilon}_{V}(T_i)}}~dT_i
		\nonumber\\ ~~~~~~~~~~~~~~~~~~~~~~~~~~~~~\displaystyle \approx\frac{ \sqrt{2q}\alpha^{'}M}{M^{2}_{p}}\int^{T_i}_{T_{i,end}}
		\frac{V(T)V(T_i)}{V^{'}(T_i)}~dT_i\,.~~~~~~ &
		\mbox{\small {\bf for {~any~arbitrary~ $q$ }}} 
	\end{array}
	\right.
\end{array}\ee
where $T_{i,end}$ characterizes the tachyonic field value at the end of inflation $t=t_{end}$ for all $i$ fields participating in assisted inflation. As in the case of assisted inflation all the $M$ fields are identical then 
one can re-express the number of e-foldings as:
\be\begin{array}{lll}\label{e-fold2v7v3}
	\displaystyle  N=\left\{\begin{array}{lll}
		\displaystyle  
		\frac{\alpha^{'}M}{M^{2}_{p}}\int^{T_i}_{T_{i,end}}
		\frac{V^2(T_i)}{V^{'}(T_i)}~dT_i=\frac{\alpha^{'}M}{M^{2}_{p}}\int^{T}_{T_{end}}
		\frac{V^2(T)}{V^{'}(T)}~dT\,,~~~~~~ &
		\mbox{\small {\bf for {$q=1/2$ }}}  \\  
		\displaystyle   
		\frac{ \sqrt{2q}\alpha^{'}M}{M^{2}_{p}}\int^{T_i}_{T_{i,end}}
		\frac{V^2(T_i)}{V^{'}(T_i)}~dT_i=\frac{ \sqrt{2q}\alpha^{'}M}{M^{2}_{p}}\int^{T}_{T_{end}}
		\frac{V^2(T)}{V^{'}(T)}~dT\,.~~~~~~ &
		\mbox{\small {\bf for {~any~arbitrary~ $q$ }}} 
	\end{array}
	\right.
\end{array}\ee
\subsubsection{Basics of tachyonic perturbations}
In this subsection we explicitly discuss about the cosmological linear perturbation theory within the framework of assisted tachyonic inflation.
Let us clearly mention that here we have various ways of characterizing cosmological perturbations in the context of inflation,
which finally depend on the choice of gauge. Let us do the computation in the longitudinal gauge, where the scalar metric perturbations of the 
FLRW background are given by the following infinitesimal line element:
\bea \label{linev1}
ds^2 &=& -\left(1+2{\bf \Phi}(t,{\bf x})\right)dt^2+a^{2}(t)\left(1-2{\bf \Psi}(t,{\bf x})\right)\delta_{ij}dx^{i}dx^{j},
\eea
where $a(t)$ is the scale factor, ${\bf \Phi}(t,{\bf x})$ and ${\bf \Psi}(t,{\bf x})$ characterizes the gauge invariant metric perturbations.
Specifically, the perturbation of the FLRW metric leads to the perturbation in the energy-momentum stress tensor via the Einstein field equation or equivalently through the Friedmann equations.
For the perturbed metric as mentioned in Eq~(\ref{linev1}), the perturbed Einstein field equations can be expressed for $q=1/2$ case of tachyonic inflationary setup as:
\bea \label{l1v1} 3H\left(H{\bf \Phi}(t,{\bf k})+\dot{\bf \Psi}(t,{\bf k})\right)+\frac{k^{2}}{a^{2}(t)}&=&-\frac{1}{2M^{2}_{p}}\delta \rho_{i},\\
\label{l2v1} \ddot{\bf \Psi}(t,{\bf k})+3H\left(H{\bf \Phi}(t,{\bf k})+\dot{\bf \Psi}(t,{\bf k})\right)+H\dot{\bf \Phi}(t,{\bf k})\nonumber~~~~\\~~~~~~~~~~+2\dot{H}{\bf \Phi}(t,{\bf k})+\frac{k^{2}}{3a^{2}(t)}
\left({\bf \Phi}(t,{\bf k})-{\bf \Psi}(t,{\bf k})\right)&=&\frac{1}{2M^{2}_{p}}\delta p_{i},\\
\label{l3v1} \dot{\bf \Psi}(t,{\bf k})+H{\bf \Phi}(t,{\bf k})&=&-\frac{\alpha^{'}V(T_i)}{\sqrt{1-\alpha^{'}\dot{T}^{2}_{i}}}\frac{\dot{T}_{i}}{M^{2}_{p}}\delta T_{i},~~~~~~~~~~\\
\label{l4v1}{\bf \Psi}(t,{\bf k})-{\bf \Phi}(t,{\bf k})&=&0.
\eea
Similarly, for any arbitrary $q$ the perturbed Einstein field equations can be expressed as:
\bea \label{l11v1} 3H\left(H{\bf \Phi}(t,{\bf k})+\dot{\bf \Psi}(t,{\bf k})\right)+\frac{k^{2}}{a^{2}(t)}&=&-\frac{1}{2M^{2}_{p}}\delta \rho_{i},\\
\label{l22v1} \ddot{\bf \Psi}(t,{\bf k})+3H\left(H{\bf \Phi}(t,{\bf k})+\dot{\bf \Psi}(t,{\bf k})\right)+H\dot{\bf \Phi}(t,{\bf k})\nonumber~~~~\\~~~~~~~~~~+2\dot{H}{\bf \Phi}(t,{\bf k})+\frac{k^{2}}{3a^{2}(t)}
\left({\bf \Phi}(t,{\bf k})-{\bf \Psi}(t,{\bf k})\right)&=&\frac{1}{2M^{2}_{p}}\delta p_{i},\\
\label{l33v1} \dot{\bf \Psi}(t,{\bf k})+H{\bf \Phi}(t,{\bf k})&=&-\frac{\alpha^{'}V(T_i)\left[1-\alpha^{'}(1-2q)\dot{T}^2_{i}\right]}{\left(1-\alpha^{'}\dot{T}^2_{i}\right)^{1-q}} \frac{\dot{T}_{i}}{M^{2}_{p}}\delta T_{i},~~~~~~~~~~\\
\label{l44v1}{\bf \Psi}(t,{\bf k})-{\bf \Phi}(t,{\bf k})&=&0.
\eea
Here ${\bf \Phi}(t,{\bf k})$ and ${\bf \Psi}(t,{\bf k})$ are the two gauge invariant metric perturbations in the Fourier space.
Additionally, it is important to note that in Eq~(\ref{l4}), the two gauge invariant metric perturbations ${\bf \Phi}(t,{\bf k})$ and ${\bf \Psi}(t,{\bf k})$ are equal in the context of minimally coupled tachyonic string field theoretic model 
with Einstein gravity sector. In Eq~(\ref{l1}) and Eq~(\ref{l2}) the perturbed energy density $\delta\rho$ and pressure $\delta p$ are given by:
\bea\label{e7aav2}
 \displaystyle \delta\rho_{i} &=&\left\{\begin{array}{lll}
                    \displaystyle  
                    \frac{V^{'}(T_{i})\delta T_{i}}{\sqrt{1-\alpha^{'}\dot{T}^2_{i}}}
                    +\frac{\alpha^{'}V(T_{i})\left(\dot{T}_{i}\delta\dot{T}_{i}+\dot{T}^2_{i}{\bf \Phi}(t,{\bf k})\right)}{\left(1-\alpha^{'}\dot{T}^2_{i}\right)^{3/2}} \,,~~~~ &
 \mbox{\small {\bf for $q=1/2$}}  \\ 
 \displaystyle  
  \frac{\left\{V^{'}(T_{i})\left[1-\alpha^{'}(1-2q)\dot{T}^2_{i}\right]\delta T_{i}-4\alpha^{'}(1-2q)V(T_{i})\dot{T}_{i}\delta \dot{T}_{i}\right\}}
  {\left(1-\alpha^{'}\dot{T}^2_{i}\right)^{1-q}}\\ \displaystyle ~~~~~
  +\frac{2\alpha^{'}(1-q)V(T_{i})\left[1-\alpha^{'}(1-2q)\dot{T}^2_{i}\right]\left(\dot{T}_{i}\delta\dot{T}_{i}+\dot{T}^2_{i}{\bf \Phi}(t,{\bf k})\right)}
  {\left(1-\alpha^{'}\dot{T}^2_{i}\right)^{2-q}} \,,~~~ &
 \mbox{\small {\bf for~any~arbitrary~$q$}}.
          \end{array}
\right.
\\ \nonumber \label{e8aav3}
 \displaystyle \delta p_{i} &=&\left\{\begin{array}{lll}
                    \displaystyle  
                    -V^{'}(T_{i})\sqrt{1-\alpha^{'}\dot{T}^2_{i}}\delta T_{i} +\frac{\alpha^{'}V(T_{i})\left(\dot{T}_{i}\delta\dot{T}_{i}+\dot{T}^2{\bf \Phi}(t,{\bf k})\right)}{\sqrt{1-\alpha^{'}\dot{T}^2}} \,,~~~~ &
 \mbox{\small {\bf for $q=1/2$}}  \\ 
 \displaystyle  
 -V^{'}(T_{i})\left(1-\alpha^{'}\dot{T}^2_{i}\right)^{q}\delta T_{i}+\frac{2q\alpha^{'}V(T_{i})\left(\dot{T}_{i}\delta\dot{T}_{i}+\dot{T}^2_{i}{\bf \Phi}(t,{\bf k})\right)}{\left(1-\alpha^{'}\dot{T}^2_{i}\right)^{1-q}} \,,~~~ &
 \mbox{\small {\bf for ~any~arbitrary~$q$}}.
          \end{array}
\right.
\eea
Similarly after the variation of the tachyoinic field equation motion
we get the following expressions for the perturbed equation of motion: 
\be\begin{array}{lll}\label{e8aav2}\tiny
 \displaystyle 0 \approx\left\{\begin{array}{lll}
                    \displaystyle  
                    \delta \ddot{T}_{i}+3H\delta\dot{T}_{i}+\frac{2\alpha^{'}\ddot{T}_{i}\left(\dot{T}_{i}\delta\dot{T}_{i}+\dot{T}^2_{i} {\bf \Phi}(t,{\bf k})\right)}{\left(1-\alpha^{'}\dot{T}^2_{i}\right)}
                    \\ \displaystyle ~~~~~~~+\frac{M_{p}\sqrt{1-\alpha^{'}\dot{T}^2_{i}}}{\alpha^{'}V(T_{i})}\left[\left(\frac{k^2}{a^2} - 3\dot{H}\right){\bf \Phi}(t,{\bf k})-\frac{2k^2}{a^2}{\bf \Psi}(t,{\bf k})
                    \right.\\ \left. \displaystyle~~~~-3\left(\ddot{\bf \Psi}(t,{\bf k})+4H\dot{\bf \Psi}(t,{\bf k})+H\dot{\bf \Phi}(t,{\bf k})+\dot{H}{\bf \Phi}(t,{\bf k})+4H^2 {\bf \Phi}(t,{\bf k})\right)\right]\\
                    \displaystyle -\left\{6H\alpha^{'}\dot{T}^3_{i} - \frac{2V^{'}(T_{i})}{\alpha^{'}V(T_{i})}\left(1-\alpha^{'}\dot{T}^2_{i}\right)\right\}{\bf \Phi}(t,{\bf k})-\left(\dot{\bf \Psi}(t,{\bf k})+3\dot{\bf \Psi}(t,{\bf k})\right)\dot{T}_{i}\\
                    \displaystyle ~~~~~~-\frac{M_{p}\left(1-\alpha^{'}\dot{T}^2_{i}\right)}{\alpha^{'}}\left(\frac{V^{''}(T_{i})}{V(T_{i})}-\frac{V^{'2}(T_{i})}{V^{2}(T_{i})}\right)\,,~~~~ &
 \mbox{\small {\bf for $q=1/2$}}  \\ 
 \displaystyle  
                    \delta \ddot{T}_{i}+3H\delta\dot{T}_{i}+\frac{2\alpha^{'}\ddot{T}_{i}\left(\dot{T}_{i}\delta\dot{T}_{i}+\dot{T}^2_{i} {\bf \Phi}(t,{\bf k})\right)}{\left(1-\alpha^{'}\dot{T}^2_{i}\right)^{2(1-q)}}
                    \\ \displaystyle ~~~~~~~+\frac{M_{p}\left(1-\alpha^{'}\dot{T}^2_{i}\right)^{1-q}}{\alpha^{'}V(T_{i})\left[1-\alpha^{'}(1-2q)\dot{T}^2_{i}\right]}\left[\left(\frac{k^2}{a^2} - 3\dot{H}\right){\bf \Phi}(t,{\bf k})-\frac{2k^2}{a^2}{\bf \Psi}(t,{\bf k})
                    \right.\\ \left. \displaystyle~~~~-3\left(\ddot{\bf \Psi}(t,{\bf k})+4H\dot{\bf \Psi}(t,{\bf k})+H\dot{\bf \Phi}(t,{\bf k})+\dot{H}{\bf \Phi}(t,{\bf k})+4H^2 {\bf \Phi}(t,{\bf k})\right)\right]\\
                    \displaystyle -\left\{6H\alpha^{'}\dot{T}^3_{i} - \sqrt{\frac{2}{q}}\frac{V^{'}(T_{i})}{\alpha^{'}V(T_{i})}\left(1-\alpha^{'}\dot{T}^2_{i}\right)^{2(1-q)}\right\}{\bf \Phi}(t,{\bf k})-\left(\dot{\bf \Psi}(t,{\bf k})+3\dot{\bf \Psi}(t,{\bf k})\right)\dot{T}_{i}\\
                    \displaystyle ~~~~~~-\frac{M_{p}\left(1-\alpha^{'}\dot{T}^2_{i}\right)^{2(1-q)}}{\sqrt{2q}\alpha^{'}}\left(\frac{V^{''}(T_{i})}{V(T_{i})}-\frac{V^{'2}(T_{i})}{V^{2}(T_{i})}\right)\,,~~~ &
 \mbox{\small {\bf for ~any~$q$}}.
          \end{array}
\right.
\end{array}\ee 
Further we will perform the following steps through out the next part of the computation:
\begin{itemize}
 \item First of all we decompose the scalar perturbations into two components-(1) {\bf entropic or isocurvature perturbations} which can be usually treated as the orthogonal projective part to the trajectory and (2) 
 {\bf adiabatic or curvature perturbations} which can be usually treated as the parallel projective part to the trajectory.
 \item Within the framework of first order cosmological perturbation theory we define a gauge invariant primordial curvature perturbation on the scales outside the horizon:
       \bea \zeta=\sum^{M}_{i=1}\left(\zeta_{r}+\zeta_{\tilde{T}_{i}}\right)= {\bf \Psi}-\sum^{M}_{i=1}\frac{H}{\dot{\rho}_{i}}\delta\rho_{i}. \eea
 \item Next we consider the uniform density hypersurface in which 
 \bea \delta \rho_{i}=0~~~~\forall i=1,2,\cdots,M.\eea
 Consequently the curvature perturbation is governed by:
  \bea \zeta=\sum^{M}_{i=1}\left(\zeta_{r}+\zeta_{\tilde{T}_{i}}\right)={\bf \Psi}.\eea
  \item Further, the time evolution of the curvature perturbation can be expressed as:
  \bea\label{pqw1} \dot{\zeta}=H\sum^{M}_{i=1}\left(\frac{\delta \bar{p}_{i}}{\rho_{i}+p_{i}}\right),\eea
  where $\delta \bar{p}_{i}$ characterizes the non-adiabatic or entropic contribution in the first order linearized cosmological perturbation. In the present context $\delta \bar{p}_{i}$ can be expressed as:
  \bea \delta \bar{p}_{i}= \Gamma \dot{p}_{i}=\left(\frac{\partial p_{i}}{\partial S}\right)_{S}\delta S,\eea
  where $\Gamma$ characterizes the relative displacement between hypersurfaces of uniform pressure and density. In case of assisted or multi-component fluid dynamical system, there are 
  two contributions to the $\delta \bar{p}$ are appearing in the computation:
  \bea \delta \bar{p}_{i}=\delta \bar{p}_{rel,\tilde{T}_{i}}+\delta \bar{p}_{int},\eea
  where $\delta \bar{p}_{rel,i}$ and $\delta \bar{p}_{int}$ characterize the relative and intrinsic contributions to the multi-component fluid system given by:
  \bea \delta \bar{p}_{rel,\tilde{T}_{i}}&=& \frac{1}{3H\dot{\rho}}\dot{\rho}_{r}\dot{\rho}_{\tilde{T}_{i}}\left(c^2_{r}-c^{2}_{S,\tilde{T}_{i}}\right){\cal S}_{r\tilde{T}_{i}}\propto \dot{\rho}_{r},\\
  \delta \bar{p}_{int}&=&\delta\bar{p}_{int,\tilde{T}_{i}}+\delta\bar{p}_{int,r}=\sum_{\alpha=r,\tilde{T}_{i}}\left(\delta p_{\alpha}-c^{2}_{\alpha}\delta \rho_{\alpha}\right).\eea
  Here $c^{2}_{S,\tilde{T}_{i}}$ and $c^2_{r}$ are the effective adiabatic sound speed and sound speed due to radiation. Also the relative entropic perturbation is defined as:
  \bea {\cal S}_{r\tilde{T}_{i}}&=& 3\left(\zeta_{r}-\zeta_{\tilde{T}_{i}}\right)\nonumber\\
  &=&-3H\left(\frac{\delta\rho_{r}}{\dot{\rho}_{r}}-\frac{\delta\rho_{\tilde{T}_{i}}}{\dot{\rho}_{\tilde{T}_{i}}}\right)\nonumber\\
  &=& \frac{\delta_{r}}{1+w_{r}}-\frac{\delta_{\tilde{T}_{i}}}{1+w_{\tilde{T}_{i}}},\eea
  where $w_{r}$ and $w_{\tilde{T}_{i}}$ signify the equation of state parameter due to radiation and $M$ number of tachyonic matter, $\delta_{r}$ and $\delta_{\tilde{T}_{i}}$ characterize the energy density contrast due to radiation and $M$ number of tachyonic matter contents, defined as:
  \bea \delta_{\alpha}&=&\frac{\delta \rho_{\alpha}}{\rho_{\alpha}}~~~~\forall \alpha=r, \tilde{T}_{i}(\forall i=1,2,\cdots, M). \eea
  In generalized prescription
  the pressure perturbation in arbitrary gauge can be decomposed into the following contributions:
  \bea \delta p=\sum^{M}_{i=1}\delta p_{i} = c^{2}_{S}\delta \rho + \frac{1}{3H\dot{\rho}}\dot{\rho}_{r}\sum^{M}_{i=1}\dot{\rho}_{\tilde{T}_{i}}\left(c^2_{r}-c^{2}_{S,\tilde{T}_{i}}\right){\cal S}_{r\tilde{T}_{i}}+\sum^{M}_{i=1}
  \sum_{\alpha=r,\tilde{T}_{i}}\left(\delta p_{\alpha}-c^{2}_{\alpha}\delta \rho_{\alpha}\right).~~~~~~~~\eea
  As $\delta\bar{p}_{rel,\tilde{T}_{i}}\propto \dot{\rho}_{r}$, the relative non-adiabatic pressure perturbation is heavily suppressed in the computation and one can safely ignore such contributions during the epoch of inflation.
  Here the relative entropic perturbation ${\cal S}_{r\tilde{T}_{i}}$ remains small and finite in the limit of small $\dot{\rho}_{r}$ due to the smallness of the curvature perturbations $\zeta_{r}$ and $\zeta_{\tilde{T}_{i}}$.
  \item Now let us consider a situation where the equation of state parameter corresponding to radiation is: 
  \be 
  w_{r}=c^2_{r}={\rm \bf Constant}.
  \ee
  Consequently the fluctuation in pressure due to radiation can be expressed in terms of the fluctuation of density can be expressed as:
  \be \delta p_{r}=w_{r}\delta \rho_{r}=c^{2}_{r}\delta \rho_{r}={\rm \bf Constant}\times\delta \rho_{r}\propto \delta \rho_{r}.\ee
  This clearly implies that:
  \be \delta\bar{p}_{int,r}=\left(\delta p_{r}-c^{2}_{r}\delta \rho_{r}\right)=0\ee
  i.e. the intrinsic non-adiabatic pressure vanishes identically. However, in terms of the effective tachyonic field $\tilde{T}$ this situation 
  is not very simple, as the equation of state parameter and the intrinsic sound speed changes with the number of component fields $M$. During inflationary 
  epoch the all $M$ number of fields obey the following equation of state:
  \be w_{\tilde{T}_{i}}=\frac{p_{\tilde{T}_{i}}}{\rho_{\tilde{T}_{i}}}\simeq -1=c^2_{S,\tilde{T}_{i}}={\rm \bf Constant}~~~~\forall i=1,2,\cdots,M.\ee
  For all $M$ number of effective tachyon fields the fluctuation in pressure can be expressed as:
  \be \delta p_{\tilde{T}_{i}} \simeq w_{\tilde{T}_{i}}\delta \rho_{\tilde{T}_{i}}=c^2_{S,\tilde{T}_{i}}\delta \rho_{\tilde{T}_{i}}\simeq -\delta \rho_{\tilde{T}_{i}}. \ee
  Consequently the intrinsic non-adiabatic pressure within each tachyonic field amongst $M$ number of fields can be expressed as:
  \be \delta p_{int,\tilde{T}_{i}}=\delta p_{\tilde{T}_{i}}-c^2_{S,\tilde{T}_{i}}\delta \rho_{\tilde{T}_{i}}\simeq 0.\ee
  Now in the present context the relative contributions to the non-adiabatic pressure between the $M$ number of tachyonic field context can be expressed as:
  \be \delta p_{rel,\tilde{T}_{i}\tilde{T}_{j}}\propto \left(c^2_{S,\tilde{T}_{i}}-c^2_{S,\tilde{T}_{j}}\right)\simeq 0,\ee
  since 
  \be c^2_{S,\tilde{T}_{i}}\simeq c^2_{S,\tilde{T}_{j}}\simeq -1~~~~\forall i,j=1,2,\cdots, M.\ee
  This implies that the total non-adiabatic pressure is negligible during inflation i.e. 
  \bea \delta \bar{p}_{int}&=&\delta\bar{p}_{int,\tilde{T}_{i}}+\delta\bar{p}_{int,r}=\sum_{\alpha=r,\tilde{T}_{i}}\left(\delta p_{\alpha}-c^{2}_{\alpha}\delta \rho_{\alpha}\right)=0.\eea
  and it is completely justified to write the fluctuation in the pressure for individual $M$ number of tachyonic field content as:
  \bea \delta \bar{p}_{i}=\delta \bar{p}_{rel,\tilde{T}_{i}}=\frac{1}{3H\dot{\rho}}\dot{\rho}_{r}\dot{\rho}_{\tilde{T}_{i}}\left(c^2_{r}-c^{2}_{S,\tilde{T}_{i}}\right){\cal S}_{r\tilde{T}_{i}}\propto \dot{\rho}_{r}.\eea
  Using these results the total fluctuation in the pressure can be re-expressed as:
  \bea \delta p=\sum^{M}_{i=1}\delta p_{i} = c^{2}_{S}\delta \rho + \sum^{M}_{i=1}\delta \bar{p}_{i}=c^{2}_{S}\delta \rho +AM\dot{\rho}_{r}\simeq c^{2}_{S}\delta \rho,~~~~~~~~\eea
  where $A$ is the proportionality constant and we have safely ignored the contribution from the isocurvatrure modes due to its smallness in the case of assisted tachyonic inflation.
  Consequently we get:
  \bea \delta \bar{p}_{i}= \Gamma \dot{p}_{i}=AM\dot{\rho}_{r}\simeq 0 ~~~~\Rightarrow~~~~\rho_{r}={\bf Constant}~~~~\Rightarrow~~~~\zeta={\bf Constant},~~~~~~~~~~~~~\eea
  which is exactly similar with the usual single field slow-roll conditions in the context of tachyonic inflationary setup discussed earlier. Finally, in the uniform density hypersurfaces, the curvature perturbation can be written in terms of the tachyonic field 
  fluctuations on spatially flat hypersurfaces as: 
  \bea \zeta=-H\sum^{M}_{i=1}\left(\frac{\delta T_{i}}{\dot{T}_{i}}\right)=-HM\left(\frac{\delta T}{\dot{T}}\right)=-HM\left(\frac{\delta \tilde{T}}{\dot{\tilde{T}}}\right).\eea
  \item But in the case of multi tachyonic inflation where the $M$ number of fields are not identical with each other the situation is not simpler like assisted inflation. In that case during the short time intervals when fields decay in a very faster rate and its corresponding equation of state charges rapidly then it would be really interesting to investigate 
  the production of isocurvature perturbations. On the scale smaller than the Hubble radius any entropy perturbation rapidly becomes adiabatic perturbation of the same amplitude, as local pressure differences, due to the local 
  fluctuations in the equation of state, re-distribute the energy density. However, this change is slightly less efficient during the epoch of radiation compared to the tachyonic matter dominated epoch and can only occur after the decoupling 
  between the photons and baryons, in the specific case of baryonic isocurvature perturbation. Causality precludes this re-distribution on scales bigger than the Hubble radius, and thus any entropy perturbation on these scales remains with constant amplitude.
  Also it is important to note that, the entropic or isocurvature perturbations are not affected by the Silk damping, which is exactly contrary to the curvature or adiabatic perturbations.  
   \end{itemize}
   
\subsubsection{Computation of scalar power spectrum}
In this subsection we will not derive the results for assisted inflation as it is exactly same as derived in the case of single tachyonic inflation. But we will state 
the results for {\bf BD} vacuum where the changes will appear due to the presence of $M$ 
identical copies of the tachyon field. One can similarly write down the detailed expressions for {\bf AV} as well. The changes will appear in the expressions for the following inflationary observables at the horizon crossing:
\begin{itemize}
 \item In the present context amplitude of scalar power spectrum can be computed as:
\begin{eqnarray} 
\label{psxcc1vv2} \Delta_{\zeta, \star}
&=&\left\{\begin{array}{lll}
                    \displaystyle 
 \displaystyle  \left\{\left[1-({\cal C}_{E}+1)\frac{\bar{\epsilon}_{V}}{M}-\frac{{\cal C}_{E}}{M}\left(3\bar{\epsilon}_{V}-\bar{\eta}_{V}\right)\right]^2 \frac{MH^{2}}{8\pi^2M^{2}_{p}c_{S}\bar{\epsilon}_{V}}\right\}_{k_{\star}=a_{\star}H_{\star}}\,,~~~~ &
 \mbox{\small {\bf for $q=1/2$}}  \\ 
 \displaystyle \left\{\left[1-({\cal C}_{E}+1)\frac{\bar{\epsilon}_{V}}{2qM}-\frac{{\cal C}_{E}}{\sqrt{2q}M}\left(3\bar{\epsilon}_{V}-\bar{\eta}_{V}\right)\right]^2 \frac{qM H^{2}}{4\pi^2M^{2}_{p}c_{S}\bar{\epsilon}_{V}}\right\}_{k_{\star}=a_{\star}H_{\star}} \,,~~~~ &
 \mbox{\small {\bf for any $q$}}.
 \end{array}
\right.\end{eqnarray}
where $H^2=MV/3M^{2}_{p}$ and ${\cal C}_{E}= -2 + \ln 2 + \gamma \approx -0.72.$  Using the slow-roll approximations one can further approximate the expression for sound speed as:
\be\begin{array}{lll}\label{rteqxcccvv2}
 \displaystyle c^2_{S} 
                  =\left\{\begin{array}{lll}
                    \displaystyle  
                  1-\frac{2}{3}\frac{\bar{\epsilon}_{V}}{M}+{\cal O}\left(\frac{\bar{\epsilon}^2_{V}}{M^2}\right)+\cdots \,,~~~~ &
 \mbox{\small {\bf for $q=1/2$}}  \\ 
 \displaystyle  
 1-\frac{(1-q)}{3q^2 M}\bar{\epsilon}_{V}+{\cal O}\left(\frac{\bar{\epsilon}^2_{V}}{M^2}\right)+\cdots \,,~~~ &
 \mbox{\small {\bf for ~any~$q$}}.
          \end{array}
\right.
\end{array}\ee
 Hence using the result in Eq~(\ref{psxcc1vv2}) we get the following simplified expression for the primordial scalar power spectrum:
\begin{eqnarray} 
\label{psxcc1v0vv2} \Delta_{\zeta, \star}
&\approx&\left\{\begin{array}{lll}
                    \displaystyle 
 \displaystyle  
                \left\{\left[1-\left({\cal C}_{E}+\frac{5}{6}\right)\frac{\bar{\epsilon}_{V}}{M}-\frac{{\cal C}_{E}}{M}
                \left(3\bar{\epsilon}_{V}-\bar{\eta}_{V}\right)\right]^2 \frac{MH^{2}}{8\pi^2M^{2}_{p}\bar{\epsilon}_{V}}\right\}_{k_{\star}=a_{\star}H_{\star}} \,,~~~~ &
 \mbox{\small {\bf for $q=1/2$}}  \\ 
 \displaystyle  
                \left\{\left[1-({\cal C}_{E}+1-\Sigma)\frac{\bar{\epsilon}_{V}}{2qM}-\frac{{\cal C}_{E}}{\sqrt{2q}M}\left(3\bar{\epsilon}_{V}-\bar{\eta}_{V}\right)\right]^2 \frac{qM H^{2}}{4\pi^2M^{2}_{p}
                \bar{\epsilon}_{V}}\right\}_{k_{\star}=a_{\star}H_{\star}} \,.~~~~ &
 \mbox{\small {\bf for any $q$}}.
 \end{array}
\right.\end{eqnarray}
\item Next one can compute the scalar spectral tilt ($n_{S}$) of the primordial scalar power spectrum as:
\begin{eqnarray} 
\label{psxcc1v2vv2} n_{\zeta, \star}-1 
&\approx&\left\{\begin{array}{lll}
                    \displaystyle 
 \displaystyle  
               \displaystyle
                \frac{1}{M}\left(2\bar{\eta}_{V}-8\bar{\epsilon}_{V}\right)-\frac{2}{M^2}\bar{\epsilon}^{2}_{V}-\frac{2}{M^2}\left(2{\cal C}_{E}+\frac{8}{3}\right)\bar{\epsilon}_{V}\left(3\bar{\epsilon}_{V}-\bar{\eta}_{V}\right)\\
               \displaystyle~~~~~~~-\frac{{\cal C}_{E}}{M^2}\left(2\bar{\xi}^{2}_{V}-5\bar{\eta}_{V}\bar{\epsilon}_{V}+36\bar{\epsilon}^{2}_{V}\right)+\cdots,~~ &
 \mbox{\small {\bf for $q=1/2$}}  \\ 
 \displaystyle  
               \displaystyle  
               \displaystyle
                \frac{1}{M}\sqrt{\frac{2}{q}}\bar{\eta}_{V}-\frac{1}{M}\left(\frac{1}{q}+3\sqrt{\frac{2}{q}}\right)\bar{\epsilon}_{V}-\frac{\bar{\epsilon}^{2}_{V}}{2q^2M^2}\\
                \displaystyle -\frac{2}{(2q)^{3/2}M^2}\left(2{\cal C}_{E}+3-2\Sigma\right)\bar{\epsilon}_{V}\left(3\bar{\epsilon}_{V}-\bar{\eta}_{V}\right)\\
               \displaystyle~~~~~~~-\frac{{\cal C}_{E}}{\sqrt{2q}M^2}\left(2\bar{\xi}^{2}_{V}-5\bar{\eta}_{V}\bar{\epsilon}_{V}+36\bar{\epsilon}^{2}_{V}\right)+\cdots\,,~~ &
 \mbox{\small {\bf for any $q$}}.
 \end{array}
\right.\end{eqnarray}
\item Next one can compute the running of the scalar spectral tilt ($\alpha_{S}$) of the primordial scalar power spectrum as:
\begin{eqnarray}
\label{psxcc1v3vv2} \alpha_{\zeta, \star}
&\approx&\small\left\{ \small\begin{array}{lll}
                    \displaystyle 
\displaystyle -\left\{\left[\frac{4}{M^2}\bar{\epsilon}_{V}\left(1+\frac{\bar{\epsilon}_{V}}{M}\right)(\bar{\eta}_{V}-3\bar{\epsilon}_{V})+\frac{2}{M^2}\left(10\bar{\epsilon}_{V}\bar{\eta}_{V}
                -18\bar{\epsilon}^2_{V}-\bar{\xi}^2_{V}\right)\right]\right.\nonumber\\ \left.
                \displaystyle -\frac{{\cal C}_{E}}{M^3}\left(2\bar{\sigma}^{3}_{V}-216\bar{\epsilon}^{3}_{V}+2\bar{\xi}^{2}_{V}\bar{\eta}_{V}-7\bar{\xi}^{2}_{V}\bar{\epsilon}_{V}
                +194\bar{\epsilon}^2_{V}\bar{\eta}_{V}-10\bar{\eta}_{V}\bar{\epsilon}_{V}
                   \right)\displaystyle\right.\\ \left.
               \displaystyle-\frac{1}{M^3}\left(2{\cal C}_{E}+\frac{8}{3}\right)\left[2\bar{\epsilon}_{V}\left(10\bar{\epsilon}_{V}\bar{\eta}_{V}-18\bar{\epsilon}^2_{V}
               -\bar{\xi}^2_{V}\right)\right.\right.\\ \left.\left.
               \displaystyle -4\bar{\epsilon}_{V}\left(3\bar{\epsilon}_{V}-\bar{\eta}_{V}\right)^2\right]\right\}\displaystyle\left(1-\frac{2}{3}\frac{\bar{\epsilon}_{V}}{M}\right)^{1/4}+\cdots,~~ &
 \mbox{\small {\bf for $q=1/2$}}  \\ 
 \displaystyle  
               -\left\{\left[\sqrt{\frac{2}{q}}\frac{\bar{\epsilon}_{V}}{qM^2}\left(1+\frac{\bar{\epsilon}_{V}}{2qM}\right)(\bar{\eta}_{V}-3\bar{\epsilon}_{V})+\frac{1}{M^2}\sqrt{\frac{2}{q}}\left(10\bar{\epsilon}_{V}\bar{\eta}_{V}
                -18\bar{\epsilon}^2_{V}-\bar{\xi}^2_{V}\right)\right]\right.\nonumber\\ \left.
                \displaystyle -\frac{{\cal C}_{E}}{\sqrt{2q}M^3}\left(2\bar{\sigma}^{3}_{V}-216\bar{\epsilon}^{3}_{V}+2\bar{\xi}^{2}_{V}\bar{\eta}_{V}-7\bar{\xi}^{2}_{V}\bar{\epsilon}_{V}
                +194\bar{\epsilon}^2_{V}\bar{\eta}_{V}-10\bar{\eta}_{V}\bar{\epsilon}_{V}
                   \right)\right.\nonumber\\ \left.
               \displaystyle-\frac{1}{M^3}\left(2{\cal C}_{E}+\frac{8}{3}\right)\left[\sqrt{\frac{2}{q}}\bar{\epsilon}_{V}\left(10\bar{\epsilon}_{V}\bar{\eta}_{V}-18\bar{\epsilon}^2_{V}
               -\bar{\xi}^2_{V}\right)\right.\right.\\ \left.\left.\displaystyle
               -\frac{4}{(2q)^{3/2}}\bar{\epsilon}_{V}\left(3\bar{\epsilon}_{V}-\bar{\eta}_{V}\right)^2\right]\right\}\displaystyle\left(1-\frac{1}{3q}\frac{\bar{\epsilon}_{V}}{M}\right)^{1/4}+\cdots\,,~~ &
 \mbox{\small {\bf for any $q$}}.
 \end{array}
\right.\end{eqnarray}
\item Finally, one can also compute the running of the running of scalar spectral tilt ($\kappa_{S}$) of the primordial scalar power spectrum as:
\begin{eqnarray}
\label{psxcc1v4vv2} \kappa_{\zeta, \star}
&\approx&\left\{ \small\begin{array}{lll}
                    \displaystyle 
\displaystyle -\frac{1}{M^3}\left[8\bar{\epsilon}_{V}(1+\bar{\epsilon}_{V})(\bar{\eta}_{V}-3\bar{\epsilon}_{V})^2+\frac{8}{M}\bar{\epsilon}^2_{V}(\bar{\eta}_{V}-3\bar{\epsilon}_{V})^2\right.\\ \left. \displaystyle 
+8\bar{\epsilon}_{V}(1+\bar{\epsilon}_{V})\left(\bar{\xi}^2_{V}-10\bar{\epsilon}_{V}\bar{\eta}_{V}
+18\bar{\epsilon}^{2}_{V}\right)\right.\\ \left. \displaystyle  +2\left(20\bar{\epsilon}_{V}\bar{\eta}_{V}
\left(\bar{\eta}_{V}-3\bar{\epsilon}_{V}\right)+10\bar{\epsilon}_{V}\left(\bar{\xi}^2_{V}-4\bar{\epsilon}_{V}\bar{\eta}_{V}\right)
                \right.\right.\\ \left.\left.\displaystyle -72\bar{\epsilon}^2_{V}\left(\bar{\eta}_{V}-3\bar{\epsilon}_{V}\right)-
                \left(\bar{\sigma}^3_{V}+\bar{\xi}^2_{V}\bar{\eta}_{V}-\bar{\xi}^2_{V}\bar{\epsilon}_{V}\right)\right)\right]\displaystyle
                \left(1-\frac{2}{3}\frac{\bar{\epsilon}_{V}}{M}\right)^{1/4}+\cdots,~~ &
 \mbox{\small {\bf for $q=1/2$}}  \\ 
 \displaystyle  
              
                \displaystyle 
\displaystyle -\frac{1}{M^3}\left[\left(\frac{2}{q}\right)^{3/2}\bar{\epsilon}_{V}\left(1+\frac{\bar{\epsilon}_{V}}{2qM}\right)(\bar{\eta}_{V}
-3\bar{\epsilon}_{V})^2+\sqrt{\frac{2}{q}}\frac{2}{q^2M}\bar{\epsilon}^2_{V}(\bar{\eta}_{V}-3\bar{\epsilon}_{V})^2\right.\\ \left. \displaystyle 
+\sqrt{\frac{2}{q}}\frac{2}{q}\bar{\epsilon}_{V}\left(1+\frac{\bar{\epsilon}_{V}}{2q}\right)\left(\bar{\xi}^2_{V}-10\bar{\epsilon}_{V}\bar{\eta}_{V}
+18\bar{\epsilon}^{2}_{V}\right)\right.\\ \left. \displaystyle  +\sqrt{\frac{2}{q}}\left(\frac{10}{q}\bar{\epsilon}_{V}\bar{\eta}_{V}
\left(\bar{\eta}_{V}-3\bar{\epsilon}_{V}\right)+10\bar{\epsilon}_{V}\left(\bar{\xi}^2_{V}-4\bar{\epsilon}_{V}\bar{\eta}_{V}\right)
                \right.\right.\\ \left.\left.\displaystyle -\frac{36}{q}\bar{\epsilon}^2_{V}\left(\bar{\eta}_{V}-3\bar{\epsilon}_{V}\right)-
                \left(\bar{\sigma}^3_{V}+\bar{\xi}^2_{V}\bar{\eta}_{V}-\bar{\xi}^2_{V}\bar{\epsilon}_{V}\right)\right)\right]\displaystyle
                \left(1-\frac{1}{3q}\frac{\bar{\epsilon}_{V}}{M}\right)^{1/4}+\cdots,~~ &
 \mbox{\small {\bf for any $q$}}.
 \end{array}
\right.\end{eqnarray}
\end{itemize}

\subsubsection{Computation of tensor power spectrum}
In this subsection we will not derive the results for assisted inflation as it is exactly same as derived in the case of single tachyonic inflation. But we will state 
the results for {\bf BD} vacuum where the changes will appear due to the presence of $M$ 
identical copies of the tachyon field. One can similarly write down the detailed expressions for {\bf AV} as well. The changes will appear in the expressions for the following inflationary observables at the horizon crossing:
\begin{itemize}
 \item In the present context amplitude of tensor power spectrum can be computed as:
\begin{eqnarray} 
\label{psxcc122vv2} \Delta_{h, \star}
&=&\left\{\begin{array}{lll}
                    \displaystyle 
 \displaystyle  \left\{\left[1-({\cal C}_{E}+1)\frac{\bar{\epsilon}_{V}}{M}\right]^2 \frac{2H^{2}}{\pi^2M^{2}_{p}}\right\}_{k_{\star}=a_{\star}H_{\star}}\,,~~~~~~~~~ &
 \mbox{\small {\bf for $q=1/2$}}  \\ 
 \displaystyle \left\{\left[1-({\cal C}_{E}+1)\frac{\bar{\epsilon}_{V}}{2qM}\right]^2 \frac{2H^{2}}{\pi^2M^{2}_{p}}\right\}_{k_{\star}=a_{\star}H_{\star}} \,,~~~~~~~~~ &
 \mbox{\small {\bf for any $q$}}.
 \end{array}
\right.\end{eqnarray}
where ${\cal C}_{E}= -2 + \ln 2 + \gamma \approx -0.72.$ 
\item Next one can compute the scalar spectral tilt ($n_{S}$) of the primordial scalar power spectrum as:
\begin{eqnarray} 
\label{psxcc133vv2} n_{h, \star} 
&\approx&\left\{\begin{array}{lll}
                    \displaystyle 
 \displaystyle  -\frac{2}{M}\bar{\epsilon}_{V}\left[1+\frac{\bar{\epsilon}_{V}}{M}+\frac{2}{M}\left({\cal C}_{E}+1\right)\left(3\bar{\epsilon}_{V}-\bar{\eta}_{V}\right)\right]+\cdots,~~ &
 \mbox{\small {\bf for $q=1/2$}}  \\ 
 \displaystyle  
               -\frac{\bar{\epsilon}_{V}}{qM}\left[1+\frac{\bar{\epsilon}_{V}}{2qM}+\frac{1}{M}\sqrt{\frac{2}{q}}\left({\cal C}_{E}+1\right)\left(3\bar{\epsilon}_{V}-\bar{\eta}_{V}\right)\right]+\cdots\,,~~ &
 \mbox{\small {\bf for any $q$}}.
 \end{array}
\right.\end{eqnarray}
\item Next one can compute the running of the tensor spectral tilt ($\alpha_{h}$) of the primordial scalar power spectrum as:
\begin{eqnarray}
\label{psxcc144vv2} \alpha_{h, \star}
&\approx&\left\{ \small\begin{array}{lll}
                    \displaystyle 
\displaystyle -\left[\frac{4}{M^2}\bar{\epsilon}_{V}\left(1+\frac{\bar{\epsilon}_{V}}{M}\right)(\bar{\eta}_{V}-3\bar{\epsilon}_{V})+\frac{4}{M^3}\bar{\epsilon}^2_{V}(\bar{\eta}_{V}-3\bar{\epsilon}_{V})
\right.\\ \left. \displaystyle ~~~~~~~~~~~-\frac{8}{M^3}\left({\cal C}_{E}+1\right)\bar{\epsilon}_{V}(\bar{\eta}_{V}-3\bar{\epsilon}_{V})^2 \right.\\ \left. \displaystyle~~~~~~~~~~~~-\frac{2}{M^3}\bar{\epsilon}_{V}\left(10\bar{\epsilon}_{V}\bar{\eta}_{V}
                -18\bar{\epsilon}^2_{V}-\bar{\xi}^2_{V}\right)\right]_{\star}\displaystyle\left(1-\frac{2}{3}\frac{\bar{\epsilon}_{V}}{M}\right)^{1/4}_{\star}+\cdots,~~ &
 \mbox{\small {\bf for $q=1/2$}}  \\ 
 \displaystyle  
               -\left[2\frac{\bar{\epsilon}_{V}}{qM^2}\left(1+\frac{\bar{\epsilon}_{V}}{2qM}\right)(\bar{\eta}_{V}-3\bar{\epsilon}_{V})+\frac{1}{q^2M^3}\bar{\epsilon}^2_{V}(\bar{\eta}_{V}-3\bar{\epsilon}_{V})
\right.\\ \left. \displaystyle ~~~~~~~~~~~-\frac{8}{(2q)^{5/2}M^3}\left({\cal C}_{E}+1\right)\bar{\epsilon}_{V}(\bar{\eta}_{V}-3\bar{\epsilon}_{V})^2 \right.\\ \left. \displaystyle~~~~~~~~~~~~-
\frac{1}{M^3}\sqrt{\frac{2}{q}}\bar{\epsilon}_{V}\left(10\bar{\epsilon}_{V}\bar{\eta}_{V}
                -18\bar{\epsilon}^2_{V}-\bar{\xi}^2_{V}\right)\right]_{\star}\displaystyle\left(1-\frac{1}{3q}\frac{\bar{\epsilon}_{V}}{M}\right)^{1/4}_{\star}+\cdots\,,~~ &
 \mbox{\small {\bf for any $q$}}.
 \end{array}
\right.\end{eqnarray}
\item Finally, one can also compute the running of the running of scalar spectral tilt ($\kappa_{S}$) of the primordial scalar power spectrum as:
\begin{eqnarray}
\label{psxcc155vv2} \kappa_{h, \star}
&\approx&\left\{ \small\begin{array}{lll}
                    \displaystyle 
\displaystyle \displaystyle -\left[\frac{8}{M^3}\bar{\epsilon}_{V}(1+\bar{\epsilon}_{V})(\bar{\eta}_{V}-3\bar{\epsilon}_{V})^2+\frac{8}{M^4}\bar{\epsilon}^2_{V}(\bar{\eta}_{V}-3\bar{\epsilon}_{V})^2
\right.\\ \left. \displaystyle-\frac{16}{M^4}\left({\cal C}_{E}+1\right)\bar{\epsilon}_{V}\left\{
(\bar{\eta}_{V}-3\bar{\epsilon}_{V})^3 -(\bar{\eta}_{V}-3\bar{\epsilon}_{V})\left(10\bar{\epsilon}_{V}\bar{\eta}_{V}
                -18\bar{\epsilon}^2_{V}-\bar{\xi}^2_{V}\right)\right\}\right.\\ 
\left. \displaystyle~~~~~~~~~~~~-\frac{4}{M^3}\bar{\epsilon}_{V}\left(1+\frac{\bar{\epsilon}_{V}}{M}\right)\left(10\bar{\epsilon}_{V}\bar{\eta}_{V}
                -18\bar{\epsilon}^2_{V}-\bar{\xi}^2_{V}\right)\right]_{\star}\displaystyle\left(1-\frac{2}{3}\frac{\bar{\epsilon}_{V}}{M}\right)^{1/4}_{\star}+\cdots,~~ &
 \mbox{\small {\bf for $q=1/2$}}  \\ 
 \displaystyle  \displaystyle 
\displaystyle -\left[\frac{4}{qM^3}\bar{\epsilon}_{V}\left(1+\frac{\bar{\epsilon}_{V}}{2qM}\right)(\bar{\eta}_{V}
-3\bar{\epsilon}_{V})^2+\frac{2}{q^2M^4}\bar{\epsilon}^2_{V}(\bar{\eta}_{V}-3\bar{\epsilon}_{V})^2\right.\\ \left. \displaystyle 
+\frac{2}{qM^3}\bar{\epsilon}_{V}\left(1+\frac{\bar{\epsilon}_{V}}{2q}\right)\left(\bar{\xi}^2_{V}-10\bar{\epsilon}_{V}\bar{\eta}_{V}
+18\bar{\epsilon}^{2}_{V}\right)\right.\\ \left. \displaystyle -\frac{8}{qM^4}\left({\cal C}_{E}+1\right)\bar{\epsilon}_{V}\left\{
(\bar{\eta}_{V}-3\bar{\epsilon}_{V})^3 -(\bar{\eta}_{V}-3\bar{\epsilon}_{V})\left(10\bar{\epsilon}_{V}\bar{\eta}_{V}
                -18\bar{\epsilon}^2_{V}-\bar{\xi}^2_{V}\right)\right\}\right]_{\star}\\
                \displaystyle~~~~~~~~~~~~~~~~~~~~~~~~~~~~\times\left(1-\frac{1}{3q}\frac{\bar{\epsilon}_{V}}{M}\right)^{1/4}_{\star}+\cdots,~~ &
 \mbox{\small {\bf for any $q$}}.
 \end{array}
\right.\end{eqnarray}
\end{itemize}

\subsubsection{Modified consistency relations}
In this subsection we derive the new (modified) consistency relations for single tachyonic field inflation:
\begin{enumerate}
 \item Next for {\bf BD} vacuum with $|k c_{S}\eta|=1$ case within slow-roll regime we can approximately write the following expression for tensor-to-scalar ratio:
\begin{eqnarray}
\label{psxcc188vv2} r_{\star}
&\approx&\left\{ \small\begin{array}{lll}
                    \displaystyle 
\displaystyle \left[\frac{16}{M}\bar{\epsilon}_{V}c_{S}\frac{\left[1-({\cal C}_{E}+1)\frac{\bar{\epsilon}_{V}}{M}\right]^2}{\left[1-({\cal C}_{E}+1)\frac{\bar{\epsilon}_{V}}{M}-\frac{{\cal C}_{E}}{M}\left(3\bar{\epsilon}_{V}-\bar{\eta}_{V}\right)\right]^2}\right]_{k_{\star}=a_{\star}H_{\star}}\\
\displaystyle =\left[\frac{16}{M}\bar{\epsilon}_{V}\frac{\left[1-({\cal C}_{E}+1)\frac{\bar{\epsilon}_{V}}{M}\right]^2}{\left[1-\left({\cal C}_{E}+\frac{5}{6}\right)\frac{\bar{\epsilon}_{V}}{M}-\frac{{\cal C}_{E}}{M}
                \left(3\bar{\epsilon}_{V}-\bar{\eta}_{V}\right)\right]^2}\right]_{k_{\star}=a_{\star}H_{\star}},~~ &
 \mbox{\small {\bf for $q=1/2$}}  \\ 
 \displaystyle  
                \left[\frac{8}{qM}\bar{\epsilon}_{V}c_{S}\frac{\left[1-({\cal C}_{E}+1)\frac{\bar{\epsilon}_{V}}{2qM}\right]^2}{\left[1-({\cal C}_{E}+1)\frac{\bar{\epsilon}_{V}}{2qM}-\frac{{\cal C}_{E}}{M\sqrt{2q}}\left(3\bar{\epsilon}_{V}-\bar{\eta}_{V}\right)\right]^2}\right]_{k_{\star}=a_{\star}H_{\star}}\\
\displaystyle =\left[\frac{8}{qM}\bar{\epsilon}_{V}\frac{\left[1-({\cal C}_{E}+1)\frac{\bar{\epsilon}_{V}}{2qM}\right]^2}{\left[1-({\cal C}_{E}+1-\Sigma)\frac{\bar{\epsilon}_{V}}{2qM}-\frac{{\cal C}_{E}}{M\sqrt{2q}}\left(3\bar{\epsilon}_{V}-\bar{\eta}_{V}\right)\right]^2}\right]_{k_{\star}=a_{\star}H_{\star}}\,,~~ &
 \mbox{\small {\bf for any $q$}}.
 \end{array}
\right.\end{eqnarray}
\item Hence the consistency relation between the tensor-to-scalar ratio $r$ and spectral tilt $n_{T}$ for tensor modes for {\bf BD} vacuum with $|k c_{S}\eta|=1$ case can be written as: 
\begin{eqnarray}
\label{psxcc199vv2} r_{\star}
&\approx& -8n_{h,\star}\times\underbrace{\left\{ \tiny\tiny\begin{array}{lll}
                    \displaystyle 
\displaystyle\left[c_{S}\frac{\left[1-({\cal C}_{E}+1)\frac{\bar{\epsilon}_{V}}{M}\right]^2}{\left[1-({\cal C}_{E}+1)\frac{\bar{\epsilon}_{V}}{M}-\frac{{\cal C}_{E}}{M}\left(3\bar{\epsilon}_{V}-\bar{\eta}_{V}\right)\right]^2
\left[1+\frac{\bar{\epsilon}_{V}}{M}+\frac{2}{M}\left({\cal C}_{E}+1\right)\left(3\bar{\epsilon}_{V}-\bar{\eta}_{V}\right)\right]}\right]_{k_{\star}=a_{\star}H_{\star}}\\
\displaystyle =\left[\frac{\left[1-({\cal C}_{E}+1)\frac{\bar{\epsilon}_{V}}{M}\right]^2}{\left[1-\left({\cal C}_{E}+\frac{5}{6}\right)\frac{\bar{\epsilon}_{V}}{M}-\frac{{\cal C}_{E}}{M}
                \left(3\bar{\epsilon}_{V}-\bar{\eta}_{V}\right)\right]^2\left[1+\frac{\bar{\epsilon}_{V}}{M}+\frac{2}{M}\left({\cal C}_{E}+1\right)\left(3\bar{\epsilon}_{V}-\bar{\eta}_{V}\right)
                \right]}\right]_{k_{\star}}, &
 \mbox{\small {\bf for $q=1/2$}}  \\ 
 \displaystyle  
                \left[c_{S}\frac{\left[1-({\cal C}_{E}+1)\frac{\bar{\epsilon}_{V}}{2qM}\right]^2}{\left[1-({\cal C}_{E}+1)\frac{\bar{\epsilon}_{V}}{2qM}-\frac{{\cal C}_{E}}
                {M\sqrt{2q}}\left(3\bar{\epsilon}_{V}-\bar{\eta}_{V}\right)\right]^2\left[1+\frac{\bar{\epsilon}_{V}}{2qM}+\frac{1}{M}\sqrt{\frac{2}{q}}\left({\cal C}_{E}+1\right)\left(3\bar{\epsilon}_{V}-\bar{\eta}_{V}\right)\right]}\right]_{k_{\star}=a_{\star}H_{\star}}\\
\displaystyle =\left[\frac{\left[1-({\cal C}_{E}+1)\frac{\bar{\epsilon}_{V}}{2qM}\right]^2}{\left[1-({\cal C}_{E}+1-\Sigma)\frac{\bar{\epsilon}_{V}}{2qM}-\frac{{\cal C}_{E}}{M\sqrt{2q}}
\left(3\bar{\epsilon}_{V}-\bar{\eta}_{V}\right)\right]^2\left[1+\frac{\bar{\epsilon}_{V}}{2qM}+\frac{1}{M}
\sqrt{\frac{2}{q}}\left({\cal C}_{E}+1\right)\left(3\bar{\epsilon}_{V}-\bar{\eta}_{V}\right)\right]}\right]_{k_{\star}}\,, &
 \mbox{\small {\bf for any $q$}}.
 \end{array}
\right.}_{Correction~factor}\end{eqnarray}
\item Next one can express the first two slow-roll parameters $\bar{\epsilon}_{V}$ and $\bar{\eta}_{V}$ in terms of the inflationary observables as:
\begin{eqnarray}
\label{psxcc1101vv2} \bar{\epsilon}_{V}
&\approx&\left\{ \begin{array}{lll}
                    \displaystyle 
\displaystyle\epsilon_{1}\approx -\frac{n_{h,\star}M}{2}+\cdots\approx \frac{r_{\star}M}{16}+\cdots,~~~~~~~~~~~~~~~~~~~~~~~~~~~~~~~~~~~~~~~~~~ &
 \mbox{\small {\bf for $q=1/2$}}  \\ 
 \displaystyle  
                2q\epsilon_{1}\approx -qn_{h,\star}M+\cdots\approx \frac{qr_{\star}M}{8}+\cdots\,,~~ &
 \mbox{\small {\bf for any $q$}}.
 \end{array}
\right. \end{eqnarray}
\begin{eqnarray}
\label{psxcc1vv2} \bar{\eta}_{V}
&\approx&\left\{ \begin{array}{lll}
                    \displaystyle 
\displaystyle 3\epsilon_{1}-\frac{\epsilon_{2}}{2}\approx \frac{M}{2}\left(n_{\zeta,\star}-1+\frac{r_{\star}}{2}\right)+\cdots\approx \frac{M}{2}\left(n_{\zeta,\star}-1-4n_{h,\star}\right)+\cdots,~~ &
 \mbox{\small {\bf for $q=1/2$}}  \\ 
 \displaystyle  
                6q\epsilon_{1}-\sqrt{\frac{q}{2}}\epsilon_{2}\approx M\sqrt{\frac{q}{2}}\left(n_{\zeta,\star}-1+\left(\frac{1}{q}+3\sqrt{\frac{2}{q}}\right)\frac{qr_{\star}}{8}\right)+\cdots\\
                \displaystyle~~~~~~~~~~~~~~~~~\approx M\sqrt{\frac{q}{2}}\left(n_{\zeta,\star}-1-q\left(\frac{1}{q}+3\sqrt{\frac{2}{q}}\right)n_{h,\star}\right)+\cdots\,,~~ &
 \mbox{\small {\bf for any $q$}}.
 \end{array}
\right.\end{eqnarray}
\item Then the connecting consistency relation between tensor and scalar spectral tilt and tensor-to-scalar ratio can be expressed as:
\begin{eqnarray} 
\label{psxcc112vv2} n_{h, \star} 
&\approx&\left\{\begin{array}{lll}
                    \displaystyle 
               \displaystyle
                -\frac{r_{\star}}{8c_{S}}\left[1-\frac{r_{\star}}{16}+\left(1-n_{\zeta,\star}\right)-{\cal C}_{E}\left\{\frac{r_{\star}}{8}
                +\left(n_{\zeta,\star}-1\right)\right\}\right]+\cdots,~~ &
 \mbox{\small {\bf for $q=1/2$}}  \\ 
 \displaystyle  
               \displaystyle
                -\frac{r_{\star}}{8c_{S}}\left[1+\left\{\left(\frac{3q}{8}\sqrt{\frac{2}{q}}
                -\left(\sqrt{\frac{2}{q}}+5\right)\frac{1}{16}\right)r_{\star}
                +\frac{\left(1-n_{\zeta,\star}\right)}{\sqrt{2q}}\right\}
                \right.\\ \left.\displaystyle+\sqrt{\frac{2}{q}}{\cal C}_{E}\left(\frac{3qr_{\star}}{8}-\frac{1}{2}\left\{n_{\zeta,\star}-1+\left(\frac{1}{q}+3\sqrt{\frac{2}{q}}\right)\frac{qr_{\star}}{8}\right\}\right)\right]+\cdots\,,~~ &
 \mbox{\small {\bf for any $q$}}.
 \end{array}
\right. \end{eqnarray}
Finally using the approximated version of the expression for $c_{S}$ in terms of slow-roll parameters one can recast this consistency condition as: 
\begin{eqnarray} 
\label{psxcc113vv2} n_{h, \star} 
&\approx&\left\{\begin{array}{lll}
                    \displaystyle 
               \displaystyle
                -\frac{r_{\star}}{8}\left[1-\frac{r_{\star}}{24}+\left(1-n_{\zeta,\star}\right)-{\cal C}_{E}\left\{\frac{r_{\star}}{8}
                +\left(n_{\zeta,\star}-1\right)\right\}\right]+\cdots,~~ &
 \mbox{\small {\bf for $q=1/2$}}  \\ 
 \displaystyle  
               \displaystyle
                -\frac{r_{\star}}{8}\left[1+\left\{\left(\frac{3q}{8}\sqrt{\frac{2}{q}}
                -\left(\sqrt{\frac{2}{q}}+5\right)\frac{1}{16}+\frac{\Sigma}{8}\right)r_{\star}
                +\frac{\left(1-n_{\zeta,\star}\right)}{\sqrt{2q}}\right\}
                \right.\\ \left.\displaystyle+\sqrt{\frac{2}{q}}{\cal C}_{E}\left(\frac{3qr_{\star}}{8}-\frac{1}{2}\left\{n_{\zeta,\star}-1+\left(\frac{1}{q}+3\sqrt{\frac{2}{q}}\right)\frac{qr_{\star}}{8}\right\}\right)\right]+\cdots\,,~~ &
 \mbox{\small {\bf for any $q$}}.
 \end{array}
\right. \end{eqnarray}
\item Next the running of the sound speed $c_{S}$ can be written in terms of slow-roll parameters as:
\be\begin{array}{lll}\label{rteqxc14vv2}
 \displaystyle S=\frac{\dot{c}_{S}}{Hc_{S}}=\frac{d\ln c_{S}}{dN}= \frac{d\ln c_{S}}{d\ln k}\\ 
                 ~~=\left\{\begin{array}{lll}
                    \displaystyle  
                   -\frac{2}{3M^2}\bar{\epsilon}_{V}\left(\bar{\eta}_{V}-3\bar{\epsilon}_{V}\right)\left(1-\frac{2}{3}\frac{\bar{\epsilon}_{V}}{M}\right)^{1/4}+\cdots \,,~~~~ &
 \mbox{\small {\bf for $q=1/2$}}  \\ 
 \displaystyle  
 -\frac{(1-q)}{3q^2M^2}\bar{\epsilon}_{V}\left(\bar{\eta}_{V}-3\bar{\epsilon}_{V}\right)\left(1-\frac{1}{3q}\frac{\bar{\epsilon}_{V}}{M}\right)^{1/4}+\cdots \,,~~~ &
 \mbox{\small {\bf for ~any~$q$}}.
          \end{array}
\right.
\end{array}\ee
which can be treated as another slow-roll parameter in the present context. One can also recast the slow-roll parameter $S$ in terms of the inflationary observables as:
\be\begin{array}{lll}\label{rteqxc15vv2}
 \displaystyle S=\left\{\begin{array}{lll}
                    \displaystyle  
                   -\frac{r_{\star}}{48}\left(n_{\zeta,\star}-1+\frac{r_{\star}}{8}\right)\left(1-\frac{r_{\star}}{24}\right)^{1/4}+\cdots \,,~~~~ &
 \mbox{\small {\bf for $q=1/2$}}  \\ 
 \displaystyle  
 -\frac{(1-q)}{24q^2}\sqrt{\frac{q}{2}}r_{\star}\left(n_{\zeta,\star}-1+\frac{r_{\star}}{8}\right)\left(1-\frac{r_{\star}}{24}\right)^{1/4}+\cdots \,,~~~ &
 \mbox{\small {\bf for ~any~$q$}}.
          \end{array}
\right.
\end{array}\ee
\item Further the running of tensor spectral tilt can be written in terms of the inflationary observables as:
\begin{eqnarray}
\label{psxcc116vv2} \alpha_{h, \star}&=&\left\{ \small\begin{array}{lll}
                    \displaystyle 
\displaystyle -\left[\frac{r_{\star}}{8}\left(1+\frac{r_{\star}}{8}\right)\left(n_{\zeta,\star}-1+\frac{r_{\star}}{8}\right)-\frac{r_{\star}}{8}\left(n_{\zeta,\star}-1+\frac{r_{\star}}{8}\right)^2
\right.\\ \left. \displaystyle ~~~~~~~~~~~-{\cal C}_{E}\frac{r_{\star}}{8}\left(n_{\zeta,\star}-1+\frac{r_{\star}}{8}\right)^2 \right]\displaystyle\left(1-\frac{r_{\star}}{24}\right)^{1/4}+\cdots, &
 \mbox{\small {\bf for $q=1/2$}}  \\ 
 \displaystyle  
               -\left[\frac{r_{\star}}{4}\sqrt{\frac{q}{2}}\left(1+\frac{r_{\star}}{8}\right)\left(n_{\zeta,\star}-1+\frac{r_{\star}}{8}\right)
\right.\\ \left. \displaystyle -\frac{1}{(2q)^{1/2}}\left({\cal C}_{E}+1\right)\frac{r_{\star}}{8}\left(n_{\zeta,\star}-1+\frac{r_{\star}}{8}\right)^2 \right]\displaystyle
\left(1-\frac{r_{\star}}{24}\right)^{1/4}+\cdots\,, &
 \mbox{\small {\bf for any $q$}}.
 \end{array}
\right.\end{eqnarray}
\item Next the scalar power spectrum can be expressed in terms of the other inflationary observables as:
\begin{eqnarray} 
\label{psxcc117vv2} \Delta_{\zeta, \star}
&\approx&\left\{\begin{array}{lll}
                    \displaystyle 
 \displaystyle  
                \left[1-\left({\cal C}_{E}+\frac{5}{6}\right)\frac{r_{\star}}{16}+\frac{{\cal C}_{E}}{2}
                \left(n_{\zeta,\star}-1+\frac{r_{\star}}{8}\right)\right]^2 \frac{2H^{2}_{\star}}{\pi^2M^{2}_{p}r_{\star}}+\cdots \,,~~~~ &
 \mbox{\small {\bf for $q=1/2$}}  \\ 
 \displaystyle  
                \left[1-({\cal C}_{E}+1-\Sigma)\frac{r_{\star}}{16}+\frac{{\cal C}_{E}}{2}\left(n_{\zeta,\star}-1
                +\frac{r_{\star}}{8}\right)\right]^2 \displaystyle\frac{2 H^{2}_{\star}}{\pi^2M^{2}_{p}
                r_{\star}}+\cdots \,,~~~~ &
 \mbox{\small {\bf for any $q$}}.
 \end{array}
\right.\end{eqnarray}
\item Further the tensor power spectrum can be expressed in terms of the other inflationary observables as:
\begin{eqnarray} 
\label{psxcc118vv2} \Delta_{h, \star}
&=&\left\{\begin{array}{lll}
                    \displaystyle 
 \displaystyle  \left[1-({\cal C}_{E}+1)\frac{r_{\star}}{16}\right]^2 \frac{2H^{2}_{\star}}{\pi^2M^{2}_{p}}+\cdots\,,~~~~~~~~~ &
 \mbox{\small {\bf for $q=1/2$}}  \\ 
 \displaystyle \left[1-({\cal C}_{E}+1)\frac{r_{\star}}{16}\right]^2 \frac{2H^{2}_{\star}}{\pi^2M^{2}_{p}}+\cdots \,,~~~~~~~~~ &
 \mbox{\small {\bf for any $q$}}.
 \end{array}
\right.\end{eqnarray}
\item Next the running of the tensor to scalar ratio can be expressed in terms of inflationary observables as:
\begin{eqnarray}
\label{psxcc119vv2} \alpha_{r,\star}&=&-8\alpha_{h, \star}+\cdots\nonumber\\
&\approx&\left\{ \small\begin{array}{lll}
                    \displaystyle 
\displaystyle \left[r_{\star}\left(1+\frac{r_{\star}}{8}\right)\left(n_{\zeta,\star}-1+\frac{r_{\star}}{8}\right)
-r_{\star}\left(n_{\zeta,\star}-1+\frac{r_{\star}}{8}\right)^2
\right.\\ \left. \displaystyle ~~~~~~~~~~~-{\cal C}_{E}r_{\star}\left(n_{\zeta,\star}-1+\frac{r_{\star}}{8}\right)^2 \right]\displaystyle\left(1-\frac{r_{\star}}{24}\right)^{1/4}+\cdots, &
 \mbox{\small {\bf for $q=1/2$}}  \\ 
 \displaystyle  
                 \left[2r_{\star}\sqrt{\frac{q}{2}}\left(1+\frac{r_{\star}}{8}\right)\left(n_{\zeta,\star}-1+\frac{r_{\star}}{8}\right)
\right.\\ \left. \displaystyle -\frac{1}{(2q)^{1/2}}\left({\cal C}_{E}+1\right)r_{\star}\left(n_{\zeta,\star}-1
+\frac{r_{\star}}{8}\right)^2 \right]\displaystyle
\left(1-\frac{r_{\star}}{24}\right)^{1/4}+\cdots\,, &
 \mbox{\small {\bf for any $q$}}.
 \end{array}
\right.\end{eqnarray}
\item Finally the scale of single field tachyonic inflation can be expressed in terms of the Hubble parameter and the other inflationary observables as: 
\begin{eqnarray} \label{p20vvv2}
H_{inf}&=& H_{\star}
\approx\left\{\begin{array}{lll}
                    \displaystyle 
 \displaystyle  
               \frac{\sqrt{\frac{\Delta_{\zeta, \star}r_{\star}}{2}}\pi M_{p}}{ \left[1-\left({\cal C}_{E}+\frac{5}{6}\right)\frac{r_{\star}}{16}+\frac{{\cal C}_{E}}{2}
                \left(n_{\zeta,\star}-1+\frac{r_{\star}}{8}\right)\right]} +\cdots \,,~~~~ &
 \mbox{\small {\bf for $q=1/2$}}  \\ 
 \displaystyle \frac{\sqrt{\frac{\Delta_{\zeta, \star}r_{\star}}{2}}\pi M_{p}}{ \left[1-({\cal C}_{E}+1-\Sigma)
 \frac{r_{\star}}{16}+\frac{{\cal C}_{E}}{2}\left(n_{\zeta,\star}-1
                +\frac{r_{\star}}{8}\right)\right]} 
                +\cdots \,,~~~~ &
 \mbox{\small {\bf for any $q$}}.
 \end{array}
\right.\end{eqnarray}
One can recast this statement in terms of inflationary potential as:
\begin{eqnarray} 
\label{psxcc12200vvv2} \sqrt[4]{V_{inf}}&=&\sqrt[4]{V_{\star}}\approx\left\{\begin{array}{lll}
                    \displaystyle 
 \displaystyle  
               \frac{\sqrt[4]{\frac{3\Delta_{\zeta, \star}r_{\star}}{2}}\sqrt{\pi} M_{p}}{ \sqrt{\left[1-\left({\cal C}_{E}+
               \frac{5}{6}\right)\frac{r_{\star}}{16}+\frac{{\cal C}_{E}}{2}
                \left(n_{\zeta,\star}-1+\frac{r_{\star}}{8}\right)\right]}} +\cdots \,, &
 \mbox{\small {\bf for $q=1/2$}}  \\ 
 \displaystyle \frac{\sqrt[4]{\frac{3\Delta_{\zeta, \star}r_{\star}}{2}}\sqrt{\pi} M_{p}}{ \sqrt{\left[1-({\cal C}_{E}+1-\Sigma)
 \frac{r_{\star}}{16}+\frac{{\cal C}_{E}}{2}\left(n_{\zeta,\star}-1
                +\frac{r_{\star}}{8}\right)\right]}} 
                +\cdots \,, &
 \mbox{\small {\bf for any $q$}}.
 \end{array}
\right.\end{eqnarray}
\end{enumerate}

\subsubsection{Field excursion for tachyon}
In this subsection we explicitly derive the expression for the field excursion for tachyonic inflation defined as:
\bea |\Delta T| &=& |T_{cmb}-T_{end}|=|T_{\star}-T_{end}| \eea
where $T_{cmb}$, $T_{end}$ and $T_{\star}$ signify the tachyon field value at the time of horizon exit, at end of inflation and at pivot scale respectively.
Here we perform the computation for both {\bf AV} and {\bf BD} vacuum.
For for the sake of simplicity the pivot scale is fixed at the horizon exit scale. To compute the expression for the field excursion we perform the following steps:
\begin{enumerate}
 \item We start with the operator identity for single field tachyon using which one can write expression for the tachyon field variation with respect to the momentum scale ($k$) or number of e-foldings ($N$) 
 in terms of the inflationary observables as:
 \be\begin{array}{lll}\label{vchj2xcvv2}
 \displaystyle  \frac{1}{H}\frac{dT}{dt}=\frac{dT}{dN}=\frac{dT}{d\ln k}\approx\left\{\begin{array}{lll}
                    \displaystyle  
                   \sqrt{\frac{r}{8MV(T)\alpha^{'}}}M_{p}\left(1-\frac{r}{24}\right)^{1/4}+\cdots\,,~~~~~~ &
 \mbox{\small {\bf for {$q=1/2$ }}}  \\ \\
 \displaystyle   
                \sqrt{\frac{qr}{4MV(T)\alpha^{'}}}\frac{M_{p}}{2q}\left(1-\frac{r}{24}\right)^{1/4}+\cdots\,.~~~~~~ &
 \mbox{\small {\bf for {~any~arbitrary~ $q$ }}} 
          \end{array}
\right.
\end{array}\ee
where the tensor-to-scalar ratio $r$ is function of $k$ or $N$.
 
 \item Next using Eq~(\ref{vchj2xc}) we can write the following integral equation:
 \be\begin{array}{lll}\label{vchj2xc22zxvv2}
 \displaystyle  \int^{T_{\star}}_{T_{end}} dT~\sqrt{V(T)} \approx\left\{\begin{array}{lll}
                    \displaystyle  
                   \int^{k_{\star}}_{k_{end}} d\ln k~\sqrt{\frac{r}{8M\alpha^{'}}}M_{p}\left(1-\frac{r}{24}\right)^{1/4}+\cdots\,\\ \displaystyle  
                  \displaystyle =\int^{N_{\star}}_{N_{end}} dN~\sqrt{\frac{r}{8M\alpha^{'}}}M_{p}\left(1-\frac{r}{24}\right)^{1/4}+\cdots\,,~~~~~~ &
 \mbox{\small {\bf for {$q=1/2$ }}}  \\ \\
 \displaystyle   
                \int^{k_{\star}}_{k_{end}} d\ln k~\sqrt{\frac{qr}{4M\alpha^{'}}}\frac{M_{p}}{2q}\left(1-\frac{r}{24}\right)^{1/4}+\cdots\,\\
                \displaystyle   
                =\int^{N_{\star}}_{N_{end}} dN~\sqrt{\frac{qr}{4M\alpha^{'}}}\frac{M_{p}}{2q}\left(1-\frac{r}{24}\right)^{1/4}+\cdots\,.~~~~~~ &
 \mbox{\small {\bf for {~any~arbitrary~ $q$ }}} 
          \end{array}
\right.
\end{array}\ee
 \item Next we use the same parametrization of the tensor-to-scalar ratio for $q=1/2$ and for any arbitrary $q$ at any arbitrary scale as mentioned earlier for the single tachyonic field case.
\item For any value of $q$ including $q=1/2$ we need to compute the following integral:
\be\begin{array}{lll}\label{raqq1vv2}\tiny\tiny
  \displaystyle \int^{k_{\star}}_{k_{end}} d\ln k~\sqrt{\frac{qr}{4M\alpha^{'}}}\frac{M_{p}}{2q}\left(1-\frac{r}{24}\right)^{1/4}\,\\ \\ \approx\small\left\{\begin{array}{ll}
                    \displaystyle  \sqrt{\frac{qr_{\star}}{4M\alpha^{'}}}\frac{M_{p}}{2q}\left(1-\frac{r_{\star}}{24}\right)^{1/4}\ln\left(\frac{k_{\star}}{k_{end}}\right) &
 \mbox{ {\bf for \underline{Case I}}}  \\ \\
         \displaystyle \frac{\sqrt{\frac{qr_{\star}}{4M\alpha^{'}}}\frac{M_{p}}{q}}{3(n_{h,\star}-n_{\zeta,\star}+1)}\left[  \left\{\, _2F_1\left[\frac{1}{2},\frac{3}{4};\frac{3}{2};\frac{r_{\star}}{24} \right]+2 \left(1-\frac{r_{\star}}{24}
         \right)^{1/4}\right\}\right.\\ \left.
         \displaystyle -\left(\frac{k_{end}}{k_{\star}}\right)^{\frac{n_{h,\star}-n_{\zeta,\star}+1}{2}} \left\{\, _2F_1\left[\frac{1}{2},\frac{3}{4};\frac{3}{2};\frac{r_{\star}}{24}
         \left(\frac{k_{end}}{k_{\star}}\right)^{n_{h,\star}-n_{\zeta,\star}+1}\right]\right.\right.\\ \left.\left.\displaystyle~~~~~~~~+2 \left(1-\frac{r_{\star}}{24}\left(\frac{k_{end}}{k_{\star}}\right)^{n_{h,\star}
         -n_{\zeta,\star}+1}
         \right)^{1/4}\right\}\right] & \mbox{ {\bf for \underline{Case II}}}\\ \\
\displaystyle  \sqrt{\frac{\pi q r_{\star}}{M\alpha^{'}(\alpha_{h,\star}-\alpha_{\zeta,\star})}}\frac{M_{p}}{48q}
e^{-\frac{3 (n_{h,\star}-n_{\zeta,\star}+1)^2}{4 (\alpha_{h,\star}-\alpha_{\zeta,\star})}}\\
\displaystyle ~~\left[12 e^{\frac{(n_{h,\star}-n_{\zeta,\star}+1)^2}{ 2(\alpha_{h,\star}-\alpha_{\zeta,\star})}}\left\{ \text{erfi}\left(\frac{n_{h,\star}-n_{\zeta,\star}+1}{2 \sqrt{\alpha_{h,\star}-\alpha_{\zeta,\star}}}
\right)\right.\right.\\ \left.\left. 
\displaystyle~~~~~-
\text{erfi}\left(\frac{n_{h,\star}-n_{\zeta,\star}+1}{2 \sqrt{\alpha_{h,\star}-\alpha_{\zeta,\star}}}+\frac{\sqrt{\alpha_{h,\star}-\alpha_{\zeta,\star}}}{2}\ln\left(\frac{k_{end}}{k_{\star}}\right)
\right)\right\}\right.\\ \left. \displaystyle~~~~~~-\frac{\sqrt{3}r_{\star}}{24} \left\{\text{erfi}
\left(\frac{\sqrt{3} (n_{h,\star}-n_{\zeta,\star}+1)}{2 \sqrt{\alpha_{h,\star}-\alpha_{\zeta,\star}}}\right)\right.\right.\\ \left.\left. 
\displaystyle-\text{erfi}
\left(\frac{\sqrt{3} (n_{h,\star}-n_{\zeta,\star}+1)}{2 \sqrt{\alpha_{h,\star}-\alpha_{\zeta,\star}}}+\frac{\sqrt{3(\alpha_{h,\star}-\alpha_{\zeta,\star})}}{2}
\ln\left(\frac{k_{end}}{k_{\star}}\right)\right)\right\}\right] & \mbox{ {\bf for \underline{Case III}}}.
          \end{array}
\right.& \mbox{ \underline{\bf for {BD}}} 
\end{array}\ee
Simalrly for {\bf AV} we get the following result:
\be\begin{array}{lll}\label{raqq2vv2}\tiny
  \displaystyle \int^{k_{\star}}_{k_{end}} d\ln k~\sqrt{\frac{qr}{4M\alpha^{'}}}\frac{M_{p}}{2q}\left(1-\frac{r}{24}\right)^{1/4}\, \approx\tiny\left\{\begin{array}{ll}
                    \displaystyle  \displaystyle  
  \sqrt{\frac{r_{\star}}{8M\alpha^{'}}}\frac{M_{p}}{2c_{S}}\left(1-\frac{qr_{\star}}{48c^{2}_{S}}\right)^{1/4}\ln\left(\frac{k_{\star}}{k_{end}}\right) &
 \mbox{ {\bf for \underline{Case I}}}  \\ \\
         \displaystyle \displaystyle  
  \sqrt{\frac{r_{\star}}{8M\alpha^{'}}}\frac{M_{p}\left|D\right|}{2c_{S}\left|C\right|}\left(1-\frac{qr_{\star}}{48c^{2}_{S}}\frac{\left|D\right|^2}
  {\left|C\right|^2}\right)^{1/4}\ln\left(\frac{k_{\star}}{k_{end}}\right) & \mbox{ {\bf for \underline{Case II}}}.
          \end{array}
\right.& \mbox{ \underline{\bf for {AV}}} 
\end{array}\ee
In terms number of e-foldings $N$ one can re-express Eq~(\ref{raqq1vv2}) and Eq~(\ref{raqq2vv2}) as:
\be\begin{array}{lll}\label{raqg1vv2}
  \displaystyle \int^{k_{\star}}_{k_{end}} d\ln k~\sqrt{\frac{qr}{4M\alpha^{'}}}\frac{M_{p}}{2q}\left(1-\frac{r}{24}\right)^{1/4}\,\\  \approx\small\left\{\begin{array}{ll}
                    \displaystyle  \sqrt{\frac{qr_{\star}}{4M\alpha^{'}}}\frac{M_{p}}{2q}\left(1-\frac{r_{\star}}{24}\right)^{1/4}\left(N_{\star}-N_{end}\right) &
 \mbox{ {\bf for \underline{Case I}}}  \\ \\
         \displaystyle \frac{\sqrt{\frac{qr_{\star}}{4M\alpha^{'}}}\frac{M_{p}}{q}}{3(n_{h,\star}-n_{\zeta,\star}+1)}\left[  
         \left\{\, _2F_1\left[\frac{1}{2},\frac{3}{4};\frac{3}{2};\frac{r_{\star}}{24} \right]+2 \left(1-\frac{r_{\star}}{24}
         \right)^{1/4}\right\}\right.\\ \left.
         \displaystyle -e^{\frac{n_{h,\star}-n_{\zeta,\star}+1}{2}(N_{end}-N_{\star})} \left\{\, _2F_1\left[\frac{1}{2},\frac{3}{4};\frac{3}{2};\frac{r_{\star}}{24}
         e^{(n_{h,\star}-n_{\zeta,\star}+1)(N_{end}-N_{\star})}\right]\right.\right.\\ \left.\left.\displaystyle~~~~~~~~+2 \left(1-\frac{r_{\star}}{24}e^{(n_{h,\star}
         -n_{\zeta,\star}+1)(N_{end}-N_{\star})}
         \right)^{1/4}\right\}\right] & \mbox{ {\bf for \underline{Case II}}}\\ \\
\displaystyle  \sqrt{\frac{\pi q r_{\star}}{M\alpha^{'}(\alpha_{h,\star}-\alpha_{\zeta,\star})}}\frac{M_{p}}{48q}
e^{-\frac{3 (n_{h,\star}-n_{\zeta,\star}+1)^2}{4 (\alpha_{h,\star}-\alpha_{\zeta,\star})}}\\
\displaystyle ~~\left[12 e^{\frac{(n_{h,\star}-n_{\zeta,\star}+1)^2}{ 2(\alpha_{h,\star}-\alpha_{\zeta,\star})}}\left\{ \text{erfi}\left(\frac{n_{h,\star}-n_{\zeta,\star}+1}{2 \sqrt{\alpha_{h,\star}-\alpha_{\zeta,\star}}}
\right)\right.\right.\\ \left.\left. 
\displaystyle~~~~~-
\text{erfi}\left(\frac{n_{h,\star}-n_{\zeta,\star}+1}{2 \sqrt{\alpha_{h,\star}-\alpha_{\zeta,\star}}}+\frac{\sqrt{\alpha_{h,\star}-\alpha_{\zeta,\star}}}{2}(N_{end}-N_{\star})
\right)\right\}\right.\\ \left. \displaystyle~~~~~~-\frac{\sqrt{3}r_{\star}}{24} \left\{\text{erfi}
\left(\frac{\sqrt{3} (n_{h,\star}-n_{\zeta,\star}+1)}{2 \sqrt{\alpha_{h,\star}-\alpha_{\zeta,\star}}}\right)\right.\right.\\ \left.\left. 
\displaystyle-\text{erfi}
\left(\frac{\sqrt{3} (n_{h,\star}-n_{\zeta,\star}+1)}{2 \sqrt{\alpha_{h,\star}-\alpha_{\zeta,\star}}}+\frac{\sqrt{3(\alpha_{h,\star}-\alpha_{\zeta,\star})}}{2}
(N_{end}-N_{\star})\right)\right\}\right] & \mbox{ {\bf for \underline{Case III}}}.
          \end{array}
\right.& \mbox{ \underline{\bf for {BD}}} 
\end{array}\ee
\be\begin{array}{lll}\label{raqg2vv2}
  \displaystyle \int^{k_{\star}}_{k_{end}} d\ln k~\sqrt{\frac{qr}{4\alpha^{'}}}\frac{M_{p}}{2q}\left(1-\frac{r}{24}\right)^{1/4}\,\approx\tiny\left\{\begin{array}{ll}
                    \displaystyle  \displaystyle  
  \sqrt{\frac{r_{\star}}{8M\alpha^{'}}}\frac{M_{p}}{2c_{S}}\left(1-\frac{qr_{\star}}{48c^{2}_{S}}\right)^{1/4}(N_{\star}-N_{end}) &
 \mbox{ {\bf for \underline{Case I}}}  \\ \\
         \displaystyle \displaystyle  
  \sqrt{\frac{r_{\star}}{8M\alpha^{'}}}\frac{M_{p}\left|D\right|}{2c_{S}\left|C\right|}\left(1-\frac{qr_{\star}}{48c^{2}_{S}}\frac{\left|D\right|^2}
  {\left|C\right|^2}\right)^{1/4}(N_{\star}-N_{end}) & \mbox{ {\bf for \underline{Case II}}}.
          \end{array}
\right.& \mbox{ \underline{\bf for {AV}}} 
\end{array}\ee

Here the two possibilities for {\bf AV} vacuum as appearing for assisted inflationary framework are:-
\begin{itemize}
                        \item \underline{\bf Case I} stands for a situation where the spectrum is characterized by the constraint i) $D_{1}=D_{2}=C_{1}=C_{2}\neq 0$, ii) $D_{1}=D_{2}$, $C_{1}=C_{2}=0$, iii) $D_{1}=D_{2}=0$,
                        $C_{1}=C_{2}$.

\item  \underline{\bf Case II} stands for a situation where the spectrum is characterized by the constraint i) $\mu\approx \nu$, $D_{1}=D_{2}=D\neq 0$ and $C_{1}=C_{2}=C\neq 0$, ii) $\mu\approx \nu$, $D_{1}=D\neq 0$, $D_{2}=0$ and $C_{1}=C\neq 0$, $C_{2}=0
$, iii) $\mu\approx \nu$, $D_{2}=D\neq 0$, $D_{1}=0$ and $C_{2}=C\neq 0$, $C_{1}=0$.
\end{itemize} 
\item Next exactly following the same steps for single field inflation and also using Eq~(\ref{raqg1vv2}), Eq~(\ref{raqg2vv2}) and Eq~(\ref{vchj2xc22zxvv2}) we get:
\be\begin{array}{lll}\label{raqggg1vv2}
  \displaystyle \frac{\Delta T}{M_{p}}\,  \approx\footnotesize\left\{\begin{array}{ll}
                    \displaystyle  \sqrt{\frac{qr_{\star}}{4M\alpha^{'}V_{0}}}\frac{1}{2q}\left(1-\frac{r_{\star}}{24}\right)^{1/4}\left(N_{\star}-N_{end}\right) &
 \mbox{ {\bf for \underline{Case I}}}  \\ \\
         \displaystyle \frac{\sqrt{\frac{qr_{\star}}{4M\alpha^{'}V_{0}}}\frac{1}{q}}{3(n_{h,\star}-n_{\zeta,\star}+1)}\left[  
         \left\{\, _2F_1\left[\frac{1}{2},\frac{3}{4};\frac{3}{2};\frac{r_{\star}}{24} \right]+2 \left(1-\frac{r_{\star}}{24}
         \right)^{1/4}\right\}\right.\\ \left.
         \displaystyle -e^{\frac{n_{h,\star}-n_{\zeta,\star}+1}{2}(N_{end}-N_{\star})} \left\{\, _2F_1\left[\frac{1}{2},\frac{3}{4};\frac{3}{2};\frac{r_{\star}}{24}
         e^{(n_{h,\star}-n_{\zeta,\star}+1)(N_{end}-N_{\star})}\right]\right.\right.\\ \left.\left.\displaystyle~~~~~~~~+2 \left(1-\frac{r_{\star}}{24}e^{(n_{h,\star}
         -n_{\zeta,\star}+1)(N_{end}-N_{\star})}
         \right)^{1/4}\right\}\right] & \mbox{ {\bf for \underline{Case II}}}\\ \\
\displaystyle  \sqrt{\frac{\pi q r_{\star}}{M\alpha^{'}(\alpha_{h,\star}-\alpha_{\zeta,\star})V_{0}}}\frac{1}{48q}
e^{-\frac{3 (n_{h,\star}-n_{\zeta,\star}+1)^2}{4 (\alpha_{h,\star}-\alpha_{\zeta,\star})}}\\
\displaystyle ~~\left[12 e^{\frac{(n_{h,\star}-n_{\zeta,\star}+1)^2}{ 2(\alpha_{h,\star}-\alpha_{\zeta,\star})}}\left\{ \text{erfi}\left(\frac{n_{h,\star}-n_{\zeta,\star}+1}{2 \sqrt{\alpha_{h,\star}-\alpha_{\zeta,\star}}}
\right)\right.\right.\\ \left.\left. 
\displaystyle~~~~~-
\text{erfi}\left(\frac{n_{h,\star}-n_{\zeta,\star}+1}{2 \sqrt{\alpha_{h,\star}-\alpha_{\zeta,\star}}}+\frac{\sqrt{\alpha_{h,\star}-\alpha_{\zeta,\star}}}{2}(N_{end}-N_{\star})
\right)\right\}\right.\\ \left. \displaystyle~~~~~~-\frac{\sqrt{3}r_{\star}}{24} \left\{\text{erfi}
\left(\frac{\sqrt{3} (n_{h,\star}-n_{\zeta,\star}+1)}{2 \sqrt{\alpha_{h,\star}-\alpha_{\zeta,\star}}}\right)\right.\right.\\ \left.\left. 
\displaystyle-\text{erfi}
\left(\frac{\sqrt{3} (n_{h,\star}-n_{\zeta,\star}+1)}{2 \sqrt{\alpha_{h,\star}-\alpha_{\zeta,\star}}}+\frac{\sqrt{3(\alpha_{h,\star}-\alpha_{\zeta,\star})}}{2}
(N_{end}-N_{\star})\right)\right\}\right] & \mbox{ {\bf for \underline{Case III}}}.
          \end{array}
\right.& \mbox{ \small\underline{\bf for {BD}}} 
\end{array}\ee
\be\begin{array}{lll}\label{raqggg2vv2}
  \displaystyle \frac{\Delta T}{M_{p}} \approx\footnotesize\left\{\begin{array}{ll}
                    \displaystyle  \displaystyle  
  \sqrt{\frac{r_{\star}}{8M\alpha^{'}V_{0}}}\frac{1}{2c_{S}}\left(1-\frac{qr_{\star}}{48c^{2}_{S}}\right)^{1/4}(N_{\star}-N_{end}) &
 \mbox{ {\bf for \underline{Case I}}}  \\ \\
         \displaystyle \displaystyle  
  \sqrt{\frac{r_{\star}}{8M\alpha^{'}V_{0}}}\frac{\left|D\right|}{2c_{S}\left|C\right|}\left(1-\frac{qr_{\star}}{48c^{2}_{S}}\frac{\left|D\right|^2}
  {\left|C\right|^2}\right)^{1/4}(N_{\star}-N_{end}) & \mbox{ {\bf for \underline{Case II}}}.
          \end{array}
\right.& \mbox{ \underline{\bf for {AV}}} 
\end{array}\ee
\item Next using Eq~(\ref{psxcc12200vvv2}) in eq~(\ref{raqggg1vv2}) and Eq~(\ref{raqggg2vv2}) we get:
\be\begin{array}{lll}\label{raqgggg1vv2}
  \displaystyle \frac{\Delta T}{M_{p}}\,  \approx\footnotesize\left\{\begin{array}{ll}
                    \displaystyle  \sqrt{\frac{qr_{\star}}{4M\alpha^{'}}}\frac{1}{2q\sqrt{\frac{3\Delta_{\zeta, \star}r_{\star}}{2}}\pi M^2_{p}}\left(1-\frac{r_{\star}}{24}\right)^{1/4}\left(N_{\star}-N_{end}\right)\\
                    \displaystyle   \left[1-({\cal C}_{E}+1-\Sigma)
 \frac{r_{\star}}{16}+\frac{{\cal C}_{E}}{2\sqrt{2q}}\left(n_{\zeta,\star}-1
                \displaystyle+\frac{r_{\star}}{8}\right)\right]&
 \mbox{ {\bf for \underline{Case I}}}  \\ 
         \displaystyle \frac{\sqrt{\frac{qr_{\star}}{4M\alpha^{'}}}\frac{1}{q\sqrt{\frac{3\Delta_{\zeta, \star}r_{\star}}{2}}\pi M^2_{p}}}{3(n_{h,\star}-n_{\zeta,\star}+1)}\left[  
         \left\{\, _2F_1\left[\frac{1}{2},\frac{3}{4};\frac{3}{2};\frac{r_{\star}}{24} \right]+2 \left(1-\frac{r_{\star}}{24}
         \right)^{1/4}\right\}\right.\\ \left.
         \displaystyle -e^{\frac{n_{h,\star}-n_{\zeta,\star}+1}{2}(N_{end}-N_{\star})} \left\{\, _2F_1\left[\frac{1}{2},\frac{3}{4};\frac{3}{2};\frac{r_{\star}}{24}
         e^{(n_{h,\star}-n_{\zeta,\star}+1)(N_{end}-N_{\star})}\right]\right.\right.\\ \left.\left.\displaystyle~~~~~~~~+2 \left(1-\frac{r_{\star}}{24}e^{(n_{h,\star}
         -n_{\zeta,\star}+1)(N_{end}-N_{\star})}
         \right)^{1/4}\right\}\right]\\
         \displaystyle   \left[1-({\cal C}_{E}+1-\Sigma)
 \frac{r_{\star}}{16}+\frac{{\cal C}_{E}}{2}\left(n_{\zeta,\star}-1
                \displaystyle+\frac{r_{\star}}{8}\right)\right]& \mbox{ {\bf for \underline{Case II}}}\\ 
\displaystyle \frac{ \sqrt{\frac{\pi q r_{\star}}{M\alpha^{'}(\alpha_{h,\star}-\alpha_{\zeta,\star})}}}{48q\sqrt{\frac{3\Delta_{\zeta, \star}r_{\star}}{2}}\pi M^2_{p}}
e^{-\frac{3 (n_{h,\star}-n_{\zeta,\star}+1)^2}{4 (\alpha_{h,\star}-\alpha_{\zeta,\star})}}\\
\displaystyle ~~\left[12 e^{\frac{(n_{h,\star}-n_{\zeta,\star}+1)^2}{ 2(\alpha_{h,\star}-\alpha_{\zeta,\star})}}\left\{ \text{erfi}\left(\frac{n_{h,\star}-n_{\zeta,\star}+1}{2 \sqrt{\alpha_{h,\star}-\alpha_{\zeta,\star}}}
\right)\right.\right.\\ \left.\left. 
\displaystyle~~~~~-
\text{erfi}\left(\frac{n_{h,\star}-n_{\zeta,\star}+1}{2 \sqrt{\alpha_{h,\star}-\alpha_{\zeta,\star}}}+\frac{\sqrt{\alpha_{h,\star}-\alpha_{\zeta,\star}}}{2}(N_{end}-N_{\star})
\right)\right\}\right.\\ \left. \displaystyle~~~~~~-\frac{\sqrt{3}r_{\star}}{24} \left\{\text{erfi}
\left(\frac{\sqrt{3} (n_{h,\star}-n_{\zeta,\star}+1)}{2 \sqrt{\alpha_{h,\star}-\alpha_{\zeta,\star}}}\right)\right.\right.\\ \left.\left. 
\displaystyle-\text{erfi}
\left(\frac{\sqrt{3} (n_{h,\star}-n_{\zeta,\star}+1)}{2 \sqrt{\alpha_{h,\star}-\alpha_{\zeta,\star}}}+\frac{\sqrt{3(\alpha_{h,\star}-\alpha_{\zeta,\star})}}{2}
(N_{end}-N_{\star})\right)\right\}\right]\\
\displaystyle   \left[1-({\cal C}_{E}+1-\Sigma)
 \frac{r_{\star}}{16}+\frac{{\cal C}_{E}}{2}\left(n_{\zeta,\star}-1
                \displaystyle+\frac{r_{\star}}{8}\right)\right]& \mbox{ {\bf for \underline{Case III}}}.
          \end{array}
\right.& \mbox{ \small\underline{\bf for {BD}}} 
\end{array}\ee
\be\begin{array}{lll}\label{raqgggg2vv2}
  \displaystyle \frac{\Delta T}{M_{p}} \approx\footnotesize\left\{\begin{array}{ll}
                    \displaystyle  \displaystyle  
  \sqrt{\frac{r_{\star}}{8M\alpha^{'}}}\frac{\left(1-\frac{qr_{\star}}{48c^{2}_{S}}\right)^{1/4}(N_{\star}-N_{end})}{2c_{S}\sqrt{\frac{3\Delta_{\zeta, \star}r_{\star}}{2}}\pi M^2_{p}}\\ \displaystyle
                 \left[1-({\cal C}_{E}+1-\Sigma)
 \frac{r_{\star}}{16}+\frac{{\cal C}_{E}}{2}\left(n_{\zeta,\star}-1
                +\frac{r_{\star}}{8}\right)\right]&
 \mbox{ {\bf for \underline{Case I}}}  \\ 
         \displaystyle \displaystyle  
  \sqrt{\frac{r_{\star}}{8M\alpha^{'}}}\frac{\left|D\right|\left(1-\frac{qr_{\star}}{48c^{2}_{S}}\frac{\left|D\right|^2}
  {\left|C\right|^2}\right)^{1/4}(N_{\star}-N_{end})}{2c_{S}\left|C\right|\sqrt{\frac{3\Delta_{\zeta, \star}r_{\star}}{2}}\pi M^2_{p}}\\ \displaystyle
                \left[1-({\cal C}_{E}+1-\Sigma)
 \frac{r_{\star}}{16}+\frac{{\cal C}_{E}}{2}\left(n_{\zeta,\star}-1
                +\frac{r_{\star}}{8}\right)\right] & \mbox{ {\bf for \underline{Case II}}}.
          \end{array}
\right.& \mbox{ \underline{\bf \small for {AV}}} 
\end{array}\ee
Further using the approximated form of the sound speed $c_{S}$ the expression for the field excursion for {\bf AV} can be re-written as:
\be\begin{array}{lll}\label{raqggggg2vv2}
  \displaystyle \frac{\Delta T}{M_{p}} \approx\small\left\{\begin{array}{ll}
                    \displaystyle  \displaystyle  
  \sqrt{\frac{r_{\star}}{8M\alpha^{'}}}\frac{\left(1-\frac{qr_{\star}}{48\left[1-\frac{(1-q)}{3q}\frac{r_{\star}}{8}\right]}\right)^{1/4}(N_{\star}-N_{end})}{2\sqrt{1-\frac{(1-q)}{3q}\frac{r_{\star}}{8}}\sqrt{\frac{3\Delta_{\zeta, \star}r_{\star}}{2}}\pi M^2_{p}}\\ \displaystyle
                 \left[1-({\cal C}_{E}+1-\Sigma)
 \frac{r_{\star}}{16}+\frac{{\cal C}_{E}}{2}\left(n_{\zeta,\star}-1
                +\frac{r_{\star}}{8}\right)\right]&
 \mbox{ {\bf for \underline{Case I}}}  \\ 
         \displaystyle \displaystyle  
  \sqrt{\frac{r_{\star}}{8M\alpha^{'}}}\frac{\left|D\right|\left(1-\frac{qr_{\star}}{48\left[1-\frac{(1-q)}{3q}\frac{r_{\star}}{8}\right]}\frac{\left|D\right|^2}
  {\left|C\right|^2}\right)^{1/4}(N_{\star}-N_{end})}{2\sqrt{1-\frac{(1-q)}{3q}\frac{r_{\star}}{8}}\left|C\right|\sqrt{\frac{3\Delta_{\zeta, \star}r_{\star}}{2}}\pi M^2_{p}}\\ \displaystyle
                \left[1-({\cal C}_{E}+1-\Sigma)
 \frac{r_{\star}}{16}+\frac{{\cal C}_{E}}{2}\left(n_{\zeta,\star}-1
                +\frac{r_{\star}}{8}\right)\right] & \mbox{ {\bf for \underline{Case II}}}.
          \end{array}
\right.& \mbox{ \underline{\bf \small for {AV}}} 
\end{array}\ee
This implies that for {\bf BD} and {\bf AV} we get roughly the following result from this analysis:
\bea \left|\frac{\Delta T}{M_p}\right|_{Assisted}&=& \frac{1}{\sqrt{M}}\times\left|\frac{\Delta T}{M_p}\right|_{Single},\eea
which means if the number of tachyonic field participating in the assisted inflation gradually increases, then the tachyonic field excursion for assisted inflation becomes more and more sub-Planckian compared to the 
single field result.
\end{enumerate}

\subsubsection{Semi analytical study and Cosmological parameter estimation}
In this subsection our prime objective are:
\begin{itemize}
 \item  To compute various inflationary observables from variants of tachyonic potentials in presence of $M$ number of identical degrees of freedom,
 
 \item  Estimate the relevant cosmological parameters from the proposed models,
 
 \item Next
to compare the effectiveness of all of these models in the light of recent Planck 2015 data along with other combined constraints.

\item Finally we will check the compatibility of all of these models with the CMB TT, TE and EE angular power spectra as observed by Planck 2015.
\end{itemize}
However instead of computing everything in detail we will not do any further computation in the context of assisted inflation using all the individual five potentials for
that we have already done the analysis in the context of single tachyonic field earlier in this paper. In this context the results are exactly same for all potentials that have already done for single field, 
provided for all the models the stringy parameter $g$, that are appearing almost every where is rescaled by the number of identical scalar fields in the present context i.e. here for the sake of simplicity 
here we define a new stringy parameter $\tilde{g}$ which given by:
\be \tilde{g}=gM= \underbrace{\frac{\alpha^{'}\lambda T^2_0 }{M^2_p}}\times M=\underbrace{\frac{M^4_s}{(2\pi)^3 g_s}\frac{\alpha^{'} T^2_0 }{M^2_p}}\times M,\ee
 where the terms pointed by the $\underbrace{}$ symbol signify the exact contribution from the single tachyonic field. Now from the observational constraints instead of constraining the parameter $g$ here we need 
 to constrain the value of $\tilde{g}$ for all five tachyonic potentials mentioned earlier. Let us mention all the constraints on the stringy parameter $\tilde{g}$ for assisted inflationary framework:
 \begin{itemize}
  \item \underline{\bf Model I: Inverse cosh potential}:\\ For $q=1/2$, $q=1$, $q=3/2$ and $q=2$ we fix $N_{\star}/\tilde{g}\sim 0.8$, which further implies that for $50<N_{\star}<70$,
the prescribed window for $\tilde{g}$ from $\Delta_{\zeta}+n_{\zeta}$ plot is given by, $63<\tilde{g}<88$. If we additionally impose the constraint from the upper bound on tensor-to-scalar ratio then also the allowed parameter 
range is lying within the almost similar window i.e. $88<\tilde{g}<100$.
  \item \underline{\bf Model II: Logarithmic potential}:\\ For $q=1/2$, and $q=1$
we fix $N_{\star}/\tilde{g}\sim 0.7$, which further implies that for $50<N_{\star}<70$,
the prescribed window for $\tilde{g}$ from $\Delta_{\zeta}+n_{\zeta}$ plot is given by, $71.4<g<100$. If we additionally impose the constraint from the upper bound on tensor-to-scalar ratio then also the allowed parameter 
range is lying within the almost similar window i.e. $71.4<\tilde{g}<90$.
  \item \underline{\bf Model III: Exponential potential-Type I}\\  For $q=1/2$, $q=1$, $q=3/2$ and $q=2$ case $\tilde{g}$ is not explicitly appearing in the various inflationary observables except the amplitude of scalar power spectrum in 
  this case. To produce the correct value of the amplitude of the scalar power spectra we fix the parameter $360<\tilde{g}<400$.
  \item \underline{\bf Model IV: Exponential potential-Type II}\\ For $q=1/2$, $q=1$, $q=3/2$ and $q=2$ we
fix $N_{\star}/g\sim 0.85$, which further implies that for $50<N_{\star}<70$,
the prescribed window for $g$ from $\Delta_{\zeta}+n_{\zeta}$ constraints is given by, $59<\tilde{g}<82.3$. If we additionally impose the constraint from the upper bound on tensor-to-scalar ratio then also the allowed parameter 
range is lying within the window i.e. $73<\tilde{g}<82.3$.

  \item \underline{\bf Model V: Inverse power-law potential}\\ For $q=1/2$, $q=1$, $q=3/2$ and $q=2$ case $\tilde{g}$ is not explicitly appearing in the various inflationary observables except the amplitude of scalar power spectrum in 
  this case. To produce the correct value of the amplitude of the scalar power spectra we fix the parameter $600<\tilde{g}<700$.
 \end{itemize}
 In table~(\ref{tab1}) we have shown the comparison between the field excursion obtained from all the tachyonic potentials from single field and assisted
inflationary framework. Here we define $y_0=T_0/M_p$, where $M_p=2.43\times 10^{18}~{\rm GeV}$. The numerics clearly implies that the assisted tachyonic inflationary framework push the field excursion value 
to sub-Planckian value for large $M$ with large amount, compared to the value obtained from single inflationary setup. Technically this implies the doing effective field theory ({\bf EFT}) with assisted framework is more safer compared 
to single field case, as it involves an additional parameter $M$. 
 
 Additionally it is important to mention here that the other conclusions and the rest of the constraints are exactly same as analyzed in case of single field case.
 Similarly the CMB TT, TE, EE spectrum for scalar modes are exactly same as obtained in the context of single tachyonic inflation and compatible with the observed Planck 2015 data. 
 
 \begin{table*}
\centering
\footnotesize
\begin{tabular}{|c|c|c|c|c|}
\hline
\hline
\hline
\hline
\footnotesize{\bf \textcolor{blue}{\bf Model}} & \footnotesize{\bf \textcolor{blue}{\bf Parameters}}& 
\footnotesize{\bf \textcolor{blue}{\bf Single field}} & \footnotesize{\bf \textcolor{blue}{\bf Assisted}}\\
 & \footnotesize{\bf ($2\sigma$~bound)}&\footnotesize$X_S:=(|\Delta T|/M_p)_{\rm Single}$ &  \footnotesize$X_A:=(|\Delta T|/M_p)_{\rm Assisted}$   \\
\hline\hline\hline
 \textcolor{red}{\bf Inverse cosh potential} & $80<g,\tilde{g}<100$ & $0.34y_0<X_S<1.00y_0$
 & $\frac{0.34y_0}{\sqrt{M}}<X_A<\frac{1.00y_0}{\sqrt{M}}$\\
  & $1/2<q<2$ & {\bf EFT:} For $y_0<1.00$ & {\bf EFT:} For I. $y_{0}<1.00$
\\
  & &  &~~~~~~~~~~II. $M\geq2$
\\
\hline
 \textcolor{red}{\bf Logarithmic potential} & $71.4<g,\tilde{g}<100$ & $0.93y_0<X_S<1.12y_0$
 & $\frac{0.93y_0}{\sqrt{M}}<X_A<\frac{1.12y_0}{\sqrt{M}}$\\
  & $1/2<q<1$ & {\bf EFT:} For $y_0<0.892$ & {\bf EFT:} For I. $y_{0}<0.892$
\\
  & $c=0.07$&  &~~~~~~~~~~II. $M\geq2$
\\
\hline
  \textcolor{red}{\bf Exponential potential-Type I} & $360<g,\tilde{g}<400$ & $4.26y_0<X_S<4.27y_0$
 & $\frac{4.26y_0}{\sqrt{M}}<X_A<\frac{4.27y_0}{\sqrt{M}}$\\
  & $1<q<2$ & {\bf EFT:} For $y_0<0.234$ & {\bf EFT:} For I. $y_{0}<0.234$
\\
  & &  &~~~~~~~~~~II. $M\geq2$
\\
\hline
 \textcolor{red}{\bf Exponential potential-Type II} & $73<g,\tilde{g}<82.3$ &$3.18y_0<X_S<6.27y_0$
& $\frac{3.18y_0}{\sqrt{M}}<X_A<\frac{6.27y_0}{\sqrt{M}}$\\
  & $1/2<q<2$ & {\bf EFT:} For $y_0<0.160$& {\bf EFT:} For I. $y_{0}<0.160$
\\
   &$6<p=\sqrt{\frac{g}{2}}<6.4$ &  &~~~~~~~~~~II. $M\geq2$
\\
\hline
 \textcolor{red}{\bf Inverse power-law potential} & $600<g,\tilde{g}<700$ & $7.2y_0<X_S<7.9y_0$
& $\frac{7.2y_0}{\sqrt{M}}<X_A<\frac{7.9y_0}{\sqrt{M}}$
\\
  & $1<q<3/2$ & {\bf EFT:} For $y_0<0.127$& {\bf EFT:} For I. $y_{0}<0.127$
\\
  & &  &~~~~~~~~~~II. $M\geq 2$
\\
\hline
\hline
\hline
\hline
\end{tabular}
\vspace{.4cm}
\caption{Comparison between the field excursion obtained from all the tachyonic potentials from single field and assisted
inflationary framework. Here we define $y_0=T_0/M_p$, where $M_p=2.43\times 10^{18}~{\rm GeV}$. The numerics clearly implies that the assisted tachyonic inflationary framework push the field excursion value 
to sub-Planckian value for large $M$ with large amount, compared to the value obtained from single inflationary setup. Technically this implies the doing effective field theory ({\bf EFT}) with assisted framework is more safer compared 
to single field case, as it involves an additional parameter $M$. 
} 
 \label{tab1}
\end{table*}
\subsection{Computation for Multi-field inflation}
\label{aa5c}
In case of multi tachyonic inflation all the tachyons are not identical. In more technical language for the most generalized prescription one can 
state that:
\bea T_{1}\neq T_{2}\neq\cdots \neq T_{M}.\eea
In the next subsections we will explore the detailed features of multi tachyonic inflation by computing the curvature, isocurvature and tensor perturbations and then we discuss the 
observational constraints and cosmological consequences from the setup. We will give all the analytical results for $M$ number of non-identical tachyonic fields and for completeness also give the results for $M$ number of non-identical 
inverse cosh separable potential.
\subsubsection{Condition for inflation}
For assisted tachyonic inflation, the prime condition for inflation is given by:
\bea
\dot{H}+H^{2}&=&\left(\frac{\ddot{a}}{{a}}\right)=-\sum^{M}_{i=1}\frac{(\rho_{i}+3p_{i})}{6 M^2_p}>0
\eea
which can be re-expressed in terms of the following constraint condition in the context of assisted tachyonic inflation:
\be\begin{array}{lll}\label{g39v2bb2}
	\displaystyle \sum^{M}_{i=1}\frac{V(T_i)}{3M^{2}_{p}\sqrt{1-\alpha^{'}\dot{T}^2_i}}\left(1-\frac{3}{2}\alpha^{'}\dot{T}^2_i\right)>0.
\end{array}\ee\\
Here Eq~(\ref{g39v2}) implies that to satisfy inflationary constraints in the slow-roll regime the following constraint always holds good:
\bea \label{wa1v2}\dot{T}_{i}&<& \sqrt{\frac{2}{3\alpha^{'}}}~~~~\forall i=1,2,\cdots,M,\\ 
\label{wa2v2}\ddot{T}_{i}&<&  3H \dot{T}_{i}<\sqrt{\frac{6}{\alpha^{'}}}H~~~~\forall i=1,2,\cdots,M.\eea
Consequently the field equations are approximated as`:
\be\begin{array}{lll}\label{e13xv2bb2c}
	\displaystyle 
	3H\alpha^{'}\dot{T}_{i}+\left(\sum^{M}_{j=1} V(T_{j})\right)^{-1}\frac{\partial}{\partial T_{i}}\left(\sum^{M}_{j=1} V(T_{j})\right)\approx 0 \,,
\end{array}\ee
Similarly, in the most generalized case, 
\be\begin{array}{lll}\label{g39v3bb2}
	\displaystyle \sum^{M}_{i=1}\frac{V(T_{i})}{3M^{2}_{p}\left(1-\alpha^{'}\dot{T}^2_{i}
		\right)^{1-q}}\left(1-(1+q)\alpha^{'}\dot{T}^2_{i}\right)>0.
\end{array}\ee
Here Eq~(\ref{g39v3bb2}) implies that to satisfy inflationary constraints in the slow-roll regime the following constraint always holds good:
\bea \dot{T}_{i}&<& \sqrt{\frac{1}{\alpha^{'}(1+q)}}~~~~\forall i=1,2,\cdots,M,\\ 
\ddot{T_{i}}&<&  3H \dot{T}_{i}<\sqrt{\frac{9}{\alpha^{'}(1+q)}}H~~~~\forall i=1,2,\cdots,M.\eea
Consequently the field equations are approximated as`:
\be\begin{array}{lll}\label{e14xxvv3bb2}
	\displaystyle 
	6q\alpha^{'}H\dot{T}_{i}+\left(\sum^{M}_{j=1} V(T_{j})\right)^{-1}\frac{\partial}{\partial T_{i}}\left(\sum^{M}_{j=1} V(T_{j})\right)\approx 0 \,
\end{array}\ee
Also for both the cases in the slow-roll regime the Friedmann equation is modified as:
\bea\label{qxcc1v3bb2} H^{2}&\approx& \sum^{M}_{i=1}\frac{V(T_{i})}{3M^{2}_{p}}.\eea
Further substituting Eq~(\ref{qxcc1v3bb2}) in Eq~(\ref{e13xv2bb2c}) and Eq~(\ref{e14xxvv3bb2}) we get:
\bea\label{e13gvxv4bb2}
\displaystyle 
\sqrt{\sum^{M}_{j=1} V(T_{j})}\frac{\sqrt{3}\alpha^{'}}{M_p}\dot{T}_{i}+\left(\sum^{M}_{j=1} V(T_{j})\right)^{-1}\frac{\partial}{\partial T_{i}}\left(\sum^{M}_{j=1} V(T_{j})\right)\approx 0 ~~~~\forall i=1,2,\cdots,M\,,~~~~~~~~\\
\label{e13gvvxbb2} \frac{6q}{\sqrt{3}M_{p}}\sqrt{\sum^{M}_{j=1} V(T_{j})}\alpha^{'}\dot{T}_{i}+\left(\sum^{M}_{j=1} V(T_{j})\right)^{-1}\frac{\partial}{\partial T_{i}}\left(\sum^{M}_{j=1} V(T_{j})\right)\approx 0 ~~~~
\forall i=1,2,\cdots,M\,~~~~~~~~~~~~~.
\eea
Finally the general solution for both the cases can be expressed in terms of the single field tachyonic potential $V(T)$ as:
\bea\label{e13gvxc1v5bb2}
\displaystyle 
\label{bw1bb2}t-t_{in,i}\approx -\frac{\sqrt{3}\alpha^{'}}{M_{p}}\int^{T_{i}}_{T_{in,i}}dT_{i}~\sqrt{\sum^{M}_{j=1} V(T_{j})}\left[\left(\sum^{M}_{j=1} V(T_{j})\right)^{-1}\frac{\partial}{\partial T_{i}}
\left(\sum^{M}_{j=1} V(T_{j})\right)\right]^{-1} \,,~~~~~~~~~~~\\
\label{bw2bb2} t-t_{in,i}\approx -\frac{6q\alpha^{'}}{\sqrt{3}M_{p}}\int^{T_{i}}_{T_{in,i}}dT_{i}~\sqrt{\sum^{M}_{j=1} V(T_{j})}\left[\left(\sum^{M}_{j=1} V(T_{j})\right)^{-1}
\frac{\partial}{\partial T_{i}}\left(\sum^{M}_{j=1} V(T_{j})\right)\right]^{-1}.~~~~~~~~~~~
\eea
Further using Eq~(\ref{bw1bb2}), Eq~(\ref{bw2bb2}) and Eq~(\ref{qxcc1v3bb2}) we get the following solution for the scale factor 
in terms of the tachyonic field for usual $q=1/2$ and for generalized value of $q$ as:
\be\begin{array}{lll}\label{scss1bb2}\small
	\displaystyle  a= a_{in,i}\times \left\{\begin{array}{lll}
		\displaystyle  
		\exp\left[-\frac{\alpha^{'}M}{M^2_p}\int^{T_{i}}_{T_{in}}dT_{i}~\sum^{M}_{j=1} V(T_{j})\left[\left(\sum^{M}_{j=1} V(T_{j})\right)^{-1}
		\frac{\partial}{\partial T_{i}}\left(\sum^{M}_{j=1} V(T_{j})\right)\right]^{-1}\right]\,,&
		\mbox{\small {\bf for {$q=1/2$ }}}  \\ 
		\displaystyle   
		\exp\left[-\frac{2q\alpha^{'}M}{M^2_p}\int^{T_{i}}_{T_{in}}dT_{i}~\sum^{M}_{j=1} V(T_{j})\left[\left(\sum^{M}_{j=1} V(T_{j})\right)^{-1}
		\frac{\partial}{\partial T_{i}}\left(\sum^{M}_{j=1} V(T_{j})\right)\right]^{-1}\right]\,. &
		\mbox{\small {\bf for {~any~$q$ }}} 
	\end{array}
	\right.
\end{array}\ee
Sometimes it is convenient to identify the inflaton field direction as the direction in field space corresponding to the evolution of the background spatially homogeneous tachyonic fields during inflation. 
To serve this purpose one can write:
\bea \sigma=\int \sum^{M}_{i=1}\hat{\sigma}_{i}\dot{T}_{i}~dt,\eea
where the inflaton direction is defined as:
\bea \hat{\sigma}_{i}\equiv \dot{T}\times \left[\sum^{M}_{j=1}\dot{T}^2_{j}\right]^{-1/2}.\eea
The $M$ number of evolution equations for the homogeneous scalar fields can be written as a evolution of single scalar field in the slow-roll regime as:
\be\begin{array}{lll}\small\displaystyle  3H\dot{\sigma}+\left(\sum^{M}_{j=1} V(T_{j})\right)^{-1}\frac{\partial}{\partial\sigma}\left(\sum^{M}_{j=1} V(T_{j})\right)=3H\dot{\sigma}+\sum^{M}_{i=1}
\hat{\sigma}_{i}\left(\sum^{M}_{j=1} V(T_{j})\right)^{-1}\frac{\partial}{\partial T_{i}}\left(\sum^{M}_{j=1} V(T_{j})\right)=0. \end{array}\ee
\subsubsection{Analysis using Slow-roll formalism}
Here our prime objective is to define slow-roll parameters for tachyon inflation in terms of the Hubble parameter and the multi tachyonic inflationary potential, where the $M$ number of tachyon fields are not identical. Using the slow-roll 
approximation one can expand various cosmological observables in terms of small dynamical quantities derived from the appropriate derivatives of the Hubble parameter and of the inflationary potential. 
To start with , in the present context the potential dependent slow-roll parameters are defined as:
\bea \label{wzq1v2} \epsilon_{V;T_{j}T_{j}}(T_{i})&=&\frac{M^{2}_{p}}{2}\left(\frac{\partial_{T_{j}}V(T_{i})}{V(T_{i})}\right)^2,\\
\label{wzq2v2} \eta_{V;T_{j}T_{j}}(T_{i})&=& M^{2}_{p}\left(\frac{\partial_{T_{j}}\partial_{T_{j}}V(T_{i})}{V(T_{i})}\right),\\
\label{wzq1v2a} \Theta_{V;T_{j}T_{k}}(T_{i})&=&M^{2}_{p}\left(\frac{\partial_{T_{j}}\partial_{T_{k}}V(T_{i})}{V(T_{i})}\right),\\
\label{wzq2v2b} \Delta_{V;T_{j}T_{k}}(T_{i})&=& \frac{M^{2}_{p}}{2}\left(\frac{\partial_{T_{j}}V(T_{i})\partial_{T_{k}}V(T_{i})}{
	V^2(T_{i})}\right),\\
\label{wzq3v2} \xi^{2}_{V;T_{j}T_{j}T_{j}T_{j}}(T_{i})&=& M^{4}_{p}\left(\frac{\partial_{T_{j}}V(T_{i})\partial_{T_{j}}\partial_{T_{j}}\partial_{T_{j}}V(T_i)}{V^2(T_{i})}\right),\\
\label{wzq3v2} \vartheta^{2}_{V;T_{j}T_{j}T_{k}T_{k}}(T_{i})&=& M^{4}_{p}\left(\frac{\partial_{T_{j}}V(T_{i})\partial_{T_{j}}\partial_{T_{k}}\partial_{T_{k}}V(T_i)}{V^2(T_{i})}\right),\\
\label{wzq3v2} \vartheta^{2}_{V;T_{k}T_{j}T_{k}T_{k}}(T_{i})&=& M^{4}_{p}\left(\frac{\partial_{T_{k}}V(T_{i})\partial_{T_{j}}\partial_{T_{k}}\partial_{T_{k}}V(T_i)}{V^2(T_{i})}\right),\\
\label{wzq3v2} \vartheta^{2}_{V;T_{j}T_{k}T_{k}T_{k}}(T_{i})&=& M^{4}_{p}\left(\frac{\partial_{T_{j}}V(T_{i})\partial_{T_{k}}\partial_{T_{k}}\partial_{T_{k}}V(T_i)}{V^2(T_{i})}\right),\\
\label{wzq3v2} \sigma^{3}_{V;T_{j}T_{j}T_{j}T_{j}T_{j}T_{j}}(T_{i})&=& M^{6}_{p}\left(\frac{\partial_{T_{j}}V(T_{i})\partial_{T_{j}}V(T_{i})\partial_{T_{j}}\partial_{T_{j}}\partial_{T_{j}}\partial_{T_{j}}V(T_i)}{V^3(T_{i})}\right),\\
\label{wzq3v2} \Upsilon^{3}_{V;T_{j}T_{j}T_{k}T_{k}T_{k}T_{k}}(T_{i})&=& M^{6}_{p}\left(\frac{\partial_{T_{j}}V(T_{i})\partial_{T_{j}}V(T_{i})\partial_{T_{k}}\partial_{T_{k}}\partial_{T_{k}}\partial_{T_{k}}V(T_i)}{V^3(T_{i})}\right),\\
\label{wzq3v2} \Upsilon^{3}_{V;T_{j}T_{k}T_{k}T_{k}T_{k}T_{k}}(T_{i})&=& M^{6}_{p}\left(\frac{\partial_{T_{j}}V(T_{i})\partial_{T_{k}}V(T_{i})\partial_{T_{k}}\partial_{T_{k}}\partial_{T_{k}}\partial_{T_{k}}V(T_i)}{V^3(T_{i})}\right),\\
\label{wzq3v2} \Upsilon^{3}_{V;T_{j}T_{j}T_{j}T_{k}T_{j}T_{k}}(T_{i})&=& M^{6}_{p}\left(\frac{\partial_{T_{j}}V(T_{i})\partial_{T_{j}}V(T_{i})\partial_{T_{j}}\partial_{T_{k}}\partial_{T_{k}}\partial_{T_{k}}V(T_i)}{V^3(T_{i})}\right),\\
\label{wzq3v2} \Upsilon^{3}_{V;T_{j}T_{j}T_{j}T_{j}T_{k}T_{k}}(T_{i})&=& M^{6}_{p}\left(\frac{\partial_{T_{j}}V(T_{i})\partial_{T_{j}}V(T_{i})\partial_{T_{j}}\partial_{T_{j}}\partial_{T_{k}}\partial_{T_{k}}V(T_i)}{V^3(T_{i})}\right),\\
\label{wzq3v2} \Upsilon^{3}_{V;T_{j}T_{j}T_{j}T_{j}T_{j}T_{k}}(T_{i})&=& M^{6}_{p}\left(\frac{\partial_{T_{j}}V(T_{i})\partial_{T_{j}}V(T_{i})\partial_{T_{j}}\partial_{T_{j}}\partial_{T_{j}}\partial_{T_{k}}V(T_i)}{V^3(T_{i})}\right),\\
\label{wzq3v2} \Upsilon^{3}_{V;T_{j}T_{k}T_{j}T_{j}T_{j}T_{j}}(T_{i})&=& M^{6}_{p}\left(\frac{\partial_{T_{j}}V(T_{i})\partial_{T_{k}}V(T_{i})\partial_{T_{j}}\partial_{T_{j}}\partial_{T_{j}}\partial_{T_{j}}V(T_i)}{V^3(T_{i})}\right),\\
\label{wzq3v2}\Upsilon^{3}_{V;T_{j}T_{k}T_{j}T_{j}T_{j}T_{k}}(T_{i})&=& M^{6}_{p}\left(\frac{\partial_{T_{j}}V(T_{i})\partial_{T_{k}}V(T_{i})\partial_{T_{j}}\partial_{T_{j}}\partial_{T_{j}}\partial_{T_{k}}V(T_i)}{V^3(T_{i})}\right),\\
\label{wzq3v2} \Upsilon^{3}_{V;T_{j}T_{k}T_{j}T_{j}T_{k}T_{k}}(T_{i})&=& M^{6}_{p}\left(\frac{\partial_{T_{j}}V(T_{i})\partial_{T_{k}}V(T_{i})\partial_{T_{j}}\partial_{T_{j}}\partial_{T_{k}}\partial_{T_{k}}V(T_i)}{V^3(T_{i})}\right),\\
\label{wzq3v2} \Upsilon^{3}_{V;T_{j}T_{k}T_{j}T_{k}T_{k}T_{k}}(T_{i})&=& M^{6}_{p}\left(\frac{\partial_{T_{j}}V(T_{i})\partial_{T_{k}}V(T_{i})\partial_{T_{j}}\partial_{T_{k}}\partial_{T_{k}}\partial_{T_{k}}V(T_i)}{V^3(T_{i})}\right).
\eea
However, for the sake of clarity here we introduce new sets of potential dependent slow-roll parameters for multi tachyonic inflation by rescaling with the appropriate powers of $\alpha^{'}\left(\sum^{M}_{k=1}V(T_{k})\right)$:
\bea \label{wzq11v2} \bar{\epsilon}_{V;T_{j}T_{j}}(T_{i})&=&\frac{\epsilon_{V;T_{j}T_{j}}(T_{i})}{\alpha^{'}\left(\sum^{M}_{k=1}V(T_{k})\right)},~~~~~~\bar{\Delta}_{V;T_{j}T_{k}}(T_{i})=\frac{\Delta_{V;T_{j}T_{k}}(T_{i})}{\alpha^{'}\left(\sum^{M}_{j=1}V(T_{j})\right)},\nonumber\\
\label{wzq22v2} \bar{\eta}_{V;T_{j}T_{j}}(T_{i})&=& \frac{{\eta}_{V;T_{j}T_{j}}(T_{i})}{\alpha^{'}\left(\sum^{M}_{j=1}V(T_{j})\right)},~~~~~~\bar{\Theta}_{V;T_{j}T_{k}}(T_{i})= \frac{{\Theta}_{V;T_{j}T_{k}}(T_{i})}{\alpha^{'}\left(\sum^{M}_{j=1}V(T_{j})\right)},\nonumber\\
\label{wzq33v2} \bar{\xi}^{2}_{V;T_{j}T_{j}T_{j}T_{j}}(T_{i})&=& \frac{\xi^{2}_{V;T_{j}T_{j}T_{j}T_{j}}(T_{i})}{\alpha^{'2}\left(\sum^{M}_{j=1}V(T_{j})\right)^2},~~~~~~\bar{\vartheta}^{2}_{V;T_{j}T_{j}T_{k}T_{k}}(T_{i})=
\frac{\vartheta^{2}_{V;T_{j}T_{j}T_{k}T_{k}}(T_{i})}{\alpha^{'2}\left(\sum^{M}_{j=1}V(T_{j})\right)^2},\nonumber\\
\label{wzq44v2} \bar{\vartheta}^{2}_{V;T_{k}T_{j}T_{k}T_{k}}(T_{i})&=& \frac{\vartheta^{2}_{V;T_{k}T_{j}T_{k}T_{k}}(T_{i})}{\alpha^{'2}\left(\sum^{M}_{j=1}V(T_{j})\right)^2},~~~~~~\bar{\vartheta}^{2}_{V;T_{j}T_{k}T_{k}T_{k}}(T_{i})= 
\frac{\vartheta^{2}_{V;T_{k}T_{k}T_{j}T_{k}}(T_{i})}{\alpha^{'2}\left(\sum^{M}_{j=1}V(T_{j})\right)^2},~~~~~~~~~~~~~~~\\
\label{wzq3v2} \bar{\sigma}^{3}_{V;T_{j}T_{j}T_{j}T_{j}T_{j}T_{j}}(T_{i})&=&\frac{{\sigma}^{3}_{V;T_{j}T_{j}T_{j}T_{j}T_{j}T_{j}}(T_{i})}{\alpha^{'3}\left(\sum^{M}_{j=1}V(T_{j})\right)^3},~~~~
~~~ \Upsilon^{3}_{V;T_{j}T_{j}T_{k}T_{k}T_{k}T_{k}}(T_{i})= \frac{\Upsilon^{3}_{V;T_{j}T_{j}T_{k}T_{k}T_{k}T_{k}}(T_{i})}{\alpha^{'3}\left(\sum^{M}_{j=1}V(T_{j})\right)^3},\\
\label{wzq3v2} \bar{\Upsilon}^{3}_{V;T_{j}T_{k}T_{k}T_{k}T_{k}T_{k}}(T_{i})&=& \frac{\Upsilon^{3}_{V;T_{j}T_{k}T_{k}T_{k}T_{k}T_{k}}(T_{i})}{\alpha^{'3}\left(\sum^{M}_{j=1}V(T_{j})\right)^3},~~~~
\bar{\Upsilon}^{3}_{V;T_{j}T_{j}T_{j}T_{k}T_{j}T_{k}}(T_{i})= \frac{\Upsilon^{3}_{V;T_{j}T_{j}T_{j}T_{k}T_{j}T_{k}}(T_{i})}{\alpha^{'3}\left(\sum^{M}_{j=1}V(T_{j})\right)^3},\\
\label{wzq3v2} \bar{\Upsilon}^{3}_{V;T_{j}T_{j}T_{j}T_{j}T_{k}T_{k}}(T_{i})&=& \frac{\Upsilon^{3}_{V;T_{j}T_{j}T_{j}T_{j}T_{k}T_{k}}(T_{i})}{\alpha^{'3}\left(\sum^{M}_{j=1}V(T_{j})\right)^3},~~~~
\bar{\Upsilon}^{3}_{V;T_{j}T_{j}T_{j}T_{j}T_{j}T_{k}}(T_{i})= \frac{\Upsilon^{3}_{V;T_{j}T_{j}T_{j}T_{j}T_{j}T_{k}}(T_{i})}{\alpha^{'3}\left(\sum^{M}_{j=1}V(T_{j})\right)^3},\\
\label{wzq3v2} \bar{\Upsilon}^{3}_{V;T_{j}T_{k}T_{j}T_{j}T_{j}T_{j}}(T_{i})&=&\frac{\Upsilon^{3}_{V;T_{j}T_{k}T_{j}T_{j}T_{j}T_{j}}(T_{i})}{\alpha^{'3}\left(\sum^{M}_{j=1}V(T_{j})\right)^3},~~~~
 \bar{\Upsilon}^{3}_{V;T_{j}T_{k}T_{j}T_{j}T_{j}T_{k}}(T_{i})= \frac{\Upsilon^{3}_{V;T_{j}T_{k}T_{j}T_{j}T_{j}T_{k}}(T_{i})}{\alpha^{'3}\left(\sum^{M}_{j=1}V(T_{j})\right)^3},\\
\label{wzq3v2} \bar{\Upsilon}^{3}_{V;T_{j}T_{k}T_{j}T_{j}T_{k}T_{k}}(T_{i})&=& \frac{\Upsilon^{3}_{V;T_{j}T_{k}T_{j}T_{j}T_{k}T_{k}}(T_{i})}{\alpha^{'3}\left(\sum^{M}_{j=1}V(T_{j})\right)^3},~~~~
 \bar{\Upsilon}^{3}_{V;T_{j}T_{k}T_{j}T_{k}T_{k}T_{k}}(T_{i})= \frac{\Upsilon^{3}_{V;T_{j}T_{k}T_{j}T_{k}T_{k}T_{k}}(T_{i})}{\alpha^{'3}\left(\sum^{M}_{j=1}V(T_{j})\right)^3}.
\eea
where in all cases $i,j,k =1,2,\cdots,M$ and $j\neq k$. For the sake of simplicity let us parametrize the flow-functions in the slow roll regime via two angular parameters,
by making use of the following transformation equations:
\bea\label{d1}
\cos\alpha_{V;T_{j}T_{j}}:&=&\sqrt{\frac{\bar{\epsilon}_{V;T_{j}T_{j}}(T_i)}{\bar{{\epsilon}}_{V}}},\\
\label{d2}\sin\alpha_{V;T_{k}T_{k}}:&=&\sqrt{\frac{\bar{\epsilon}_{V;T_{k}T_{k}}(T_i)}{\bar{{\epsilon}}_{V}}},\\
\label{d3}\cos\beta_{V;T_{k}}:&=&M_p ~\bar{{\epsilon}}_{V;T_{k}},\\
\label{d4}\sin\beta_{V;T_{j}}:&=&M_p ~\bar{\epsilon}_{V;T_{k}}.
\eea
For the two field set set up this can be visualized in a better way. In that case we need to fix $j=1$, $k=2$ or $j=2$, $k=1$. and $i$ is the free index which can take values $i=1,2$ depending on the field derivative, $\partial_{T_{1}}$
or $\partial_{T_{2}}$ acting on it. In the present context these two sets of angular parameters physically represent 
the angle between the adiabatic perturbation, the tangent of the first slow-roll parameter and the field contents. 
Here also we define the following reduced parameters for multi tachyonic inflation: 
\bea\label{hub}
\bar{\epsilon}_{V}&=&\sum^{M}_{i=1}\sum^{M}_{j=1}\bar{\epsilon}_{V;T_{j}T_{j}}(T_{i})+\sum^{M}_{i=1}\sum^{M}_{j=1}\sum^{M}_{k=1}\left(\bar{\Delta}_{V;T_{j}T_{k}}(T_{i})+\bar{\Delta}_{V;T_{k}T_{j}}(T_{i})\right),\\
\bar{\eta}_{V}&=&\sum^{M}_{i=1}\sum^{M}_{j=1}\bar{\eta}_{V;T_{j}T_{j}}(T_{i})+\sum^{M}_{i=1}\sum^{M}_{j=1}\sum^{M}_{k=1}\left(\bar{\Theta}_{V;T_{j}T_{k}}(T_{i})+\bar{\Theta}_{V;T_{k}T_{j}}(T_{i})\right),\\
\bar{\xi}^2_{V}&=&\sum^{M}_{i=1}\sum^{M}_{j=1}\bar{\xi}^{2}_{V;T_{j}T_{j}T_{j}T_{j}}(T_{i})+\sum^{M}_{i=1}\sum^{M}_{j=1}\sum^{M}_{k=1}\left(\bar{\vartheta}^{2}_{V;T_{j}T_{j}T_{k}T_{k}}(T_{i})+\bar{\vartheta}^{2}_{V;T_{k}T_{j}T_{k}T_{k}}(T_{i})\nonumber\right.\\&& \left.~~~
~~~~~~~~~~~~~~~~~~~~~~~~~~~~~~~~~+\bar{\vartheta}^{2}_{V;T_{j}T_{k}T_{k}T_{k}}(T_{i})+\bar{\vartheta}^{2}_{V;T_{j}T_{k}T_{j}T_{k}}(T_{i})\nonumber\right.\\&& \left.~~~
~~~~~~~~~~~~~~~~~~~~~~~~~~~~~~~~~+\bar{\vartheta}^{2}_{V;T_{j}T_{k}T_{k}T_{j}}(T_{i})+\bar{\vartheta}^{2}_{V;T_{k}T_{k}T_{j}T_{k}}(T_{i})\nonumber\right.\\&& \left.~~~
~~~~~~~~~~~~~~~~~~~~~~~~~~~~~~~~~+\bar{\vartheta}^{2}_{V;T_{k}T_{k}T_{k}T_{j}}(T_{i})\right),\\
\bar{\sigma}^3_{V}&=&\sum^{M}_{i=1}\sum^{M}_{j=1}\bar{\sigma}^{3}_{V;T_{j}T_{j}T_{j}T_{j}T_{j}T_{j}}(T_{i})+\sum^{M}_{i=1}\sum^{M}_{j=1}\sum^{M}_{k=1}\left(\Upsilon^{3}_{V;T_{j}T_{j}T_{k}T_{k}T_{k}T_{k}}(T_{i})
+\bar{\Upsilon}^{3}_{V;T_{j}T_{k}T_{k}T_{k}T_{k}T_{k}}(T_{i})\nonumber
\right.\\&& \left.~+\bar{\Upsilon}^{3}_{V;T_{j}T_{j}T_{j}T_{j}T_{k}T_{k}}(T_{i})
+\bar{\Upsilon}^{3}_{V;T_{j}T_{j}T_{j}T_{j}T_{j}T_{k}}(T_{i})+\bar{\Upsilon}^{3}_{V;T_{j}T_{k}T_{j}T_{j}T_{j}T_{j}}(T_{i})\nonumber\right.\\&& \left.~
+\bar{\Upsilon}^{3}_{V;T_{j}T_{k}T_{j}T_{j}T_{j}T_{k}}(T_{i})+\bar{\Upsilon}^{3}_{V;T_{j}T_{k}T_{j}T_{j}T_{k}T_{k}}(T_{i})
+\bar{\Upsilon}^{3}_{V;T_{j}T_{k}T_{j}T_{k}T_{k}T_{k}}(T_{i})+\bar{\Upsilon}^{3}_{V;T_{j}T_{j}T_{k}T_{j}T_{k}T_{k}}(T_{i})\nonumber\right.\\&& \left.~
+\bar{\Upsilon}^{3}_{V;T_{j}T_{j}T_{k}T_{k}T_{j}T_{k}}(T_{i})+\bar{\Upsilon}^{3}_{V;T_{j}T_{j}T_{k}T_{k}T_{k}T_{j}}(T_{i})
+\bar{\Upsilon}^{3}_{V;T_{j}T_{j}T_{j}T_{k}T_{j}T_{k}}(T_{i})+\bar{\Upsilon}^{3}_{V;T_{j}T_{j}T_{j}T_{k}T_{k}T_{j}}(T_{i})\nonumber\right.\\&& \left.~
+\bar{\Upsilon}^{3}_{V;T_{j}T_{j}T_{k}T_{j}T_{k}T_{j}}(T_{i})+\bar{\Upsilon}^{3}_{V;T_{j}T_{j}T_{k}T_{j}T_{j}T_{k}}(T_{i})
+\bar{\Upsilon}^{3}_{V;T_{j}T_{j}T_{j}T_{j}T_{k}T_{j}}(T_{i})+\bar{\Upsilon}^{3}_{V;T_{j}T_{j}T_{j}T_{k}T_{j}T_{j}}(T_{i})\nonumber\right.\\&& \left.~~~~~~~~~~~~~~~~~
+\bar{\Upsilon}^{3}_{V;T_{j}T_{j}T_{k}T_{j}T_{j}T_{j}}(T_{i})+\bar{\Upsilon}^{3}_{V;T_{k}T_{j}T_{j}T_{j}T_{j}T_{j}}(T_{i})\nonumber\right.\\&& \left.~~~~~~~~~~~~~~~~~
+\bar{\Upsilon}^{3}_{V;T_{j}T_{k}T_{j}T_{j}T_{k}T_{j}}(T_{i})+\bar{\Upsilon}^{3}_{V;T_{j}T_{k}T_{j}T_{k}T_{j}T_{j}}(T_{i})\nonumber\right.\\&& \left.~~~~~~~~~~~~~~~~~
+\bar{\Upsilon}^{3}_{V;T_{j}T_{k}T_{k}T_{j}T_{j}T_{j}}(T_{i})+\bar{\Upsilon}^{3}_{V;T_{k}T_{j}T_{j}T_{j}T_{j}T_{k}}(T_{i})\nonumber\right.\\&& \left.~~~~~~~~~~~~~~~~~
+\bar{\Upsilon}^{3}_{V;T_{k}T_{j}T_{j}T_{j}T_{k}T_{j}}(T_{i})+\bar{\Upsilon}^{3}_{V;T_{k}T_{j}T_{j}T_{k}T_{j}T_{j}}(T_{i})\nonumber\right.\\&& \left.~~~~~~~~~~~~~~~~~
+\bar{\Upsilon}^{3}_{V;T_{k}T_{j}T_{k}T_{j}T_{j}T_{j}}(T_{i})+\bar{\Upsilon}^{3}_{V;T_{j}T_{k}T_{j}T_{k}T_{j}T_{k}}(T_{i})\nonumber\right.\\&& \left.~~~~~~~~~~~~~~~~~
+\bar{\Upsilon}^{3}_{V;T_{j}T_{k}T_{k}T_{j}T_{j}T_{k}}(T_{i})+\bar{\Upsilon}^{3}_{V;T_{j}T_{k}T_{j}T_{k}T_{k}T_{j}}(T_{i})\nonumber\right.\\&& \left.~~~~~~~~~~~~~~~~~
+\bar{\Upsilon}^{3}_{V;T_{j}T_{k}T_{k}T_{j}T_{k}T_{j}}(T_{i})+\bar{\Upsilon}^{3}_{V;T_{k}T_{j}T_{j}T_{j}T_{k}T_{k}}(T_{i})\nonumber\right.\\&& \left.~~~~~~~~~~~~~~~~~
+\bar{\Upsilon}^{3}_{V;T_{k}T_{j}T_{j}T_{k}T_{j}T_{k}}(T_{i})+\bar{\Upsilon}^{3}_{V;T_{k}T_{j}T_{k}T_{j}T_{j}T_{k}}(T_{i})\nonumber\right.\\&& \left.~~~~~~~~~~~~~~~~~
+\bar{\Upsilon}^{3}_{V;T_{k}T_{j}T_{j}T_{k}T_{k}T_{j}}(T_{i})+\bar{\Upsilon}^{3}_{V;T_{k}T_{j}T_{k}T_{j}T_{k}T_{j}}(T_{i})\nonumber\right.\\&& \left.~~~~~~~~~~~~~~~~~
+\bar{\Upsilon}^{3}_{V;T_{j}T_{k}T_{k}T_{j}T_{k}T_{k}}(T_{i})+\bar{\Upsilon}^{3}_{V;T_{j}T_{k}T_{k}T_{k}T_{j}T_{k}}(T_{i})\nonumber\right.\\&& \left.~~~~~~~~~~~~~~~~~
+\bar{\Upsilon}^{3}_{V;T_{j}T_{k}T_{k}T_{k}T_{k}T_{j}}(T_{i})+\bar{\Upsilon}^{3}_{V;T_{k}T_{j}T_{j}T_{k}T_{k}T_{k}}(T_{i})\nonumber\right.\\&& \left.~~~~~~~~~~~~~~~~~
+\bar{\Upsilon}^{3}_{V;T_{k}T_{j}T_{k}T_{j}T_{k}T_{k}}(T_{i})+\bar{\Upsilon}^{3}_{V;T_{k}T_{j}T_{k}T_{k}T_{j}T_{k}}(T_{i})\nonumber\right.\\&& \left.~~~~~~~~~~~~~~~~~
+\bar{\Upsilon}^{3}_{V;T_{k}T_{j}T_{k}T_{k}T_{k}T_{j}}(T_{i})+\bar{\Upsilon}^{3}_{V;T_{j}T_{j}T_{j}T_{k}T_{k}T_{k}}(T_{i})\right).~~~~
\eea 
Now simplify the results let us mention the symmetries appearing due to various permutation of indices:
\bea \bar{\Delta}_{V;T_{j}T_{k}}(T_{i})&=&\bar{\Delta}_{V;T_{k}T_{j}}(T_{i}),\\
\bar{\Theta}_{V;T_{j}T_{k}}(T_{i})&=&\bar{\Theta}_{V;T_{k}T_{j}}(T_{i}),\\
\bar{\vartheta}^{2}_{V;T_{j}T_{j}T_{k}T_{k}}(T_{i})&=&\bar{\vartheta}^{2}_{V;T_{j}T_{k}T_{j}T_{k}}(T_{i})=\bar{\vartheta}^{2}_{V;T_{j}T_{k}T_{k}T_{j}}(T_{i}),\\
\bar{\vartheta}^{2}_{V;T_{k}T_{j}T_{k}T_{k}}(T_{i})&=&\bar{\vartheta}^{2}_{V;T_{k}T_{k}T_{j}T_{k}}(T_{i})=\bar{\vartheta}^{2}_{V;T_{k}T_{k}T_{k}T_{j}}(T_{i}),\\
\bar{\Upsilon}^{3}_{V;T_{j}T_{j}T_{j}T_{k}T_{k}T_{k}}(T_{i})&=& \bar{\Upsilon}^{3}_{V;T_{j}T_{j}T_{k}T_{j}T_{k}T_{k}}(T_{i})=
\bar{\Upsilon}^{3}_{V;T_{j}T_{j}T_{k}T_{k}T_{j}T_{k}}(T_{i})=
\bar{\Upsilon}^{3}_{V;T_{j}T_{j}T_{k}T_{k}T_{k}T_{j}}(T_{i}),~~~~~~~~~\\
\bar{\Upsilon}^{3}_{V;T_{j}T_{j}T_{j}T_{j}T_{k}T_{k}}(T_{i})&=& \bar{\Upsilon}^{3}_{V;T_{j}T_{j}T_{j}T_{k}T_{j}T_{k}}(T_{i})=
\bar{\Upsilon}^{3}_{V;T_{j}T_{j}T_{j}T_{k}T_{k}T_{j}}(T_{i})=
\bar{\Upsilon}^{3}_{V;T_{j}T_{j}T_{k}T_{j}T_{k}T_{j}}(T_{i})\nonumber\\
&&~~~~~~~~~~~~~~~~~~~~~~~~~~~~~~~~=\bar{\Upsilon}^{3}_{V;T_{j}T_{j}T_{k}T_{j}T_{j}T_{k}}(T_{i}),~~~~~~~~~\\
\bar{\Upsilon}^{3}_{V;T_{j}T_{j}T_{j}T_{j}T_{j}T_{k}}(T_{i})&=& \bar{\Upsilon}^{3}_{V;T_{j}T_{j}T_{j}T_{j}T_{k}T_{j}}(T_{i})=
\bar{\Upsilon}^{3}_{V;T_{j}T_{j}T_{j}T_{k}T_{j}T_{j}}(T_{i})=
\bar{\Upsilon}^{3}_{V;T_{j}T_{j}T_{k}T_{j}T_{j}T_{j}}(T_{i}),~~~~~~~~~\\
\bar{\Upsilon}^{3}_{V;T_{j}T_{k}T_{j}T_{j}T_{j}T_{j}}(T_{i})&=& \bar{\Upsilon}^{3}_{V;T_{k}T_{j}T_{j}T_{j}T_{j}T_{j}}(T_{i}),~~~~~~~~~~\\
\bar{\Upsilon}^{3}_{V;T_{j}T_{k}T_{j}T_{j}T_{j}T_{k}}(T_{i})&=& \bar{\Upsilon}^{3}_{V;T_{j}T_{k}T_{j}T_{j}T_{k}T_{j}}(T_{i})=
\bar{\Upsilon}^{3}_{V;T_{j}T_{k}T_{j}T_{k}T_{j}T_{j}}(T_{i})=
\bar{\Upsilon}^{3}_{V;T_{j}T_{k}T_{k}T_{j}T_{j}T_{j}}(T_{i}),~~~~~~~~~\\
\bar{\Upsilon}^{3}_{V;T_{k}T_{j}T_{j}T_{j}T_{j}T_{k}}(T_{i})&=& \bar{\Upsilon}^{3}_{V;T_{k}T_{j}T_{j}T_{j}T_{k}T_{j}}(T_{i})=
\bar{\Upsilon}^{3}_{V;T_{k}T_{j}T_{j}T_{k}T_{j}T_{j}}(T_{i})=
\bar{\Upsilon}^{3}_{V;T_{k}T_{j}T_{k}T_{j}T_{j}T_{j}}(T_{i}),~~~~~~~~~\\
\bar{\Upsilon}^{3}_{V;T_{j}T_{k}T_{j}T_{j}T_{k}T_{k}}(T_{i})&=& \bar{\Upsilon}^{3}_{V;T_{j}T_{k}T_{j}T_{k}T_{j}T_{k}}(T_{i})=
\bar{\Upsilon}^{3}_{V;T_{j}T_{k}T_{k}T_{j}T_{j}T_{k}}(T_{i})=
\bar{\Upsilon}^{3}_{V;T_{j}T_{k}T_{j}T_{k}T_{k}T_{j}}(T_{i})\nonumber\\
&&~~~~~~~~~~~~~~~~~~~~~~~~~~~~~~~~=\bar{\Upsilon}^{3}_{V;T_{j}T_{k}T_{k}T_{j}T_{k}T_{j}}(T_{i}),~~~~~~~~~\\
\bar{\Upsilon}^{3}_{V;T_{k}T_{j}T_{j}T_{j}T_{k}T_{k}}(T_{i})&=& \bar{\Upsilon}^{3}_{V;T_{k}T_{j}T_{j}T_{k}T_{j}T_{k}}(T_{i})=
\bar{\Upsilon}^{3}_{V;T_{k}T_{j}T_{k}T_{j}T_{j}T_{k}}(T_{i})=
\bar{\Upsilon}^{3}_{V;T_{k}T_{j}T_{j}T_{k}T_{k}T_{j}}(T_{i})\nonumber\\
&&~~~~~~~~~~~~~~~~~~~~~~~~~~~~~~~~=\bar{\Upsilon}^{3}_{V;T_{k}T_{j}T_{k}T_{j}T_{k}T_{j}}(T_{i}),~~~~~~~~~\\
\bar{\Upsilon}^{3}_{V;T_{k}T_{j}T_{j}T_{k}T_{k}T_{k}}(T_{i})&=& \bar{\Upsilon}^{3}_{V;T_{k}T_{j}T_{k}T_{j}T_{k}T_{k}}(T_{i})=
\bar{\Upsilon}^{3}_{V;T_{k}T_{j}T_{k}T_{k}T_{j}T_{k}}(T_{i})=
\bar{\Upsilon}^{3}_{V;T_{k}T_{j}T_{k}T_{k}T_{k}T_{j}}(T_{i}),~~~~~~~~~\\
\bar{\Upsilon}^{3}_{V;T_{j}T_{k}T_{j}T_{k}T_{k}T_{k}}(T_{i})&=& \bar{\Upsilon}^{3}_{V;T_{j}T_{k}T_{k}T_{j}T_{k}T_{k}}(T_{i})=
\bar{\Upsilon}^{3}_{V;T_{j}T_{k}T_{k}T_{k}T_{j}T_{k}}(T_{i})=
\bar{\Upsilon}^{3}_{V;T_{j}T_{k}T_{k}T_{k}T_{k}T_{j}}(T_{i}).
\eea
Using these results one can finally re-express the reduced slow-roll parameters as:
\bea\label{hub}
\bar{\epsilon}_{V}&=&\sum^{M}_{i=1}\sum^{M}_{j=1}\bar{\epsilon}_{V;T_{j}T_{j}}(T_{i})+2\sum^{M}_{i=1}\sum^{M}_{j=1}\sum^{M}_{k=1}\bar{\Delta}_{V;T_{j}T_{k}}(T_{i}),\\
\bar{\eta}_{V}&=&\sum^{M}_{i=1}\sum^{M}_{j=1}\bar{\eta}_{V;T_{j}T_{j}}(T_{i})+2\sum^{M}_{i=1}\sum^{M}_{j=1}\sum^{M}_{k=1}\bar{\Theta}_{V;T_{j}T_{k}}(T_{i}),\\
\bar{\xi}^2_{V}&=&\sum^{M}_{i=1}\sum^{M}_{j=1}\bar{\xi}^{2}_{V;T_{j}T_{j}T_{j}T_{j}}(T_{i})+\sum^{M}_{i=1}\sum^{M}_{j=1}\sum^{M}_{k=1}\left(3\bar{\vartheta}^{2}_{V;T_{j}T_{j}T_{k}T_{k}}(T_{i})\right.\\&& \left.~~~~~~~~~~~~~~~~~~~~~~~~~~+3\bar{\vartheta}^{2}_{V;T_{k}T_{j}T_{k}T_{k}}(T_{i})
+\bar{\vartheta}^{2}_{V;T_{j}T_{k}T_{k}T_{k}}(T_{i})\right),~~~~~~~~~~\\
\bar{\sigma}^3_{V}&=&\sum^{M}_{i=1}\sum^{M}_{j=1}\bar{\sigma}^{3}_{V;T_{j}T_{j}T_{j}T_{j}T_{j}T_{j}}(T_{i})+\sum^{M}_{i=1}\sum^{M}_{j=1}\sum^{M}_{k=1}\left(\Upsilon^{3}_{V;T_{j}T_{j}T_{k}T_{k}T_{k}T_{k}}(T_{i})
+\bar{\Upsilon}^{3}_{V;T_{j}T_{k}T_{k}T_{k}T_{k}T_{k}}(T_{i})\nonumber
\right.\\&& \left.+5\bar{\Upsilon}^{3}_{V;T_{j}T_{j}T_{j}T_{j}T_{k}T_{k}}(T_{i})
+4\bar{\Upsilon}^{3}_{V;T_{j}T_{j}T_{j}T_{j}T_{j}T_{k}}(T_{i})+2\bar{\Upsilon}^{3}_{V;T_{j}T_{k}T_{j}T_{j}T_{j}T_{j}}(T_{i})+8\bar{\Upsilon}^{3}_{V;T_{j}T_{k}T_{j}T_{j}T_{j}T_{k}}(T_{i})\nonumber\right.\\&& \left.
+10\bar{\Upsilon}^{3}_{V;T_{j}T_{k}T_{j}T_{j}T_{k}T_{k}}(T_{i})
+8\bar{\Upsilon}^{3}_{V;T_{j}T_{k}T_{j}T_{k}T_{k}T_{k}}(T_{i})+4\bar{\Upsilon}^{3}_{V;T_{j}T_{j}T_{j}T_{k}T_{k}T_{k}}(T_{i})
\right).~~~~
\eea
Now in our computation we take separable potentials having same structural form for all M number of non identical tachyons. In such case, for an example if take:
\bea V(T_1)&=& A_{1}\exp(-T_{1}/T_{01}),\\
V(T_2)&=& A_{2}\exp(-T_{2}/T_{02}),\\
\cdots\cdots \cdots&&\cdots\cdots \cdots \cdots\cdots \cdots\\
V(T_M)&=& A_{M}\exp(-T_{M}/T_{0M}),\eea which implies the structural form, we get the following simplified expression:
\bea\label{hubg5}
\label{zb1g5}\bar{\epsilon}_{V}&=&\sum^{M}_{i=1}\sum^{M}_{j=1}\bar{\epsilon}_{V;T_{j}T_{j}}(T_{i}),\\
\label{zb2g5}\bar{\eta}_{V}&=&\sum^{M}_{i=1}\sum^{M}_{j=1}\bar{\eta}_{V;T_{j}T_{j}}(T_{i}),\\
\bar{\xi}^2_{V}&=&\sum^{M}_{i=1}\sum^{M}_{j=1}\bar{\xi}^{2}_{V;T_{j}T_{j}T_{j}T_{j}}(T_{i}),~~~~~~~~~~\\
\bar{\sigma}^3_{V}&=&\sum^{M}_{i=1}\sum^{M}_{j=1}\bar{\sigma}^{3}_{V;T_{j}T_{j}T_{j}T_{j}T_{j}T_{j}}(T_{i}).~~~~
\eea
The cross terms only appears when the structural form of the $M$ number of tachyons are different. For an example, if we choose:
\bea V(T_{1})&=&A_{1}\exp(-T_{1}/T_{01}),\\
V(T_{2})&=& B_{2}{\rm cosh}(T_{2}/T_{02}),\\
\cdots\cdots \cdots&&\cdots\cdots \cdots \cdots\cdots \cdots\eea then all the cross terms in slow-roll vanish. Also for non-separable potentials this explanations works.
\subsubsection{The $\delta N$ formalism for Multi tachyons}
\begin{figure}[t]
{\centerline{\includegraphics[width=16.5cm, height=14cm] {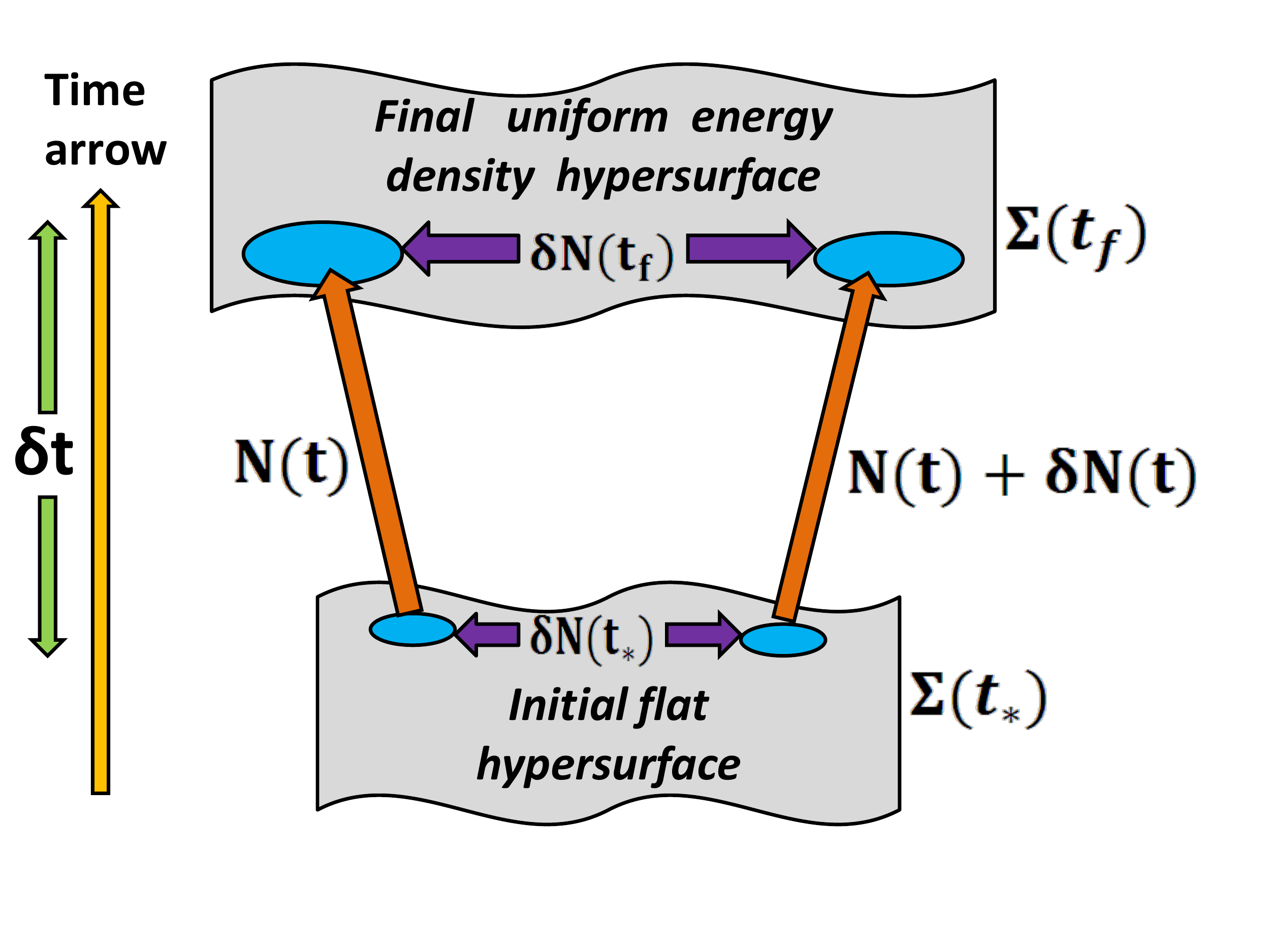}}}
\caption{Diagrammatic representation of $\delta N$ formalism. In this schematic picture $\Sigma(t_{i})$ and $\Sigma(t_{f})$ represent the initial and final hypersurface
where time arrow flows from $t_{i}\rightarrow t_{f}$.} \label{fig3ccz}
\end{figure}

\begin{figure}[t]
	{\centerline{\includegraphics[width=15.5cm, height=12cm] {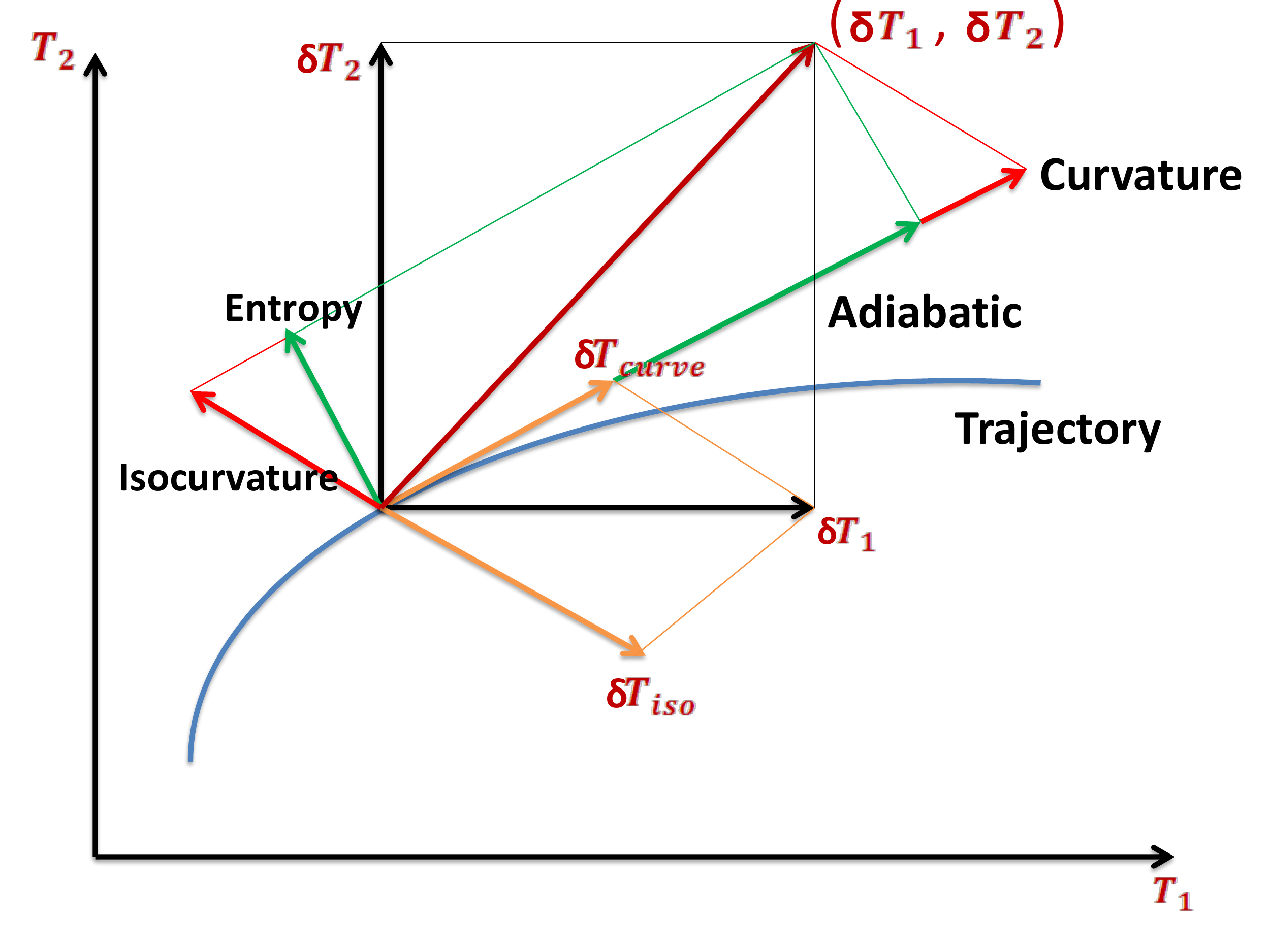}}}
	\caption{Diagrammatic representation of trajectories and perturbation decompositions in the field space.
 Here it is explicitly shown that any perturbation including field
perturbations can be decomposed to a curvature component ``curve'' and an isocurvature component
``iso''.} \label{fig3cczvc}
\end{figure}

In this section we have used the $\delta N$ formalism 
to compute the inflationary observables, for the multi tachyonic field setup. Here $N$ signifies the number of e-foldings which can be expressed in the multi tachyonic set up as:
\be\begin{array}{lll}\label{e-fold2v7v3} N= \int^{t_{end}}_{t} Hdt=\left\{\begin{array}{lll}
		\displaystyle  
		\frac{\alpha^{'}}{M^{2}_{p}}\sum^{M}_{i=1}\int^{T_i}_{T_{i,end}}
		\frac{V^2(T_i)}{V^{'}(T_i)}~dT_i\,,~~~~~~ &
		\mbox{\small {\bf for {$q=1/2$ }}}  \\ \\ 
		\displaystyle   
		\frac{ \sqrt{2q}\alpha^{'}}{M^{2}_{p}}\sum^{M}_{i=1}\int^{T_i}_{T_{i,end}}
		\frac{V^2(T_i)}{V^{'}(T_i)}~dT_i\,.~~~~~~ &
		\mbox{\small {\bf for {~any~arbitrary~ $q$ }}} 
	\end{array}
	\right.
\end{array}\ee
In the non-attractor regime, the $\delta N$
formalism shows various non trivial features which has to be taken into account during explicit calculations. Once the solution
reaches the attractor behaviour, the dominant contribution comes from only on the
perturbations of the scalar-field
trajectories with respect to the tachyon field value at the initial
hypersurface, $T_{i}$, as the velocity, $\dot T_{i}$, is uniquely
determined by $T_{i}$ where $i=1,2,\cdots,M$. However, in the non-attractor regime of solution, 
both the information from the field value $T_{i}$ and also $\dot T_{i}$ are required to determine the 
trajectory. This can be understood by providing two initial conditions on $T_i$ and $\dot T_i$ on the
initial hypersurface. During the computation of the trajectories let us assume here that the universe has already arrived 
at the adiabatic limit via attractor phase by this epoch,
or equivalently it can be stated that a typical phase transition phenomena appears to an attractor phase at the time $t=t_*$. 
More specifically, in the present context, we have assumed that the evolution of the universe is unique
after the value of the scalar field arrived at $T=T_*$ where it is mimicking the role of standard clock, 
irrespective of the value of its velocity $\dot T_*$. 
Let me mention that only in this case $\delta N$ is equal to
the final value of the comoving curvature perturbation $\zeta$ which
is conserved at $t\geq t_*$. In figure~(\ref{fig3ccz}) and figure~(\ref{fig3cczvc}), we have shown the schematic picture of the $\delta N$ formalism and the
trajectories and perturbation decompositions in the field space for multi-field tachyon inflation. 

On sufficiently large scales for a given suitable assumptions about the dynamical behaviour which permit us to ignore time
derivatives of the perturbations, we expect that each horizon volume will evolve in such a manner that it
were a self contained universe, and consequently the curvature perturbation can be written
beyond liner order in cosmological perturbation theory as \cite{Choudhury:2014uxa,Sugiyama:2012tj}:
\bea \zeta &=& \delta N\nonumber\\ &=& \sum^{M}_{i=1}N_{,i}\delta T^{i}+\frac{1}{2!}\sum^{M}_{i=1}\sum^{M}_{j=1}N_{,ij}\delta T^{i}\delta T^{j}+\frac{1}{3!}\sum^{M}_{i=1}\sum^{M}_{j=1}\sum^{M}_{k=1}N_{,ijk}\delta T^{i}\delta T^{j}\delta T^{k}+\cdots,~~~~~~~~~~~~~~~~ \eea
where we use the following short hand notations:
\bea N_{,i}&=&\partial_{T_{i}}N,\\
N_{,ij}&=&\partial_{T_{i}}\partial_{T_{j}}N,\\
N_{,ijk}&=&\partial_{T_{i}}\partial_{T_{j}}\partial_{T_{k}}N.
\eea
More precisely here $\delta T_{i}\forall i=1,2,\cdots,M$ represent the deviations of the fields from their unperturbed values in some specified region of the universe. 

When the potential is sum-separable, the derivatives of $N$  can be simplified into the following expressions:
\bea N_{,i}&=&\frac{1}{\sqrt{2\bar{\epsilon}_{V}(T^{\star}_{i})}}\frac{V(T^{\star}_{i})+Z^{c}_{i}}{\left(\sum^{M}_{j=1}V(T^{\star}_{j})\right)},\\
N_{,ij}&=&\delta_{ij}\left[1-\frac{\bar{\eta}_{V}(T^{\star}_{i})}{2\bar{\epsilon}_{V}(T^{\star}_{i})}\frac{V(T^{\star}_{i})+Z^{c}_{i}}{\left(\sum^{M}_{j=1}V(T^{\star}_{j})\right)}\right]+\frac{1}{\sqrt{2\bar{\epsilon}_{V}(T^{\star}_{j})}\left(\sum^{M}_{m=1}V(T^{\star}_{m})\right)}Z^{c}_{j,i},~~~~~~~~~~~
\eea
where $Z^{c}_{i}$ and $Z^{c}_{i,j}$ is defined as:
\bea Z^{c}_{i}&=& \left(\sum^{M}_{m=1}V(T^{c}_{m})\right)\frac{\bar{\epsilon}_{V}(T^{c}_{i})}{\bar{\epsilon}^{c}_{V}}-V(T^{c}_{i}),\\
Z^{c}_{i,j}&=& -\frac{\left(\sum^{M}_{m=1}V(T^{c}_{m})\right)^2}{\sum^{M}_{p=1}V(T^{\star}_{p})}\sqrt{\frac{2}{\bar{\epsilon}_{V}(T_{j})}}\nonumber\\
&& ~~~~~~~~~~~~~~\times\left[\sum^{M}_{k=1}\bar{\epsilon}_{V}(T_{k})\left(\frac{\bar{\epsilon}_{V}(T_{i})}{\bar{\epsilon}_{V}}-\delta_{ik}\right)
\left(\frac{\bar{\epsilon}_{V}(T_{j})}{\bar{\epsilon}_{V}}-\delta_{jk}\right)\left(1-\frac{\bar{\eta}_{V}(T_{k})}{\bar{\epsilon}_{V}}\right)\right]_{c}~~~~~~~~~~~~\nonumber\\
&=&\sqrt{\frac{2}{\bar{\epsilon}_{V}(T_{j})}}\left(\sum^{M}_{m=1}V(T^{\star}_{m})\right){\cal K}_{ij},\eea
where $\star$ indicates the horizon crossing and $c$ denotes the constant density surface. In this context additionally we get:
\bea dN&=&\sum^{M}_{j=1}\left[\left(\frac{V(T_{j})}{\partial_{j}V(T_{i})}\right)-\sum^{M}_{i=1}\frac{\partial T^{c}_{i}}{\partial T^{\star}_{j}}\left(\frac{V(T_{i})}{\partial_{i}V(T_{k})}\right)\right]dT^{\star}_{j},\eea
where 
\bea \frac{\partial T^{c}_{i}}{\partial T^{\star}_{j}} &=&-\frac{\sum^{M}_{k=1}V(T^{c}_{k})}{\sum^{M}_{m=1}V(T^{\star}_{m})}\sqrt{\frac{\bar{\epsilon}_{V}(T^{c}_{i})}{\bar{\epsilon}_{V}(T^{\star}_{j})}}\left(\frac{\bar{\epsilon}_{V}(T^{c}_{i})}{\sum^{M}_{p=1}\bar{\epsilon}_{V}(T_{p})}-\delta_{ij}\right).
\eea Further using this one can write the following differential operator identity which is very useful to compute various inflationary observables mentioned in the next section: 
\bea \frac{d}{dN}&=&\sum^{M}_{j=1}\left[\left(\frac{V(T_{j})}{\partial_{j}V(T_{i})}\right)-\sum^{M}_{i=1}\frac{\partial T^{c}_{i}}{\partial T^{\star}_{j}}\left(\frac{V(T_{i})}{\partial_{i}V(T_{k})}\right)\right]^{-1}\frac{d}{dT^{\star}_{j}},\eea
In the next section we will write all the inflationary observables in terms of the components of $\delta N$ by including all the effects of non-linearities in cosmological perturbations for multi tachyons. 
\subsubsection{Computation of scalar power spectrum}
In this subsection we start with two point function for multi tachyonic scalar modes. Implementing the $\delta N$ procedure we get:
\bea  \langle \zeta_{\bf k} \zeta_{\bf k^{'}} \rangle  &=& \sum^{M}_{i=1}\sum^{M}_{j=1}\langle \zeta^{i}_{\bf k} \zeta^{j}_{\bf k^{'}} \rangle=(2\pi)^3 \delta^{3}({\bf k}+{\bf k^{'}})\frac{2\pi^2}{k^3}\Delta_{\zeta}(k),\eea
where the primordial power spectrum for scalar modes at any arbitrary momentum scale $k$ can be expressed as:
\bea \Delta_{\zeta}(k)&=& \sum^{M}_{i=1}\sum^{M}_{j=1}N_{,i}N_{'j}G^{ij}\frac{k^3 P_{\zeta}(k)}{2\pi^2}.\eea
 Sometimes it is convenient to express everything in terms of the normalized adiabatic curvature power spectrum in the following way:
\bea  \langle \bar{\zeta}_{\bf k} \bar{\zeta}_{\bf k^{'}} \rangle  &=&(2\pi)^3 \delta^{3}({\bf k}+{\bf k^{'}})\frac{2\pi^2}{k^3}\bar{\Delta}_{\zeta}(k),\eea
where 
\bea \bar{\zeta}&=& -\frac{H}{\dot{T}_0}\sum^{M}_{i=1}\sum^{M}_{j=1}\Theta_{i}\delta T_{j}G^{ij},\\
 \bar{\Delta}_{\zeta}(k)&=& \frac{1}{2\bar{\epsilon}_{V}}\sum^{M}_{i=1}\sum^{M}_{j=1}\Theta_{i}\Theta_{j}G^{ij}\frac{k^3 P_{\zeta}(k)}{2\pi^2}.\eea
 Here $\Theta_{i}=\frac{\dot{T}_{i}}{\dot{T}_{0}}$ is a basis vector that projects $\delta T_{i}$ along the direction of classical background trajectory. The vector ${\bf \Theta}$ and a complementary set of $(M-1)$ mutually orthonormal basis vectors ${\bf s_{p}}$ form the kinematic basis. Applying the Gram-Schmidt orthogonalization technique one can determine ${\bf s_{p}}$. 
 
 On the other hand for the multi tachyonic scenario the isocurvature perturbations ${\cal S}_{p}$ is the orthogonal projection along the ${\bf s_{K}}$ directions:
 \bea {\cal S}_{p}&=& -\frac{H}{\dot{T}_{0}}\sum^{M}_{j=1}\sum^{M}_{m=1}s_{pj}G^{jm}\delta T_{m}.\eea
  Using this one can write down the expression for the normalized isocurvature perturbation:
  \bea \langle {\cal S}_{{\bf k}} {\cal S}_{{\bf k^{'}}}\rangle&=&\sum^{M-1}_{p=1}\sum^{M-1}_{q=1}\langle {\cal S}_{p{\bf k}} {\cal S}_{q{\bf k^{'}}}\rangle =(2\pi)^3\delta^{3}({\bf k}+{\bf k^{'}})\frac{2\pi^2}{k^3}\bar{\Delta}_{\cal S}(k),\eea
  where 
 \bea \bar{\Delta}_{\cal S}(k)&=& \frac{1}{2\bar{\epsilon}_{V}}\sum^{M-1}_{p=1}\sum^{M-1}_{q=1}\sum^{M}_{j=1}\sum^{M}_{m=1}s_{j}^{p}s_{m}^{q}G^{jm}\frac{k^3 P_{\zeta}(k)}{2\pi^2},\eea
 where we have explicitly shown all the summations to indicate that the isocurvature basis vectors are $(M-1)$ dimensional.	For the sake of simplicity applying Gram-Schmidt orthogonalization technique one can choose a basis where $G^{ij}$ is diagonal and given by $G^{ij}=\delta^{ij}$.	
 
 Similarly, for the completeness one can define the adiabatic-isocurvature cross spectra by the following fashion: 
 \bea \langle \bar{\zeta}_{{\bf k}} {\cal S}_{{\bf k^{'}}}\rangle&=&\sum^{M-1}_{p=1}\langle \bar{\zeta}_{{\bf k}} {\cal S}_{p{\bf k^{'}}}\rangle =(2\pi)^3\delta^{3}({\bf k}+{\bf k^{'}})\frac{2\pi^2}{k^3}\bar{\Delta}_{\zeta{\cal S}}(k),\eea
 where 
 \bea \bar{\Delta}_{\zeta{\cal S}}(k)&=& \frac{1}{2\bar{\epsilon}_{V}}\sum^{M-1}_{p=1}\sum^{M}_{j=1}\sum^{M}_{m=1}\Theta_{j}s_{m}^{p}\left(G^{jm}+G^{mj}\right)\frac{k^3 P_{\zeta}(k)}{2\pi^2}.\eea	
Cross correlations are generically expected if the background trajectory is curved as modes of interest leave the horizon. 

 Now we write down the the expressions for all the inflationary observables computed from the multi tachyonic set up at horizon crossing:
\begin{itemize}
	\item In the present context amplitude of scalar power spectrum can be computed as:
	\begin{eqnarray} 
	\label{psxcc1vv5} \Delta_{\zeta, \star}
	&=&\left\{\begin{array}{lll}
	\displaystyle 
	\displaystyle \sum^{M}_{i=1}\sum^{M}_{j=1}N_{,i}N_{,j}G^{ij} \left\{\left[1-({\cal C}_{E}+1)\bar{\epsilon}_{V}-{\cal C}_{E}\left(3\bar{\epsilon}_{V}-\bar{\eta}_{V}\right)\right]^2 \frac{H^{2}}{4\pi^2c_{S}}\right\}_{\star}\,, &
	\mbox{\small {\bf for $q=1/2$}}  \\ 
	\displaystyle \sum^{M}_{i=1}\sum^{M}_{j=1}N_{,i}N_{,j}G^{ij}\left\{\left[1-({\cal C}_{E}+1)\frac{\bar{\epsilon}_{V}}{2q}-\frac{{\cal C}_{E}}{\sqrt{2q}}\left(3\bar{\epsilon}_{V}-\bar{\eta}_{V}\right)\right]^2 \frac{q H^{2}}{2\pi^2c_{S}}\right\}_{\star}\,, &
	\mbox{\small {\bf for any $q$}}.
	\end{array}
	\right.\end{eqnarray}
	where $H^2=V/3M^{2}_{p}=\sum^{M}_{k=1}V(T_{k})/3M^2_p$ and ${\cal C}_{E}= -2 + \ln 2 + \gamma \approx -0.72.$  Using the slow-roll approximations one can further approximate the expression for sound speed as:
	\be\begin{array}{lll}\label{rteqxcccvv5}
		\displaystyle c^2_{S} 
		=\left\{\begin{array}{lll}
			\displaystyle  
			1-\frac{2}{3}\bar{\epsilon}_{V}+{\cal O}\left(\bar{\epsilon}^2_{V}\right)+\cdots \,,~~~~ &
			\mbox{\small {\bf for $q=1/2$}}  \\ 
			\displaystyle  
			1-\frac{(1-q)}{3q^2 }\bar{\epsilon}_{V}+{\cal O}\left(\bar{\epsilon}^2_{V}\right)+\cdots \,,~~~ &
			\mbox{\small {\bf for ~any~$q$}}
		\end{array}
		\right.
	\end{array}\ee
	where $\bar{\epsilon}_{V}$ and $\bar{\eta}_{V}$ are the cumulative or total contribution to the slow-roll parameter as defined earlier.
	Hence using the result in Eq~(\ref{psxcc1vv5}) we get the following simplified expression for the primordial scalar power spectrum:
	\begin{eqnarray} 
	\label{psxcc1v0vv5} \Delta_{\zeta, \star}
	&=&\left\{\begin{array}{lll}
	\displaystyle 
	\displaystyle \sum^{M}_{i=1}\sum^{M}_{j=1}N_{,i}N_{,j}G^{ij} \left\{\left[1-\left({\cal C}_{E}+\frac{5}{6}\right)\bar{\epsilon}_{V}-{\cal C}_{E}
	\left(3\bar{\epsilon}_{V}-\bar{\eta}_{V}\right)\right]^2 \frac{H^{2}}{4\pi^2}\right\}_{\star}\,, &
	\mbox{\small {\bf for $q=1/2$}}  \\ 
	\displaystyle \sum^{M}_{i=1}\sum^{M}_{j=1}N_{,i}N_{,j}G^{ij}\left\{\left[1-({\cal C}_{E}+1-\Sigma)\frac{\bar{\epsilon}_{V}}{2q}-\frac{{\cal C}_{E}}{\sqrt{2q}}\left(3\bar{\epsilon}_{V}-\bar{\eta}_{V}\right)\right]^2 \frac{q H^{2}}{2\pi^2}\right\}_{\star}\,, &
	\mbox{\small {\bf for any $q$}}.
	\end{array}
	\right.\end{eqnarray}
	Similarly using the normalized adiabatic curvature power spectrum we get:
	\begin{eqnarray} 
	\label{psxcc1v0vv5ad} \bar{\Delta}_{\zeta, \star}
	&=&\left\{\begin{array}{lll}
	\displaystyle 
	\displaystyle \sum^{M}_{i=1}\sum^{M}_{j=1}\Theta_{i}\Theta_{j}G^{ij} \left\{\left[1-\left({\cal C}_{E}+\frac{5}{6}\right)\bar{\epsilon}_{V}-{\cal C}_{E}
	\left(3\bar{\epsilon}_{V}-\bar{\eta}_{V}\right)\right]^2 \frac{H^{2}}{8\pi^2\bar{\epsilon}_{V}}\right\}_{\star}\,, &
	\mbox{\small {\bf for $q=1/2$}}  \\ 
	\displaystyle \sum^{M}_{i=1}\sum^{M}_{j=1}\Theta_{i}\Theta_{j}G^{ij}\left\{\left[1-({\cal C}_{E}+1-\Sigma)\frac{\bar{\epsilon}_{V}}{2q}-\frac{{\cal C}_{E}}{\sqrt{2q}}\left(3\bar{\epsilon}_{V}-\bar{\eta}_{V}\right)\right]^2 \frac{q H^{2}}{4\pi^2\bar{\epsilon}_{V}}\right\}_{\star}\,, &
	\mbox{\small {\bf for any $q$}}.
	\end{array}
	\right.\end{eqnarray}
	
	\item In the present context the normalized amplitude of isocurvature power spectrum can be computed as:
	\begin{eqnarray} 
	\label{psxcc1vv5iso} \bar{\Delta}_{{\cal S}, \star}
	&=&\left\{\begin{array}{lll}
	\displaystyle 
	\displaystyle \sum^{M-1}_{p,q=1}\sum^{M}_{j,m=1}s_{j}^{p}s_{m}^{q}G^{jm} \left\{\left[1-({\cal C}_{E}+1)\bar{\epsilon}_{V}-{\cal C}_{E}\left(3\bar{\epsilon}_{V}-\bar{\eta}_{V}\right)\right]^2 \frac{H^{2}}{8\pi^2c_{S}\bar{\epsilon}_{V}}\right\}_{\star}\,, &
	\mbox{\small {\bf for $q=1/2$}}  \\ 
	\displaystyle \sum^{M-1}_{p,q=1}\sum^{M}_{j,m=1}s_{j}^{p}s_{m}^{q}G^{jm}\left\{\left[1-({\cal C}_{E}+1)\frac{\bar{\epsilon}_{V}}{2q}-\frac{{\cal C}_{E}}{\sqrt{2q}}\left(3\bar{\epsilon}_{V}-\bar{\eta}_{V}\right)\right]^2 \frac{q H^{2}}{4\pi^2c_{S}\bar{\epsilon}_{V}}\right\}_{\star}\,, &
	\mbox{\small {\bf for any $q$}}.
	\end{array}
	\right.\end{eqnarray}
	where $H^2=V/3M^{2}_{p}=\sum^{M}_{k=1}V(T_{k})/3M^2_p$ and ${\cal C}_{E}= -2 + \ln 2 + \gamma \approx -0.72.$  Further using approximated expression for $c_{S}$ in the slow-roll regime and also using
	the result in Eq~(\ref{psxcc1vv5iso}) we get the following simplified expression for the power spectrum:
		\begin{eqnarray} 
		\label{psxcc1vv5iso2} \bar{\Delta}_{{\cal S}, \star}
		&=&\left\{\begin{array}{lll}
		\displaystyle 
		\displaystyle \sum^{M-1}_{p,q=1}\sum^{M}_{j,m=1}s_{j}^{p}s_{m}^{q}G^{jm} \left\{\left[1-\left({\cal C}_{E}+\frac{5}{6}\right)\bar{\epsilon}_{V}-{\cal C}_{E}
		\left(3\bar{\epsilon}_{V}-\bar{\eta}_{V}\right)\right]^2 \frac{H^{2}}{8\pi^2\bar{\epsilon}_{V}}\right\}_{\star}\,, &
		\mbox{\small {\bf for $q=\frac{1}{2}$}}  \\ 
		\displaystyle \sum^{M-1}_{p,q=1}\sum^{M}_{j,m=1}s_{j}^{p}s_{m}^{q}G^{jm}\left\{\left[1-({\cal C}_{E}+1-\Sigma)\frac{\bar{\epsilon}_{V}}{2q}-\frac{{\cal C}_{E}}{\sqrt{2q}}\left(3\bar{\epsilon}_{V}-\bar{\eta}_{V}\right)\right]^2 \frac{q H^{2}}{4\pi^2\bar{\epsilon}_{V}}\right\}_{\star}\,, &
		\mbox{\small {\bf for any $q$}}.
		\end{array}
		\right.\end{eqnarray}
		
			\item Similarly the normalized amplitude of adiabatic-isocurvature cross power spectrum can be computed as:
			\begin{eqnarray} \tiny
			\label{psxcc1vv5isoad} \bar{\Delta}_{\zeta{\cal S}, \star}
			&=&\left\{\footnotesize\begin{array}{lll}\tiny
			\displaystyle 
			\displaystyle \sum^{M-1}_{p=1}\sum^{M}_{j,m=1}\Theta_{j}s_{m}^{p}\left(G^{jm}+G^{mj}\right) \left\{\left[1-({\cal C}_{E}+1)\bar{\epsilon}_{V}-{\cal C}_{E}\left(3\bar{\epsilon}_{V}-\bar{\eta}_{V}\right)\right]^2 \frac{H^{2}}{8\pi^2c_{S}\bar{\epsilon}_{V}}\right\}_{\star}\,, &
			\mbox{\small {\bf for $q=1/2$}}  \\ 
			\displaystyle \sum^{M-1}_{p=1}\sum^{M}_{j,m=1}\Theta_{j}s_{m}^{p}\left(G^{jm}+G^{mj}\right)\left\{\left[1-({\cal C}_{E}+1)\frac{\bar{\epsilon}_{V}}{2q}-\frac{{\cal C}_{E}}{\sqrt{2q}}\left(3\bar{\epsilon}_{V}-\bar{\eta}_{V}\right)\right]^2 \frac{q H^{2}}{4\pi^2c_{S}\bar{\epsilon}_{V}}\right\}_{\star}\,, &
			\mbox{\small {\bf for any $q$}}.
			\end{array}
			\right.\end{eqnarray}
			where $H^2=V/3M^{2}_{p}=\sum^{M}_{k=1}V(T_{k})/3M^2_p$ and ${\cal C}_{E}= -2 + \ln 2 + \gamma \approx -0.72.$  Further using approximated expression for $c_{S}$ in the slow-roll regime and also using
			the result in Eq~(\ref{psxcc1vv5isoad}) we get the following simplified expression for the power spectrum:
			\begin{eqnarray} 
			\label{psxcc1vv5iso2} \bar{\Delta}_{\zeta{\cal S}, \star}
			&=&\left\{\footnotesize\begin{array}{lll}
			\displaystyle 
			 \sum^{M-1}_{p=1}\sum^{M}_{j,m=1}\Theta_{j}s_{m}^{p}\left(G^{jm}+G^{mj}\right) \left\{\left[1-\left({\cal C}_{E}+\frac{5}{6}\right)\bar{\epsilon}_{V}-{\cal C}_{E}
			\left(3\bar{\epsilon}_{V}-\bar{\eta}_{V}\right)\right]^2 \frac{H^{2}}{8\pi^2\bar{\epsilon}_{V}}\right\}_{\star}\,, &
			\mbox{\small {\bf for $q=\frac{1}{2}$}}  \\ 
			\displaystyle \sum^{M-1}_{p=1}\sum^{M}_{j,m=1}\Theta_{j}s_{m}^{p}\left(G^{jm}+G^{mj}\right)\left\{\left[1-({\cal C}_{E}+1-\Sigma)\frac{\bar{\epsilon}_{V}}{2q}-\frac{{\cal C}_{E}}{\sqrt{2q}}\left(3\bar{\epsilon}_{V}-\bar{\eta}_{V}\right)\right]^2 \frac{q H^{2}}{4\pi^2\bar{\epsilon}_{V}}\right\}_{\star}\, &
			\mbox{\small {\bf for any $q$}}.
			\end{array}
			\right.\end{eqnarray}

	\item Next one can compute the scalar spectral tilt ($n_{\zeta}, n_{\cal S},n_{\zeta{\cal S}}$) of the primordial adiabatic, isocurvature and cross power spectrum as:
	\begin{eqnarray} 
	\label{psxcc1v2vv5sc} n_{\zeta, \star}-1 
	&\approx&\left\{\footnotesize\begin{array}{lll}
	\displaystyle 
	\displaystyle  
	\displaystyle
	\left(2\bar{\eta}_{V}-8\bar{\epsilon}_{V}\right)-\frac{2}{\sum^{M}_{i=1}\sum^{M}_{j=1}N_{,i}N_{,j}G^{ij}}\\
	\displaystyle +
	\frac{2}{\sum^{M}_{m=1}V(T_{m})}\frac{\sum^{M}_{n=1}\sum^{M}_{i=1}\sum^{M}_{j=1}\sum^{M}_{l=1}\sum^{M}_{k=1}V_{,ij}(T_{n})N_{,l}N_{,k}G^{il}G^{jk}}{\sum^{M}_{i^{'}=1}\sum^{M}_{j^{'}=1}N_{,i^{'}}N_{,j^{'}}G^{i^{'}j^{'}}}+\cdots, &
	\mbox{\small {\bf for $q=1/2$}}  \\ 
	\displaystyle  
	\displaystyle  
	\displaystyle
	\sqrt{\frac{2}{q}}\bar{\eta}_{V}-\left(\frac{1}{q}+3\sqrt{\frac{2}{q}}\right)\bar{\epsilon}_{V}-\frac{1}{q\sum^{M}_{i=1}\sum^{M}_{j=1}N_{,i}N_{,j}G^{ij}}\\
	\displaystyle +
	\frac{1}{q\sum^{M}_{m=1}V(T_{m})}\frac{\sum^{M}_{n=1}\sum^{M}_{i=1}\sum^{M}_{j=1}\sum^{M}_{l=1}\sum^{M}_{k=1}V_{,ij}(T_{n})N_{,l}N_{,k}G^{il}G^{jk}}{\sum^{M}_{i^{'}=1}\sum^{M}_{j^{'}=1}N_{,i^{'}}N_{,j^{'}}G^{i^{'}j^{'}}}+\cdots\,, &
	\mbox{\small {\bf for any $q$}}
	\end{array}
	\right.\end{eqnarray}
	and additionally we have:
	\bea n_{\cal S,\star}-1&\approx& n_{\zeta,\star}-1\approx n_{\zeta{\cal S},\star}-1.\eea
	\item One can also compute the expression for the running and running of the running by following the same procedure as mentioned above.

\end{itemize}
For more completeness let us mention the behaviour of the cosmological perturbation at later times for multi Gtachyonic inflation. 
To start with it is important to mention here that in a more generalized physical prescription the time dependence of adiabatic
and entropy perturbations in the large cosmological scale limit can always be written in the following
simplified form for multi Gtachyonic inflationary paradigm as:
\bea \label{evol1}
\frac{d\zeta}{dt}=\dot{\zeta}&=&\Upsilon H {\cal S},\\
\label{evol2}\frac{d{\cal S}}{dt}=\dot{\cal S}&=&\Sigma H {\cal S},
\eea
where $\Upsilon$ and $\Sigma$ physically represent the generalized time-dependent dimensionless functions in cosmological
perturbation theory of multi Gtachyonic fields.
Now to extract more informations form Eq.~(\ref{evol1}) and Eq.~(\ref{evol2}), we further integrate both the equations over the specified cosmological time scale and finally following this prescription 
one can easily compute the expression for the generalized form
of the transfer matrix which relate the curvature and entropic perturbations generated
during the situation where the fluctuating mode is stretched outside the
expansion (Hubble) scale during the epoch of inflation to the
curvature and entropic perturbations at later time via the following simplified form
of the matrix equation as:
\be
\label{matrix}
\left(\begin{array}{c}
{\zeta}(t) \\ {{\cal S}(t)}
\end{array}\right) = \left(
\begin{array}{ccccccccc}
1 & & & & & &\hat{T}_{\zeta{\cal S}} \\ 0 & & & & & &\hat{T}_{{\cal S}{\cal S}}
\end{array}
\right) \left(\begin{array}{c}\zeta(t_{\star}) \\ {\cal S}(t_{\star})
\end{array}
\right),
\ee
where the transfer functions are represented by the following equations in the context
of Multi tachyonic inflation from effective GTachyon setup:
\bea\label{tgst}
\hat{T}_{\zeta{\cal S}}(t_{\star},t) &=& \int^t_{t_{\star}} \Upsilon(t^{''}) \hat{T}_{{\cal S}{\cal S}}(t_{\star},t^{''}) H(t^{''})
dt^{''},\\
\label{tsst}
\hat{T}_{{\cal S}{\cal S}}(t_{\star},t) &=& \exp \left( \int^t_{t_{\star}} \Sigma(t^{'}) H(t^{''})
dt^{''} \right).
\eea
It is also important to note that the evolution in the large-scale limit is independent
of the cosmological scale under consideration in the present context and consequently
the derived generalized form of the transfer functions $\hat{T}_{\zeta{\cal S}}$ and
$\hat{T}_{{\cal S}{\cal S}}$ are implicit functions
of cosmological scale due to their dependence upon the cosmic time scale $t_{\star}(k)$ as appearing in the argument of the transfer functions. The scale-dependence of the
transfer functions are governed by the following sets of evolution equations:
\bea
\label{htgs}
\left(H_{\star}^{-1} {\partial_{t_{\star}}} +\Sigma_{\star}\right) \hat{T}_{\zeta{\cal S}}+\Upsilon_{\star}&=& 0, \\
\label{htss}
\left(H_{\star}^{-1} \partial_{t_{\star}}+\Sigma_{\star}\right) \hat{T}_{{\cal S}{\cal S}} &=&0.
\eea
where we introduce a notation $\partial_{t_{\star}}=\partial/\partial t_{\star}$ to define the partial differentiation in a simplified way.
In the present context, the cosmological (momentum) scale dependence of the generalized transfer functions for
multi tachyonic inflationary paradigm is explicitly determined
by the two factors, $\Upsilon_{\star}$ and $\Sigma_{\star}$, which physically represent 
the cosmic time scale evolution of the curvature and entropic fluctuations at the horizon crossing during
the epoch of multi GTachyonic inflation.

Now one can apply the above mentioned generalized transfer matrix as stated in Eq.~(\ref{matrix}) to the primordial
scalar power spectra explicitly computed in the previous subsection. After doing the
detailed analysis one can finally compute the resulting curvature and entropic primordial
power spectra at the beginning point of the radiation dominated epoch as:
\bea
\label{powg}
\bar{\Delta}_{\zeta} &=& \left( 1 + \hat{T}_{\zeta{\cal S}}^2 \right) \bar{\Delta}_{\zeta,\star},\\
\label{pows}
\bar{\Delta}_{{\cal S}} &=& \hat{T}_{{\cal S}{\cal S}}^2 \bar{\Delta}_{\zeta,\star},\\
\label{powgs}
\bar{\Delta}_{\zeta{\cal S}} &=& \hat{T}_{\zeta{\cal S}} \hat{T}_{{\cal S}{\cal S}} \bar{\Delta}_{\zeta,\star}.
\eea
For the sake of simplicity let us define a dimensionless cosmological measure of the correlation
function in terms of a correlation angle $\theta$ in the following simplified way:
\bea
\label{cd1} \cos\theta &\equiv& {\bar{\Delta}_{\zeta{\cal S}} \over \sqrt{\bar{\Delta}_{\zeta}
		\bar{\Delta}_{\cal S}}} = {\hat{T}_{\zeta{\cal S}} \over \sqrt{1+\hat{T}_{\zeta{\cal S}}^2}} \,,\\
\label{cd2} \sin\theta &\equiv& \sqrt{1-\left( {\bar{\Delta}_{\zeta{\cal S}} \over \sqrt{{\Delta}_{\zeta}
			\bar{\Delta}_{\cal S}}}\right)^2} = \frac{1}{\sqrt{1+\hat{T}_{\zeta{\cal S}}^2}} \,,\\
\label{cd3} \theta &=&  \cot^{-1}(\hat{T}_{\zeta{\cal S}}) \,,
\eea
which are surely very very useful for further computation in the present context.
Further using the Eq.~(\ref{powg}) in Eq.~(\ref{cd1}), finally we get the following expression for the scalar metric perturbation at
the horizon crossing which is expressed in terms of the observed curvature perturbation
at later times in the cosmic time scale and also in terms of the the cross-correlation angle as:
\bea
\label{delau}
\bar{\Delta}_{\zeta,\star} \simeq \bar{\Delta}_{\zeta} \sin^2\theta \,.
\eea
Next using Eqs.~(\ref{powg}-\ref{powgs}), the spectral indices of the primordial power spectrum
at later stage of times in cosmological time scale can be expressed by the following simplified expressions:
\bea
n_\zeta-1 &=& n_{\zeta,\star}-1 + H_{\star}^{-1} \left(\partial_{t_{\star}}{\hat{T}}_{\zeta{\cal S}}\right) \sin 2\theta,\\
n_{\cal S}-1 &=& n_{\zeta,\star}-1 + 2 H_{\star}^{-1} \left(\partial_{t_{\star}}{\hat{T}}_{{\cal S}{\cal S}}\right),\\
n_{\cal C}-1 &=& n_{\zeta,\star}-1 + H_{\star}^{-1} \left\{ \left(\partial_{t_{\star}}{\hat{T}}_{\zeta{\cal S}}\right)
\tan\theta + \left(\partial_{t_{\star}}{\hat{T}}_{{\cal S}{\cal S}}\right) \right\}.
\eea
which are very very useful to study the scale dependent behaviour of the primordial power spectra in the present context.
Following the same procedure one can also compute the expressions for the running and running of the running of the spectral indices.
Additionally, it is important to mention here that, the overall amplitude
of the generalized transfer functions $\hat{T}_{\zeta{\cal S}}$ and $\hat{T}_{{\cal S}{\cal S}}$ are dependent on the time scale evolution
after horizon crossing via reheating phenomena and into the radiation dominated
epoch. But in spite of this important fact, the spectral tilts of the resulting primordial perturbation spectra can be finally
expressed in terms of the slow-roll parameters at the horizon crossing during the
epoch of inflation and also in terms of the cross-correlation angle $\theta$, which we have already introduced earlier in Eq~(\ref{cd3}).

\subsubsection{Computation of tensor power spectrum}

In this subsection we will not derive the crucial results for multi Gtachyonic inflation. But we will state 
the results for {\bf BD} vacuum where the changes will appear due to the presence of $M$ 
number of different tachyon field. One can similarly write down the detailed expressions for {\bf AV} (including $\alpha$ vacuum) as well. 

\bea  \langle h_{\bf k} h_{\bf k^{'}} \rangle  &=& \sum^{M}_{i=1}\sum^{M}_{j=1}\langle h^{i}_{\bf k} h^{j}_{\bf k^{'}} \rangle=(2\pi)^3 \delta^{3}({\bf k}+{\bf k^{'}})\frac{2\pi^2}{k^3}\Delta_{h}(k),\eea
where the primordial power spectrum for scalar modes at any arbitrary momentum scale $k$ can be expressed as:
\bea \Delta_{h}(k)&=&8 \frac{\sum^{M}_{m=1}\sum^{M}_{n=1}N_{,m}N_{,n}G^{mn}}{\sum^{M}_{i=1}\sum^{M}_{j=1}N_{,i}N_{'j}G^{ij}}\frac{k^3 P_{\zeta}(k)}{2\pi^2}=\frac{4k^3 P_{\zeta}(k)}{\pi^2}.\eea

The changes will appear in the expressions for the following inflationary observables at the horizon crossing:
\begin{itemize}
	\item In the present context amplitude of tensor power spectrum can be computed as:
	\begin{eqnarray} 
	\label{psxcc122vv2g4} \Delta_{h, \star}
	&=&\left\{\begin{array}{lll}
	\displaystyle 
	\displaystyle  \left\{\left[1-({\cal C}_{E}+1)\bar{\epsilon}_{V}\right]^2 \frac{2H^{2}}{\pi^2M^{2}_{p}}\right\}_{\star}\,, &
	\mbox{\small {\bf for $q=1/2$}}  \\ 
	\displaystyle \left\{\left[1-({\cal C}_{E}+1)\frac{\bar{\epsilon}_{V}}{2q}\right]^2 \frac{2H^{2}}{\pi^2M^{2}_{p}}\right\}_{\star} \,, &
	\mbox{\small {\bf for any $q$}}.
	\end{array}
	\right.\end{eqnarray}
	where ${\cal C}_{E}= -2 + \ln 2 + \gamma \approx -0.72.$  
	\item Next one can compute the tensor spectral tilt ($n_{h}$) of the primordial scalar power spectrum as:
	\begin{eqnarray} 
	\label{psxcc133vv2g4} n_{h, \star} 
	&\approx&\left\{\begin{array}{lll}
	\displaystyle 
	\displaystyle  -2\bar{\epsilon}_{V}\left[1+\bar{\epsilon}_{V}+2\left({\cal C}_{E}+1\right)\left(3\bar{\epsilon}_{V}-\bar{\eta}_{V}\right)\right]+\cdots,~~ &
	\mbox{\small {\bf for $q=1/2$}}  \\ 
	\displaystyle  
	-\frac{\bar{\epsilon}_{V}}{q}\left[1+\frac{\bar{\epsilon}_{V}}{2q}+\sqrt{\frac{2}{q}}\left({\cal C}_{E}+1\right)\left(3\bar{\epsilon}_{V}-\bar{\eta}_{V}\right)\right]+\cdots\,,~~ &
	\mbox{\small {\bf for any $q$}}.
	\end{array}
	\right.\end{eqnarray}
	\item Finally the tensor to scalar ratio for multi tachyonic set up can be expressed as:
	\begin{eqnarray} 
	\label{psxcc133vv2g444} r_{ \star} 
	&\approx&\left\{\begin{array}{lll}
	\displaystyle 
	\displaystyle  \frac{8}{\sum^{M}_{i}\sum^{M}_{j=1}N_{,i}N_{,j}G^{ij}}+\cdots,~~ &
	\mbox{\small {\bf for $q=1/2$}}  \\ 
	\displaystyle  
	 \frac{4}{q\sum^{M}_{i}\sum^{M}_{j=1}N_{,i}N_{,j}G^{ij}}+\cdots\,,~~ &
	\mbox{\small {\bf for any $q$}}.
	\end{array}
	\right.\end{eqnarray}
\end{itemize}
On the contrary, compared to the scalar (curvature and entropic) part of the cosmological perturbations, the tensor perturbations
remain frozen in on the large cosmological scales and finally decoupled from the scalar (curvature and entropic) part of the cosmological
perturbations at the linear order of the cosmological perturbation theory. Consequently, in the present context, the primordial cosmological perturbation
spectrum for gravitational waves is given by the following simplified expression:
\bea
\label{gravitywaves} {\Delta}_h &=& {\Delta}_{h,\star},\\
n_h &=& n_{h,\star}.
\eea
Finally, it also important to note that, the consistency condition for the tensor-to-scalar amplitudes of the primordial power spectrum at the
Hubble-crossing be rewritten using 
Eq.~(\ref{delau}) and Eq.~(\ref{gravitywaves}), as a model independent consistency relation
between the tensor-to-scalar amplitudes of the primordial power spectrum
at late times in cosmological scale for GTachyon as:
\be
\label{xcxcxcqq} r={{\Delta}_{h}\over {\Delta}_{\zeta}} \simeq - 8 n_h \sin^2\theta+\cdots = - 8 n_{h,\star} \sin^2\theta+\cdots = r_{\star} \sin^2\theta+\cdots.
\ee
In the present context, the scale dependence of the final scalar (curvature and entropic) power spectra depends on 
both of the cosmological scale dependence of the initial spectral index ($n_{\zeta,\star}$)
and on the explicit form of the generalized transfer functions $T_{\zeta{\cal S}}$ and $T_{{\cal S}{\cal S}}$.
Here we have not explicitly discussed the cosmological parameter estimation and numerical estimations from a specific class of multi-tachyonic potentials.
But to understand more, one can carry forward the results obtained in this section to test various models of multi-tachyonic potential.

\subsubsection{Analytical study for Multi-field model }
Here we compute the expression for Inverse cosh potential. One can repeat the computation for the other proposed models of multi-field inflation as well.
For multi field case the Inverse cosh potential is given by:
\be\label{webbbb1g5}
V(T_{j})=\frac{\lambda_{j}}{{\rm cosh}\left(\frac{T_{j}}{T_{0j}}\right)}\forall j=1,2,\cdots,M,\ee
and the total effective potential is given by:
\be\label{webbbb1g6}
V=\sum^{M}_{j=1}V(T_{j})=\sum^{M}_{j=1}\frac{\lambda_{j}}{{\rm cosh}\left(\frac{T_{j}}{T_{0j}}\right)},\ee
where $\lambda_{j}$ characterize the scale of inflation in each branch and $T_{0j}$ are $j$ number of different parameter of the model. Next using specified form of the potential the potential dependent slow-roll parameters for $i$ th species are computed as:
\bea \bar{\epsilon}_{V}(T_i)&=&\frac{1}{\sum^{M}_{m=1}\lambda_{m}{\rm sech}\left(\frac{T_m}{T_{0m}}\right)}\left[\frac{\lambda_{i}}{2g_{i}}\frac{\sinh^2\left(\frac{T_{i}}{T_{0i}}\right) }{\cosh\left(\frac{T_i}{T_{0i}}\right) }\right],~~~~~~~~~~~~\\
\bar{\eta}_{V}(T_{i})&=&\frac{1}{\sum^{M}_{m=1}\lambda_{m}{\rm sech}\left(\frac{T_m}{T_{0m}}\right)}\left\{\frac{\lambda_{i}}{g_{i}}\left[\tanh^2\left(\frac{T_{i}}{T_{0i}}\right)-{\rm sech}^2\left(\frac{T_i}{T_{0i}}\right)\right]\right\}.\eea
Also the total contribution in the slow-roll is expressed through the reduced slow-roll parameters:
\bea \bar{\epsilon}_{V}&=&\frac{1}{\sum^{M}_{m=1}\lambda_{m}{\rm sech}\left(\frac{T_m}{T_{0m}}\right)}\left[\sum^{M}_{i=1}\frac{\lambda_{i}}{2g_{i}}\frac{\sinh^2\left(\frac{T_{i}}{T_{0i}}\right) }{\cosh\left(\frac{T_i}{T_{0i}}\right) }\right],~~~~~~~~~~~~\\
\bar{\eta}_{V}&=&\frac{1}{\sum^{M}_{m=1}\lambda_{m}{\rm sech}\left(\frac{T_m}{T_{0m}}\right)}\left\{\sum^{M}_{i=1}\frac{\lambda_{i}}{g_{i}}\left[\tanh^2\left(\frac{T_{i}}{T_{0i}}\right)-{\rm sech}^2\left(\frac{T_i}{T_{0i}}\right)\right]\right\}.\eea
Additionally here we introduce a new factor $g_{j}$ for each non identical $j$ number of multi tachyonic fields, defined as:
\be g_{j}= \frac{\alpha^{'}\lambda_{j} T^2_{0j}}{M^2_p}=\frac{M^4_{sj}}{(2\pi)^3 g_{sj}}\frac{\alpha^{'} T^2_{0j}}{M^2_p}.\ee
Next we compute the number of e-foldings from this model:
\be\begin{array}{lll}\label{e-fold5}
	\displaystyle  N=\left\{\begin{array}{lll}
		\displaystyle  
		\sum^{M}_{j=1}g_j~\ln\left[\frac{\tanh\left(\frac{T_{end,j}}{2T_{0j}}\right)}{\tanh\left(\frac{T_j}{2T_{0j}}\right)}\right]\,,~~~~~~ &
		\mbox{\small {\bf for {$q=1/2$ }}}  \\ 
		\displaystyle   
		\sum^{M}_{j=1}\sqrt{2q}~g_j~\ln\left[\frac{\tanh\left(\frac{T_{end,j}}{2T_{0j}}\right)}{\tanh\left(\frac{T_j}{2T_{0j}}\right)}\right]\,.~~~~~~ &
		\mbox{\small {\bf for {~any~arbitrary~ $q$ }}} 
	\end{array}
	\right.
\end{array}\ee
Further using the condition to end inflation:
\bea \bar{\epsilon}_{V}(T_{end,i})=1,\\
|\bar{\eta}_{V}(T_{end,i})|=1,\eea
we get the following field value at the end of inflation:
\be T_{end,i}=T_{0i}~{\rm sech}^{-1}(g_{i}).\ee
Next using $N_i=N_{cmb,i}=N_{\star,i}$ and $T_{i}=T_{cmb,i}=T^{\star}_{i}$ at the horizon crossing we get,
\be\begin{array}{lll}\label{e-fold6}
	\displaystyle  T^{\star}_{i}\approx 2T_{0,i}\times\left\{\begin{array}{lll}
		\displaystyle  
		\tanh^{-1}\left[\exp\left(-\frac{N_{\star,i}}{g_i}\right)\right]\,,~~~~~~ &
		\mbox{\small {\bf for {$q=1/2$ }}}  \\ 
		\displaystyle   
		\tanh^{-1}\left[\exp\left(-\frac{N_{\star,i}}{\sqrt{2q}g_i}\right)\right]\,.~~~~~~ &
		\mbox{\small {\bf for {~any~arbitrary~ $q$ }}} 
	\end{array}
	\right.
\end{array}\ee
Using these results one can compute: 
\bea N_{,i}&=&\frac{1}{\sqrt{2\bar{\epsilon}_{V}(T^{\star}_{i})}}\frac{V(T^{\star}_{i})-V(T^{c}_{i})+\left(\sum^{M}_{m=1}V(T^{c}_{m})\right)\frac{\bar{\epsilon}_{V}(T^{c}_{i})}{\bar{\epsilon}^{c}_{V}}}{\left(\sum^{M}_{j=1}V(T^{\star}_{j})\right)}\nonumber\\
&=&\frac{\lambda_{i}{\rm sech}\left(\frac{T^{\star}_i}{T_{0i}}\right)-\lambda_{i}{\rm sech}\left(\frac{T^{c}_i}{T_{0i}}\right)+\left(\sum^{M}_{m=1}\lambda_{m}{\rm sech}\left(\frac{T^{c}_m}{T_{0m}}\right)\right)\frac{\bar{\epsilon}_{V}(T^{c}_{i})}{\bar{\epsilon}^{c}_{V}}}{\sqrt{\left[\frac{\lambda_{i}}{g_{i}}\frac{\sinh^2\left(\frac{T_{i}}{T_{0i}}\right) }{\cosh\left(\frac{T_i}{T_{0i}}\right) }\right]\left(\sum^{M}_{j=1}\lambda_{j}{\rm sech}\left(\frac{T^{\star}_j}{T_{0j}}\right)\right)}}
\eea
which is very very useful to further compute the inflationary observables from multi-tachyonic setup. Here $\star$ indicates the horizon crossing and $c$ denotes the constant density surface. 

Using these results
finally we compute the following inflationary observables:
\begin{eqnarray} 
\label{psxcc1v0vv5vv5}\footnotesize \Delta_{\zeta, \star}
&=&\sum^{M}_{i=1}\left(\frac{\lambda_{i}{\rm sech}\left(\frac{T^{\star}_i}{T_{0i}}\right)-\lambda_{i}{\rm sech}\left(\frac{T^{c}_i}{T_{0i}}\right)+\left(\sum^{M}_{m=1}\lambda_{m}{\rm sech}\left(\frac{T^{c}_m}{T_{0m}}\right)\right)\frac{\bar{\epsilon}_{V}(T^{c}_{i})}{\bar{\epsilon}^{c}_{V}}}{\sqrt{\left[\frac{\lambda_{i}}{g_{i}}\frac{\sinh^2\left(\frac{T_{i}}{T_{0i}}\right) }{\cosh\left(\frac{T_i}{T_{0i}}\right) }\right]\left(\sum^{M}_{j=1}\lambda_{j}{\rm sech}\left(\frac{T^{\star}_j}{T_{0j}}\right)\right)}}\right)^2\nonumber\\
&&~~~~~~~~~~~~~~~~~~~~~~~~~~~~~~~~\times\left\{\begin{array}{lll}\footnotesize
\displaystyle 
\displaystyle  \left\{ \frac{\left(\sum^{M}_{j=1}\lambda_{j}{\rm sech}\left(\frac{T^{\star}_j}{T_{0j}}\right)\right)}{12\pi^2M^2_p}\right\}_{\star}\,, &
\mbox{\small {\bf for $q=1/2$}}  \\ 
\displaystyle \left\{ \frac{q \left(\sum^{M}_{j=1}\lambda_{j}{\rm sech}\left(\frac{T^{\star}_j}{T_{0j}}\right)\right)}{6\pi^2M^2_p}\right\}_{\star}\,, &
\mbox{\small {\bf for any $q$}}.
\end{array}
\right.\\
	\label{psxcc1v0vv5vv5cc}\footnotesize \Delta_{\zeta,c}
	&=&\left[\sum^{M}_{i=1}\frac{\lambda_{i}}{2g_{i}}\frac{\sinh^2\left(\frac{T_{i}}{T_{0i}}\right) }{\cosh\left(\frac{T_i}{T_{0i}}\right) }\right]^{-1}_{c}\times\left\{\begin{array}{lll}\footnotesize
		\displaystyle 
		\displaystyle  \left\{ \frac{\left(\sum^{M}_{j=1}\lambda_{j}{\rm sech}\left(\frac{T^{\star}_j}{T_{0j}}\right)\right)^2}{12\pi^2M^2_p }\right\}_{c}\,, &
		\mbox{\small {\bf for $q=1/2$}}  \\ 
		\displaystyle \left\{ \frac{q \left(\sum^{M}_{j=1}\lambda_{j}{\rm sech}\left(\frac{T^{\star}_j}{T_{0j}}\right)\right)^2}{6\pi^2M^2_p}\right\}_{c}\,, &
		\mbox{\small {\bf for any $q$}}.
	\end{array}
	\right.\end{eqnarray}
\begin{eqnarray} 
\label{psxcc1v2vv5scxxxc} n_{\zeta, \star}-1 
&\approx&\left\{\footnotesize\begin{array}{lll}
\displaystyle 
\displaystyle  
\displaystyle
\left(2\bar{\eta}_{V}-8\bar{\epsilon}_{V}\right)-\frac{2}{\sum^{M}_{i=1}\left(\frac{\lambda_{i}{\rm sech}\left(\frac{T^{\star}_i}{T_{0i}}\right)-\lambda_{i}{\rm sech}\left(\frac{T^{c}_i}{T_{0i}}\right)+\left(\sum^{M}_{m=1}\lambda_{m}{\rm sech}\left(\frac{T^{c}_m}{T_{0m}}\right)\right)\frac{\bar{\epsilon}_{V}(T^{c}_{i})}{\bar{\epsilon}^{c}_{V}}}{\sqrt{\left[\frac{\lambda_{i}}{g_{i}}\frac{\sinh^2\left(\frac{T_{i}}{T_{0i}}\right) }{\cosh\left(\frac{T_i}{T_{0i}}\right) }\right]\left(\sum^{M}_{j=1}\lambda_{j}{\rm sech}\left(\frac{T^{\star}_j}{T_{0j}}\right)\right)}}\right)^2}+\cdots, &
\mbox{\small {\bf for $q=1/2$}}  \\ 
\displaystyle  
\displaystyle  
\displaystyle
\sqrt{\frac{2}{q}}\bar{\eta}_{V}-\left(\frac{1}{q}+3\sqrt{\frac{2}{q}}\right)\bar{\epsilon}_{V}\\
\displaystyle-\frac{1}{q\sum^{M}_{i=1}\left(\frac{\lambda_{i}{\rm sech}\left(\frac{T^{\star}_i}{T_{0i}}\right)-\lambda_{i}{\rm sech}\left(\frac{T^{c}_i}{T_{0i}}\right)+\left(\sum^{M}_{m=1}\lambda_{m}{\rm sech}\left(\frac{T^{c}_m}{T_{0m}}\right)\right)\frac{\bar{\epsilon}_{V}(T^{c}_{i})}{\bar{\epsilon}^{c}_{V}}}{\sqrt{\left[\frac{\lambda_{i}}{g_{i}}\frac{\sinh^2\left(\frac{T_{i}}{T_{0i}}\right) }{\cosh\left(\frac{T_i}{T_{0i}}\right) }\right]\left(\sum^{M}_{j=1}\lambda_{j}{\rm sech}\left(\frac{T^{\star}_j}{T_{0j}}\right)\right)}}\right)^2}+\cdots\,, &
\mbox{\small {\bf for any $q$}}
\end{array}
\right.\end{eqnarray}
\begin{eqnarray} 
	\label{psxcc1v2vv5scxxxc} n_{\zeta,c}-1 
	&\approx&\left\{\footnotesize\begin{array}{lll}
		\displaystyle 
		\displaystyle  
		\displaystyle
		-6\bar{\epsilon}_{V}+2\frac{\sum^{M}_{i=1}\bar{\epsilon}_{V}(T_i)\bar{\eta}_{V}(T_i)}{\bar{\epsilon}_{V}}+\cdots, &
		\mbox{\small {\bf for $q=1/2$}}  \\ 
		\displaystyle  
		\displaystyle  
		\displaystyle
		-3\sqrt{\frac{2}{q}}\bar{\epsilon}_{V}+\frac{\sum^{M}_{i=1}\bar{\epsilon}_{V}(T_i)\bar{\eta}_{V}(T_i)}{q\bar{\epsilon}_{V}}\cdots\,, &
		\mbox{\small {\bf for any $q$}}
	\end{array}
	\right.\end{eqnarray}
\begin{eqnarray} 
\label{psxcc133vv2g444dfr} r_{ \star} 
&\approx&\left\{\begin{array}{lll}
\displaystyle 
\displaystyle  \frac{8}{\sum^{M}_{i=1}\left(\frac{\lambda_{i}{\rm sech}\left(\frac{T^{\star}_i}{T_{0i}}\right)-\lambda_{i}{\rm sech}\left(\frac{T^{c}_i}{T_{0i}}\right)+\left(\sum^{M}_{m=1}\lambda_{m}{\rm sech}\left(\frac{T^{c}_m}{T_{0m}}\right)\right)\frac{\bar{\epsilon}_{V}(T^{c}_{i})}{\bar{\epsilon}^{c}_{V}}}{\sqrt{\left[\frac{\lambda_{i}}{g_{i}}\frac{\sinh^2\left(\frac{T_{i}}{T_{0i}}\right) }{\cosh\left(\frac{T_i}{T_{0i}}\right) }\right]\left(\sum^{M}_{j=1}\lambda_{j}{\rm sech}\left(\frac{T^{\star}_j}{T_{0j}}\right)\right)}}\right)^2}+\cdots,~~ &
\mbox{\small {\bf for $q=1/2$}}  \\ 
\displaystyle  
\frac{4}{q\sum^{M}_{i=1}\left(\frac{\lambda_{i}{\rm sech}\left(\frac{T^{\star}_i}{T_{0i}}\right)-\lambda_{i}{\rm sech}\left(\frac{T^{c}_i}{T_{0i}}\right)+\left(\sum^{M}_{m=1}\lambda_{m}{\rm sech}\left(\frac{T^{c}_m}{T_{0m}}\right)\right)\frac{\bar{\epsilon}_{V}(T^{c}_{i})}{\bar{\epsilon}^{c}_{V}}}{\sqrt{\left[\frac{\lambda_{i}}{g_{i}}\frac{\sinh^2\left(\frac{T_{i}}{T_{0i}}\right) }{\cosh\left(\frac{T_i}{T_{0i}}\right) }\right]\left(\sum^{M}_{j=1}\lambda_{j}{\rm sech}\left(\frac{T^{\star}_j}{T_{0j}}\right)\right)}}\right)^2}+\cdots\,,~~ &
\mbox{\small {\bf for any $q$}}.
\end{array}
\right.\end{eqnarray}
For Inverse cosh potential we get the following consistency relations:
\begin{eqnarray} 
\label{psxcc1v2vv5scxxxc} n_{\zeta, \star}-1 +\frac{r_{\star}}{4}
&\approx&\left\{\footnotesize\begin{array}{lll}
\displaystyle 
\displaystyle  
\displaystyle
\left(2\bar{\eta}_{V}-8\bar{\epsilon}_{V}\right)+\cdots, &
\mbox{\small {\bf for $q=1/2$}}  \\ 
\displaystyle  
\displaystyle  
\displaystyle
\sqrt{\frac{2}{q}}\bar{\eta}_{V}-\left(\frac{1}{q}+3\sqrt{\frac{2}{q}}\right)\bar{\epsilon}_{V}+\cdots\,, &
\mbox{\small {\bf for any $q$}}
\end{array}
\right.\end{eqnarray}

\section{Conclusion}
\label{aa6}
In this paper, we have explored various cosmological consequences from GTachyonic field. We start with the basic introduction of tachyons in the context
of non-BPS string theory, where we also introduce the GTachyon
field, in presence of which the tachyon action is getting modified and one can
quantify the amount of the modification via a superscript $q$ instead of $1/2$.
This modification exactly mimics the role of effective field theory operators and
studying the various cosmological features from this theory our one of the final
objectives is to constrain the index $q$ and a specific combination ($\propto \alpha^{'}M^4_s/g_s$) of Regge slope parameter $\alpha^{'}$, string coupling constant $g_{s}$ and mass scale of 
tachyon $M_s$, from the recent Planck 2015 and Planck+BICEP2/Keck
Array joint data. To serve this purpose, we introduce various types of tachyonic
potentials-Inverse cosh, Logarithmic, Exponential and Inverse polynomial, using
which we constraint the index $q$. To explore this issue in detail, we start with
the detailed characteristic features of the each potentials.
Next we discuss the dynamics of GTachyon as well as usual tachyon for single,
assisted and multi-field scenario. We also derive the dynamical solutions
for various phases of Universe, including two situations- $T<<T_0$ and $T>>T_0$, where
$T_0$ is interpreted to be the minimum or the mass scale of the tachyon.
Next we have explicitly studied the inflationary paradigm from single field, assisted
field and multi-field tachyon set up. Specifically for single field and assisted field
case we have derived the results in the quasi-de-Sitter background in which we have
utilized the details of- (1) cosmological perturbations and quantum fluctuations
for scalar and tensor modes, (2) Slow-roll prescription up to all orders. In this context we have
derived the expressions for all inflationary observables using any arbitrary
vacuum and also for Bunch-Davies vacuum by exactly solving the Mukhanov-Sasaki equation as obtained from the fluctuation of scalar and tensor modes.
For single field and assisted field
case in presence of GTachyon we have derived-the inflationary
Hubble flow and potential dependent flow equations, new sets of consistency
relations, which are valid in the slow-roll regime and also derived the expression for the field excursion formula for tachyon
in terms of inflationary observables from both of the solutions obtained from arbitrary and Bunch-Davies initial conditions for inflation. 
Particularly the derived formula for the field excursion for GTachyon can be treated as a one of the probes through which one can -
(1) test the validity 
and the applicability of the effective field theory prescription within the present setup, 
(2) distinguish between various classes of models from effective field theory point of view.
Also we have shown that in case of assisted tachyon inflation validity of effective field theory prescription is much more better compared to the 
single field case. This is because of the fact that 
the field excursion for the assisted inflation is expressed as in terms of the field excursion for single field, provided the multiplicative scaling factor is 
$1/\sqrt{M}$, where $M$ is the number of identical tachyons participating in assisted inflation. This derived formula also suggests that if $M$ is very large number then 
it is very easy to validate effective theory techniques, as in that case $|\Delta T|_{Assisted}<<M_p$.
Next using the explicit for of the
tachyonic potentials we have studied the inflationary constraints and quantify
the allowed range of the generalized index $q$ for each potentials. Hence
using the each specific form of the tachyonic potentials in the context of
single filed scenario, we have studied the features of CMB Angular power
spectrum from TT, TE and EE correlations from scalar fluctuations within
the allowed range of $q$ for each potentials. We also put the constraints
from the Planck temperature anisotropy and polarization data, which
shows that our analysis confronts well with the data. 
We have additionally studied the features of tensor contribution in the
CMB Angular power spectrum from TT, BB, TE and EE correlations, which
will give more interesting information in near future once one can
detect the signature of  primordial B-modes. Further, using
$\delta N$ formalism we have derived the expressions for inflationary
observables in the context of multi-field tachyons. We have also
demonstrated the results for two-field tachyonic case to understand
the cosmological implications of the results in a better way.   

The future prospects of our work are appended below:
\begin{itemize}
	\item  Through this analysis contains each an every details, but one can further study
	the features of primordial non-Gaussianity from single field, assisted field and multi-field 
	case \cite{Maldacena:2002vr,Maldacena:2011nz,Arkani-Hamed:2015bza,Mata:2012bx,Ghosh:2014kba,Kundu:2014gxa,Kundu:2015xta}. Due to the presence of an generalized index $q$ one would expect that the consistency
	relations, which connect non-Gaussian parameters with the inflationary observables, are
	getting modified in the slow-roll regime of inflation and consequently one can generate
	large amount of primordial non-Gaussianity from the GTachyonic set up. Also our
	aim is generalize the results as well as the consistency relations in all order of cosmological
	perturbations by incorporating the effect of sound speed $c_{S}$ and the generalized parameter $q$. 
	Using setup one can also compute the cosmological Ward identities for inflation \cite{Kundu:2014gxa,Kundu:2015xta,Hinterbichler:2013dpa}, through which one can write down the recursion relation between correlation functions. This will 
	help to derive the modified consistency relations in the present context. Additionally, one can also explore the possibility of devising cosmological observables which violate Bell's
inequalities recently pointed in ref.~\cite{Maldacena:2015bha}. Such observables could be used to argue that cosmic scale features
were produced by quantum mechanical effects in the very early universe, specifically in the context of inflation.
	We are planning to report on these issues explicitly very soon. 
	
	\item One
	would also like to ask what happened when we add additional higher derivatives in the gravity sector within tachyonic set up, which are important around the string scale $M_s$.
	How the higher derivative gravity sector will change the cosmological dynamics and detailed study of 
	primordial non-Gaussianity in presence of non-canonical GTachyonic
	sector is completely unknown to all. We have also a plan to extend this project in that direction.
	
	\item The generation of the seed (primordial) magnetic field from inflationary sector \cite{Choudhury:2015pqa,Choudhury:2014hua} is a long standing issue in primordial cosmology. 
	      To address this well known issue one can also study the cosmological consequences from the effective field theory of 
	      inflationary magnetogenesis from GTachyon within the framework 
	      of type IIA/IIB string theory.
	
	\item Also, our present analysis have
	been performed for five different potentials motivated from tachyonic string theory, specifically in the context of single field and assisted field
	inflation. For multi-field we have quoted the results only for a specific kind of separable 
	potential. It would be really very interesting to
	check whether we can study the behaviour of non-separable potentials in the present context \cite{Vernizzi:2006ve,Senatore:2010wk}. Also it is interesting to 
	reconstruct any general form of a tachyonic inflationary potential, using which one can study the applicability of the effective field theory
	framework within this present set up. 
	
	\item One can also carry forward our analysis in the context of non-minimal set up where the usual Einstein Hilbert term in the gravity sector
	is coupled with the GTachyonic field. By applying the conformal transformation in the metric one can construct
	an equivalent representation of the gravity sector~\footnote{In cosmological literature this frame is known as the Einstein frame.} in which the 
	gravity sector is decoupled from the tachyonic sector and the usual tachyonic matter sector modified in presence of an extra conformal factor. This clearly implies that
	the tachyonic potentials that we have studied in this paper all getting modified via the additional conformal factor. Consequently it is expected 
	that the cosmological dynamics will be modified in this context. Also one can check the applicability of slow-roll prescription, which helps further 
	to understand the exact behaviour of cosmological perturbations in presence of conformally rescaled modified tachyonic potentials.
	
	\item The detailed study of the generation of the dark matter using effective field
theory framework~\footnote{For the completeness, it is important to note that, very recently in ref.~\cite{Choudhury:2015eua} we have explicitly
studied the effective field theory framework from membrane inflationary paradigm with Randall
Sundrum single brane setup. We also suggest the readers to study the refs.~\cite{Busoni:2013lha,Busoni:2014sya}, in which the effective theory of dark matter have already been studied very well.}, other crucial particle physics issues i.e. reheating phenomena,
leptogenesis and baryogenesis from GTachyonic setup is also an unexplored
issue till date.
	
\end{itemize}

\section*{Acknowledgments}
SC would like to thank Department of Theoretical Physics, Tata Institute of Fundamental
Research, Mumbai for providing me Visiting (Post-Doctoral) Research Fellowship.  SC take this opportunity to thank sincerely to Sandip P. Trivedi, Shiraz Minwalla,
 Soumitra SenGupta, Varun Sahni, Sayan Kar and Supratik Pal
for their constant support and inspiration. SC take this opportunity to thank all the active members and the
regular participants of weekly student discussion meet ``COSMOMEET'' from Department of Theoretical Physics and Department of Astronomy and Astrophysics, Tata Institute of Fundamental
Research for their strong support. SC also thank Sandip Trivedi and Shiraz Minwalla for giving the opportunity to be the part of String Theory and
Mathematical Physics Group. SC also thank the other post-docs and doctoral students from String Theory and
Mathematical Physics Group for providing an excellant academic ambience during the research work. SC also thanks Institute of Physics (IOP), Bhubaneswar and specially Prof. Sudhakar Panda
for arranging the official visit where the problem have been formulated and the initial part have been done.  SC also thank the organizers of COSMOASTRO, 2015, Institute
of Physics (IOP), Bhubaneswar for giving the opportunity to give an invited talk
on related issues of effective field theory and primordial non-Gaussianity. 
SC additionally thank Inter University Centre for Astronomy and Astrophysics (IUCAA), Pune and specially Varun Sahni for 
extending hospitality during the work. Additionally SC take this opportunity to thank the organizers of STRINGS, 2015, International Centre for Theoretical Science, Tata Institute of Fundamental Research (ICTS,TIFR)
and Indian Institute of Science (IISC)
for providing the local hospitality during the work and give a chance to discuss with Prof. Nima Arkani-Hamed on related issues, which finally helped us to improve the qualitative and quantitative discussion in the paper. 
Last but not the least, we would all like to acknowledge our debt to the people of
India for their generous and steady support for research in natural sciences, especially for string theory and cosmology.



\end{document}